\documentclass[11pt, a4paper,twoside]{book}

\usepackage{StyleFile}
\usepackage{colordvi,fancybox,latexsym,amstext,latexsym}
\usepackage[english,ngerman]{babel}
\usepackage{a4wide}
\usepackage{graphicx}
\usepackage{psfrag}
\usepackage{amsmath, amssymb}
\usepackage{amsfonts}
\usepackage{epic}
\usepackage{eepic}
\usepackage{epsfig}
\usepackage{boxedminipage}
\usepackage{verbatim}
\usepackage{textcomp}
\usepackage{afterpage}
\usepackage{multirow}
\usepackage{array}
\usepackage{marvosym}
\usepackage{axodraw}
\usepackage[T1]{fontenc} 

\newcommand{\eff}{e\hspace{-0.4mm}f\hspace{-0.6mm}f}
\newcommand{\f}{\hspace{-0.6mm}f}
\newcommand{\auf}{au\hspace{-0.4mm}f}
\newcommand{\rcg}{$\vec{r}_{\text{CG}}$}
\newcommand{\rcga}{$\vec{r}_{\text{CGA}}$}

\newcommand{\nnu}{0.70}

\newcommand{\nnuup}{0.35}
\newcommand{\ncalpha}{$10.7^{\circ}$}
\newcommand{\ncalphacut}{$3.5^{\circ}$}
\newcommand{\ncElog}{0.17}
\newcommand{\ncE}{1.5}
\newcommand{\ncEcutlog}{0.14}
\newcommand{\ncEcut}{1.4}
\newcommand{\ccalpha}{$9.5^{\circ}$}
\newcommand{\ccalphacut}{$3.0^{\circ}$}
\newcommand{\ccElog}{0.19}
\newcommand{\ccE}{1.5}
\newcommand{\ccEcutlog}{0.16}
\newcommand{\ccEcut}{1.4}
\newcommand{\alphabest}{$2^{\circ}$}
\newcommand{\sensitivity}{1.7 \times 10^{-7}}
\newcommand{\sensitivup}{3.6 \times 10^{-7}}

\begin{document}
\frontmatter
\pagenumbering{Roman}

\begin{titlepage}
\begin{flushright}
FAU-PI1-DISS-06-001
\end{flushright}
\begin{center}
\normalsize \hspace*{9.5cm} $\strut$ \\
\vspace*{10mm} \huge Reconstruction of Neutrino-Induced\\ Hadronic and Electromagnetic Showers
\\with the ANTARES Experiment

\vspace*{35mm} \large Den Naturwissenschaftlichen Fakult\"aten \\ der
Friedrich-Alexander-Universit\"at Erlangen-N\"urnberg \\ zur \\
Erlangung des Doktorgrades \\

\vspace*{25mm} \epsfig{file=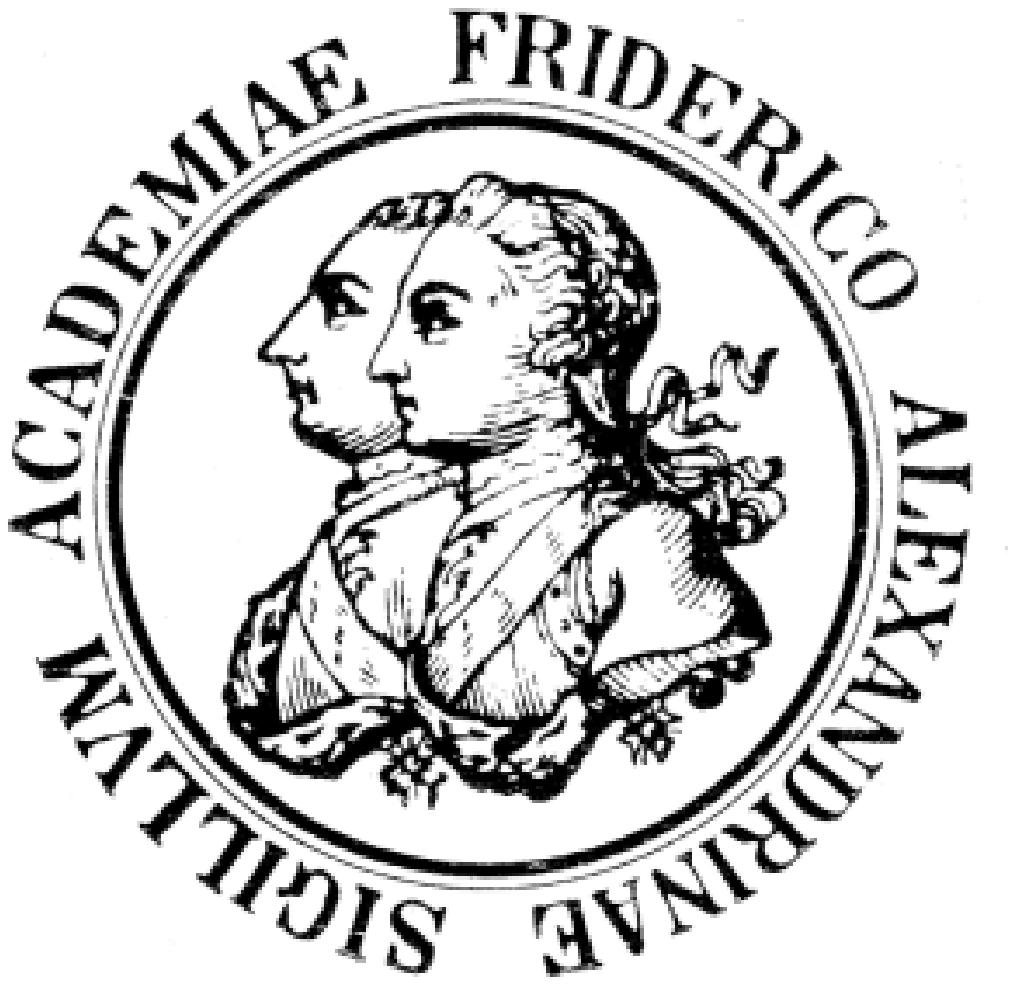,width=5cm}
\vspace{25mm}

vorgelegt von \\
Bettina Diane Hartmann \\ 
\vspace{1cm}
aus Stuttgart

\end{center}
\end{titlepage}

\thispagestyle{empty}
\begin{center}

\begin{tabular}{ll}
& \\
& \\
& \\
& \\
& \\
\end{tabular}

\large Als Dissertation genehmigt \\ von den Naturwissenschaftlichen Fakult\"aten \\ 
der Universit\"at Erlangen-N\"urnberg. 

\vspace{10cm}

\begin{tabular}{ll}
Tag der m\"undlichen Pr\"ufung: & 2.\,Juni 2006\\ \\
Vorsitzender der Pr\"ufungskommission: & Prof. Dr. D.-P. H\"ader \\ \\
Erstberichterstatter: & Prof. Dr. U. Katz \\ \\
Zweitberichterstatter: & Prof. Dr. E. Steffens
\end{tabular}

\end{center}

\selectlanguage{ngerman}
\chapter*{Zusammenfassung}
\thispagestyle{plain}

\begin{figure}
\begin{minipage}[b]{7.2cm}
  \centering \epsfig{figure=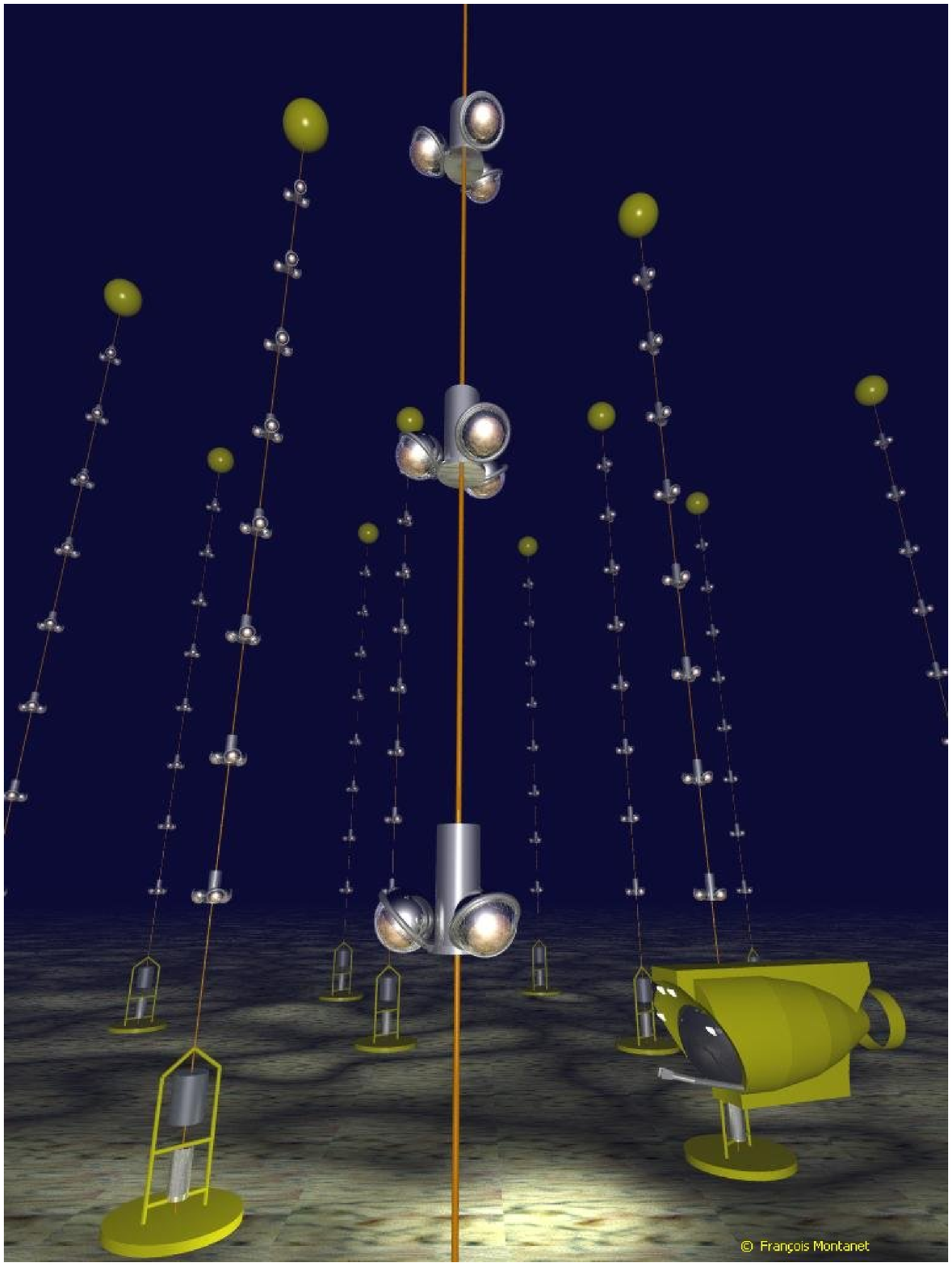, width=7.2cm}
\caption[K\"unstlerische Darstellung des ANTARES-Detektors]{K\"unstlerische Darstellung des
  ANTARES-Detektors~\cite{montanet}. Statt der eigentlichen 25 sind zur besseren \"Ubersichtlichkeit nur
  8 Stockwerke je String gezeigt.} 
\label{fig:Z:ant_pub}
\end{minipage}
\hspace{0.4cm}
\begin{minipage}[b]{7.2cm}
  \centering \epsfig{figure=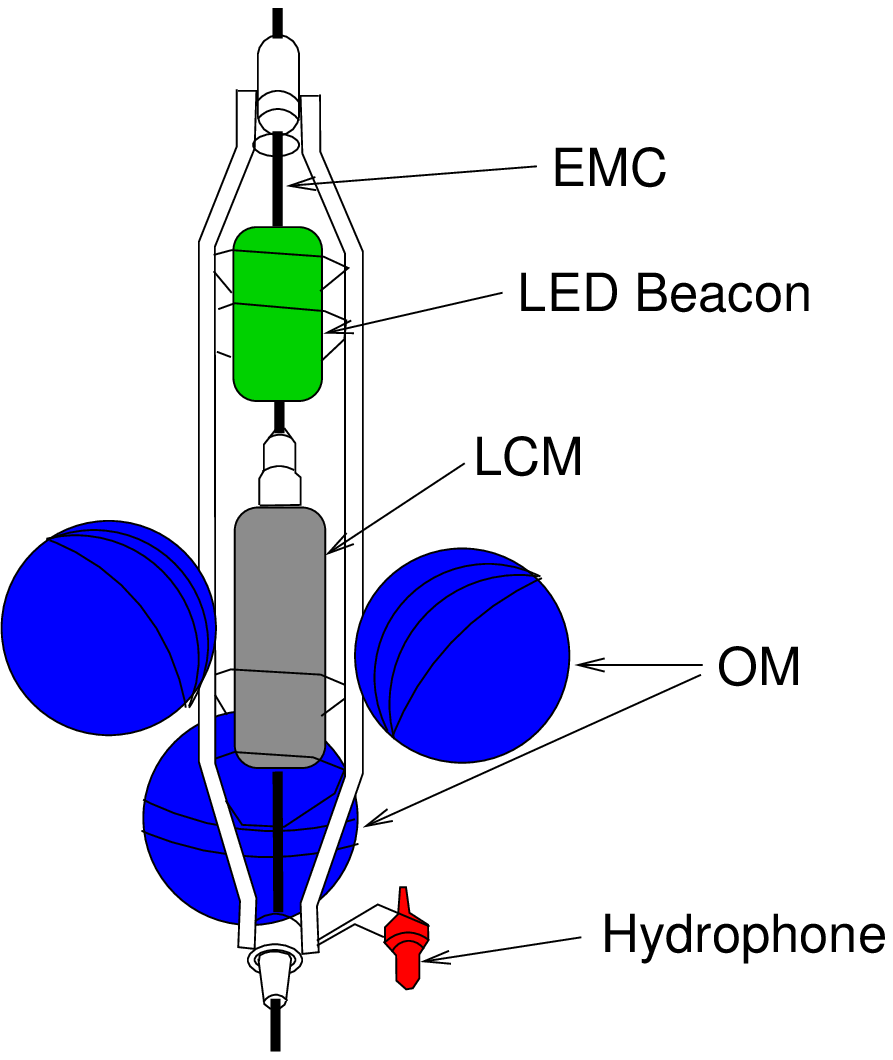, width=5cm}
\caption[Schematische Ansicht eines ANTARES-Stockwerks]{Schematische Ansicht eines
  ANTARES-Stockwerks. Das elektromechanische Verbindungskabel (EMC) ist in schwarz
  eingezeichnet. W\"ahrend das Lokale Kontrollmodul (LCM) und die OMs auf allen Stockwerken 
  installiert werden, sind das f\"ur die akustische Positionsmessung vorgesehene Hydrophon und der zur
  optischen Kalibration gedachte LED-Strahler (LED beacon) nur einmal alle f\"unf Stockwerke
  vorgesehen.} 
\label{fig:Z:storey}
\end{minipage} 
\end{figure}

Die Geschichte der Neutrinophysik begann im Jahr 1930, als Wolfgang Pauli, auf der 
Suche nach einer L\"osung des R\"atsels um den Betazerfall, ein leichtes, neutrales Teilchen
postulierte. Er nannte dieses Teilchen zun\"achst {\it Neutron}, ehe es dann von Fermi den Namen {\it
  Neutrino} erhielt. Pauli h\"atte sich wohl damals nicht tr\"aumen lassen, dass 75 Jahre sp\"ater
f\"ur den Nachweis dieser Teilchen Detektoren mit einem instrumentierten Volumen von bis zu einem
Kubikkilometer existieren bzw. sich im Bau befinden w\"urden. Eines dieser riesigen {\it
  Neutrinoteleskope} ist das 
ANTARES-Experiment~\cite{proposal}. Dieses 0.03\,km$^3$ gro\ss e Neutrinoteleskop wird derzeit vor
der Franz\"osischen Mittelmeerk\"uste in 2400\,m Meerestiefe aufgebaut. \\ 
Wie der Name Teleskop bereits andeutet, haben Neutrinoteleskope die Vermessung von aus gro\ss en
Entfernungen kommenden Signalen zum Ziel, in diesem Fall von Neutrinos, die im Kosmos
erzeugt werden. Man geht davon aus, dass die meisten Quellen kosmischer Gamma-Strahlung, wie
z.B.~Gammastrahlenblitze (engl. Gamma Ray Bursts), Aktive Galaktische Kerne oder
Supernova-\"Uberreste, neben Gammastrahlen auch hochenergetische Neutrinos, d.h. Neutrinos mit
Energien $> 1$\,GeV, in gro\ss er Zahl produzieren. Neutrinos haben wegen ihres sehr kleinen
Wirkungsquerschnittes die vorteilhafte Eigenschaft, nur sehr schwach mit dem interstellaren Medium
oder stellarem Staub zu wechselwirken, und k\"onnen daher praktisch aus beliebig gro\ss en
kosmischen Entfernungen ohne Abschw\"achung zu uns auf die Erde gelangen. Der extrem kleine
Wirkungsquerschnitt birgt jedoch gleichzeitig auch den gro\ss en Nachteil, dass Neutrinos nur mit
Hilfe sehr gro\ss er Targetmassen nachweisbar sind. Aus diesem Grund benutzen Neutrinoteleskope
wie ANTARES nat\"urlich vorkommende gro\ss e, optisch transparente Volumina, wie z.B.~das Meer, als
Detektormedium. \\ 
Neutrinos k\"onnen nur indirekt, \"uber ihre
Reaktionsprodukte, nachgewiesen werden. Experimente wie ANTARES nutzen f\"ur den Nachweis der
Reaktionsprodukte den so genannten Cherenkov-Effekt~\cite{cherenkov}: Wenn geladene Teilchen ein
Medium mit einer Geschwindigkeit durchfliegen, die gr\"o\ss er als die Lichtgeschwindigkeit in
diesem Medium ist, emittieren sie entlang ihrer Bahn unter einem festen, vom Brechungsindex des Mediums
abh\"angigen Winkel Photonen. F\"ur Meerwasser liegt dieser Winkel bei ca.~42$^{\circ}$. Die
Cherenkov-Photonen k\"onnen mit Hilfe von Photomultipliern nachgewiesen werden.
Aus den gemessenen Photonensignalen werden dann wiederum die Energie und Richtung des
prim\"aren Neutrinos rekonstruiert. \\ 
Der ANTARES-Detektor wird in seiner endg\"ultigen Form aus 900 Optischen Modulen (OMs) bestehen, die
je einen Photomultiplier enthalten, der \"uber ein Verbindungskabel mit der
Ausleseelektronik im Lokalen Kontrollmodul (LCM) verbunden ist. Die OMs sind in {\it Stockwerken} zu
je drei St\"uck an vertikalen Kabelstrukturen, den so genannten {\it Strings}, befestigt. Der
gesamte Detektor wird aus 12 solcher Strings bestehen, mit je 25 Stockwerken in einem vertikalen Abstand
von 14.5\,m. Abbildung~\ref{fig:Z:ant_pub} zeigt eine k\"unstlerische Darstellung des
ANTARES-Detektors~\cite{montanet}, mit 8 statt der eigentlichen 25 Stockwerke. Ein einzelnes
Stockwerk als Detailansicht ist in Abbildung~\ref{fig:Z:storey} gezeigt. 

\begin{figure}[h] \centering
\includegraphics[width=13.5cm]{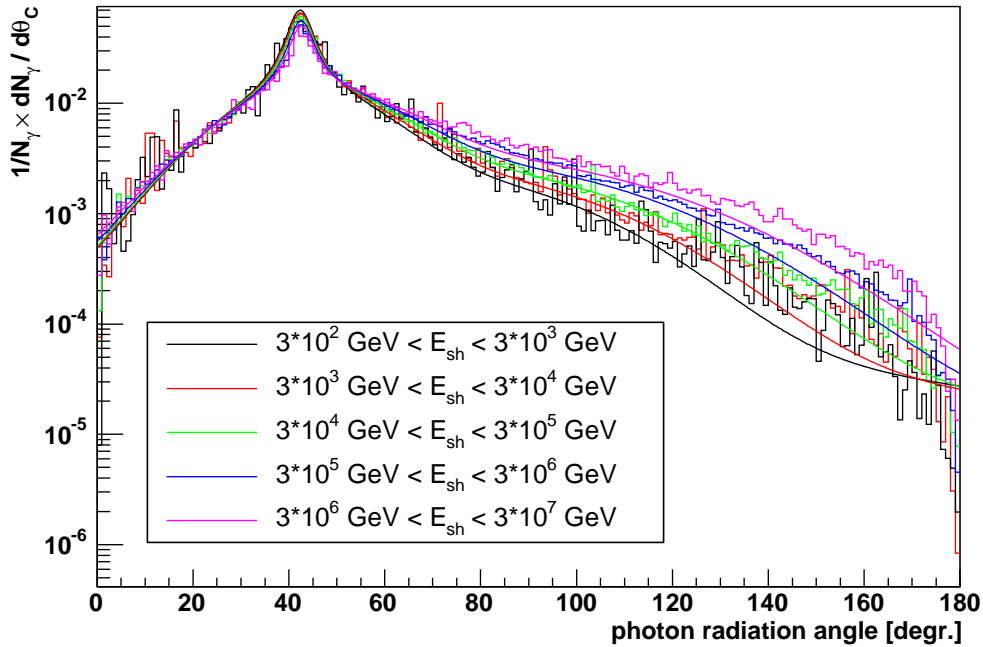}
\caption[Winkelverteilung der Cherenkov-Photonen, und Parameterisierung]
{Polarwinkelverteilung der Cherenkov-Photonen in Bezug auf die Schauerachse f\"ur je eine Dekade
  in der Schauerenergie, von 300\,GeV bis 30\,PeV, und die zugeh\"orige Parameterisierungsfunktion,
  die dem mittleren logarithmischen Energiewert des jeweils betrachteten Energiebereichs entspricht.}
\label{fig:Z:cheren_fit}
\end{figure}

Die Rekonstruktion der Neutrinorichtung und -energie aus den in den Optischen Modulen gemessenen
Signalen gestaltet sich, je nach Ereignisart,
unterschiedlich kompliziert. \"Ublicherweise sind Neutrinoteleskope auf die Rekostruktion von
{\it Myon-Ereignissen} optimiert, d.h.~auf Ereignisse, bei denen das Neutrino unter Austausch des
geladenen Stroms mit einem Nukleon aus dem es umgebenden Medium wechselwirkt und ein Myon und einen
hadronischen Schauer erzeugt. Da das Myon bei den betrachteten Energien eine sehr viel gr\"o\ss ere
Reichweite als der Schauer hat, ist es in den meisten F\"allen das 
einzige Teilchen, das den Detektor erreicht. Dabei produziert es Cherenkov-Photonen unter einem
festen Winkel entlang seiner geraden Spur, sodass die Richtung des Myons, und damit auch die des
Neutrinos, auf einige Zehntelgrad genau bestimmt werden k\"onnen. Ein Nachteil f\"ur diese Art von
Ereignissen ist jedoch, dass die Energie schwierig zu rekonstruieren ist, da unbekannt ist, welcher
Anteil der Myonenspur au\ss erhalb des Detektors verlief. \\
Das Myon ist jedoch nicht das h\"aufigste Endprodukt in einer Neutrinoreaktion. Dies sind
vielmehr die {\it hadronischen Schauer}, wobei angemerkt werden muss, dass bei ANTARES aufgrund der
im Verh\"altnis zu den Ausdehnungen eines Schauers groben Instrumentierung nicht zwischen
hadronischen und {\it elektromagnetischen} Schauern unterschieden werden kann, zumal bei den f\"ur
diese Studie relevanten Energien im TeV-Bereich und dar\"uber auch der \"uberwiegende Anteil der
Teilchen im hadronischen Schauer aus elektromagnetischen Wechselwirkungen stammt. Hadronische
Schauer werden sowohl in Neutrinoreaktionen mit geladenem Strom als auch in solchen mit neutralem
Strom erzeugt. In letzteren sind sie sogar die einzigen nachweisbaren Bestandteile des Endzustandes,
da als weiteres auslaufendes Teilchen ein Neutrino erzeugt wird, welches nicht beobachtet werden
kann. Die Rekonstruktion hadronischer Schauer ist daher von gro\ss em Interesse, da sie die
Untersuchung zus\"atzlicher Ereignisklassen erm\"oglicht und somit die durch den kleinen
Wirkungsquerschnitt bedingt geringe Ereignisrate erh\"oht. 
\\
Nachteilig auf die Rekonstruierbarkeit von Schauern wirkt sich jedoch deren
Richtungs- und Abstrahlcharakteristik aus. Typische Schauerl\"angen bei den betrachteten Energien
zwischen 100\,GeV und 100\,PeV betragen um die 10\,m, was im Verh\"altnis zu den Abst\"anden der
einzelnen Detektorstrings, die zwischen 60\,m und 75\,m liegen, klein ist. Schauer k\"onnen
somit als quasi punktf\"ormige Ereignisse im Detektor betrachtet werden. Sie werden auch nur dann
detektiert, wenn die Neutrinoreaktion im instrumentierten Volumen oder innerhalb eines Bereiches von
$\sim 100$\,m um den Detektorrand erfolgte,
w\"ahrend Myonen noch Kilometer von ihrem Entstehungsort entfernt registriert werden k\"onnen, wenn
sie in Richtung des Detektors fliegen. Die Sensitivit\"at des Detektors ist daher f\"ur
Schau\-er\-er\-eig\-nis\-se erheblich kleiner als f\"ur Myonereignisse. Die kurze Schauerl\"ange ist
auch daf\"ur verantwortlich, dass der R\"uckschluss vom Schauersignal auf die Neutrinorichtung
ungenauer als im Fall des Myons ist. Aufgrund der gro\ss en Zahl von Sekund\"arteilchen im Schauer,
die keineswegs alle in Richtung der Schauerachse erzeugt werden, haben die von diesen
Sekund\"arteilchen erzeugten Cherenkov-Photonen verschiedene Winkel in Bezug auf die
Schauerachse. Somit entsteht hier auch kein klarer, scharfer Kegel wie beim Myon, sondern eine
breite Verteilung, die lediglich ihr Maximum im Bereich des Cherenkov-Winkels von $42^{\circ}$
hat. Abbildung~\ref{fig:Z:cheren_fit} zeigt diese Polarwinkelverteilung f\"ur verschiedene Energien
in logarithmischer Darstellung, sowie eine im Rahmen dieser Arbeit erstellte Parameterisierung
dieser Verteilung.  

\begin{figure}[h] \centering
\includegraphics[width=7.4cm]{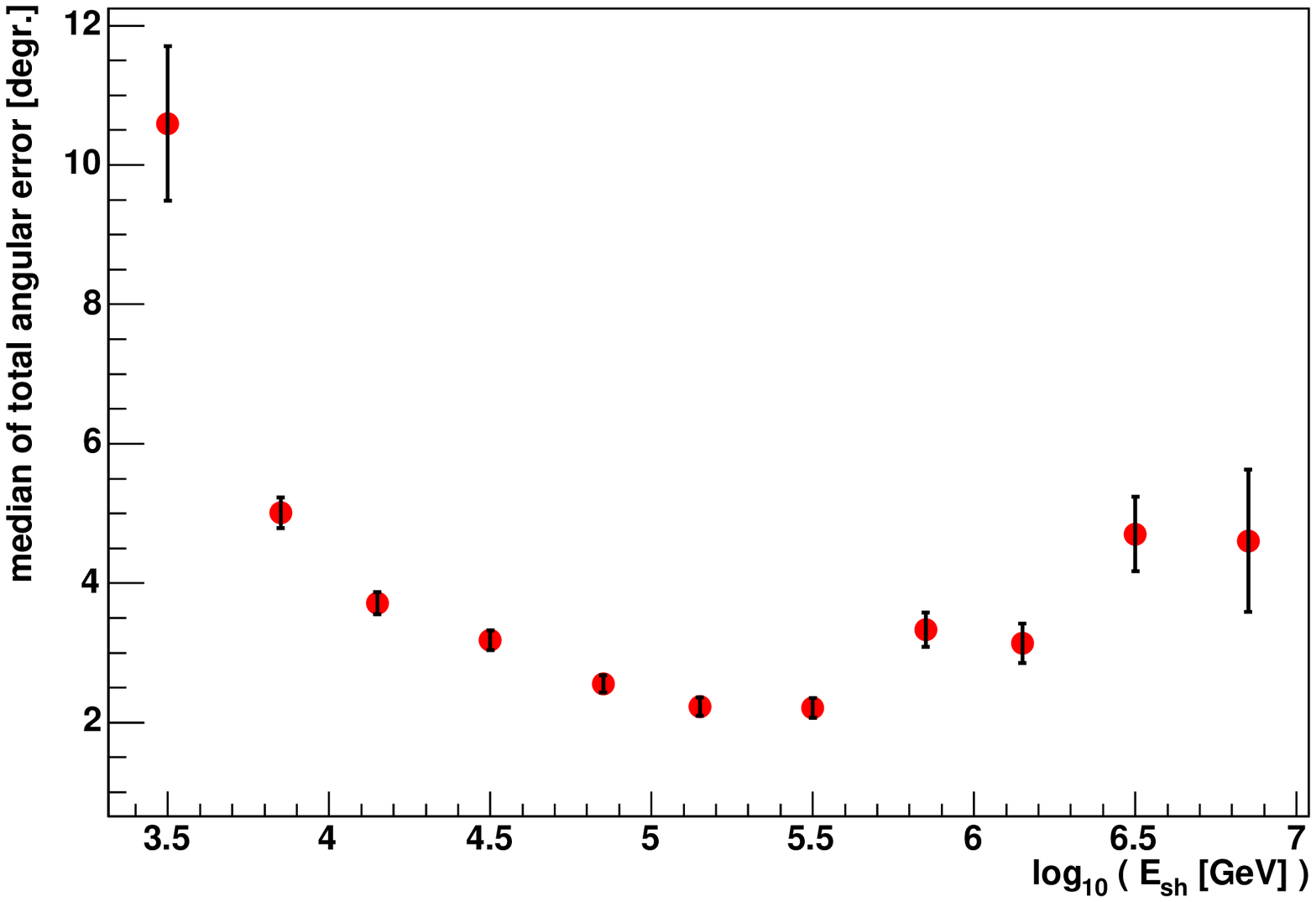}
\includegraphics[width=7.4cm]{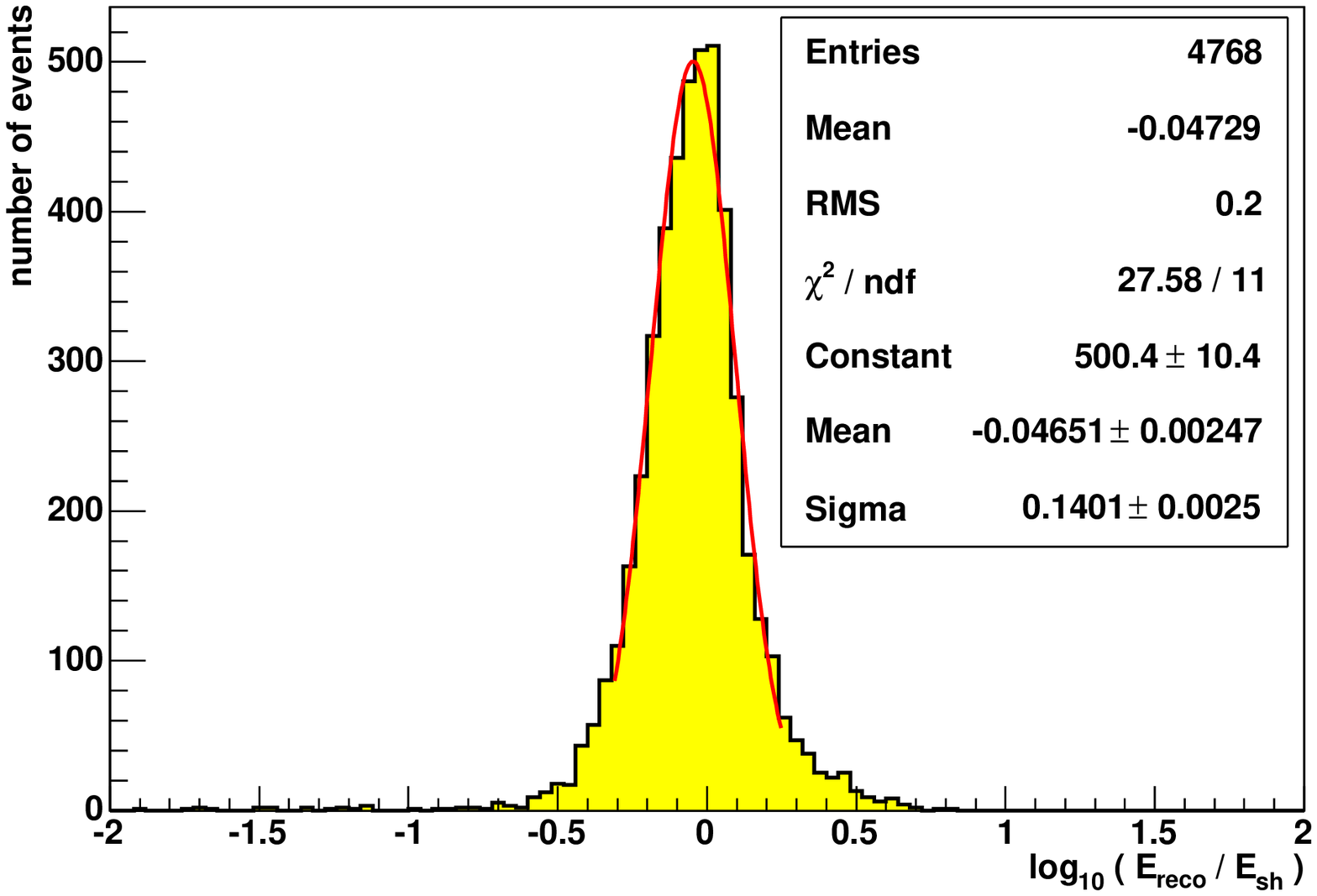}
\caption[Gesamtwinkelfehler und Energieaufl\"osung nach Schnitten]{Median des Gesamtwinkelfehlers,
  aufgetragen \"uber der wahren Schauerenergie (links), und
  Energieaufl\"osung in Bezug auf die Schauerenergie (rechts), f\"ur Neutralstromereignisse, nach
  Schnitten.}
\label{fig:Z:result}
\end{figure}

\clearpage
W\"ahrend bei der Richtungsrekonstruktion von Schauern also mit schlechteren Ergebnissen als bei
Myonen zu rechnen ist, erwartet man ein deutlich besseres Ergebnis f\"ur die Energierekonstruktion
des Schauers, da dessen gesamte Energie innerhalb eines verh\"altnism\"a\ss ig kleinen Volumens
deponiert wird und man von der Annahme ausgehen kann, dass die ausgesandte Lichtmenge proportional
zur Energie des Schauers ist. Die Lichtmenge pro Ereignis l\"asst sich aus den in den Optischen
Modulen gemessenen Amplituden berechnen, indem diese auf die Entfernung zum Reaktionsort und die
Winkelakzeptanz der Optischen Module in Bezug auf die Photonrichtung korrigiert wurden. Zus\"atzlich
wird die in Abbildung~\ref{fig:Z:cheren_fit} gezeigte Polarwinkelverteilung der Photonen
verwendet, um eine Hochrechnung auf die Photondichte des gesamten Raumwinkels vorzunehmen. Da f\"ur
die genannten Berechnungen die Schauerrichtung ben\"otigt wird, bietet sich eine kombinierte
Rekonstruktion von Schauerrichtung und -energie an, wobei beide Gr\"o\ss en gleichzeitig variiert
werden. Das Auffinden der idealen Werte erfolgt \"uber den Abgleich der f\"ur einen momentan
angenommenen Wert von Schauerrichtung und -energie erwarteten Amplitude in jedem einzelnen 
Optischen Modul mit der tats\"achlich gemessenen Amplitude. Mit Hilfe eines Log-Likelihood-Fits
werden Richtung und Energie dann variiert, bis die maximale
\"Ubereinstimmung gefunden ist. \\
Der Algorithmus erlaubt die Rekonstruktion der Schauerrichtung mit einer Aufl\"osung von
$\sim 10^{\circ}$ im Median, f\"ur Ereignisse mit einer rekonstruierten Schauerenergie $> 5$\,TeV. 
Durch weitere geeignete Schnitte kann der Gesamtwinkelfehler f\"ur einzelne Energiebereiche auf 
Werte bis \alphabest~reduziert werden, wie aus Abbildung~\ref{fig:Z:result} (links) ersichtlich 
wird. Hier wurde f\"ur die jeweils gezeigten 
Bereiche in der wahren Schauerenergie der Median des Gesamtwinkelfehlers berechnet. W\"ahrend die
Winkelaufl\"osung bis zu einer Schauerenergie von ca.~300\,TeV stetig besser wird, steigt sie f\"ur
noch gr\"o\ss ere Energien wieder leicht an, was daran liegt, dass dann f\"ur die meisten Optischen
Module das S\"attigungsniveau der Ausleseelektronik erreicht ist. \\
Zur Darstellung der erzielten Aufl\"osung f\"ur die Schauerenergie wird auf der rechten Seite von
Abbildung~\ref{fig:Z:result} der Logarithmus des Quotienten von rekonstruierter und wahrer
Schauerenergie gezeigt. Der Gro\ss teil der Ereignisse liegt nahe Null, was einer 
\"Ubereinstimmung der beiden Werte entspricht. Aus der Standardabweichung von \ncElog\ ergibt sich, dass
die Schauerenergie bis auf einen Faktor $10^{\ncElog} \approx \ncE$ genau bestimmt werden kann. Dieser
Wert wird durch einige Ereignisse bei zu klein rekonstruierten Energien noch verf\"alscht; die wahre
Breite des Maximums, erkennbar durch die an die Verteilung angepasste, in rot eingezeichnete Gau\ss
kurve, liegt bei $\sim \ncEcutlog$, was einem Faktor von $\ncEcut$ in der Energie entspricht. 

\begin{figure}[h] \centering
\includegraphics[width=7.4cm]{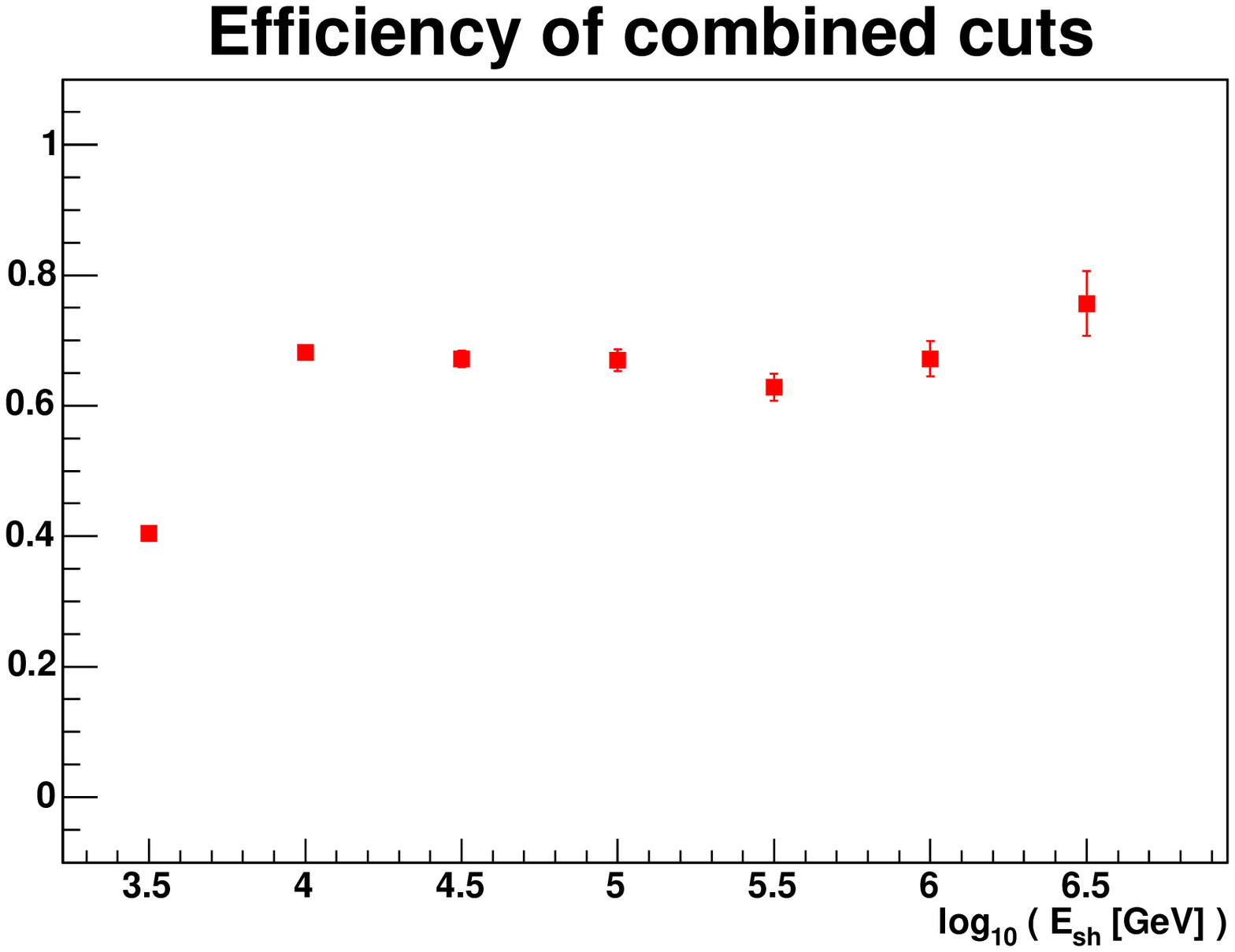}
\includegraphics[width=7.4cm]{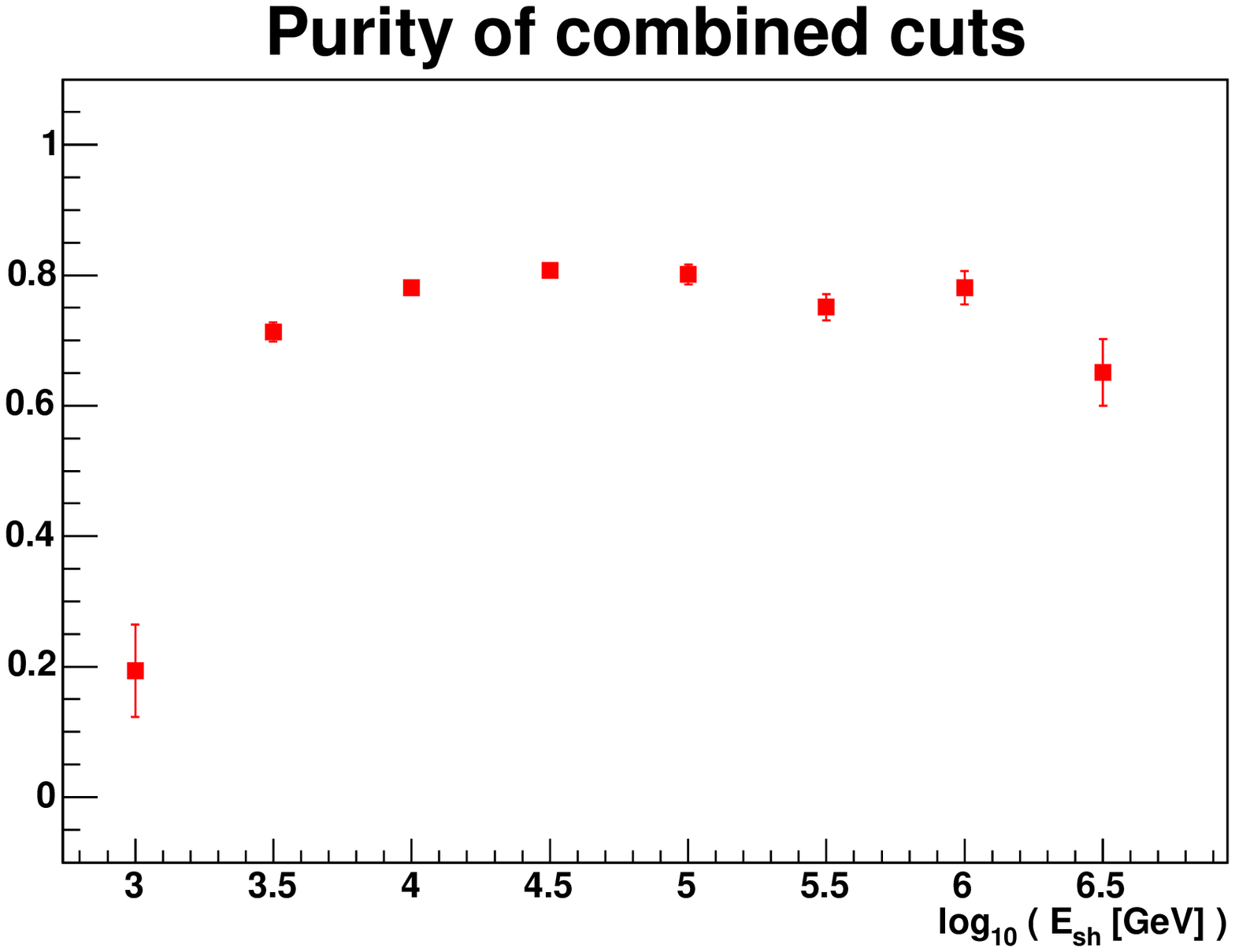}
\caption[Effizienz und Reinheit der Schnitte]{Effizienz (links) und Reinheit (rechts) der
  vorgenommenen Schnitte, aufgetragen \"uber der wahren Schauerenergie.}
\label{fig:Z:effi}
\end{figure}

\clearpage
Die {\it Effizienz} der Schnitte, d.h.~der Anteil der den Schnitt passierenden Ereignisse mit gutem
Ergebnis (Winkelfehler $< 10^{\circ}$), liegt f\"ur Schauerenergien oberhalb von 10\,TeV bei
ca.~70\%, wie aus Abbildung~\ref{fig:Z:effi} (links) ersichtlich wird. In derselben Abbildung
rechts ist die {\it Reinheit} der Schnitte, d.h.~der Anteil der Ereignisse mit gutem Ergebnis
(Winkelfehler $< 10^{\circ}$) nach den Schnitten an der Gesamtzahl der den Schnitt passierenden
Ereignisse, gezeigt. Die Reinheit erreicht einen Wert von ca.~80\% zwischen 10\,TeV und 1\,PeV.  

\begin{figure}[h] \centering
\includegraphics[width=10cm]{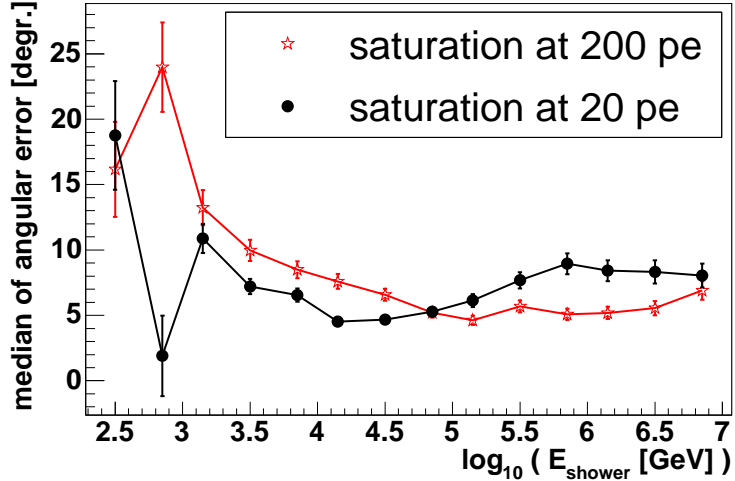}
\caption[Gesamtwinkelfehler bei verschiedenen S\"attigungsniveaus]{Median des Gesamtwinkelfehlers in
  Abh\"angigkeit der Schauerenergie, f\"ur S\"attigung bei 200\,pe (WF-Modus, rote Sterne) und bei
  20\,pe (SPE-Modus, schwarze Kreise), vor allen Schnitten.}
\label{fig:Z:SPE_WF}
\end{figure}

Bei den gezeigten Ergebnissen handelt es sich um rekonstruierte Neutralstromereignisse. Bei diesem
Ereignistyp ist der hadronische Schauer der einzige detektierbare Bestandteil des Endzustandes, und
da das Neutrino im Endzustand einen unbekannten Energieanteil tr\"agt, kann f\"ur die
Prim\"arenergie nur eine untere Grenze in H\"ohe der Schauerenergie angegeben werden. Anders sieht
es bei der Reaktion eines {\it Elektron-Neutrinos} \"uber den 
geladenen Strom aus. Hier geht die gesamte Energie des Prim\"arneutrinos in einen
elektromagnetischen und einen hadronischen Schauer \"uber, die mit dem vorliegenden Algorithmus
gemeinsam rekonstruiert werden k\"onnen. Die ermittelte Schauerenergie entspricht dann der
Neutrinoenergie, sodass f\"ur diesen Ereignistyp nach Schnitten eine Aufl\"osung erzielt werden kann,
die einem Faktor \ccEcut\ in der Neutrinoenergie entspricht. \\ 
Die hier gezeigten Ergebnisse wurden unter der Annahme einer S\"attigung der Elektronik der 
Photomultiplier bei 200\,Photoelektronen (pe) ermittelt. Dieses S\"attigungsniveau
entspricht der Aufnahme von {\it Wellenformen (WF)}, welche sehr bandbreiten- und speicherintensiv
ist. Die Datennahme in diesem Modus ist mit der Elektronik der Optischen Module zwar m\"oglich,
jedoch ist unklar, in wie weit er im fertigen Detektor wirklich genutzt werden wird. Der alternative
Aufnahmemodus w\"are der so genannte {\it Einzelelektronenmodus (SPE-Modus)}, bei dem jeweils
Amplitude und Zeitpunkt eines Signals aufgezeichnet werden, ohne weitere Informationen \"uber die
Wellenform. Dieser Modus verbraucht nur etwa $\frac{1}{40}$ der f\"ur Wellenformen ben\"otigten
Bandbreite, hat jedoch im Hinblick auf die Schauerrekonstruktion den Nachteil, dass das
S\"attigungsniveau hier bereits bei etwa 20\,pe liegt, und daher der Vergleich zwischen 
berechneter und gemessener Amplitude bei hohen Energien ungenauer wird. Dies wird in
Abbildung~\ref{fig:Z:SPE_WF} deutlich, wo f\"ur beide Datennahmemoden der Median des
Gesamtwinkelfehlers, vor allen Schnitten, f\"ur rekonstruierte Neutralstromereignisse \"uber der
Schauerenergie  aufgetragen ist. \\
Die erreichte Aufl\"osung im Bereich bis $\sim 70$\,TeV ist f\"ur das niedrigere S\"attigungsniveau sogar
besser als f\"ur das h\"ohere, da in diesem Bereich hohe Fluktuation in der Photonenverteilung der
einzelnen Ereignissen auftreten k\"onnen, die durch die niedrigere S\"attigung besser unterdr\"uckt
werden. Oberhalb von 100\,TeV zeigt das niedrigere S\"attigungsniveau ein leichte Verschlechterung 
der Richtungsrekonstruktion um ca.~3$^{\circ}$ im Vergleich zu einer S\"attigung bei 200\,pe. Die
erreichte Aufl\"osung ist jedoch immer noch gut genug, um eine Richtungsrekonstruktion zu erm\"oglichen. 
Zudem zeigt sich in der Energieaufl\"osung kein nennenswerter Unterschied zwischen beiden Moden.

\begin{figure}[h] \centering
\includegraphics[width=14.cm]{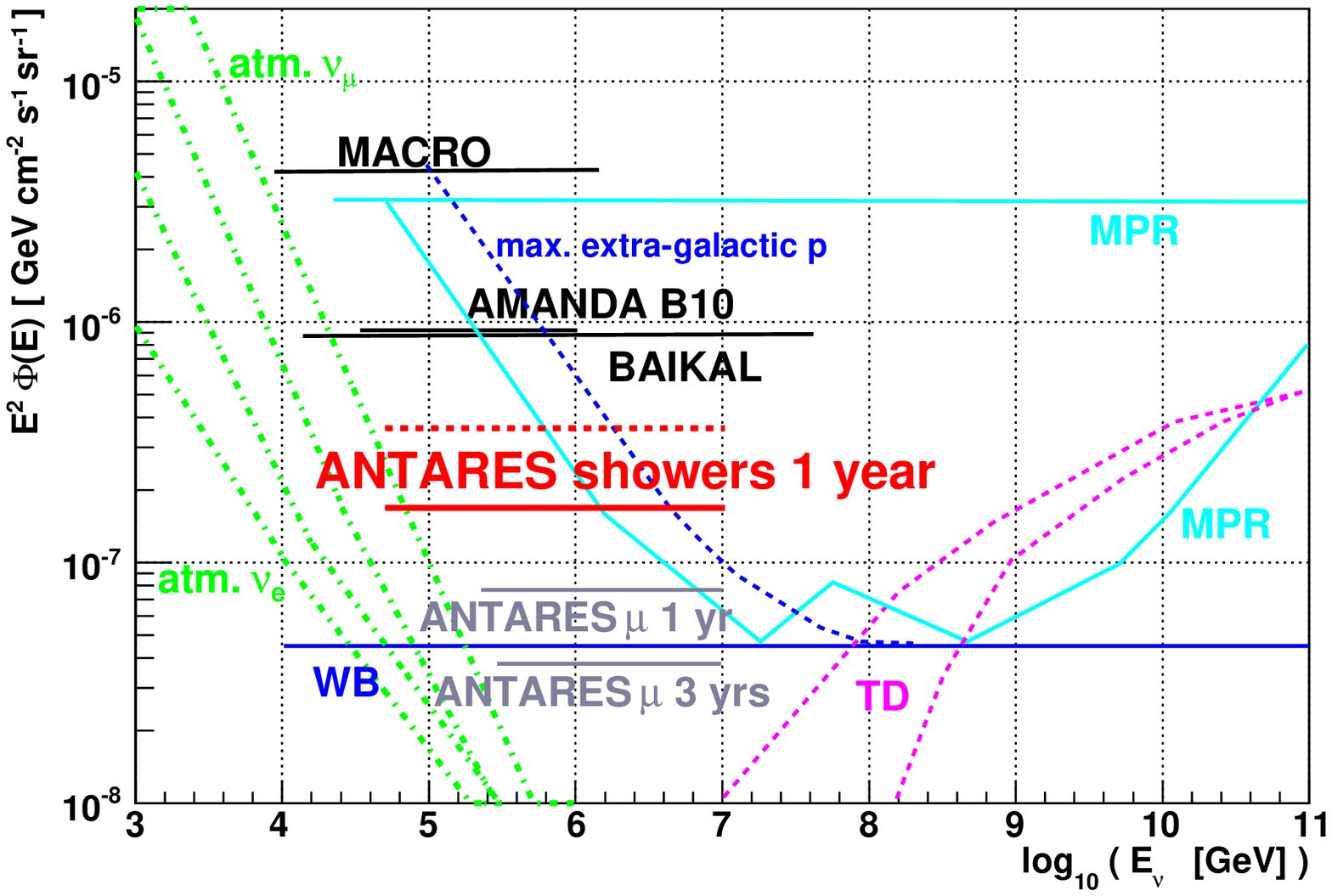}
\caption[Sensitivit\"at f\"ur Schauerereignisse]{Sensitivit\"at f\"ur Schauer in ANTARES, die von
  isotrop einfallenden (durchgezogene rote Linie), bzw. von von unten kommenden Neutrinos
  (gestrichelte rote Linie) erzeugt wurden, f\"ur ein Jahr Messdauer, und
  im Vergleich dazu Ergebnisse und Erwartungen anderer Ereignisarten und Experimente, welche im Text
  genauer erl\"autert werden.}
\label{fig:Z:Sensitivitaet}
\end{figure}

Verschiedene Untergrundquellen sind zu ber\"ucksichtigen: Zum einen erzeugen radioaktive Zerf\"alle
und Mikroorganismen in der Tiefseeumgebung ein zeitlich und r\"aumlich nur langsam variierendes {\it
  optisches Rauschen}, welches zu unkorrelierten, einzelne Photoelektronen erzeugenden Signalen in
den Optischen Modulen f\"uhrt. Diese Untergrundsignale 
k\"onnen durch entsprechend gew\"ahlte Kausalit\"atsbedingungen f\"ur die Signalzeitpunkte in
verschiedenen Optischen Modulen, sowie die Forderung einer Mindestamplitude, zu gro\ss en Teilen
unterdr\"uckt werden. \\ 
Weitere Untergrundquellen sind atmosph\"arische Myonen und atmosph\"arische
Neutrinos. {\it Atmosph\"arische Myonen} werden durch die Wechselwirkung kosmischer Wasserstoff- oder
anderer Kerne mit der Erdatmosph\"are erzeugt und stellen einen gef\"ahrlichen, von oben kommenden
Untergrund f\"ur Schauerereignisse dar, wenn sie nicht eindeutig als Myonen identifiziert werden
k\"onnen. Dies ist dann der Fall, wenn die Myonen durch starke Bremsstrahlungsverluste einen
elektromagnetischen Schauer im Detektor induzieren, oder wenn mehrere Myonen gleichzeitig, als so
genanntes {\it Myonenb\"undel}, den Detektor passieren. Durch Qualit\"atsschnitte lassen sich die 
Myonen, die die Schauerrekonstruktion \"uberlebt haben, zu \"uber 99\% unterdr\"ucken. Der durch 
atmosph\"arische Myonen erzeugte Untergrund kann zus\"atzlich reduziert werden, indem nur von 
unten kommende Ereignisse betrachtet werden. \\ 
{\it Atmosph\"arische Neutrinos} werden wie atmosph\"arische Myonen durch die Wechselwirkung
geladener kosmischer Strahlung mit der Erdatmosph\"are erzeugt, k\"onnen jedoch, anders als erstere,
auch von unten, durch die Erde hindurch, den Detektor erreichen. Da atmosph\"arische Neutrinos nicht
von kosmischen Neutrinos unterscheidbar sind, kann dieser
Untergrund durch einfache Ereignisselektion nicht unterdr\"uckt werden; kosmische Neutrinos k\"onnen
lediglich als \"Uberschuss \"uber dem erwarteten atmosph\"arischen Neutrinofluss detektiert werden. \\
Der atmosph\"arische Myonen-, bzw.~Neutrinountergrund dominiert das Signal der kosmischen Neutrinos
unterhalb von 20\,TeV bzw.~50\,TeV. Die Messung eines isotropen diffusen Flusses ist somit nur oberhalb 
dieser Energien m\"oglich. F\"ur den verbleibenden betrachteten Energiebereich bis ca.~10\,PeV werden 
noch \nnu\ atmosph\"arische Neutrinos pro Jahr erwartet. Nimmt man an, dass tats\"achlich
w\"ahrend einer einj\"ahrigen Messperiode in ANTARES ein einziges Ereignis detektiert worden ist und dass der
kosmische Neutrinofluss proportional zu $E_{\nu}^{-2}$ ist, so ergibt sich daraus, mit einem
Konfidenzniveau von 90\%, eine enegieunabh\"angige Obergrenze des kosmischen Neutrinoflusses von 

\begin{equation*}
E_{\nu}^2 {\Phi_{90\%}} = \sensitivity\,\textrm{GeV cm}^{-2}\,\textrm{s}^{-1}\,\textrm{sr}^{-1}
\end{equation*}

f\"ur den betrachteten Energiebereich zwischen 50\,TeV und 10\,PeV.
Dieser Wert ist in Abbildung~\ref{fig:Z:Sensitivitaet} als durchgezogene rote Linie, im Vergleich zu
den f\"ur Schauer gemessenen Obergrenzen der Neutrinoexperimente AMANDA~\cite{amanda_cascades} und
BAIKAL~\cite{wischnewski}, sowie zu den bei MACRO gemessenen~\cite{macro}, bzw. f\"ur ANTARES
berechneten~\cite{zornoza} (graue Linien) Obergrenzen f\"ur Myon-Ereignisse, gezeigt. Die in gr\"un
gezeigten atmosph\"arischen Neutrinofl\"usse entsprechen dem Modell von Bartol~\cite{bartol} f\"ur
verschiedene Einfallwinkel. Die weiteren Eintragungen zeigen verschiedene theoretische Obergrenzen
des Neutrinoflusses nach Modellen von Waxman und Bahcall~\cite{waxman-bahcall1,waxman-bahcall2}
(blaue, mit {\it WB} und {\it max.~extra-galactic p} beschriftete Linien; Mannheim, Protheroe und
Rachen~\cite{MPR} (t\"urkisfarbene, mit {\it MPR} beschriftete Kurven); sowie \glqq
top-down\rq\rq-Modelle~\cite{topdown} (violette, mit {\it TD} beschriftete Linien). \\ 
Betrachtet man nur von unten kommende Ereignisse, so dominiert oberhalb von $\sim 5$\,TeV die
Neutrinorate \"uber der Rate f\"alschlicherweise als von unten kommend rekonstruierter 
atmosph\"arischer Myonen. Der kosmische Neutrinofluss \"ubersteigt den
atmosph\"arischen bei 50\,TeV, und oberhalb dieser Energie ist die erwartete Untergrundrate noch
0.35 pro Jahr. Die insgesamt pro Jahr erwartete Untergrundrate ist also in etwa mit Null
vertr\"aglich. Unter der Annahme, dass nach einj\"ahriger Messung in ANTARES kein einziges Ereignis
gefunden wurde, ergibt sich die energieunabh\"angige Obergrenze f\"ur den kosmischen Neutrinofluss
mit einem Konfidenzniveau von 90\% zu 

\begin{equation*}
E^2 {\Phi_{90\%}}^{\auf} = \sensitivup\,\textrm{GeV cm}^{-2}\,\textrm{s}^{-1}\,\textrm{sr}^{-1}.
\end{equation*}

Dieser Wert ist in Abbildung~\ref{fig:Z:Sensitivitaet} als gestrichelte rote Linie eingezeichnet.

Die vorgelegte Rekonstruktionsstrategie schlie\ss t eine gro\ss e L\"ucke in der
Ereignisrekonstruktion bei ANTARES, da erst durch sie die Rekonstruktion von Ereignissen mit
Schauern erm\"oglicht wird. Damit ist nun prinzipiell die Rekonstruktion aller bei ANTARES
auftretenden Ereignistypen m\"oglich. Aufgrund der ohnehin geringen Fl\"usse der kosmischen
Neutrinos ist jede zus\"atzlich rekonstruierbare Ereignisklasse von gro\ss er Bedeutung f\"ur die
Sensitivit\"at des Experiments. \\ 
In einer Weiterentwicklung kann durch die Verbindung dieses
Rekonstruktionsalgorithmus mit demjenigen f\"ur die Myonrekonstruktion auch eine Verbesserung in der
Energierekonstruktion von Myonereignissen erzielt werden, sofern diese so nahe am Detektor
stattgefunden haben, dass der hadronische Schauer mit detektiert wurde. In diesem Fall k\"onnen
z.B.~die Energien von Myon und Schauer mit den verschiedenen Strategien getrennt rekonstruiert und
das Ergebnis dann kombiniert werden. 

\selectlanguage{english}


\tableofcontents
\listoffigures
\listoftables 

\mainmatter

\selectlanguage{english}
\chapter{Introduction}

When Wolfgang Pauli postulated the neutrino in 1930, he probably would not have imagined that today, 75 
years later,  giant instruments for the detection of what he had called a \lq\lq desperate way 
out\rq\rq\, of the beta decay puzzle would exist, let alone in such hostile surroundings as the deep 
sea, or Antarctica. One of these experiments is ANTARES~\cite{proposal}, a neutrino telescope that
is currently under construction at a depth of 2400\,m in the Mediterranean Sea. \\ 
The goal of ANTARES is the detection of high-energy neutrinos, i.e.~neutrinos with an energy
$\gtrsim 50$\,GeV, from the cosmos. While neutrinos generated in nuclear power plants and particle
accelerators, in the atmosphere of the Earth, and also inside the Sun, have all long been
detected in large numbers, the only neutrinos from outside our solar system that have been measured
so far are a handful of events from Supernova
1987A~\cite{SN1987_baksan,SN1987_IMB,SN1987_kamiokande} with energies in the 10 MeV range. \\
Strong evidence exists, however, that high-energy cosmic neutrinos are generated in
powerful cosmic particle accelerators, like Supernova Remnants or Gamma Ray Bursts:
Air shower experiments on Earth measure high rates of high-energy charged cosmic rays, for which
these stellar accelerators are possible sources. As these sources can be associated with dense
matter concentrations, it is expected that a part of the accelerated cosmic rays interacts with this
dense matter to produce secondary photons and neutrinos. Photons at TeV energies from
these sources have already been measured, and astroparticle physicists are therefore
convinced that it is only a matter of time until high-energy neutrinos
will be detected as well. Theoretical models suggest that the flux of these neutrinos is small,
with predicted values around $10^{-6}$\,m$^{-2}$\,s$^{-1}$\,sr$^{-1}$ for energies above 1\,TeV. 
Also, the cross section of neutrinos is very small, because they only interact through the weak
force. \\
The detection efficiency therefore depends crucially on the size of the detector. To enable an
experiment to measure a statistically relevant neutrino rate, huge target masses have  
to be instrumented. This is the reason why neutrino detectors use natural targets like
the sea or the Antarctic ice. The transparency of these targets is an important factor for the
neutrino detection, because neutrinos can only be detected indirectly, through secondary, charged
particles. When travelling faster than the speed of light in the medium, these charged particles
produce light, the so called Cherenkov radiation~\cite{cherenkov}, which is detected by
photomultipliers of the experiment. \\ 
Generally, the inelastic neutrino-nucleon interaction cross section exceeds that of the neutrino-electron
interaction by several orders of magnitude\footnote{with the exception of the $W$-resonance at 6.3\,PeV in
the channel ${\bar{\nu}_e + e^- \to W \to anything}$.}.
When a neutrino interacts inelastically with a nucleon, a hadronic shower and a lepton are produced,
the type of the latter depending on the flavour of the incident neutrino ($\nu_{e}, \nu_{\mu}$ or
$\nu_{\tau}$), and the type of interaction: In charged current reactions, a charged lepton
corresponding to the neutrino flavour is produced; in neutral current reactions, the final state
lepton equals the incident neutrino. The interaction channels for anti-neutrinos are equivalent, and
in the following, if a neutrino channel is mentioned, the respective anti-neutrino channel is always
meant as well. \\ 
From the detection point of view, the most favourable secondary lepton from a neutrino interaction
is the muon, because it can travel over distances up to several km in water, emitting Cherenkov
light at a fixed angle along a straight track. This allows for the reconstruction of the muon
direction with sub-degree resolution. Therefore, experiments like ANTARES have been optimised
for muon detection, and most of the studies conducted so far have specialised on muon
reconstruction. This implies, however, the loss of those event classes which are not characterised
by an isolated muon track, but instead, by cascades: {\it Hadronic cascades} occur in all
neutrino-nucleon interactions --- in neutral current reactions, the hadronic cascade is even the
only detectable part of the interaction; {\it electromagnetic cascades} are generated from secondary
electrons in the charged current interactions.  \\
This thesis presents the first full and detailed reconstruction strategy inside ANTARES for this class of
cascade-, or shower-type events\footnote{In 2000, F.~Bernard~\cite{bernard} has conducted a
simplified study on showers in ANTARES, which did however not become 
part of the official ANTARES software.}. For the reconstruction of these events, a pattern
matching algorithm has been developed. The basic feature of this algorithm is the matching between
the amplitudes measured in the photomultipliers of the detector, and their expected values which 
are calculated assuming a starting value for the energy proportional to the
amount of light that is measured in the shower. Under the assumption of a certain position and
direction of the shower, and considering the photon directions distributed according to a
parameterisation derived from simulations, the expected amplitude for each photomultiplier is
calculated. The photon attenuation in water and the angular efficiency of the photomultipliers are
taken into account. The matching of the calculated and the measured amplitudes in each
photomultiplier is then tested by a likelihood function. Shower direction and energy are varied
until the likelihood, and therefore the agreement between expectation and measurement, becomes
maximal. \\ 
With this algorithm it is possible to reconstruct the direction of a shower in ANTARES
with a resolution as good as \alphabest. The resolution for the reconstruction of the
shower energy is usually obtained by fitting a Gaussian to the distribution of
${\log_{10}(E_{reco}/E_{MC})}$. The width $\sigma_E$ of the Gaussian describes the logarithmic
energy resolution. In this study, a resolution of $\sigma_E = \ccEcutlog$ is reached, 
which corresponds to a factor of $10^{\ccEcutlog} \approx \ccEcut$~in the shower energy. In the case of
charged current interactions of electron neutrinos, the shower energy is equivalent to the neutrino
energy; for neutral current interactions, the neutrino carries away part of the energy, and the
reconstructed shower energy only provides a lower limit on the primary neutrino energy, which
introduces an additional bias and error. In comparison, for muon events, resolutions as good as a
few tenths of a degree are reached for the reconstruction of the direction, but the muon energy can
only be determined within a factor of 2 -- 2.5, and provides, as for neutral current events, only a
lower limit of the primary neutrino energy. \\
The content of this thesis is the following: In Chapter~\ref{ch:CR}, a general introduction to the
sources and fluxes of cosmic rays is provided, together with a short overview of the history of
neutrino physics. Generation mechanisms for high-energy cosmic particles are explained, and a list
of the possible or known sources of high-energy neutrinos is given. Chapter~\ref{sec:interactions}
illustrates neutrino interactions and the detection of the secondaries and discusses the expected
neutrino fluxes. Chapter~\ref{ch:experiment} gives an overview of the ANTARES detector. In
Chapter~\ref{ch:showers}, possible event types in ANTARES are presented, and characteristics of
electromagnetic and hadronic showers are discussed. In Chapter~\ref{sec:background}, various
background sources, both from atmospheric particles and from optical noise in the deep sea, are
described. Different possible algorithms for an individual calculation of the shower position,
direction and energy are discussed in Chapter~\ref{sec:minor_strategies}. The final strategy for the
shower reconstruction is explained in Chapter~\ref{sec:shower_fitter}. A set of cuts to separate
well reconstructed events from poorly reconstructed ones is described in Chapter~\ref{sec:cuts}
which also covers the suppression of atmospheric muon background. The results for different data
samples, both neutral current and charged current $\nu_e$ events, before and after the cuts, are
presented in Chapter~\ref{sec:results}, together with effective areas and a sensitivity estimate for 
diffuse neutrino flux. Finally, in Chapter~\ref{ch:conclusion}, a summary and an outlook to further
developments is given.

\chapter{Cosmic High-Energy Particles}\label{ch:CR} 

The cosmic high-energy particles which arrive at Earth can be divided into three
classes: Charged particles, denoted {\it cosmic rays} for historical reasons,
gamma rays, i.e.~high-energy photons, and neutrinos. \\
The detection of cosmic rays, almost 100\,years ago, opened the new window of non-optical
astronomy and consequently led to the development of stellar acceleration models which 
predict the generation of high-energy neutrinos. It is therefore
appropriate to start this chapter with a  short historical introduction on cosmic rays in 
Section~\ref{sec:cr}, together with a description of the measured energy spectrum. Thereafter, a
brief overview of the history of neutrino physics follows in
Section~\ref{sec:nu_history}. Section~\ref{sec:generation} describes some models for the generation
of high-energy particles in the cosmos, while Section~\ref{sec:nu_sources} comments on known and
possible sources, with a focus on high-energy neutrinos. 

\section{Cosmic Rays: History and Measured Spectrum}\label{sec:cr}

\subsection{History of Cosmic Ray Detection}\label{sec:cr_history}
The Earth's exposure to radiation from space was discovered as early as 1912 by the Austrian
physicist {\it Victor Hess (1883 - 1964)}. At that time ionising radiation on Earth had already been
detected, but the source of it was still uncertain. It was (correctly) believed that this radiation
came from radioactive decays in rocks or other ground matter, and therefore it was expected that
with increasing altitudes the radiation would decrease and finally vanish. In an attempt to study this
hypothesis, Hess measured the ionisation during several balloon experiments, and found that instead
of vanishing, the radiation increased with increasing altitude, which led him to the conclusion
that the Earth is exposed to ionising radiation from outside the atmosphere. Hess was awarded the
Nobel Prize for this discovery in 1936. \\
In the decades that followed, cosmic rays were studied mainly with balloon experiments,
and later with satellites. However, also ground-based experiments were constructed, the first one by 
{\it Pierre Auger (1899 - 1993)}, who discovered extensive air showers, caused by the
interaction of high-energy charged primaries with the atmosphere, in 1938. The energy contained
in these showers turned out to be several orders of magnitude larger than the energy of the cosmic rays
measured with balloons. Starting from 1946, arrays of interconnected detectors to study extended air
showers were constructed by groups in the USSR and the USA. It became clear that the cosmic ray flux
decreases with increasing energy, according to a power law (cf.~Section~\ref{sec:cr_fluxes} and
Figure~\ref{fig:cr_spectrum}). Cosmic ray experiments have therefore been built on larger and larger
scales, in order to detect particles at highest energies. The largest air shower array 
to date, the Pierre Auger Observatory~\cite{auger} which is under construction in Argentina, will
cover an area of 3000\,km$^2$. Its size will allow for the measurement of about 30 cosmic ray
events with energies above $10^{20}$\,eV a year~\cite{auger}. The completion of the experiment is
expected for mid-2006. 

\subsection{The Cosmic Ray Spectrum}\label{sec:cr_fluxes}
\begin{figure}[h]\centering
\includegraphics[width=15cm]{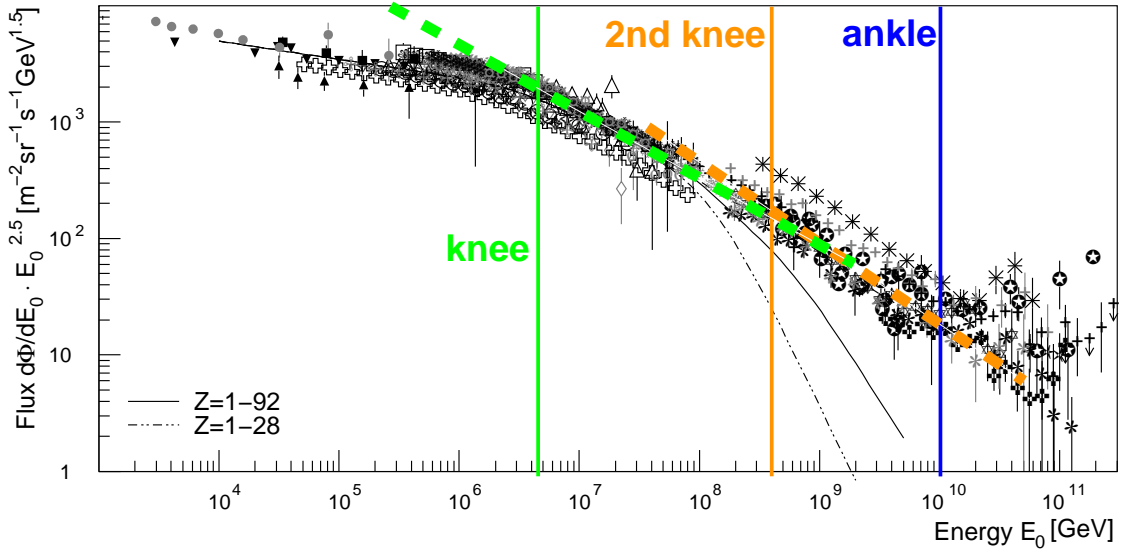}
\caption[All-particle energy spectrum]
{All-particle energy spectrum for cosmic rays as measured by different experiments, and according to
  model calculations (lines marked $Z = 1-92$ and $Z=1-28$). Note that the spectrum has been
  multiplied by $E^{2.5}$. The energies of the knees and the ankle, together with the approximate
  slope of the distribution, have been marked for better visualisation. After~\cite{hoerandel05}.}  
\label{fig:cr_spectrum}
\end{figure}

Cosmic rays consist of ionised nuclei, where, for particle energies below some TeV, protons account
for the largest fraction, about 90\%; helium nuclei make up for about 9\% of all cosmic rays, and
the rest consists of heavier nuclei up to iron. \\ 
The flux of the all-particle spectrum can be described by a power law,
\begin{equation*}
\frac{dN}{dE} \propto E^{-\gamma},
\end{equation*}
where $\gamma \approx 2.7$ for energies $\lesssim 4 \times 10^{15}$\,eV.
Figure~\ref{fig:cr_spectrum} shows the energy spectrum for the all-particle cosmic radiation
in an energy range between $10^6$\,eV and $10^{21}$\,eV. Note that the flux shown in this figure was  
multiplied by $E^{2.5}$, and therefore the spectrum appears less steep. Above about ${4 \times
10^{15}}$\,eV, the cosmic ray spectrum steepens to $\gamma \approx 3$; because of this change in the
slope, the region is referred to as the {\it knee}. This effect can be explained phenomenologically 
by assigning a cut-off energy to the cosmic ray components, proportional to their charge or
mass~\cite{hoerandel}. This would also explain why at around $4 \times 10^{17}$\,eV, the slope becomes
even steeper, an effect that is called the {\it second knee}. This steepening could be caused by the
cut-off of heavier elements in the cosmic rays. There are theoretical models~\cite{gaisser} which
support this explanation by assigning an upper energy limit to some of the cosmic ray sources,
so that above a certain energy the proton flux from this source is cut off. Heavier nuclei have
larger charges and must therefore be accelerated to larger energies $E$ to achieve the same rigidity
$R$ as the protons, ${R = E / (Ze B)}$, where $Ze$ is the charge of the nucleus and $B$ the magnetic
field which the nucleus propagates through. Consequently, for heavier elements, the cutoff lies
at higher energies. Connected with this suggestion is the assumption that for energies above the
knee, heavier nuclei start to dominate over the protons. The exact composition of the cosmic rays at
these high energies is very difficult to measure, as experiments on Earth only detect the
secondaries of the cosmic rays, after their interaction with the atmosphere. Nevertheless, many
important improvements have been achieved during the last years, e.g.~by the KASCADE air shower
array~\cite{kascade03,kascade}. \\
The cosmic ray spectral index remains stable up to $\sim 10^{19}$\,eV. Then the spectrum seems to
become harder again, an effect which is called the {\it ankle}. A possible
explanation~\cite{peters} for this effect is that an extra-galactic component begins to dominate the
spectrum at this energy. However, statistics becomes very small above $10^{19}$\,eV, and
different experiments are no longer in agreement with each other (see below). From the theoretical
point of view, it is expected that protons with an energy of some $10^{19}$\,eV start interacting
with the Cosmic Microwave Background by formation of $\Delta$-resonances, which would limit their
range to $\sim 50$\,Mpc. This effect is called the Greisen-Zatsepin-Kuz'min-cutoff
(GZK-cutoff)~\cite{greisen,zatsepin} and applies to heavier nuclei as well, at the corresponding
resonance energies. Additionally, these heavier nuclei lose energy by inducing photo-pair production
on the Cosmic Microwave Background.
If the GZK-mechanism holds, the sources of the highest-energy cosmic rays measured
on Earth must lie within the range of the primary nuclei. As currently no potential source for
particles of such high energies is known in the vicinity of our galaxy, a drop in the cosmic ray flux
should be observed at the GZK-energy. However, only the measurement of the Fly's Eye
Observatory~\cite{flyseye} supports this expectation, whereas the AGASA Experiment~\cite{agasa}
observes a flattening of the spectrum, back to smaller values of $\gamma$. The statistics of
both experiments are limited; clarification is expected from the new, much larger
Pierre Auger Observatory~\cite{auger} within the next years.  

\section{Neutrino Physics: Historical Overview}\label{sec:nu_history}

\subsection{Postulation of Neutrinos and First Detections}
Neutrinos have been postulated by {\it Wolfgang Pauli (1900 - 1958)} in 1930 to solve the energy
conservation problem in the beta decay of nuclei. In beta decay, the fundamental laws of the
conservation of energy, momentum and angular momentum seemed to be violated. Pauli made quite a
reckless suggestion to solve the problem, he invented an additional, hitherto unknown particle which
would be created in the beta decay and which would carry the missing energy, momentum and spin. This
particle would have to be neutral and very light. Pauli suggested to call it neutron; to avoid
confusion with the nucleon of the same name, which was discovered two years later (1932) by {\it James 
Chadwick (1891 - 1974)}, the name {\it neutrino} (\lq\lq small neutron\rq\rq ) was invented by {\it
  Enrico Fermi (1901 - 1954)}. Fermi also formulated a theory for the {\it weak force}, based on the
neutrino hypothesis. \\  
It took another 14 years, however, until neutrinos, to be more precise, electron
anti-neutrinos $\bar{\nu}_e$, were finally detected for the first time 
by {\it Frederick Reines (1918 - 1998)} and {\it Clyde L. Cowan, Jr. (1919 - 1974)} at the Savannah
River Plant. Reines received the Nobel Prize for the detection of the neutrino in 1995. 
The muon neutrino, $\nu_{\mu}$, was detected only a few years after the electron neutrino,
in 1962; it took another 38\,years until in 2000 the final direct evidence for the tau
neutrino, $\nu_{\tau}$, was found at the Fermilab. Experimental results had indicated before that
besides $\nu_e, \nu_{\mu}$ and $\nu_{\tau}$, no other light neutrinos should exist (here, light
means a mass of less than half the mass of the $Z$ boson). The $Z$ boson has been studied
extensively at the LEP experiments and its decay width fits very well to the 
hypothesis of three light neutrino generations~\cite{zmass}.

\subsection{Solar Neutrinos}
The dominant nuclear fusion reaction inside the Sun is the $pp$-chain. Neutrinos are produced in
this and several other reaction chains at energies ranging from less than 0.1\,MeV to 19\,MeV. In 1968
the first solar neutrinos were detected by the Homestake experiment~\cite{davis68,davis,homestake};
for a long time they were the only non-terrestrial neutrinos to be measured. A number of large
experiments like (Super)Kamiokande~\cite{kamiokande,superk} and SNO~\cite{SNO} has
been dedicated to the study of solar neutrinos; in 2002, {\it Raymond Davis Jr. (*1914)}, the
initiator of the Homestake experiment, received a Nobel Prize for his studies of the solar neutrinos
together with the initiator of the Kamiokande and SuperKamiokande experiments, {\it Masatoshi
  Koshiba (*1926)}. Solar neutrino experimentalists were confronted with a big puzzle during long
years of thorough studies: The measured flux was significantly smaller than expected from
theoretical predictions. Although the surprising solution, neutrino oscillation, was already proposed
in the late 1960s~\cite{gribov}, the final direct evidence for neutrino oscillations could only be
presented in 2001~\cite{nu-osc}. 
\\ 
Very detailed and precise reviews on solar neutrinos and neutrino oscillations have been
written by one of the pioneers of the field, {\it John N. Bahcall (1934 -
  2005)}~\cite{bahcall_sun,bahcall_neutrino}.  

\subsection{Cosmic Neutrinos}
\begin{figure}[h]\centering
\includegraphics[width=10cm]{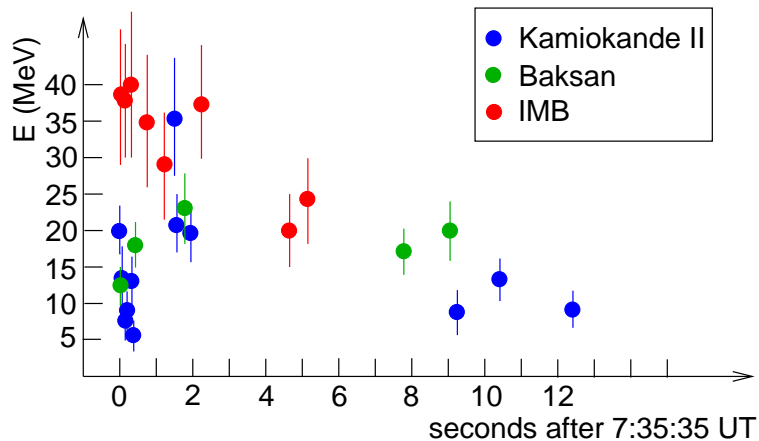}
\caption[Neutrinos from SN 1987A]
{Energy and arrival time of the neutrinos from Supernova 1987A~\cite{SN1987_kamiokande, SN1987_IMB,
    SN1987_baksan}. The signals of the three experiments 
  have been shifted so that they all start at the same time.}
\label{fig:SN_neutrinos}
\end{figure}

Even though neutrinos from the Sun have been studied for several decades, and also detectors
for cosmic neutrinos have been planned and built for more than 20\,years, cosmic neutrino
research is a young and still growing field of studies.  An overview of neutrino
telescopes is provided in Section~\ref{sec:other_ex}. However, none of these
neutrino telescopes has yet been able to identify any cosmic neutrinos. In fact there exists only
one proven source of non-terrestrial neutrinos besides the Sun, the Supernova 1987A:
On February 23, 1987, around 7:35 UT, the neutrino observatories Kamiokande II, IMB and Baksan have
detected a total of 25 neutrinos, with energies in the MeV range, from this
Supernova~\cite{SN1987_kamiokande, SN1987_IMB, SN1987_baksan}. The energies and arrival times of
the measured signals are shown in Figure~\ref{fig:SN_neutrinos}. The experiments had different
energy thresholds of 6\,MeV (Kamiokande II), 10\,MeV (Baksan) and 20\,MeV (IMB). There are also
uncertainties up to 1 minute in the absolute timing of the experiments; therefore the signals in the
plot have been shifted to the same starting point. \\
Even though the number of detected neutrinos from Supernova 1987A was not large, their detection led
to huge activities in the field and a number of new insights, e.g.~for the theory of stellar
evolution; it was also possible to derive a direct limit of $m_{\nu_e} \leq 20$\,eV on the
electron neutrino mass, from the length of the pulse and the measured neutrino
energies~\cite[p.79]{longair}.  

\section{Generation of High-Energy Particles in the Cosmos}\label{sec:generation}

\subsection{Charged Particles}\label{sec:fermi}

High-energy cosmic rays can be produced either in the \lq\lq bottom-up\rq\rq\ way, by 
acceleration, or in the \lq\lq top-down\rq\rq\ way, by the decay of super-massive particles.
The following explanation of acceleration through the Fermi Mechanism follows the book of 
Gaisser~\cite[pp.150-155]{gaisser_book}. 

\subsubsection{Acceleration: The Fermi Mechanism}

\begin{figure} [h] \centering
\includegraphics[width=7cm]{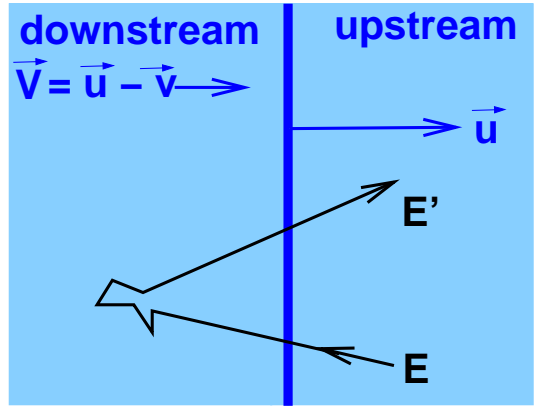}
\caption[Schematic illustration of the Fermi Mechanism]{Schematic illustration of the First Order
  Fermi Mechanism: A shock front moves with a velocity $\vec{u}$ through an accumulation of gas; the
  velocity of the shocked (\lq\lq downstream\rq\rq) gas relative to the shock front is
  $-\vec{v}$. The energy a particle gains when moving from upstream to downstream and being
  reflected in the magnetic field is proportional to the velocity of the downstream gas in the
  laboratory frame, $\vec{V} = \vec{u}-\vec{v}$. The cycle between up- and downstream has to be
  encountered many times for an acceleration to TeV energies. After~\cite[p.153]{gaisser_book}.} 
\label{fig:fermi_acc}
\end{figure}

The original {\it Second Order Fermi Mechanism}, a theory on the acceleration of cosmic rays, was
already proposed in the 1940s~\cite{fermi}; it implied, however, the drawback of being too
inefficient to allow for an acceleration to high energies. An alternative version of the model was
formulated by several authors in the late 1970s~\cite{fermi1,fermi2,fermi3,fermi4}: One considers an
accumulation of gas through which a shock front moves (caused by a star explosion, for example) with
a velocity $\vec{u}$, see Figure~\ref{fig:fermi_acc}, where the gas is indicated as the light-blue
background. The gas in front of the shock is denoted as \lq\lq upstream\rq\rq\ and is considered to
be at rest, whereas the shocked gas is denoted \lq\lq downstream\rq\rq\ and has a velocity
$-\vec{v}$ relative to the shock front. Therefore, its velocity in the laboratory frame is $\vec{V}
= \vec{u}-\vec{v}$.  It can be shown that the net energy gain which a particle can receive by moving
from upstream to downstream and being reflected by irregularities of the magnetic field in the gas,
$\frac{E^{\prime} - E}{E} = \frac{\Delta E}{E}$, is proportional to the relative velocity between
shocked and unshocked gas, $\frac{V}{c} = \beta$:

\begin{equation*}
\frac{\Delta E}{E} \propto \beta. 
\end{equation*}

The acceleration mechanism is therefore called {\it First Order Fermi Mechanism}. The particle has
to run through the cycle between upstream and downstream gas several 100 times until it is
accelerated to TeV energies. \\
It turns out that the spectral index $\gamma$ for particles accelerated by this mechanism is
independent of the absolute magnitude of the gas velocities, but depends instead only on the ratio
of the upstream and downstream velocities, and can be approximated as $\gamma \approx
2.1$. The flux at the source should therefore have the same energy dependence for all sources
inside which this mechanism holds. \\ 
The maximum energy which an accelerated particle can reach through this mechanism depends on the
lifetime of the shock front and on the strength of the magnetic fields involved; however,
an acceleration to energies above $\sim 100$\,TeV -- 1\,PeV is hard to predict applying the model to the 
known, mainly galactic, sources. 

\subsubsection{Top-Down Scenarios}

A different approach to explain the detection of cosmic particles with energies $>10^{20}$\,eV are 
so called \lq\lq top-down\rq\rq\, scenarios. In these theories, instead of being accelerated by cosmic
objects, the cosmic rays are decay products of super-heavy big-bang relics, which would have
masses at the GUT-scale of $10^{24} - 10^{25}$\,eV. These theories have neither been proven nor
disproven, but a fact that seems to contradict them is that one would expect a large number of ultra
high-energy gamma rays to be produced in these decays, while studies on the composition of the 
highest energy cosmic particles conclude that these do not contain of a large fraction of
photons~\cite{agasa2,ave}. 

\subsection{Generation of Neutral Particles}\label{sec:beam_dump}

Both photons and neutrinos have the advantage compared to cosmic rays that they are electrically
neutral and therefore do not undergo deflection in the galactic or extragalactic magnetic
fields. Thus, they point back directly to the source where they have been produced. \\
There are two different mechanism for the production of high-energy gamma rays: Via the decay of
neutral pions produced in hadronic interactions, or by electromagnetic interactions, via inverse
Compton scattering or bremsstrahlung. Neutrinos, on the other hand, can only be generated in
hadronic interactions, and therefore their detection would be a direct proof for hadron acceleration,
e.g.~according to the Fermi mechanism.   

\subsubsection{Hadronic Interactions: The Beam Dump Model}

The generation of high-energy neutral particles, both neutrinos and photons, in hadronic
interactions can be explained by the {\it beam dump model}. This model borrows its name from
accelerator physics, where the particle beam is \lq\lq turned off\rq\rq\ by deflecting it
into a massive target, the beam dump. Hitting the dump, the particles interact with the target
matter and produce a large number of secondaries, most of which are absorbed in the dump. \\ 
The particles in the cosmic \lq\lq beam\rq\rq\ are the high-energy charged cosmic rays,
generated as explained in Section~\ref{sec:fermi}. The cosmic beam dump, and
this is the crucial difference to a terrestrial beam dump, consists of a diffuse gas or
plasma. Therefore, the range of the mesons produced in the interactions of the \lq\lq beam\rq\rq\ hadrons
is long enough to allow them to decay before they are absorbed, yielding 
neutrinos or photons as decay products. \\ 
The most common mesons produced in such a beam dump are charged and neutral pions. The
charged $\pi^{\pm}$ produce neutrinos in their decay chains, while the neutral $\pi^0$ produces
photons (or, to a very small fraction, electrons). The production and decay chains yielding
neutrinos (left) and photons (right) are shown in equation~(\ref{eq:p_decay}). No distinction has
been made here between particles and anti-particles.

\begin{alignat}{6}\label{eq:p_decay}
p + p / \gamma \to & \, \pi^{\pm} + & X & \hspace{4cm} p + p / \gamma \to & \pi^0 + & X &   \\
   & \hookrightarrow \mu & + & \nu_{\mu} & \hspace{0.5mm} \hookrightarrow \gamma & + & \gamma \notag \\
   & & \hookrightarrow & \,e + \nu_e + \nu_{\mu}  & \qquad & & \notag
\end{alignat} 

A charged pion decays in 99.99\% of all cases into a muon and a muon
neutrino; the muon itself decays into an electron, an electron neutrino and another
muon neutrino. The $\pi^0$ decays in 98.8\% of all cases into 2 photons, or else, into an
electron-positron pair and a photon. \\
For the neutrino production chain, it can be seen that for each decaying $\pi$ meson, three 
neutrinos are produced, with a ratio $\nu_e : \nu_{\mu} : \nu_{\tau} = 1 : 2 : 0$. Due to neutrino 
oscillations, however, one expects a ratio of $\nu_e : \nu_{\mu} : \nu_{\tau} = 1 : 1 : 1$ on
Earth~\cite{nu_osc}.  

\subsubsection{Electromagnetic Interactions}

\begin{figure}[h] \centering
\includegraphics[width=6cm]{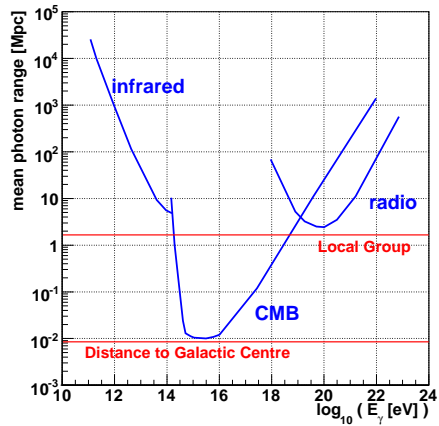}
\caption[Mean photon range over energy]
{Mean photon range over photon energy, and the cosmic interaction partners. As an orientation
of the length scale, the distance to the Galactic Centre and the size of the Local Group are shown
as well. After~\cite{kouchner}.} 
\label{fig:photon}
\end{figure}

Contrary to neutrinos, which are only produced in hadronic interactions, high-energy gamma rays can
also be generated in electromagnetic interactions: Electrons (and po\-si\-trons) can be accelerated to
very high energies, by the Fermi mechanism described above in Section~\ref{sec:fermi} or in
electromagnetic fields; those electrons produce high-energy gamma rays via bremsstrahlung in the
medium that surrounds the source or by inverse Compton scattering, when they transfer a
part of their energy to an ambient photon, which then leaves the source as a high-energy gamma
ray. \\ 
However, the fact that gamma rays interact electromagnetically also limits their range, because 
they interact with photons in the interstellar medium. The main interaction
partners of high-energy gamma rays are photons in the infrared or radio region, and photons from the
Cosmic Microwave Background. The latter have a mean energy of the order of $10^{-4}$\,eV, so that the
cross section for the production of an electron-positron-pair is maximal when the high-energy
gamma has an energy of $\sim10^{15}$\,eV. For this energy the mean photon range reaches its
minimum of only $\sim10$\,kpc, which is approximately the distance from Earth to the 
Galactic Centre. Figure~\ref{fig:photon} displays the correlation between the photon
range and the photon energy. \\
Neutrinos, on the other hand, do not interact electromagnetically, and their range is not limited
by any interaction partners in the interstellar medium. They could thus be used as messengers even 
for very distant sources. 

\section{Sources of High-Energy Cosmic Particles}\label{sec:nu_sources}

Generally speaking, all galactic or extragalactic objects where large amounts of energies
are released are possible sources of high-energy cosmic particles. The candidate sources listed here
have been observed by the photons they emit, which can have all possible wavelengths from 
radio over visible and X-ray up to wavelengths of $10^{-21}$\,m for gamma rays with energies in the 
range of around 1\,PeV, the highest energies which can be measured by ground based gamma ray
telescopes using the Imaging Atmospheric Cherenkov Technique, like HESS~\cite{hess2}.
Gamma rays of even higher energies might have been detected already by air shower experiments, but
at these energies, there exists no possibility to clearly identify the type of primary which induced
the air shower. \\  
Gamma rays with energies in the GeV range or above cannot be produced any more by thermal processes, 
they must originate from one of the acceleration mechanisms described above in 
Section~\ref{sec:beam_dump}. The sources of such gamma rays are therefore of special interest for 
neutrino telescopes, as they could be sources for high-energy neutrinos as well, if the acceleration
mechanism is of hadronic nature. Recently, the gamma ray experiment HESS has found a TeV gamma ray 
source with a spectrum favouring a hadronic acceleration process~\cite{hess}. That this is really
the case can be proven directly only by measuring neutrinos from the same source. \\
The most important cosmic accelerators, from the point of view of neutrino detection, are described
in the following subsections. 

\subsection{Supernova Remnants}

Supernova Remnants are expected to be the main sources of cosmic neutrinos below 1\,PeV. A
Supernova Remnant is the leftover of a Supernova, the explosion of a massive star at the end of its
life. For a short time, the light of this explosion can outshine a whole galaxy. Depending on the
mass of the star, the residues of the explosion can either form a neutron star or a black hole. 
In the common shell-type Supernova Remnants, photons from radio to TeV gamma rays are
emitted from an expanding shell. As the recently measured photon energy spectra seem to favour
hadronic acceleration processes~\cite{hess}, Supernova Remnants are very strong candidates for
high-energy neutrinos.  \\ 
In many cases, a pulsar emerges from a Supernova explosion. Pulsars are stars with a strong magnetic 
field; they are sources of high-energy gamma rays but the acceleration mechanism is not yet 
understood. If protons are accelerated inside the magnetic field, neutrinos are produced as well. \\
It should be mentioned that during the Supernova explosion itself, a large number of neutrinos are
produced, but with an energy range of only a few MeV, and thus not
detectable for a neutrino telescope like ANTARES. 

\subsection{Gamma Ray Bursts}

Gamma Ray Bursts (GRBs) are probably the brightest flashes that exist in the universe. They last only
a few seconds and produce an amount of gamma rays which outshines all other gamma sources in
the universe for the duration of the explosion. \\
At present there exist several different GRB models. Some GRBs, but not all, can be associated with
an extreme Supernova, a so called Hypernova. A Hypernova is the explosion of a rapidly rotating,
very massive star, which collapses into a rapidly rotating black hole surrounded by an accretion
disk. Bursts of gamma rays are produced perpendicular to the accretion disc; this
is what is detected as the GRB. \\
GRBs which cannot be associated with a Supernova might be caused by the merging of a binary system 
of a neutron star and a black hole, two neutron stars or two black holes. This merging would again
lead to the formation of a black hole, and an accretion disc with bursts, in the same way as for the
Hypernova. \\ 
The exact nature and mechanism of the acceleration of particles inside the bursts are still objects of
speculations; it is however expected that, associated with the gamma rays, a large amount of 
neutrinos is produced. If these were detected by a neutrino telescope, a lot of questions 
on the mechanisms which drive the GRBs could be clarified.

\subsection{Active Galactic Nuclei}

Active Galactic Nuclei (AGN) is the collective term for Seyfert galaxies, radio galaxies, quasars and
other high-energy astrophysical objects. These objects have in common that they consist of a galaxy 
with a super-massive black hole in its centre, and an accretion disc which builds up around the black 
hole. Highly relativistic jets, which can be up to one Mpc long, are produced perpendicular to 
the disc. In the special case of one of the jets pointing towards the observer,
the AGN is called blazar. \\ 
AGNs have been identified as emitters of high-energy gamma rays. If these gamma rays
originate from the interactions of accelerated protons in the accretion disc, neutrinos are 
produced along with them. Production of neutrinos in the jets is expected as well. 
Theoretical models of AGN are still highly speculative, and therefore the detection of neutrinos 
from an AGN would be a great step forward in the examination of these objects. 

\subsection{Microquasars}

Microquasars are interpreted as galactic binary systems emitting gamma rays in a pattern very 
similar to that of quasars, but with the scale of emission six orders of 
magnitude smaller (1\,ly, compared to $\sim 10^6$\,ly for quasars); this is where the name of these 
objects originates from. Unlike quasars, Microquasars do not consist of a whole galaxy with a black 
hole or neutron star in its centre, but of a black hole or neutron star of about a solar mass,
accompanied by a single star from which it permanently accretes mass. Therefore the
relativistic jet which builds up perpendicular to the accretion disc has a much shorter length. 
Recent calculations~\cite{distefano, antares_rates} have shown that Microquasars are very promising
candidate sources for high-energy neutrinos. Numbers for the expected neutrino rates of
selected Microquasars are provided in Section~\ref{sec:nu_fluxes}.

\chapter{Neutrino Interactions and Detection}\label{sec:interactions}

While the previous chapter was dedicated to the sources and production mechanisms of cosmic neutrinos, 
this chapter deals with the interactions of neutrinos at (or rather inside) the Earth and the
detection of these interactions. \\
Section~\ref{sec:cross_sections} gives an overview of the possible interactions and the cross
sections involved. Section~\ref{sec:nu_detection} explains the principle of a neutrino telescope,
and Section~\ref{sec:nu_fluxes} presents some predictions of neutrino fluxes and equations for the
calculation of variables defining the sensitivity of the experiment.

\section{Neutrino Interactions}\label{sec:cross_sections}

Neutrinos are electrically neutral, very light leptons\footnote{The combination of results from
direct and indirect measurements leads to mass limits of $\lesssim 2$\,eV for all three
flavours~\cite{pdg}.}. They 
interact only through the weak force (and, as all matter, through gravitation). Neutrino cross
sections are very small; for example, the $\nu_{\mu}$ on isoscalar nucleon charged current total
cross section is $6.77\pm0.14$\,fb/GeV~\cite{pdg}\footnote{The energy-dependence of the
neutrino-nucleon cross section can be considered as linear in the GeV region.}. Therefore, huge
target masses are needed for the detection, and the detection can only be indirect, i.e.~through the
detection of the interaction products. 

\subsection{Kinematic Variables of the Interaction}\label{sec:nu_variables}

\begin{figure} [h] \centering
  \begin{picture}(100,160)(-60,-105)
    \ArrowLine(-70,35)(0,0)
    \ArrowLine(0,0)(70,35)
    \Vertex(0,0){2}
    \Photon(0,0)(0,-50){5}{5}
    \Vertex(0,-50){2}
    \ArrowLine(-70,-50)(0,-50)
    \ArrowLine(0,-50)(70,-35)
    \ArrowLine(-70,-55)(0,-55)
    \ArrowLine(-70,-60)(0,-60)
    \ArrowLine(0,-55)(70,-55)
    \ArrowLine(0,-60)(70,-60)
    \Text(-105,15)[l]{$p = ( E, \vec{p} )$}
    \Text(-85,35)[l]{$\nu$}
    \Text(55,15)[l]{$p^{\prime} = ( E^{\prime}, \vec{p}^{\,\prime} )$}
    \Text(75,35)[l]{$l^{\prime}$}
    \Text(-30,-20)[l]{$-Q^2$}
    \Text(15,-25)[l]{$W, Z$}
    \Text(-130,-55)[l]{$P = ( M, \vec{0} )$}
    \Text(80,-45)[l]{$P^{\prime} = ( M^{\prime}, \vec{P}^{\prime} )$}
  \end{picture}
  \caption[Kinematic situation of deep inelastic scattering]{The kinematic situation of the
  interaction: a neutrino with four-momentum ${p = ( E, \vec{p}\,) \approx ( |\vec{p}|, \vec{p}\,)}$ 
  interacts with a nucleon with
  mass $M$ which is at rest before the interaction; the final products are a lepton $l^{\prime}$ and
  a hadronic shower of momentum $\vec{P}^{\prime}$ and mass $M^{\prime} > M$. } 
  \label{fig:kinematics}
\end{figure}
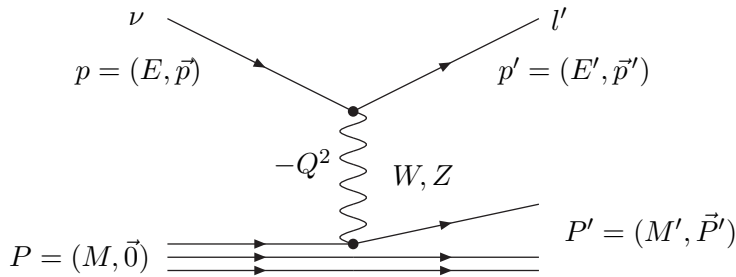

The reaction of main interest for the neutrino detection in water or other large volumes is the
deep inelastic scattering of a neutrino with a target matter nucleon. Figure~\ref{fig:kinematics}
gives a schematic view of the kinematic situation of the interaction. 

To describe the interaction mathematically, one generally uses the kinematic variables $Q^2, x$ and
$y$. $Q^2$ is the negative squared four-momentum transfer between the incoming and the outgoing
lepton:  
 
\begin{equation}
Q^2 := - (p - p^{\prime})^2.
\end{equation}
For high-energy neutrinos, the interaction partner can be considered to be at rest, so that
the rest frame of this particle is equivalent to the laboratory system. For this case, the {\it
  Bjorken variable} $x$ is defined as 

\begin{equation}
x := \frac{Q^2}{2M(E-E^{\prime})} \quad .  
\end{equation}
For elastic interactions, $Q^2 = 2M(E-E^{\prime})$, so that $x \equiv 1$. For inelastic
reactions, however, $0 < x < 1$, because in this case ${Q^2 = 2M(E-E^{\prime}) + M^2 -
M^{\prime2}}$. $x$ is therefore a measure for the inelasticity of the interaction~\cite[p.91]{rith}.
The hypothetical case of $x \equiv 0$ would be reached if all energy of the neutrino
went into the hadronic shower, such that $E^{\prime} = 0$. $M^{\prime}$ would then reach its
maximum value of ${M^{\prime} = \sqrt{M^2 + 2ME}}$, and $Q^2 = 0$. Therefore, the larger 
$M^{\prime}$, the smaller $x$. The high-energy neutrino interactions which are subject of this
study are {\it deep inelastic}, and characterised by $x \ll 1$. \\ 
The {\it Bjorken variable} $y$ is defined in the laboratory system as 

\begin{equation}
y := \frac{E-E^{\prime}}{E} \quad .
\end{equation}
$y$ is therefore the relative energy transfer from the neutrino to the hadronic system. \\

\subsection{Interaction Types and Cross Sections}\label{sec:interaction_types}

Neutrino interactions with matter are to a large extent dominated by the inelastic scattering of the
neutrino on a target nucleon, for which the cross section is generally several orders of magnitude
larger than for the interaction of a neutrino with an electron. An exception to this, the Glashow
resonance, is discussed below. \\ 
Neutrinos can interact with a nucleon by exchanging a charged $W^{\pm}$ or a neutral $Z^0$
boson. The first interaction type is called {\it Charged-Current interaction},
abbreviated CC interaction, while the second type is called {\it Neutral-Current
  interaction}, abbreviated NC interaction. Because of the electro-weak coupling terms which must be
taken into account for the neutral current interactions,  the NC cross sections are only about one
third of the CC cross sections. \\ 
There exist several software packages for the calculation of neutrino cross sections
which have been compared in detail by~\cite{gandhi}. The authors find that the different models
agree well up to $\sim 10^7$\,GeV, while for the highest considered energies of $E_{\nu} =
10^{20}$\,eV, the uncertainty between the different models is cited as a factor of $2\pm1$. \\ 
The differential cross section for the CC interaction ${\nu_{\mu} + N \to \mu^- + anything}$, where
$N$ is an isoscalar target nucleon\footnote{i.e.~the target contains the same number of protons and
neutrons.} can be expressed in leading order as~\cite{gandhi} 

\begin{equation}\label{eq:cs_cc}
\frac{d^2\sigma}{dxdy} = \frac{2 G_F^2 M E_{\nu}}{\pi} \left( \frac{M_W^2}{Q^2 + M_W^2} \right)^2 
[ xq(x,Q^2) + x\bar{q}(x,Q^2)(1 - y)^2]
\end{equation}

where $G_F = 0.117 \cdot 10^{-4}$\,GeV$^{-2}$ is the Fermi coupling constant (with $\hbar c=1$), $M$
is the mass of the target, $E_{\nu}$ the neutrino energy, $M_W$ the mass of the $W$ boson and 

\begin{align}\label{eq:q_cc}
q(x,Q^2) = \frac{1}{2} \left( u_v(x,Q^2) + d_v(x,Q^2) \right) 
+ \frac{1}{2} \left( u_s(x,Q^2) + d_s(x,Q^2) \right) \notag \\
 + s_s(x,Q^2) + b_s(x,Q^2)
\end{align}

and

\begin{align}\label{eq:antiq_cc}
\bar{q}(x,Q^2) = & \frac{1}{2} \left( u_s(x,Q^2) + d_s(x,Q^2) \right) 
+ c_s(x,Q^2) + t_s(x,Q^2) 
\end{align}

are the quark and anti-quark distribution functions, respectively, consisting of the distributions of
the valence (indexed $v$) and sea (indexed $s$) quark flavours of the target. \\
For the NC process ${\nu_{\mu} + N \to \nu_{\mu} + anything}$, the expression for the differential 
cross section is~\cite{gandhi}

\begin{equation}\label{eq:cs_nc}
\frac{d^2\sigma}{dxdy} = \frac{G_F^2 M E_{\nu}}{2 \pi} \left( \frac{M_Z^2}{Q^2 + M_Z^2} \right)^2 
[ xq^0(x,Q^2) + x\bar{q}^0(x,Q^2)(1 - y)^2]
\end{equation}

where $M_Z$ is the mass of the $Z$ boson, and the NC quark and anti-quark distribution functions are,
respectively, 

\begin{alignat}{1}\label{eq:q_nc}
q^0(x,Q^2) & = \left( \frac{1}{2} \left( u_v(x,Q^2) + d_v(x,Q^2) \right) 
+ \frac{1}{2} \left( u_s(x,Q^2) + d_s(x,Q^2) \right)\right) \cdot ( L_u^2 + L_d^2 ) \notag \\
& + \frac{1}{2} \left( u_s(x,Q^2) + d_s(x,Q^2) \right) \cdot ( R_u^2 + R_d^2 ) \notag \\
& + \left( s_s(x,Q^2) + b_s(x,Q^2) \right) \cdot( L_d^2 + R_d^2 ) \notag \\
& + \left( c_s(x,Q^2) + t_s(x,Q^2) \right) \cdot( L_u^2 + R_u^2 ) 
\end{alignat}

and

\begin{alignat}{1}\label{eq:antiq_nc}
\bar{q}^0(x,Q^2) & = \left( \frac{1}{2} \left( u_v(x,Q^2) + d_v(x,Q^2) \right) 
+ \frac{1}{2} \left( u_s(x,Q^2) + d_s(x,Q^2) \right)\right) \cdot ( R_u^2 + R_d^2 ) \notag \\
& + \frac{1}{2} \left( u_s(x,Q^2) + d_s(x,Q^2) \right) \cdot ( L_u^2 + L_d^2 ) \notag \\
& + \left( s_s(x,Q^2) + b_s(x,Q^2) \right) \cdot ( L_d^2 + R_d^2 ) \notag \\
& + \left( c_s(x,Q^2) + t_s(x,Q^2) \right) \cdot ( L_u^2 + R_u^2 ).
\end{alignat}

\begin{figure} \centering
\includegraphics[width=10cm]{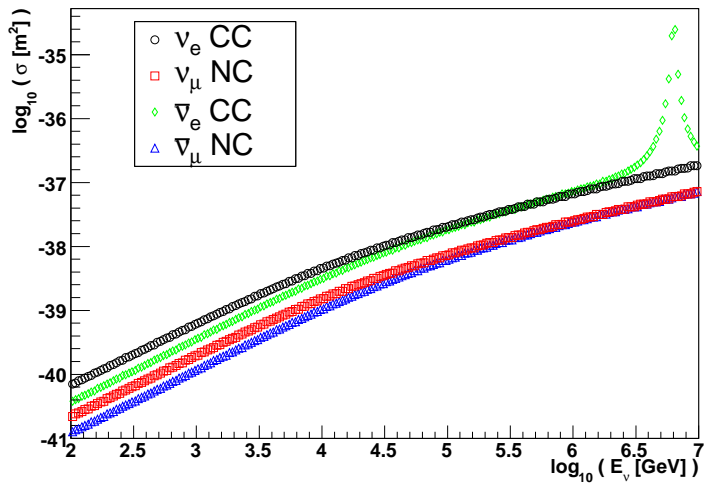}
\caption[Neutrino cross sections]{Cross sections for neutrino interactions producing showers:
  neutral current interaction of $\nu_{\mu}$ and $\bar{\nu}_{\mu}$ and charged current interaction
  of $\nu_e$ and $\bar{\nu}_e$. The resonance for $\bar{\nu}_e$ at 6.3\,PeV can be clearly seen. The
  plot was generated using the values of the CTEQ6 parton distribution functions inside the ANTARES 
  simulation software (see Appendix~\ref{sec:genhen}).} 
\label{fig:cross_sections}
\end{figure}

The chiral couplings $L_{u,d}$ and $R_{u,d}$ are defined as

\begin{alignat}{3}
& L_u = 1 - \frac{4}{3} \sin^2 \theta_W, \qquad & L_d = & -1 + \frac{2}{3} \sin^2 \theta_W \notag \\
& R_u =   - \frac{4}{3} \sin^2 \theta_W, \qquad & R_d = & \frac{2}{3} \sin^2 \theta_W
\end{alignat}

where $\sin^2 \theta_W = 0.231$ is the weak-mixing angle. From equations (\ref{eq:cs_cc}) and
(\ref{eq:cs_nc}), the total cross sections are determined using standard parton distribution sets like 
CTEQ6~\cite{cteq}.\\
For $E_{\nu} > 10^6$\,GeV, the largest contribution to the cross section by far comes from the sea
quarks, and the role of the valence quarks becomes small. The limit of $E_{\nu} \to \infty$
corresponds to setting the valence terms in 
equations (\ref{eq:q_cc}),(\ref{eq:antiq_cc}),(\ref{eq:q_nc}) and (\ref{eq:antiq_nc}) to zero; the
distribution functions for quarks and anti-quarks are then equivalent, and the cross sections of
the anti-neutrino interactions coincide with those of the neutrino interactions.
\\[0.5cm]
The total neutrino cross sections are dominated to a very large extent by the deep inelastic
scattering reactions of the neutrino on the nucleon. There is, however, one exception: 
For a $\bar{\nu}_e$ with an energy of around 6.3\,PeV, the cross section is dominated by the 
{\it Glashow resonance}. At this energy, a $W$ boson is produced resonantly by the CC interaction
of the $\bar{\nu}_e$ with an electron of one of the target molecules, in the channel
$\bar{\nu}_e + e^- \to W \to anything$.\\
The cross sections for some selected shower-producing channels ($\nu_{\mu}$ NC,
$\bar{\nu}_{\mu}$ NC, $\nu_e$ CC and $\bar{\nu}_e$ CC) are shown in
Figure~\ref{fig:cross_sections}. The neutral current interaction of $\nu_e$ is expected to be
equal to the one of $\nu_{\mu}$.  

\begin{figure}[h] \centering
\includegraphics[width=7.4cm]{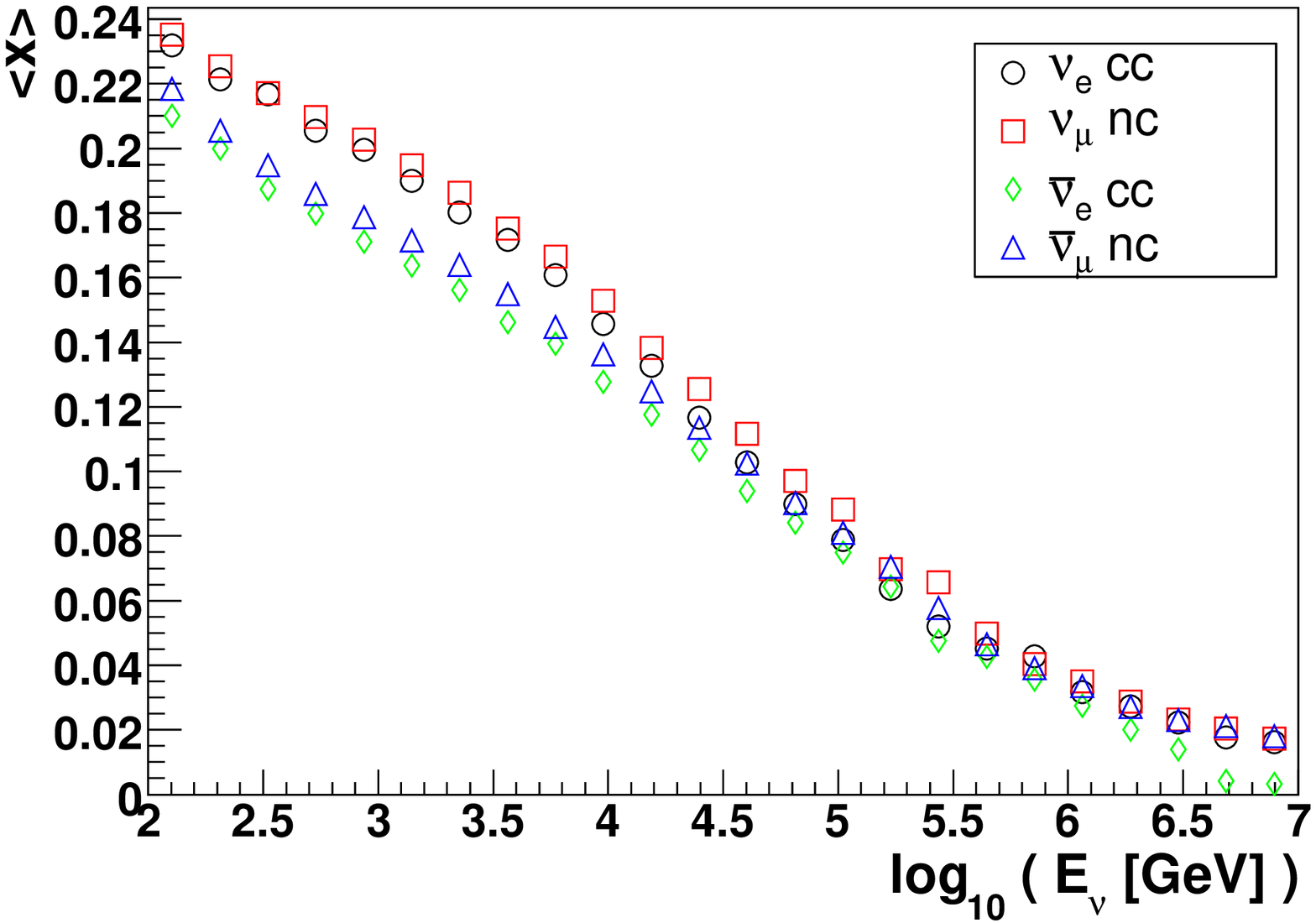}
\includegraphics[width=7.4cm]{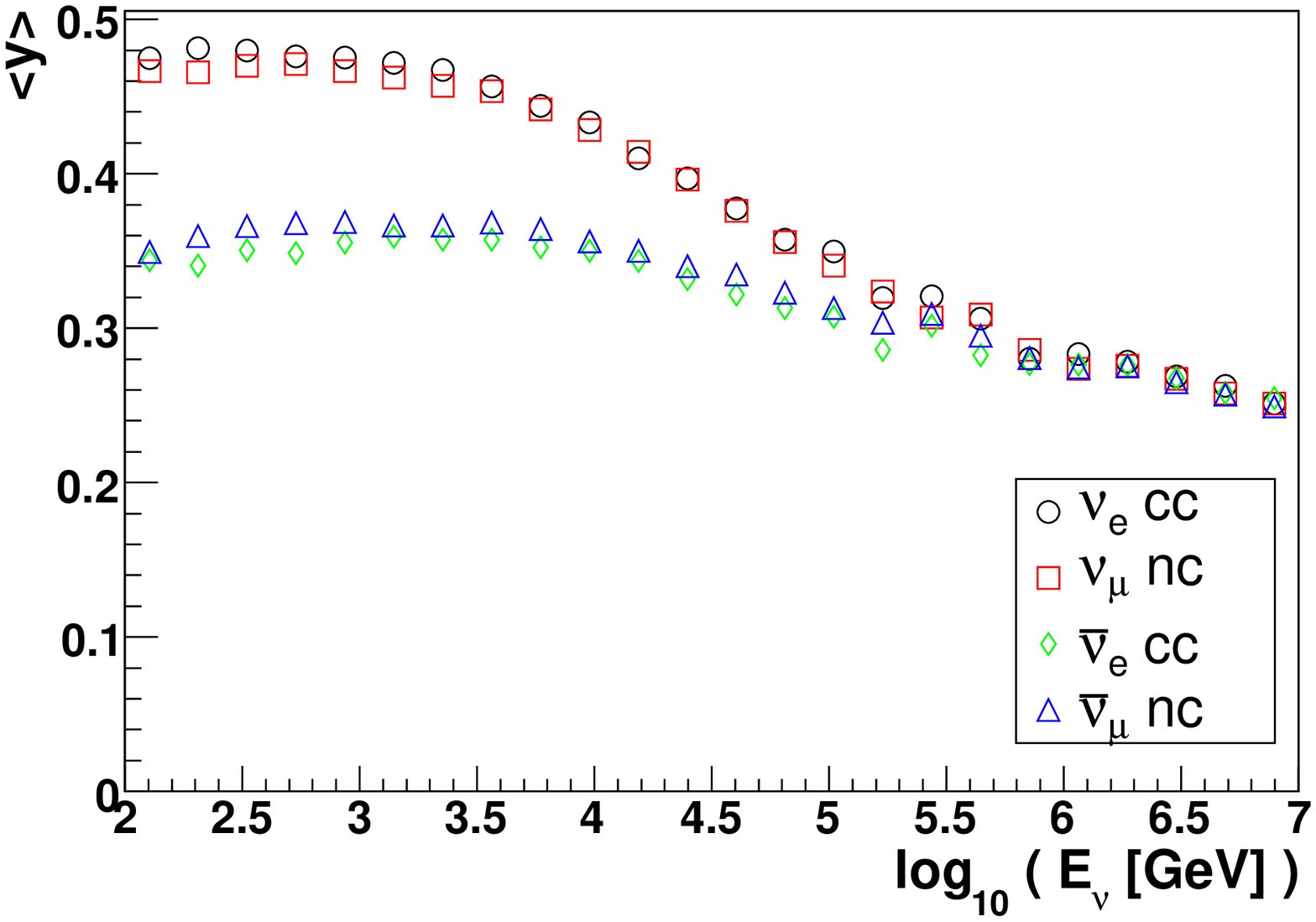}
\caption[Mean values of $x$ and $y$]{Mean values of the Bjorken variables $x$ and
  $y$ as a function of the neutrino energy. The values were calculated with the ANTARES simulation
  (see Appendix~\ref{sec:genhen}).}   
\label{fig:bjorken}
\end{figure}

The mean values of $x$ and $y$ are shown in Figure~\ref{fig:bjorken} as functions of the neutrino
energy, for different neutrino interaction types. The values were calculated with the ANTARES
simulation software (see Appendix~\ref{sec:genhen}) using the CTEQ6 parton distribution function
set~\cite{cteq}. One can see that the mean value of $x$ decreases for higher energies; this poses a 
challenge to the prediction of neutrino interactions at very high energies, because of the
increasing uncertainty of the parton distributions extracted from accelerator experiments. In the case
of the Glashow resonance, $x$ is set to zero by the simulation software. \\
An overview of the interaction products and their characteristics is given in 
Section~\ref{sec:event_types}.

\section{Neutrino Detection}\label{sec:nu_detection}

\subsection{The Cherenkov Effect}
\begin{figure}[h] \centering
\includegraphics[width=15cm]{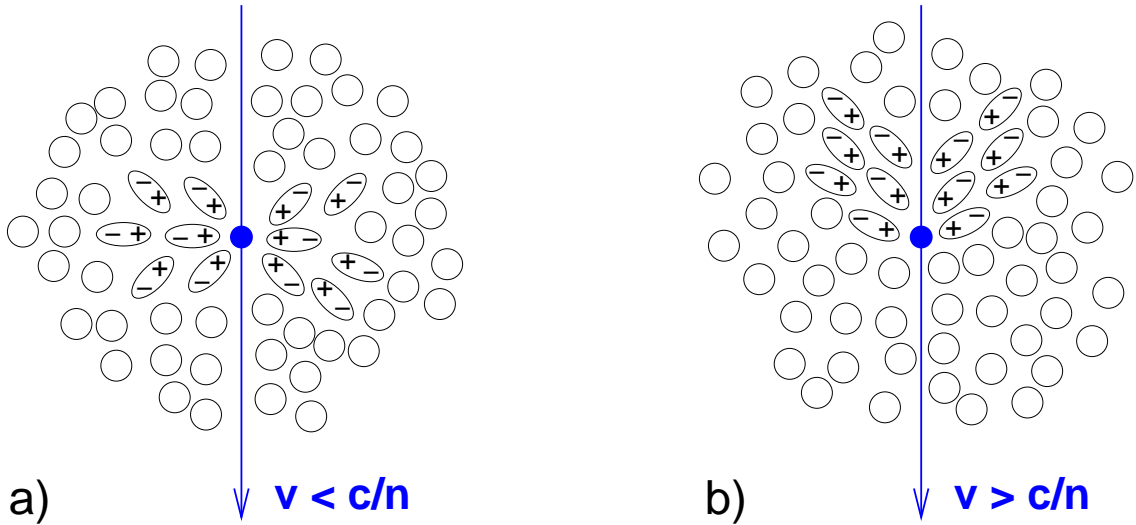}
\caption[Schematic explanation of the Cherenkov effect]{a) If the speed $v$ of a charged particle 
  traversing a medium is smaller than the speed of light in the medium, $c/n$, the dipoles induced 
  by polarisation are distributed symmetrically around the particle, and the overall
  dipole moment is zero. b) If $v > c/n$, the dipole symmetry is broken, and radiation is emitted
  along the Cherenkov cone. After~\cite[p.239]{grupen}.}
\label{fig:ch_effect}
\end{figure}

As neutrinos are electrically neutral, they can only be detected by their interaction
products. Because of the small neutrino cross sections, neutrino telescopes use a large target mass
to register as many neutrino interactions as possible inside, or close to, the detector. \\
The first generation of neutrino experiments (see Section~\ref{sec:nu_history}),
aiming at the detection of low energy neutrinos, used liquid targets of a precisely
known chemical composition, and literally counted the number of molecules which had undergone an
interaction with a neutrino. A famous example for this type of detector is the Homestake
experiment~\cite{homestake}. \\
High-energy neutrino telescopes like ANTARES, AMANDA or others described in
Chapter~\ref{ch:experiment}, along with other, lower-energy threshold experiments dedicated to the 
detection of solar and atmospheric neutrinos, like SuperKamiokande~\cite{superk} or SNO~\cite{SNO}, 
are based on a different detection principle. This type of experiment detects the
interaction products making use of the {\it Cherenkov effect}~\cite{cherenkov}: \\  
When a charged particle moves through a medium, it polarises the atoms along
its trajectory, turning them into electric dipoles, as shown symbolically in
Figure~\ref{fig:ch_effect}a. As long as the particle's speed $v$ is smaller than the speed of light
in the medium, $c/n$ ($n$ being the refraction index of the medium) the dipoles are orientated
symmetrically around the particle track, so that the overall dipole moment is zero and no
radiation is emitted. However, if the speed of the particle is larger than $c/n$,
the symmetry is broken, see Figure~\ref{fig:ch_effect}b, and dipole radiation is emitted along the
so-called {\it Cherenkov cone}. This radiation can be measured in photon detectors. 

The angle of the Cherenkov cone depends on the refraction index $n$ of the medium and the speed ${v =
\beta c}$ of the particle and can be calculated geometrically: During a time $t$, the particle will
move a distance $d_1 = \beta c t$. Light, however, will only move a distance $d_2 = (c/n) t$. The
relation between $d_1$ and $d_2$ determines the Cherenkov angle $\vartheta_C$:

\begin{equation}
\cos \vartheta_C = \frac{d_2}{d_1} = \frac{1}{\beta n}.
\end{equation}

For reactions of high-energy neutrinos, the interaction products have a
velocity $v \approx c$ and thus $\cos \vartheta_C \approx 1 / n$. 

\subsection{Neutrino Detection Principle}\label{sec:nu_passage}

\begin{figure}[h] \centering
\includegraphics[width=14cm]{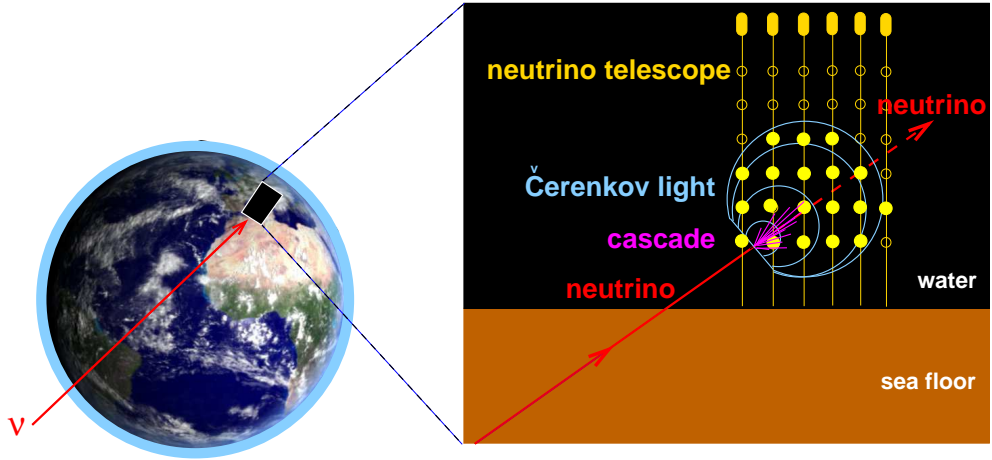}
\caption[Neutrino detection principle]{Detection principle in a neutrino telescope, shown for the
  example of a neutral current interaction: The neutrino traverses Earth and interacts somewhere
  close to the detector. The charged interaction products of the hadronic cascade produce
  superimposed Cherenkov cones, while the secondary neutrino leaves the detector without producing
  any signal.}
\label{fig:detection}
\end{figure}

Neutrino telescopes are generally situated deep undersea, underground or under ice, to suppress
the high background caused especially by secondary muons from cosmic rays, which produce
Cherenkov light in the target material as well. For the ANTARES experiment, the photomultipliers
are additionally orientated towards the ground, so that the sensitivity for particles coming from 
below, which can only be neutrinos, is enhanced. The detection principle is shown schematically in
Figure~\ref{fig:detection}. 

\begin{figure}[h]\centering
\begin{minipage}[b]{7.9cm}
\includegraphics[width=7.9cm]{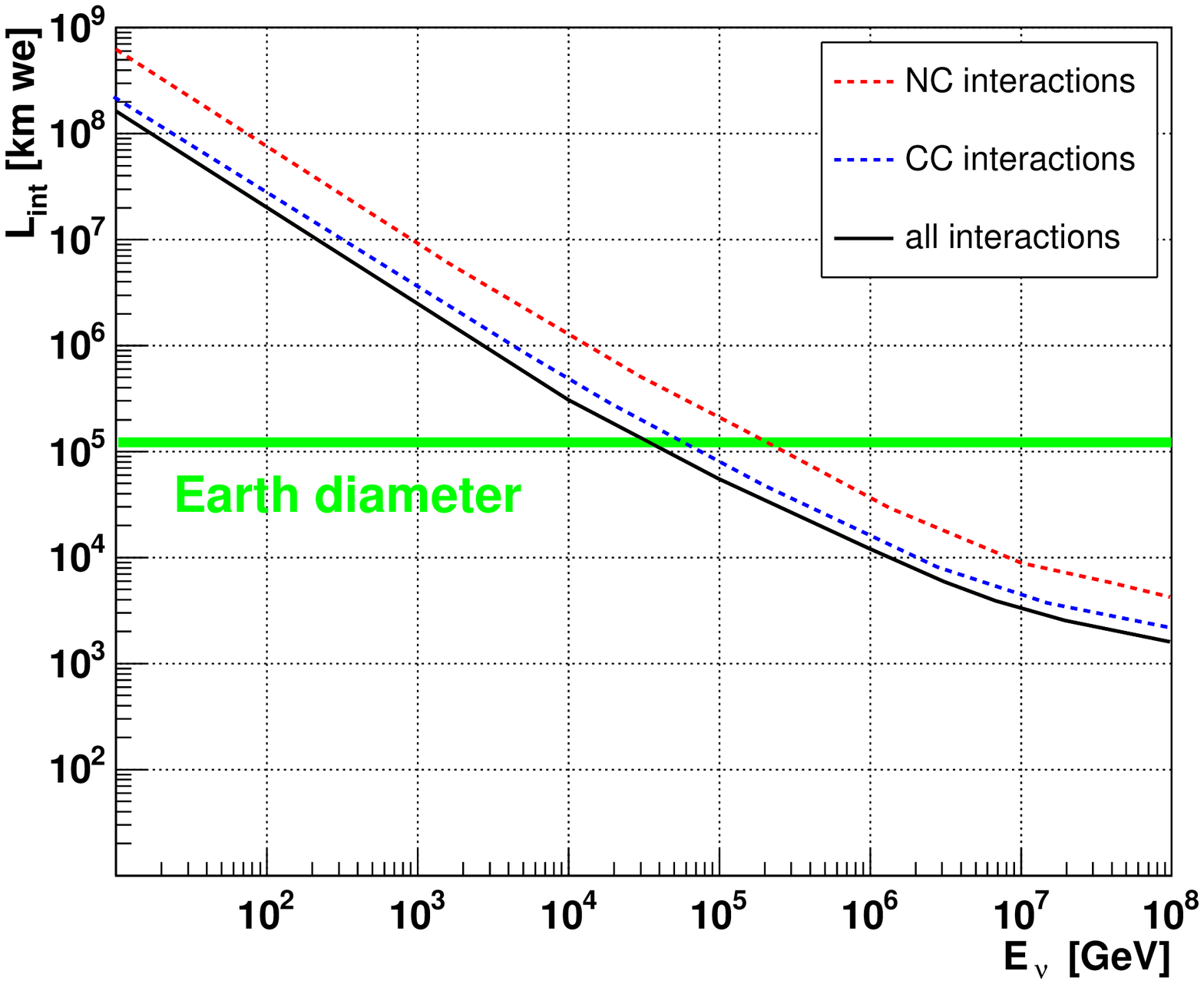}
\caption[Neutrino interaction lengths]{Interaction lengths for CC and NC neutrino-nucleon
  interactions, and total interaction length, given in km water equivalent. The diameter of the
  Earth, according to the parameterisation in Figure~\ref{fig:earth_density}, is also
  shown. After~\cite{gandhi}.} 
\label{fig:opacity}
\end{minipage}
\hspace{0.5cm}
\begin{minipage}[b]{6.3cm}
\includegraphics[width=6.3cm]{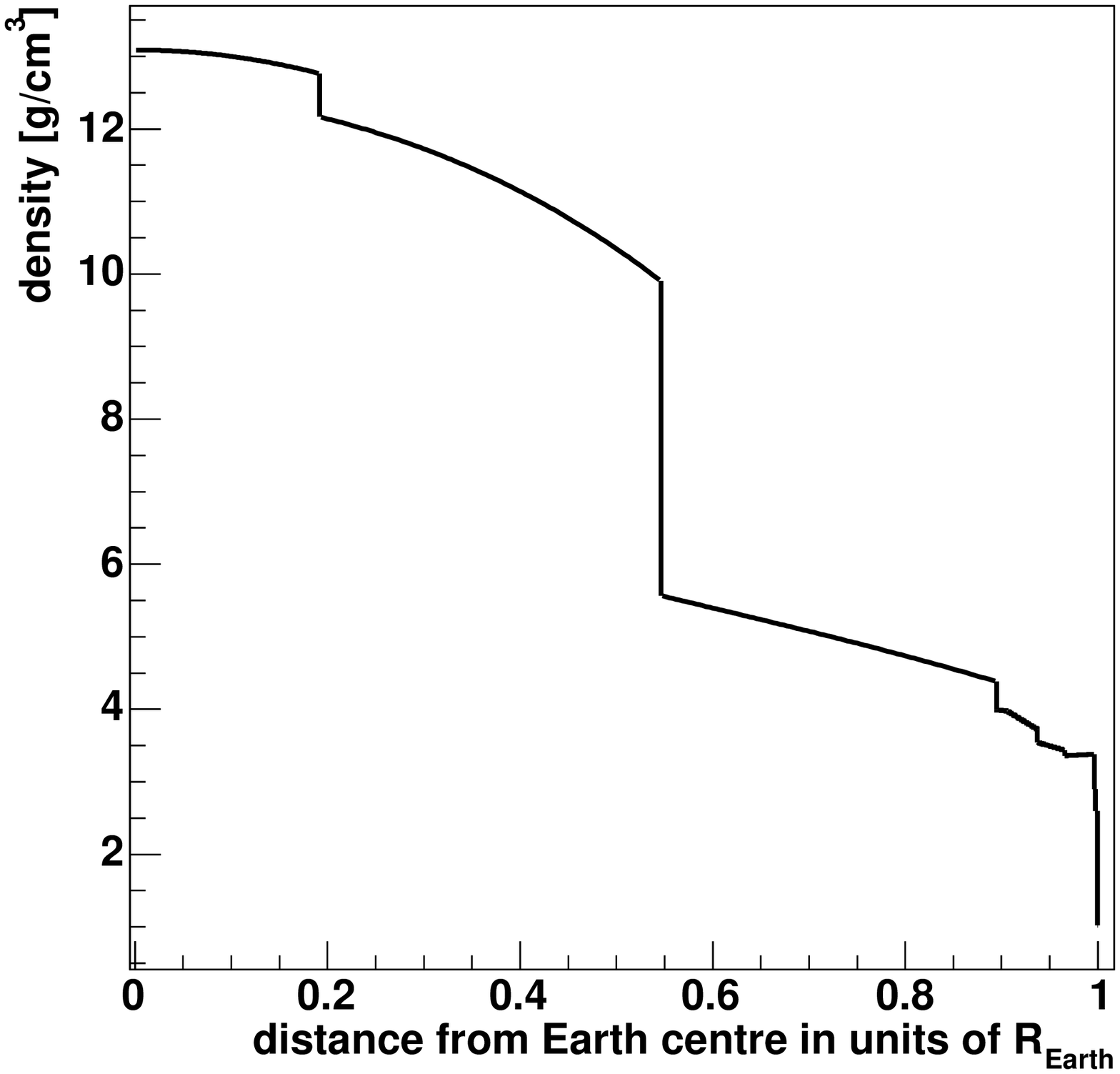}
\caption[Earth density profile]{Earth density profile, in units of the Earth radius, as
  parameterised in~\cite{earth}.}
\label{fig:earth_density}
\end{minipage}
\end{figure}

Even though the cross section of neutrino interactions is very small, for neutrino energies above
about 1\,PeV it becomes large enough to make the Earth opaque for
neutrinos. Figure~\ref{fig:opacity} shows the interaction length for neutrino-nucleon interactions 
in terms of km water equivalent, as a function of the neutrino energy. The Earth radius, in the same
units, was calculated according to the parameterised Earth density profile~\cite{earth} shown in
Figure~\ref{fig:earth_density}. One can see that for energies
above $\sim 40$\,TeV, the neutrino interaction length is smaller than the diameter of the Earth,
so that neutrinos with higher energies preferentially enter the detector at larger zenith angles. \\
Due to the smaller cross section of NC interactions, the corresponding interaction length is 
larger than the one corresponding to the CC interactions. 

\afterpage{\clearpage}

\section{Cosmic Neutrino Fluxes}\label{sec:nu_fluxes}

In general, one distinguishes between two classes of neutrino fluxes: {\it Diffuse fluxes}, 
i.e.~the inseparable superposition of neutrino fluxes from all known or hypothetical sources 
disregarding their position in space, and fluxes expected from so-called {\it point sources}, 
individual sources of neutrinos which, despite the misleading denomination, may also be 
extended. Before discussing the two types of fluxes, the variables describing the sensitivity 
of the experiments are defined.

\subsection{Effective Volume and Effective Neutrino Area}\label{sec:eff_area}

The {\it effective volume} and the {\it effective neutrino area} are variables which provide
objective measures of the sensitivity of an experiment for a selected reconstruction strategy. \\
The effective volume $V_{\eff}$ is defined as the volume within which the events have been generated,
multiplied by the fraction of events that are successfully reconstructed:

\begin{equation}
V_{\eff} = V_{gen} \cdot \frac{N_f}{N_{gen}}
\end{equation}

Here, $V_{gen}$ is the generation volume, $N_{gen}$ is the number of events which have been
generated inside $V_{gen}$ and $N_f$ is the final number of events after the reconstruction or the
quality cuts, if any have been applied. The effective volume is generally smaller for shower-type 
events than for muon-type events, because of the much shorter path length of the showers which
limits the interaction volume for detectable event. \\
The effective neutrino area is the area that a neutrino effectively \lq\lq sees\rq\rq\, when
traversing the instrumented volume. It is calculated by multiplying the effective volume $V_{\eff}$
with the number density of molecules in the target matter times the (energy dependent) neutrino
interaction cross section and the Earth penetration probability:

\begin{alignat}{3}
A_{\eff} & = & \frac{N_f}{N_{gen}} \cdot V_{gen} & \cdot \rho \cdot N_A \cdot \sigma(E) \cdot
         P_E(E,\theta) \\ 
         & = &  V_{\eff} \quad & \cdot \rho \cdot N_A \cdot \sigma(E) \cdot P_E(E,\theta).
\end{alignat}

\begin{figure}[h]\centering
\includegraphics[width=8cm]{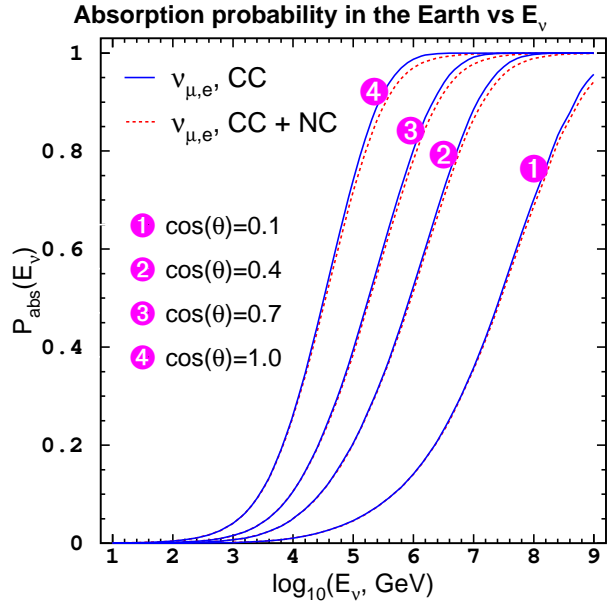}
\caption[Absorption probability in the Earth]{Absorption probability for $\nu_e$ and $\nu_{\mu}$
  when propagating through the Earth, as a function of the incident neutrino energy, for different
  incident angles. From~\cite{genhenv6}.}
\label{fig:nu_abs}
\end{figure}

In this formula, $\rho$ is the density of the medium inside which the reaction takes place,
$N_A$ is the Avogadro number (so that $\rho \cdot N_A$ is the number of molecules per unit
volume in the target), $\sigma(E)$ is the total neutrino-target interaction cross section and
$P_E(E,\theta)$ is the Earth penetration probability which depends on the zenith angle of the
incident neutrino and the neutrino energy, as demonstrated in Figure~\ref{fig:nu_abs} taken
from~\cite{genhenv6}. The Earth absorption was simulated by~\cite{genhenv6} according to the Earth
density profile shown in Figure~\ref{fig:earth_density} in the previous section. 

\subsection{Sensitivity Estimates}\label{sec:calc_diffuse_flux}

From the effective neutrino area, the sensitivity for neutrino fluxes from a given source, or
for the isotropic diffuse neutrino flux can be predicted.  Assuming a flux $\Phi$
which is constant in time and isotropic within the observed angular range, one can
calculate the expected number of events $N$ within a given energy range $\{E,E+\Delta E\}$, during
the observation period $T$, and within the angular element $\Omega$ as  

\begin{equation}
N(E,E+\Delta E) = T \Omega \int_{E}^{E+\Delta E} \Phi(E^{\prime}) A_{\eff}(E^{\prime}) dE^{\prime}.
\end{equation}

One can invert this equation by assuming that the neutrino flux follows a specific energy
spectrum. For cosmic neutrinos, one often assumes that $\Phi(E) = \phi_0 \cdot E^{-2}$, following the
predicted spectrum for the charged particle acceleration in the Fermi mechanism (see
Section~\ref{sec:fermi}). The proportionality constant $\phi_0$ can then be calculated as 

\begin{gather}\label{eq:sensitivity}
\phi_0 = \frac{N}{T \Omega \int E^{-2} A_{\eff}(E) dE} \quad .
\end{gather}

The sensitivity is then described as the limit on the neutrino flux for a number of events $N$ which
is chosen according to the Poissonian statistics for small numbers described
in~\cite{feldman}: Depending on the expected background rate, one selects the highest number of
events which, at the desired confidence level, is still compatible with the assumption of a
background-only detection. The experiment is consequently sensitive to the detection of any neutrino
flux causing a higher event rate than the assumed $N$.

\subsection{Neutrino Flux from Point Sources}\label{sec:point_sources}

\begin{figure} \centering
\includegraphics[width=7.4cm]{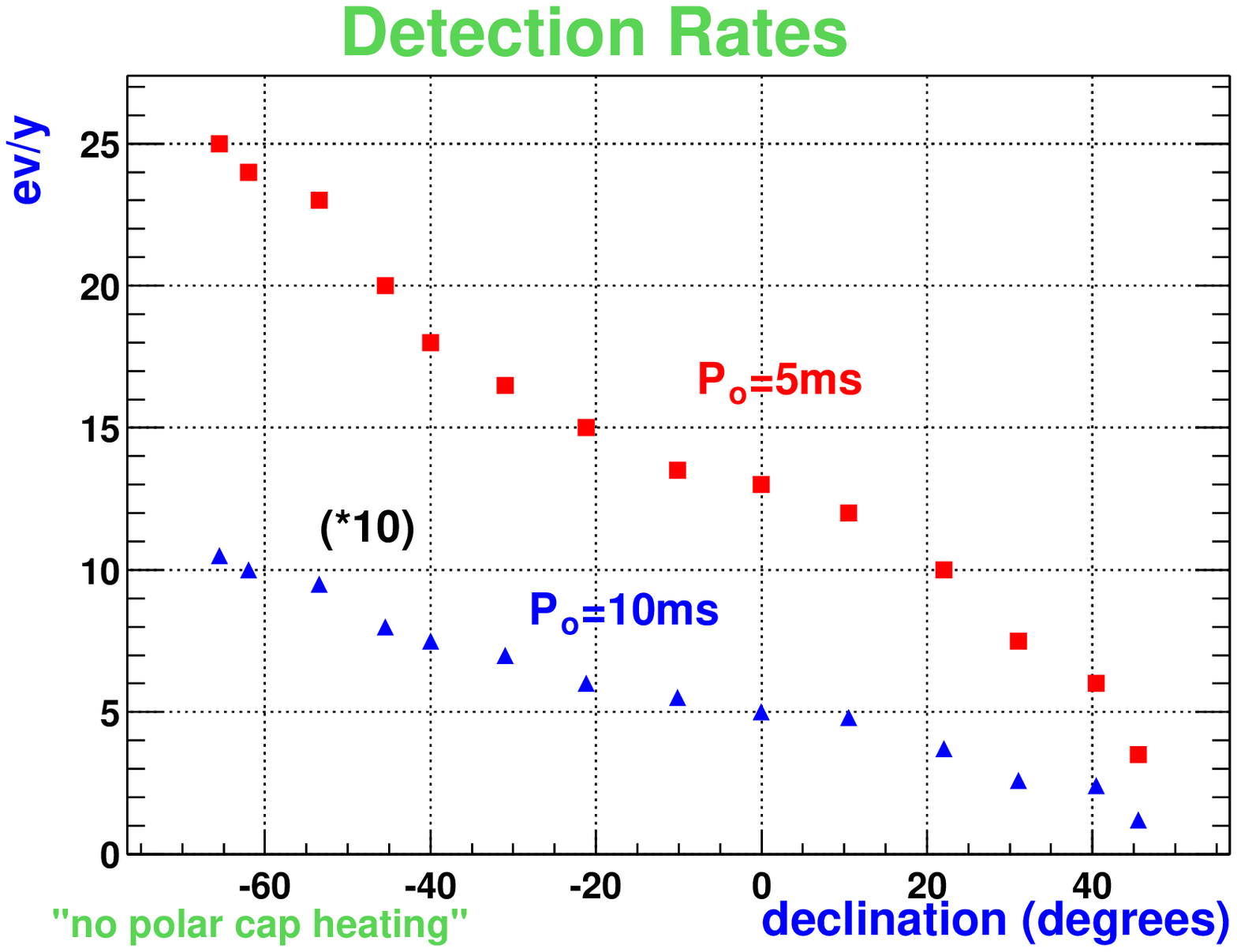}
\includegraphics[width=7.4cm]{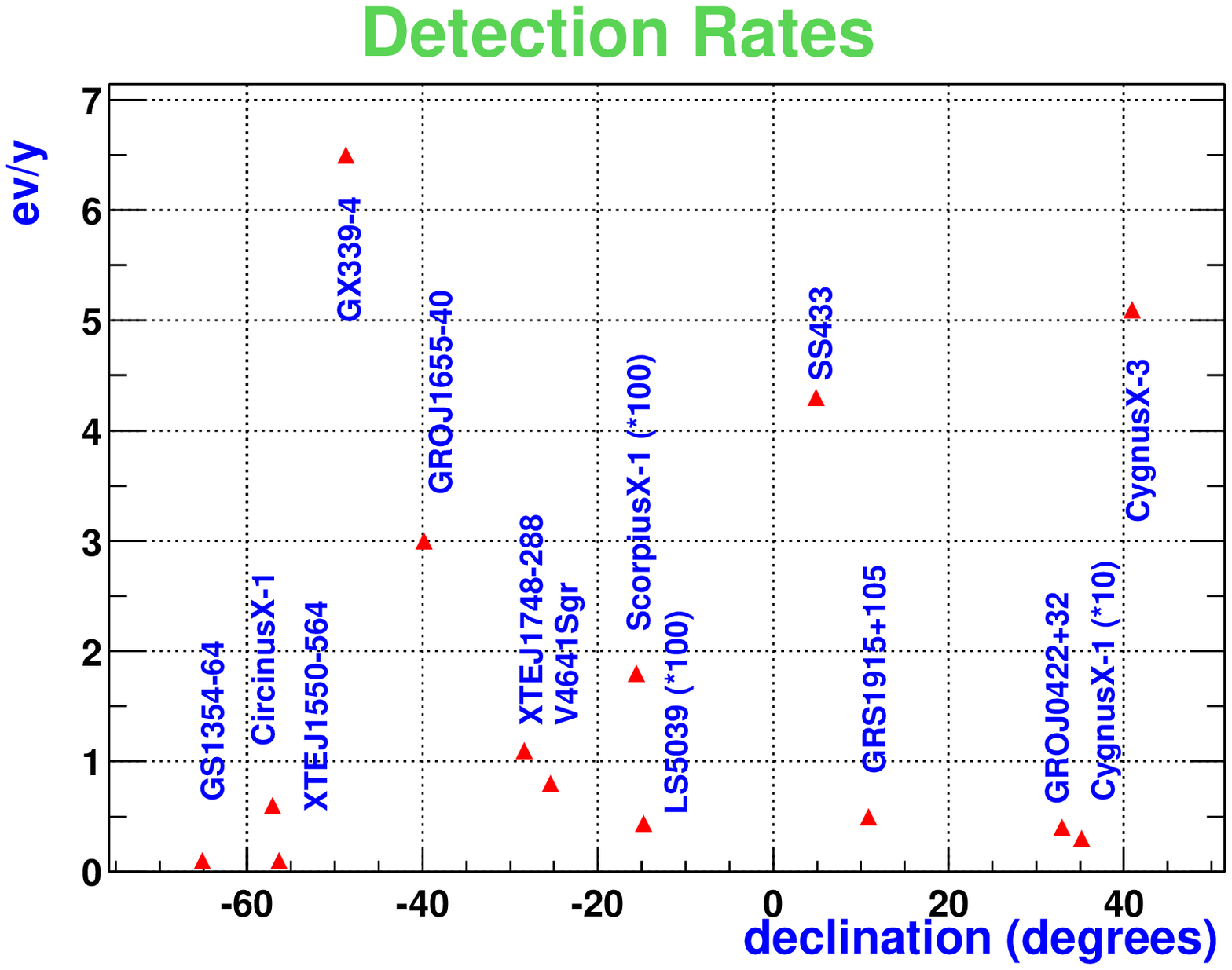}
\caption[Theoretical event rates in ANTARES for selected point sources]{Theoretical event rates per 
  year as functions of the absolute source declination in equatorial coordinates, for muons from
  galactic neutrinos, as calculated for the ANTARES experiment. Left: Event rates from a young
  Supernova Remnant at 10\,kpc distance, for a model with no polar cap heating (see text), and
  initial pulsar periods of $P_0=5$\,ms and $P_0=10$\,ms (the rates for the latter have been
  multiplied by 10). Right: Event rates for a list of known Microquasars. The rates for ScorpiusX-1 
  and LS5039 have been multiplied by 100, the rate for CygnusX-1 by 10. From~\cite{antares_rates}.} 
\label{fig:point_sources}
\end{figure}

A good angular resolution is the crucial factor for the detection of point sources, for two reasons:
Firstly, to be able to locate, or even resolve, an individual source as precisely as possible,
and secondly, to keep the background as low as possible; as the minimum size of the angular bin increases 
quadratically with the resolution, so does the rate of background collected within this angular bin.
For the ANTARES experiment, theoretical event rates for muons from galactic neutrinos have been 
estimated for selected source types in~\cite{antares_rates}. In Figure~\ref{fig:point_sources} 
taken from this article, the expected number of events from the two types of sources which are found 
to be the most promising ones in that article, namely young Supernova Remnants and Microquasars, are shown 
as examples. 
\\
The rates for the young Supernova Remnants were calculated for a theoretical source at 10\,kpc distance, 
0.1\,years after the Supernova explosion. The model used assumes that the whole star has a temperature
equal to the surface temperature (\lq\lq no polar cap heating\rq\rq ); rates for initial pulsar periods of 
$P_0=5$\,ms and $P_0=10$\,ms were calculated. For this theoretical model, the authors arrive at
predicted rates of up to 25 detected events per year. For the most promising Microquasar 
GX339-4, the authors of~\cite{antares_rates} expect a rate of 6.5 per year. \\
These predictions seem promising; however, much smaller rates are
derived from the measurements of the TeV energy gamma ray experiment HESS~\cite{hess2}:
For the two strongest TeV gamma ray sources found so far, RXJ0852 and RXJ1317, the detection rates
expected for one year of data taking in ANTARES are 0.3 and 0.1~\cite{alex}, assuming that the
neutrino rates at the source are equivalent to the photon rates. \\
In this context it should also be noted that in a recent study~\cite{aart_point_sources} on the
potential of detecting point sources in ANTARES using the reaction $\nu_{\mu},\bar{\nu}_{\mu} \to
\mu^{\pm}$ it was found that in order to arrive at 50\% probability for a $3\sigma$ discovery of an
individual source after 2 years of data taking, between 4 (for a source declination of 40$^{\circ}$)
and 8 events (for a source declination of -80$^{\circ}$) have to be detected from that source.

As the angular resolution for showers is generally poorer than for muon events, the applicability of
shower events for point source searches seems doubtful. Exact predictions on the point source
sensitivity are beyond the scope of this study and will not be discussed any further. We will only 
present a very rough estimate on the perspectives, taking into account the following points:
\begin{itemize}
\item {\bf Angular resolution:} \\
The angular resolution can be decreased
down to $2^{\circ}$ with the help of some quality cuts, as will be shown in
Chapter~\ref{sec:results}. This is still about ten times the angular resolution reached in
the reconstruction of muon events. The observational bin would therefore be 100 times
larger than for muon events, which means that {\it a priori} a correspondingly higher background
rate has to be expected. On the other hand, in a study on the expected muon neutrino fluxes from
young Supernova Remnants~\cite{protheroe}, it was found that even if the resolution is only
$10^{\circ}$, the muon neutrino flux of a 0.1\,year young Supernova Remnant at a distance of
10\,kpc, with an initial pulsar period $P_0 = 5$\,ms, lies above the atmospheric muon neutrino flux
in the ${10^{\circ}\times 10^{\circ}}$ angular bin, for neutrino energies above 10\,TeV~\cite{protheroe}.  
\item {\bf Detectable event classes:} \\
As will be explained in more detail in Section~\ref{sec:event_types}, the number of event classes
that generate a shower-type event in the detector is larger than the number of event classes which
generate a muon event. Taking into account that the cross section for NC interactions is about one third
of that of CC interactions (see Section~\ref{sec:interaction_types}), one can estimate that all 6 NC
channels, plus the 2 channels $\nu_e,\bar{\nu}_e + N \to e^{\pm} + X$ produce approximately twice
the event rate of the two CC channels $\nu_{\mu},\bar{\nu}_{\mu} + N \to \mu^{\pm} + X$.
\item {\bf Effective area:} \\
The effective area tends to be 10 -- 15 times smaller for shower events than for muon events. The
reason for this is that the much shorter length of the shower (see Figure~\ref{fig:pathlength} in
Section~\ref{sec:event_types}) leads to a decrease of the effective volume, because only events that
are within an absorption length of the instrumented volume can be detected, whereas muons can travel
through several km of water towards the detector.   
\end{itemize}
Combining the two latter factors, the sensitivity for shower events deteriorates approximately by a
factor of $(2 \cdot 0.1)^{-1} = 5$. To achieve the same sensitivity as for muons, 5 times more
events would be needed; and, taking the first point of the list into account as well, one would
expect a 100 times higher background at the same time. \\
More precise calculations would have to be conducted to retrieve exact predictions on the perspectives 
of point source searches with shower events in ANTARES; at least from the estimations presented above, 
the feasibility of such a search seems uncertain.

\subsection{Diffuse Neutrino Flux}\label{sec:diffuse_flux}

\begin{figure}[h] \centering
\includegraphics[width=12cm,height=7.5cm]{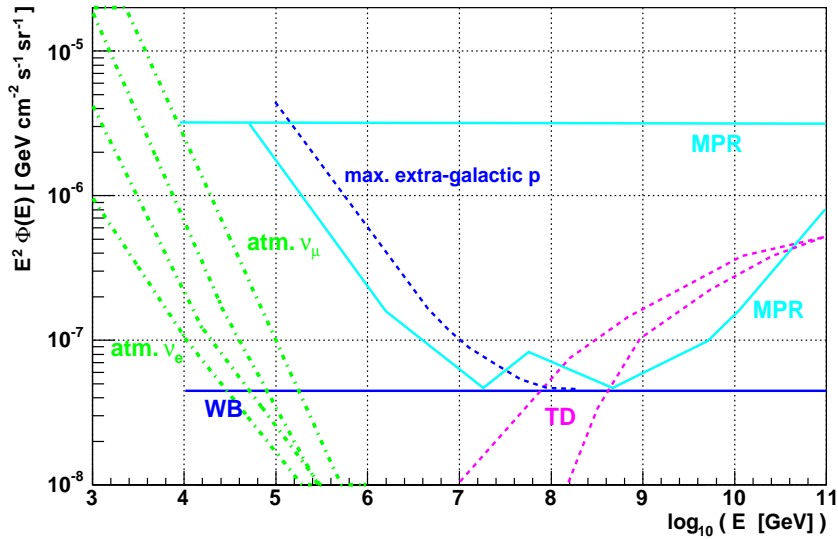}
\caption[Diffuse neutrino fluxes]{Diffuse neutrino fluxes as predicted by different models:
  Ranges for conventional~\cite{bartol} and prompt atmospheric neutrinos~\cite{naumov}: Green,
  dashed-dotted lines; 
  Waxman-Bahcall upper limit~\cite{waxman-bahcall1}: Solid blue line marked WB; maximum excess due 
  to extra-galactic protons~\cite{waxman-bahcall2}: Dashed blue line marked max.~extra-galactic $p$; 
  Mannheim-Protheroe-Rachen predictions~\cite{MPR}: Solid cyan lines marked MPR, upper line for
  optically thick, lower curve for optically thin sources; top down models~\cite{topdown}: Dashed,
  magenta lines, with (left) and without (right) SUSY. See text for details.}
\label{fig:theo_fluxes}
\end{figure}

Figure~\ref{fig:theo_fluxes} shows a collection of theoretical predictions for the diffuse neutrino
flux, the superimposed flux of all neutrinos from a certain type of sources. The same plot, with
experimental limits added, is shown again at the end of Chapter~\ref{ch:experiment} as 
Figure~\ref{fig:flux_limits}. \\
The solid blue line marked WB, the upper bound for diffuse neutrino flux calculated by Waxman and
Bahcall~\cite{waxman-bahcall1}, from now on abbreviated WB bound, is based on the fluxes of cosmic
rays measured at Earth at energies from $10^{16} - 10^{20}$\,eV and on the assumption of a
cosmic ray spectrum of $E^{-2}$ at the source, as predicted by the Fermi mechanism, see 
Section~\ref{sec:fermi}. The predictions are valid for sources which are optically thin 
for proton photo-meson interactions, in the sense that cosmic rays or photons escape the sources and can 
be measured at Earth. The WB bound is denoted to be a conservative upper bound by the authors, because 
it was assumed that the entire energy of the proton is transfered to pions in the photo-meson 
production, while realistic is a transfer of 20\% of the energy. In calculating the upper bound, WB have
assumed that only a very small fraction of the cosmic ray flux in the considered energy
region is composed of protons and that most of the cosmic ray flux in this energy range comes from
heavy nuclei, for which the photo-dissociation cross section is higher than the photo-meson
production cross section, so that they cannot account for a large neutrino flux. Assuming that
extra-galactic protons yield a higher contribution to the cosmic ray flux than expected, one can
exceed the WB bound in the way shown by the dashed blue curve marked \lq\lq max.~extra-galactic
$p$\rq\rq~\cite{waxman-bahcall2}, which has however already been partially excluded 
experimentally, see Figure~\ref{fig:flux_limits} in Section~\ref{sec:other_ex}. \\ 
Mannheim, Protheroe and Rachen~\cite{MPR} (abbreviated from now on as MPR) do not assume a fixed $E^{-2}$
cosmic ray spectrum at the sources, but take into account source characteristics, like the opacity to 
neutrons which determines the rate of neutrons which may escape the source. These neutrons would then 
decay to produce protons which would consequently be measured in the cosmic
ray flux. The authors arrive at somewhat higher flux limits, as shown in the cyan coloured lines 
marked MPR in Figure~\ref{fig:theo_fluxes}. The lower line refers to optically thin sources,
i.e.~sources that are transparent to neutrons (under the assumption that part of the cosmic rays measured 
at Earth originate from neutrons that have escaped the source and have then decayed into protons). If one
assumes that there exist a lot of optically thick sources, this would mean that the neutrino flux
can be much higher than expected from the measured cosmic ray flux because a lot of neutrino sources
remain unseen by cosmic ray experiments. The upper, straight line shows the case of optically very
thick sources. The actual upper limit on the neutrino flux is expected to be somewhere in between
the two lines. Whether the neutrino flux of the lower curve rises again for energies above
$10^9$\,GeV or not depends on the nature of the cosmic rays which have been measured beyond the GZK
cutoff. If they are caused by extragalactic sources, their flux is strongly damped by the GZK
cutoff, such that the corresponding flux of neutrinos, which do not suffer from attenuation by the
GZK mechanism, is expected to be much higher; if, on the other hand, the ultra high-energy cosmic
rays are caused by a single strong nearby source, no rise in the neutrino flux is expected; it would
stay flat instead, at the limit predicted by WB.  \\
The dashed magenta lines marked TD show the predictions for some top-down models~\cite{topdown}. The
limits refer to a model of a GUT particle with an energy of $10^{16}$\,GeV, a high universal radio
background and a relatively high extra-galactic magnetic field of $10^{-10}$\,G. The flux limit marked
by the left curve includes supersymmetry, while the right one does not. \\
The dashed-dotted green curves show the atmospheric neutrino flux, including prompt
neutrinos. The two upper lines mark the range of the atmospheric muon neutrino flux, depending on
the meson incident angle, while the two lower lines mark the range of the atmospheric electron
neutrino flux. The shown flux was simulated inside the ANTARES neutrino interaction simulation
(see Appendix~\ref{sec:genhen}); the Bartol model~\cite{bartol} was used to retrieve the
conventional flux, and the results of Naumov~\cite{naumov}, using the recombination
quark-parton model (RQPM)~\cite{RQPM}, to obtain the prompt neutrino flux. See
Section~\ref{sec:atm_nus} for more details on this.  

\chapter{The ANTARES Experiment}\label{ch:experiment}

\begin{figure} \centering
\includegraphics[width=10cm]{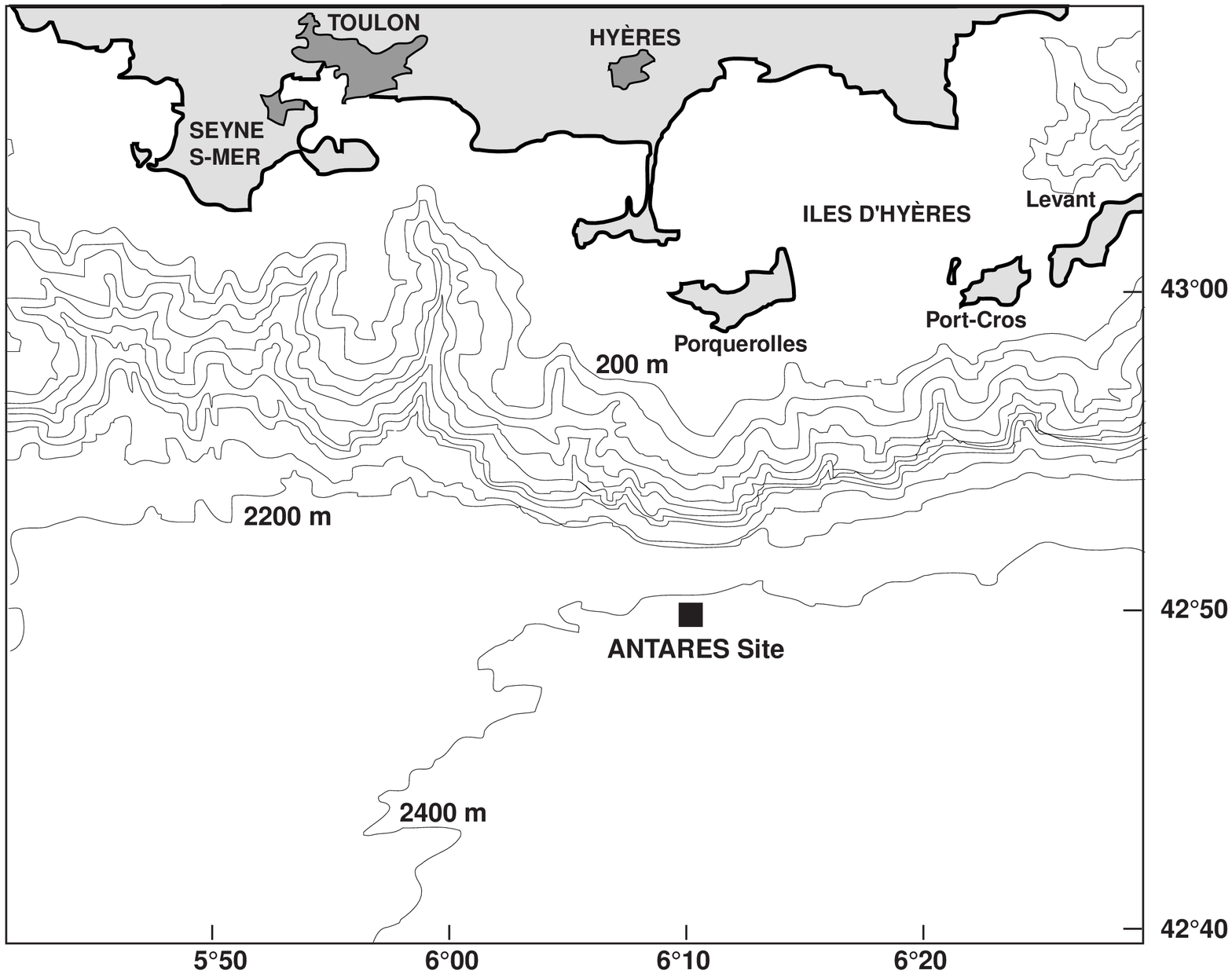}
\caption[Map of the ANTARES site location]
{Map showing the location of the ANTARES site in the Mediterranean Sea, including the French coast area (in
  grey) and sea depths. From~\cite{proposal}.}
\label{fig:location}
\end{figure}

The ANTARES\footnote{{\bf{A}}stronomy with a {\bf{N}}eutrino {\bf{T}}elescope and {\bf{A}}byss
environmental {\bf{RES}}earch.} collaboration was formed in 1996 with the objective to construct
and operate a neutrino telescope in the Mediterranean Sea. Currently the collaboration consists of
around 150 members from particle physics, astronomy and sea science institutes in 6 European
countries. Though the main purpose of the experiment is the detection of high-energy cosmic
neutrinos, it is also intended to be used as an experimental platform for
studies of the deep-sea environment. \\
ANTARES is being built in a depth of 2400\,m in the Mediterranean Sea, about 40 km South-East of the
French coastal city of Toulon. The location of the site is shown in Figure~\ref{fig:location}. As will
be discussed in more detail in Chapter~\ref{sec:background}, the location in the deep sea has the
advantage of suppressing to a large extend the background of atmospheric muons produced by cosmic
rays. On the other hand, the high water pressure of 240\,bar, as well as the 
aggressive salt water environment, impose strong requirements on the detector components which are
designed to have a lifetime of at least 10 years. \\
A description of the detector components and an overview of the detector layout is
given in Section~\ref{sec:layout}. The \lq\lq eyes\rq\rq\, of the detector, i.e.~the Optical
Modules which actually detect the Cherenkov photons emitted by the charged secondaries produced in neutrino
interactions with matter, are described in Section~\ref{sec:OM}. The digitisation of the
signal is explained in Section~\ref{sec:digitisation}. The efficiency of the detector also depends
strongly on the optical properties of the environment, i.e.~the absorption and scattering length, and the
refraction index, all three of them functions of the wavelength. These properties are described
in Section~\ref{sec:properties}; they are constantly monitored {\it in situ}.
\\
During the writing of this thesis, the first out of 12 large detector units (the so-called strings
or lines) has been deployed; the completion of the detector is forseen for the year 2007. In
Section~\ref{sec:status}, important milestones in the development and construction of the detector are
presented, and the current status is reported in more detail. \\ 
ANTARES is just one of several experiments aiming at the detection of high-energy cosmic
neutrinos. Some of these experiments are already partly or fully operational, others are under
construction or in the planning stage. Section~\ref{sec:other_ex} gives an overview of other
neutrino telescopes. 

\section{Layout of the Experiment} \label{sec:layout}

\begin{figure}
\begin{minipage}[b]{7.2cm}
  \centering \epsfig{figure=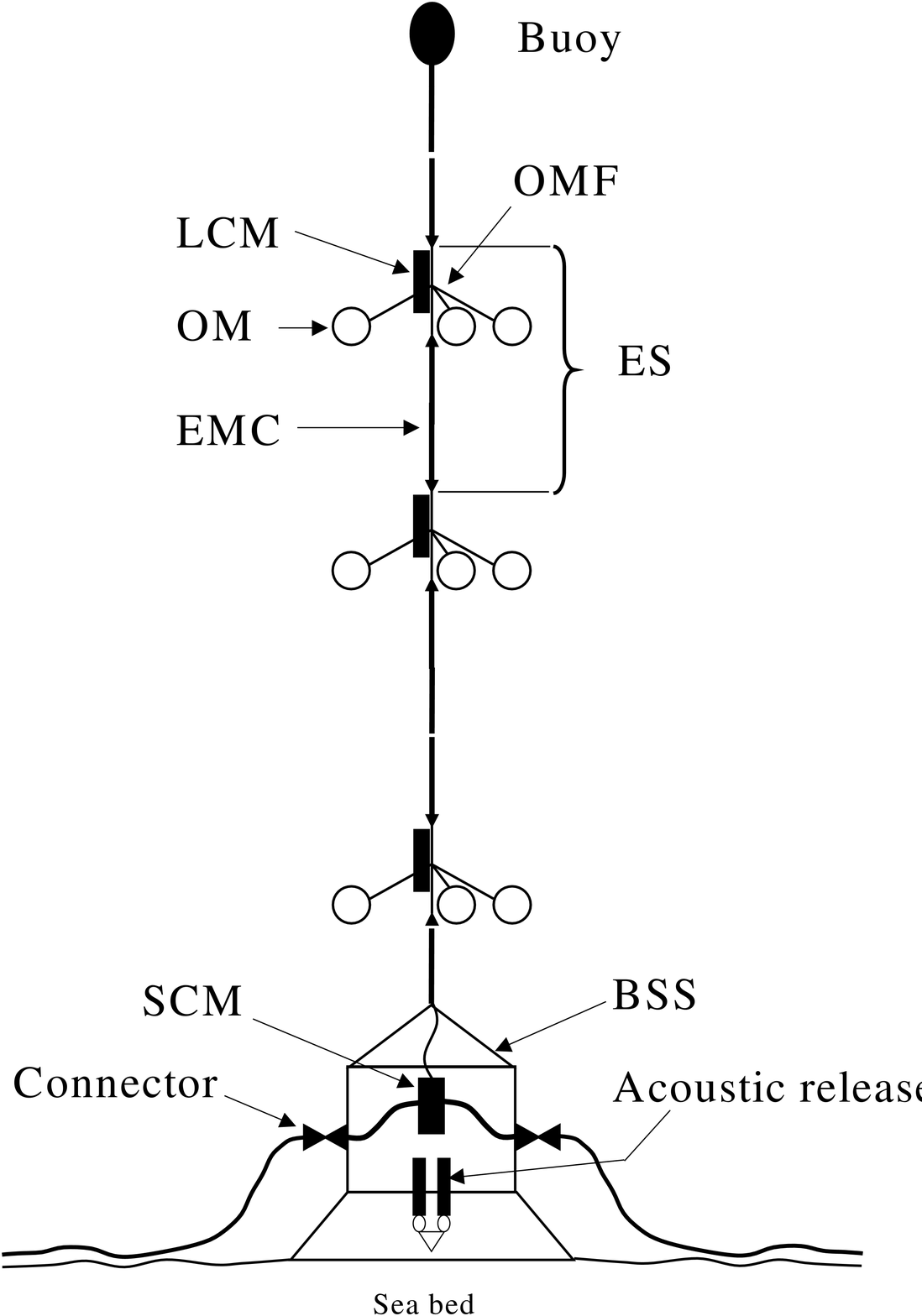, width=7.2cm}
\caption[Schematic view of an ANTARES string]
{Schematic view of an ANTARES string~\cite{proposal}. A full string contains 25 elementary segments
  (ES). The abbreviations are explained in the text.}
\label{fig:ant_string}
\end{minipage} 
\hspace{0.4cm}
\begin{minipage}[b]{7.2cm}
  \centering \epsfig{figure=antares_pub.eps, width=7.2cm}
\caption[Artist's view of the ANTARES detector]{Artist's view~\cite{montanet} of the ANTARES
 detector (not to scale).} 
\label{fig:ant_view}
\end{minipage}
\end{figure}

\begin{figure}[h] \centering
\includegraphics[width=7cm]{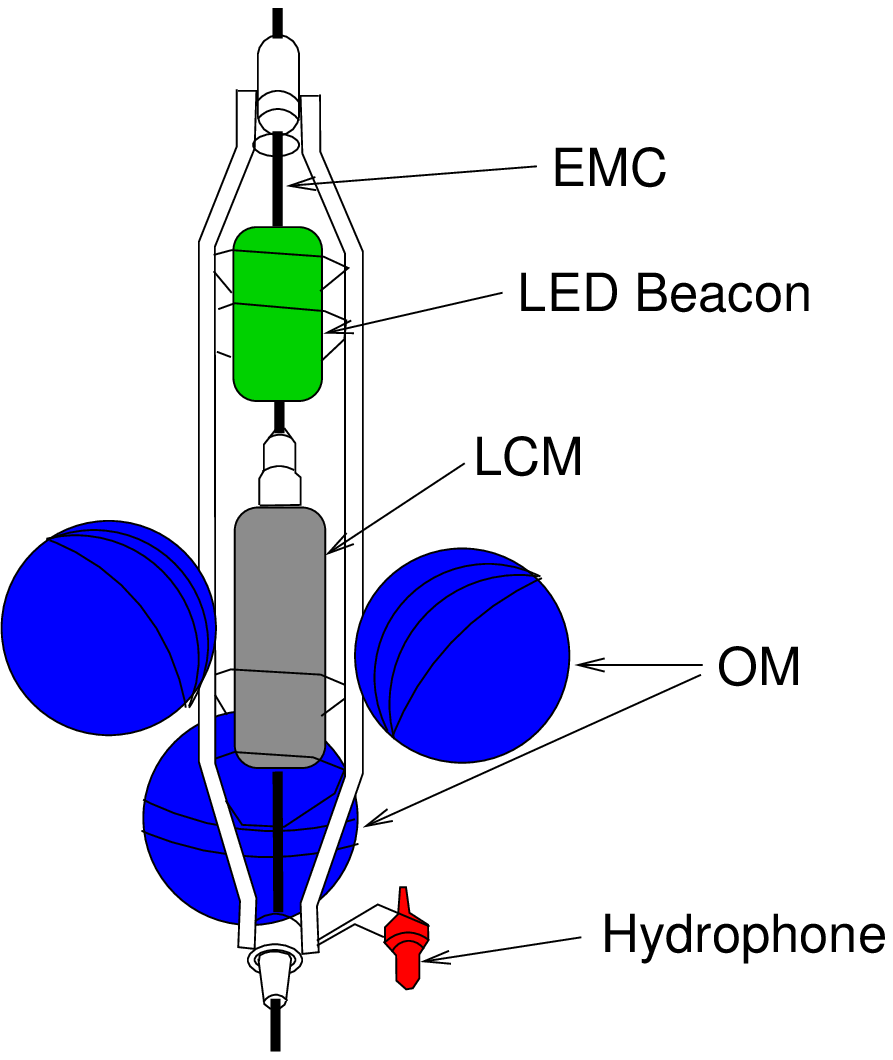}
\caption[Schematic view of an ANTARES storey]{Schematic view of an ANTARES storey. Mechanical parts
  are coloured in white, the LCM in grey, the cable in black and the OMs in blue. While the LCM and
the OMs are installed on all storeys, the hydrophone (red) and the LED beacon (green) are planned
  approximately once every five storeys, on separate storeys.}
\label{fig:storey}
\end{figure}

\begin{figure}[h] \centering
\includegraphics[width=10cm]{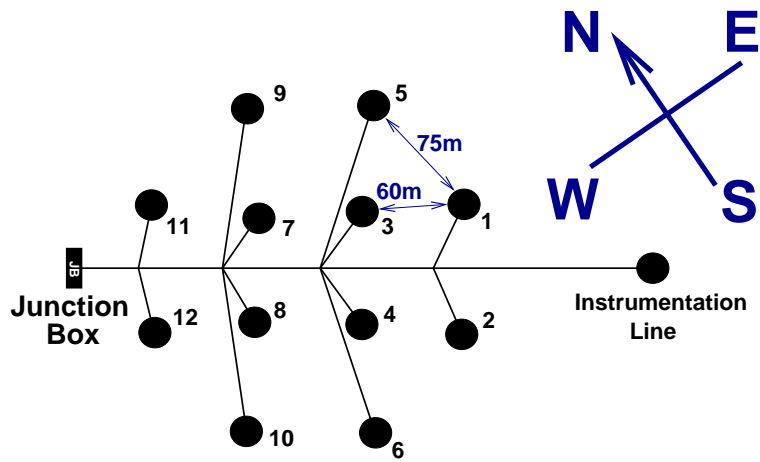}
\caption{Top view of the ANTARES sea floor layout.}
\label{fig:ant_layout}
\end{figure}

The ANTARES Optical Modules are mounted on 12 {\it strings}, long cables which are fixed to the sea
bed by the {\it Bottom String Socket (BSS)} and are kept vertically straight by a buoy. The BSS
consists of a dead weight, connectors for the cables, an acoustic device to release the string, 
and the electronics of the {\it String Control Module (SCM)}. The single 
elements of the string are both mechanically supported and electrically interconnected by a total of
460\,m of {\it Electro-Mechanical Cables (EMC)}. \\
One string consists of five {\it sectors}, each containing five {\it storeys}, so that there are 25
storeys per string. The distance between two storeys is 14.5\,m. 
The unit storey plus corresponding EMC section is called {\it Elementary Segment (ES)}. 
Each storey is made up of three {\it Optical Modules (OM)} which are supported by the {\it Optical
  Module Frame (OMF)} (see Section~\ref{sec:OM} for a more detailed description of the OMs). The {\it
  Local Control Module (LCM)} is a titanium container placed in the middle of the OMF, housing the
data read-out, clock and control electronics for the storey. A schematic view of the 
string design is given in Figure~\ref{fig:ant_string}, while Figure~\ref{fig:ant_view} shows an
artist's view of the arrangement of the ANTARES strings~\cite{montanet}. \\
Figure~\ref{fig:storey} illustrates the setup of an ANTARES storey in more detail. The figure also
shows the hydrophone for the acoustic positioning system and the LED beacon for the time
calibration of the photomultipliers. One hydrophone per sector is forseen in each line, and an LED
beacon in four of the five sectors of every line. Sound velocimeters required for the acoustic
positioning are also planned to be installed on 3 lines. They are not shown in the drawing. \\
In addition to the 12 detector strings, a 13th string, the {\it Instrumentation Line}, is foreseen for
the monitoring of environmental parameters. For the first strings, it is planned to
use the MILOM, which is already deployed at the ANTARES site, for this task. A more detailed
description of the MILOM is given in Section~\ref{sec:status}. \\ 
The detector strings are arranged on the sea floor as an octagon, as shown in
Figure~\ref{fig:ant_layout}. Each string is connected to the {\it Junction Box (JB)} by an
electro-optical cable. The Junction Box supplies the strings with electrical power and with slow-control
and clock signals, and it receives the data taken by the OMs; it contains power converters
and the electro-optical interface for data, slow-control and clock transmission. The Junction
Box is connected to the shore station located at La Seyne sur Mer by the main electro-optical
cable. The shore station houses the control room for the data-taking and the slow-control instruments.  

\afterpage{\clearpage}

\section{The Optical Modules} \label{sec:OM}
\begin{figure}
\begin{minipage}[b]{7cm}
  \centering \epsfig{figure=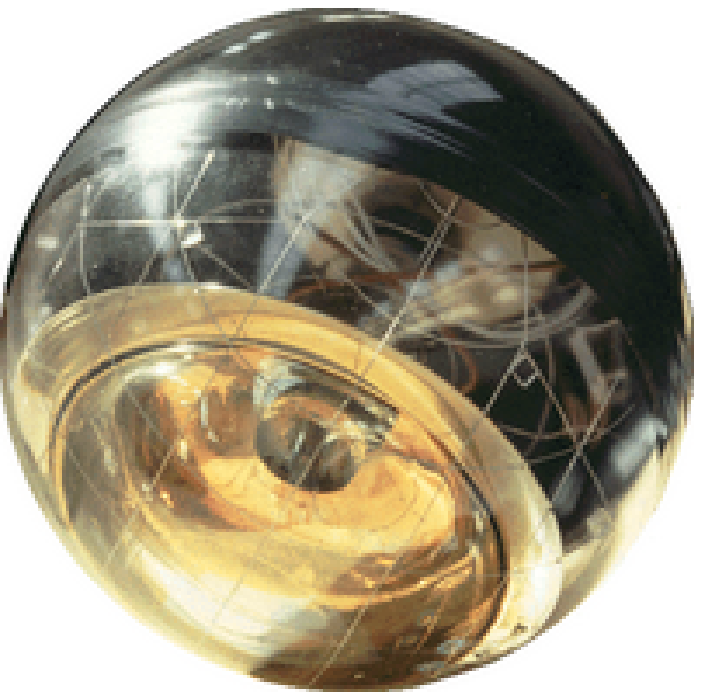, width=5cm}
  \caption[Photograph of an OM]{Photograph of an OM, taken from~\cite{antares OM}.}
  \label{fig:OM_photo}
\end{minipage} 
\hspace{0.5cm}
\begin{minipage}[b]{7cm}
  \centering \epsfig{figure=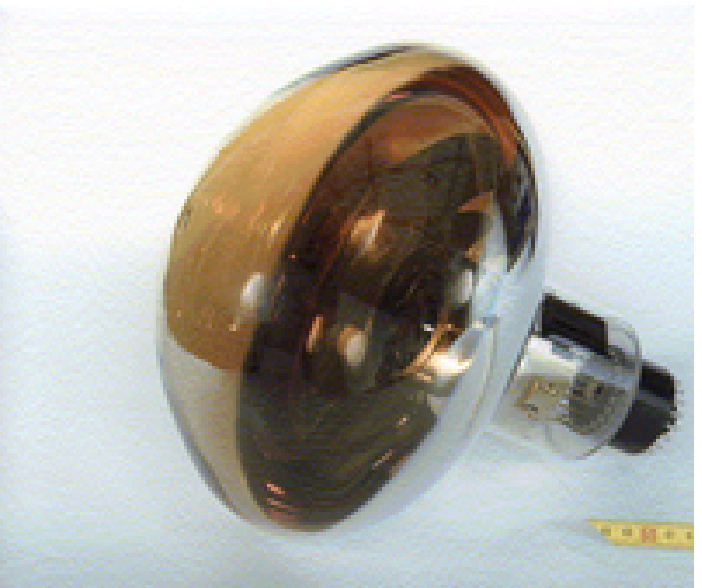, width=5.5cm}
  \caption[Photograph of the $10''$ Hamamatsu PMT]{Photograph of the $10''$ Hamamatsu
  PMT. From~\cite{antares OM}.} 
  \label{fig:PMT_photo}
\end{minipage}
\end{figure}

\begin{figure} \centering
\includegraphics[width=5cm]{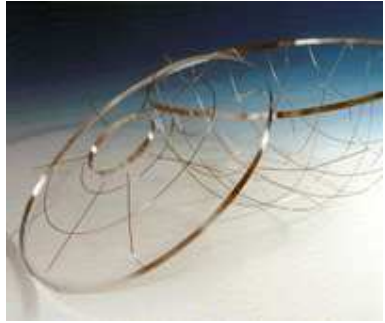}
\caption[The $\mu$-metal cage]{The $\mu$-metal cage which is used to shield the
  Earth's magnetic field. From~\cite{antares OM}.} 
\label{fig:cage}
\end{figure}

\begin{figure} \centering
\includegraphics[width=10cm]{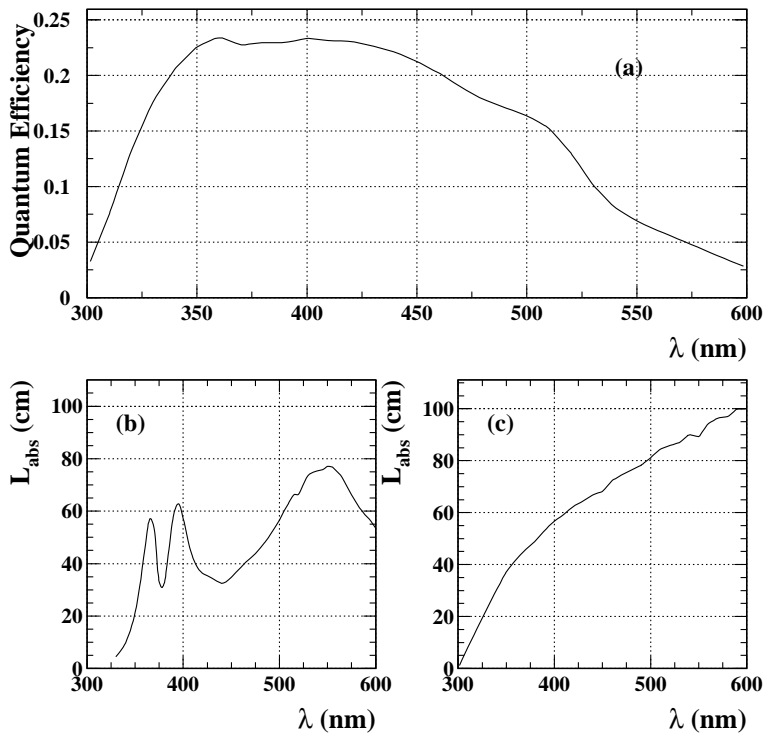}
\caption[PMT characteristics]{Characteristics of the PMTs: quantum efficiency (a),
  measured absorption length of the glass sphere (b) and of the silicone gel (c). All curves have
  been plotted as a function of the incident light wavelength. From~\cite{antares OM}.}
\label{fig:OM_effi}
\end{figure}

The Optical Modules (OMs) are the key elements of the detector. They consist of a
pressure-resistant glass sphere of about 43\,cm diameter and 15\,mm thickness, housing a
photomultiplier tube. A photograph of an OM is shown in Figure~\ref{fig:OM_photo}. This picture, as
well as all other figures in this section, has been taken from~\cite{antares OM}, where a more
detailed description of the OMs can be found. \\ 
The photomultiplier tubes (PMTs) ($10''$ R7081-20 from Hamamatsu) are glued to the glass spheres
with a thin layer of optical silicone gel. A photograph of a PMT is shown in
Figure~\ref{fig:PMT_photo}. A cage made of $\mu$-metal, a nickel-iron alloy with very 
high magnetic permeability at low field strengths, surrounds the PMT to shield the magnetic
field of the Earth. A picture of this cage is shown in Figure~\ref{fig:cage}.  \\
The back side of the sphere is painted black to avoid scattered light to reach the PMT from that
direction. It houses a penetrator which connects the photomultiplier electronics to the cable
leading to the LCM. \\  
The three OMs of one storey are fixed to the OMF with an azimuthal spacing of 120$^{\circ}$ with
respect to each other. In order to increase the sensitivity for upgoing neutrinos, the OMs are
looking downward in a 45$^{\circ}$ angle. The deployment of upward looking OMs would also bear the
problem that a layer of sediment material quickly covers the glass spheres; this affects
mainly the upper half-sphere, so that a downward looking OM barely encounters any deterioration of
sensitivity. The average transmission loss after one year, at 45$^{\circ}$ from the horizontal, was
measured to be around 2\%~\cite{fouling}.\\
Figure~\ref{fig:OM_effi} shows some characteristics of the OM: The quantum efficiency of the PMTs
(a) and the measured absorption length of the glass sphere (b) and of the silicone gel (c), all
as a function of the wavelength.

\section{Data Taking in ANTARES}
\subsection{Data Digitisation and Data-Taking Modes}\label{sec:digitisation}

\begin{figure}[h] \centering
\includegraphics[width=10cm]{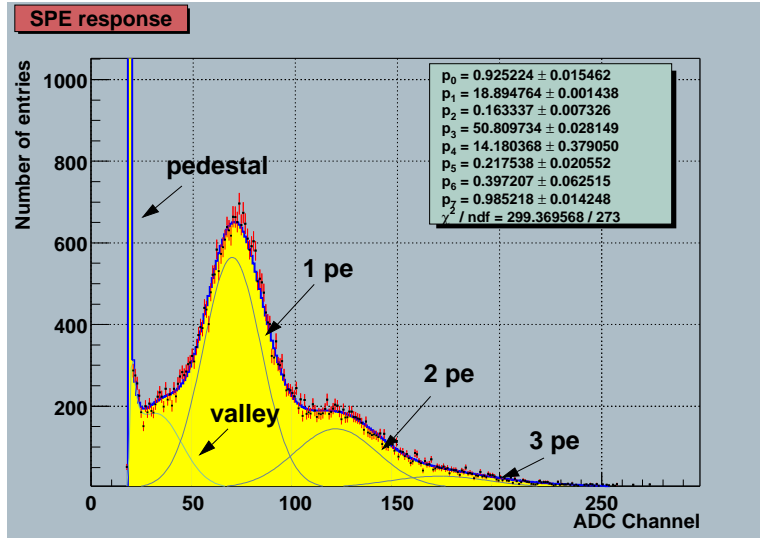}
\caption[Single photo-electron signal]{PMT response for a single photo-electron signal. The $x$-axis
  shows the Analogue to Digital Converter (ADC) channel which is proportional to the anode charge;
  each channel stands for about 0.25\,pC charge. The pedestal is the intrinsic noise of the PMT when
  the signal approaches zero; because of this intrinsic noise, no measurement is possible below a
  certain threshold, which is usually about 0.3\,pe to 0.5\,pe (for this study, a minimum pe
  threshold of 0.3\,pe was used). The measured charge for a given number of pe follows a Gaussian
  probability distribution. The prominent peak in the figure, marked 1\,pe, comes from the most
  probable case that one pe was produced from the photon pulse. The cases where two or three pe were
  produced can be seen as smaller peaks marked 2\,pe and 3\,pe. The parameters of a fit function
  describing the total shape of the histogram are also shown. This figure was taken
  from~\cite{zornoza_pmt}.} 
\label{fig:SPE_response}
\end{figure}

When a photon hits the PMT surface, it produces, with a probability according to the PMT quantum
efficiency, a secondary electron which then travels towards the anode; cascades of secondary
electrons are produced in the 14 dynodes in-between the photocathode and the anode, yielding an
amplification of $10^7 - 10^9$ between the cathode and the anode~\cite{zornoza_pmt}. The connection
between the anode signal, i.e.~charge, and the original number of single photo-electrons (pe) can be
found in calibration measurements. Figure~\ref{fig:SPE_response} shows such a 
measurement~\cite{zornoza_pmt} for a number of single photo-electron signals (which means that laser
pulses with an average number of one photon per pulse were sent to the PMT; with Poissonian probability,
also two, three or more photons are present in one pulse. The photo-electrons from the photon
pulses are produced according to the quantum efficiency of the PMT.). 

The conversion of the PMT pulses to digital signals which can be sent to the shore station for
further processing and storage is done with the {\it Analogue Ring Sampler (ARS)} chip. Each OM
is read out by two ARSs, allowing for data-taking in two different modes: The {\it Single Photo
  Electron (SPE) mode} and the {\it Waveform (WF) mode}. In SPE mode, charge and time of a single
pulse are converted to a digital signal and sent to shore. In WF mode more complex pulse shapes can
be recorded; 128 signal samples in 1\,ns windows are recorded for one signal in this mode. It
therefore requires a much higher band-width, about a factor forty more than the SPE mode. A {\it
  Pulse Shape Discriminator (PSD)} in the ARS is used to switch the data-taking to the WF mode,
e.g.~in the case of a particularly wide pulse. As will be discussed later, for the detection of
shower events, the WF mode is more suitable than the SPE mode, because it allows for a larger
dynamical range in the amplitude measurement. In order to obtain a reasonable charge resolution, the
ADC settings are chosen such that the maximum number of pe measurable with one ARS in the SPE mode
is about 20. For the WF mode, the maximum number of measurable pe, i.e.~the saturation level is
$\sim$ 200, ten times higher than for the SPE mode. \\ 
While registering a signal in SPE mode, an ARS is active for 25\,ns. Therefore, if
several photons reach the PMT within this time period, they are all integrated as one {\it
  hit}. After these 25\,ns, the ARS needs 250\,ns to process the data. If another
signal reaches the same OM within the dead time, it is recorded by the second ARS. Only the case that
both ARSs are in processing mode means dead time for the OM; data-taking is then impossible. \\ 
For this study, a saturation of 200\,pe was assumed in most cases, because the showers studied
in this work generally produce larger amplitudes in the PMTs than muons and are therefore more
likely to trigger pulse shapes which would be recorded in WF mode. However, no WF implementation 
exists in the software at the moment. It was therefore assumed that the signals recorded in the WF
mode are taken with the same integration and dead time than in the SPE mode, but at a higher 
amplitude saturation level.  

\subsection{The Software Trigger}

The ANTARES data taking follows the philosophy of \lq\lq all data to shore\rq\rq, i.e.~all signals
recorded in the PMTs are sent to the shore station. The amount of recorded data from the whole
detector is about 1\,GB per second, depending on the background rate. This data rate is too high to
be permanently stored; the aim is therefore to reduce it by several orders of magnitude, selecting those
time windows containing physics signals and discarding those containing optical background. This
is done by a software algorithm mimicking a hardware trigger, which is 
thus called {\it Software Trigger}~\cite{brams}. The trigger conditions are based on the assumption
that the hits induced in different OMs by a muon or particle shower are correlated, whereas the hits
caused by optical background are not. The conditions for correlations are explained in detail in
Section~\ref{ch:trigger}, where they are used as {\it filter conditions} for the suppression of the
optical background. Correlated hits are collected in so-called {\it clusters}. When the cluster has
reached its adjustable minimal size (e.g.~4 hits for a background rate of 70\,kHz), an {\it event} is
built by selecting all hits within a time window of 2000\,ns around the cluster. If another cluster lies
within this time window, the two events are merged. The events are then written to disk. The trigger
conditions can be adjusted for a variable background, such that the final storage rate is always
about 1\,MB/s.

\section{Optical Properties of the ANTARES Site}\label{sec:properties}

\subsection{Absorption and Scattering}\label{sec:absorption}

The efficiency of Cherenkov light detection and the accuracy of the muon or shower direction
reconstruction depends on the absorption and scattering of the Cherenkov light in the water. These
effects are described by the {\it absorption length $\lambda_{abs}$} and the {\it scattering length
  $\lambda_{scatt}$}. $\lambda_{abs}$ and $\lambda_{scatt}$ can be combined to the total {\it
  attenuation length $\lambda_{att}$}, a measure of the overall attenuation of light in water:

\begin{equation}
\frac{1}{\lambda_{att}} = \frac{1}{\lambda_{abs}} + \frac{1}{\lambda_{scatt}}.
\label{eq:attenuation}
\end{equation}

\begin{figure}[h] \centering
\vspace{1cm}
\includegraphics[width=10cm]{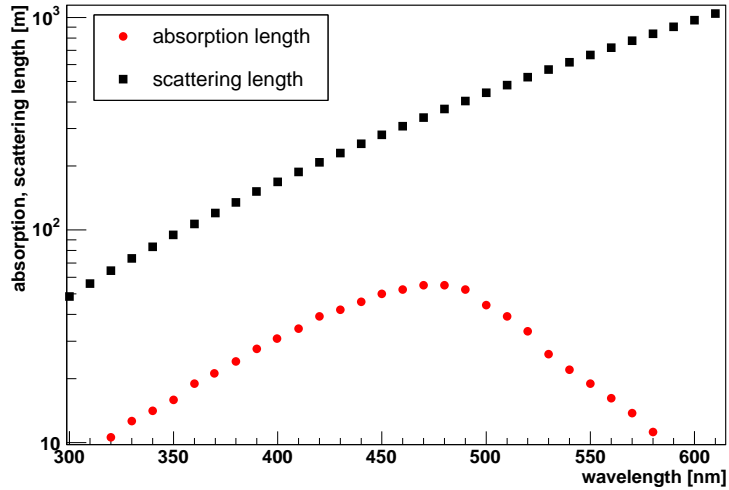}
\caption[Absorption and scattering length at the ANTARES site]{Absorption length and scattering
  length in m at the ANTARES site, as a function of the wavelength, as parameterised in~\cite{geasim}.}
\label{fig:absorption}
\end{figure}

$\lambda_{att}$ is a function of the wavelength of the light; the relevant wavelength region for the
detection of Cherenkov light is between 320\,nm and 620\,nm, 
according to the characteristics of the photomultipliers and the glass spheres (see
Figure~\ref{fig:OM_effi} in Section~\ref{sec:OM}). Measurements in 
the deep sea have shown that the attenuation length in deep salt water is largest (and therefore,
the attenuation itself smallest) at $\sim 460 - 470$\,nm~\cite{price}. This is in the region of blue
light, and therefore a number of test measurements at the ANTARES site using blue LEDs at a
wavelength of 466\,nm have been performed~\cite{palanque99,palanque01}. For a detector like ANTARES, the
scattered photons are not necessarily lost, and therefore the scattering length is replaced by an
{\it effective scattering length} which depends on the scattering angle. For the ANTARES site, an
effective scattering length between 230\,m and 300\,m, and an absorption length between 50\,m and
69\,m for blue light were measured in various sea campaigns~\cite{transmission}. 
Figure~\ref{fig:absorption} shows absorption length and scattering length, as parameterised in the
ANTARES software~\cite{geasim} following the results of the test measurements and a theoretical
model for the scattering~\cite{kopelevich}. Scattering effects are, however, not fully simulated in
the software. \\
For this thesis, attenuation effects were considered at a fixed wavelength of 475\,nm, using only 
the respective absorption length, $\lambda_{abs} \equiv \tau = 55$\,m.

\subsection{Refraction Index and Velocity of Light}\label{sec:refraction}

\begin{figure}[h] \centering
\includegraphics[width=10cm]{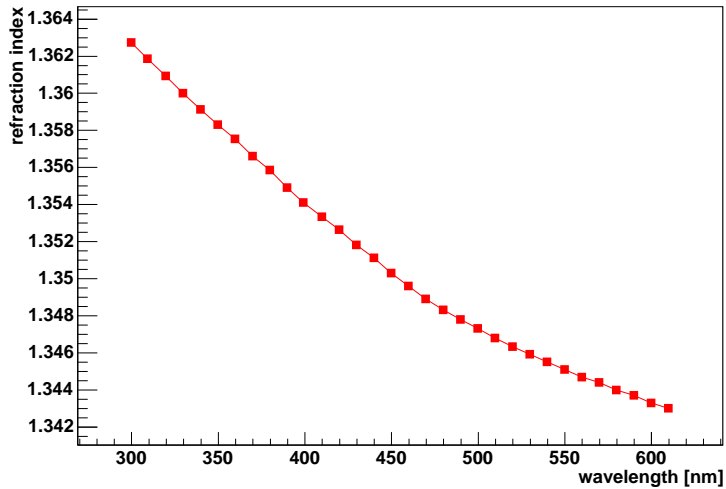}
\caption[Refraction index at the ANTARES site]{Refraction index at the ANTARES site, depending on
  the wavelength of the Cherenkov radiation, for a depth of 2000\,m~\cite{geasim,brunner}.}
\label{fig:refraction}
\end{figure}

The refraction index naturally depends on the composition of the medium and hence on the salinity of
the water (higher refraction index at higher salinity), its temperature (higher refraction index at
lower temperature), and on the water depth (higher refraction index at larger depth, i.e.~higher
pressure). Typical values of these  parameters at the ANTARES site are:
\begin{itemize}
\item salinity: $38.44$ \textperthousand;
\item temperature: $13.1^\circ$C;
\item pressure: 200 -- 240\,bar.
\end{itemize}
As it connects directly to the velocity of light in water and the Cherenkov angle, a profound
knowledge of the refraction index is required. The velocity of light as derived from refraction index
measurements is calculated for the ANTARES site in~\cite{palanque}; the value that was found for
blue light ($\lambda = 466$\,nm) is $(0.21755 \pm 2 \cdot 10^{-5})$\,m/ns. The dependence of the
refraction index on the various parameters has also been discussed in~\cite{brunner} and is
encoded in the Cherenkov light propagation software~\cite{geasim} as a function of the
wavelength. Its dependence on the wavelength, for a depth of 2000\,m, is shown in
Figure~\ref{fig:refraction}. For this study, a refraction index of 1.3499 was used, which
corresponds to a Cherenkov angle of ${\vartheta_{C} = \arccos(1/1.3499) \approx 42.2^{\circ}}$. 

\section{Status of the Experiment} \label{sec:status}
\begin{figure}[h]
\centering
\includegraphics[width=8cm]{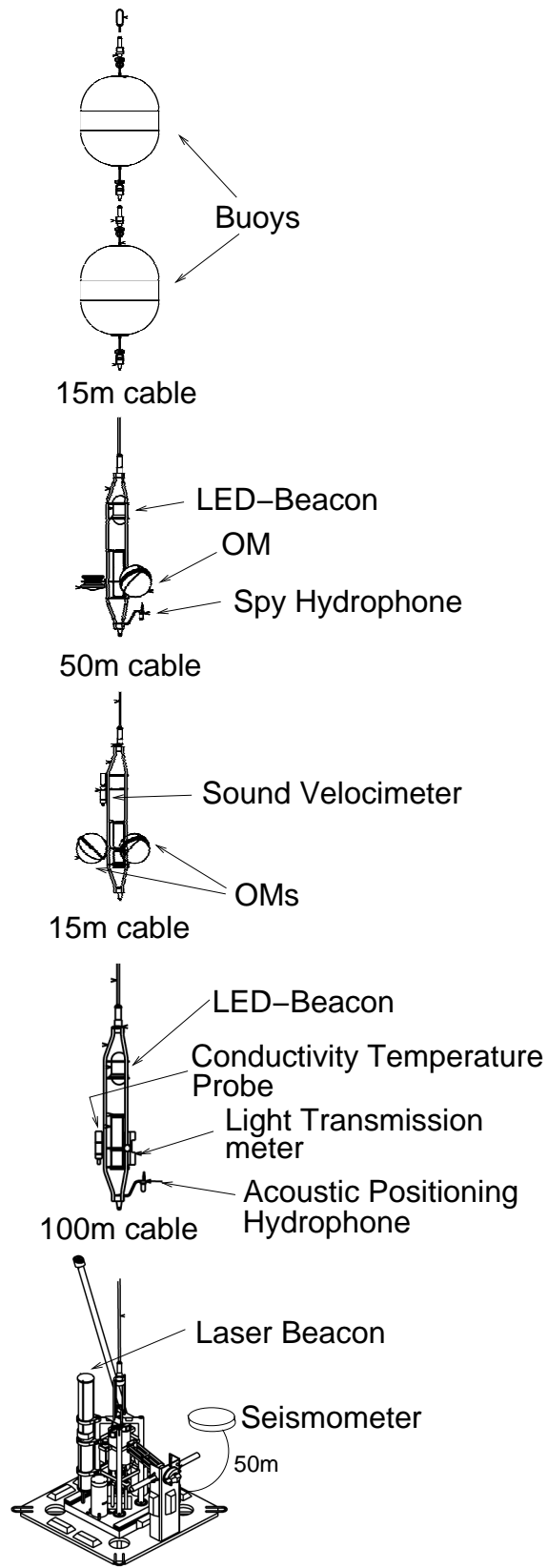}
\caption[The MILOM]{Schematic view of the MILOM (Mini-Instrumentation-Line with OMs). See text for
  more details.}
\label{fig:milom}
\end{figure}

Between 1996, the beginning of the ANTARES activities, and 2000, several autonomous lines have been
deployed at various sites to measure the optical properties of the water and test the detector
components. The most recent of these test lines, recovered in June 2000, also measured 
signals from atmospheric muons~\cite{test_muons}. \\ 
In October 2001 the main electro-optical cable leading from the ANTARES site to the shore was
deployed, followed by the Junction Box, about one year later. Shortly after the Junction Box, a line
prototype with five storeys, called Prototype Sector Line (PSL), and the Mini-Instrumentation Line
(MIL) were deployed. Both the PSL and the MIL were connected to the Junction Box in March 2003. The MIL
was recovered in May 2003 whereas the PSL stayed connected until July 2003. \\
During that four-month period the PSL took a large amount of data, and the feasibility
of the deployment and recovery sea operations was proven. There were some problems, though: Due 
to a damage of the optical fibre distributing the clock signal, no nanosecond timing was available;
also, there was a water leak in the electronics container of the MIL. These problems were thoroughly
investigated and eliminated by design modifications. To assure the functionality of the system, two
additional test strings, Line\,0 and MILOM, were built. Line\,0 is a full sized 
line containing all the mechanical elements of a string, but no electronics. It is used to intensely
test the integrity of the mechanical parts of a string during deployment, operation and recovering.
MILOM stands for Mini-Instrumentation Line with Optical Modules. The MILOM is an improved
version of the MIL; it contains LED beacons for OM calibrations, a hydrophone for acoustic
positioning, a sound velocimeter, a conductivity temperature probe, a light transmission meter and a
laser beacon. A seismometer is also deployed at 50\,m distance, with a connection to the
MILOM. Besides these monitoring and control instruments, the MILOM also contains four OMs, three
of them arranged on a normal OMF and one of them alone on a separate storey. The smaller
buoyancy of the MILOM, as compared to a normal string, is compensated by two buoys. The whole setup
is shown in Figure~\ref{fig:milom}. 

\afterpage{\clearpage}

Both Line\,0 and MILOM were deployed in March 2005. With the MILOM, a timing resolution better than 
1\,ns and a positioning resolution of the order of 10\,cm was proven~\cite{calibration}. An example
for a measurement of the timing resolution is shown in Figure~\ref{fig:time_res}.  
While MILOM is at present still taking data and
is foreseen to stay at the ANTARES site to serve as an environmental control instrument for the first
complete lines, Line\,0 was recovered in May 2005. Leakage tests showed that there were no water
leaks in the electronic containers; there were however some optical transmission losses due to an
interface problem between EMC penetrators and the LCM. These problems have been understood and
solved, and the first complete string has deployed in February 2006 and connected on March, 2nd. The
line has been fully functional from the first moment of power-up and the first downgoing muons have
been reconstructed. The successful connection of Line\,1 is a huge step forward in the completion of
the detector planned for the year 2007. \\ 

\section{Other Experiments} \label{sec:other_ex}
The planning of the first neutrino telescopes already started as early as 1975. This section lists
high-energy neutrino telescopes of the past, the present and the future, which use the Cherenkov
detection technique in water or ice.  

\begin{itemize}
\item {\bf DUMAND}\\
DUMAND was the pioneering experiment in underwater neutrino detection. It was started in 1975 with
the objective to build a neutrino telescope in the deep sea close to Hawaii, USA. Unfortunately, after
encountering some technical problems, the experiment was cancelled in 1995 without having
taken data. However, the expertise gained in this experiment, theoretically as well as
experimentally, has been an extremely useful basis for all later experiments.
\end{itemize}
\afterpage{\clearpage}

\begin{figure}[h] \centering
\includegraphics[width=12cm,height=7.5cm]{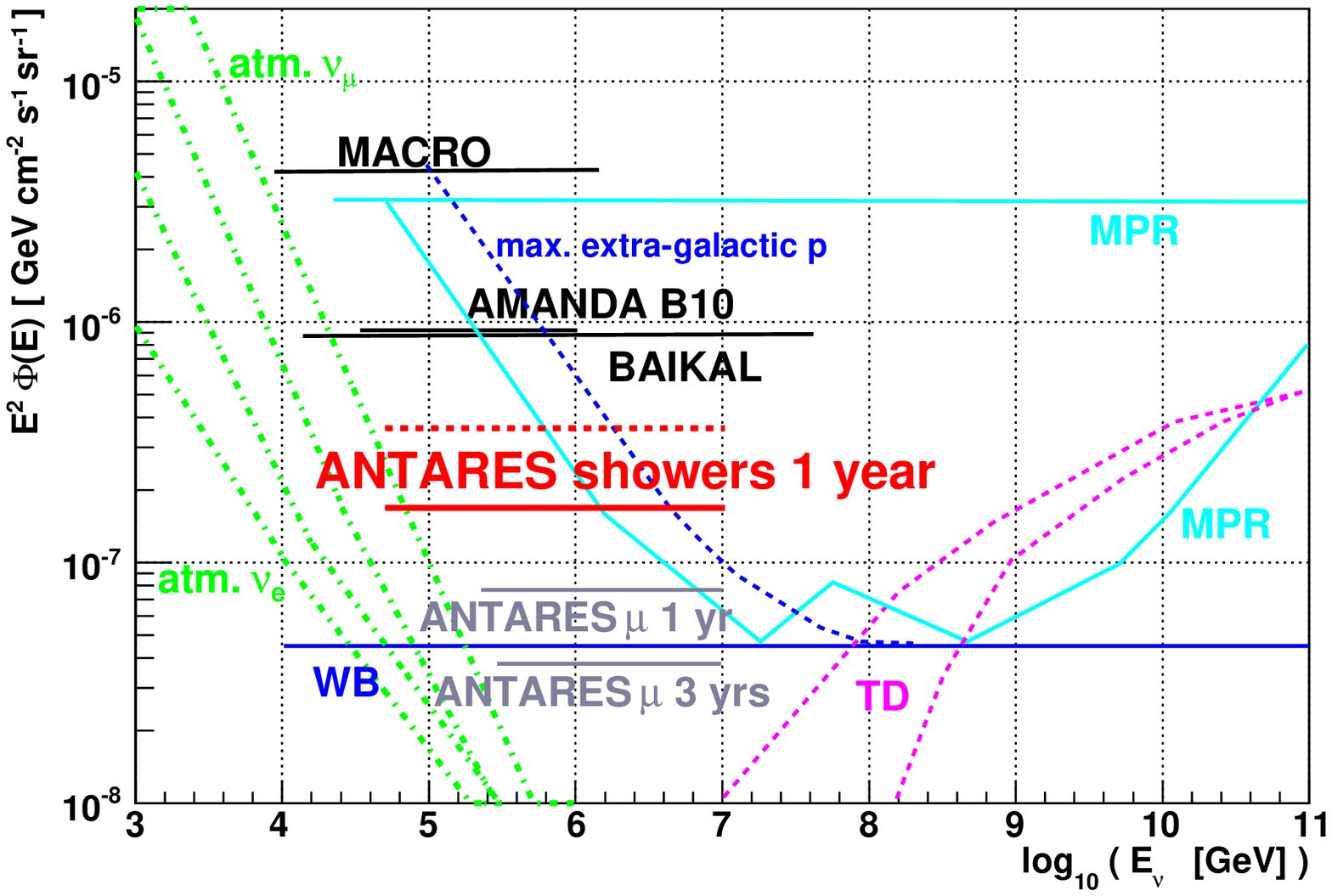}
\caption[Neutrino flux limits]
{Neutrino flux limits reached or expected with different neutrino experiments, and predicted
  bounds~\cite{waxman-bahcall1,waxman-bahcall2,MPR,topdown}. ANTARES expectations
  from~\cite{zornoza}, Macro limit from~\cite{macro}, BAIKAL limit from~\cite{wischnewski}, AMANDA
  limit (for cascades) from~\cite{amanda_cascades}. See Figure~\ref{fig:theo_fluxes} in
  Section~\ref{sec:nu_fluxes} for details on the theoretical predictions.}    
\label{fig:flux_limits}
\end{figure}

\begin{figure}[h] \centering
\includegraphics[width=12cm]{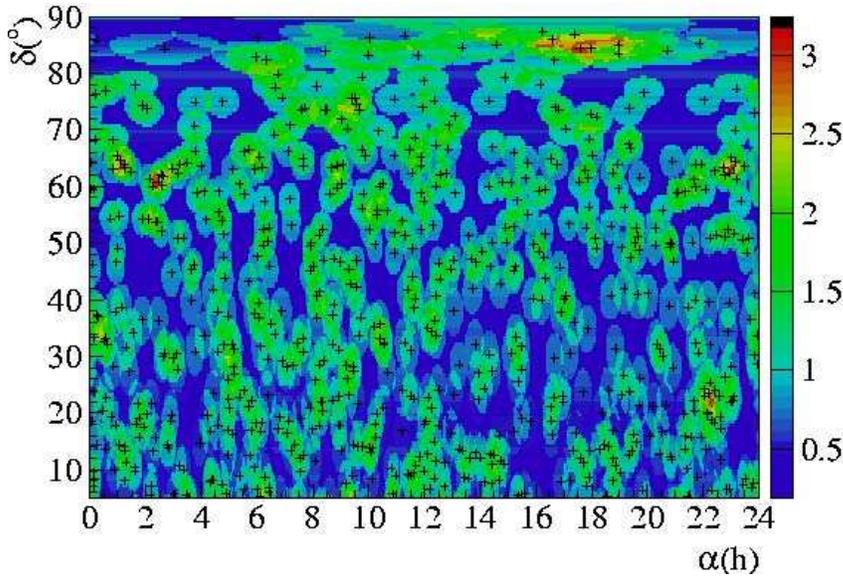}
\caption[AMANDA upper limits on neutrino point sources]
{AMANDA upper limits (90\% C.L.) on the neutrino flux integrated above 10\,GeV in equatorial
  coordinates, declination $\delta > 5^{\circ}$ over right ascension $\alpha$ (in hours), in units
  of $10^{-8}$\,cm$^{-2}$\,s$^{-1}$ (colour-coding on the right). The data were taken in the
  years 2000-2002, and an $E^{-2}$ spectrum has been assumed. The cross symbols represent the
  observed events. From~\cite{amanda_point}.}
\label{fig:amanda_point}
\end{figure}

\begin{itemize}
\item {\bf BAIKAL}\\
The BAIKAL experiment~\cite{baikal} is located at 1070\,m depth in Lake Baikal (Russia). It has been
the first running high-energy neutrino telescope, with site tests and research and development going
on since 1980. The oldest components of the current detector, NT 200+, have been in use since 1993. 
NT 200+  consists of 8 closely spaced strings plus 3 additional strings in a distance of 100\,m to
the others. Recently published flux limits of four years of data-taking~\cite{wischnewski} set a
limit of ${E^2 \Phi < 8.1 \cdot 10^{-7} \textrm{GeV cm}^{-2}\,\textrm{s}^{-1}\,\textrm{sr}^{-1}}$
for cosmic neutrinos between 10\,TeV and 10\,PeV (see Figure~\ref{fig:flux_limits}).\\
The advantage of BAIKAL compared to the other neutrino telescopes is the
relatively easy access to the detector. The strings can be deployed and recovered during winter when
the surface of the lake is frozen, by melting holes into the ice. It is more easily accessed
than the South Pole, and no ship is needed, other than for the experiments in the sea. On the other
hand, the optical properties of the water are relatively poor compared to the deep-sea sites in the
Mediterranean, with a significantly shorter attenuation length of $\sim 20$\,m. 

\begin{figure}[t]
\centering
\includegraphics[width=11.5cm]{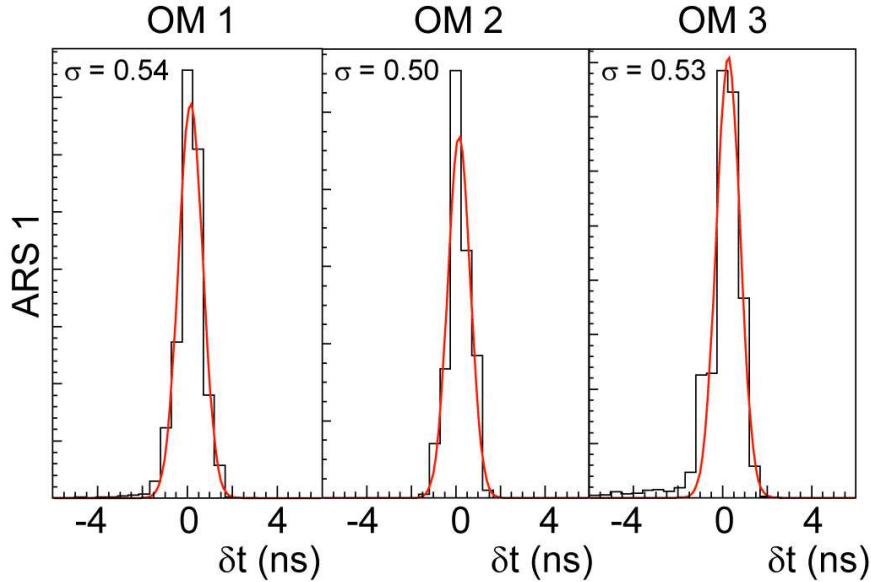}
\caption[Timing resolution with the MILOM]{Time distribution of a signal in three OMs of the MILOM,
  relative to the reference signal of the LED beacon, after subtracting an offset for the light
  propagation. From~\cite{MILOM}.}
\label{fig:time_res}
\end{figure}

\afterpage{\clearpage}

\item {\bf AMANDA}\\
The AMANDA experiment~\cite{amanda} has been built at the South Pole and is operational since
1997. The last stage, AMANDA-II, consists of 677 optical modules mounted on 19 strings which are
arranged in concentric circles. The telescope is installed in the glacial ice in depths between 1500
and 2000\,m. Using ice instead of water as detector medium has both advantages and disadvantages. In
ice, the optical background is very low ($< 1$\,kHz per optical module) which allows for a lower
energy threshold than in water. On the other hand, due to inhomogeneities in the ice, scattering
effects are much larger than in water, which deteriorates the resolution, especially for the
reconstruction of the neutrino direction. Nevertheless, many important results on neutrino astronomy
come from the AMANDA experiment, like stringent limits on diffuse neutrino flux, see
Figure~\ref{fig:flux_limits}, and for point sources. No evidence for individual sources has been
found yet, as can be seen from the recently published results~\cite{amanda_point} shown in
Figure~\ref{fig:amanda_point}. Note that all areas in the plot showing an enhancement of the flux 
can be explained as statistical fluctuations.  \\ 
In spring 2005, the AMANDA experiment has officially become a part of the km$^3$-sized detector
IceCube, which is under construction at the same site (see below).

\item {\bf NESTOR} \\
NESTOR~\cite{nestor} is a neutrino telescope which is being constructed in 4000\,m depth off the Greek
coast. Other than ANTARES, it uses a rigid structure of towers, each consisting of 12 floors which
carry 12 OMs each. The OMs are symmetrically orientated upwards and downwards, which allows 
for a uniform angular acceptance, such that the background from above can be studied. 
A first test floor of reduced size has been successfully deployed and operated in
2003~\cite{nestor2005a,nestor2005b}.\\
\end{itemize}

The size of the neutrino telescopes mentioned above, about 0.01\,km$^3$, will
only allow to measure a few high-energy events per year, because the fluxes are so
small. As will be explained in Section~\ref{sec:event_types}, some types of events, e.g.~the tau
double bang events, will even be difficult to detect at all, due to their large extension in comparison
to the detector size. \\
To be able to collect data with a higher statistical significance, larger volumes are needed. 
Therefore, km$^3$-sized detectors are being planned or under construction:

\begin{itemize} 

\item {\bf NEMO} \\ 
The NEMO collaboration~\cite{nemo} is doing research and development with regard to a future km$^3$
detector in the Mediterranean. A candidate site for the km$^3$ detector, 70\,km off the Sicilian 
coast in a depth of 3500\,m, has been identified and exhibits good optical properties.

\item {\bf KM3NeT} \\
KM3NeT is a consortium formed by the members of the three neutrino experiments in the Mediterranean,
ANTARES, NEMO and NESTOR, with the objective to build a km$^3$-sized detector in the
Mediterranean. The location on the Northern Hemisphere is complementary to the site of Antarctic
IceCube detector (see below) to allow for the observation of the Southern sky, including a large
part of the Galactic disc and the Galactic Centre, which is barely visible with IceCube. \\
The experiment is now in its R\&D phase; a design study funded by the European Union is currently
being conducted. In the course of this, the technical design will be worked out. 

\item {\bf IceCube} \\
IceCube~\cite{icecube} is the km$^3$-sized extension to the AMANDA neutrino telescope in
the Antarctic Ice. Nine strings have been deployed since the Antarctic summer 2004/2005,
and the completion of the detector is expected for 2011. The detector will then consist of 4800
photomultiplier tubes on 80 lines, each of them over 1000\,m long, with a horizontal spacing of
125\,m to each other. On the surface, an air shower array, IceTop, is 
being built, which will serve for calibration purposes and as a veto for atmospheric muons.

\end{itemize}

\chapter{Signal Events in ANTARES}\label{ch:showers}

This chapter is devoted to the different classes of signal events that occur in the ANTARES detector,
with a particular weight on shower-type events. \\ 
Section~\ref{sec:event_types} provides a closer look at the signatures of the different event classes. 
In Section~\ref{sec:em_showers}, details on the characteristics of electromagnetic cascades are
given. Section~\ref{sec:hadronic_showers} provides similar information for hadronic cascades. \\
The Monte Carlo results shown in this chapter are mostly obtained using ANTARES simulation
software. See Appendix~\ref{sec:software} for a more detailed description of this software.

\section{The Different Event Classes in ANTARES}\label{sec:event_types}

In an electromagnetic field-free environment, charged particles travel along straight lines through
the medium until they either decay or interact with the interaction medium. The mean length of the
distance travelled is called the {\it path length} of the particle and depends on its amount of
energy loss in the medium. If the path length exceeds the spatial resolution of the
detector, so that the trajectory of the particle can be resolved, one calls the trajectory a
particle {\it track}. In a neutrino telescope, one 
can distinguish between two main event classes: events with a track, and events without a track. 

\subsection{Event Classes with a Track}\label{sec:ev_track}

Event classes with a track are

\begin{itemize}
\item $\nu_{\mu},\bar{\nu}_{\mu}$ CC: \quad $(\nu_{\mu},\bar{\nu}_{\mu}) + N \to \mu^{\mp} + $
  hadronic shower 
\item $\nu_{\tau},\bar{\nu}_{\tau}$ CC:\\[1mm] 
\quad $(\nu_{\tau},\bar{\nu}_{\tau}) + N \hspace{0.45mm} \to $ hadronic shower $ \hspace{0.4mm} +
\, ( \tau^{\mp} \to (\nu_{\tau},\bar{\nu}_{\tau}) + (\bar{\nu}_{\mu},\nu_{\mu}) +
\mu^{\mp})$ \\ 
\quad $(\nu_{\tau},\bar{\nu}_{\tau}) + N \to $ hadronic shower $ + \, ( \tau^{\mp} \to
  (\nu_{\tau},\bar{\nu}_{\tau}) + (\bar{\nu}_e,\nu_e) \, + \, (e^{\pm} \to $ em.~shower))
\quad $(\nu_{\tau},\bar{\nu}_{\tau}) + N \hspace{0.45mm} \to $ hadronic shower
  $ \hspace{0.4mm} + \, ( \tau^{\mp} \to (\nu_{\tau},\bar{\nu}_{\tau}) \hspace{1mm} + $\,hadronic
shower). \\ 
\end{itemize}

\begin{figure}[h] \centering
\includegraphics[width=6.8cm]{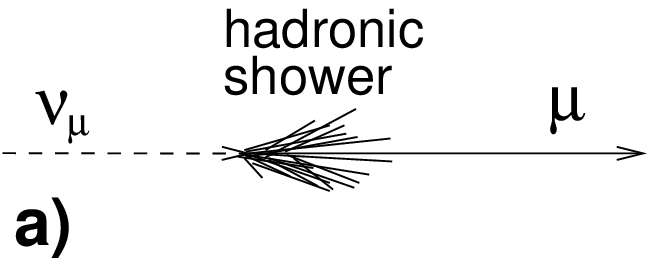}
\includegraphics[width=7.4cm]{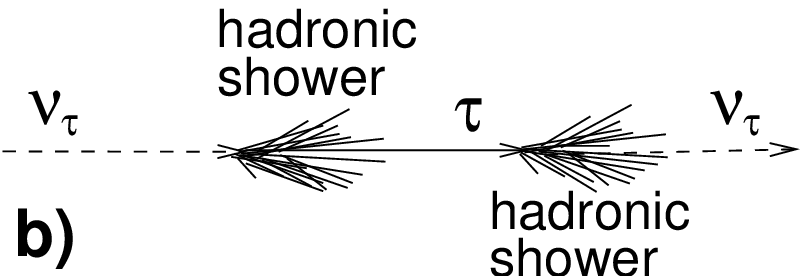}
\vspace{0.5cm}
\includegraphics[width=5.6cm]{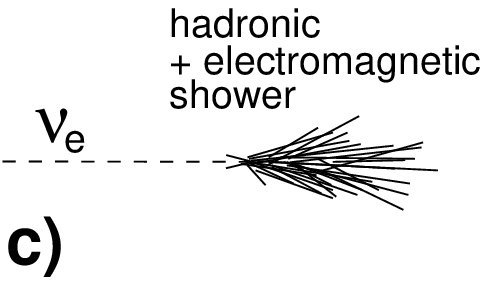}
\includegraphics[width=6.8cm]{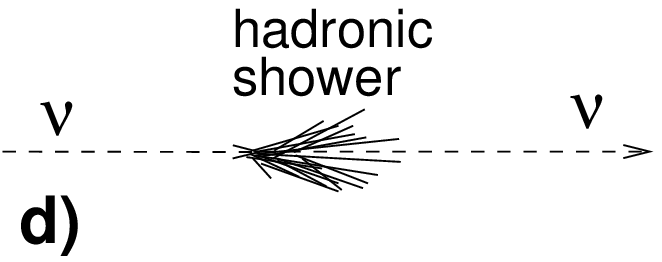}
\caption[Different event classes in ANTARES]{Schematic view of the most important event classes in
  ANTARES, as described in the text: $\nu_{\mu}$ CC (a), $\nu_{\tau}$ CC (\lq\lq double bang
  event\rq\rq) (b), $\nu_e$ CC (c) and $\nu$ NC (d).}
\label{fig:event_classes}
\end{figure}

\begin{figure}[h] \centering
\includegraphics[width=9.4cm]{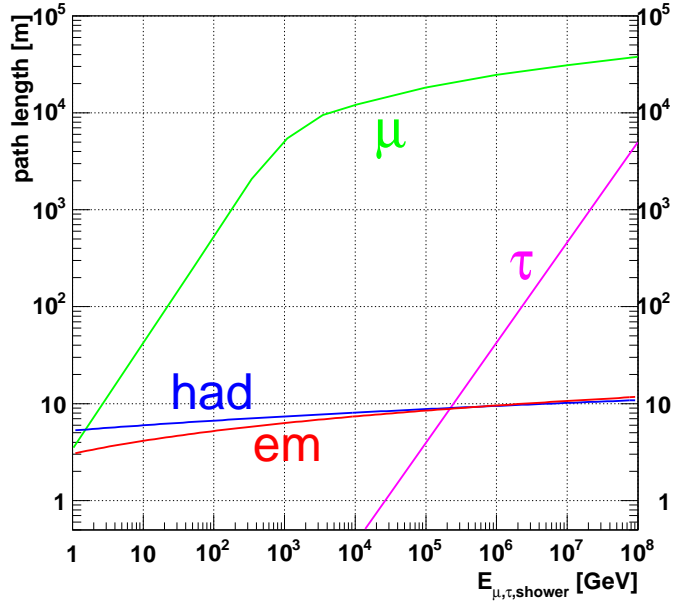}
\caption[Path length of neutrino interaction products]{Path length of neutrino interaction
  products in water: muons, taus, electromagnetic and hadronic showers, over their respective
  energy. The shower lengths are calculated using a shower profile parameterisation and the results
  of~\cite{niess}, see Section~\ref{sec:hadronic_showers} for details. Muons and taus
  after~\cite{muons_taus}. } 
\label{fig:pathlength}
\end{figure}

The last two channels only produce tracks within certain energy ranges, see below. 
In the list, \lq\lq em.~shower\rq\rq\ stands for electromagnetic shower. Schematic views of a
$\nu_{\mu}$ CC event and of the last of the three $\nu_{\tau}$ CC channels are shown in
Figure~\ref{fig:event_classes}. Neutrino and anti-neutrino reactions are
not distinguishable; thus, no differentiation between particles and anti-particles is made. Showers
occur in all event classes shown in Figure~\ref{fig:event_classes}. However, for $\nu_{\mu}$ and
$\bar{\nu}_{\mu}$ CC, often only the muon track is detected, as 
the path length of a muon in water exceeds that of a shower by more than 3 orders of magnitude for
energies above $\sim 2$\,TeV (see Figure~\ref{fig:pathlength}). Therefore, such an event might very
well be detected even if the interaction has taken place several km outside the instrumented volume,
provided that the muon traverses the detector. \\
Because of its highly relativistic velocity, the muon undergoes a large number of interactions
before it decays. For energies above $\sim 2$\,TeV, these interactions are dominated by radiation
losses.  \\ 
For the $\tau$, the situation is different: due to the much shorter lifetime, it travels only a few m
to a few km, depending on its energy, before it decays again according to one of the possible decay
modes given in the list above. Radiation losses, on the other hand, play a much smaller role than
for the muon, because of the 17 times larger mass of the $\tau$. Most of the possible $\tau$ decay modes
include the generation of a hadronic or electromagnetic cascade. Thus, if the track of the $\tau$ is
long enough to distinguish between the primary interaction of the $\nu_{\tau}$ and the decay of the
tau, i.e.~for a $\tau$ energies above $\sim 1$\,PeV (see Figure~\ref{fig:pathlength}), the expected
signatures for the $\nu_{\tau}$ and $\bar{\nu}_{\tau}$ CC events are that of a shower, a track and a
shower; this signature is called \lq\lq double bang event\rq\rq. Alternatively, if the $\tau$ starts or
ends outside the instrumented volume, a track and a shower, a \lq\lq lollipop event\rq\rq, are
detected. Most probably, if events at these high energies are observed in ANTARES at all (the
expected event rate above 1\,PeV is far below 1 event per year), this latter signature is the only
one observable in ANTARES, because the 
tau path length rapidly exceeds the dimensions of the detector for increasing energies. \\
If the $\tau$ decays into a muon, a $\nu_{\tau}$ and a $\nu_{\mu}$, the event is presumably not
distinguishable from an original $\nu_{\mu}$ CC interaction. 

\subsection{Event Classes without a Track}

Event classes without a track are

\begin{itemize}
\item $(\nu_e,\bar{\nu}_e)$ CC: \quad $(\nu_e,\bar{\nu}_e) + N \to $ hadronic shower $ + \, ( e^{\mp}
  \to $ electromagnetic shower) 
\item $(\nu,\bar{\nu})$ NC: \,\, \quad $(\nu_l,\bar{\nu}_l) + N \to (\nu_l,\bar{\nu}_l) + $ hadronic shower
\end{itemize}

Both event classes are depicted schematically in Figure~\ref{fig:event_classes}.  
Due to its small mass, the electron has a high probability for energy loss via bremsstrahlung, so
that it does not produce a resolvable track in the detector, but interacts immediately after its
generation, producing an electromagnetic shower. \\
The NC channel gives the same signature for all neutrino flavours. In this channel, a part of the
interaction energy is always carried away unobserved by the outgoing neutrino, and therefore the
error on the reconstructed energy of the primary neutrino increases accordingly. Even though
electromagnetic and hadronic showers are different from each other in principle (see
Sections~\ref{sec:em_showers} and~\ref{sec:hadronic_showers}), the $\nu_e / \bar{\nu}_e$ CC and the
$\nu$ NC channels are not distinguishable in reality, because the detector is too sparsely
instrumented.\\
Below $\sim 1$\,PeV, also the $\nu_{\tau}$ CC channels, except for the case where the $\tau$ produces a
muon, belong to the class of events without a track, because the $\tau$ track cannot be resolved
at these energies (see Figure~\ref{fig:pathlength}).

\subsection{Other Event Classes}

The resonance channel of $\bar{\nu}_e$ at 6.3\,PeV, the Glashow resonance, is a special case which is
mentioned here for completeness. The following signatures are possible for this resonance:

\begin{itemize}
\item $\bar{\nu}_e + e^- \to \bar{\nu}_e + ( e^- \to $ electromagnetic shower)
\item $\bar{\nu}_e + e^- \to \bar{\nu}_{\mu} + \mu^-$
\item $\bar{\nu}_e + e^- \to \bar{\nu}_{\tau} + ( \tau^- \to \tau^-$ decay modes (see
  Section~\ref{sec:ev_track}) ) 
\item $\bar{\nu}_e + e^- \to q + \bar{q\,}^{\prime} \to $ hadronic shower
\end{itemize}

For the energy range examined in this study, between $\sim 100$\,GeV and $\sim
100$\,PeV, the resonance channel constitutes only a small portion to the overall cross section. It
will also not be identifiable as such in the experiment: In the case of $\bar{\nu}_e + e^-$ or
$q\bar{q\,}^{\prime}$ final states, the interaction has the same characteristics as a 
\lq\lq normal\rq\rq\ shower event without a track. As it is presumably not possible to
determine between a track traversing the detector and a track starting inside the
detector, neither of the two other channels will be identifiable: In the case of $\bar{\nu}_{\tau} +
\tau^-$, the interaction would be regarded as a $\bar{\nu}_{\tau}$ CC with the primary neutrino
interaction outside  the detector and in the case of $\bar{\nu}_{\mu} + \mu^-$, the only event type
without any shower, it would be interpreted as a $\nu_{\mu}$ CC event happening outside the
detector. In a larger, km$^3$ sized detector, this would be different, because it would be possible
to determine whether a particle track starts within the instrumented volume or outside of it. For
ANTARES, where the majority of the strings are situated at the edge of the detector (see
Figure~\ref{fig:ant_layout} in Section~\ref{sec:layout}), this seems very improbable.

\subsection{Event Classification}

\begin{figure}[h] \centering
\includegraphics[width=7.cm]{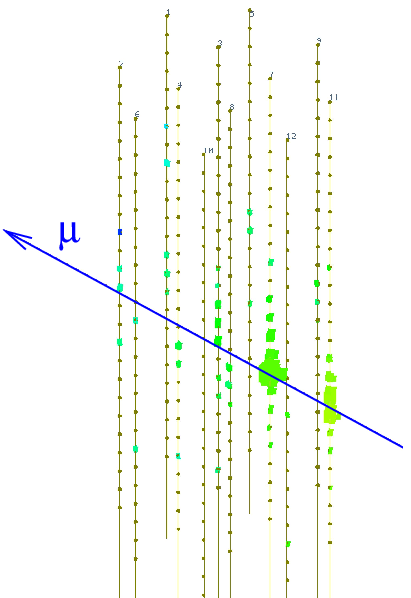}
\includegraphics[width=7.cm]{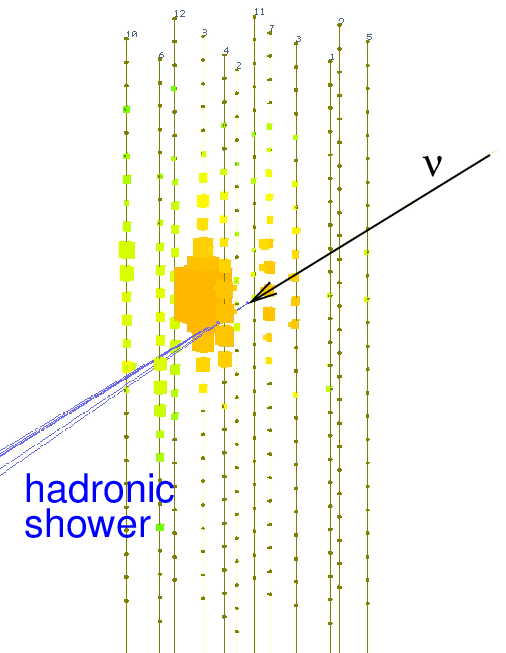}
\caption[Simulated ANTARES Events]{Passage of a muon through the ANTARES detector (left) and an NC 
  interaction inside the detector (right), as visualised by the ANTARES event display 
  A3D~\cite{a3d}. The shower length is not to scale. See text for details.} 
\label{fig:event_display}
\end{figure}

Figure~\ref{fig:event_display} shows two different Monte Carlo events in the ANTARES detector in
comparison. On the left, the passage of a muon (blue) through the
detector is shown. The muon is travelling from the lower right to the upper left in the shown
perspective; its initial energy was 19\,TeV. On the right, an NC interaction is shown; the neutrino,
marked in black, comes from the upper right and interacts in the middle of the detector, generating
a hadronic shower (blue). The energy of the hadronic shower was also 19\,TeV.  Note that the shower
constituents have been extrapolated according to their directions; the length of the lines is not to
scale. The events have been visualised using the ANTARES event display 
A3D~\cite{a3d}. The storeys in the strings are drawn as small dots; the photomultiplier signals,
integrated over 25\,ns, are shown as squares, with a size proportional to the signal amplitude,
and a colour coding according to the arrival time of the photon signal in the
photomultipliers (yellow to green to light blue). \\ 
One can clearly see that the shower deposits its entire energy in a relatively small volume, while
the muon loses only a part of its energy inside the detector.

\section{Electromagnetic Cascades}\label{sec:em_showers}

The evolution of an electromagnetic cascade can be described in a very simple way: An electron
suffers bremsstrahlung and produces photons. Each photon produces an electron-positron pair via
pair production. The electron and the positron again produce photons via bremsstrahlung, and so
on, until the energy of the constituents falls below the critical energy\footnote{for electrons, the
  critical energy is usually defined as the energy at which the rates of bremsstrahlung and
  ionisation losses are equal~\cite{pdg}.} and 
the shower production stops; the remaining energy is then dissipated by ionisation and
excitation. For electromagnetic showers it can be assumed that all Cherenkov light is emitted
isotropically in azimuth $\phi$ from the shower axis, as the lateral extension of an electromagnetic
shower is of the order of 10\,cm~\cite{pdg} and therefore negligible in comparison to the
longitudinal one. Thus, the description  of an electromagnetic shower simplifies to describing the
longitudinal profile of the shower and the angular profile of the emitted Cherenkov light. 

\subsection{Longitudinal Profile}
The longitudinal profile of the energy deposition in the shower can be
parameterised as~\cite{pdg} 

\begin{equation}\label{eq:long_shower_profile}
\frac{dE}{dt} = E_0 b \frac{(bt)^{a-1}e^{-bt}}{\Gamma (a)},
\end{equation}

where $E_0$ is the initial shower energy, $a$ and $b$ are parameters depending on the initial energy
and particle type, and $t$ is the distance from the interaction vertex along the shower axis in
units of radiation lengths, $t = z/X_0$, with $X_0 = 35$\,cm in salt water at the ANTARES
site~\cite{brunner_cherenkov}. One finds that the maximum of distribution
(\ref{eq:long_shower_profile}) is at 

\begin{equation}\label{eq:shower_max}
t_{max} = (a-1) / b.
\end{equation}

The parameters $a$ and $b$ for electron- and photon-induced interactions in salt water have been
determined by various authors~\cite{wiebusch,brunner_cherenkov,niess}. Their results are
listed in Table~\ref{tab:shower_a}, showing good compatibility within the
typical precision of around 10\%, except for the slightly larger constant offset
of~\cite{niess}. Note that~\cite{niess} uses a different form of equation 
(\ref{eq:shower_max}), $t_{max} = D \ln(E/E_c) + C$, from  which the expression for $a$ can be
calculated using the values cited for $b$ and a critical energy of $E_c = 54.27$\,MeV. The numbers
given here are valid up to $\sim 1$\,PeV; above this energy, the LPM effect, a suppression of pair
production and bremsstrahlung, has to be taken into account. \cite{niess} uses the
GEANT4~\cite{geant4} simulation software to simulate the particle interactions in water, while the
other two have done their simulations with the older version GEANT3.21~\cite{geant}.  

\begin{table}[h] \centering
\begin{tabular}{|c||c|c||c|c|}
\hline
 & \multicolumn{2}{c||}{shower parameter $a$} & \multicolumn{2}{c|}{shower parameter $b$ } \\ \hline
author & $e^-$ & $\gamma$ & $e^-$ & $\gamma$  \\ \hline
\cite{wiebusch} &  $2.0 + 0.60 \ln(E/\textrm{GeV})$ & (not calculated) & 0.63 & (not calculated) \\ \hline
\cite{brunner_cherenkov} & $1.9 + 0.64 \ln (E/\textrm{GeV})$ 
& $2.6 + 0.64 \ln (E/\textrm{GeV})$ & 0.66 & 0.66 \\ \hline
\cite{niess} & $2.6 + 0.69 \ln(E/\textrm{GeV})$ 
& $3.4 + 0.74 \ln(E/\textrm{GeV}) $  & 0.69 & 0.74 \\ 
\hline
\end{tabular}
\caption[Electromagnetic shower parameters]{Mean value of the longitudinal shower profile
  parameters $a$ and $b$ for electrons and photons, as determined by various
  authors. The value of $b$ cited for~\cite{brunner_cherenkov} was obtained by calculating the mean
  from the values given for different energies; the author himself used $b = 0.64$ to obtain 
  parameter $a$. $b$ is expected to be constant in energy and identical for electrons and photons.}
\label{tab:shower_a}
\end{table}

\begin{figure}[h] \centering
\includegraphics[width=10cm]{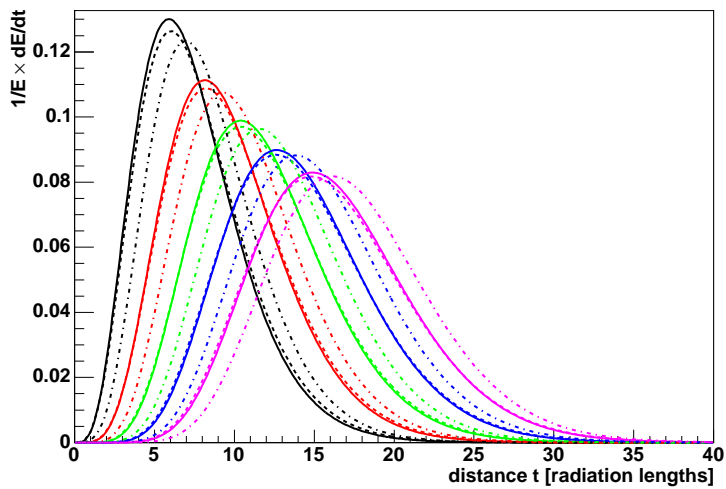}
\caption[Longitudinal profiles of electron-induced showers]{Longitudinal profiles of
  electron-induced showers at energies of 100\,GeV (black), 1\,TeV (red), 10\,TeV (green), 100\,TeV
  (blue) and 1\,PeV (magenta), as parameterised
  by~\cite{wiebusch} (dashed lines),~\cite{brunner_cherenkov} (solid lines) and~\cite{niess}
  (dashed-dotted lines).} 
\label{fig:long_prof_em}
\end{figure}

As an example, the longitudinal profiles for electron-induced showers at 100\,GeV, 1\,TeV, 10\,TeV,
100\,TeV and 1\,PeV, using the values from Table~\ref{tab:shower_a}, are shown in
Figure~\ref{fig:long_prof_em}. The shapes of all distributions are very similar to each other, but
the positions of the curves derived from~\cite{niess} are shifted to higher values of $t$. 

The longitudinal profiles are used to parameterise the total shower length as a function of the initial
shower energy. The shower length is defined here as the distance within which 95\% of the total
shower energy has been deposited. Integrating the curves shown in Figure~\ref{fig:long_prof_em}
numerically, and parameterising the results as linear functions in $\log_{10} (E)$,
one obtains the expressions for the shower length listed in Table~\ref{tab:em_shower_length};
again, the agreement of the results is within 10\%.

\begin{table}[h] \centering
\begin{tabular}{|c|c|}
\hline
author & path length $L$ [m] \\ \hline
\cite{wiebusch} & $2.93 + 1.04 \log_{10} ( E /$GeV ) \\ \hline
\cite{brunner}  & $2.79 + 1.06 \log_{10} ( E /$GeV ) \\ \hline
\cite{niess}    & $3.04 + 1.09 \log_{10} ( E /$GeV ) \\ \hline
\end{tabular}
\caption[Length of electron-induced showers]{Path length $L$ for electron-induced showers as a function
  of the shower energy, parameterised using the results given by the different authors.}
\label{tab:em_shower_length}
\end{table}

For this study, the values of~\cite{niess} have been used to retrieve the shower length and the
position of the maximum, because they are the most recent ones; the new GEANT4 simulation uses new
results for the cross sections at high energies, which makes its output more reliable than that
obtained with the older version, GEANT3.21. See Figure~\ref{fig:pathlength} in
Section~\ref{sec:event_types} for a graphical representation of the electromagnetic shower length as a
function of the shower energy.

\subsection{Angular Profile}

\begin{figure}[h] \centering
\includegraphics[width=7.4cm]{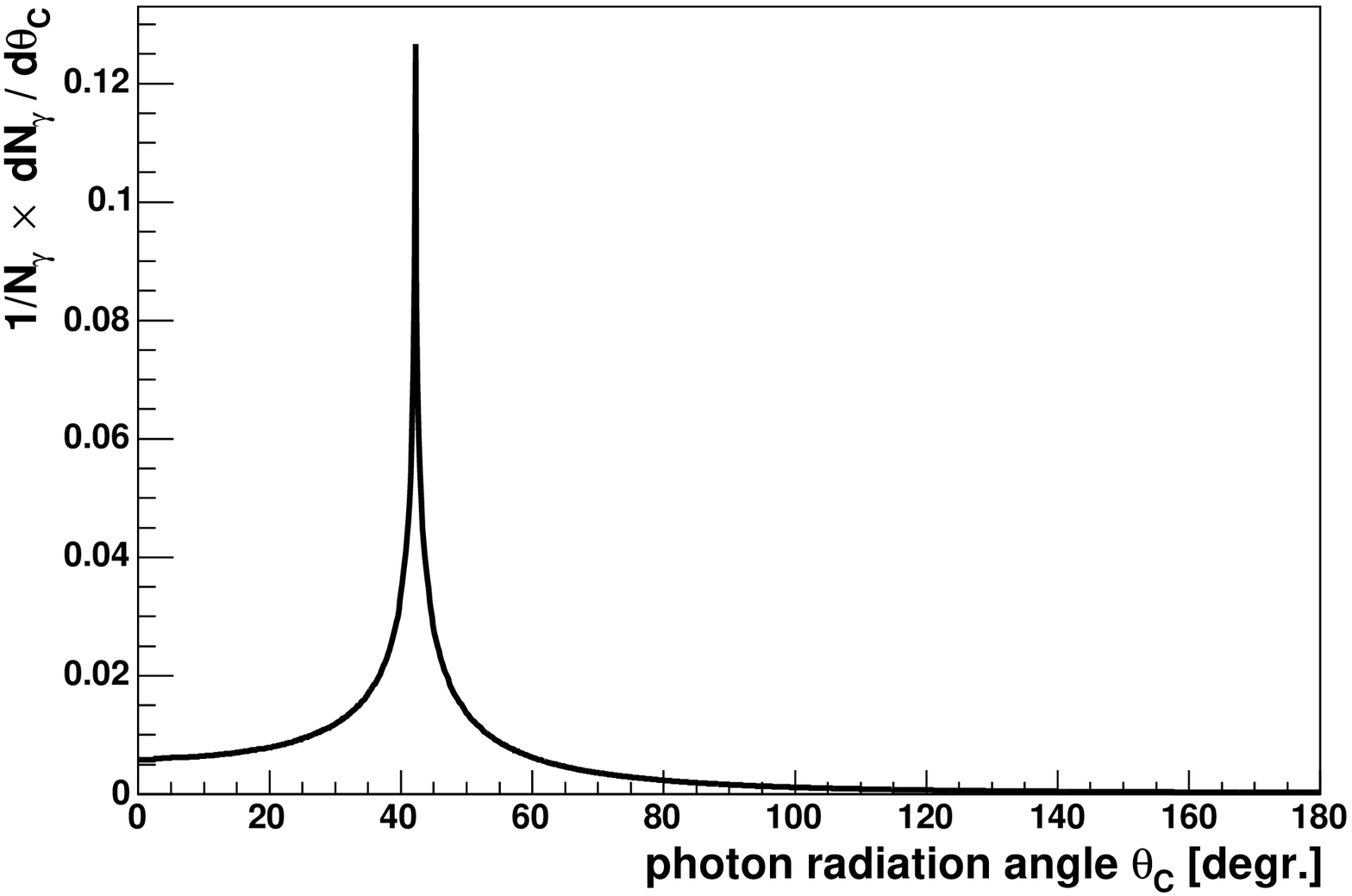}
\includegraphics[width=7.4cm]{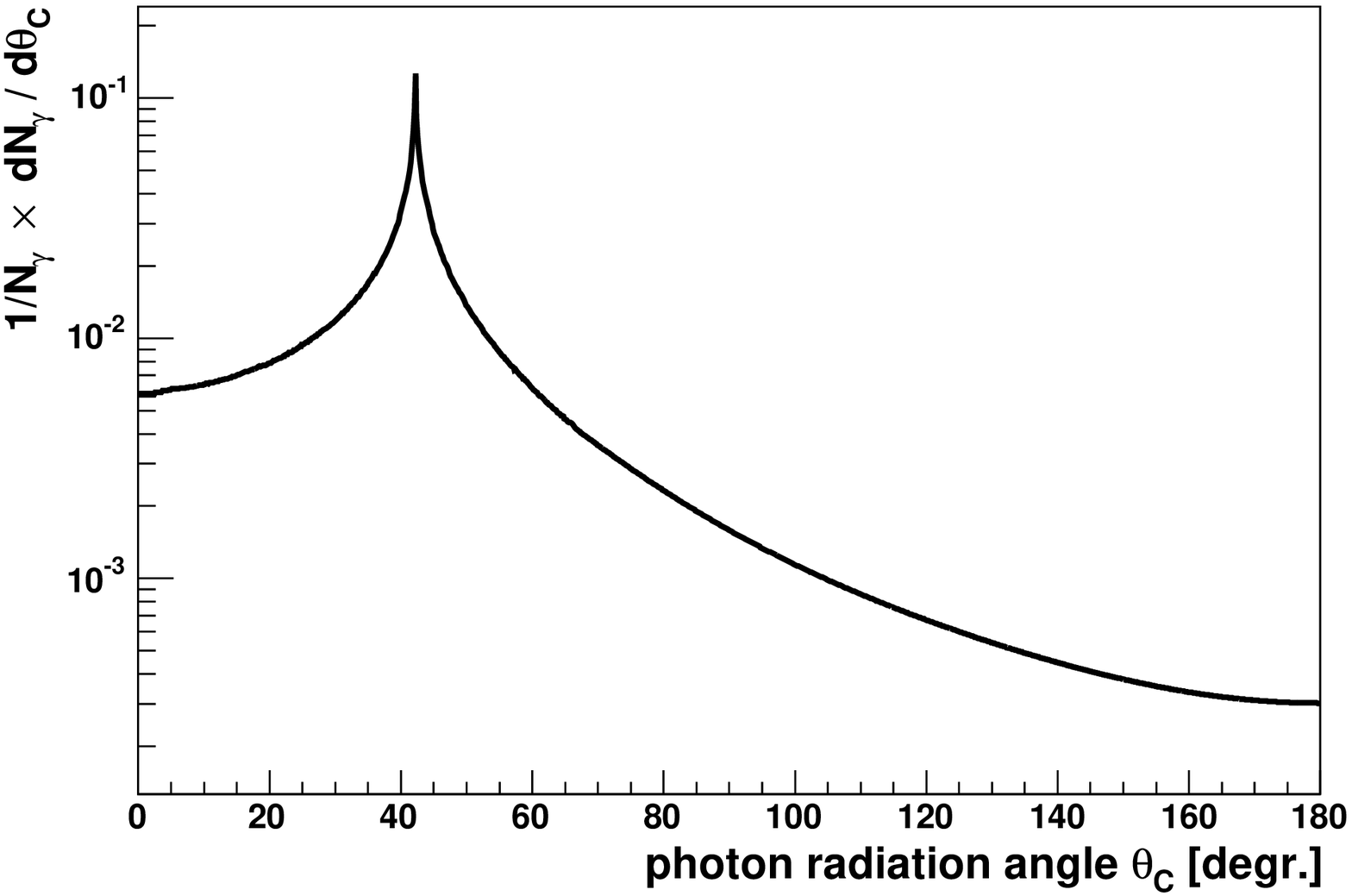}
\caption[Angular distribution for electromagnetic showers]{Angular distribution of photons in an 
  electromagnetic shower with respect to the shower axis, as parameterised within the ANTARES
  simulation software (see Appendix~\ref{sec:geasim}), following the studies
  of~\cite{brunner_cherenkov}. Left: linear $y$-scale, right: logarithmic $y$-scale. } 
\label{fig:cherenkov_em_all}
\end{figure}

It has been found both by~\cite{wiebusch} and by~\cite{brunner_cherenkov} that above 1\,GeV, the 
distribution of the Cherenkov photon radiation angle $\vartheta_C$ with respect to the shower axis is
energy-independent for electromagnetic showers, and very well reproduceable from event to event. It
is therefore possible to save computing time during the shower simulation by calculating the total
number of emitted Cherenkov photons according to the shower energy and assuming a fixed angular profile for
the photons, instead of producing the photons separately for each shower particle. In the ANTARES
simulation software (see Appendix~\ref{sec:geasim}), this has been done by parameterising the
angular distribution in the region of $\cos \vartheta_C < 0.4$, i.e.~$\vartheta_C \gtrsim 66^{\circ}$,
following the results of~\cite{brunner_cherenkov} for the simulation of electromagnetic showers. The
region of smaller angles is hard-coded in the simulation package, with a bin width of $1 \times
10^{-3}$ in $\cos \vartheta_C$. The overall distribution obtained that way is shown in
Figure~\ref{fig:cherenkov_em_all}. 

\section{Hadronic Cascades}\label{sec:hadronic_showers}

\begin{figure}[h] \centering
\includegraphics[width=10cm]{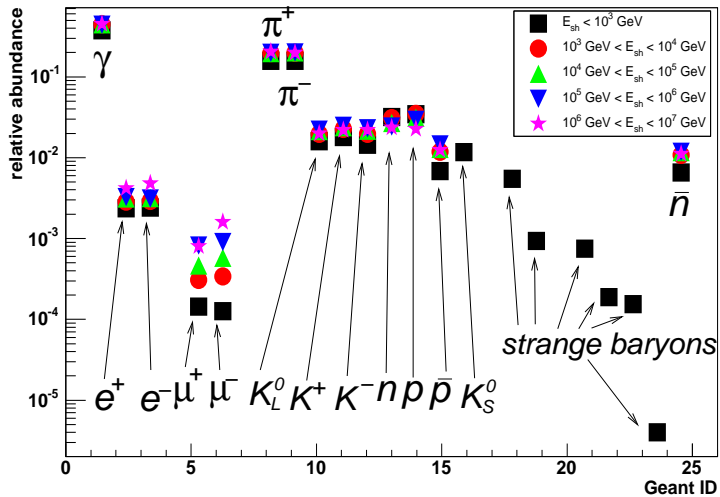}
\caption[Composition of a hadronic shower]{Composition of hadronic showers: relative abundance of
  secondary particles in hadronic showers with different primary energies, as simulated by
  the ANTARES event simulation software genhen (see Appendix~\ref{sec:genhen}). Note that only 
  long-lived particles are displayed by the simulation software. }
\label{fig:shower_composition}
\end{figure}

The description of hadronic cascades is less straight-forward than that of electromagnetic
ones, the main reason being that event-to-event fluctuations are much more important in
hadronic showers, since they do not consist only of two particle types like electromagnetic
cascades, but of many, and the fraction of the different particle types depends on the shower
energy. The dominant secondary particles in a hadronic shower are pions; 
other hadrons like kaons, protons or neutrons occur in variable fractions. A number of muons can be
present, as well; as these usually leave the shower producing long tracks, they contribute
significantly to the fluctuations. For increasing shower energy, the electromagnetic component in
the hadronic shower increases, because the number of $\pi^0$ increases significantly, due to the
interactions of the charged pions; at lower energies, these charged pions decay before they can
interact, producing mainly muons.
$\pi^0$ have a much shorter lifetime than charged pions and decay into two
photons; these photons are then the origin of an electromagnetic cascade. The percentage of
electrons or positrons on the total track length of a hadronic shower exceeds 90\% for a shower
energy of 1\,TeV~\cite{brunner_cherenkov}, which means that above this energy, the largest part of
the Cherenkov light in the shower is generated by electromagnetic sub-showers. \\
Figure~\ref{fig:shower_composition} shows the relative abundance of shower particles for different
shower energies, as simulated by the ANTARES neutrino interaction package (see
Appendix~\ref{sec:genhen}). The output of this simulation consists of all long-lived particles,
i.e.~particles with a lifetime $\gtrsim 10^{-10}$\,s, so that e.g.~$\pi^0$ occur only through
the photons that have been produced in their decays. Consequently, the photon
abundance in the shower is very high. Note also that only the primary shower particles are displayed here;
e.g.~secondary electrons produced during cascading, or Cherenkov photons, are not displayed, as they
are not generated at this simulation step, but only in the propagation that follows. 

\subsection{Longitudinal Profile}

\begin{figure}[h] \centering
\includegraphics[width=10cm]{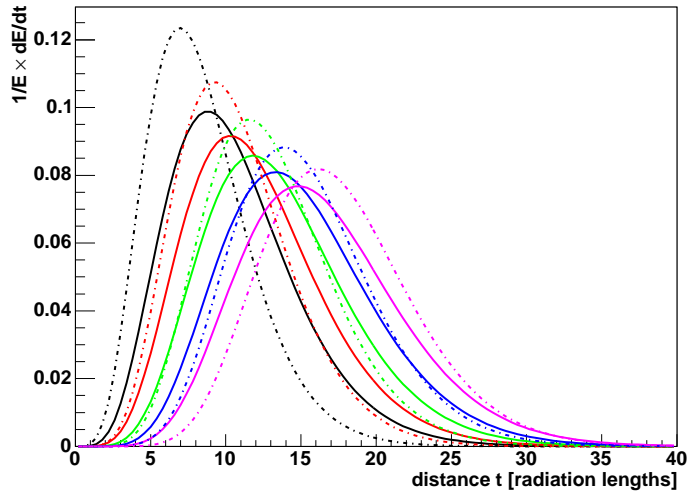}
\caption[Longitudinal profiles for hadronic showers]{Longitudinal profiles for a pion-induced
  hadronic shower at energies of 100\,GeV (black), 1\,TeV (red), 10\,TeV (green), 100\,TeV (blue)
  and 1\,PeV (magenta) (solid lines), as parameterised by~\cite{niess}. The longitudinal profiles for
  an electron-induced shower as parameterised by~\cite{niess} at the same energies are shown as
  dashed-dotted lines, for comparison.} 
\label{fig:long_prof_had}
\end{figure}

One of the authors mentioned in the previous section,~\cite{niess}, has
used his parameterisation for the electromagnetic showers, equation~(\ref{eq:long_shower_profile}),
also to determine the longitudinal profile of hadronic showers. He simulated the hadronic cascade
generated by a primary charged pion at different energies. 
The values he finds for the parameters $a$ and $b$ are given in Table~\ref{tab:shower_a_had}. 
The resulting longitudinal profiles from this parameterisation are shown 
in Figure~\ref{fig:long_prof_had} for shower energies of 100\,GeV, 1\,TeV, 10\,TeV, 100\,TeV and
1\,PeV, in units of the electromagnetic radiations length in salt water, $X_0 = 35$\,cm, to allow
for a comparison with the electromagnetic showers. The profiles for electron-induced showers at the
same energies, following the results of~\cite{niess}, are shown as dashed-dotted lines in the same
figure. Below 100\,TeV, the electromagnetic shower has a shorter length than the hadronic shower. 
While the shower maximum for electromagnetic showers with energies $\geq 100$\,TeV lies at
a larger distance from the interaction vertex than for hadronic showers, the overall shower length
becomes approximately equivalent for both shower types (compare Figure~\ref{fig:pathlength}). \\
Due to the increasing electromagnetic component, fluctuations in the hadronic shower shape are less
dominant at higher energies. \cite{niess} finds that the fluctuations with respect to the
longitudinal profile are around 10\% at some 10\,TeV; they exceed 30\% in the region of some hundred
GeV. The shower parameter $b$ showed no constant behaviour below $\sim 5$\,TeV. For shower energies
in this energy region, the parameterisations presented here should therefore be used with care. 

\begin{table}[h] \centering
\begin{tabular}{|c||c|}
\hline
parameter & $\pi^\pm$ \\ \hline
$a$ & $4.26 + 0.364 \ln(E/\textrm{GeV})$  \\ 
$b$ & 0.56 ($E \ge 10$\,TeV) \\
\hline
\end{tabular}
\caption[Hadronic shower parameters]{Mean value of the longitudinal shower profile
  parameters $a$ and $b$ for charged pions, as determined by~\cite{niess}. The parameter $b$ is
  expected to be constant in energy.}
\label{tab:shower_a_had}
\end{table}

One can again use the longitudinal profile, as explained in Section~\ref{sec:em_showers}, to
parameterise the total shower length. The shower length $L$ was again defined as the distance within
which 95\% of the total shower energy is deposited. It can be described as

\begin{gather}
L/\textrm{m} = 5.28 + 0.70 \log_{10} ( E /\textrm{GeV} ).
\label{eq:had_shower_length}
\end{gather}

\begin{figure}[h] \centering
\includegraphics[width=7.4cm]{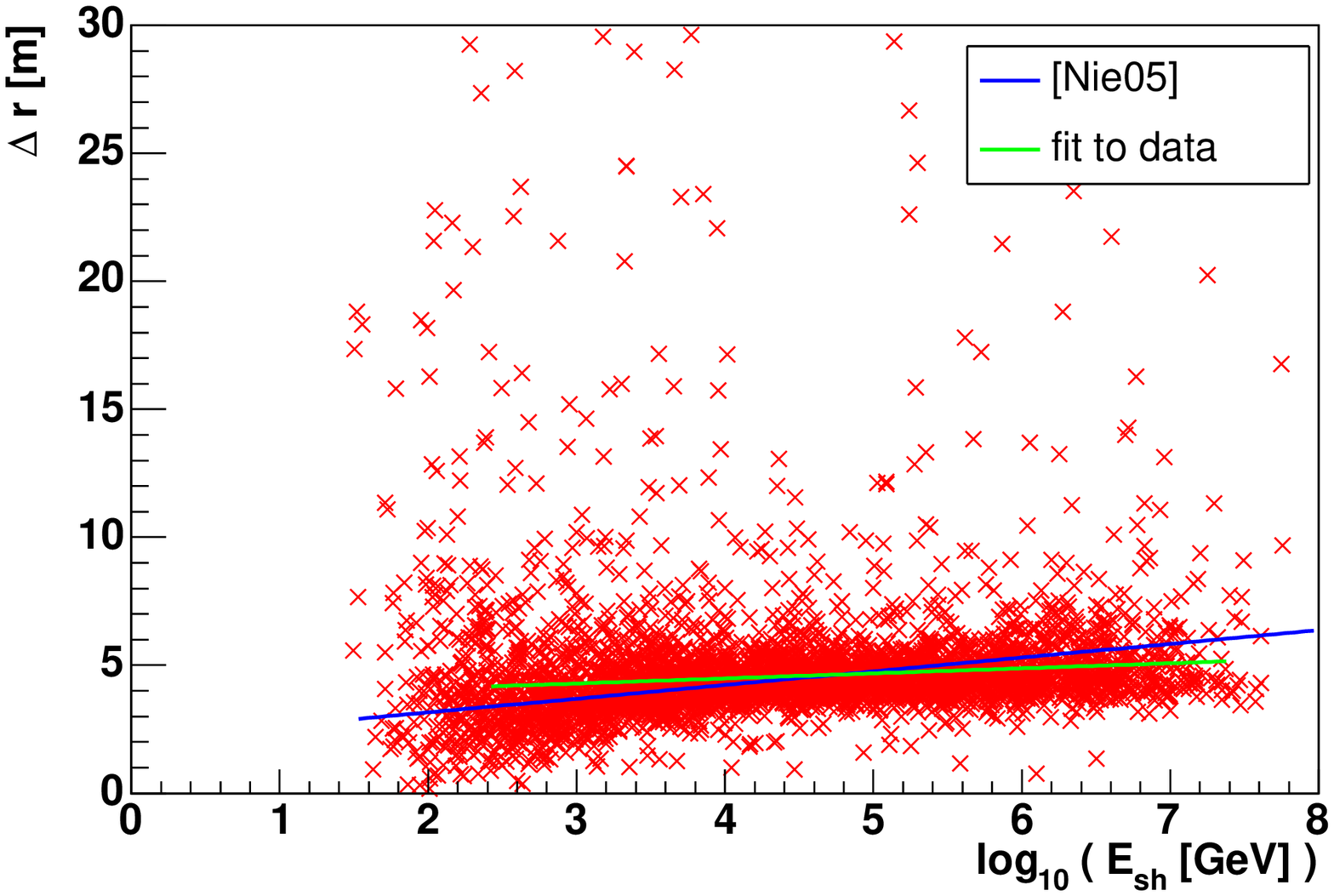}
\includegraphics[width=7.4cm]{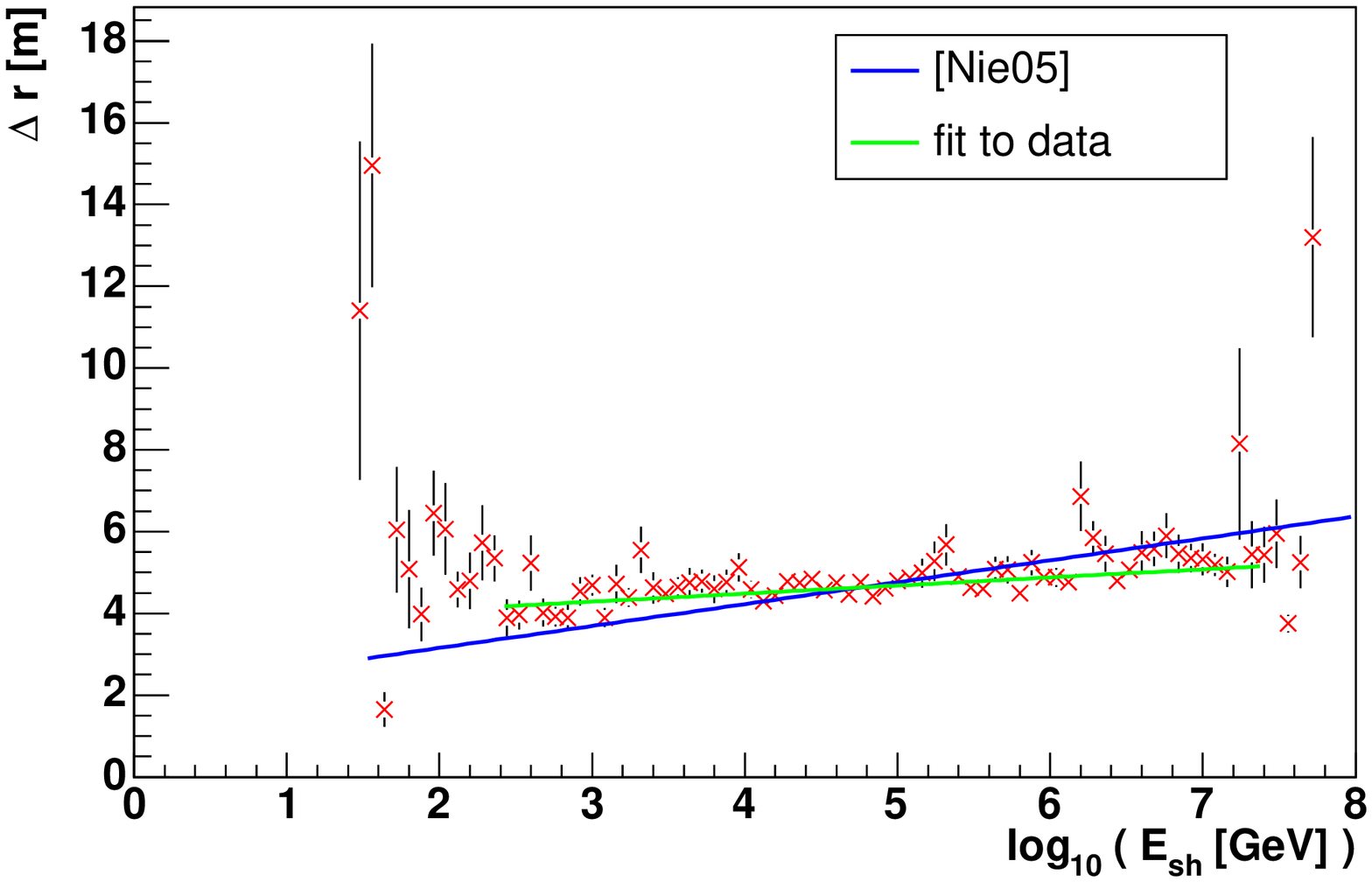}
\caption[Calculated shower maximum for hadronic showers]{Distance between the MC interaction vertex
  and the amplitude-weighted centre-of-gravity \rcga\ as calculated for the NC event sample A, a fit
  to the data (green), and distance between the MC interaction vertex and the position of the shower
  maximum as predicted by~\cite{niess} (blue), all as a function of the shower energy. See text for
  details.}  
\label{fig:shower_maximum}
\end{figure}

The length of a hadronic shower using equation~(\ref{eq:had_shower_length}), in comparison to the
electromagnetic shower length and the muon and $\tau$ ranges is shown in Figure~\ref{fig:pathlength}
(Section~\ref{sec:event_types}).  \\
The maximum of the shower retrieved from Table~\ref{tab:shower_a_had} was compared to the shower
maximum as calculated in a study that was conducted within the context of this thesis.  For this
study, the ANTARES simulation tools (see Appendix~\ref{sec:software}) were used to generate a sample of
NC events with shower energies between $\sim 40$\,GeV and $\sim 100$\,PeV. In the following, this
event sample will be denoted event sample A (Appendix~\ref{sec:data_sample} gives a detailed
description of the different event samples that were used for this work). \\
The output of the Monte Carlo simulation contains no information about the longitudinal extension of
the shower and thus, no information about the position of the shower maximum is
available. However, from the timings and positions of the measured {\it hits}, one can
reconstruct the point from where the largest fraction of Cherenkov photons is emitted, according to
an algorithm described in Section~\ref{sec:pos}. Here and in the following, the term {\it hit} is
used to denote a set of photons which reach a photomultiplier within the 25\,ns
integration time that follow the first photon signal in the photomultiplier (see
Section~\ref{sec:digitisation}). The number of photo-electrons (pe) in a hit determines the {\it hit
  amplitude}. Taking into account the hit amplitudes for the calculation, by weighting each hit
position and time by its amplitude, a {\it centre-of-gravity} of the shower, \rcga, can be
determined as an estimate of the central point of photon emission. This centre-of-gravity is
approximately equivalent to the shower maximum, as can be seen from Figure~\ref{fig:shower_maximum}:
On the left hand side, the distance $\Delta r$ between the interaction vertex and \rcga,
calculated as described above, is shown in red for each event. On the right hand side, the profile
of the calculated data points is shown. A straight line that was fitted to this profile is shown in
green in both graphs. The length of the line marks the validity range of the fit. For energies below
$\sim 1$\,TeV, the event-to-event fluctuations begin to dominate the distribution. The fit function
is given as  

\begin{equation}\label{eq:shower_max_amp}
\Delta r(E)/\textrm{m} = 3.7 +  0.20 \log_{10} ( E /\textrm{GeV} ).
\end{equation}

The position of the shower maximum with respect to the interaction vertex, as expected from the
simulations of~\cite{niess}, is shown in blue in both graphs. The agreement between the two lines is
relatively good, considering the different methods that were used, though the increase of the fit
line with energy is not as strong as predicted by~\cite{niess}. It should also be noted that the
shower maximum may not necessarily be equivalent to the point of maximum photon emission.

\subsection{Angular Profile}

\begin{figure}[h] \centering
\includegraphics[width=12cm]{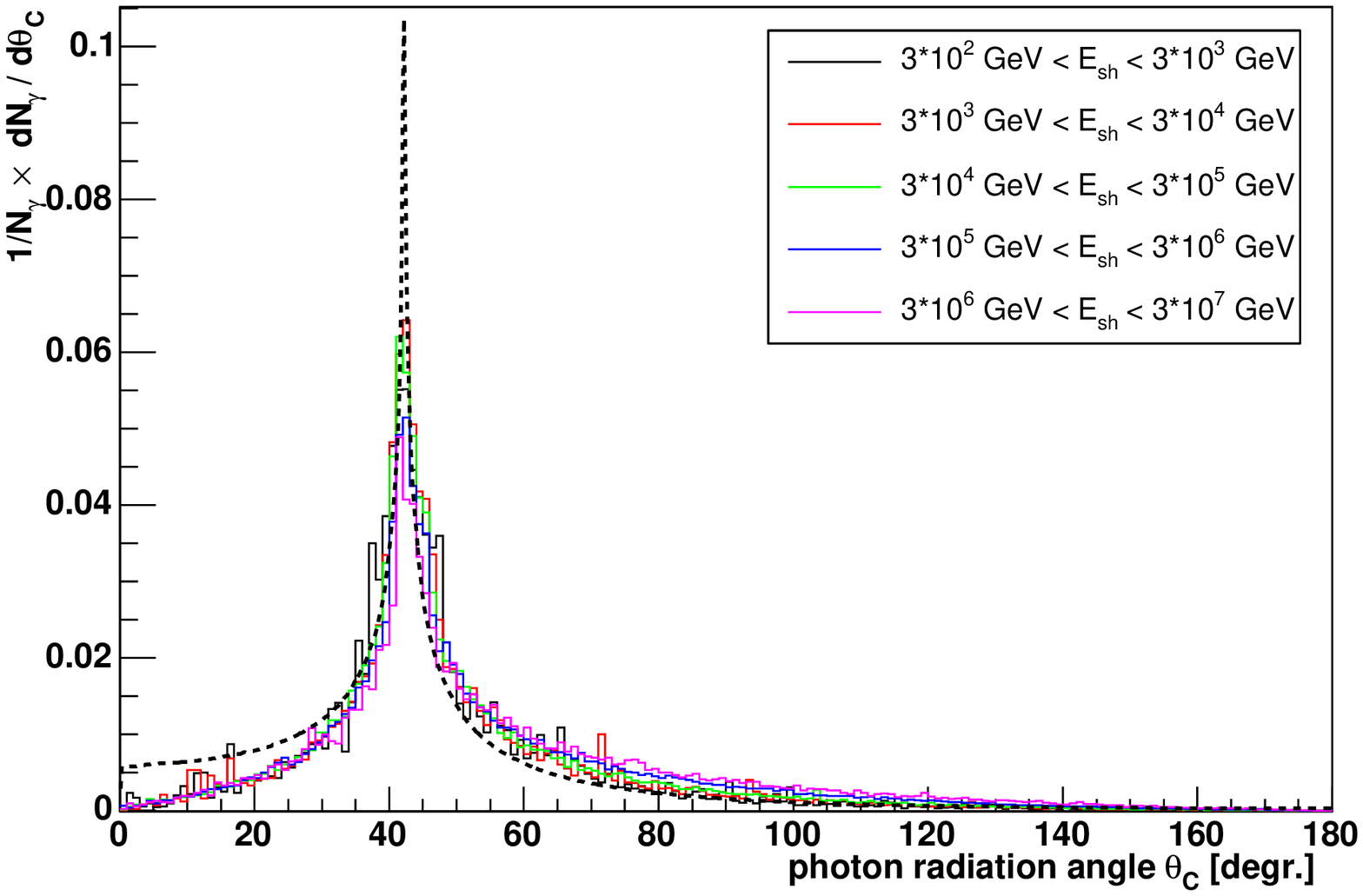}
\includegraphics[width=12cm]{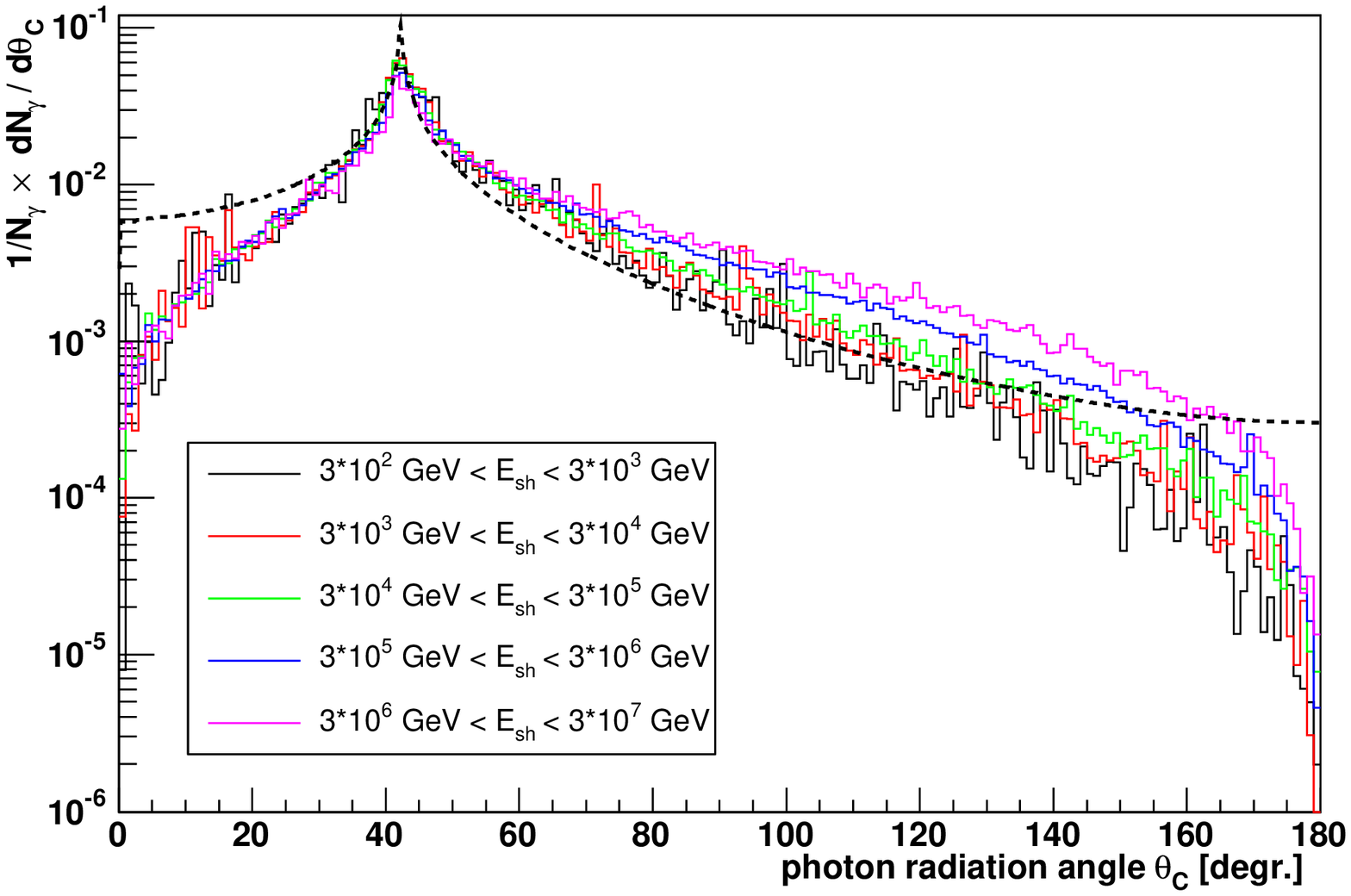}
\caption[Angular distribution for hadronic showers]{Angular distribution of photons from a 
  hadronic shower with respect to the shower axis for different energies.
  Top: linear vertical scale, bottom: logarithmic vertical scale. The angular distribution for
  electromagnetic showers as parameterised by~\cite{brunner_cherenkov} is shown as a dashed line for
  comparison. See text for details.} 
\label{fig:cherenkov_all}
\end{figure}

Same as the longitudinal shower profile, also the angular profile of the Cherenkov photons suffers
from large fluctuations for hadronic showers, in particular for low energies. Considering different
energies, it also becomes obvious that the angular distribution slightly broadens with increasing
energy, which might by due to the increasing total number of particles in the
shower. Figure~\ref{fig:cherenkov_all} shows the 
angular distribution for showers between $3 \times 10^2$ and $3 \times 10^7$\,GeV from event sample
A; each histogram contains events within one decade in energy. Shown is the angle between the
photons which produced a hit in the ANTARES detector and the shower axis. The angle was calculated
under the assumption that all photons are emitted from one point in the shower. It will be shown
later in Section~\ref{sec:pos} that the position reconstructed by the algorithm used for this study
does not refer exactly to the shower maximum, but to a distance to the interaction vertex that can
be parameterised using equation~(\ref{eq:shower_max_corr}) from Section~\ref{sec:pos}. The point of
the photon emission was therefore calculated by moving the reference point from the Monte Carlo (MC)
interaction vertex according to ${2.8 + 0.48 \cdot \log_{10} ( E /\textrm{GeV} )}$ along the shower
axis. Effects due to detector geometry characteristics can not be fully excluded but are expected to
cancel out due to large statistics and an isotropic event sample.  \\  
In Figure~\ref{fig:cherenkov_all}, the parameterisation of the angular profile for electromagnetic
showers (dashed line) as determined by~\cite{brunner_cherenkov} (see Section~\ref{sec:em_showers}),
is also shown for comparison. Close to the Cherenkov peak, the distribution does not differ much the
one for hadronic showers; it lies below the histogram of the hadronic showers with the lowest
energy, for angles between the maximum and $\sim 100^{\circ}$. This agrees with the expectation
that the angular distribution is narrower for an electromagnetic shower than for a hadronic shower,
as no energy dependence was found for the distribution of the electromagnetic shower. However, for
angles below $\sim 35^\circ$ and above $\sim 130^\circ$ (depending on the shower energy), the photon
yield for the hadronic shower is found to lie below the photon yield for the electromagnetic shower
and is therefore probably underestimated with the method used here. This effect is possibly caused
by the fact that especially low Cherenkov thresholds of $< 1$\,MeV were used for electrons and photons
in~\cite{brunner_cherenkov}, while for the angular distributions of the hadronic showers retrieved
in this study, the nominal value in the ANTARES software~\cite{geasim} for the high-energy mode,
500\,MeV, was used to save computing time. One should also be aware that what is compared here is on
the one side a photon distribution of electromagnetic showers which has been retrieved from the
Monte Carlo truth of the Cherenkov photon simulation, and on the other side distributions of
photo-electrons from hadronic showers as measured in the detector according to the ANTARES
simulation software.

\subsection{Angle Between Neutrino and Hadronic Shower}\label{sec:shower_angle}

\begin{figure}[h] \centering
\includegraphics[width=10cm]{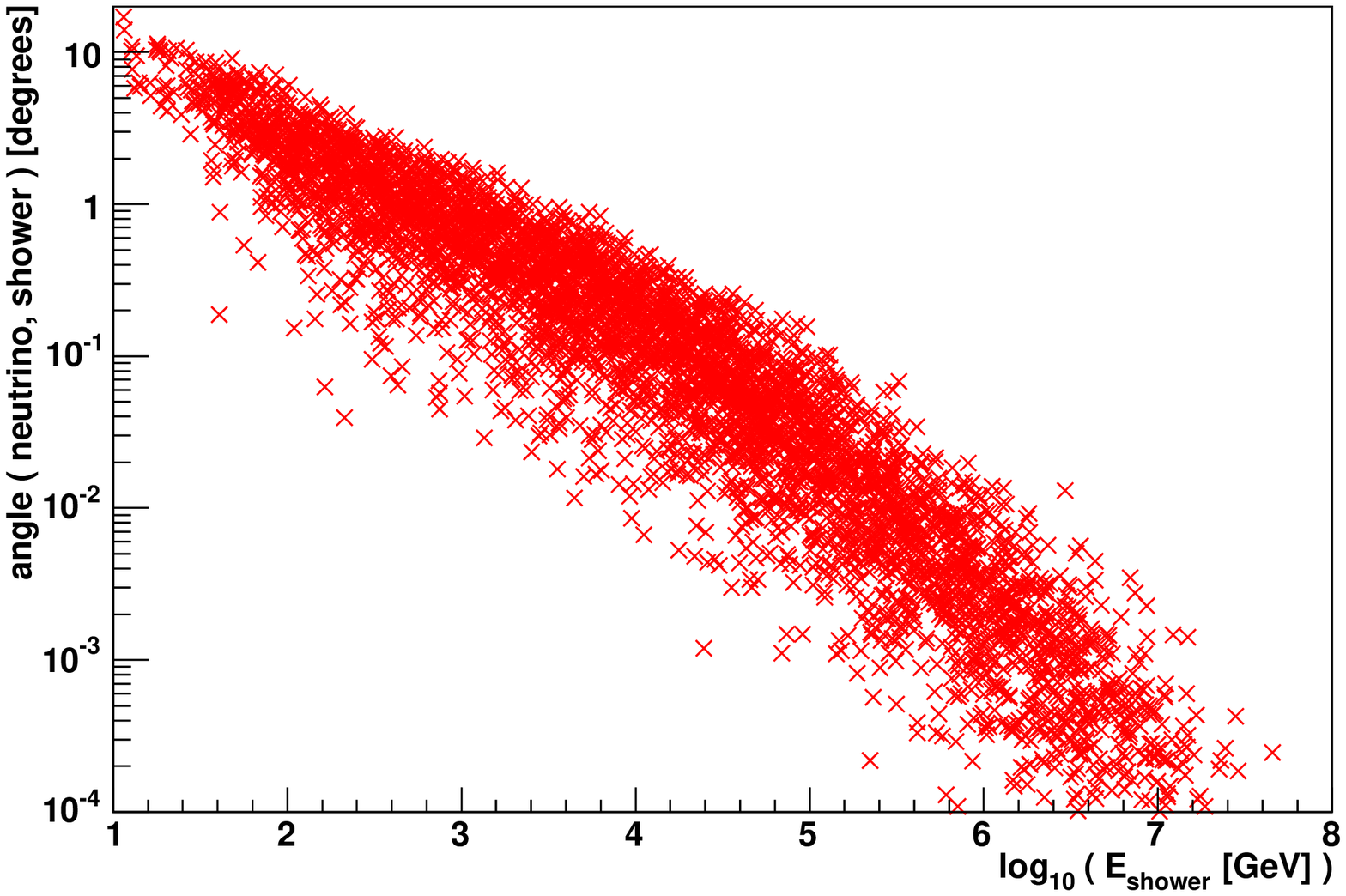}
\includegraphics[width=10cm]{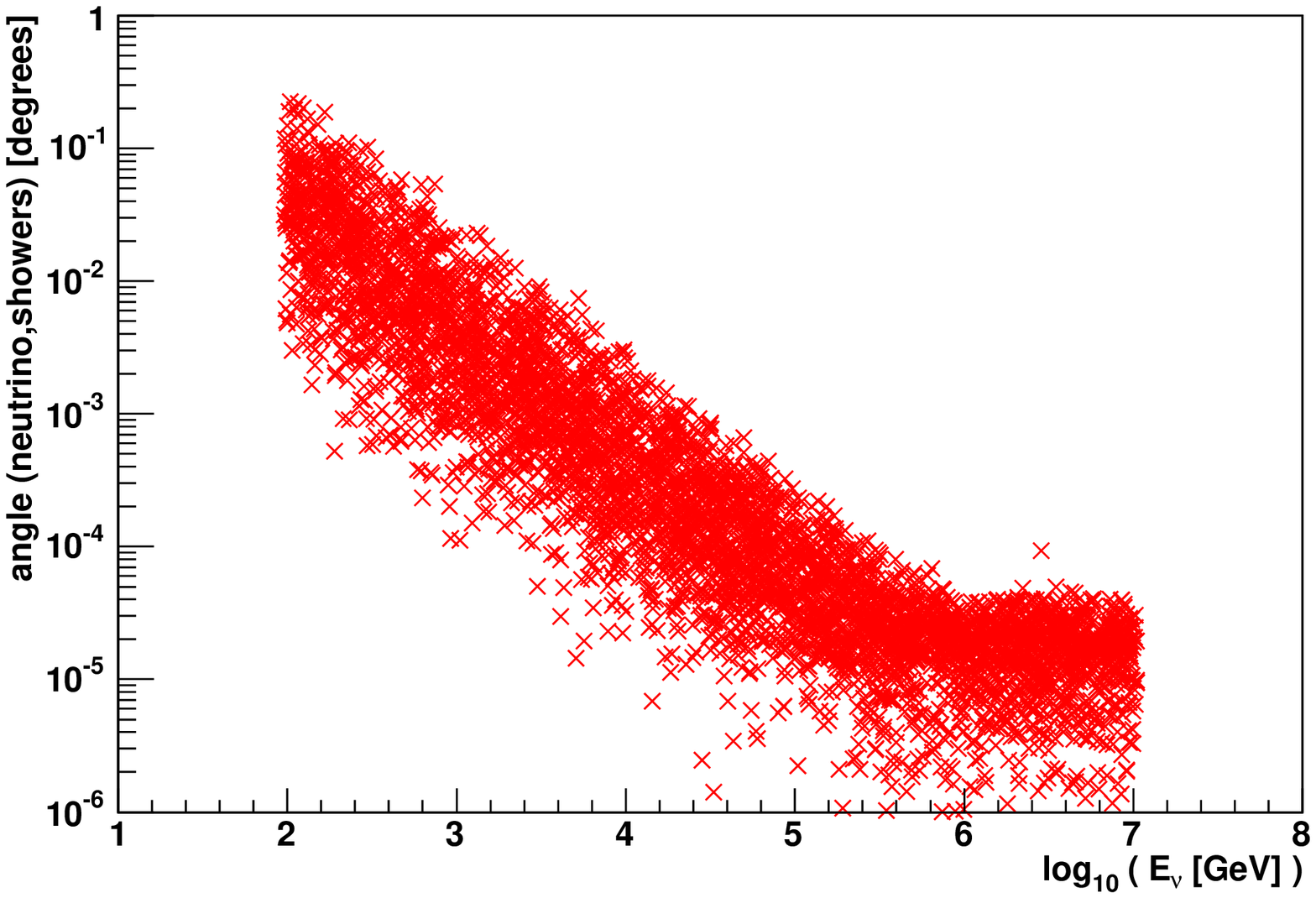}
\caption[Angle between neutrino and shower]{
Top: Angle between the neutrino and the hadronic shower, in degrees, as a function of the shower
energy, for NC interactions. Bottom: Angle between the neutrino and the common axis of
electromagnetic and hadronic shower, in degrees, as a function of the neutrino energy, for $\nu_e$
CC interactions. See text for details.} 
\label{fig:angle_shower_nu}
\end{figure}

One of the important objectives in reconstructing a neutrino event is to determine
the direction of the neutrino. In case of the NC events, one has no information about the outgoing
neutrino, so that the only object which can be reconstructed is the hadronic shower. However, the
direction of the hadronic shower is very similar to the direction of the incoming neutrino, as
can be seen in the upper plot of Figure~\ref{fig:angle_shower_nu} for events from event sample
A. Here, the direction of the shower was calculated by summing over the directions of all particles
in the shower as given by the simulation (see Figure~\ref{fig:shower_composition}), each
particle weighted with its respective MC energy. One can see that above $\sim 1$\,TeV the angular
difference between the shower and the neutrino falls below $2^{\circ}$, which is the best value that
the resolution for the direction reconstruction of a shower can reach (see
Chapter~\ref{sec:results}; for energies smaller than 1\,TeV, the intrinsic resolution is
worse). Therefore, within the acquired precision, one can safely regard the shower 
direction as identical to the neutrino direction. \\
In the lower plot of Figure~\ref{fig:angle_shower_nu}, the same is shown for the case of
$\nu_e$ CC events, for a subsample of the event sample described in
Appendix~\ref{sec:nue_sample}. In this event class (see Section~\ref{sec:event_types}), an
electromagnetic and a hadronic shower are produced. However, these two are not separable in the
detection, and therefore, a common shower axis from all particles in the two showers has been
calculated by the method described above. This axis is almost identical to the neutrino axis;
for the lower energies, inaccuracies of simulation and method cause the small difference between the
neutrino and the common shower axis, while above 1\,PeV, the size of the calculated angle reaches the 
precision of the directions in the event sample, and therefore the distribution flattens. 
The cutoffs at 100\,GeV and 10\,PeV mark the energy region within which the neutrinos were produced.

\chapter{Background}\label{sec:background}

The ANTARES experiment is constructed in the deep sea to suppress as well as possible the background
caused by cosmic rays. However, atmospheric muons and neutrinos are present as background at the
ANTARES site, and the deep-sea surroundings add additional background sources. \\
The most dangerous background for $\nu_{\mu}$ CC events (which are reconstructed using the
neutrino-induced muon) are atmospheric muons; they can be distinguished well from shower events, so
that the latter are much less affected by this background source than $\nu_{\mu}$ CC events, which
can only be identified as such if the neutrino comes from below. Atmospheric muons are a potential
background for shower events only in two cases: If they experience catastrophic bremsstrahlung
losses during their passage through the detector, or if they occur in bundles of 
several muons crossing the detector at the same time. Atmospheric muons are covered in
Section~\ref{sec:atm_muons}. \\
Atmospheric neutrinos, on the other hand, are an irreduceable background for the detection of 
high-energy cosmic neutrinos, for all event types. Characteristics of atmospheric neutrino fluxes
are given in Section~\ref{sec:atm_nus}. \\ 
A third type of background is the noise from the detector medium itself, the deep sea. Decays
of $^{40}$K nuclei in the salt water produce a constant background rate of one-photon signals; 
multi-cellular organisms add high bursts erratically. The different types of this optical noise are
discussed in Section~\ref{sec:noise}, whereas methods for its suppression are introduced in
Section~\ref{ch:trigger}. 

\section{Atmospheric Muons} \label{sec:atm_muons}

\begin{figure}[h]\centering
\includegraphics[width=10cm]{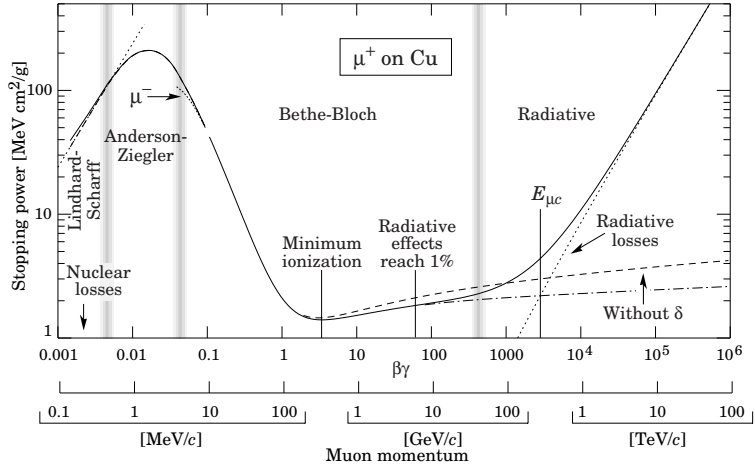}
\caption[Energy loss of muons in copper]{Stopping power (i.e.~avarage energy loss) of muons in
  copper. The shape curve is equivalent for other media but shifted in energy; the critical energy
  $E_{\mu c}$ lies at 1\,TeV for water. The dashed line marked \lq\lq without $\delta$\rq\rq~ refers
  to the omission of density effect corrections to the ionisation energy loss. From~\cite{pdg}.}
\label{fig:muonpassage}
\end{figure}

\begin{figure}
\begin{minipage}[h]{7.2cm}
  \centering \epsfig{figure=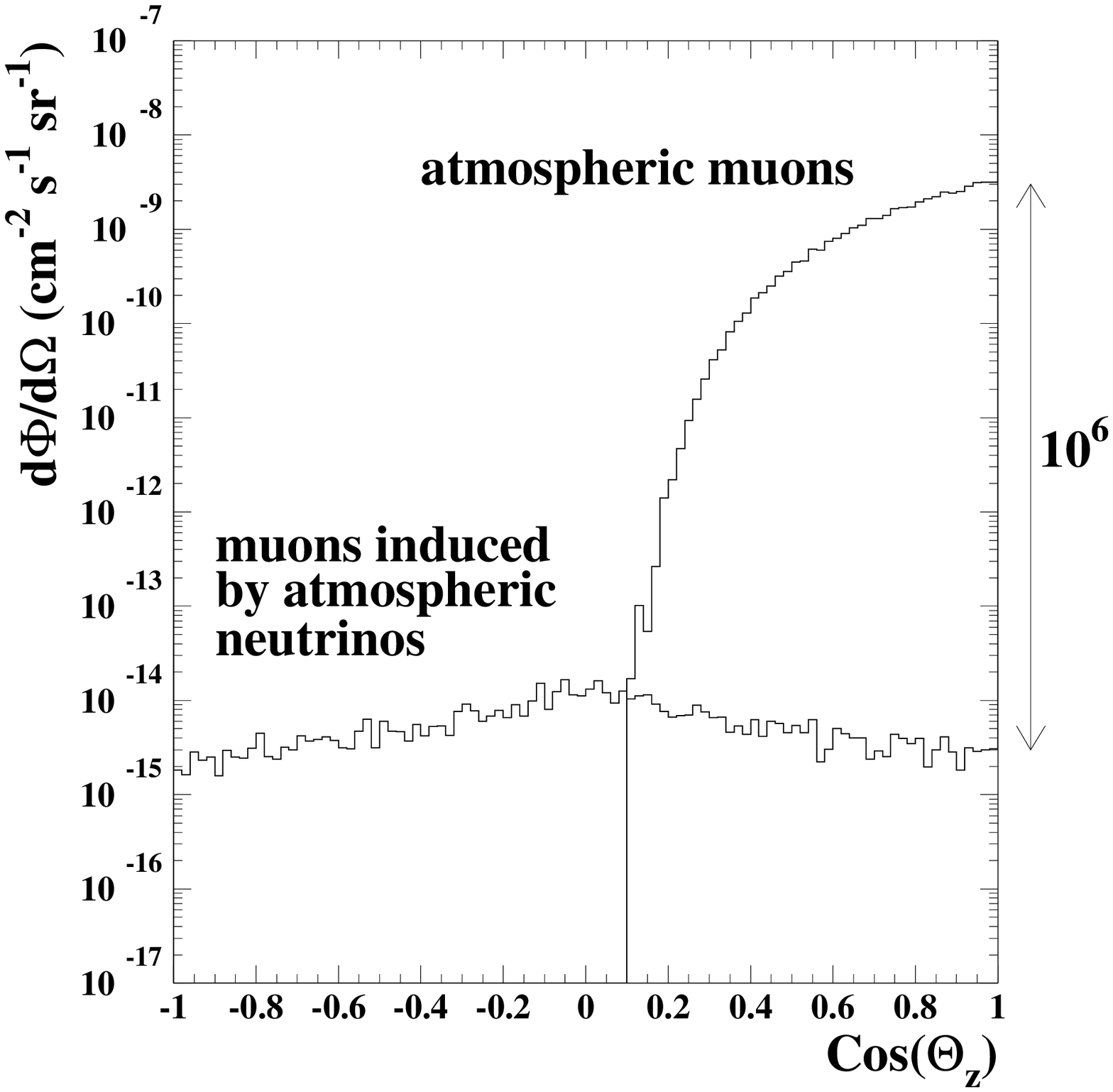, width=7.2cm}
  \caption[Muon background in ANTARES]{Flux of atmospheric muons and muons induced by atmospheric
  neutrinos as a function of the zenith angle $\theta$. This plot was taken from~\cite{proposal}.}
\label{fig:atm_muons}
\end{minipage} 
\hspace{0.4cm}
\begin{minipage}[h]{7.2cm}
  \centering \epsfig{figure=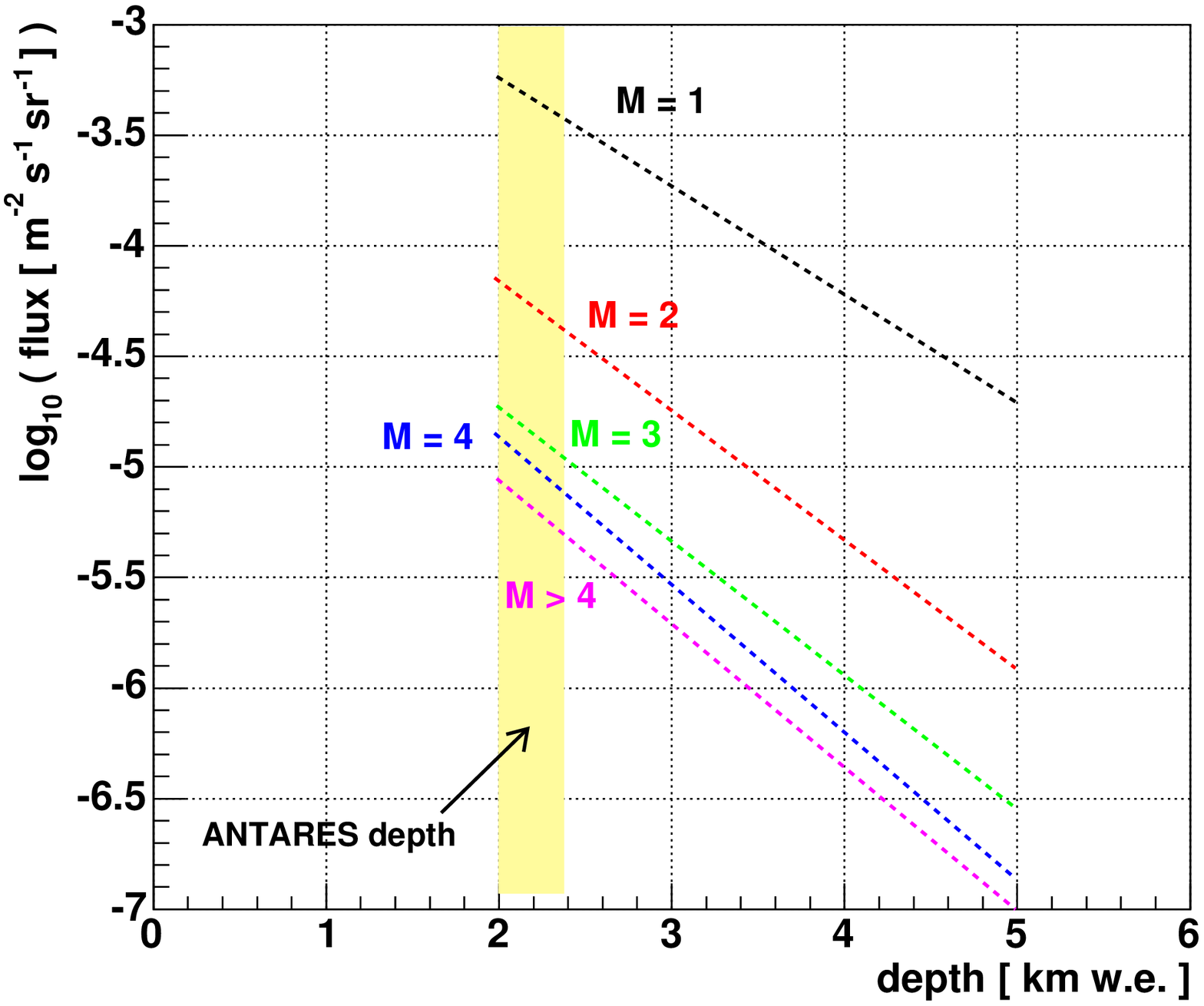, width=7.2cm}
  \caption[Flux of muon bundles]{Vertical flux of muon bundles for different muon multiplicities $M$,
  after~\cite{muon_param}.}
  \label{fig:atm_mu_flux}
\end{minipage}
\end{figure}

Atmospheric muons are produced in interactions of high-energy cosmic rays in the atmosphere,
through the decay of secondary mesons. The production chain is the same as the one discussed
for the production of neutrinos (see equation (\ref{eq:p_decay}) in Section~\ref{sec:beam_dump}). 
Atmospheric muons are produced abundantly in the atmosphere, and at energies above $\sim 1$\,TeV
their free path in water is long enough to reach the detector in 2400\,meters depth (see
Figure~\ref{fig:pathlength} in Section~\ref{sec:event_types}). They present a serious background to
the detection of muons from $\nu_{\mu}$ CC interactions, as they exceed the muon rate from atmospheric
neutrinos by several orders of magnitude (see Figure~\ref{fig:atm_muons}). For the detection of
neutrino-induced muons it is therefore necessary to restrict the analysis to events coming from
below. \\
Owing to the different topology of muon events and shower events (see
Section~\ref{sec:event_types}), the situation is different for showers. It is, however, possible
that an atmospheric muon suffers a strong radiative loss, a so-called {\it catastrophic energy
  loss}, which means that it radiates a substantial fraction of its energy into one secondary
photon, which in turn produces an electromagnetic cascade. If the energy of the cascade exceeds a
few 100\,GeV, it can be detected and hence be confused with a cascade from a neutrino interaction. \\
The average energy loss of muons is shown in Figure~\ref{fig:muonpassage} for copper as target
medium. One can see that for energies in the TeV region, radiative losses are by far the dominating
mechanism. 

The second background type caused by atmospheric muons are {\it multi-muon events}, where a bundle
of muons generated by one primary cosmic ray passes the detector, so that their signals in the
detector are causally connected. These events could be misinterpreted as shower-type events as well,
especially if the muon multiplicity is high, because in this case a large amount of electromagnetic
radiation could be produced, which could then be misinterpreted as a neutrino-induced shower. For
protons as primary cosmic rays, the muon multiplicity $M$ at sea level depends on the proton energy
$E$ approximately according to a power-law: ${M \propto E^{0.83}}$~\cite{sukowski}. The more
energetic the cosmic rays are, the larger is thus the probability that they produce a muon 
bundle. A similar correlation between multiplicity and primary energy was also found for iron
primaries, but with a multiplicity that is about an order of magnitude higher than for the primary
protons. It is assumed that the multiplicity distributions for the other primaries with masses
between protons and iron are in between these two distributions. The decrease of the
muon flux with increasing water depth can be seen in Figure~\ref{fig:atm_mu_flux}, for different
muon multiplicities, as calculated by~\cite{muon_param}. The fluxes shown in the figure refer to
vertical muon bundles from a combined simulation of five different types of
primaries\footnote{protons, He, the CNO group, primaries from Mg to Si, and Fe.}. Note that 
increasing water depth leads to a suppression of the flux, but only to a weak change in the
multiplicity ratios. At the depth of the ANTARES detector, marked by the yellow area, the flux of
single muons is about a factor of 80 higher than that of bundles with more than four muons. \\
As these two event types imitate neutrino-induced showers, they will not be recognised as background
in the reconstruction, but must be suppressed afterwards. This suppression, together with event
rates, will be discussed in Section~\ref{sec:atm_mu}. It will be shown there that the relation
between single muons and muon bundles is modified significantly in the reconstruction, as the events
with a high muon multiplicity are much more likely to pass the event filter and survive the
reconstruction. Even after reconstruction and quality cuts, the neutrino rate exceeds the
atmospheric muon rate only above $\sim 50$\,TeV (see Section~\ref{sec:result_flux}).  

\section{Atmospheric Neutrinos} \label{sec:atm_nus}

Atmospheric neutrinos are produced when high-energy cosmic particles interact in the atmosphere
of the Earth and produce charged mesons, which generate neutrinos in their decay. Due to the nature
of the meson decay chains (see equation~(\ref{eq:p_decay}) in Section~\ref{sec:beam_dump}), the flux
of atmospheric muon neutrinos is about twice as high as that of electron neutrinos, and it also
extends to higher energies. Below a meson energy of about 500\,TeV, the neutrino production is
dominated by the decay of pions and kaons. This is called the {\it conventional atmospheric neutrino
  flux}. Above this energy, the lifetimes of the light mesons become large enough to allow them to
interact before they can decay. Short-lived charm particles are the dominant source of atmospheric
neutrino production at these higher energies; neutrinos produced in that way are called {\it 
  prompt atmospheric neutrinos}.  \\
The atmospheric neutrino flux as a function of the neutrino energy is shown in
Figure~\ref{fig:theo_fluxes} (Section~\ref{sec:diffuse_flux}). The marked areas in that plot show
flux limits for $\nu_{\mu}$ (upper two lines) and $\nu_e$ (lower two lines) for different incident
angles, as calculated by the Bartol group~\cite{bartol} (conventional flux) and Naumov~\cite{naumov}
(prompt neutrinos), using the recombination quark-parton model (RQPM)~\cite{RQPM}. It should 
be noted that there exist several different models and predictions both for the conventional and the
prompt neutrino flux; authors often cited for the conventional flux besides the one mentioned above
are~\cite{honda} and~\cite{volkova}. While the differences between the conventional models are up
to $\sim 40\%$ at TeV energies~\cite{atm_flux_antares}, the models for prompt neutrinos can differ
more than one order of magnitude~\cite{prompt} due to the lack of high-energy measurements of 
charm  production cross sections in hadron-nucleus collisions. \\
{\it A priori}, atmospheric neutrinos are an irreduceable background in the detector; for the diffuse
neutrino flux, one can differentiate between cosmic and atmospheric neutrinos by detecting the
cosmic neutrinos as an excess of events in the energy spectrum of the atmospheric neutrinos. The
diffuse cosmic neutrino flux is expected to exceed the atmospheric neutrino flux above some 10 --
100\,TeV. Searching for individual sources, the usage of a small search window ($\lesssim 1^{\circ}
\times 1^{\circ}$) reduces the background from atmospheric neutrinos significantly. On the other
hand, the detection of atmospheric neutrinos is an important tool for the calibration of the
detector and the fine-tuning of the event classification and reconstruction. 

\section{Optical Noise in the Deep Sea}\label{sec:noise}

\subsubsection{The Baseline Rate}

\begin{figure}[h] \centering
\includegraphics[width=10cm]{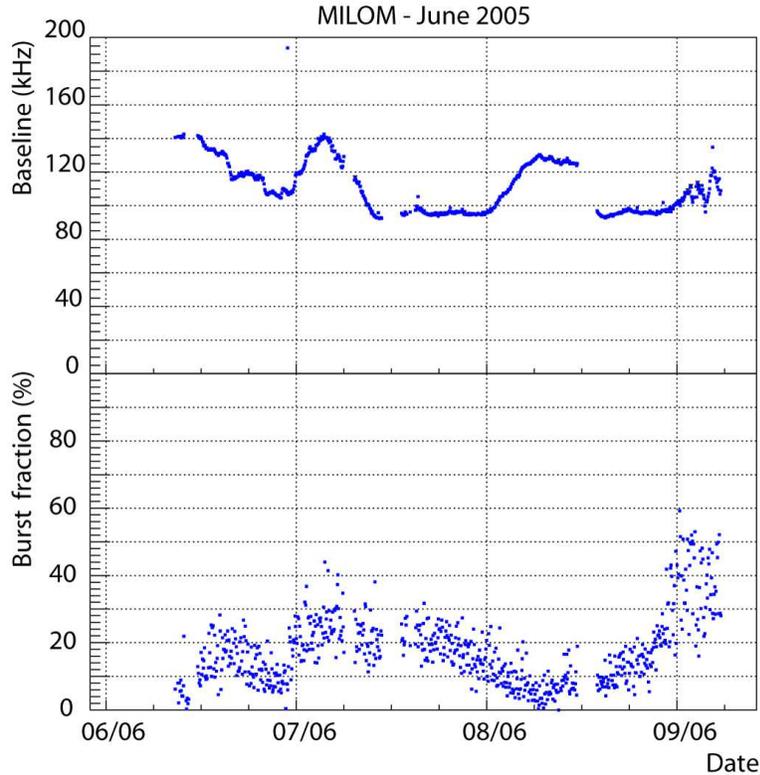}
\caption[Baseline rate and burst fraction]{Baseline rate and burst fraction as
  measured by the MILOM between June 6th and 9th, 2005. From~\cite{MILOM}.}
\label{fig:fraction_base_burst}
\end{figure}

The noise rate in the deep sea can be divided into two components: the {\it baseline} and the {\it 
bursts}. The baseline is a slowly varying rate in each photomultiplier. It is produced partly by the
radioactivity of the $^{40}$K contained in the salt water, which produces a constant noise of $\sim
40$\,kHz for the 10'' photomultipliers used in ANTARES, and partly by the {\it bioluminescence} of
micro-organisms which varies with the environmental parameters, like deep-sea current or weather
conditions on the surface. This background causes mainly single, uncorrelated photon signals in
individual photomultipliers, so that it can be removed to a large extent by a software filter  
(see Section~\ref{ch:trigger}).  \\
The {\it baseline rate} is defined as the mean of a Gaussian which is fitted to the counting rate
distribution measured in a period of 5\,minutes~\cite{escoffier}. The ANTARES test
strings PSL and MILOM (see Section~\ref{sec:status}) have observed baseline rates between 40 and
almost 200\,kHz; the long-term average is about 60--70\,kHz. The baseline rate for a period of
3\,days is shown in Figure~\ref{fig:fraction_base_burst}, in the upper graph. It should be noted
that also the photomultipliers themselves contribute to the baseline by their internal noise; this
contribution is about 3\,kHz~\cite{escoffier}.  

\subsubsection{Bursts}

\begin{figure}[h] \centering
\includegraphics[width=10cm]{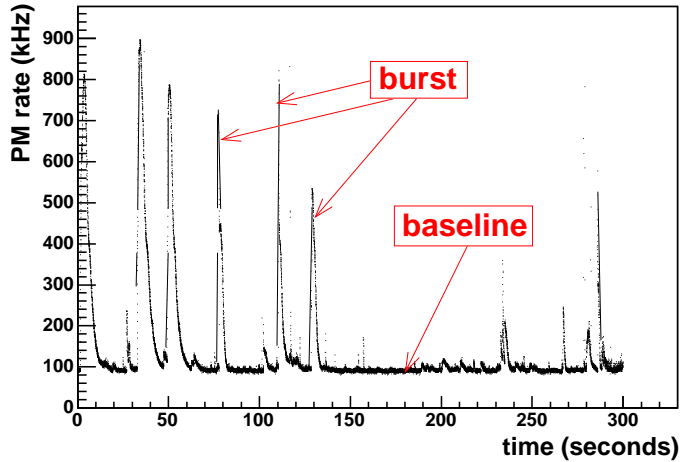}
\caption[Photomultiplier signal: baseline and bursts]{Photomultiplier counting rate
  within a 5-minute time window, as measured with the PSL. One can clearly distinguish the baseline
  at about 80\,kHz and the bursts. The figure was taken from~\cite{escoffier} and slightly
  modified. }
\label{fig:baseline}
\end{figure}

Bursts are caused by all kinds of multi-cellular organisms which emit light: Fluorescent squids,
crustaceans or fish. Contrary to the baseline rate, burst signals are very bright flashes lasting up
to a few seconds, which cause counting rates up to several MHz in the nearby photomultipliers, but are
aperiodic and localised, so that they do not affect the whole detector. The affected
photomultipliers, however, have to be excluded from data taking during the burst. \\
The {\it burst fraction} in the optical background is defined as the fraction of time in a 5-minute window
during which the counting rate in the photomultiplier exceeds the baseline by more than
20\%~\cite{escoffier}. As for the baseline, the burst fraction also varies with time; there are
periods where it is almost zero, whereas during other periods it rises to more than 40\%. There is
no direct correlation between a high baseline rate and a high burst fraction. An example for the
burst fraction measured by the MILOM during a 3\,day period is shown in the lower graph of
Figure~\ref{fig:fraction_base_burst}. \\ 
Figure~\ref{fig:baseline} shows the counting rate of one photomultiplier of the PSL, in a 5-minute
window. In this example the baseline rate is about 70\,kHz, and a number of smaller bursts
occur in irregular intervals. 

\section{Suppression of Optical Background}\label{ch:trigger}

The optical noise which causes the baseline rate can be suppressed by using the fact that the hits
caused by this background are randomly distributed and uncorrelated. To prevent noise
contamination of an event sample, the optical background must be filtered out {\it before} the
reconstruction (contrary to atmospheric muon background which is removed by quality cuts after the
reconstruction). This filtering is done for every single hit of an event by checking its correlation
with the other hits. If a hit does not fulfill the {\it filtering conditions} listed in the
following sections, it will be removed from the event. The first two conditions are taken from the
ANTARES Software Trigger~\cite{brams} where they are used as trigger conditions to determine whether a
physics event has been observed in the detector and should be written to disk, whereas the third
condition was developed in the context of this thesis.

\subsection{Filter Conditions for a Signal Hit}\label{sec:trigger}

\subsubsection{Condition 1: Global Causality}

Every hit must be causally connected to the largest hit of the event. As the optical background hits
are usually single photon hits, the largest hit is considered not to be caused by noise.
The difference between the arrival time $t_i$ of the considered hit and the arrival time $t_0$ of
the largest hit in the event must fulfill the condition 

\begin{equation}
|t_i - t_0| \overset{!}{<} d / v + \delta t,
\end{equation}

where $d$ is the distance of the two photomultipliers that were hit, $v$ is the speed of light in
water and $\delta t = 100$\,ns is an extra time window to account for time measurement inaccuracies
and delays due to scattering of the light.

\subsubsection{Condition 2: Local Coincidences or Large Amplitude}

Every hit must either have an amplitude larger than the adjustable $A_{min}$, or be in
local coincidence within $\Delta t = 20$\,ns to at least one other hit {\it on the same
  storey}. $A_{min}$ was set to 3\,photoelectrons (pe) for this study.

\subsubsection{Condition 3: Background Coincidence Suppression}

To avoid accidental coincidences of two background hits, those hits which have passed Condition 2
because they are in coincidence with another hit (and not because of their large amplitude) must
fulfill an additional condition: A second pair of coincident hits {\it on the same string} is
required. If there are no other coincident hits on the same string, the single pair of coincident
hits can still pass the filter condition if at least one of the hits has an amplitude larger than
$A_{min,2}$. $A_{min,2}$ is set to 1.5\,pe, smaller than $A_{min}$, because the hits which this
condition is applied to have already passed the other two conditions and the background has
therefore already been reduced. \\[1cm] 
Conditions 1 and 2 are used in a very similar way in the ANTARES Software Trigger as trigger
conditions for physics events; they are further described in~\cite{brams}. ${A_{min} = 3}$\,pe of
Condition 2 is above the default value of 2.5 used for the Software Trigger, as in general higher
amplitudes for showers than for muon events are expected in the individual photomultipliers. \\
Condition 3 has been implemented additionally in the context of this thesis, in order to prevent
accidental coincidences between two background hits on the same storey. These coincidences are not
so rare, as one can see from the following consideration: \\
The $k$-fold coincidence rate $R_k$ of $n$ photomultipliers within a coincidence window
$\Delta t$ and for a background rate $r$ measured by each photomultiplier, is~\cite{rates}

\begin{equation}
R_k = \frac{n!}{(n-k)!} \cdot r^2 \cdot \Delta t,
\end{equation}

where $\frac{n!}{(n-k)!} = k! \begin{pmatrix} n \\ k \end{pmatrix}$ is the number of {\it ordered
sequences} of $k$ samples taken from a total of $n$ samples, without repetition. For $n=3$ and $k=2$
(coincidence in two out of three photomultipliers on the same storey), and for $r = 60$\,kHz and 
$\Delta t = 20$\,ns, ${R_{k=2} = 432}$\,s$^{-1}$. The number of coincidences $N$ in the whole
detector (300 storeys), for a typical event duration $T = 2000$\,ns, then becomes

\begin{equation}
N = 300 \cdot R_{k=2} \cdot T = 0.26, 
\end{equation}

approximately one background coincidence in every four events. The number of coincidences grows
quadratically with increasing background rate. \\ 
Only hits of an event which have passed all the conditions are used. These hits are required to be
distributed on at least 3 different strings to generate a sufficiently large lever-arm for the
reconstruction algorithm. If this is not the case, the event is discarded. 

\subsection{Event-wise Efficiency and Purity}\label{sec:event_purity}

One can define an event-wise efficiency and purity for the filtered events. The {\it efficiency}
is the ratio of the number of events which have passed the filter conditions to the original number of
events. The {\it purity} is the ratio of the number of {\it good events} which have passed the filter to
the total number of events which have passed the filter. Here, good events are defined as events
that have passed the filter purely because of signal hits. Therefore, as an estimate for the purity
we consider the number of events of a background-free sample which have passed the filter, divided
by the number of events of a background-contaminated sample which have passed the filter. \\
Figure~\ref{fig:eff_trig} shows the efficiency and purity for event sample A (see
appendix~\ref{sec:data_sample}), a sample of NC events with interactions within the instrumented
volume and neutrino energies between 100\,GeV and 100\,PeV, for different rates of optical background
per photomultiplier.  
For shower energies above 25\,TeV the efficiency is practically 100\%, whereas for energies smaller
than 10\,TeV it decreases strongly. At low energies, also many signal hits are cut away by the
filter conditions, so that events fail to pass the filter; one can also see that the efficiency in
these first three bins increases for higher background rates, which means that some of the
background hits contribute to the coincidences and make the event pass the filter. This also shows
up in the purity plot: For the lowest energy bin, the purity for the highest background rate of
140\,kHz per photomultiplier is 0.65, so that 35\% of the events are background-contaminated. \\ 
One can conclude that above 10\,TeV, the background filter works well.

\begin{figure}[h] \centering
\includegraphics[width=7.4cm]{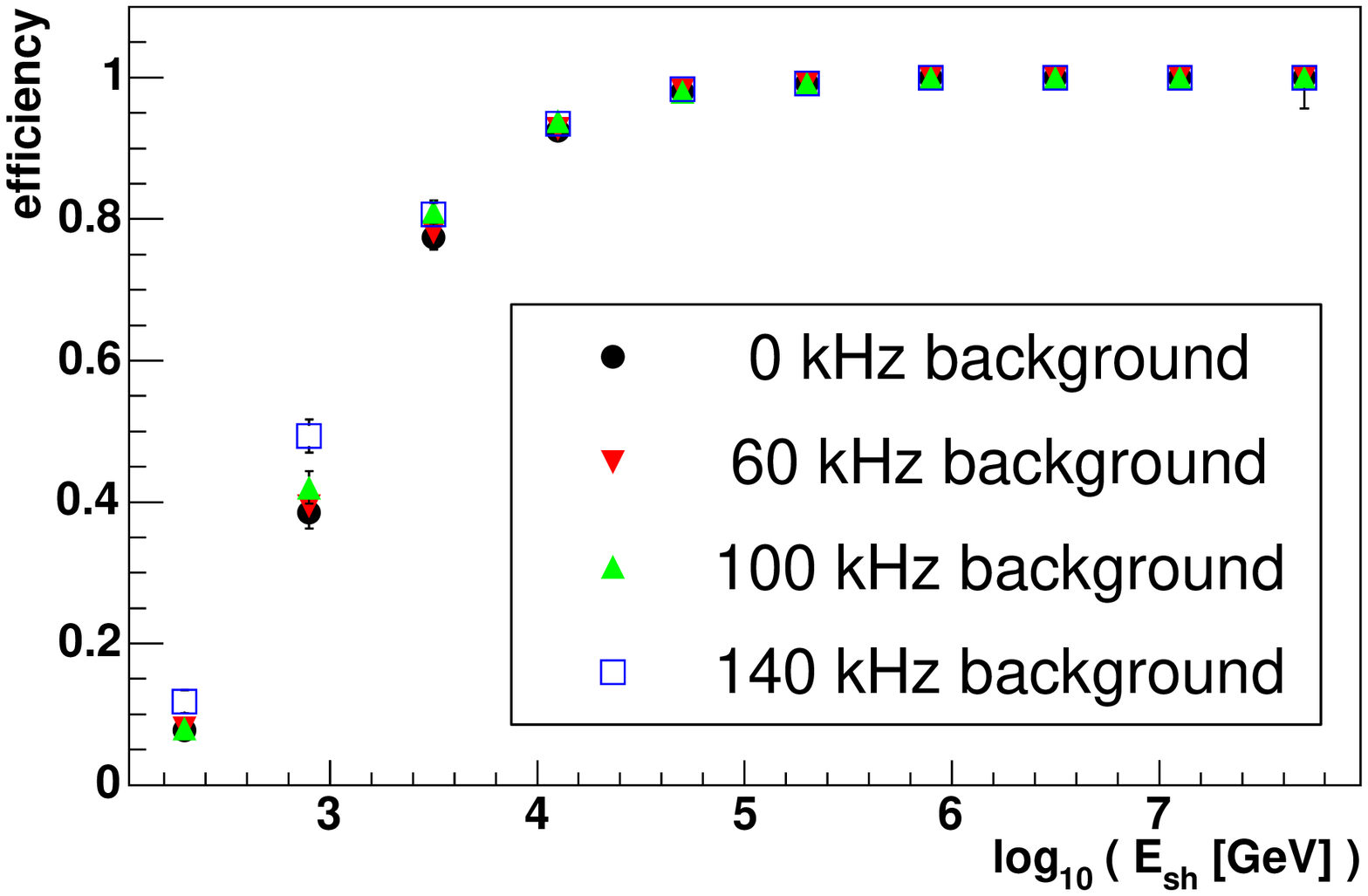}
\includegraphics[width=7.4cm]{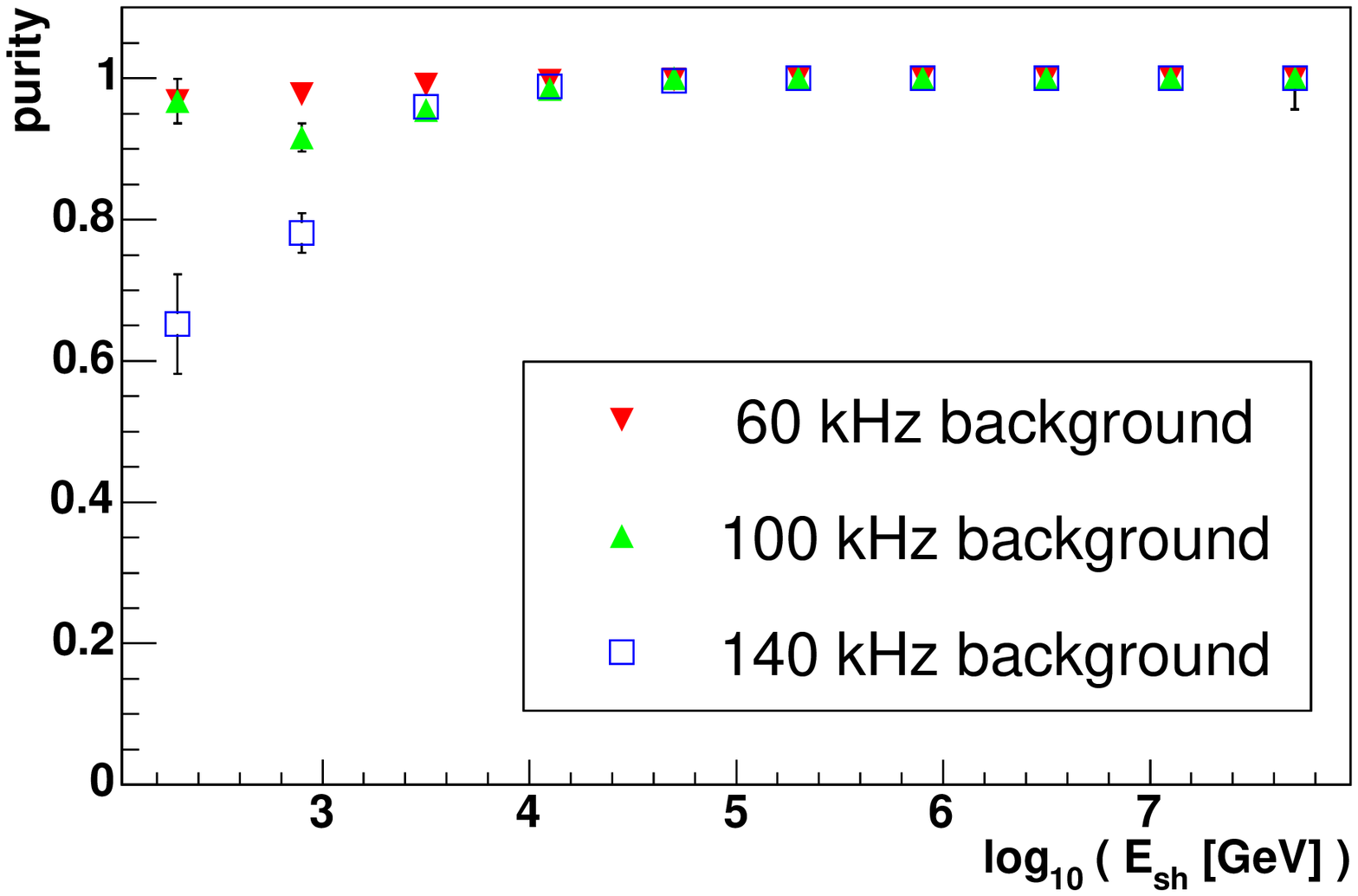}
\caption[Event-wise efficiency and purity of the filter]{Efficiency (left) and purity (right) of
  the filter conditions described in the text, over shower energy, for event sample A. The error
  bars are merely statistical, and for most energy bins they are smaller than the points
  themselves. The calculation of these statistical errors is described separately in
  Appendix~\ref{sec:stat_errors}.}  
\label{fig:eff_trig}
\end{figure}

\subsection{Hit-wise Efficiency and Purity}\label{sec:hit_purity}

\begin{figure}[h] \centering
\includegraphics[width=7.4cm]{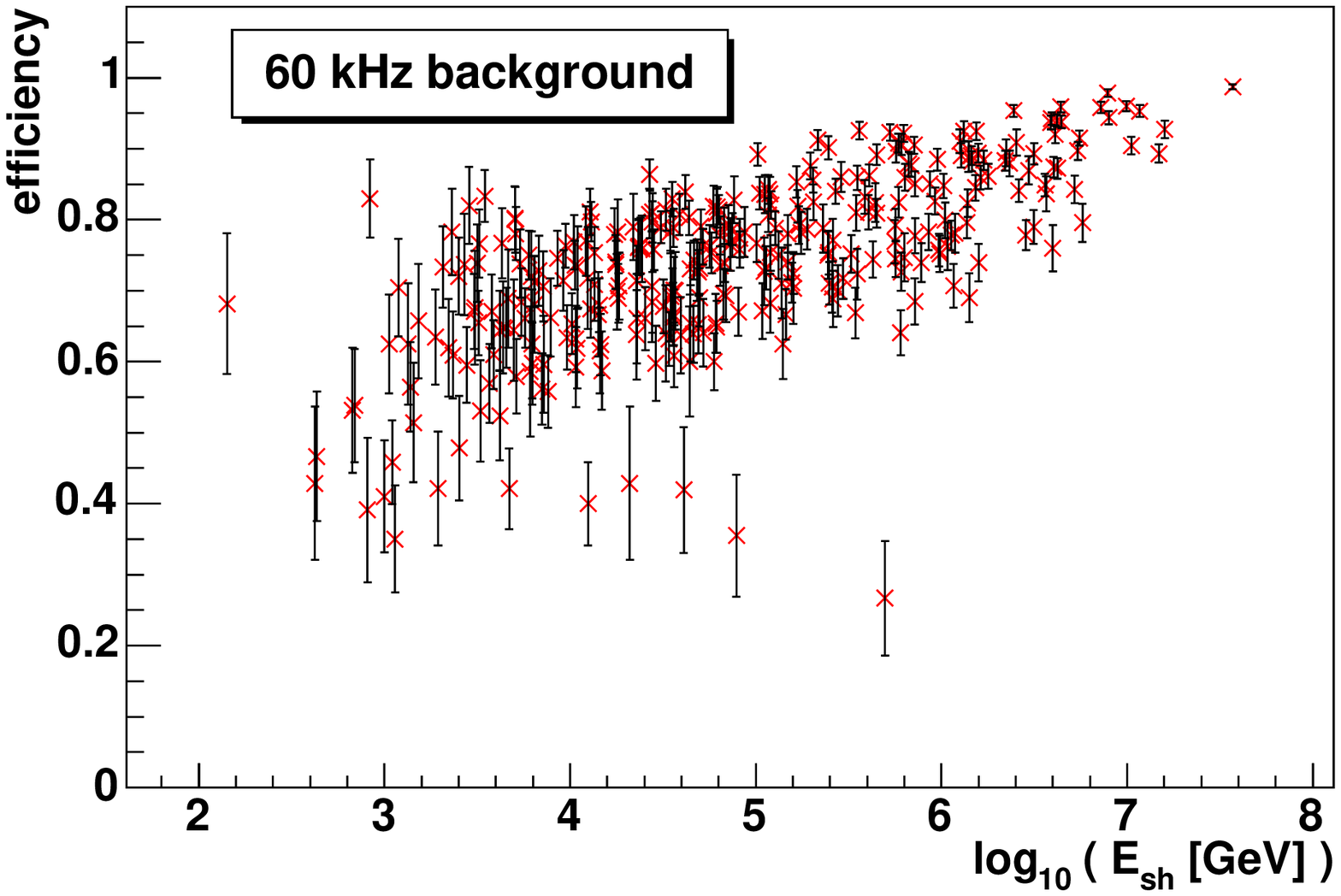}
\includegraphics[width=7.4cm]{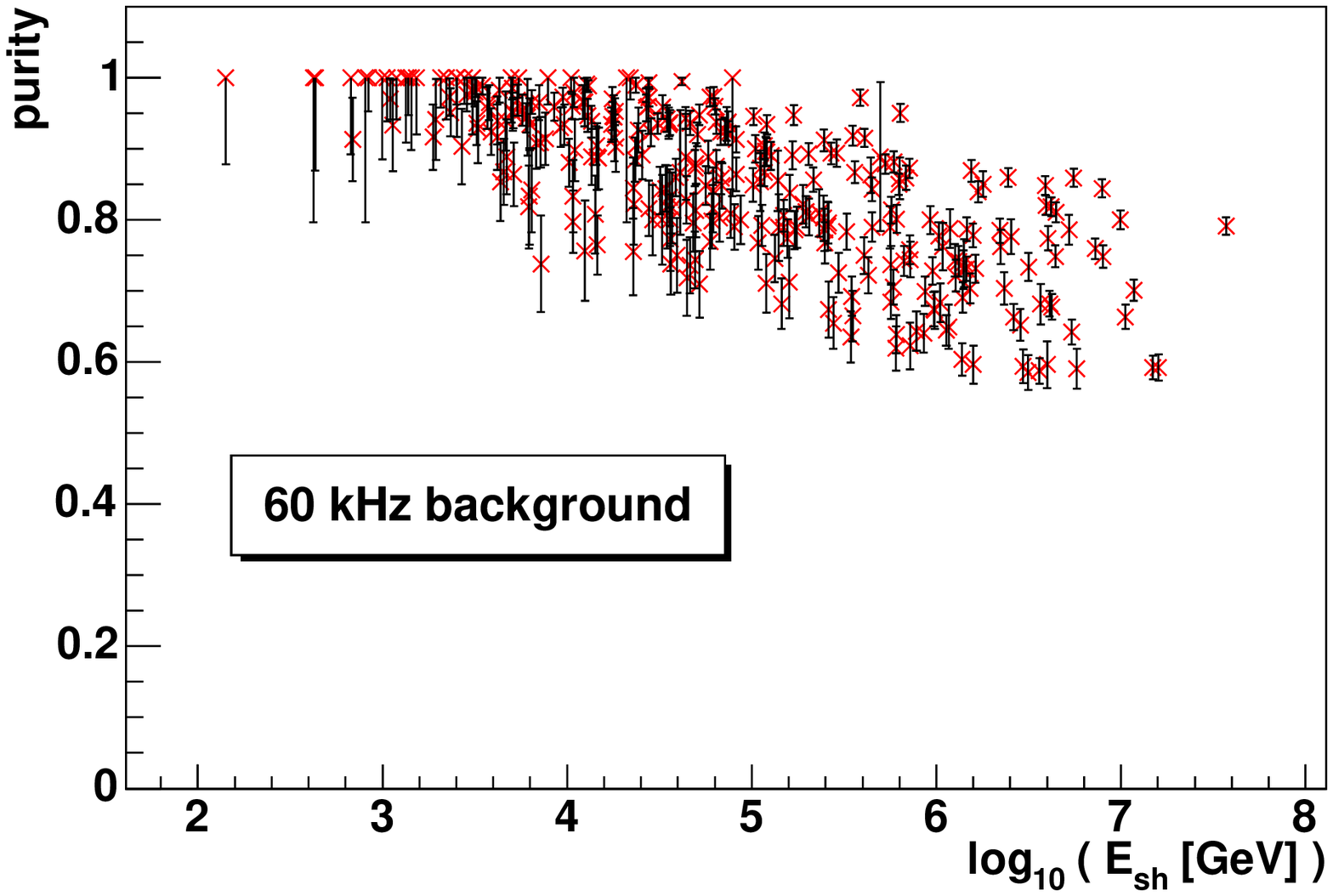}
\caption[60\,kHz hit-wise efficiency and purity]{Hit-wise efficiency (left) and purity (right) of the
  filter conditions described in the text vs.~the shower energy $E_{sh}$, for events with 60\,kHz
  background per photomultiplier. Each cross stands for one event. Statistical 
  errors calculated according to Appendix~\ref{sec:stat_errors} are also shown.}
\label{fig:hit_eff_trig}
\end{figure}

\begin{figure}[h] \centering
\includegraphics[width=7.4cm]{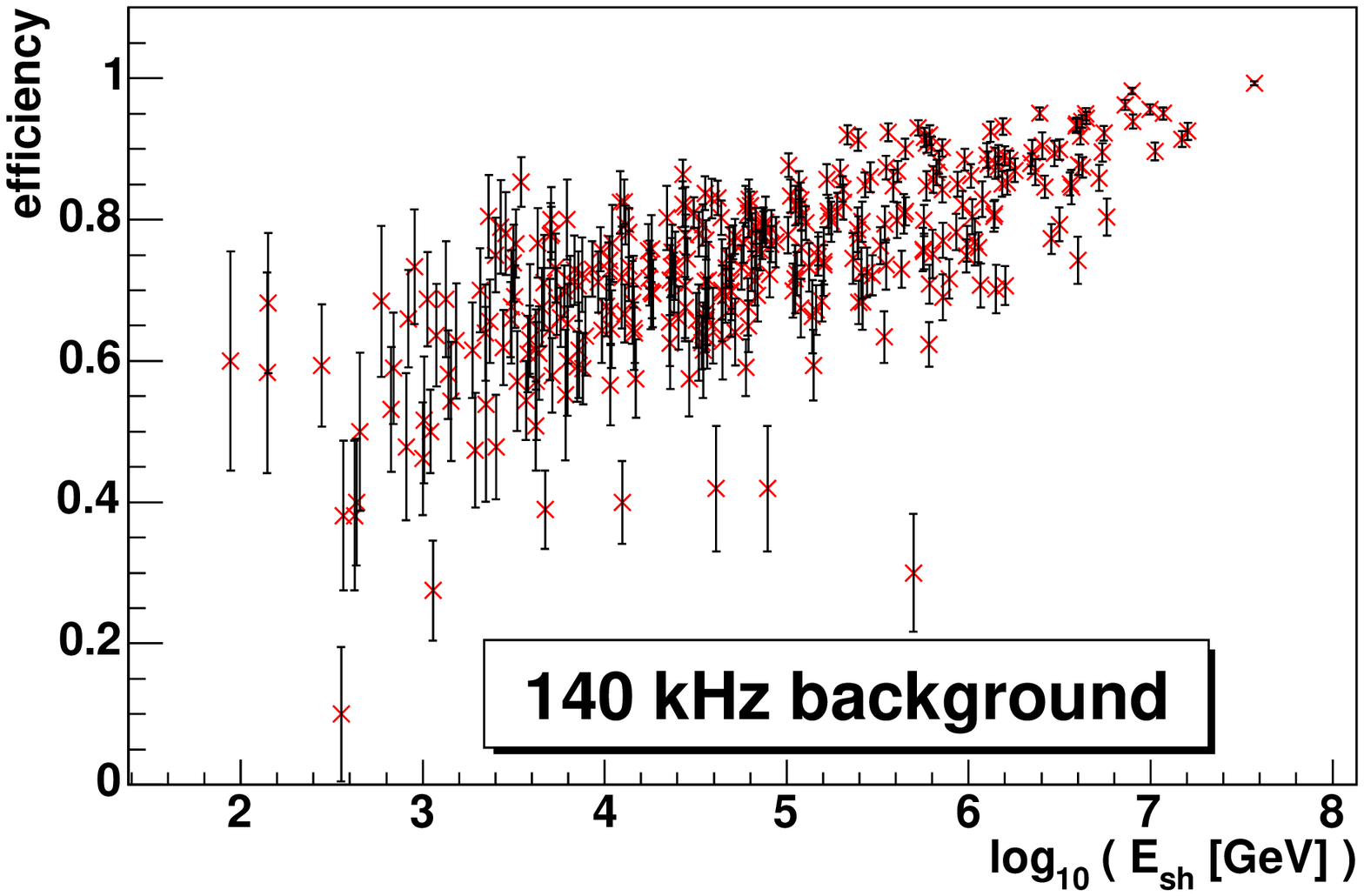}
\includegraphics[width=7.4cm]{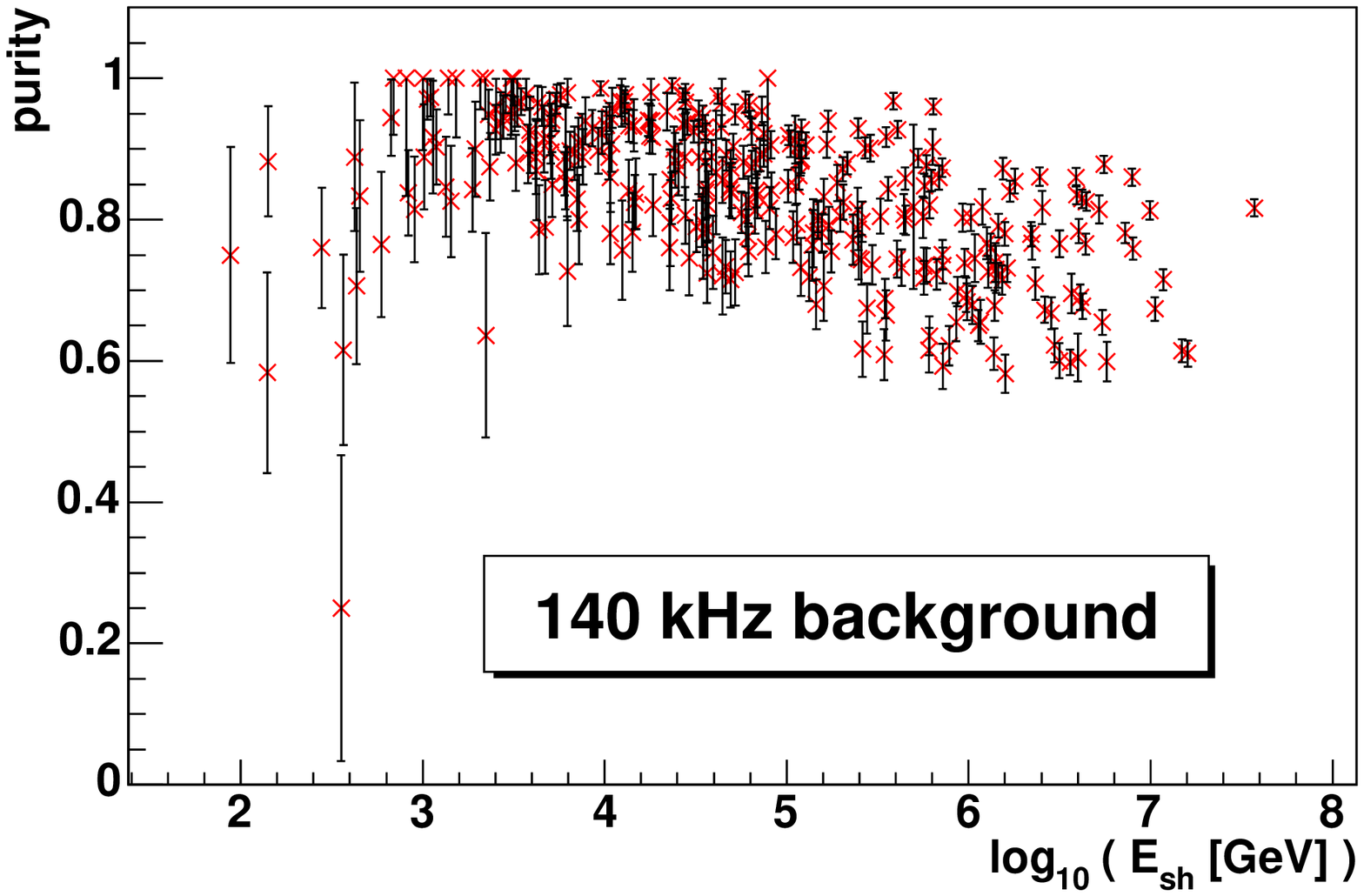}
\caption[140\,kHz hit-wise efficiency and purity]{Hit-wise efficiency (left) and purity (right) of
  the filter conditions described in the text vs.~the shower energy $E_{sh}$, for events with
  140\,kHz background per photomultiplier. Each cross stands for one event. Statistical errors 
  calculated according to Appendix~\ref{sec:stat_errors} are also shown.}
\label{fig:hit_eff_trig140}
\end{figure}

\begin{figure}[h] \centering
\includegraphics[width=10cm]{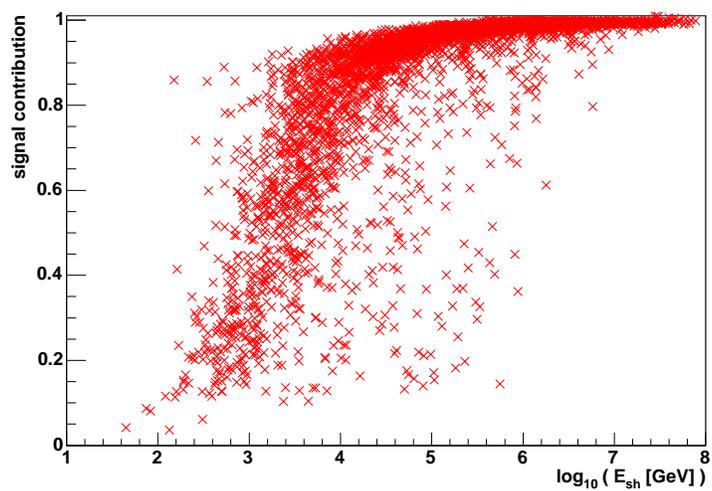}
\caption[Signal contribution to total amplitude]{Contribution of signal hits to the total amplitude
  of an event as a function of the shower energy, for 140\,kHz background, before the filter.}  
\label{fig:amplitude_contr}
\end{figure}

The efficiencies and purities shown in the previous section state whether an event has passed the
filter conditions or not; they do not indicate whether the hits remaining in the event after the
filter contain only signal, or also a fraction of background. Note that a hit is generated by
the photons, both from Cherenkov light and from optical noise, which arrive at a photomultiplier within
25\,ns (see Section~\ref{sec:digitisation}). We therefore define a {\it signal hit} as a hit with a
background contribution to the amplitude below 1\%. To examine the quality of the hits which pass
the filter, the hit-wise efficiency and purity are introduced: The hit-wise efficiency is defined as
the ratio of the number of signal hits {\it after} the filter to that {\it before} the filter.
The hit-wise purity is defined as the ratio of the number of signal hits remaining after the filter
to the {\it total number of hits} that are present after the filter. Note that a purity of e.g.~60\% 
implies that in 40\% of the hits a fraction $> 1\%$ of the hit amplitude was caused by background.
Efficiency and purity have been calculated on an event-by-event basis for a sample of $\sim 330$
events, selecting every 10th event of event sample A, both for 60\,kHz and for 140\,kHz optical
background. The results are shown in figures~\ref{fig:hit_eff_trig}
and~\ref{fig:hit_eff_trig140}. The numbers vary from event to event because the number of hits in an 
event does not only depend on the shower energy, but also on the position and direction of the
shower with respect to the detector (note that the events used all had an interaction vertex inside
the instrumented volume). \\  
For the sample contaminated with 60\,kHz noise, the hit-wise efficiency for events with a shower
energy below 10\,TeV is about 60\%, increasing to about 90\% for the higher energies. 
The purity for the 60\,kHz sample is close to 100\% up to $\sim 40$\,TeV and decreases to about 70\%
for higher energies. This is due to the longer duration of the events, which causes a larger total
number of background hits.  \\
For the events with 140\,kHz background, the results are very similar. There are, however, some 
additional events below 1\,Tev, which have only passed the filter conditions because of the higher
background rate. \\
It should be noted that the decrease in the purity for the highest energies (for both background
rates shown) has no impact on the reconstructibility of these events, because the signal amplitudes at
these energies are so high that the background only contributes a small fraction to the total hit
amplitude. One can see this from Figure~\ref{fig:amplitude_contr}, where the amplitude fraction
caused by signal hits with respect to the total amplitude measured in an event is calculated for
140\,kHz background contamination, before the filter. The plots was retrieved without simulation of
the detector electronics, i.e.~the shown hits are the real photon signals, not the signals
integrated over the ARS data taking time window (see Section~\ref{sec:digitisation}). It is
therefore possible to take into account the origin of the single photon signals, from Cherenkov
radiation or optical background. One can see that for events above 100\,TeV almost 100\% of the
amplitude is due to signal hits.

The studies show that the filter assures a satisfying suppression of the background above
10\,TeV, while for lower energies the majority of events is discarded. The pattern
matching algorithm for the shower reconstruction (see Chapter~\ref{sec:shower_fitter}) yields best
results above 10\,TeV (see also Chapter~\ref{sec:results}, where the results for the shower
reconstruction are presented). The reconstruction of low-energy events in the region below 10\,TeV
requires the use of other methods of background suppression, and different reconstruction
algorithms, and is outside the scope of this thesis. 

\chapter[Reconstruction of Individual Shower Parameters]{Strategies for the Reconstruction of Individual Shower Parameters}\label{sec:minor_strategies}

This chapter addresses strategies for the separate reconstruction of individual parameters of a shower:
the position, the direction of the shower axis and the shower energy. Some of the 
algorithms described here are part of the shower reconstruction package {\it ShowerFitter} which was
created within the context of this thesis\footnote{In the {\it ShowerFitter}, direction and energy
  are correlated and can therefore not be reconstructed separately.}. Other algorithms have been
developed to check the shower characteristics and are not intended for the use in a more
sophisticated analysis. \\
Section~\ref{sec:pos} describes an algorithm for the reconstruction of the interaction vertex, 
needed as an auxiliary variable for the reconstruction of direction and energy. The interaction time
is reconstructed in the same step but not used for further analysis\footnote{Even for
  astrophysical phenomena with a very short duration or periodicity, it is sufficient to use the
  time of the first trigger instead of the interaction time; this allows for a timing better than
  30\,ns.}. \\
In Section~\ref{sec:prefit_dir}, strategies for the reconstruction of the shower direction are
described. These algorithms are intended to provide a first estimate of the shower direction; the
resolution which results from these algorithms is not good enough to qualify them for a precise
physics analysis. They can, however, be used to check the consistency of the final fit. \\
Section~\ref{sec:e_pre} describes a method to reconstruct the energy of a shower, if its direction
and the interaction vertex are known. This method is not used in the final reconstruction, but
it provides a good consistency check for the combined reconstruction of direction and energy. \\
It should be noted that within the context of this work, the shower direction is considered to be
identical with the neutrino direction, which is a very good approximation within the precision of a
few degrees that can be reached in the direction reconstruction (see Section~\ref{sec:shower_angle}).
For the angular resolution, the reconstructed shower direction will therefore be directly compared
to the MC neutrino direction. 

\section{Reconstruction of the Interaction Vertex}\label{sec:pos}

\subsection{Method}

\begin{figure}[h] \centering
\includegraphics[width=7cm]{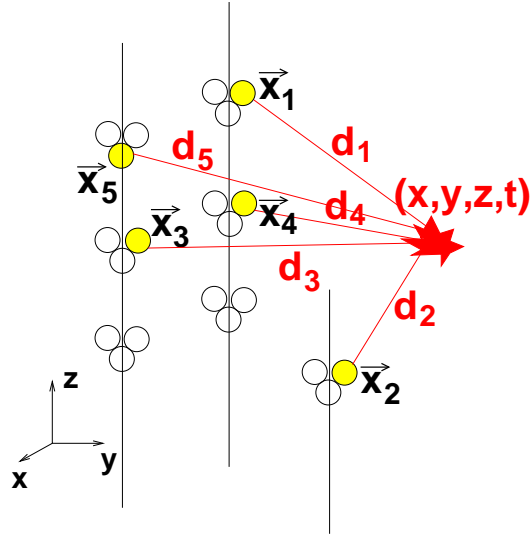}
\caption[Geometric situation for the vertex reconstruction]
{Schematic view of the geometric situation for the reconstruction of the shower position $(x,y,z)$
  and interaction time $t$.}
\label{fig:reco_position_scheme}
\end{figure}

The reconstruction of the interaction vertex is based on the simple assumption that a shower is 
a point-like object, as compared to the granularity of the detector. Therefore, all the light which
is produced in a shower is assumed to be emitted from one point. The point which is reconstructed
under this assumption is in fact not equivalent to the interaction vertex of the neutrino, but
should rather be denoted as {\it OM-centre-of-gravity} \rcg\ (the centre-of-gravity of all
Optical Modules (OMs) which have been hit in the event). To calculate \rcg, the point of
photon emission is calculated from the time and position of all hits, but neglecting the hit
amplitudes. To retrieve the true {\it centre-of-gravity of a shower} \rcga, as defined in
Section~\ref{sec:hadronic_showers}, the amplitudes of the hits are also taken into account. The
relation between these two positions will be discussed below. \\
Under the assumption of a point-like light source, the photon emission point can be reconstructed
via a simple triangulation. 
For an event with hits in $N$ OMs with space-time coordinates $\vec{x}_i = (x_i,y_i,z_i,t_i)$ ($i = 
1, ... , N$) and an unknown space-time position of the shower $\vec{x} = (x,y,z,t)$, the distance
$d_i$ between $\vec{x}$ and $\vec{x}_i$ can be written as 

\begin{equation}\label{eq:dist}
d_i = \sqrt{(x-x_i)^2 + (y-y_i)^2 + (z-z_i)^2} = c/n \cdot (t-t_i)
\end{equation}

where $c/n$ is the velocity of light in water with the refraction index $n$.
The situation is shown schematically for $N = 5$ in Figure~\ref{fig:reco_position_scheme}. Note that
in the ANTARES coordinate system the spacial point (0,0,0) lies in the centre of the detector, with
the $x$ axis pointing towards East, the $y$ axis towards North and the $z$ axis vertically upward
(see Figure~\ref{fig:ant_layout}).  

Equation~(\ref{eq:dist}) yields a system of $N$ equations which are quadratic in the unknown
$\vec{x}$. By pairwise subtraction of these equations, the quadratic terms of the components of
$\vec{x}$ cancel out and one is left with an $(N-1)$--dimensional linear equation system of the form 
\begin{alignat}{1}\label{eq:LGS}
& x_{i+1}^2 - x_i^2 + y_{i+1}^2 - y_i^2 + z_{i+1}^2 - z_i^2 - (c/n)^2t_{i+1}^2 + (c/n)^2t_i^2 \notag \\
= & 2x\cdot(x_{i+1} - x_i) + 2y\cdot(y_{i+1} - y_i) + 2z\cdot(z_{i+1} - z_i) - 2(c/n)^2t\cdot(t_{i+1} - t_i). 
\end{alignat}
This equation system can be written as 
\begin{equation}\label{eq:system}
A \cdot \vec{x} = \vec{b}, 
\end{equation}
where $A$ is a $(N-1) \times 4$ matrix with the components 
\begin{alignat}{1}
a_{j1} & = 2 \, ( x_{j+1} - x_j ) \notag \\
a_{j2} & = 2 \, ( y_{j+1} - y_j ) \notag \\
a_{j3} & = 2 \, ( z_{j+1} - z_j ) \notag \\
a_{j4} & = - 2 \, (c/n)^2(t_{j+1} - t_j),
\end{alignat}
and $\vec{b}$ is a vector with the
components 
\begin{equation}
b_j = x_{j+1}^2 - x_j^2 + y_{j+1}^2 - y_j^2 + z_{j+1}^2 - z_j^2 - (c/n)^2\cdot(t_{j+1}^2 -
t_j^2),
\end{equation}
with $j = 1, ..., N-1$. \\
The system (\ref{eq:system}) will normally not have an exact solution; a residual vector ${\vec{r} =
A\vec{x} - \vec{b}}$, ${\vec{r} \neq \vec{0}}$ will remain. The best solution $\vec{x}$ of the equation
system is given by the condition
\begin{equation}\label{eq:res}
\sum_j r_j^2 = \vec{r}^T\vec{r} = (A\vec{x} - \vec{b})^T (A\vec{x} - \vec{b}) = \text{minimal}.
\end{equation}
Differentiating equation (\ref{eq:res}) by $\vec{x}$ results in
\begin{gather}
2 A^T A \vec{x} - 2 A^T \vec{b} = 0 \leadsto \vec{x} = (A^TA)^{-1}A^T\vec{b}.
\end{gather}
This solution for $\vec{x}$ is calculated analytically.  There are four unknowns in the
$(N-1)$--dimensional equation system (\ref{eq:system}), so $N$ must be at least 5 to constrain
$\vec{x}$ unambiguously. As a further condition, it is required that hits occur on at least 3
different strings. This is necessary to ensure that the equation system (\ref{eq:system}) is
solvable for all geometries. To ensure numerical stability, the pairwise subtractions in
equation~(\ref{eq:LGS}) are arranged such that they preferentially involve hits from different
strings. 

\subsection{Calculation of the Centre-of-Gravity of a Shower}

As mentioned above, the {\it centre-of-gravity of a shower} \rcga, as introduced in
Section~\ref{sec:hadronic_showers}, is defined as the point of photon emission calculated from the
{\it amplitude-weighted} positions and times of the hits. To calculate \rcga\
from the physics information that is available after the detector simulation, one has to
follow the algorithm introduced in the last section, but the contributions of the respective OMs
have to be weighted according to the hit amplitude that the OMs have measured.  \\
However, the resolution achieved when calculating \rcga\ is not better than the
resolution achieved for the calculation of the {\it OM-centre-of-gravity} \rcg.
In the combined reconstruction of direction and energy, for which the shower position is needed as an
auxiliary parameter, it makes no difference whether the interaction vertex, \rcg\ or \rcga\ is used as
shower position, as long as the same variable is used consistently throughout the reconstruction. It
was therefore chosen to use \rcg\ throughout this thesis. 

\subsection{Results for Events without Background}

The distance between the interaction vertex and \rcg\ is larger than the distance between the
interaction vertex and \rcga. This can be seen in Figure~\ref{fig:ortsfehler} for event sample A
(see Appendix~\ref{sec:data_sample}), without optical background and without the usage of the filter
for background suppression (see Section~\ref{ch:trigger}): On the left, the total distance
between \rcg\ and the true interaction vertex is shown. The average offset is $\sim 5.4$\,m. On the
right, the correlation between that distance and the shower energy is shown. 
A linear function in $\log_{10} E$ (shown in red) has been fitted to the data within the region
marked by the endpoints of the line. The numerical values are:

\begin{equation}
\Delta r(E)/\textrm{m} = 2.8 + 0.48 \cdot \log_{10} ( E /\textrm{GeV} ),
\label{eq:shower_max_corr}
\end{equation}

with $\Delta r(E) = |$\rcg$ - \vec{r}_{\text{MC}}|$, the distance between \rcg\ and the MC interaction
vertex. The green line marks the distance between \rcga\ and the MC interaction vertex, equation
(\ref{eq:shower_max_amp}) (cf.~Section~\ref{sec:hadronic_showers}). The distance between the
interaction vertex and \rcg\ grows more steeply with increasing energy than the distance between the
interaction vertex and \rcga. This is possibly due to the fact that for increasing energy, a larger
amount of Cherenkov light is radiated in a transverse direction (see
Figure~\ref{fig:cherenkov_all}), and therefore, the hit amplitude in OMs positioned at larger polar
angles with respect to the shower axis increases. This is not taken into account in the calculation
of \rcg, as all OMs are treated in the same way and information regarding the hit amplitude is not
used, and therefore the OMs in forward direction have too large an impact on the reconstructed
position. \\  
\rcg and \rcga are at the same position at $\sim 40$\,TeV, and there is a discrepancy of $\sim
1.5$\,m for the highest energies studied.

\begin{figure}[h] \centering
\includegraphics[width=7.4cm,height=5.3cm]{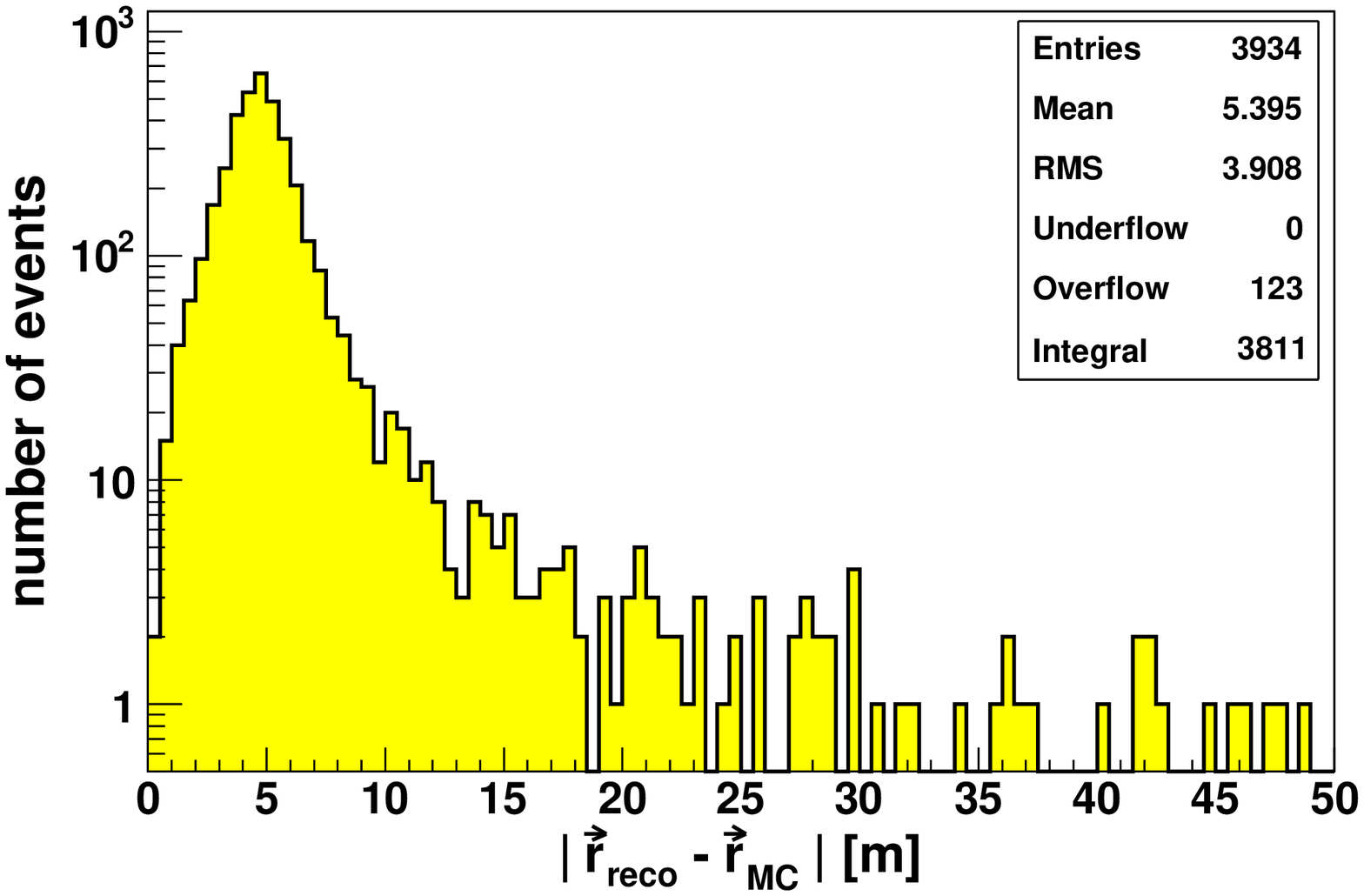}
\includegraphics[width=7.4cm]{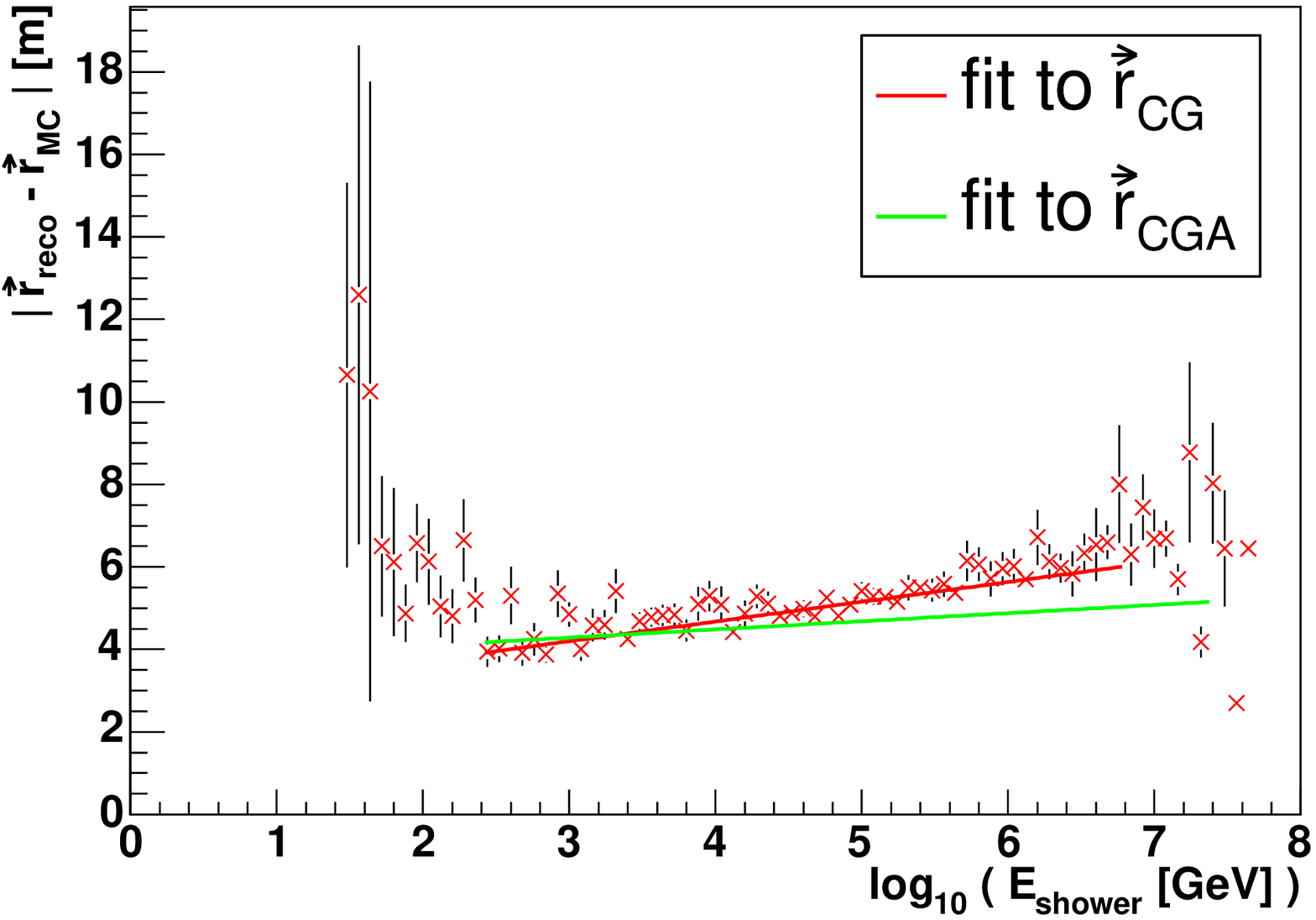}
\caption[Difference between reconstructed and MC vertex]
{Left: Total distance between reconstructed position and Monte Carlo (MC) vertex. Right: Profile of
  the distance, plotted as a function of the MC shower energy. The green line refers to equation
  (\ref{eq:shower_max_amp}) from Section~\ref{sec:hadronic_showers}, the red line is a fit to the
  data shown; the endpoints of the line mark the range of the fit.}
\label{fig:ortsfehler}
\end{figure}

\begin{figure}[h] \centering
\includegraphics[width=7.4cm]{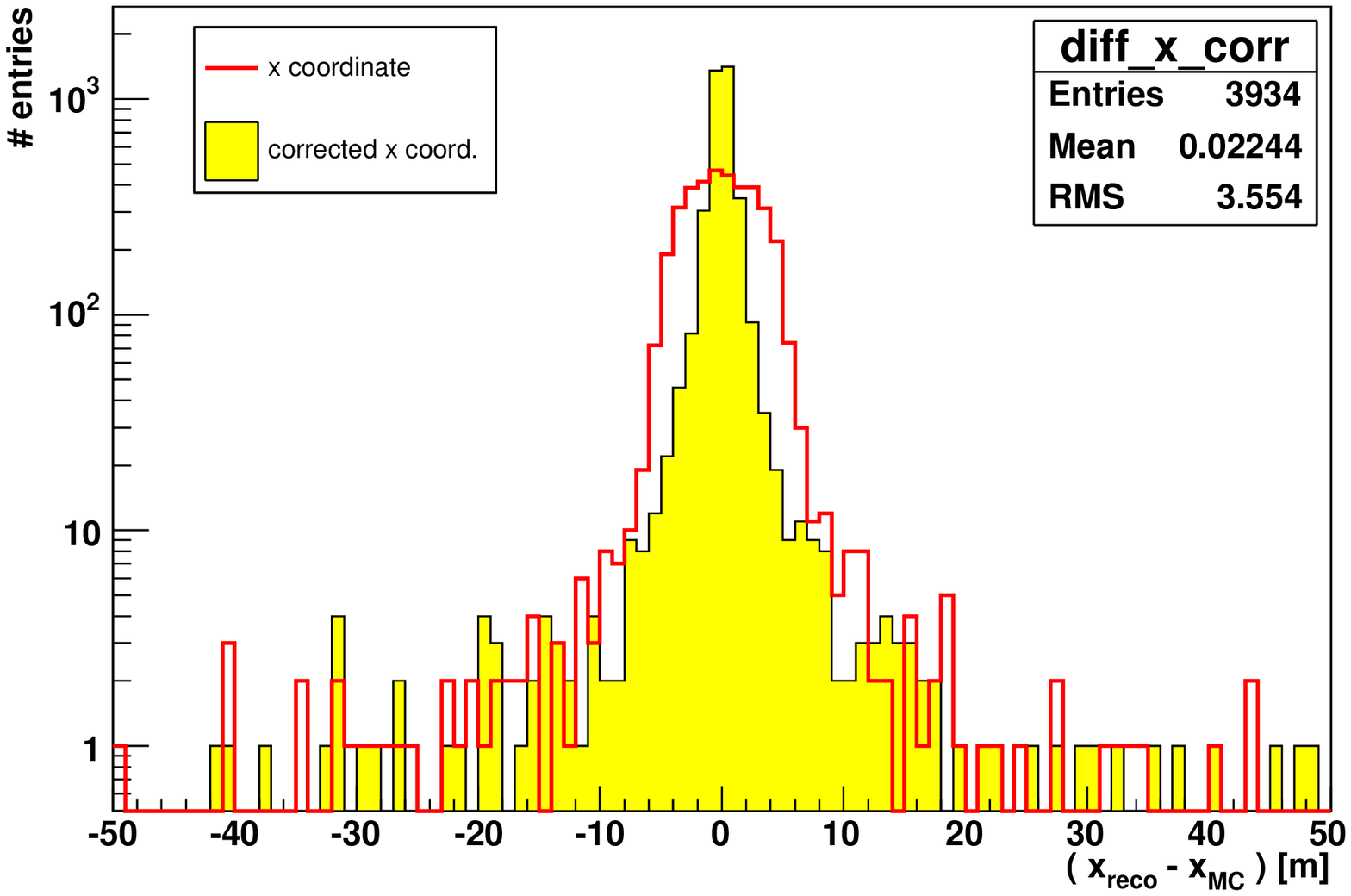}
\includegraphics[width=7.4cm]{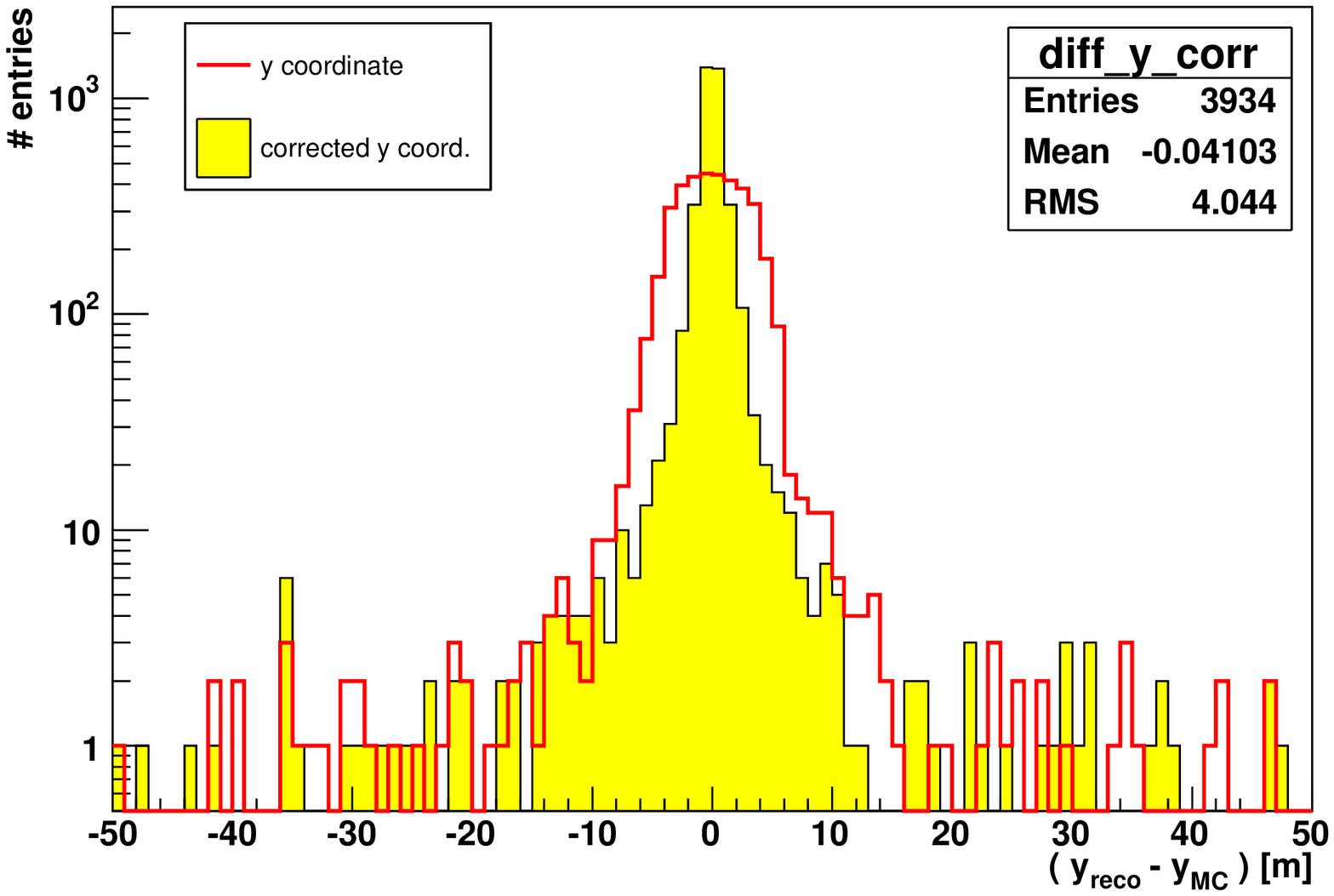}
\includegraphics[width=7.4cm]{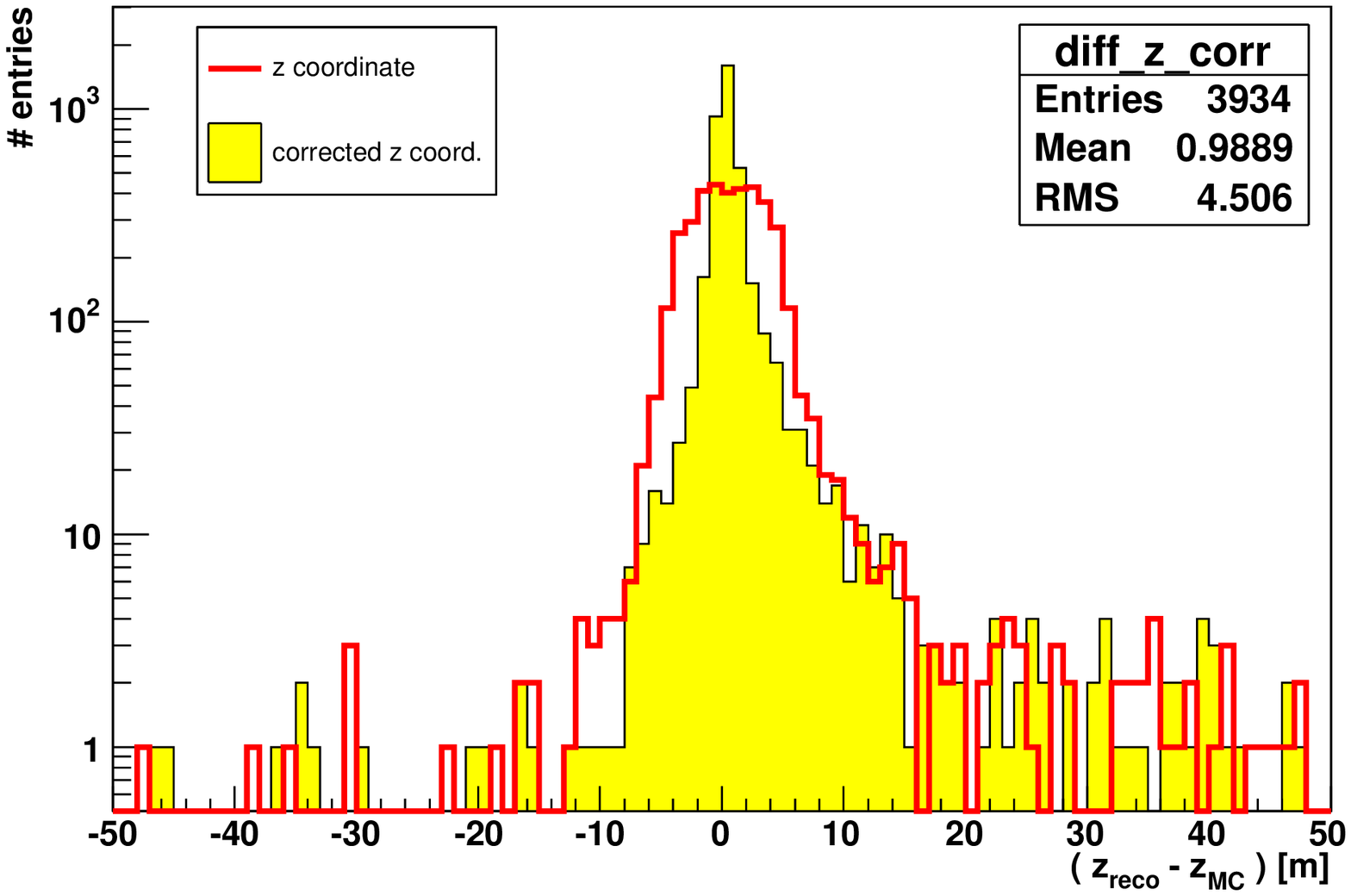}
\includegraphics[width=7.4cm]{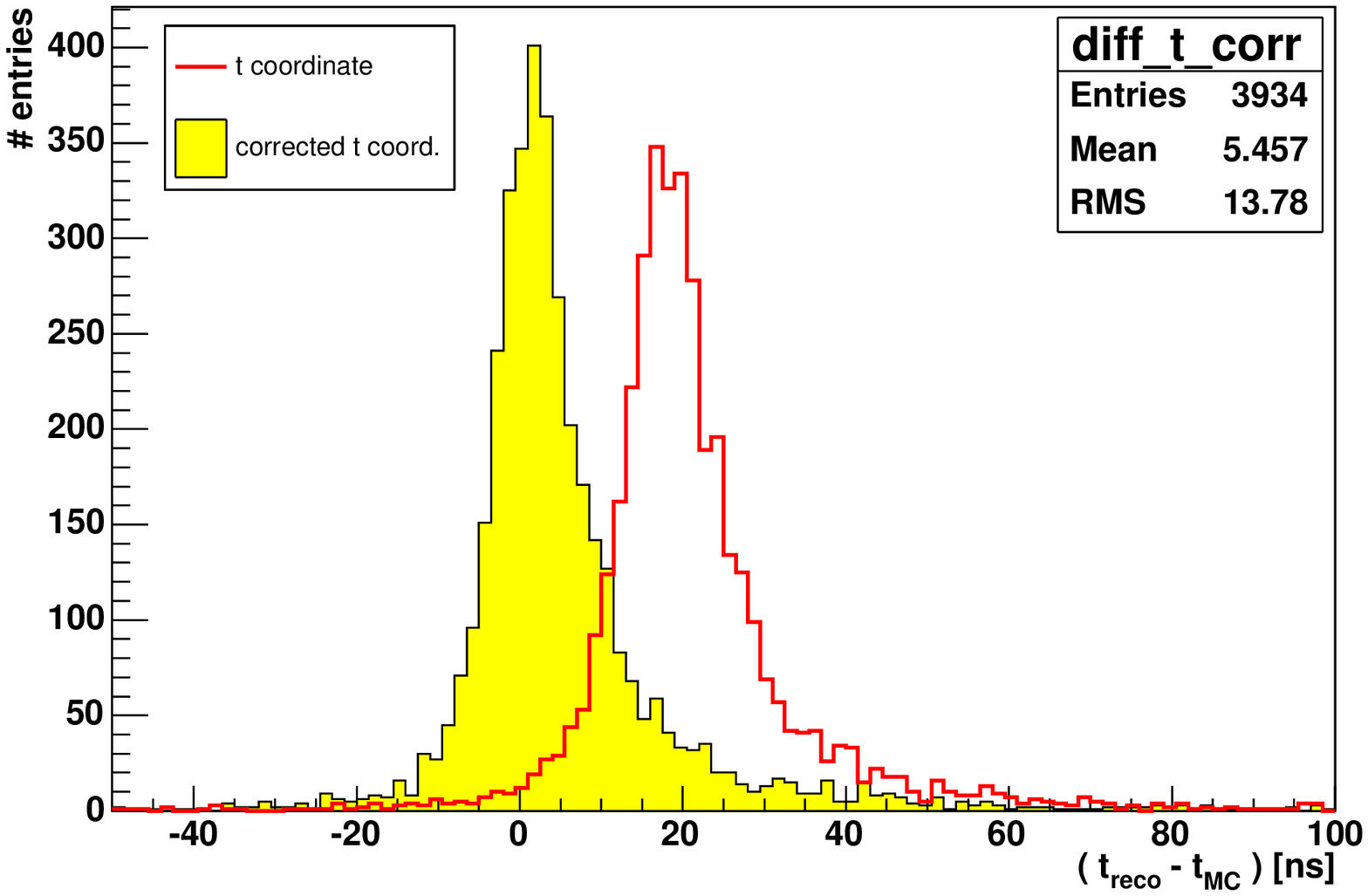}
\caption[Results of the reconstruction of the interaction point]
{Results of the reconstruction of the interaction point. Shown are the differences between
  the reconstructed spacial coordinates $x, y ,z$ and those from the MC, and the difference between
  the reconstructed and the MC interaction time. The red line shows the results for the
  reconstruction of \rcg, the yellow filled histograms those after correcting \rcg\ towards the
  interaction vertex, using equation (\ref{eq:shower_max_corr}).} 
\label{fig:ort}
\end{figure}

Figure~\ref{fig:ort} shows the results of the reconstruction of \rcga\ for the individual spacial
coordinates and the interaction time. The plots show as red lines the resolution of the three
spacial coordinates $x, y, z$ after the reconstruction. \\
As mentioned above, \rcg\ is a suitable auxiliary variable for the
reconstruction of direction and energy. If a knowledge of the interaction vertex is desired, its
coordinates can be calculated from \rcg\ following equation~(\ref{eq:shower_max_corr}), if direction
and energy are known (from reconstruction or MC). The results of this calculation, using MC values,
is shown as yellow filled histograms in the same figure. The individual coordinates are now in good
agreement with the MC values. The resolution reaches values between 3.5\,m and 4.5\,m. \\
The slight trend to positive values in the $z$ coordinate, both before and after the correction
towards the interaction vertex, is caused by the non-isotropic angular acceptance of the OMs which
was not corrected for here. As the OMs are orientated looking downward at a 45$^{\circ}$ angle (see
Section~\ref{sec:OM}), they are less sensitive for light coming from above than for light coming
from below; however, the position reconstruction assumes an isotropic light detection efficiency,
which leads to a slight overestimate of the $z$ coordinate. \\ 
Figure~\ref{fig:ort} also shows the resolution of the reconstruction of the interaction time, again
before (red lines), and after the correction towards the interaction vertex (yellow areas). The
offset $\Delta t \approx 20$\,ns before the correction corresponds to the distance $\Delta r$
between the interaction vertex and \rcg\ and is corrected accordingly, assuming that the particle is
travelling with the speed of light. For this correction, again the MC values of energy and direction
were used. The mean of the time distribution is closer to zero after the correction, though still
slightly shifted to times reconstructed too large. \\ 
It can therefore be concluded that it is possible to reconstruct \rcg\ and the interaction vertex of
the shower with a resolution of a few metres by this method. As mentioned above, the interaction
vertex is not needed for the final reconstruction of direction and energy, as \rcg\ is the reconstructed
position which is used throughout all following reconstruction steps of the {\it ShowerFitter} (see
following chapter). \\
The reconstructed position is an important parameter for the reconstruction of direction and energy,
as the fit for direction and energy is likely to fail, if a wrong position is used (see
Section~\ref{sec:topologies}).   

\subsection{Results for Events with Optical Background}

\begin{figure}[h] \centering
\includegraphics[width=7.4cm]{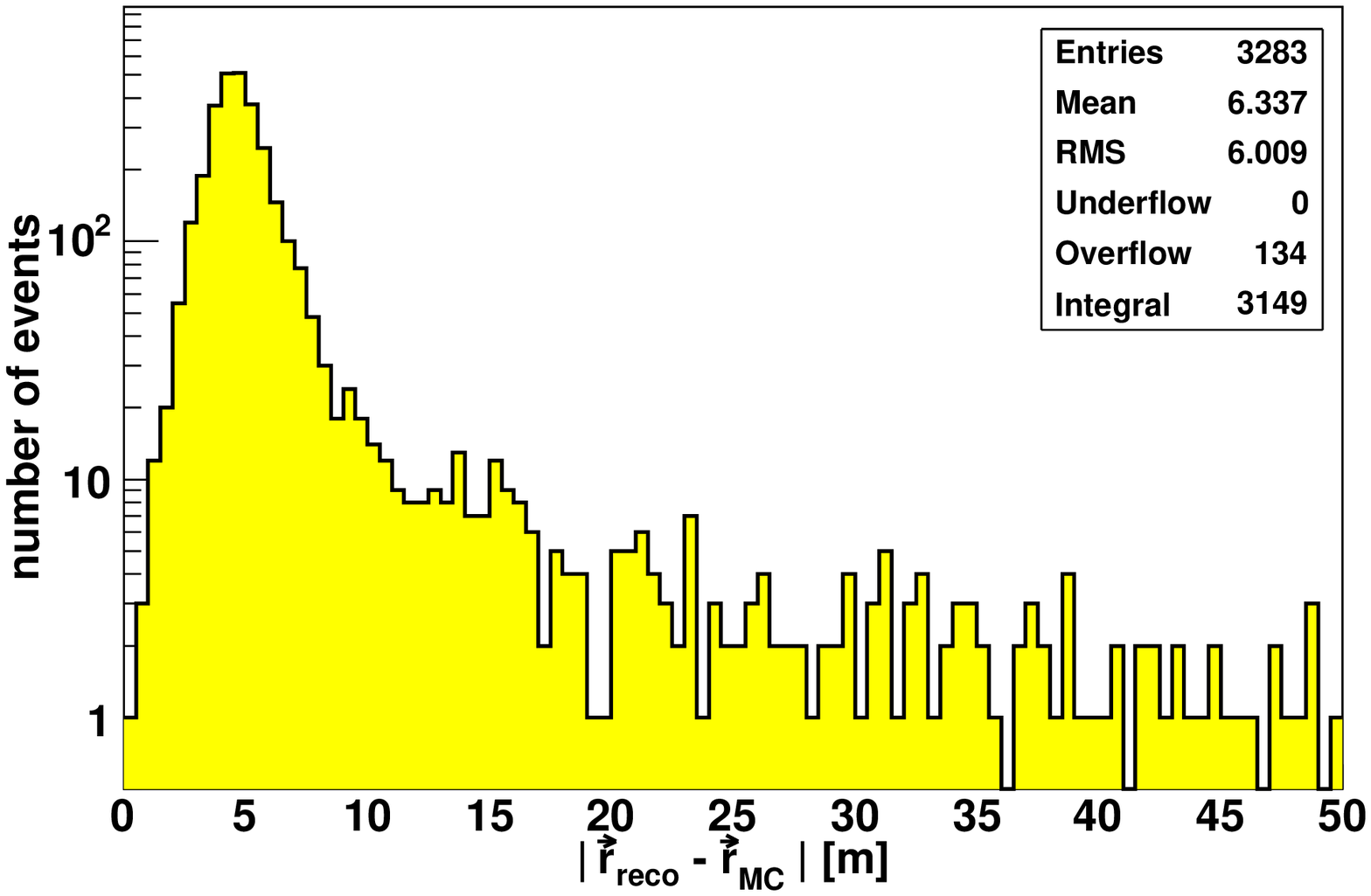}
\includegraphics[width=7.4cm]{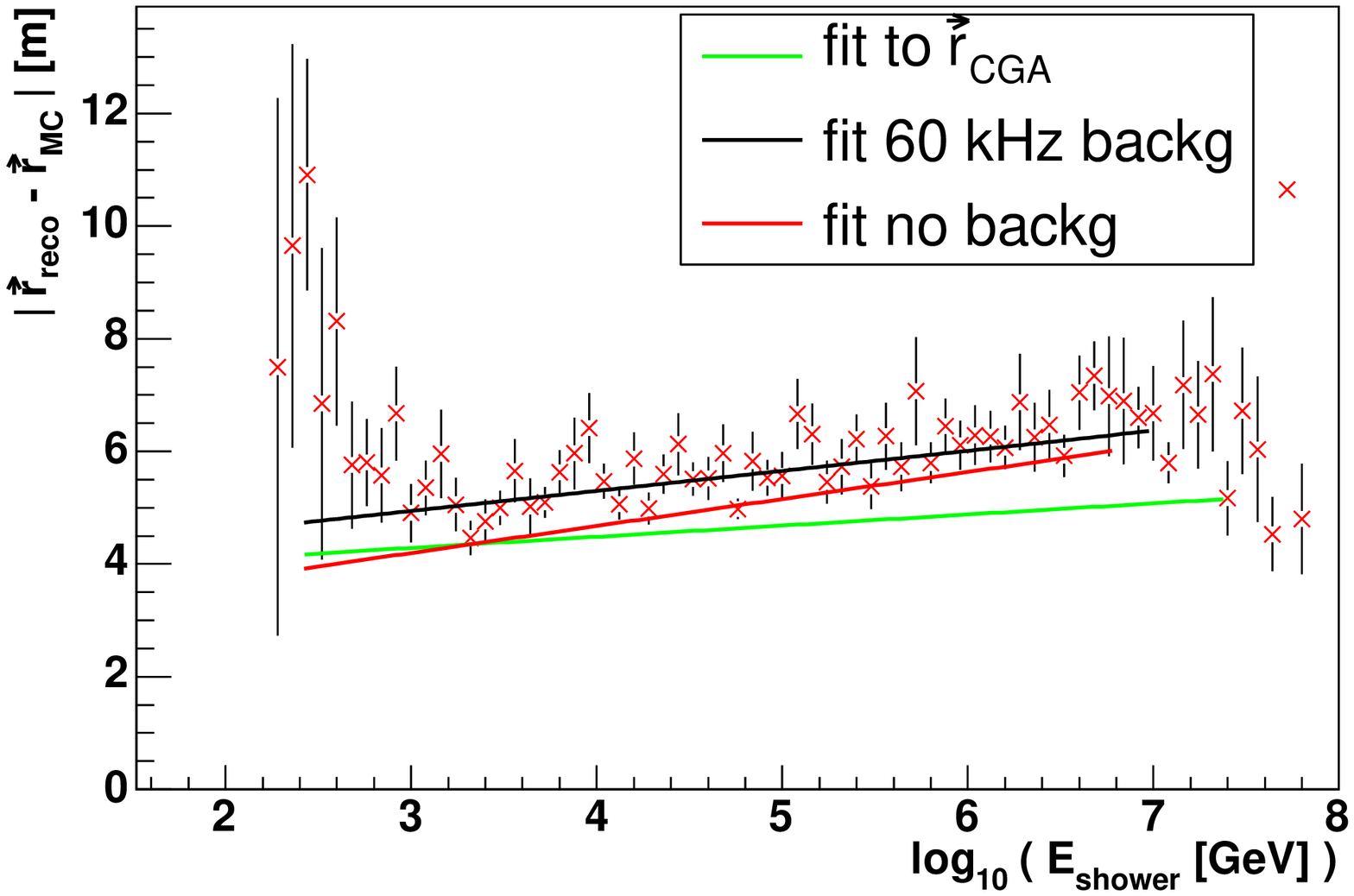}
\caption[Difference between reconstructed and MC vertex, with noise]
{Left: Total distance between reconstructed and MC vertex, for an event sample with 60\,kHz
  background. Right: Profile of the distance, plotted as a function of the MC shower energy. The
  green line again refers to equation (\ref{eq:shower_max_amp}) from
  Section~\ref{sec:hadronic_showers}, the red line is the fit to the background-free data shown
  in Figure~\ref{fig:ortsfehler}, and the black line is a fit to the data shown in this figure.}
\label{fig:ortsfehler_60k}
\end{figure}

Figures~\ref{fig:ortsfehler} and~\ref{fig:ort} are the results of the position reconstruction for a
background-free event sample. However, at the ANTARES site, signal events always contain optical
noise from the deep sea (see Section~\ref{sec:noise}). A filter strategy to suppress this background
has been described in Section~\ref{ch:trigger}. The strategy is based on the distinction between
{\it physics hits} caused by Cherenkov radiation and optical noise hits. However, it is unavoidable
that some of the physics hits are also cut away in the filter or that some background hits remain in
the sample. Therefore, the results of the position reconstruction will be slightly different for 
events with background and using the filter conditions. \\ 
Figure~\ref{fig:ortsfehler_60k} shows the results of the position reconstruction for the same data
sample as used above, with neutrino energies between 100\,GeV and 100\,PeV, but with 60\,kHz
optical noise added per OM. A fit to the data is shown in black, given by

\begin{equation}
|\vec{r}_{\text{CG}} - \vec{r}_{\text{MC}}|/\textrm{m} = \Delta r(E)/\textrm{m} = 3.9 + 0.36 \cdot \log_{10} ( E /\textrm{GeV} ).
\end{equation}

\begin{figure}[h] \centering
\includegraphics[width=10cm]{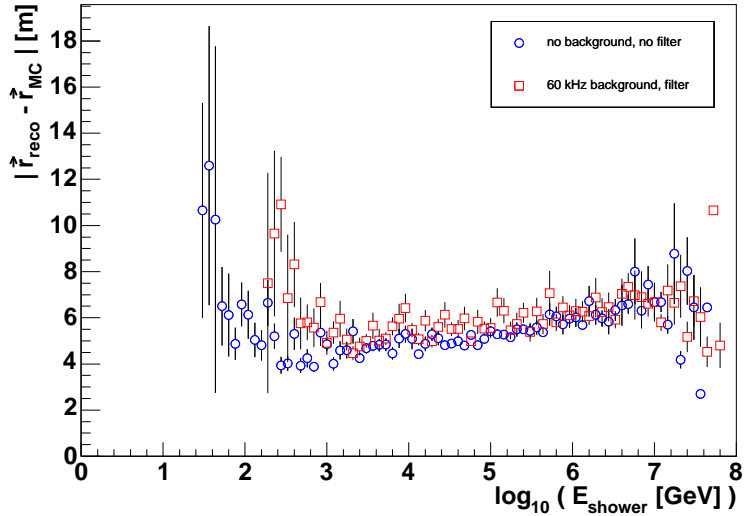}
\caption[Comparison of position with and without noise]{Comparison of the results of the position
  reconstruction for background-free events (blue circles) and the results for the same events 
  with 60\,kHz background added and after the filter conditions (red squares).}
\label{fig:ortsfehler_vergleich}
\end{figure}

The left plot indicates that the resolution has deteriorated, and also that some 660 events, mainly
with very small shower energies below 100\,GeV (see Figure~\ref{fig:ortsfehler_vergleich}), are
missing in the event sample because they did not pass the filter conditions. From
Figure~\ref{fig:ortsfehler_vergleich} it can be seen that events with a shower energy above $\sim
40$\,TeV show very similar results with and without background, which is due to the larger overall
number of hits in these events. For events with a lower energy, on the other hand, the offset
between the reconstructed and the true position is slightly larger for the sample with than for the
sample without background. 

\section{Reconstruction of the Shower Direction}\label{sec:prefit_dir}
\subsection{Method 1: Using the Cherenkov Angle} \label{sec:prefit_dir1}

\begin{figure}[h] \centering
\includegraphics[width=6cm]{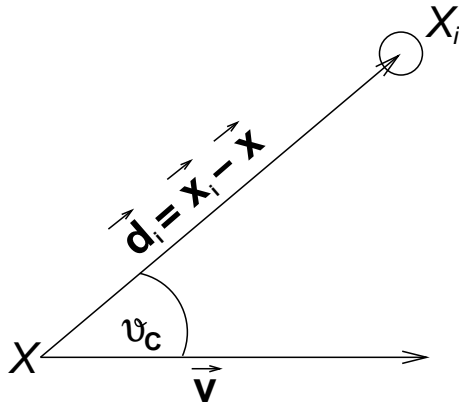}  
\caption[Geometric situation for the direction reconstruction]
{Schematic view of the geometric situation for the reconstruction of the
  direction. $\vec{v}$ is the direction of the shower, $X$ and $X_i$ mark the position of the
  shower and OM$_i$, respectively, and $\vartheta_C$ is the Cherenkov angle, $42.2^{\circ}$.}
\label{fig:reco_dir_scheme}
\end{figure}

An isolated charged particle (e.g.~a muon) traversing the detector medium produces Cheren\-kov light
on a cone along its track. The opening angle of this cone depends on the refraction index of the
medium and is around $42.2^{\circ}$ at the ANTARES site (see Section~\ref{sec:refraction}).
This is also true for the individual charged particles in a shower; however, with respect to the
shower axis, the angular distribution of the photons is rather broad, with a peak around the
Cherenkov angle (see Section~\ref{sec:hadronic_showers}). \\
To get a rough estimate of the shower direction, e.g.~as a starting value for a later
fit, one can assume that all photons are produced on a Cherenkov cone under a fixed angle with
respect to the shower axis. One can then develop an algorithm to calculate the shower direction from
the position of the hits. This algorithm is very similar to the one described in
Section~\ref{sec:pos}. For shower reconstruction in ANTARES it has first been described
in~\cite{naumann}. Its advantage is that it does not need the reconstructed interaction point as
input information. \\ 
Consider Figure~\ref{fig:reco_dir_scheme}, where $\vec{x} = (x,y,z)$ represents the position vector
of the interaction and $\vec{v} = (v_x, v_y, v_z)$ the direction vector of the shower. Here, the
direction of the shower is approximated by the direction of the neutrino (see
Section~\ref{sec:hadronic_showers}). 

\begin{figure}[h] \centering
\includegraphics[width=15cm]{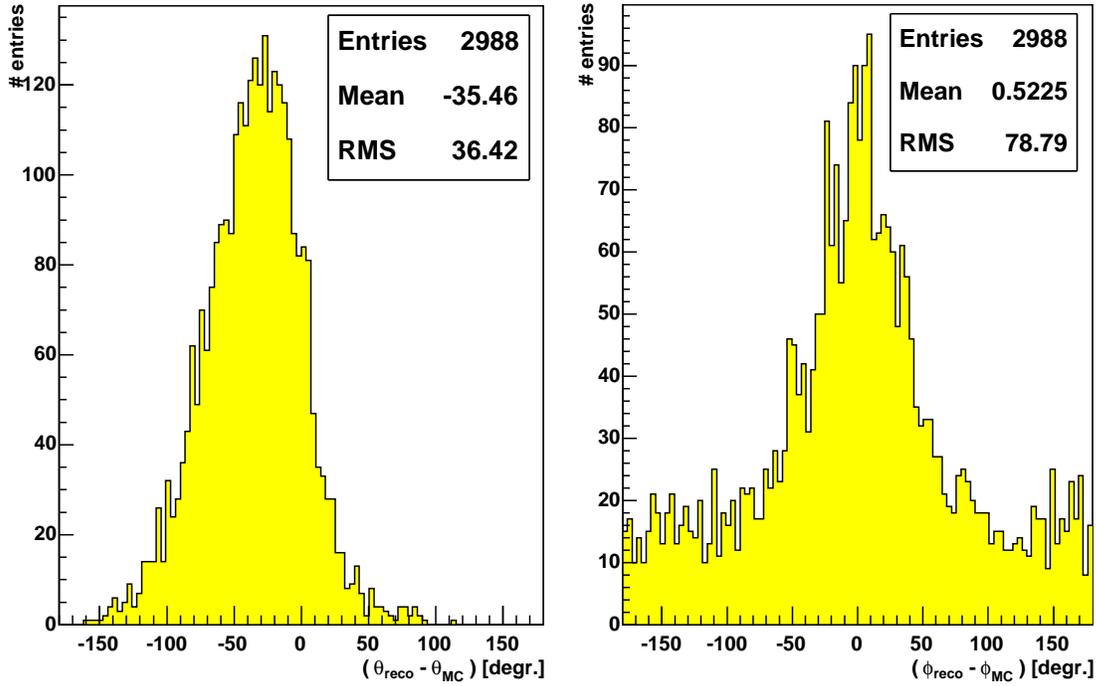}  
\caption[Reconstruction using the Cherenkov cone: results]
{Results of the reconstruction of the direction, using the Cherenkov cone condition, for the zenith
  angle $\theta$ and the azimuth angle $\phi$. Shown are the differences between the reconstructed
  angles and the true MC angles.}
\label{fig:theta_phi_pre}
\end{figure}

\begin{figure}[h]
\begin{minipage}{7.0cm}
\centering \epsfig{figure=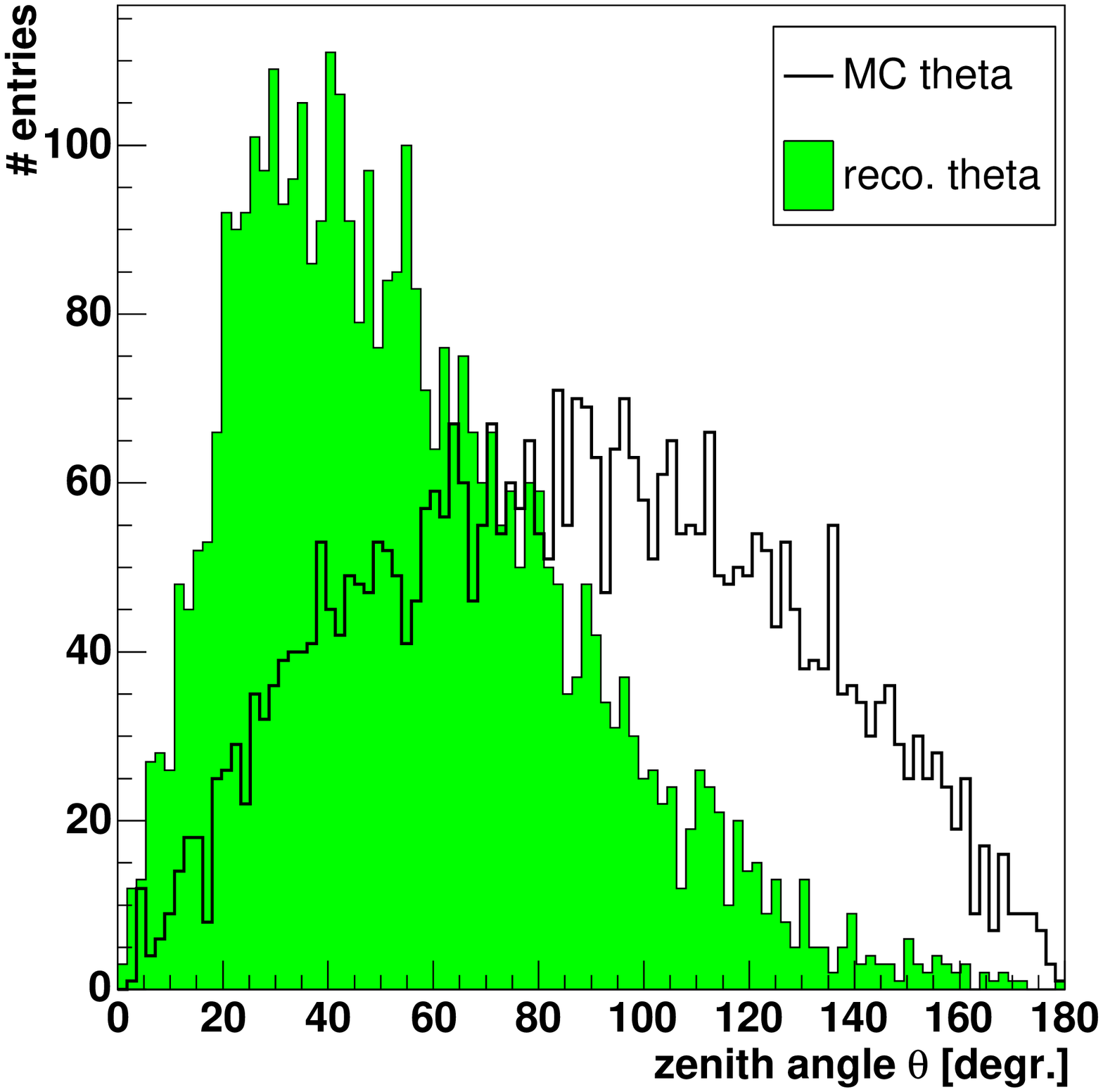, width=7cm}  
\caption[MC and reconstructed zenith angle]{MC and reconstructed zenith angle $\theta$. It is
  clearly visible how the reconstruction tends to misreconstruct downgoing events ($\theta >
  90^{\circ}$) as upgoing.}
\label{fig:theta_theta_pre}
\end{minipage}
\hspace{3mm}
\begin{minipage}{7.5cm}
\centering \epsfig{figure=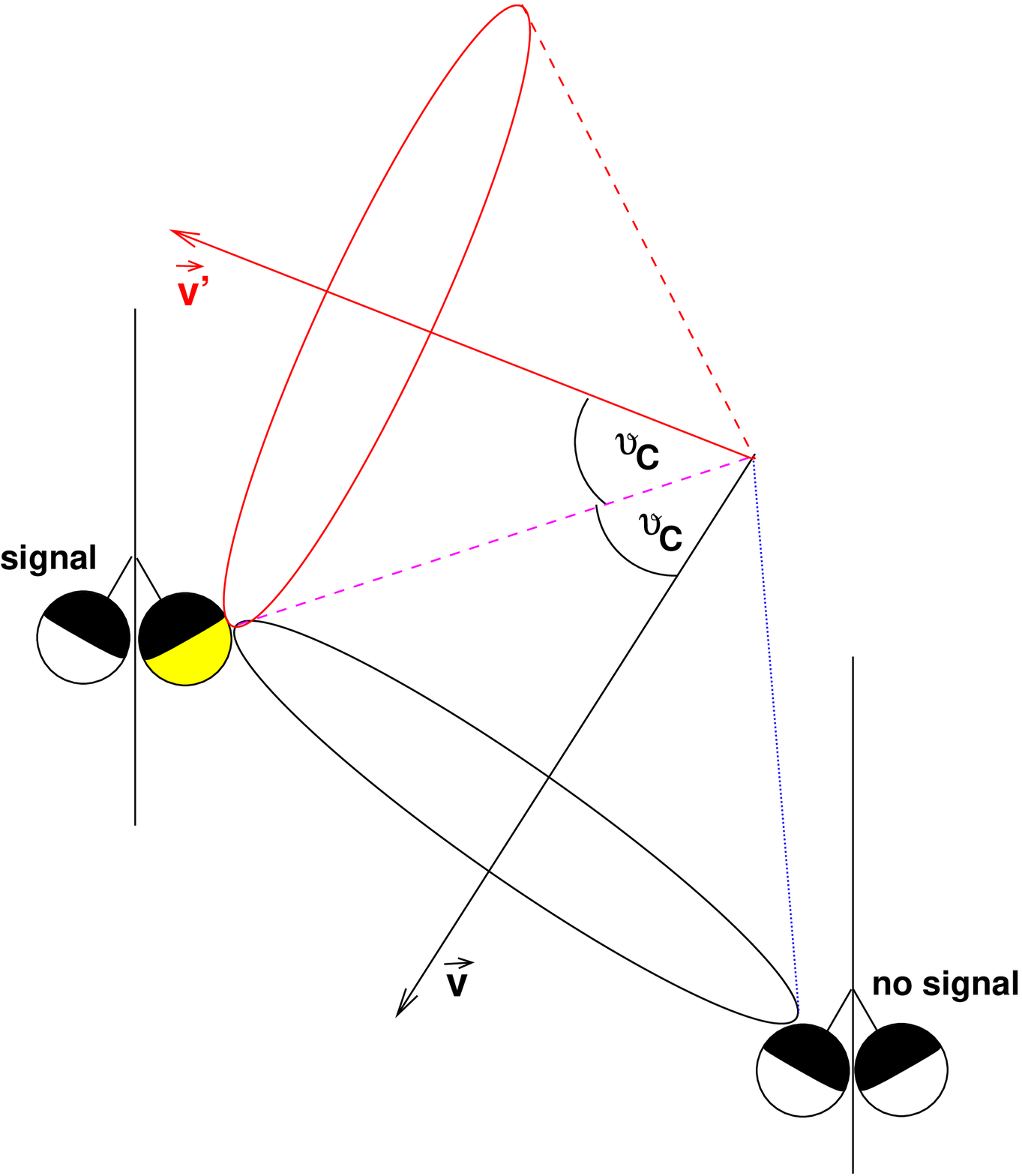, width=7cm}  
\caption[Schematic view of a possible misreconstruction]{Schematic view of a possible
  misreconstruction: Instead of the true direction $\vec{v}$, the reconstruction produces
  $\vec{v}^{\prime}$ as a result, due to the unfavourable angle with respect to the OM orientation.}
\label{fig:misreco}
\end{minipage}
\end{figure}

A photon emitted under the Cherenkov angle $\vartheta_C$ hits OM$_i$, which has the coordinates
$\vec{x}_i = (x_i,y_i,z_i)$. The vector $\vec{d}_i$ connecting the points $X$
and $X_i$ is then defined by ${\vec{d}_i = (x_i - x, y_i - y, z_i - z)}$, with

\begin{equation}
\cos \vartheta_C = \frac{\vec{v}\cdot\vec{d}_i}{|\vec{v}|\cdot|\vec{d}_i|}
\label{eq:theta}
\end{equation}

for all hits $i$. The norm of $\vec{d}_i$ can be written using the speed of light and the travel
time: $|\vec{d_i}| = (c/n) (t_i - t)$. Suppose $|\vec{v}| \equiv 1$; then equation (\ref{eq:theta})
can be rewritten as

\begin{equation*}
\cos \vartheta_C\cdot (c/n)\cdot(t_i - t) = v_x\cdot(x_i - x) + v_y\cdot(y_i - y) + v_z\cdot(z_i - z)
\end{equation*}

for all hits $i$ in the event. Again one can subtract these equations pairwise, thus eliminating the
contributions of $\vec{x}$. For $N$ hits one obtains $(N-1)$ equations of the form 

\begin{equation*}
\cos \vartheta_C\cdot (c/n)\cdot(t_i - t_{i+1}) = v_x\cdot(x_i - x_{i+1}) + v_y\cdot(y_i - y_{i+1}) + v_z\cdot(z_i -z_{i+1})
\end{equation*}

where $v_x, v_y$ and $v_z$ are the unknown quantities. This system of equations can be solved as
described in Section~\ref{sec:pos}. \\
This algorithm yields a rough estimate of the shower direction, as shown in
Figure~\ref{fig:theta_phi_pre} for event sample A, without background. The resolution is
about 36$^{\circ}$ in zenith angle $\theta$ and 79$^{\circ}$ in azimuth angle $\phi$. In $\theta$,
there is a systematic shift towards small angles which is due to the fact that downgoing events
($\theta > 90^{\circ}$) are often misreconstructed as upgoing. This is illustrated in 
Figure~\ref{fig:theta_theta_pre}, where the reconstructed (in green) and the MC values (bold line) of
$\theta$ are plotted for each event. While the MC distribution is flat in $\cos \theta$, the
reconstructed one shows a clear trend towards smaller values of $\theta$ and therefore tends to
reconstruct downgoing events as upgoing. The effect can be explained as schematically shown in
Figure~\ref{fig:misreco}: One side (marked as a blue, dotted line) of the Cherenkov cone cannot be
detected because the light hits the OM at an angle where the OM's angular acceptance is zero. With
the information from only one side of the cone, the direction of the shower might as well be the
vector marked $\vec{v}^{\prime}$ in the plot. In this case, the reconstructed direction is $2
\vartheta_C$ off the real direction and is upgoing instead of downgoing. In a real event with hits
in more than two OMs, the effect of the missing information is less prominent, and the shift in
$\theta$ is about $-35^{\circ}$ on average.  \\
The result of this method depends also on the shower energy. This is demonstrated in
Figure~\ref{fig:delta_over_E}, where the differences between the reconstructed and the MC angles, as
shown in Figure~\ref{fig:theta_phi_pre}, are plotted versus the MC shower energy. Especially for
$\phi$, the resolution deteriorates very strongly for increasing shower energies. This deterioration
can be understood from the fact that the distribution of the photon radiation angle with
respect to the shower axis becomes broader and broader for higher shower energies 
(cf.~Figure~\ref{fig:cherenkov_all}).

\afterpage{\clearpage}

\subsection{Method 2: Amplitude Ratio}\label{sec:prefit_dir2}
\begin{figure}[h] \centering
\includegraphics[width=8cm]{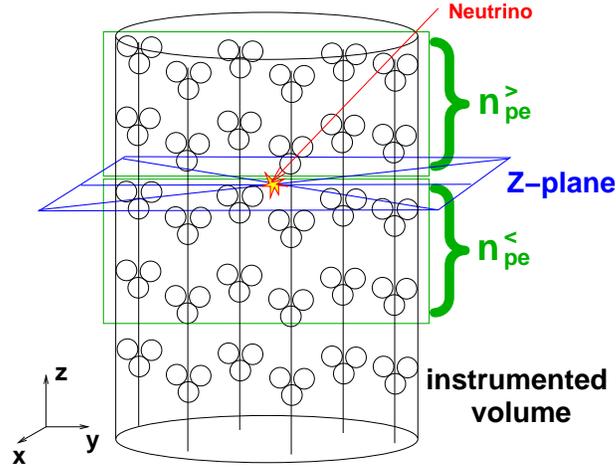}
\caption[Geometric principle of the amplitude ratio method]{Geometric principle of the amplitude
  ratio method: The instrumented volume is symmetrically divided into two parts by a horizontal
  plane through the $z$-coordinate of the interaction vertex.} 
\label{fig:ratio_scheme}
\end{figure}

As shown in the previous subsection, the reconstruction of the shower direction with the Cherenkov
angle method is not always successful. The knowledge of the neutrino orientation, i.e.~whether a
neutrino is upgoing or downgoing, is, however, a significant quantity in the reconstruction. Most
importantly, it has to be excluded that the event was caused by a downgoing atmospheric muon
imitating a shower (see Section~\ref{sec:atm_muons}), instead of an upgoing neutrino. \\  
Therefore, a method was developed to be able to quickly distinguish between upgoing and downgoing
events. The algorithm is simple: The detector is divided into two regions by placing a
horizontal plane, the so-called {\it $Z$-plane}, at the $z$-coordinate of the interaction vertex. 
The assumption is then that for {\it upgoing} events, the largest part of the emitted light is
measured in OMs which are positioned {\it above} the $Z$-plane, while for {\it downgoing} events, a
larger fraction is measured {\it below} the $Z$-plane. A schematic view of the geometric principle
is shown in Figure~\ref{fig:ratio_scheme}. \\ 
To discriminate between upgoing and downgoing events, the ratio $r_{pe}$ of the amplitudes measured
in the two regions, i.e.~of the total number of photo-electrons in OMs above the $Z$-plane,
$n_{pe}^{>}$, and below the $Z$-plane, $n_{pe}^{<}$, is calculated: 

\begin{figure}[t] \centering
\includegraphics[width=13.5cm]{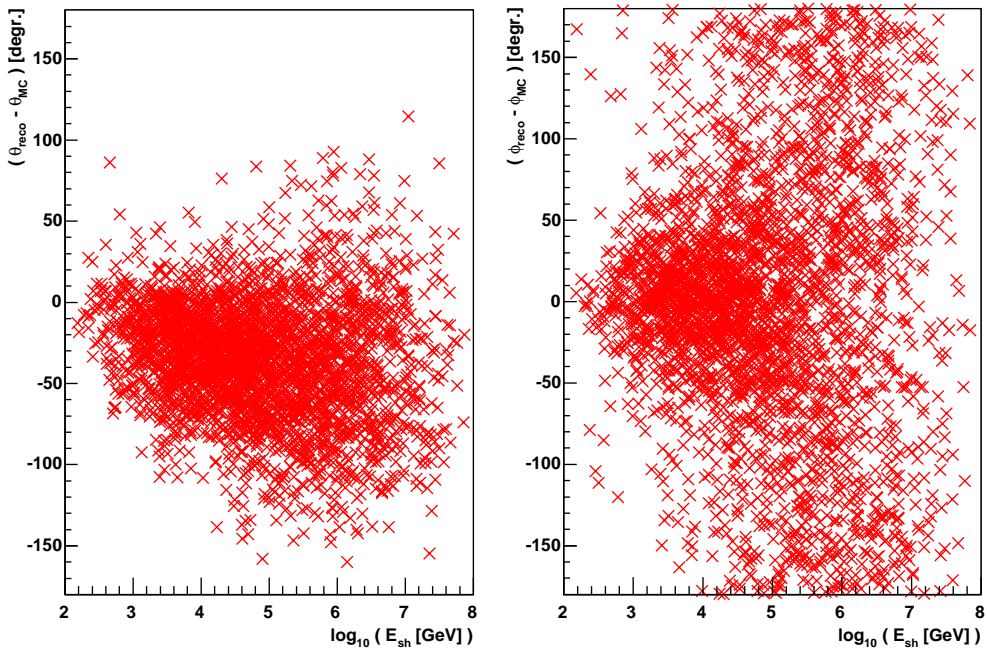}  
\caption[Reconstruction results over shower energy]{Resolution of the zenith angle $\theta$ (left)
  and the azimuth angle $\phi$ (right) as shown in Figure~\ref{fig:theta_phi_pre}, plotted vs.~the MC
  shower energy. }
\label{fig:delta_over_E}
\end{figure}

\begin{equation}
r_{pe} = \frac{n_{pe}^{>}}{n_{pe}^{<}}.
\end{equation}

\begin{figure}[h] \centering
\includegraphics[width=7.4cm,height=5cm]{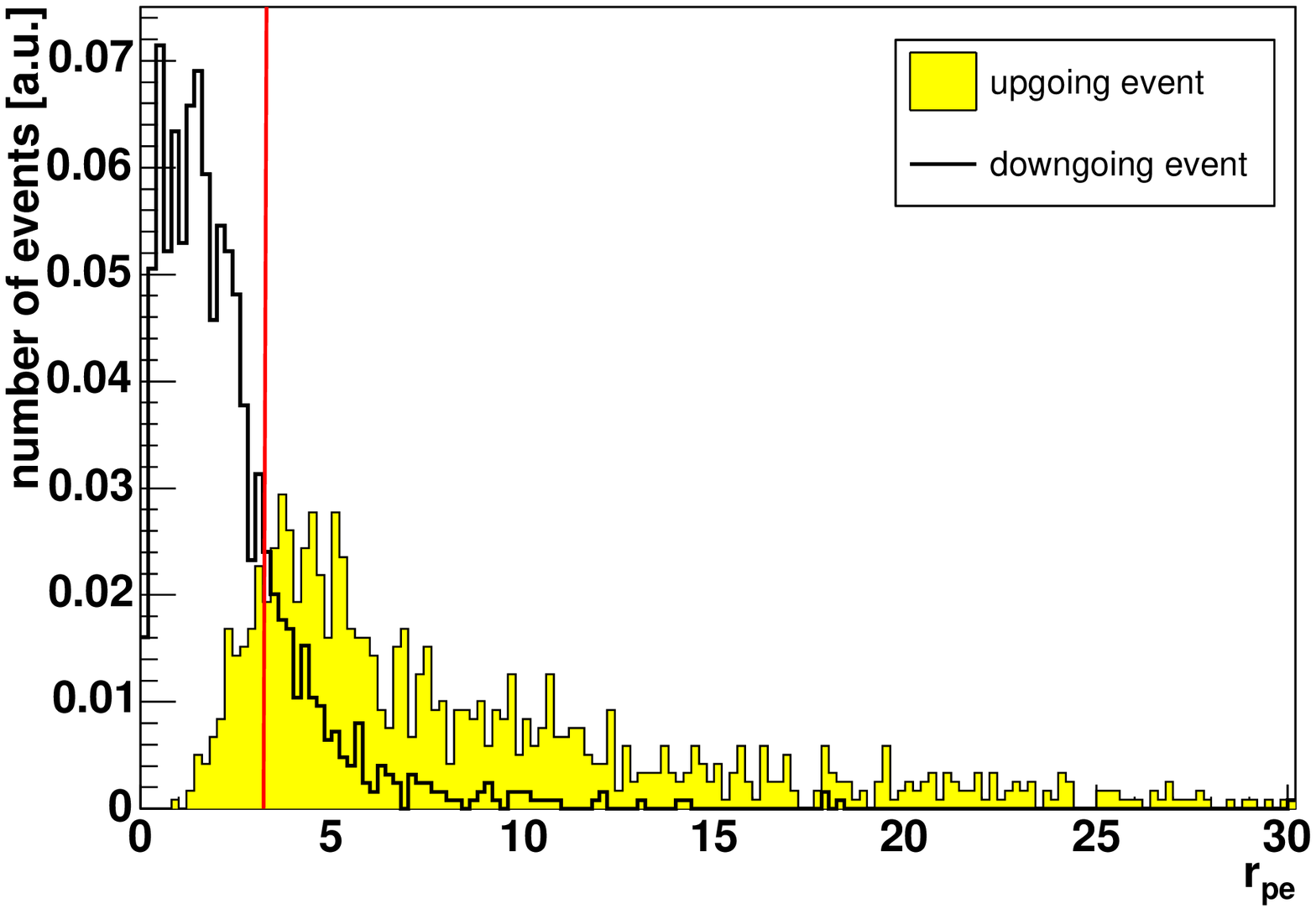}  
\includegraphics[width=7.4cm,height=5cm]{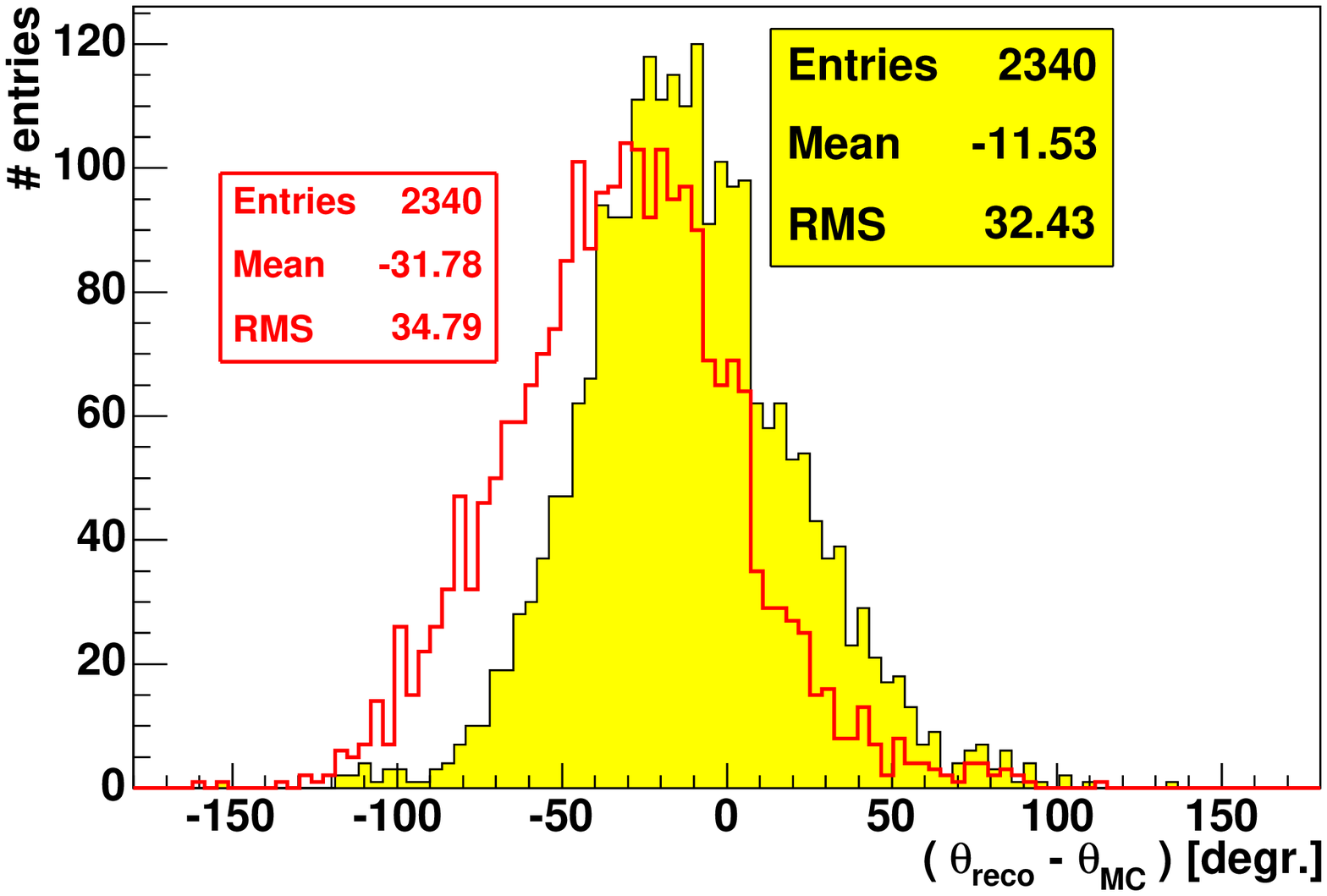}  
\caption[Ratio of pe above and below the shower position]
{Left: Ratio of pe above and below the shower position, for upgoing and downgoing events; both
  histograms have been normalised to 1. The red line marks the cut proposed at a ratio of 3.3.
Right: Distribution of the difference between the reconstructed and the MC $\theta$ angles, before
the amplitude ratio correction (red lines) and afterwards (yellow histogram).}
\label{fig:npe_ratio}
\end{figure}

To avoid edge effects, the regions inside which $n_{pe}^{>}$ and $n_{pe}^{<}$ are evaluated are
chosen symmetrically, i.e.~the number of storeys used for the calculation is the same for both sides
of the $Z$-plane and defined by the minimum number of storeys between the $Z$-plane and the detector
top or bottom. By this method, the probability that $r_{pe}$ is affected by edge effects is reduced.
Note that $r_{pe}$ can only be calculated if the event occurs well inside the detector, such that 
$n_{pe}^{<} \neq 0$. \\
$r_{pe}$ as calculated for MC events from event sample A without background is shown in
Figure~\ref{fig:npe_ratio} on the left hand side. As expected, $r_{pe}$ is much smaller for
downgoing than for upgoing events. By applying a cut on $r_{pe}$, events are classified as downgoing
or upgoing, respectively. In the example shown here, the cut was applied at $r_{pe} = 3.3$.
88.1\% of the upgoing events are above that value, so that the probability to misinterpret an upgoing
neutrino as downgoing is 11.9\%; on the other hand, 80.7\% of the downgoing events have a
ratio smaller than 3.3, so that the probability to misinterpret a downgoing neutrino as upgoing is
19.3\%.

The right hand side of Figure~\ref{fig:npe_ratio} shows the improvement which is possible if this
method is combined with the one from Section~\ref{sec:prefit_dir1}: The red histogram shows the
equivalent distribution as in Figure~\ref{fig:theta_phi_pre} (left). The reconstructed values of
$\theta$ were then corrected by using $r_{pe}$; if the orientation did not agree with the
reconstructed $\theta$, $\theta$ was shifted to $\pi - \theta$. The comparison of the reconstructed
and the MC direction after this correction is shown on the right hand side of
Figure~\ref{fig:npe_ratio} in yellow. Both the mean and the RMS of the distribution are improved by
the correction. The number of events shown in Figure~\ref{fig:npe_ratio} is smaller than in
Figure~\ref{fig:theta_phi_pre}, as the events with undetermined $r_{pe}$ (due to $n_{pe}^{<} = 0$)
have been removed from the sample. 

\section{Energy Reconstruction}\label{sec:e_pre}
\begin{figure}[h] \centering
\includegraphics[width=5cm]{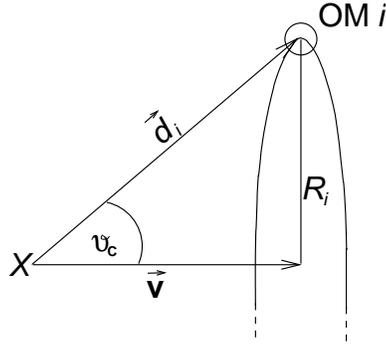}
\caption[Geometric situation for the reconstruction of the energy]
{Schematic view of the geometric situation for the reconstruction of the energy: $\vec{v}$ is the
  direction of the shower axis, while $\vec{d_i}$ is the direction of the photon hitting OM $i$.}
\label{fig:reco_e_scheme}
\end{figure}

The algorithm for the reconstruction of the shower energy which is described in this section uses
the fact that the total hit amplitude, i.e.~the sum of the amplitudes of all hits in an
event, corrected according to distance and angle of each hit with respect to the shower, is
correlated to the shower energy. The method described here is not used in the final energy  
determination, but is useful for consistency checks or for the analysis of data with known shower
direction and interaction vertex. It uses the simplified assumption that all photons are emitted
from one point, $\vec{r}_{\text{CG}}$ (see Section~\ref{sec:pos}), marked $X$ in
Figure~\ref{fig:reco_e_scheme}, and under a fixed polar angle $\vartheta_C$ with respect to the
shower axis, i.e.~along a cone (see Figure~\ref{fig:reco_e_scheme}). $\vartheta_C$ is defined as: 

\begin{equation}
\cos \vartheta_C = \frac{\vec{d_i} \cdot \vec{v}}{|\vec{d_i}| \cdot |\vec{v}|}
\end{equation}

Assuming that the photons are emitted isotropically in the azimuthal angle $\varphi$, the photon
density (in units of length$^{-1}$) at a distance $d = |\vec{d}|$ from $\vec{r}_{\text{CG}}$ is
proportional to the inverse of the circumference of the cone, and thus proportional to $1/R = 1 / (d
\cdot \sin(\vartheta_C))$ (see Figure~\ref{fig:reco_e_scheme}). The measured amplitude $a_i$ in 
OM $i$ can therefore be regarded as the photon density times a unit segment of the circle,
$a_{i,0} = \rho_i \cdot 2 \pi R_0$; the total number of photons on the circle is then $a_i = \rho_i
\cdot 2 \pi R_i = a_{i,0} \cdot R_i/R_0$. The attenuation in water, $\exp(-d/\tau)$, is also taken into 
account; as scattering is not fully simulated in the light propagation software, the attenuation
length $\tau$ is set to 55\,m, which corresponds to the absorption length at 475\,nm, the wavelength
of maximum absorption at the ANTARES site (see Section~\ref{sec:absorption}). Thus, the corrected
number of photons $a_i^{corr}$ emitted along the cone hitting OM $i$ is $a_i^{corr} = a_i
\exp(d_i/\tau)$. The total number of photons $\mathcal{A}$ in an event with $N$ hits is therefore 

\begin{equation}
\mathcal{A} = \sum_{i=1}^N a_i^{corr} = \sum_{i=1}^N a_{i,0} \cdot \frac{R_i}{R_0} \cdot \exp(d_i/\tau).
\end{equation}

\begin{figure}[h] \centering
\includegraphics[width=7.4cm]{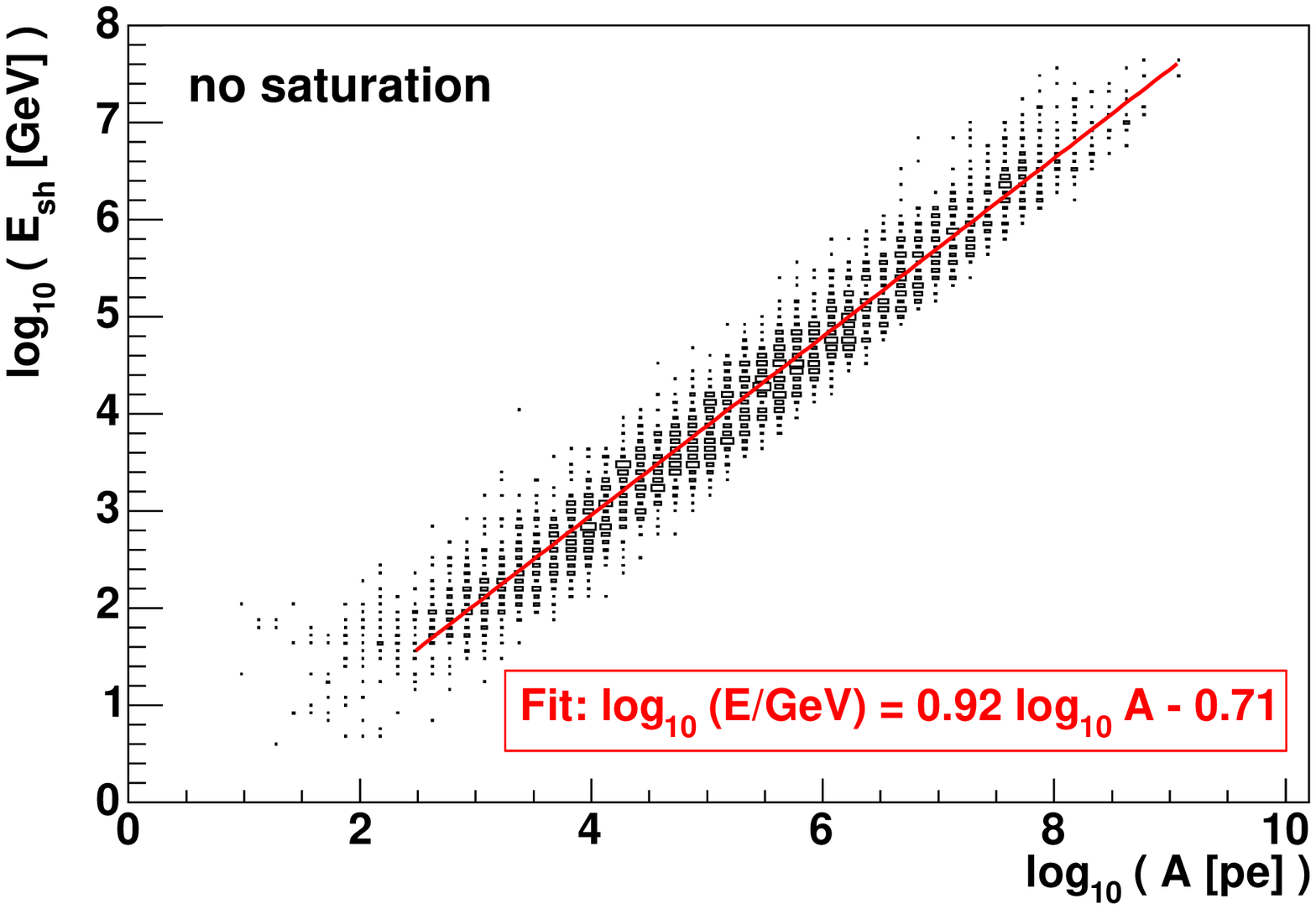}
\includegraphics[width=7.4cm]{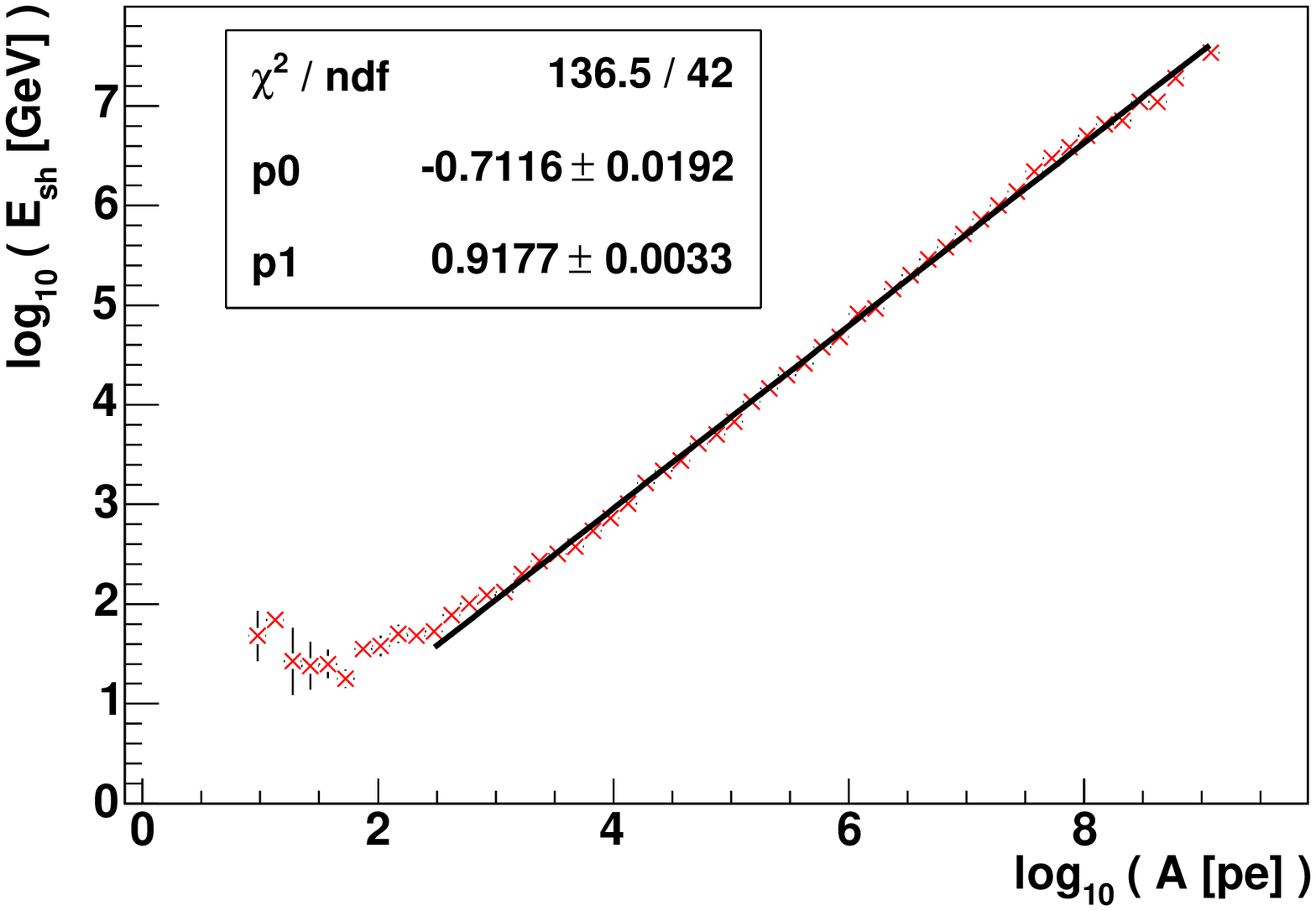}
\caption[Shower energy over total amplitude]{Left: Shower energy $E_{sh}$ in GeV vs.~total
  corrected amplitude $\mathcal{A}$ per event. The linear function which has been fitted to the
  distribution is also shown. Right: Profile of the distribution shown on the left, and fit.}  
\label{fig:npe_E}
\end{figure}

For event sample A, without optical background, $\mathcal{A}$ was calculated using the MC direction
and \rcg\ calculated according to equation (\ref{eq:shower_max_corr}), using MC values. The
distribution of $\mathcal{A}$ versus the MC shower energy $E_{sh}$ is shown in
Figure~\ref{fig:npe_E} on the left hand side. 

To find a fit curve to this distribution, the profile of the MC shower energy $E_{sh}$ with respect
to $\mathcal{A}$ has been considered, and a polynomial has been fit to that profile, as shown
on the right hand side of Figure~\ref{fig:npe_E}. This yields the following relation between
averaged total amplitude and shower energy:

\begin{equation}
  \log_{10}(E_{sh}/{\textrm{GeV}}) = 0.92 \cdot \log_{10} \mathcal{A} - 0.71.
\label{eq:e_pre}
\end{equation}

\begin{figure}[h] \centering
\includegraphics[width=7.4cm]{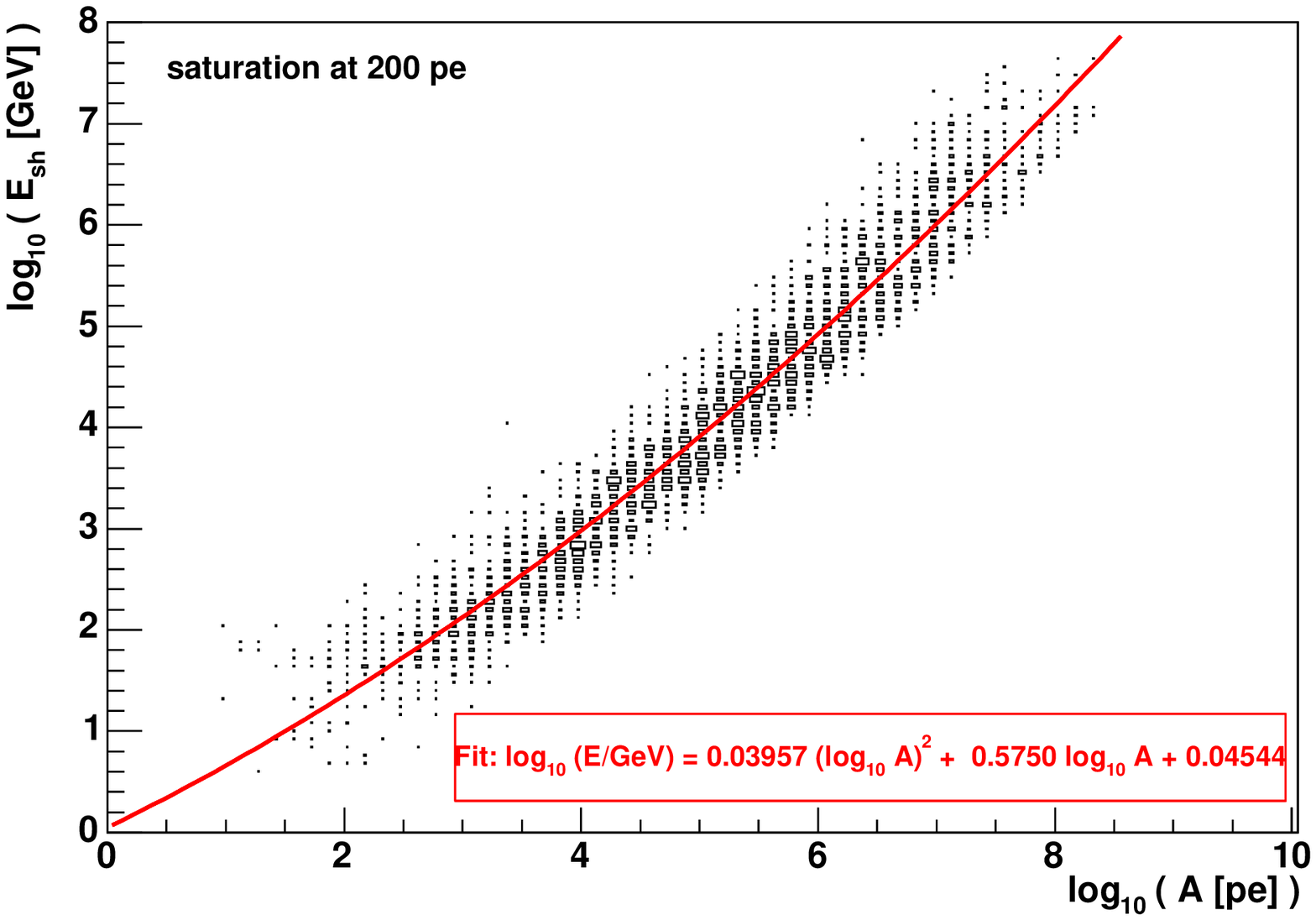}
\includegraphics[width=7.4cm]{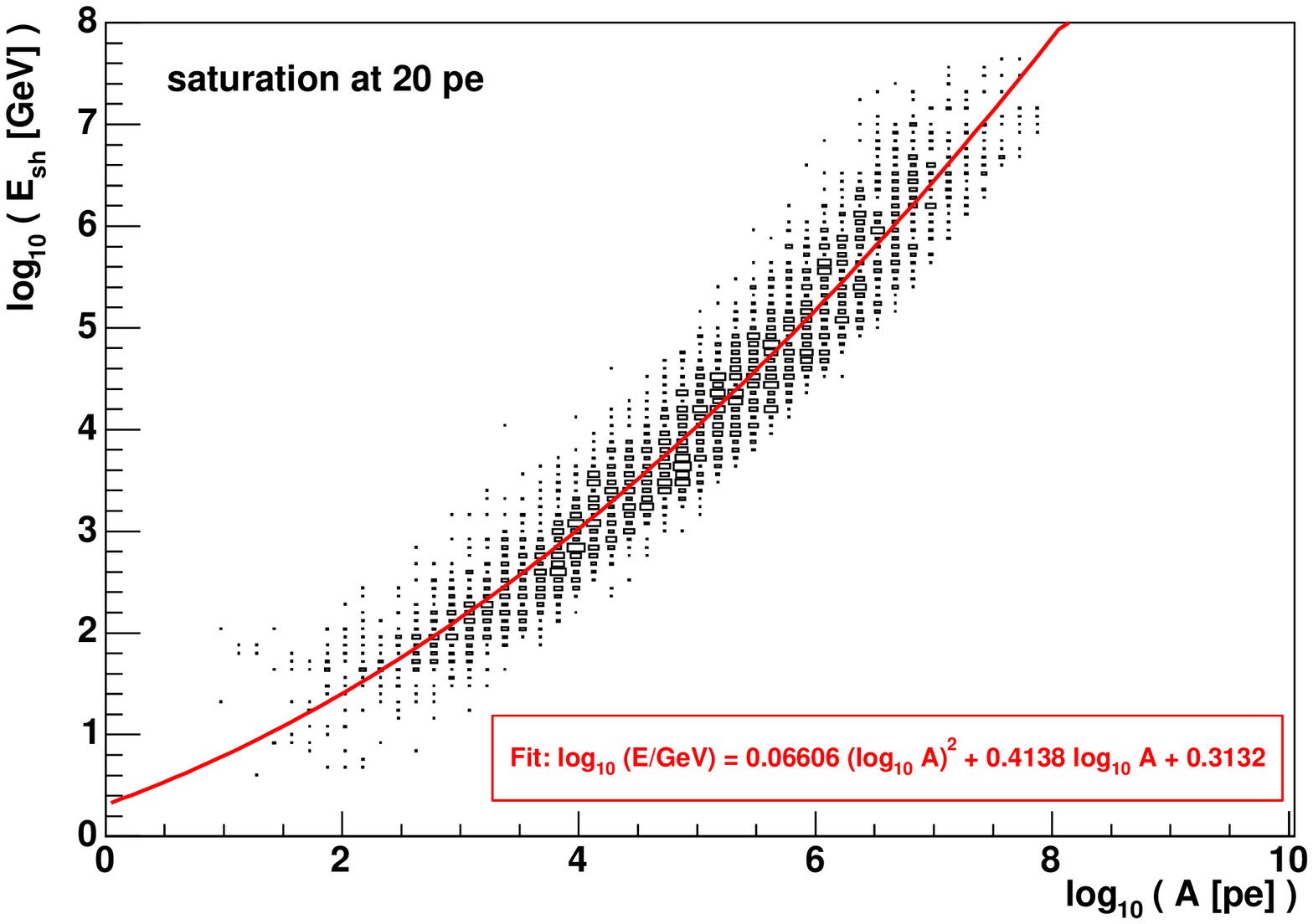}
\caption[Shower energy over total amplitude including saturation]{Shower energy $E_{sh}$ in GeV
  vs.~total corrected amplitude $\mathcal{A}$ per event, for a saturation at 200\,pe (left) and 20\,pe
  (right). The polynomials which have been fitted to the distributions are also shown.}  
\label{fig:npe_E_WF_SPE}
\end{figure}

Something which has not been taken into account for the calculation of the total amplitude
$\mathcal{A}$ is the fact that the photomultiplier electronics saturate at a certain amplitude,
depending on the data taking mode (see Section~\ref{sec:digitisation}). To study the
effect of saturation on the energy reconstruction, two values for saturation, 200\,pe (equivalent to
the WF mode) and 20\,pe (equivalent to the SPE mode) have been studied in addition to the
non-saturation case shown before. The resulting distributions are shown in
Figure~\ref{fig:npe_E_WF_SPE}. 

For the saturation at 200\,pe, one finds the following parameterisation for the shower energy:
\begin{equation}
  \log_{10}(E_{sh}/{\textrm{GeV}}) = 0.040 \cdot (\log_{10} \mathcal{A})^2 + 0.58 \cdot \log_{10} \mathcal{A} + 0.045;
\label{eq:e_pre_WF}
\end{equation}

for the saturation at 20\,pe, the parameterisation becomes
\begin{equation}
  \log_{10}(E_{sh}/{\textrm{GeV}}) = 0.066 \cdot (\log_{10} \mathcal{A})^2 + 0.41 \cdot \log_{10} \mathcal{A} + 0.31.
\label{eq:e_pre_SPE}
\end{equation}

\begin{figure}[h] \centering
\includegraphics[width=7.4cm]{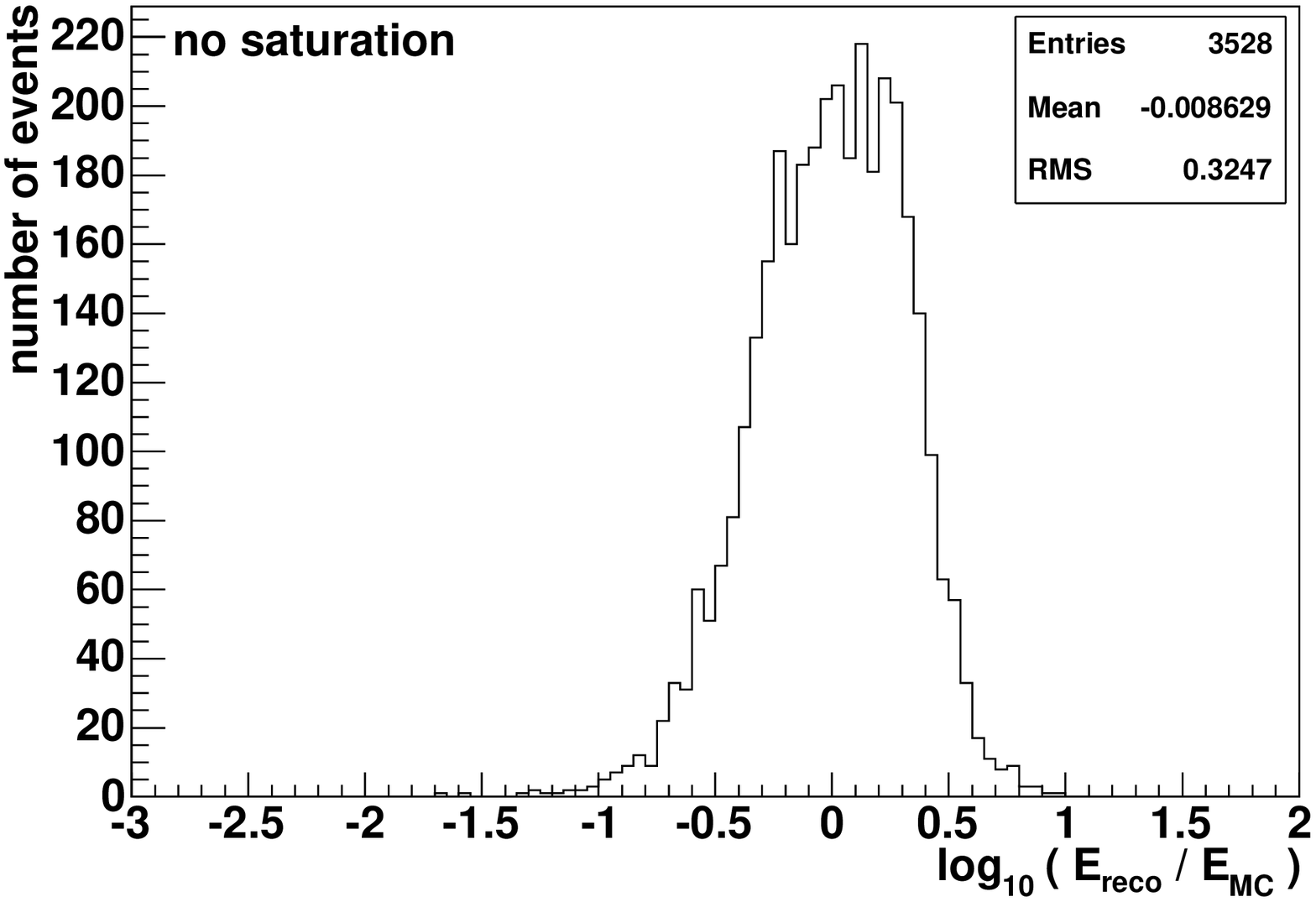}
\includegraphics[width=7.4cm]{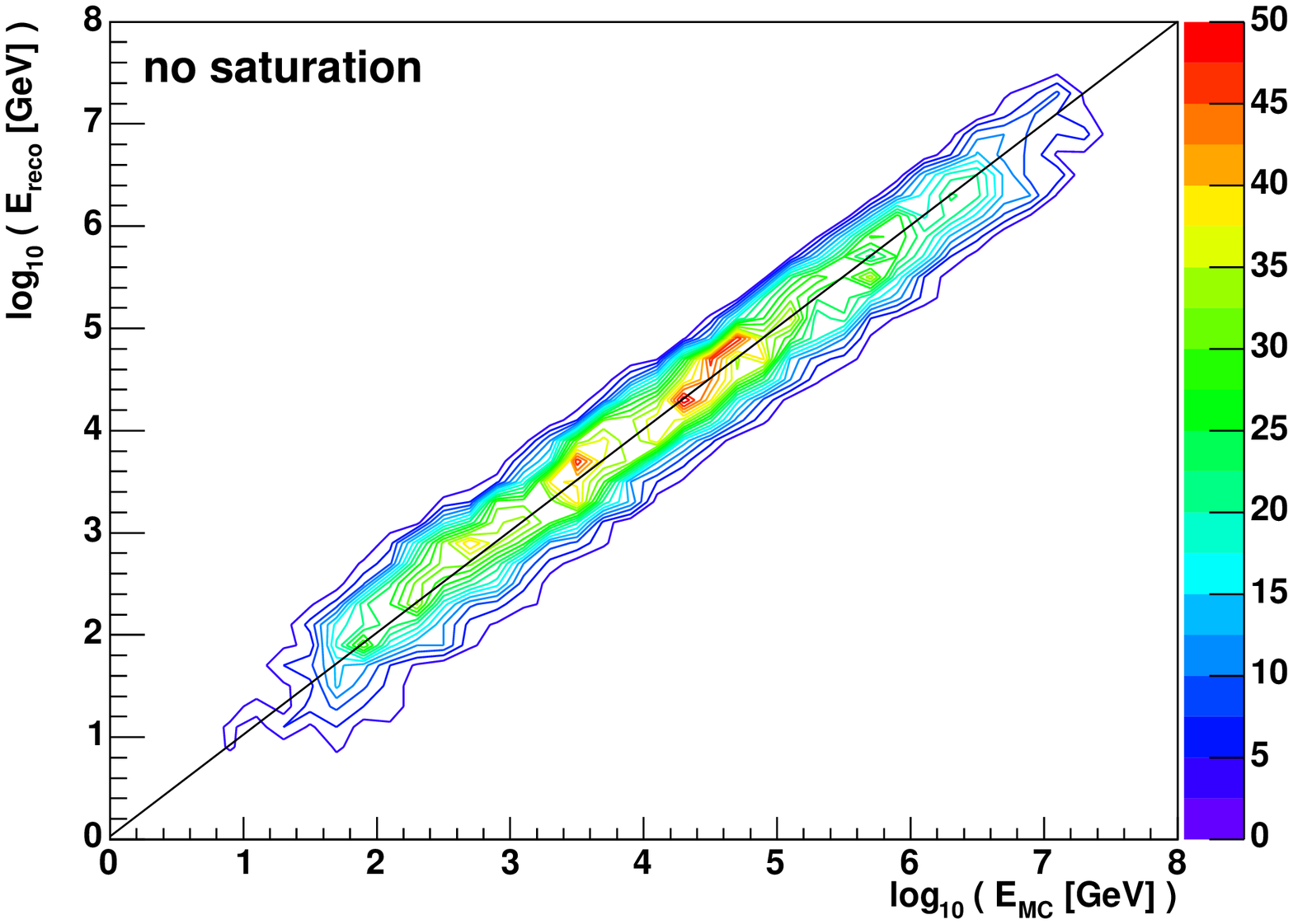}
\caption[Energy reconstruction results, no saturation]
{Results of the energy reconstruction using the corrected total amplitude, for the case of no
  saturation: Distribution of the logarithmic quotient of reconstructed and MC shower energy
  (left), and reconstructed versus MC shower energy as contour plot (right).}
\label{fig:e_pre}
\end{figure}

\begin{figure}[h] \centering
\includegraphics[width=7.4cm]{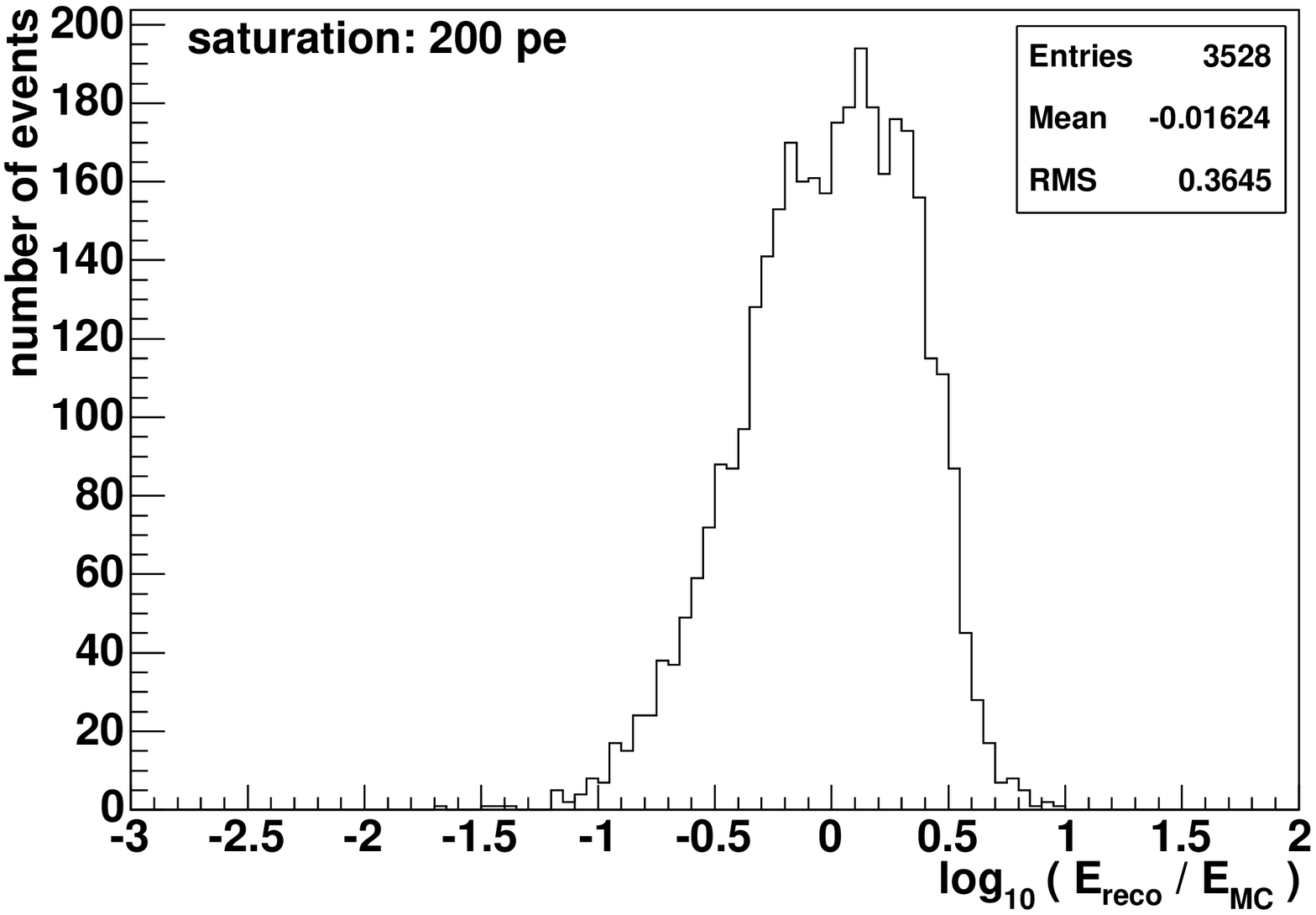}
\includegraphics[width=7.4cm]{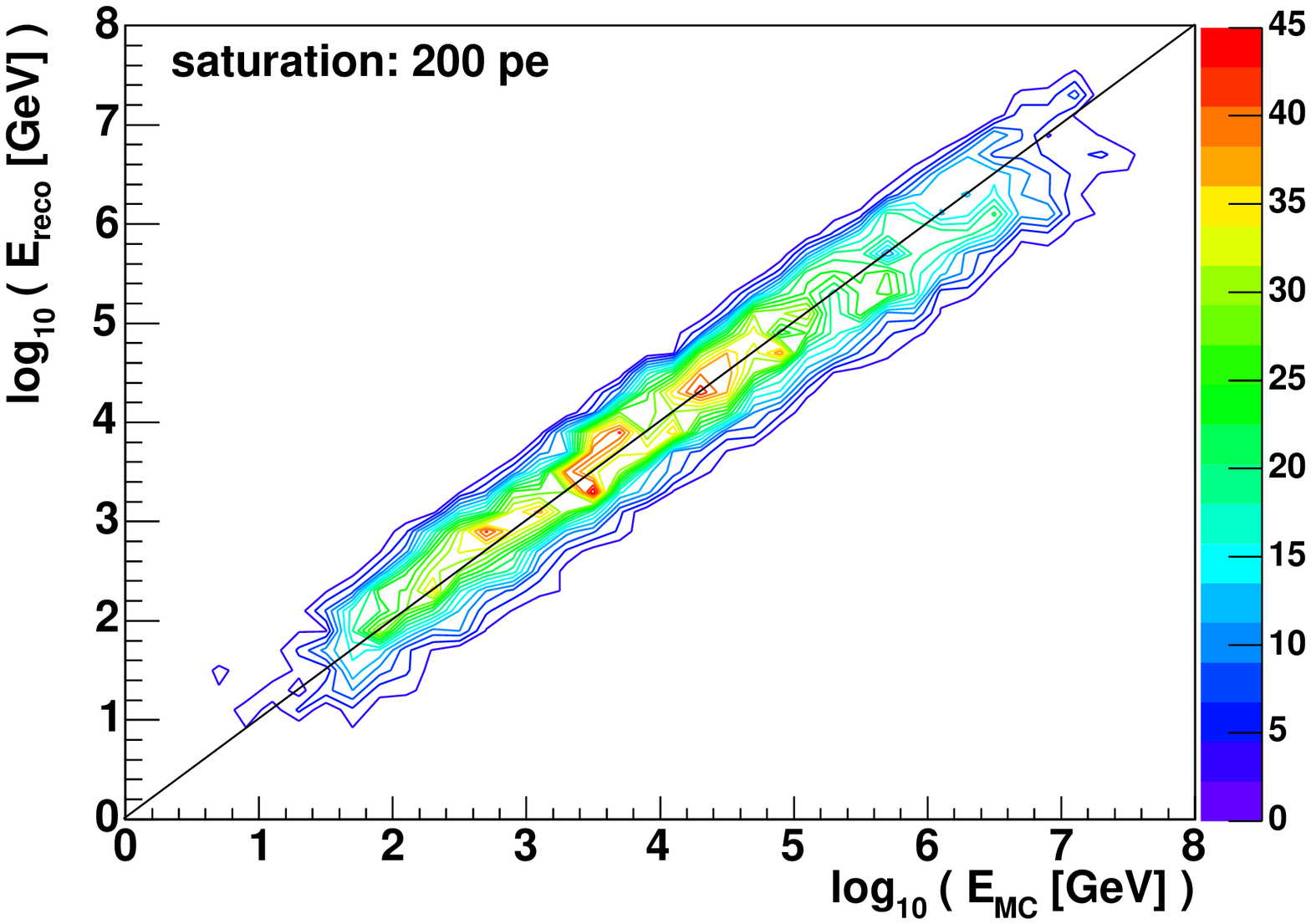}
\caption[Energy reconstruction results, WF mode]
{Results of the energy reconstruction using the corrected total amplitude, for the WF mode
    case, saturation at 200\,pe: Distribution of the logarithmic quotient of reconstructed and MC
    shower energy (left), and reconstructed versus MC shower energy as contour plot (right).} 
\label{fig:e_pre_WF}
\end{figure}

\begin{figure}[h] \centering
\includegraphics[width=7.4cm]{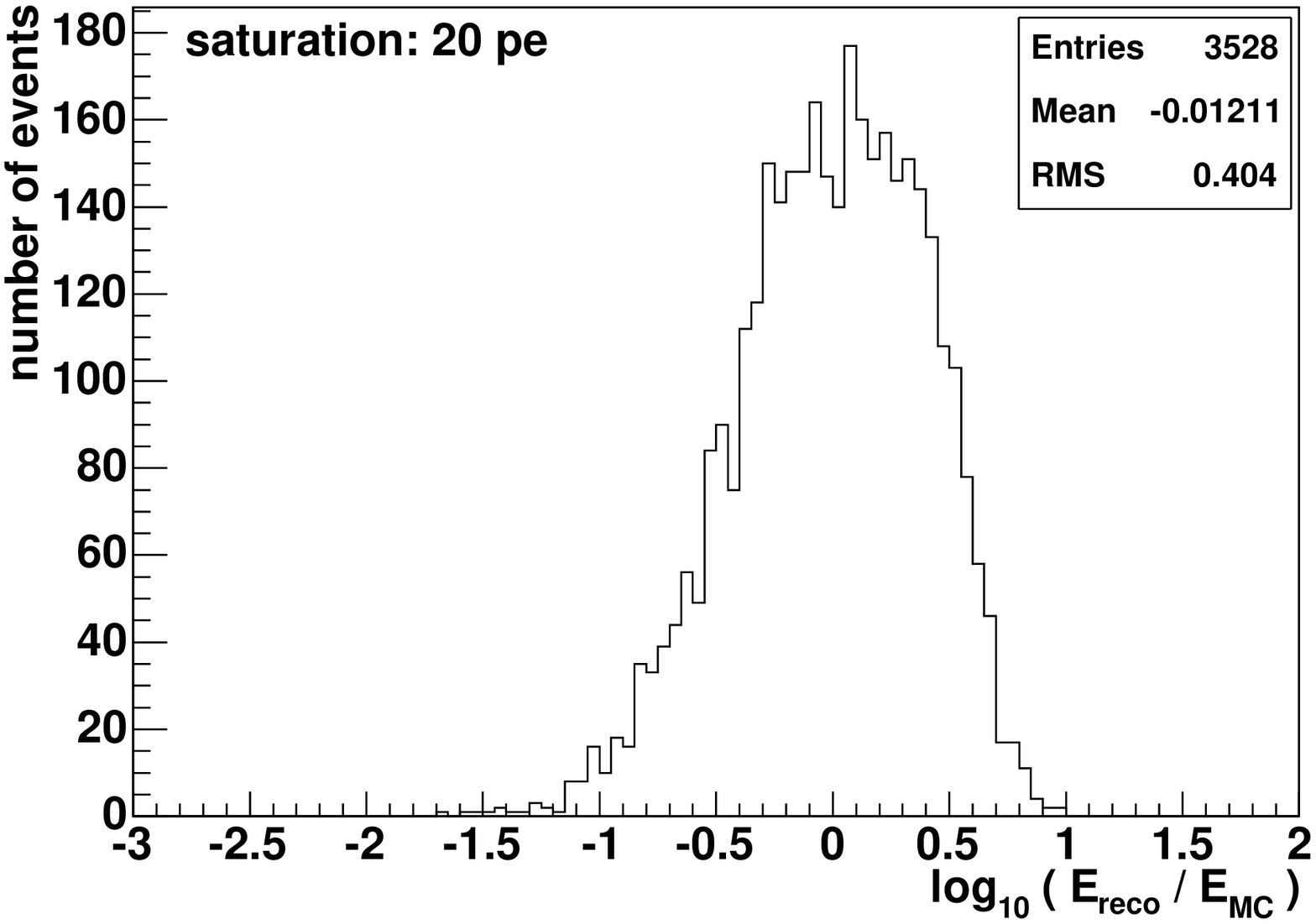}
\includegraphics[width=7.4cm]{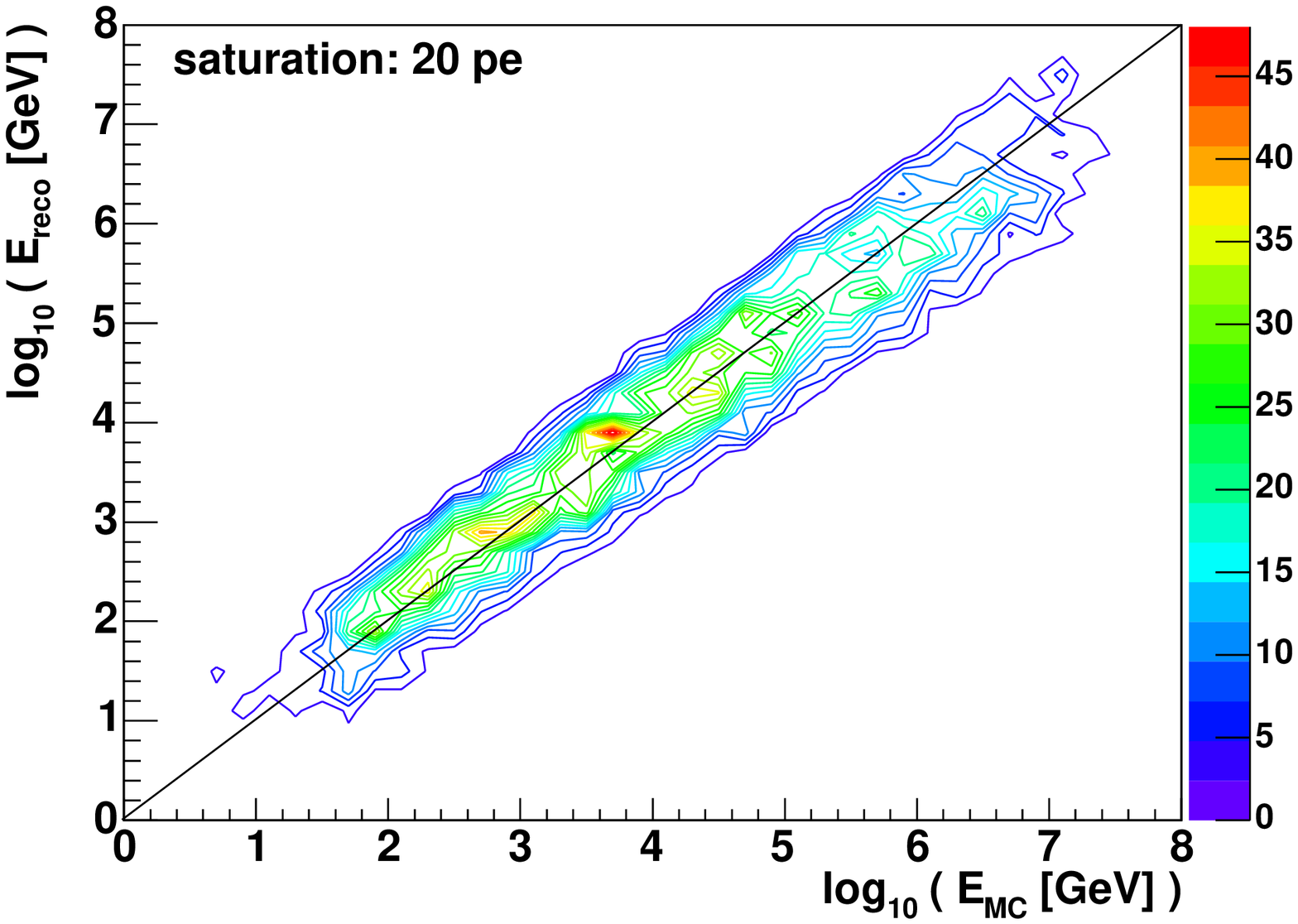}
\caption[Energy reconstruction results, SPE mode]
{Results of the energy reconstruction using the corrected total amplitude, for the SPE mode
    case, saturation at 20\,pe: Distribution of the logarithmic quotient of reconstructed and MC
    shower energy (left), and reconstructed versus MC shower energy as contour plot (right).}
\label{fig:e_pre_SPE}
\end{figure}

As this algorithm makes use of the shower position and direction, it can only be used
if these two values are at least approximately known. \\ 
Results for this energy reconstruction method using the MC values for shower position and
direction are shown in figures~\ref{fig:e_pre},~\ref{fig:e_pre_WF} and~\ref{fig:e_pre_SPE} for no
saturation, saturation at 200\,pe, and saturation at 20\,pe, respectively. All plots show
distributions of the logarithmic difference of the reconstructed and the MC shower energy on the
left hand side, and the logarithm of the reconstructed energy versus that of the MC shower energy on
the right hand side. The ideal result of a complete agreement is marked by a diagonal line. No event or hit
selection of any kind was used in this reconstruction, and the events were background-free. \\
The best results are obtained in case of no saturation. The overall resolution
is $\sim 0.32$, which corresponds to a factor of $10^{0.32} \approx 2.1$ in energy. One can see on
the right hand side of Figure~\ref{fig:e_pre} that the resolution remains approximately constant for
all energies. \\
In the case of saturation at 200\,pe, the WF mode, the resolution deteriorates slightly to $0.36$,
corresponding to a factor of $10^{0.36} \approx 2.3$ in energy, and for saturation at 20\,pe, the
SPE mode, the resolution becomes $0.40$, which corresponds to $10^{0.40} \approx 2.5$. The
deterioration is not surprising, since the relation between the measured amplitude and the shower
energy is smeared by the saturation effects. \\
For this comparably simple method and keeping in mind the simplifications that have been used in
describing the correlation between total amplitude and energy, these results are satisfying. The
resolution is already in the same order of magnitude as the resolution achieved for the
reconstructed muon energy in ANTARES~\cite{muon_energy}.

\chapter[Combined Reconstruction of Shower Parameters]{Combined Reconstruction of Shower Parameters using a Maximum Likelihood Fit}\label{sec:shower_fitter}

The major reconstruction package {\it ShowerFitter}, which was developed in the context of this
thesis, reconstructs simultaneously the direction and energy of a neutrino-induced shower, using a
maximum likelihood fit. This chapter describes the reconstruction algorithm applied, i.e.~the
matching between the measured and the calculated hit amplitudes in each Optical Module (OM). \\
In Section~\ref{sec:pattern_matching} the calculation of the hit amplitude in an OM is discussed, 
while Section~\ref{sec:likelihood} provides further details of the fit, like the determination of
the initial parameter values, and a closer examination of the likelihood parameter space with
examples for different event topologies.

\section{Pattern Matching Algorithm}\label{sec:pattern_matching}

The reconstruction of a hadronic shower in a sparsely instrumented detector like AN\-TAR\-ES is a
challenging task, due to fluctuations in the Cherenkov light emission between different showers, and
the fact that most of the light is lost in the sparsely instrumented detector. It is therefore
important that all relevant information available from the measurements is used. As a
consequence, it is not only important for the reconstruction which OMs have been hit (or not hit),
but also how large the hit amplitudes are. \\ 
The reconstruction strategy presented here is based on pattern matching: For an assumed shower
direction and energy, what would be the expected hit amplitude in each OM? And how does this hit 
pattern compare to the actual measurement? \\
The variables which are to be determined are the shower direction, i.e.~the azimuth angle $\phi$ and
the zenith angle $\theta$, and the shower energy. As the pattern matching deals with hit amplitudes,
and as the total hit amplitude of an event is directly correlated to the shower energy (see
e.g.~Figure~\ref{fig:nphot_fit_E}), the {\it total hit amplitude}, in other words, {\it number of
  (measured) photo-electrons} will be used as third fit parameter. The conversion between the number
of photo-electrons and the shower energy will be discussed at the end of this section.

\subsection{Geometrical Situation, OM Efficiency and Attenuation}\label{sec:geometry}

\begin{figure}[h]
\begin{minipage}{7cm}
\centering \epsfig{figure=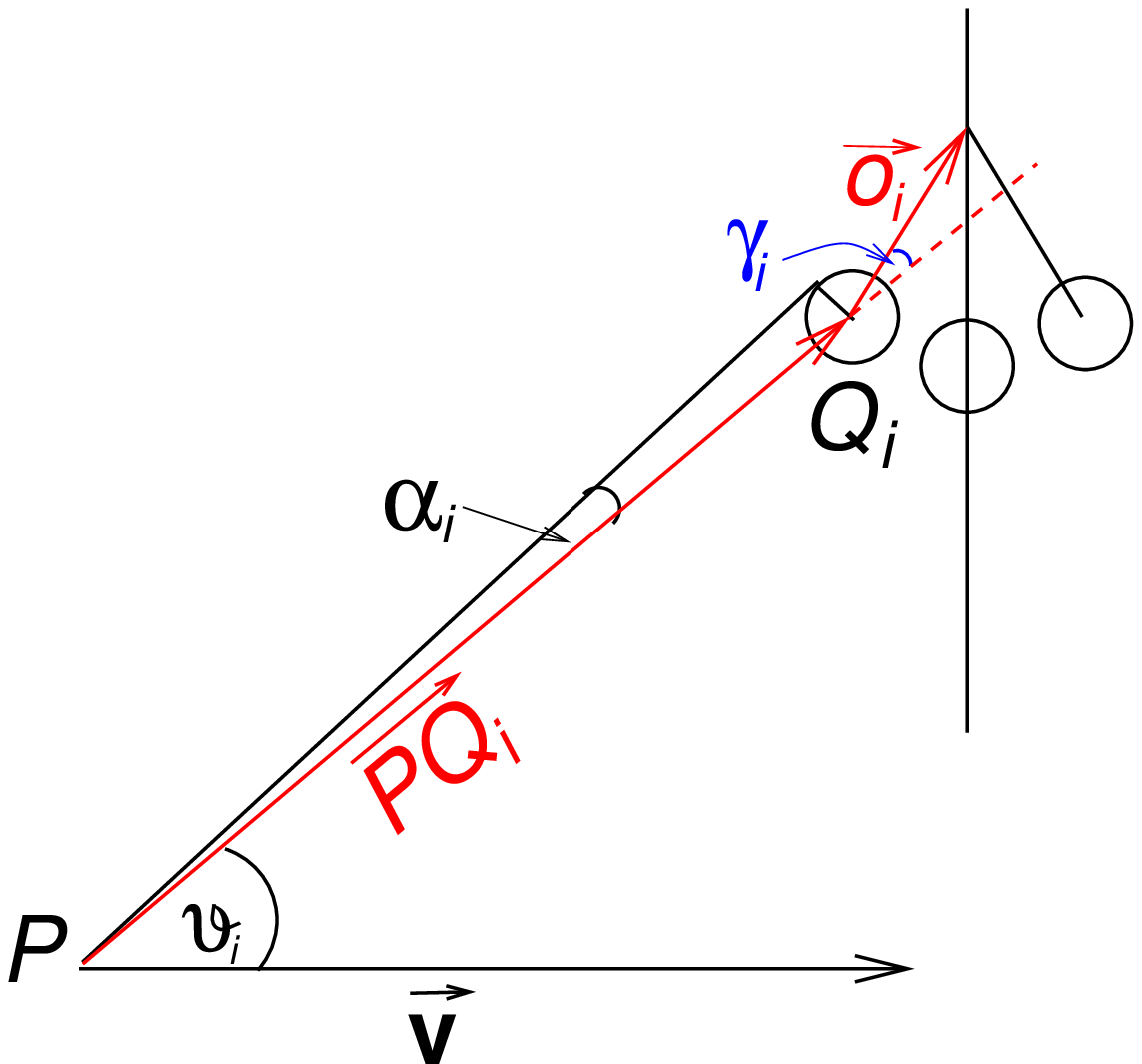, width=6cm}
\caption[Variables used to calculate the total photo-electron number]
{Definition of the variables used to calculate the total photo-electron number from the shower
  direction and the hits in the single OMs.}
\label{fig:fit_scheme}
\end{minipage}
\hspace{2mm}
\begin{minipage}{7cm} 
\centering \epsfig{figure=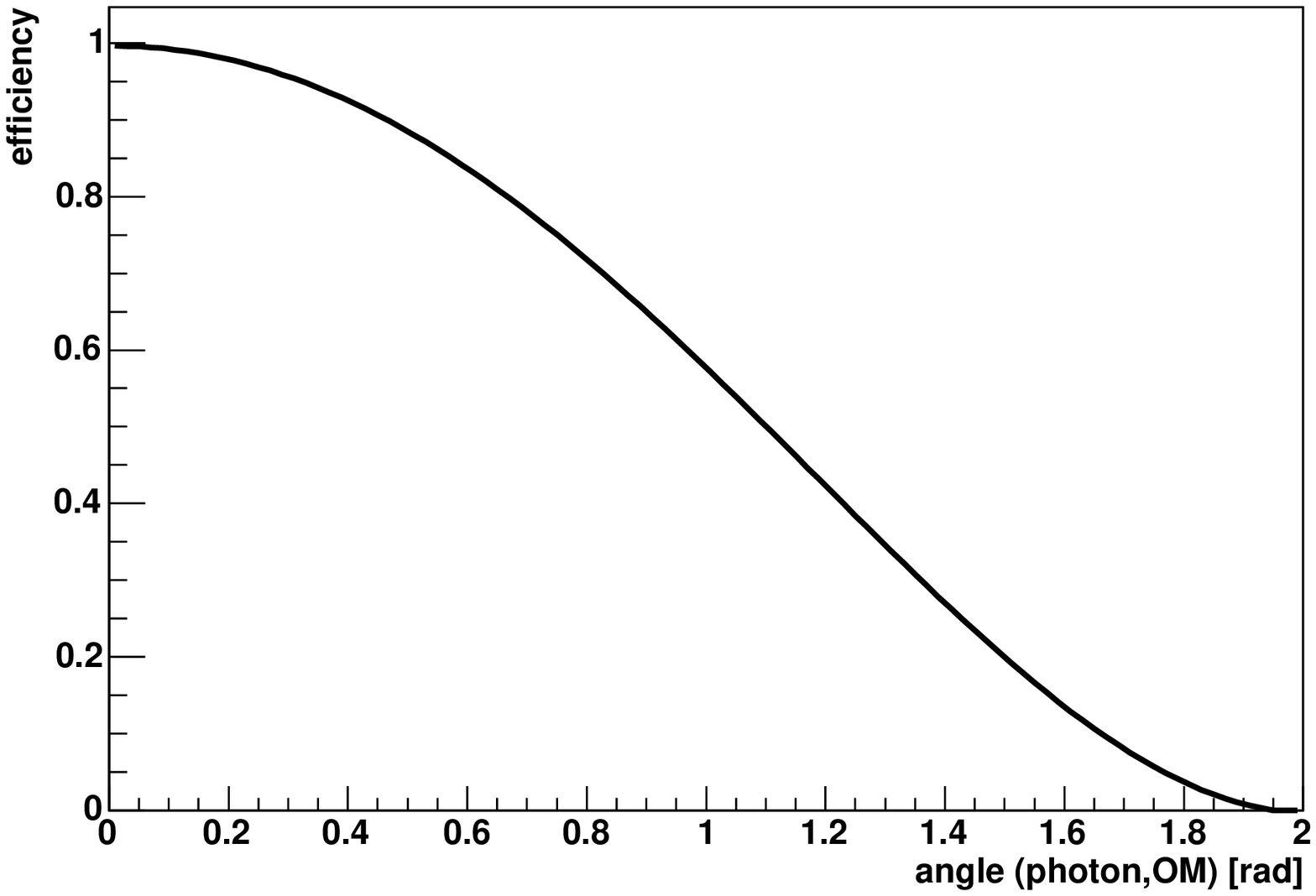, width=7cm}
\caption[Parameterised angular efficiency]{Parameterised angular efficiency $a_i$ of the ANTARES
  OMs, from~\cite{antcc}.}
\label{fig:ang_eff}
\end{minipage}
\end{figure}

To correlate the shower direction with the measured hit amplitudes, some geometrical considerations
are needed. The situation is explained schematically in Figure~\ref{fig:fit_scheme}; the following
notations are used: 

\begin{itemize}
\item $\vec{v}$ = assumed direction of the shower
\item $P$ = point from which a photon generating hit $i$ was emitted from the shower
\item $Q_i$ = position of OM$_i$ which has measured the corresponding hit $i$
\item $\overrightarrow{PQ_i}$ = direction of the photon
\item $\vartheta_i$ = angle between $\vec{v}$ and $\overrightarrow{PQ_i}$
\item $\alpha_i$ = opening angle of OM$_i$, approximated as $\tan \alpha_i = R_{OM} /
  |\overrightarrow{PQ_i}|$, with the OM radius $R_{OM} = 21.7$\,cm
\item $\vec{o}_i$ = orientation of OM$_i$
\item $\gamma_i$ = angle between $\overrightarrow{PQ_i}$ and $\vec{o}_i$. 
\end{itemize}

In order to obtain the total number of photo-electrons $N$ in the event from the hit amplitude $n_i$
in one OM, the relative {\it angular efficiency} $a_i(\gamma_i)$ of the OM has to be taken into
account. A parameterisation from the ANTARES software~\cite{antcc} has been used to describe the
angular efficiency. The parameterised function $a_i$ is shown in Figure~\ref{fig:ang_eff}.   

One also has to take into account the {\it attenuation} $\Lambda_{att}$ of light in water:
\begin{equation*}
\Lambda_{att} = e^{-|\overrightarrow{PQ_i}| / \tau}.
\end{equation*}
Here, $\tau$ is the attenuation length and has been set to 55\,m, corresponding to the maximum
absorption length at a wavelength of 475\,nm (see Section~\ref{sec:absorption}). Scattering effects
are not fully simulated by the light propagation software and are therefore not taken into account
here; however, the effective scattering length at the ANTARES site is measured to be about 300\,m,
such that scattering would only slightly increase the photon attenuation. \\
Note that for these considerations, the Cherenkov light is assumed to be monochromatic.
Because of this, also the quantum efficiency of the OMs is not taken into account as it is equal for
all OMs and hence yields only an additional global normalisation factor. The goal of the
calculations presented here is not to determine the true number of photons emitted from the shower,
but to find a measure for the energy of the shower. 

\subsection{Angular Photon Distribution}\label{sec:ang_distrib}

\begin{figure}[h] \centering
\includegraphics[width=6.4cm]{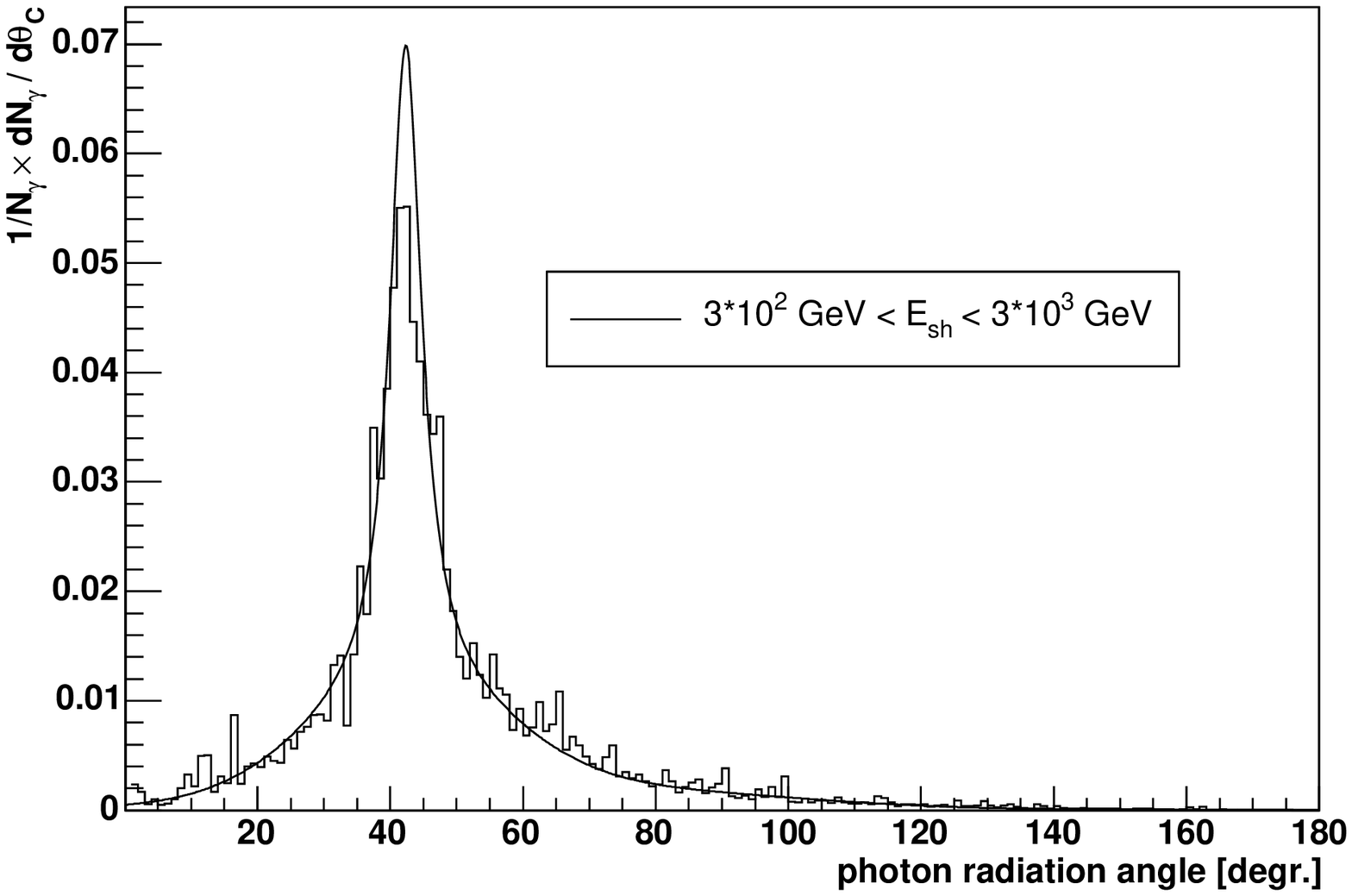}
\includegraphics[width=6.4cm]{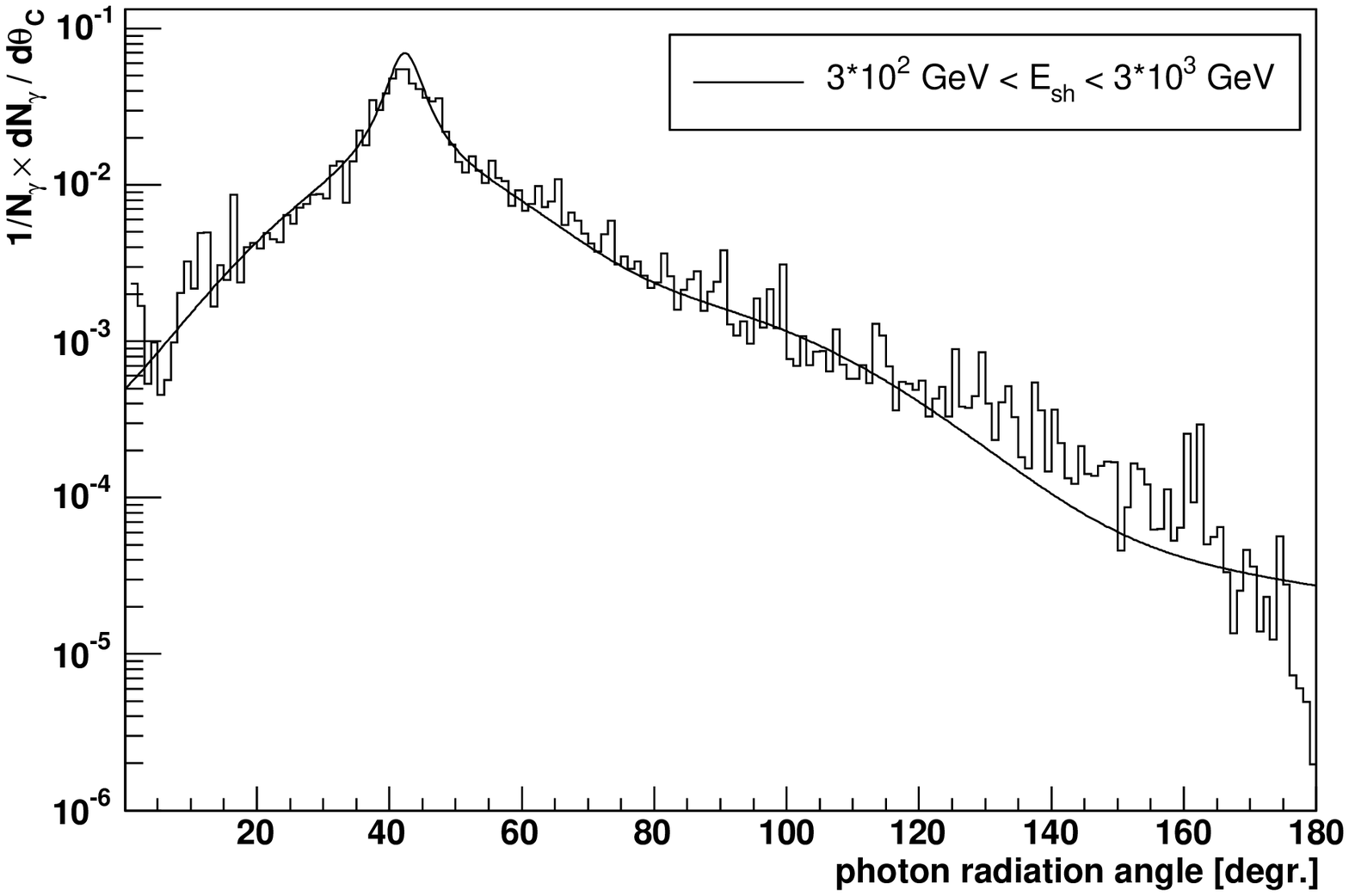}
\includegraphics[width=6.4cm]{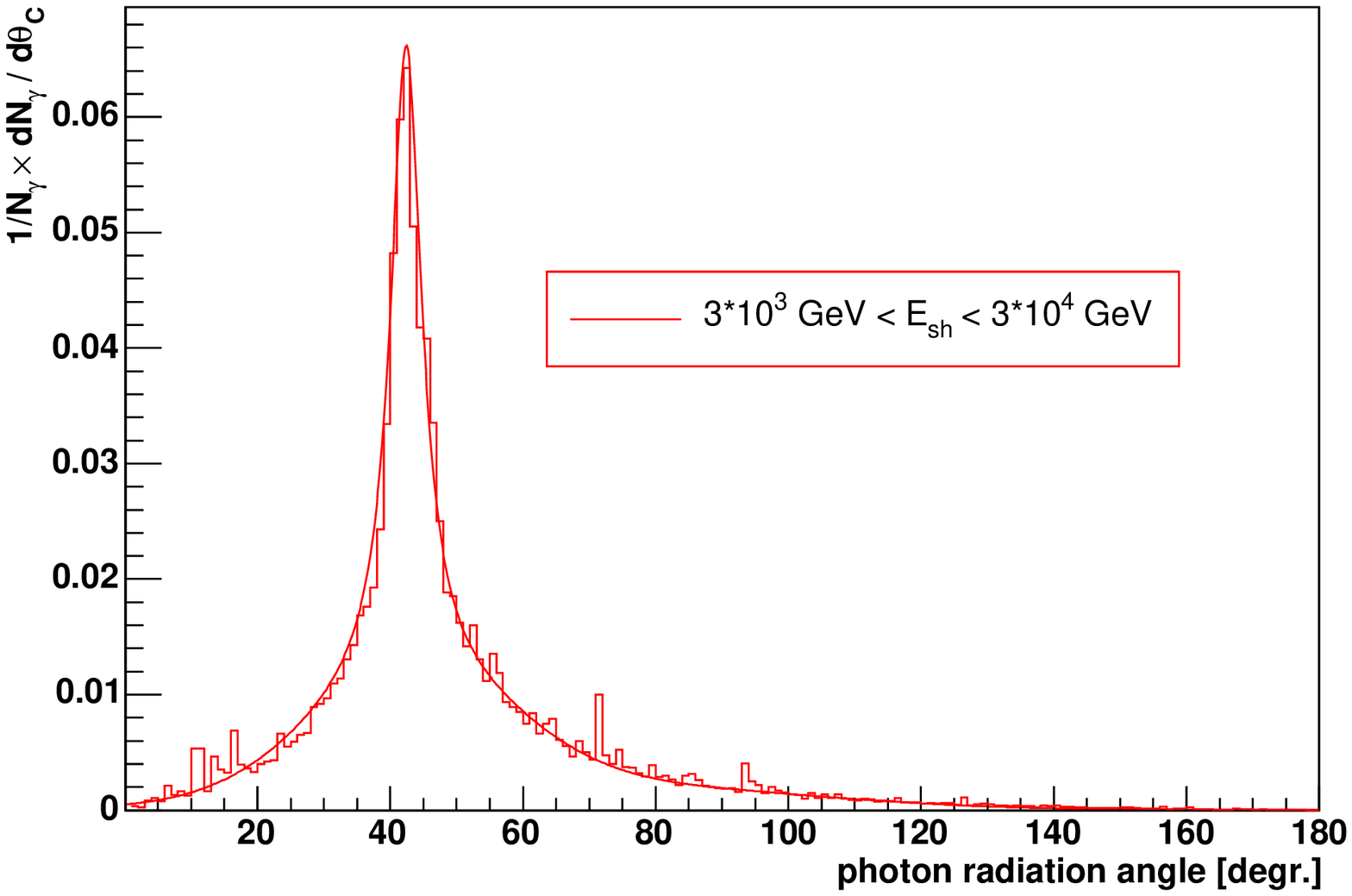}
\includegraphics[width=6.4cm]{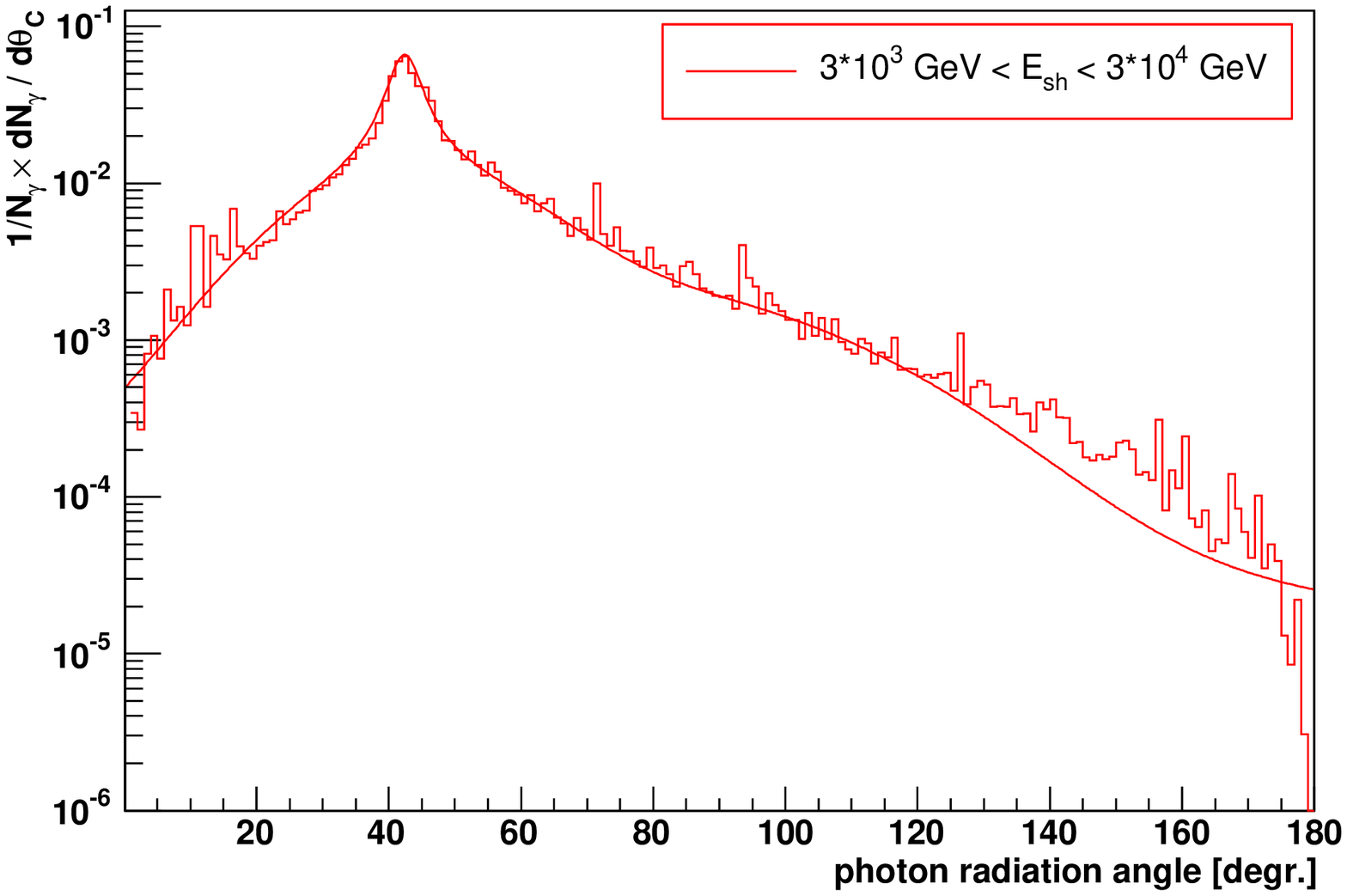}
\includegraphics[width=6.4cm]{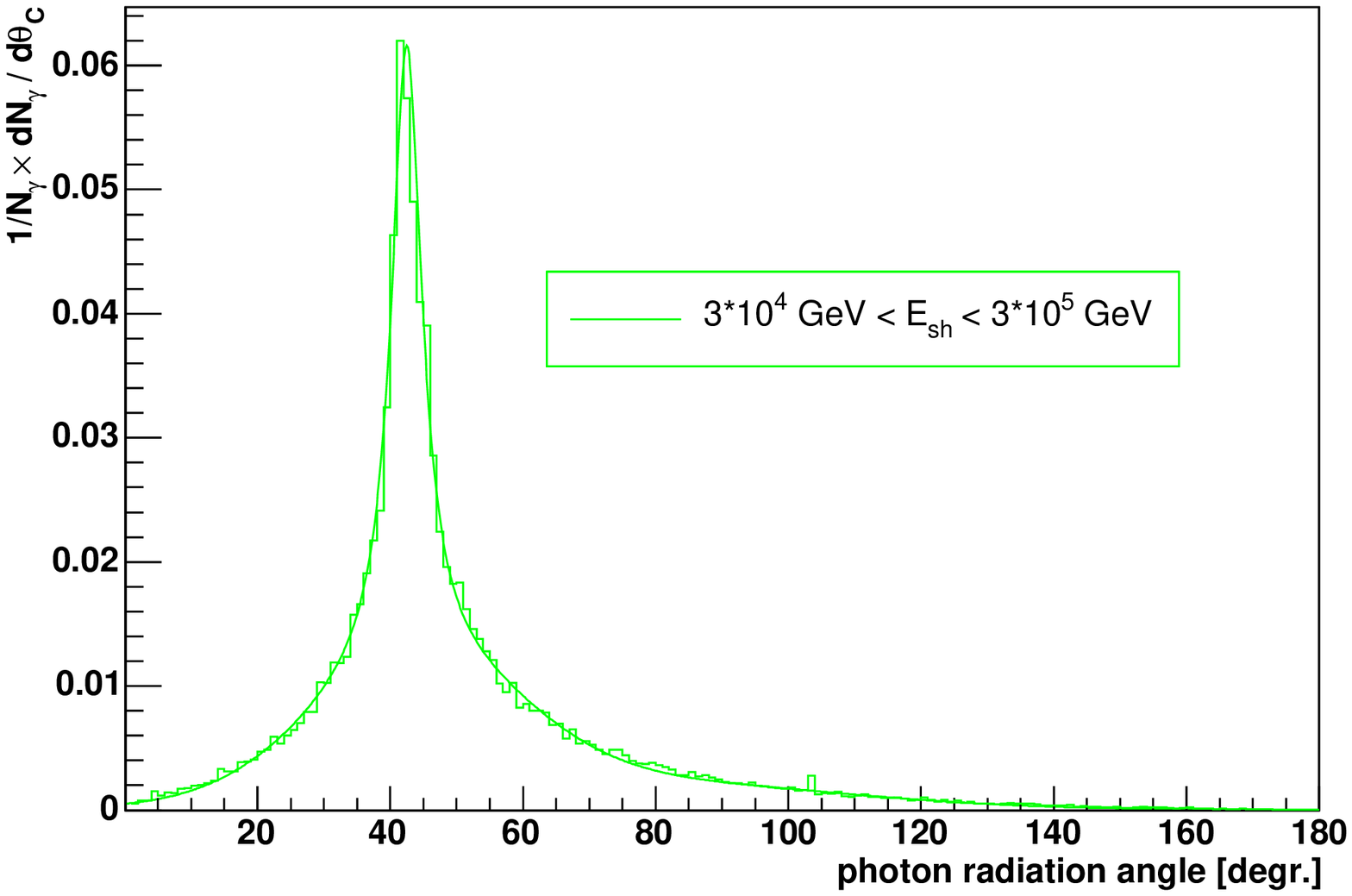}
\includegraphics[width=6.4cm]{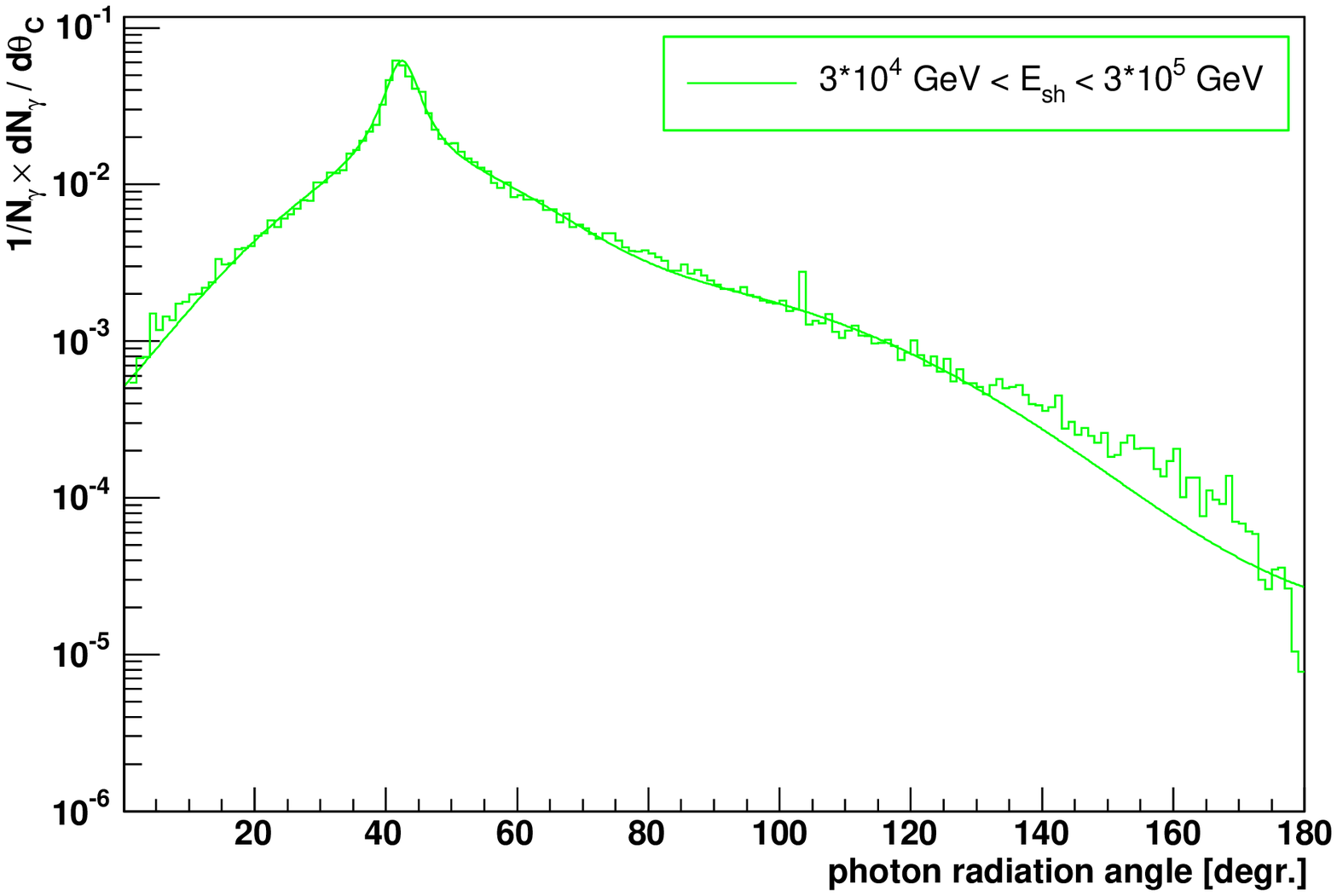}
\includegraphics[width=6.4cm]{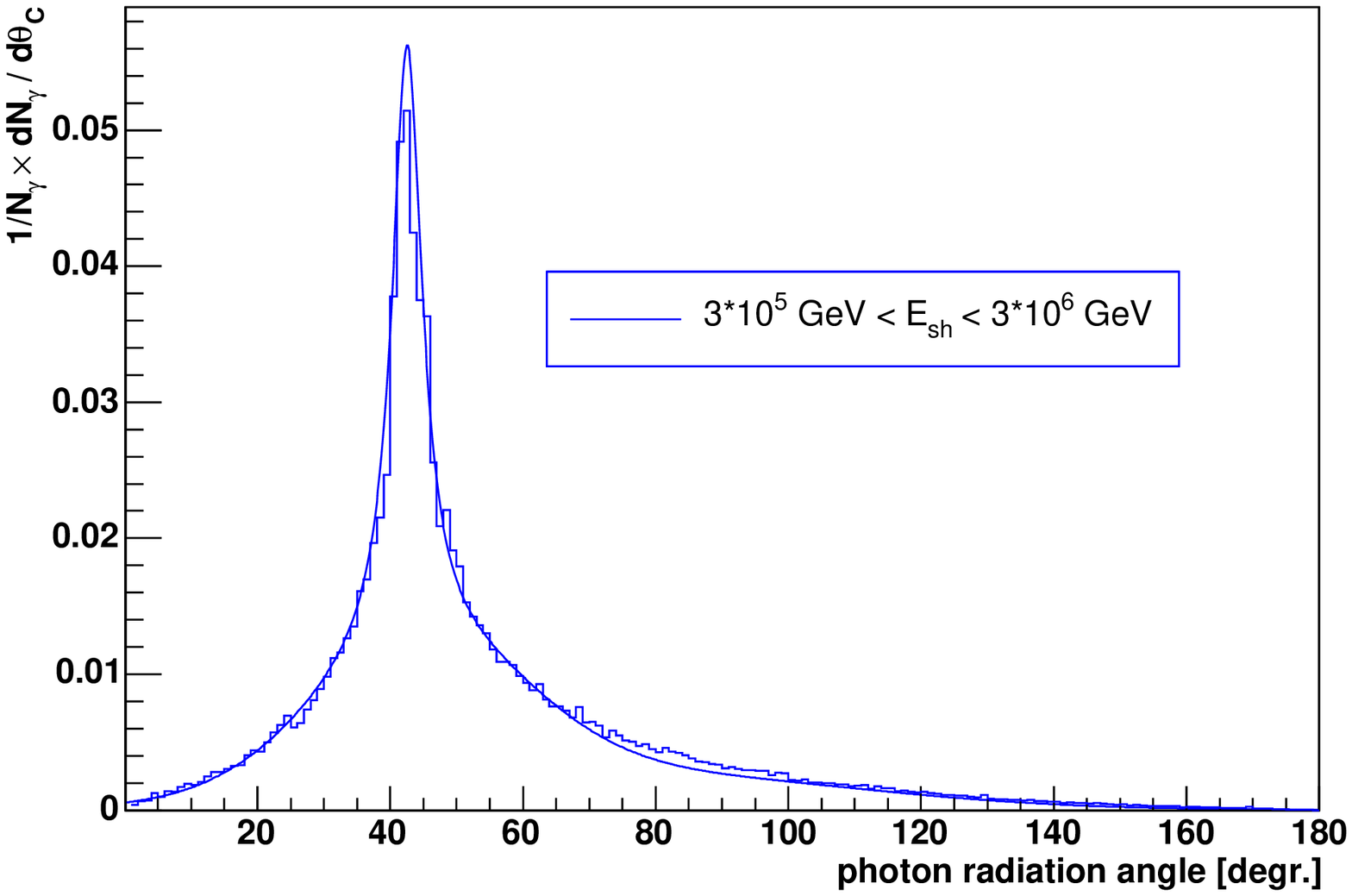}
\includegraphics[width=6.4cm]{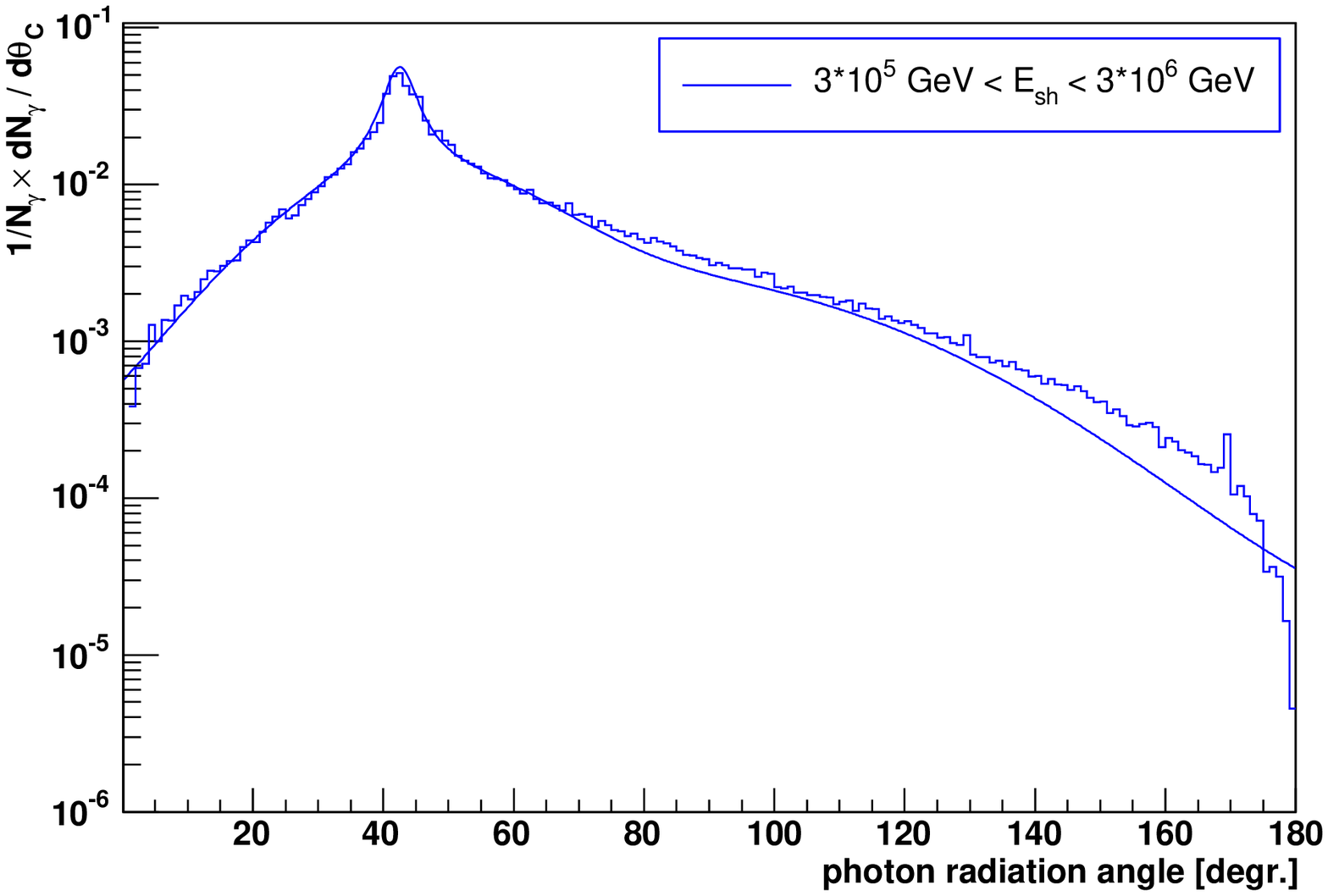}
\includegraphics[width=6.4cm]{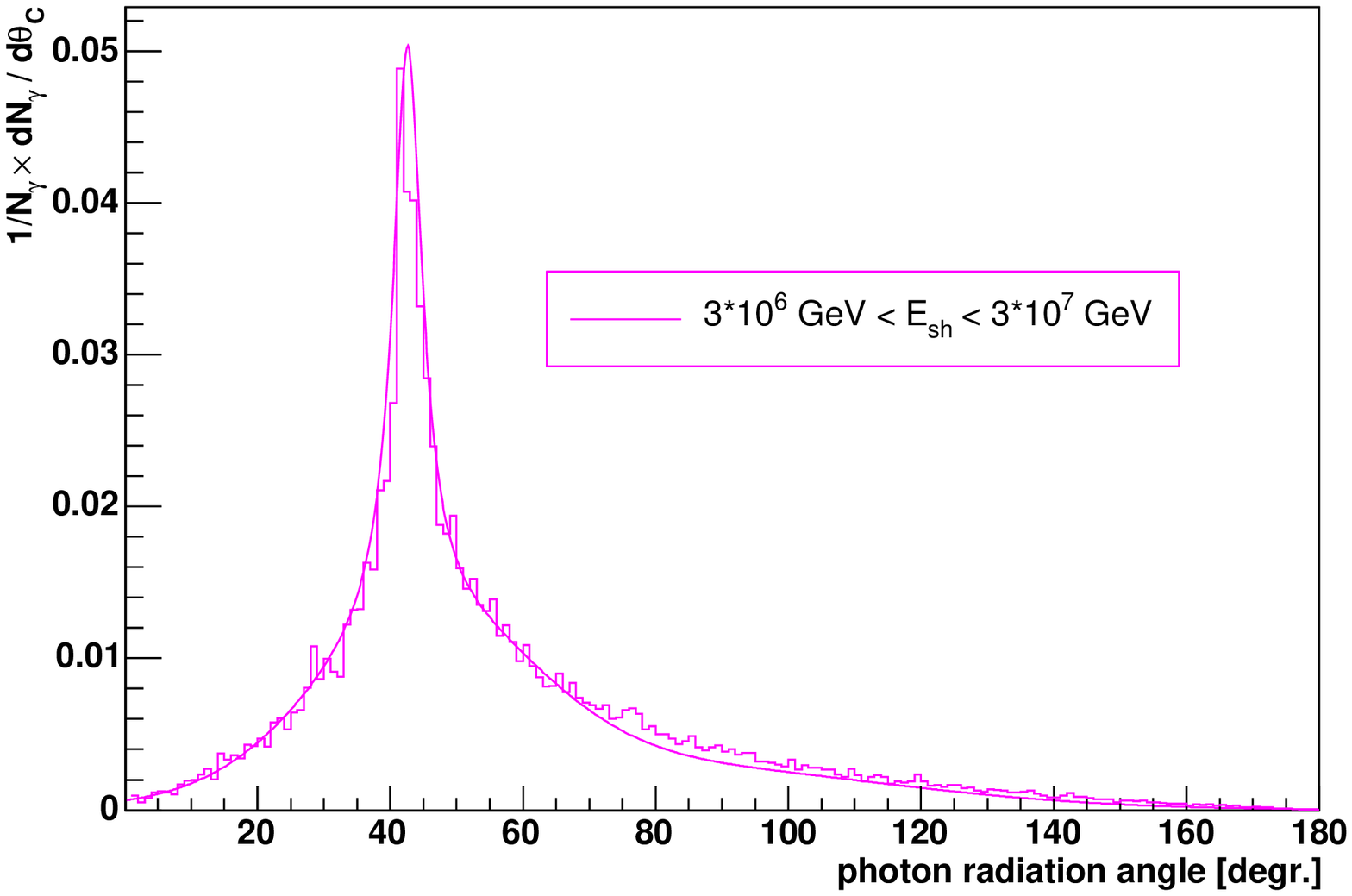}
\includegraphics[width=6.4cm]{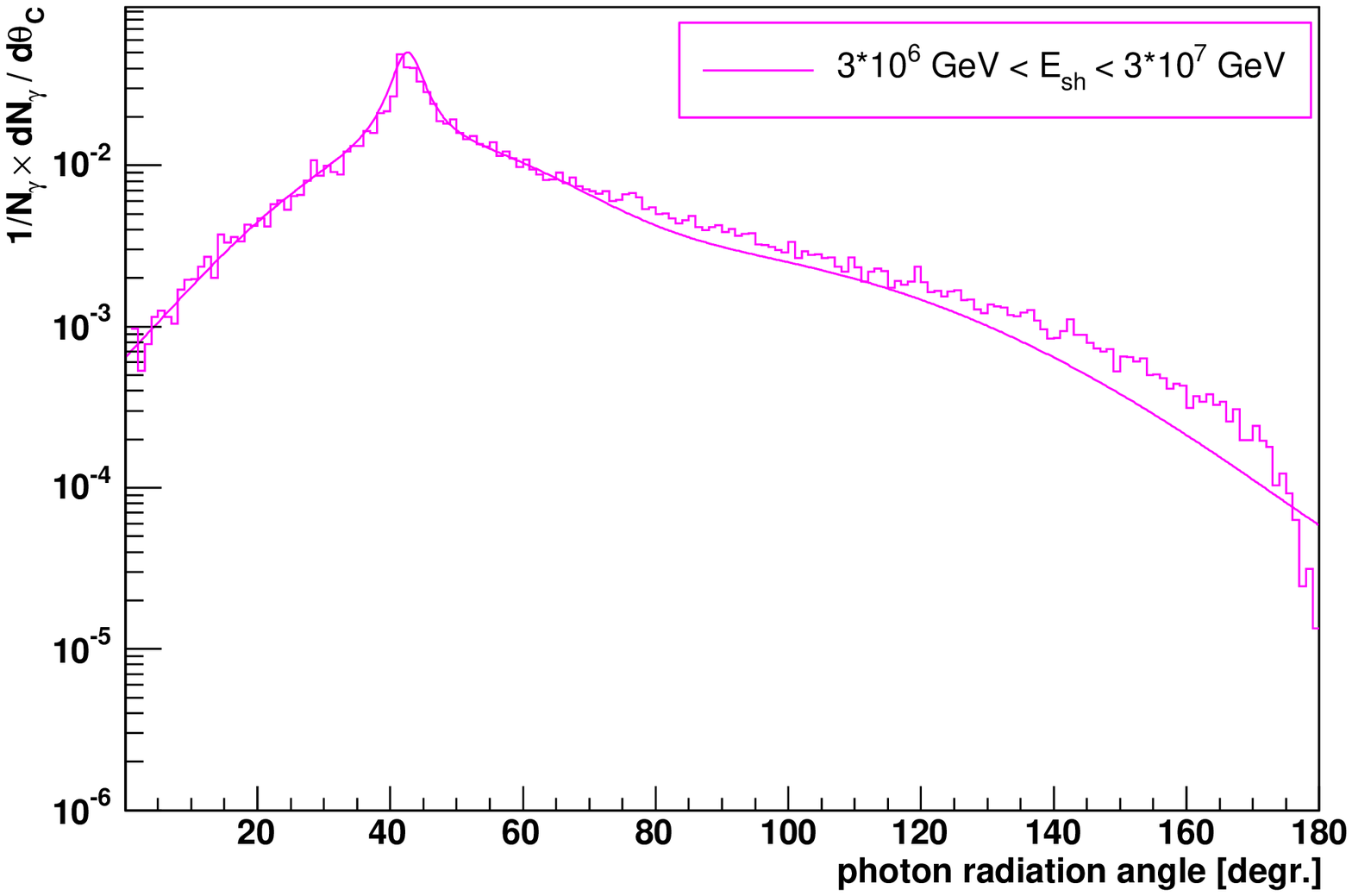}
\caption[$\vartheta$ distribution of the Cherenkov photons, and parameterisation]
{Angular distribution of the Cherenkov photons with respect to the shower axis for decades of energy
between 300\,GeV and 30\,PeV, together with the parameterisation function for an energy in the
logarithmic centre of the
respective energy range (solid lines).}
\label{fig:cheren_fit}
\end{figure}

It is assumed that the photons emitted by the shower are distributed uniformly in $\varphi$ and
according to a distribution in $\vartheta$ obtained from Monte Carlo studies which has been
shown in Figure~\ref{fig:cherenkov_all}. The distribution is shown again, together with the
parameterisation presented in the following, in Figure~\ref{fig:cheren_fit}. The distribution is
characterised by a large, narrow peak at about 42$^{\circ}$ and is slightly asymmetric, with a
larger tail towards larger angles. It has therefore been parameterised by a
Breit-Wigner-distribution with its maximum at the peak position and two Gaussians, one with its mean
value approximately at the peak position and one with its mean at about 79$^{\circ}$ to account for
the tail; the parameterisation function $D(\vartheta,E_{sh})$ is given by:

\begin{equation} \label{eq:d_theta}
  D(\vartheta,E_{sh}) =  p \cdot \left( \frac{p_0}{\sqrt{2\pi} p_1} \cdot e^{ -(\vartheta - p_2)^2 / (2p_1^2)} 
                       + \frac{p_3}{\sqrt{2\pi} p_4} \cdot e^{ -(\vartheta - p_5)^2 / (2p_4^2) } 
                       + \frac{p_6}{2\pi} \frac{p_7}{(\vartheta - p_8)^2 + \frac{p_7^2}{4}} \right)
\end{equation}

\begin{figure}[h] \centering
\includegraphics[width=4.9cm,height=4cm]{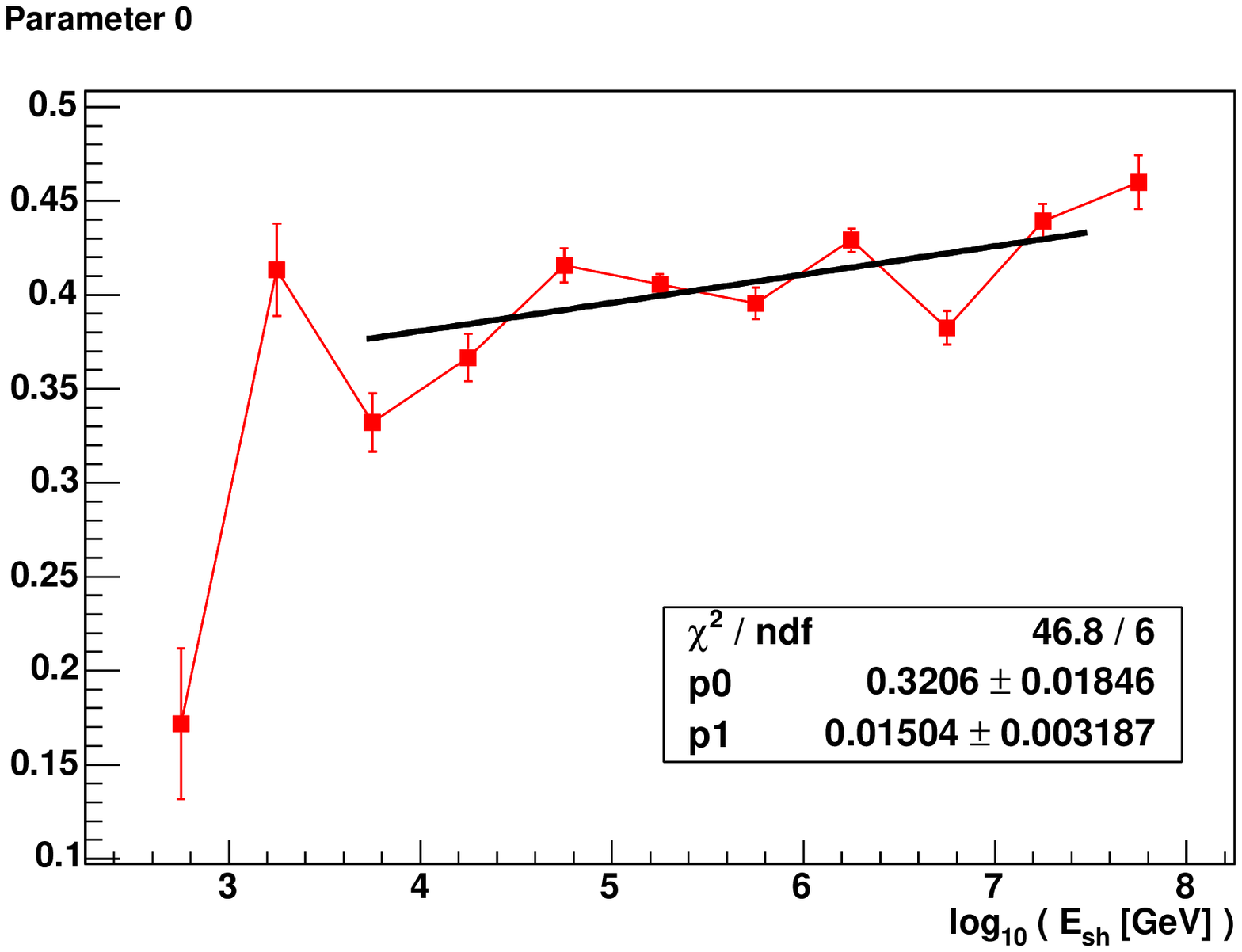}
\includegraphics[width=4.9cm,height=4cm]{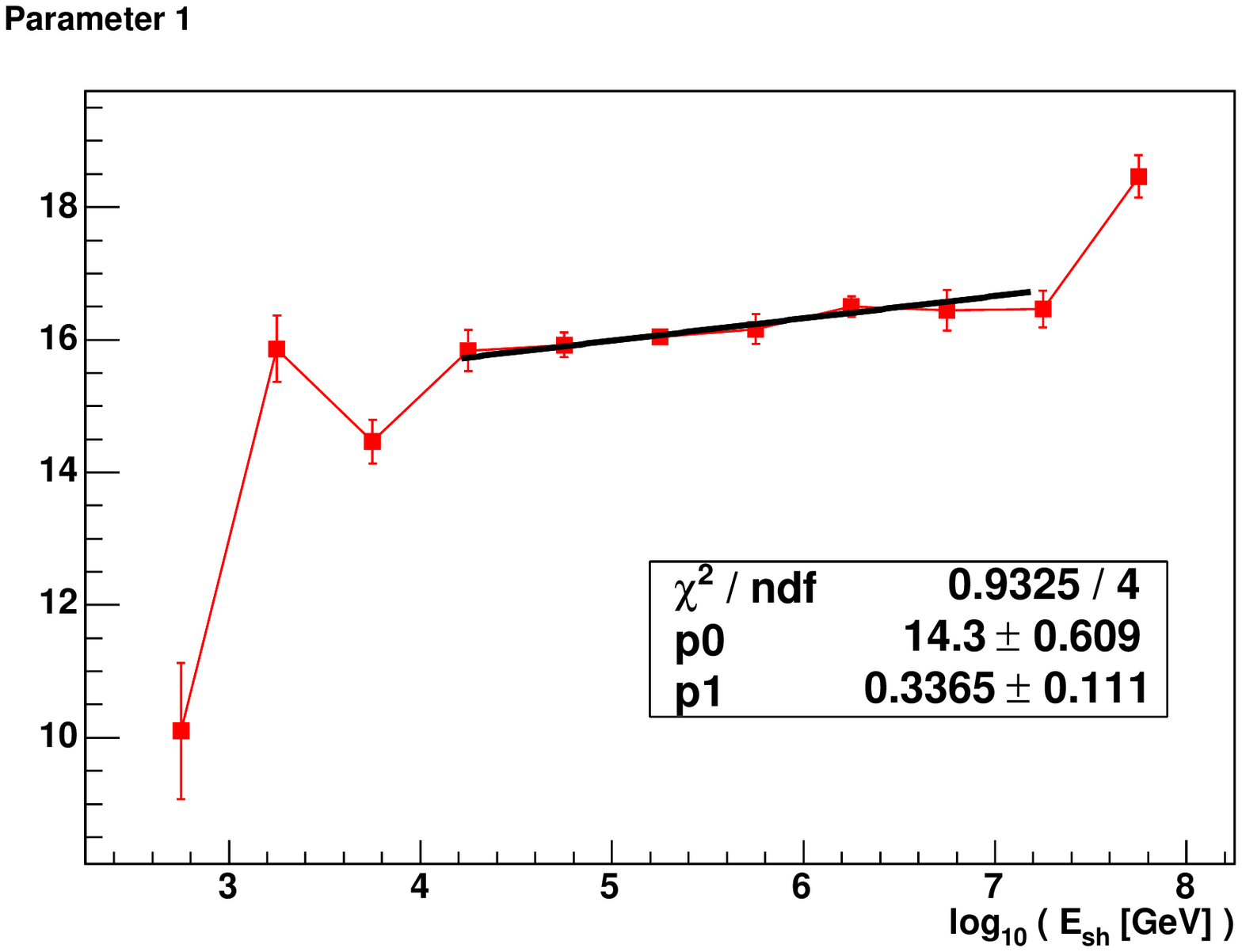}
\includegraphics[width=4.9cm,height=4cm]{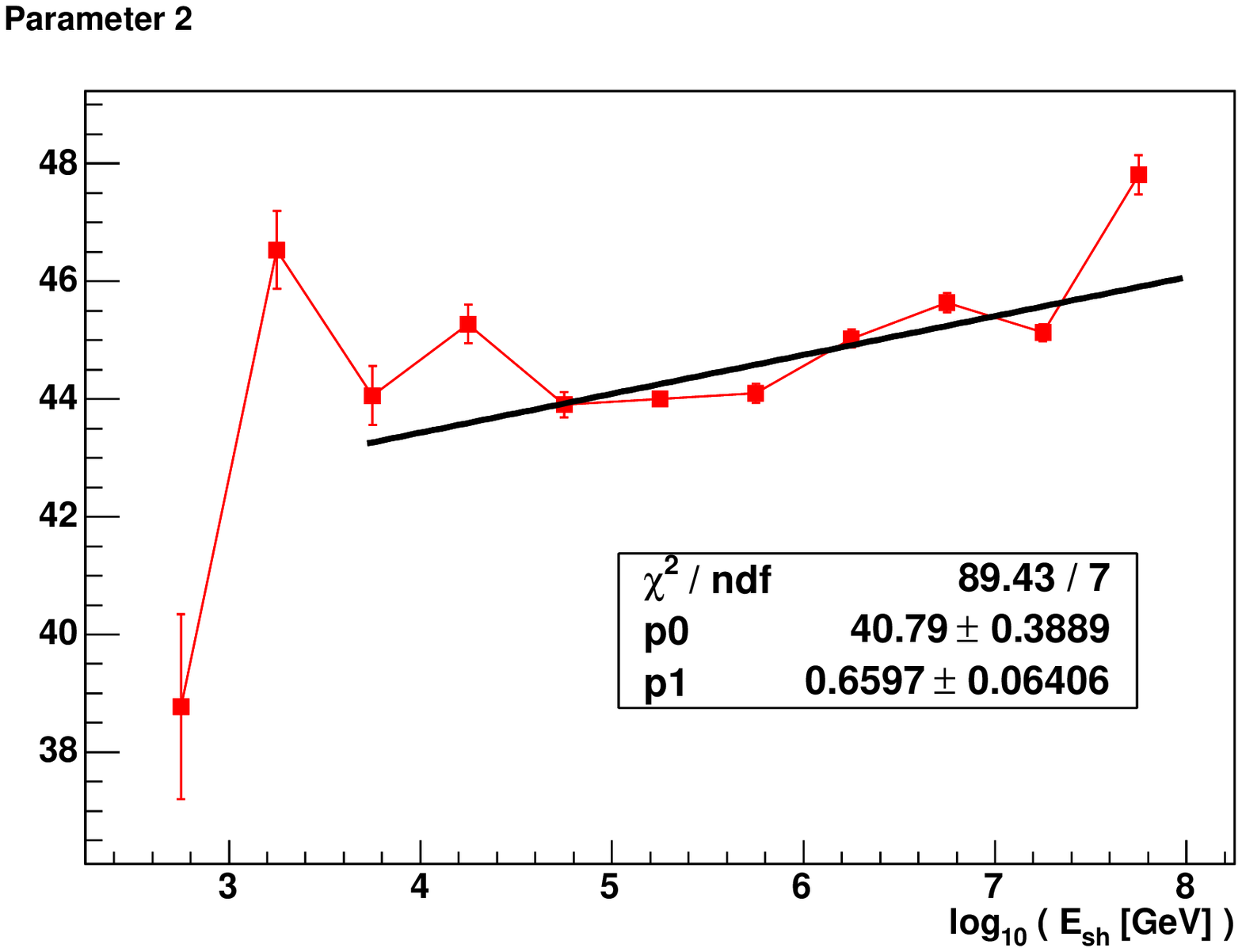}
\includegraphics[width=4.9cm,height=4cm]{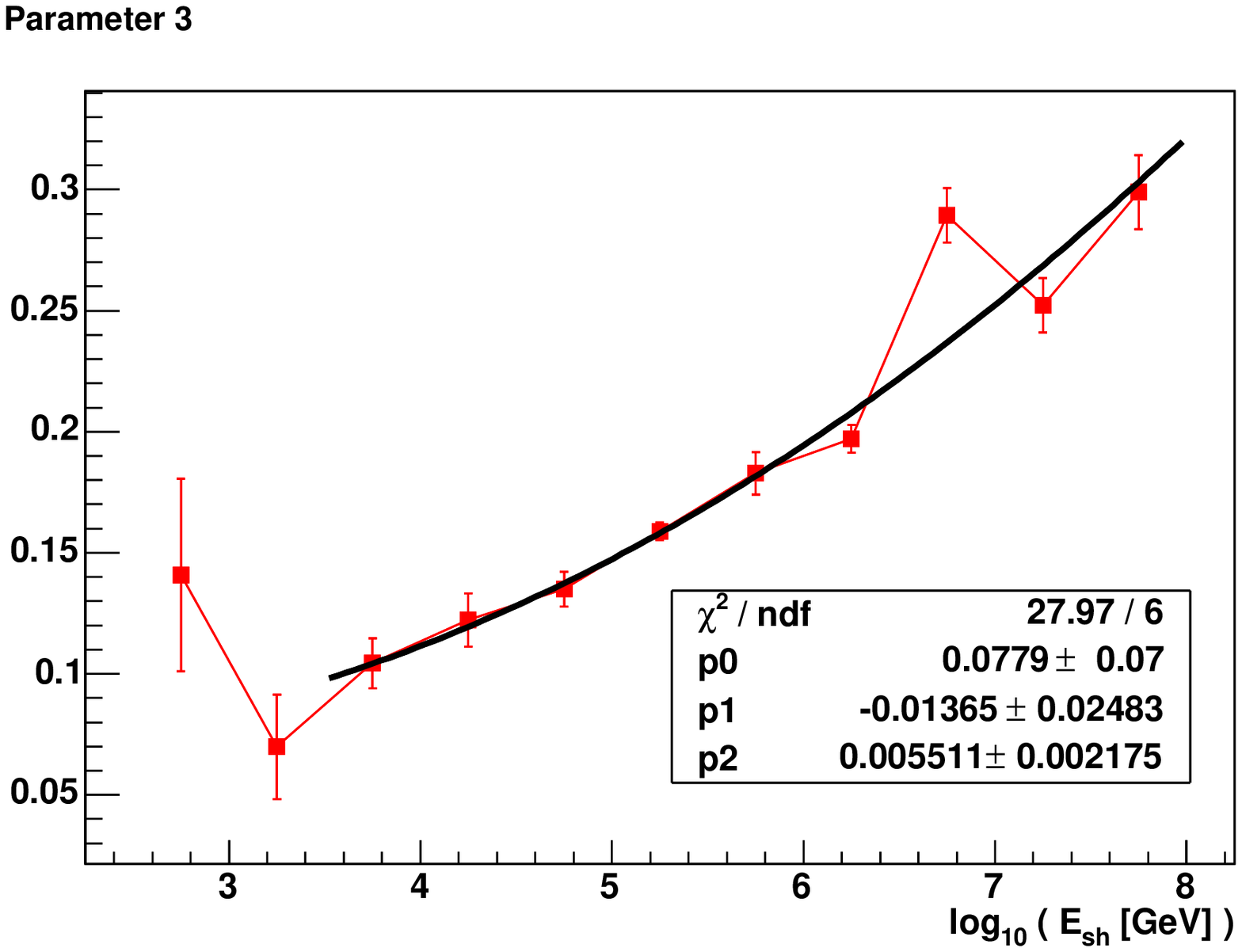}
\includegraphics[width=4.9cm,height=4cm]{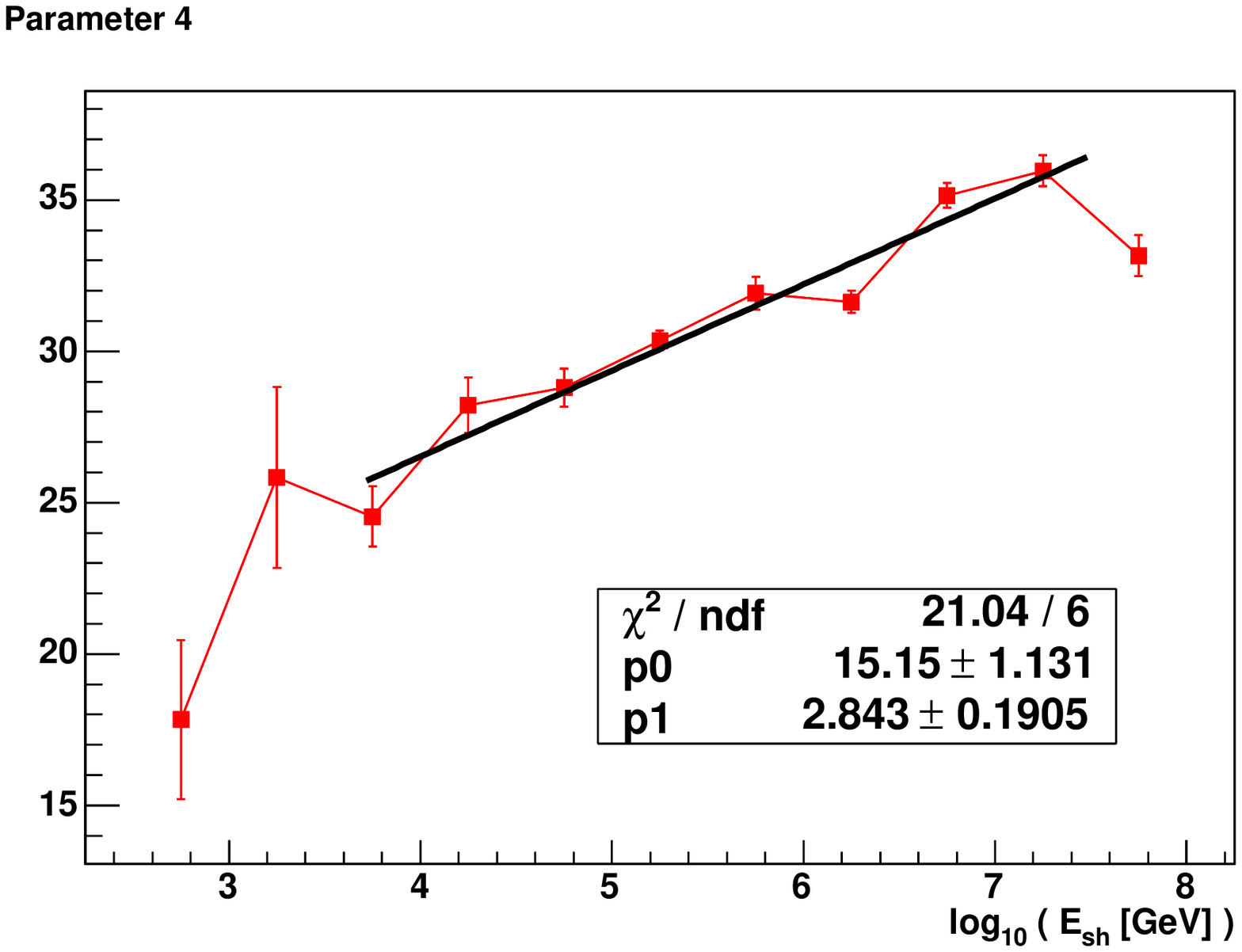}
\includegraphics[width=4.9cm,height=4cm]{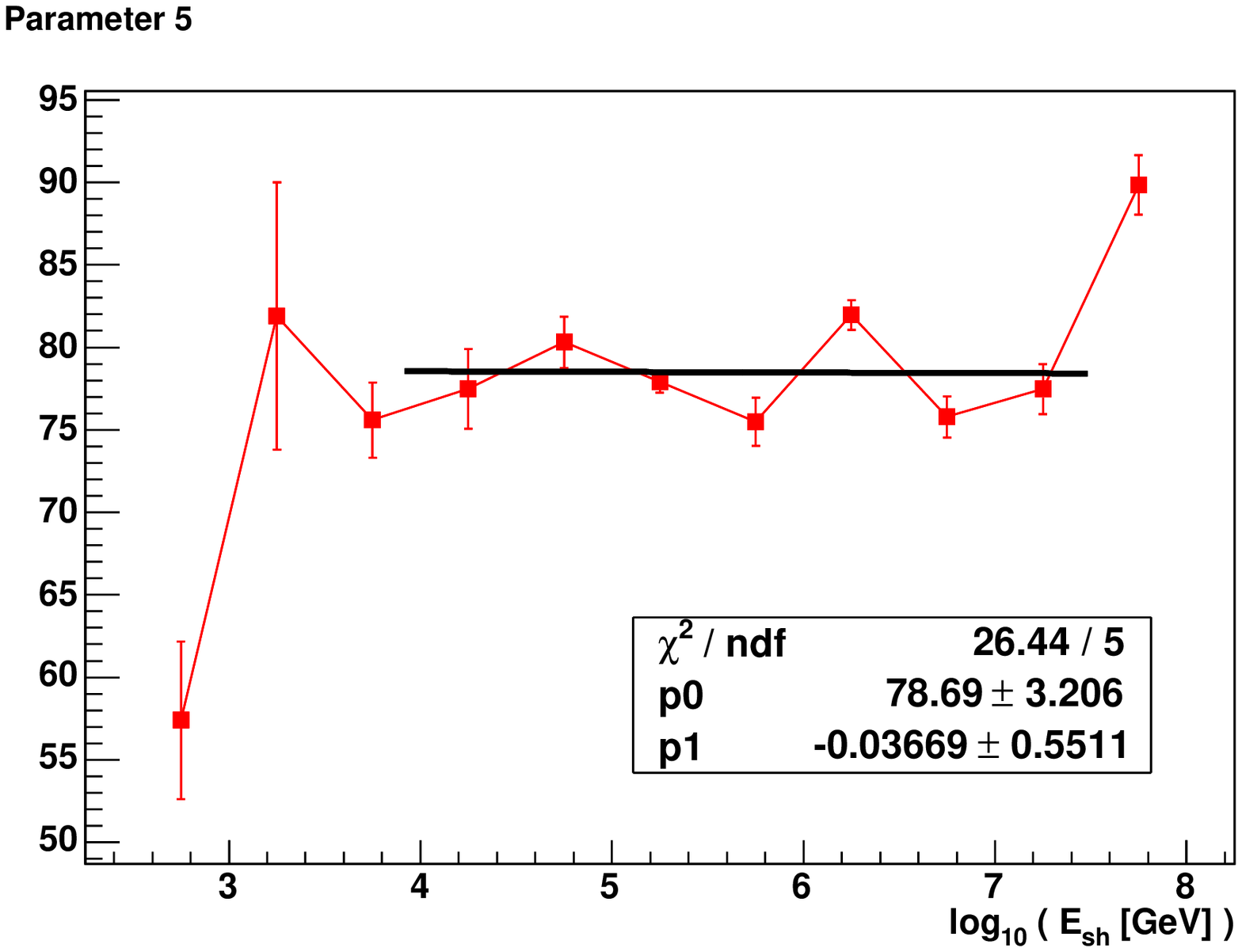}
\includegraphics[width=4.9cm,height=4cm]{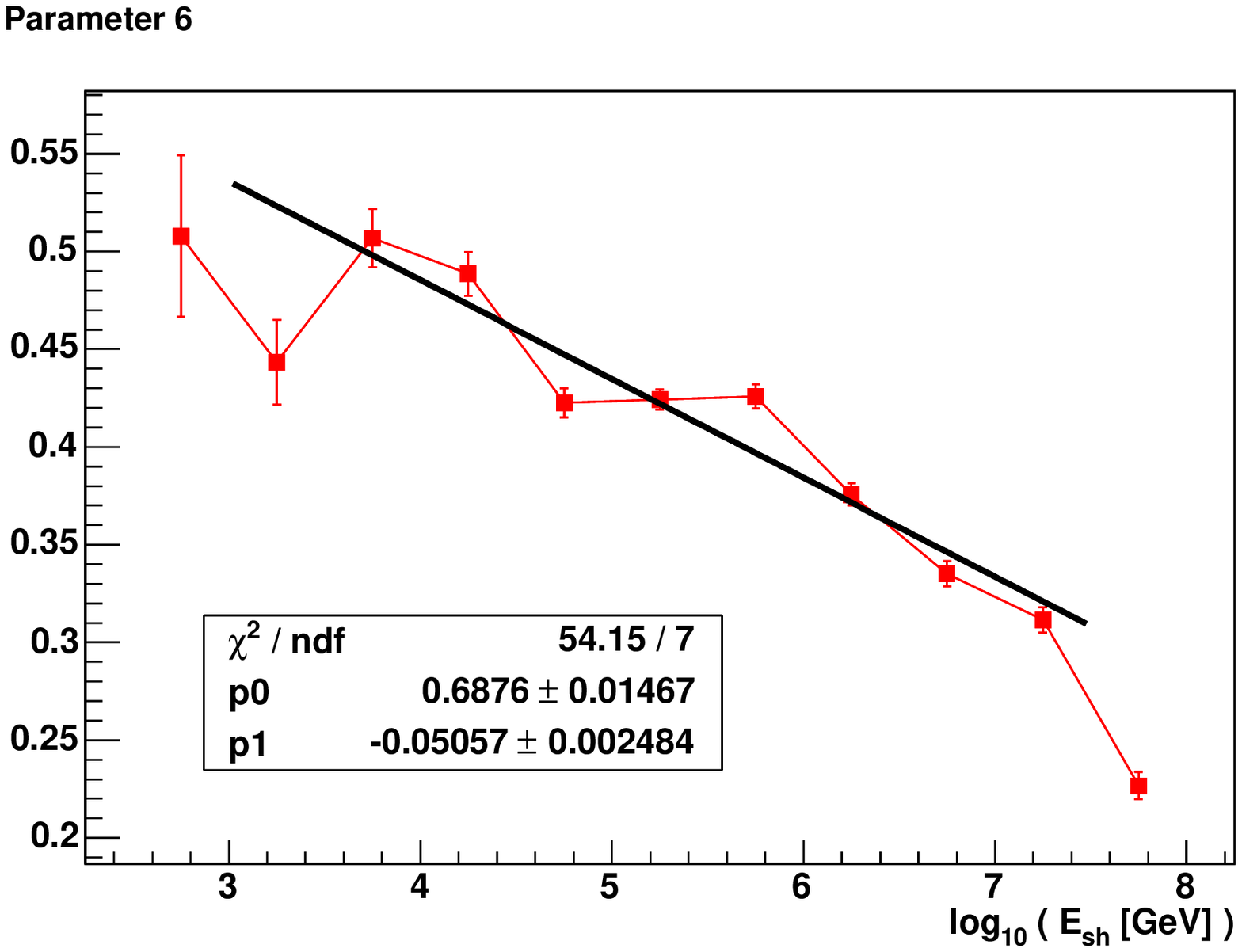}
\includegraphics[width=4.9cm,height=4cm]{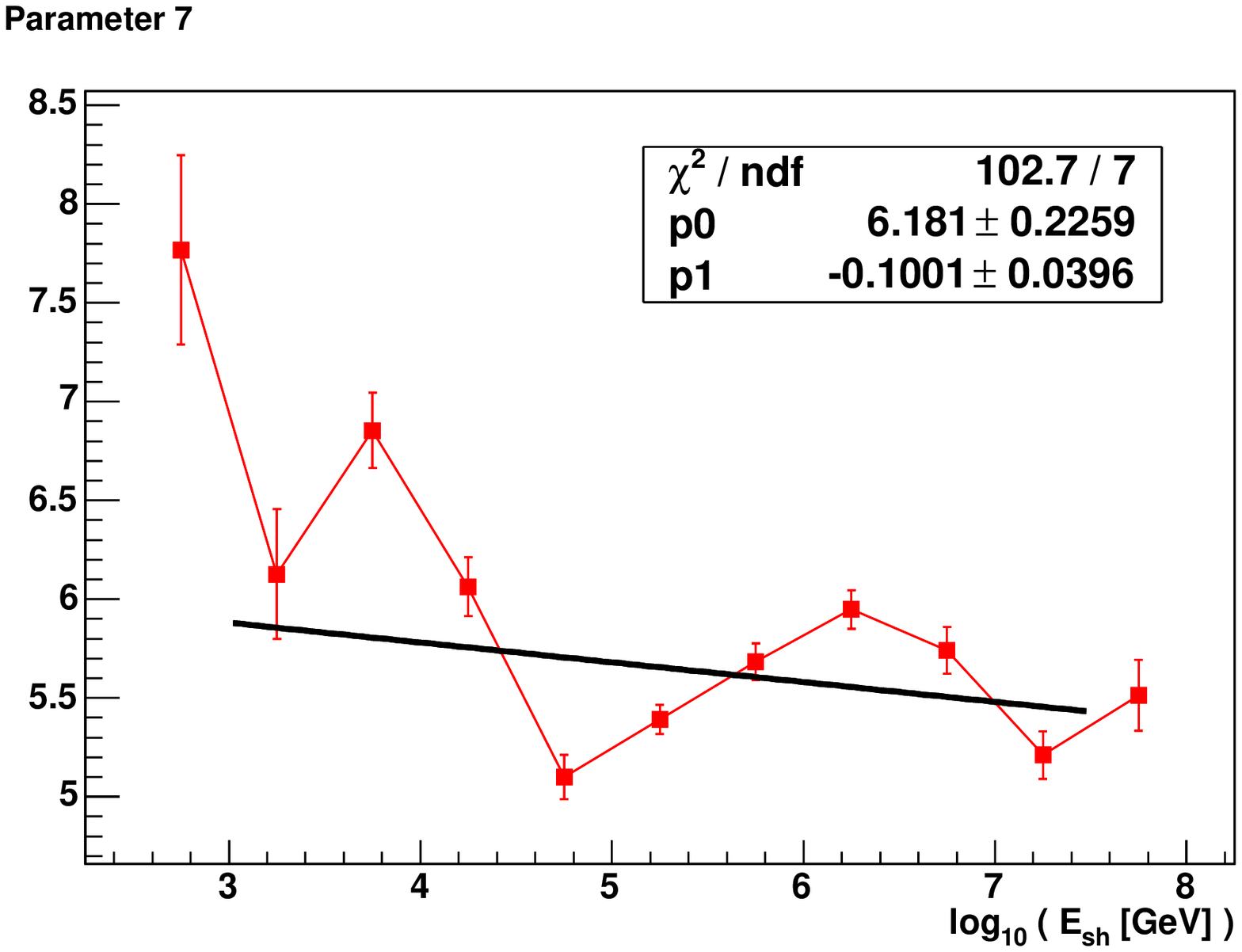}
\includegraphics[width=4.9cm,height=4cm]{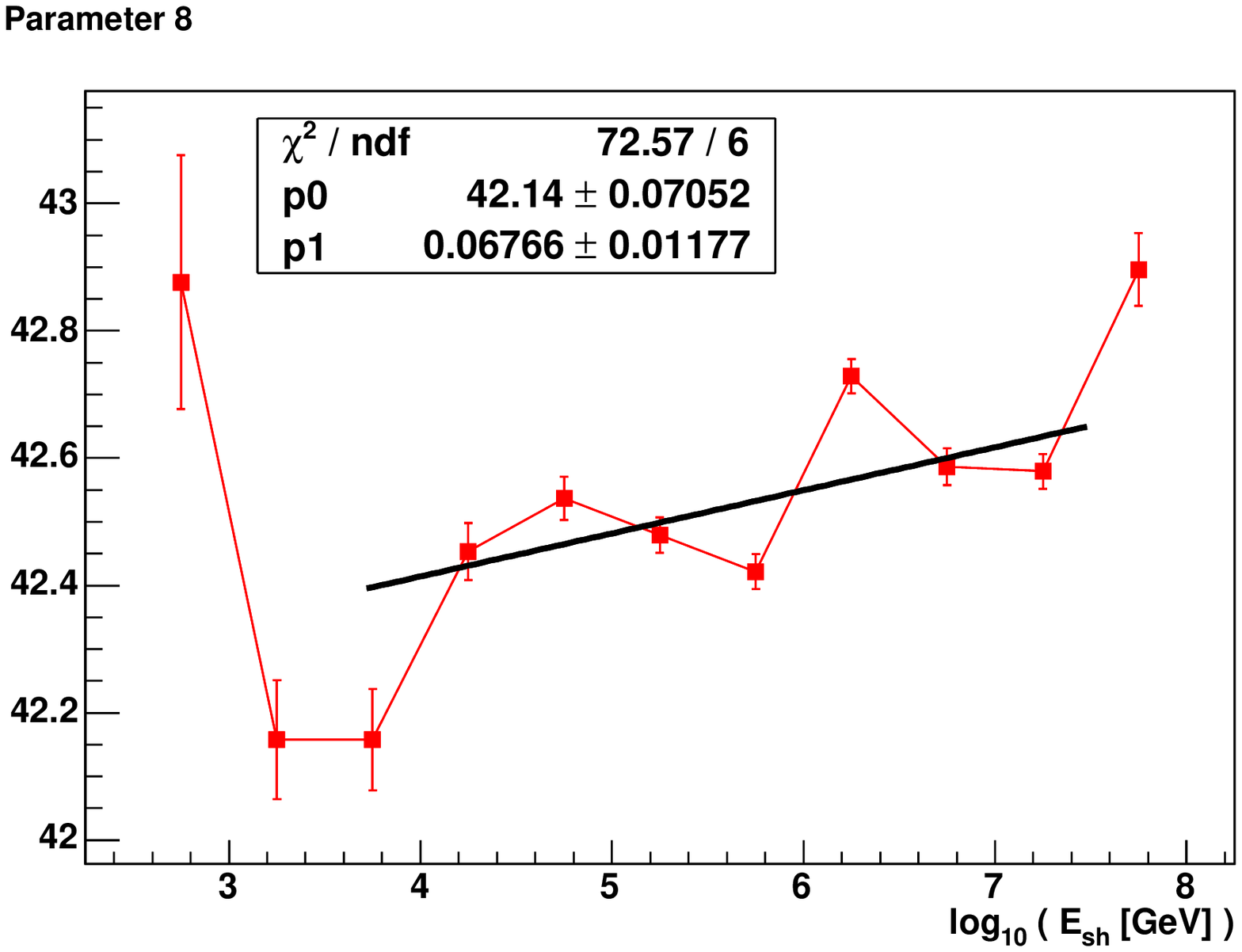}
\caption[$D(\vartheta,E_{sh})$ parameterisation]{Energy-dependent values of the nine parameters
  in the function describing the $\vartheta$ distribution, $D(\vartheta,E_{sh})$, and fit curves to
  the values.}
\label{fig:params}
\end{figure}

Here and in the following, $\vartheta$ is measured in degrees. The distribution implicitly depends
on the shower energy $E_{sh}$, through the parameters $p_0$ to $p_8$. These parameters have been
determined by fitting equation~(\ref{eq:d_theta}) to the photon distribution, at 11 different shower
energy values between $\sim 560$\,GeV and $\sim 56$\,PeV, equally distributed in $\log_{10} E_{sh}$. The
parameters have than been interpolated as functions of $\log_{10} E_{sh}$ as shown in
Figure~\ref{fig:params}. Due to event-to-event fluctuations, it was not always possible to find a
fit function for the whole energy range. The values for the parameters found at the lowest energies
apparently do not fit to the overall shape of the parameter distributions and were left out in the
fits. It will however be shown below that nevertheless, $D(\vartheta,E_{sh})$ describes the photon
distribution at these energies well. \\  
The parameterisations for the parameters are: 

\begin{alignat*}{6}
  p_0 = & \, 0.321 & + & \, 0.0150 & \cdot \log_{10}(E_{sh}/\textrm{GeV}) & \\
  p_1 = & \,14.304 & + & \, 0.336 & \cdot \log_{10}(E_{sh}/\textrm{GeV}) & \\
  p_2 = & \,40.791 & + & \, 0.660 & \cdot \log_{10}(E_{sh}/\textrm{GeV}) & \\
  p_3 = & \, 0.0779 & - & \, 0.0137 & \cdot \log_{10}(E_{sh}/\textrm{GeV}) & 
  + 0.0055 \cdot (\log_{10}(E_{sh}/\textrm{GeV}))^2 \\
  p_4 = & \,15.150 & + & \, 2.843 & \cdot \log_{10}(E_{sh}/\textrm{GeV}) & \\
  p_5 = & \,78.694 & - & \, 0.0367 & \cdot \log_{10}(E_{sh}/\textrm{GeV}) & \\
  p_6 = & \, 0.688 & - & \, 0.0506 & \cdot \log_{10}(E_{sh}/\textrm{GeV}) & \\
  p_7 = & \, 6.181 & - & \, 0.100 & \cdot \log_{10}(E_{sh}/\textrm{GeV}) & \\
  p_8 = & \,42.144 & + & \, 0.0677 & \cdot \log_{10}(E_{sh}/\textrm{GeV}) &
\end{alignat*}

$D(\vartheta,E_{sh})$ is normalised to 1. The normalisation parameter $p$ is calculated analytically by
solving the integral 
\begin{alignat}{1}
\int_{0^{\circ}}^{180^{\circ}} \frac{D(\vartheta,E_{sh})}{p} \textrm{d} \vartheta  
\approx & \int_{-\infty}^{\infty} \left( \frac{p_0}{\sqrt{2\pi} p_1} \cdot e^{ -(\vartheta - p_2)^2 / (2p_1^2) }
        + \frac{p_3}{\sqrt{2\pi} p_4} \cdot e^{ -(\vartheta - p_5)^2 / (2p_4^2) } \right) \textrm{d}
	\vartheta \notag \\
& + \int_{0^{\circ}}^{180^{\circ}} \left( \frac{p_6}{2\pi} \frac{p_7}{(\vartheta - p_8)^2 + \frac{p_7^2}{4}} \right)
	\textrm{d} \vartheta  \qquad = \frac{1}{p} 
\end{alignat}
which gives
\begin{equation}
p = (p_0 + p_3 + p_6 / \pi \cdot (\arctan((360^{\circ}-2 p_8)/p_7) - \arctan(-2 p_8/p_7)))^{-1}.
\end{equation}

Here, the fact that the two Gaussian distributions in $D(\vartheta,E_{sh})$ have their maxima in the
middle of the $\vartheta$-parameter space, at $\sim 43^{\circ}$ and $\sim 79^{\circ}$, and fall to very
small values beyond $0^{\circ}$ and $180^{\circ}$, is used
to allow for the approximate analytical calculation of the normalisation integral.

The main purpose of the parameterisation of the parameters was to find a fit which is in the ideal
case a first degree polynomial, to keep equation~(\ref{eq:d_theta}) as simple as possible. The
parameterisations presented here suffice to describe the distribution of $\vartheta$ for different
energies, as can be seen from Figure~\ref{fig:cheren_fit}, where the distribution of $\vartheta$ is
displayed for the event sample A, together with the parameterisation $D(\vartheta, E_{sh})$, for
energies between 300\,GeV and 30\,PeV (one decade per histogram). Here, $D(\vartheta, E_{sh})$ was
calculated at the logarithmic centre of the respective energy bin. For angles $\leq 120^{\circ}$, the
agreement between $D(\vartheta,E_{sh})$ and the photon distribution is very good. The slight
deviations between $D(\vartheta,E_{sh})$ and the photon distribution at larger angles are
insignificant for the calculation of the total photo-electron number. \\
It should be noted that for these studies the angle $\vartheta$ was calculated with respect to \rcg,
the {\it OM-centre-of-gravity}, as calculated in equation (\ref{eq:shower_max_corr}), not the
interaction vertex. The reason for this is that this position is the outcome of the position
reconstruction, as discussed in Section~\ref{sec:pos}. Note also that during the fit, the 
momentary value of $E_{sh}$ is used for the determination of $D(\vartheta,E_{sh})$.

\afterpage{\clearpage}

$D(\vartheta,E_{sh})$ has a similar shape also for energies larger or smaller than the ones studied here,
so that also in the case of very large or very small energies, the photon distribution is still
described approximately. However, it has to be ensured that none of the parameters $p_0$--$p_8$
becomes negative during the fit, which is achieved here by setting the lower limit of all parameters
to $10^{-10}$.

\subsection{Calculation of the Photo-Electron Number from the Hit Amplitudes}\label{sec:n_phot}

All the variables that are needed to calculate the photo-electron number $N_i$ in the event from the hit
amplitude $n_i$ of hit $i$ in OM$_i$ have now been collected. Starting point is the correction for
the angular efficiency $a_i(\gamma_i)$ of the OM: 

\begin{equation*}
n_i^{\prime} = \frac{n_i}{a_i(\gamma_i)}.
\end{equation*}

Next, the photon attenuation is taken into account: 

\begin{equation*}
n_i^{\prime\prime} = \frac{n_i^{\prime}}{\Lambda_{att}}.
\end{equation*}

In the last step, the photons are distributed isotropically in $\varphi$ and according to
$D(\vartheta,E_{sh})$ in $\vartheta$: 

\begin{equation*}
n_i^{\prime\prime\prime} = n_i^{\prime\prime} \cdot \frac{360^{\circ} \cdot \sin \vartheta_i}{2\alpha_i} \cdot 
\frac{1}{D(\vartheta_i,E_{sh}) \cdot 2\alpha_i} \equiv N_i,
\end{equation*}

where $N_i$ is the total number of photo-electrons in the event according to hit $i$. \\
Altogether, one obtains:

\begin{equation}
N_i = \frac{n_i}{a_i(\gamma_i) \cdot \Lambda_{att}} \cdot \frac{360^{\circ}\cdot \sin \vartheta_i}{2\alpha_i}
\cdot \frac{1}{D(\vartheta_i,E_{sh}) 2\alpha_i}.
\label{eq:n_calc_total}
\end{equation}

\begin{figure}[h]\centering
\includegraphics[width=11cm]{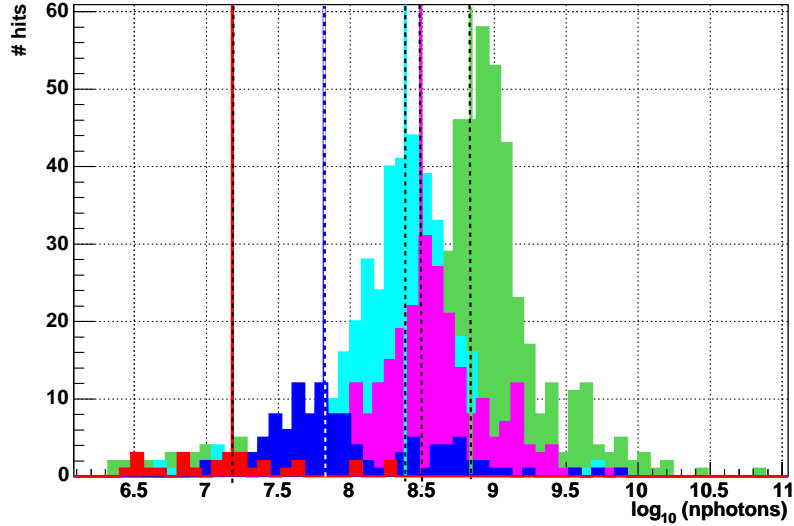}
\caption[Photo-electron number calculated from hit amplitude]{Photo-electron number in 5 events as
  calculated from the hit amplitudes in every hit OM; the values for each event are marked by
  different colours. The respective values of the medians are also shown as dashed lines.}
\label{fig:photon_number}
\end{figure}

\begin{figure}[h] \centering
\includegraphics[width=11cm]{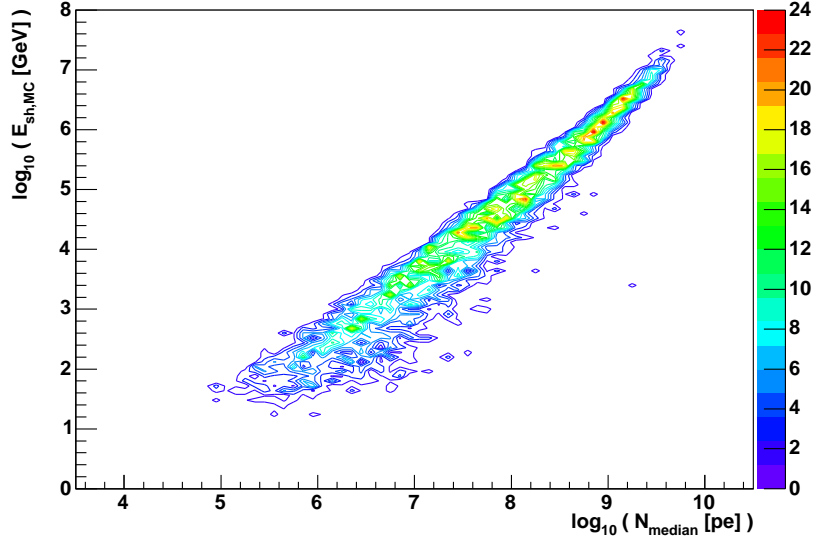}
\caption[Monte Carlo shower energy over median of photo-electron number]
{Monte Carlo shower energy over the median of the photo-electron numbers $N_i$, calculated from the hit
  amplitude $n_i$ in all OMs, as described in the text. }
\label{fig:nphot_E}
\end{figure}

$N_i$ is calculated independently for all hits $i$ in the event. After this, the median of all $N_i$ is
calculated; this value is then used as a starting value for the number of photo-electrons produced
by this shower, which is directly proportional to the shower energy. Figure~\ref{fig:photon_number}
shows exemplarily the distribution of the photo-electron numbers calculated once for every hit in an
event, for a selection of five events from event sample A with different energies. 
Though for most events there is a clear trend towards one favourable value for the
photo-electron number, the values of the photo-electron numbers calculated from each hit can be very
widely spread, mainly due to the angular acceptance $a_i(\gamma_i)$ which decreases to very small
values for incident angles larger than $\sim 115^{\circ}$ (see Figure~\ref{fig:ang_eff}).   
Therefore, to obtain one single and reliable value for the
photo-electron number in the event, the median was chosen; it is more stable against wide spreads of
the values than the arithmetic mean. The MC shower energy plotted versus the median of the
photo-electron numbers calculated in the described way is shown in Figure~\ref{fig:nphot_E} for
event sample A.  

\subsection{Comparison of the Measured and the Calculated Hit Amplitudes}

To calculate the expected amplitude $n_{i,c}$ in OM$_i$ from the total number of photo-electrons $N$,
position, direction and energy of the shower are needed. The position is calculated following the
algorithm from Section~\ref{sec:pos}, and for the shower direction, a starting value is assumed from
a scan of the parameter space (see Section~\ref{sec:scan}). The shower energy can be calculated from
the total number of photo-electrons $N$ (see Section~\ref{sec:calc_nu_energy}). By inverting
equation~(\ref{eq:n_calc_total}), $n_{i,c}$ is calculated as

\begin{equation}
n_{i,c} = N \cdot a_i(\gamma_i) \cdot \Lambda_{att} \cdot \frac{2\alpha_i}{360^{\circ}\cdot \sin \vartheta_i} \cdot
D(\vartheta_i,E_{sh}) \cdot 2\alpha_i.
\label{eq:n_calc_inv}
\end{equation}

$n_{i,c}$ can then be compared with $n_{i,meas}$, the actually measured amplitude in OM$_i$. The
lower limit of $n_{i,meas}$ is the SPE threshold (see Section~\ref{sec:digitisation}); measured
amplitudes below the threshold are considered as noise. For this study, an SPE threshold of 
0.3\,pe was used. The measurement process in the
photomultiplier smears the number of photo-electrons, so that the probability to measure a signal
$n_{i,meas}$ when $j$ photo-electrons were generated in the photomultiplier is 

\begin{equation}\label{eq:p_gauss}
  P_g(n_{i,meas},j) = \frac{1}{\sigma_j \cdot \sqrt{ 2\pi }} \cdot e^{-(n_{i,meas} - j)^2 /
  (2\sigma_j^2)}, \qquad j>0.
\end{equation}

\begin{figure}[h] \centering
\includegraphics[width=10cm]{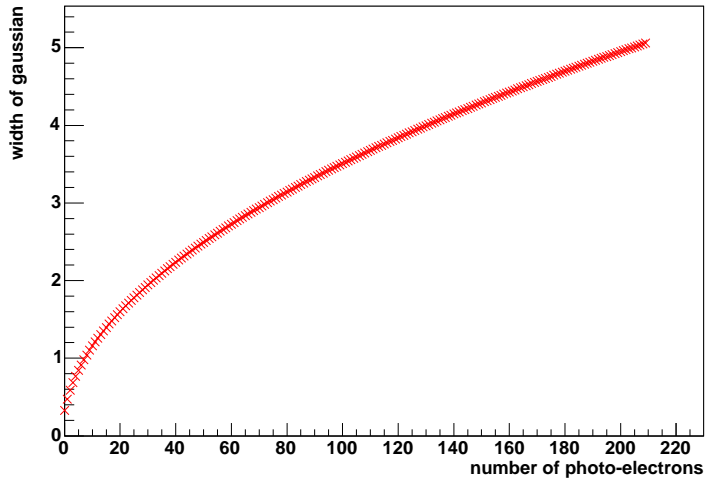}
\caption[Width of the Gaussian spread of the signal]{Width $\sigma$ of the Gaussian distribution
 (\ref{eq:p_gauss}) for signals between 1 and 210 photo-electrons.}
\label{fig:sigma_gauss}
\end{figure}

The width $\sigma_j$ depends on the number of photo-electrons $j$. It has been calculated for ${1
\le j \le 210}$ as proposed by~\cite{kopper}: For each $j$, a measurement is simulated $5 \cdot
10^8$ times using the ANTARES photomultiplier simulation~\cite{antcc}. A Gaussian fit is the applied
to the resulting distribution of $n_{i,meas}$; the width of the fit is the 
desired $\sigma_j$. The values of $\sigma_j$ calculated that way are as shown in 
Figure~\ref{fig:sigma_gauss}. 

The probability $P_g(n_{i,meas},0)$ to measure an amplitude $n_{i,meas}$ if no photo-electrons 
at all been produced in the OM, i.e.~to measure electronics noise, must be taken into account as
well. For the analysis of real data, the individual calibration data for each OM would 
have to be used to determine the noise pedestal (see Figure~\ref{fig:SPE_response}). The 
minimum threshold would have to be adjusted for each event, as is is set according to the
current optical noise. For this MC based study, the noise term is estimated as an exponentially
decreasing function, with ${P_g(n_{i,meas},1) \geq P_g(n_{i,meas},0)}$ above the minimum threshold
$n_{i,meas} = 0.3$: 

\begin{equation}
  P_g(n_{i,meas},0) = 0.99 \cdot e^{- 10 \cdot n_{i,meas}}.
\end{equation}

The factor 0.99 is used to ensure $P_g(n_{i,meas},0) < 1$. \\ [0.5cm]
The relation between the expected amplitude $n_{i,c}$ and the actual photo-electron number
$j$ can be described by a Poissonian distribution: 

\begin{equation}
P_p(n_{i,c},j) = \frac{n_{i,c}^j}{j!} \cdot \textrm{e}^{-n_{i,c}}.
\end{equation}

If the expected amplitude is $n_{i,c}$, the probability $l_{i,j}$, that $j$ photo-electrons 
have been produced which generate a measured amplitude $n_{i,meas}$, is the product of the two
probabilities $P_p(n_{i,c},j)$ and $P_g(n_{i,meas},j)$:  

\begin{equation}
l_{i,j} = P_p(n_{i,c},j) \cdot P_g(n_{i,meas},j).
\label{eq:likeli1}
\end{equation}

Summing over all $j$ gives the probability to measure $n_{i,meas}$ if $n_{i,c}$ is expected:  

\begin{equation}
L_i(n_{i,meas},n_{i,c})  = \sum^{j_{max}}_{j = j_{min}} l_{i,j} 
  = \sum^{210}_{j = 0} P_p(n_{i,c},j) \cdot P_g(n_{i,meas},j). 
\end{equation}

The lower limit of the summation represents the case of no photo-electrons being produced in 
OM$_i$, while the upper limit of 210 is 10\,pe above the saturation of the photomultiplier 
electronics of 200\,pe. In principle, both in simulation and in measurement, a signal of 
200\,pe can also be caused by more than 200 photo-electrons, such that the upper limit of $j$
could be chosen arbitrarily large. However, both $P_g(n_{i,meas},j)$ and $P_p(n_{i,c},j)$ decrease
rapidly for large $j$, and it is therefore possible to set the upper summation limit 
slightly above the saturation limit, at 210. \\
$L_i$ is the function which determines the likelihood of the combination $n_{i,meas}$ and $n_{i,c}$
in OM$_i$. Therefore, the overall probability $\mathcal{P}$ for a hit constellation in the whole
detector is given by multiplying the probabilities of the single OMs:

\begin{equation} \label{eq:prob}
\begin{split}
\mathcal{P} &= \prod_{i=1}^{900} L_i 
= \prod_{i=1}^{900} \sum^{210}_{j = 0} P_p(n_{i,c},j) \cdot P_g(n_{i,meas},j)  \\
&= \prod_{i=1}^{900} \sum^{210}_{j = 0}  \frac{n_{i,c}^j}{j!} \cdot \textrm{e}^{-n_{i,c}} \cdot
P_g(n_{i,meas},j)  \\
&=  \prod_{i=1}^{900} \sum^{210}_{j = 0} \left( \frac{(N_i \cdot a_i(\gamma_i) \cdot \Lambda_{att} 
\cdot \frac{2\alpha_i}{360^{\circ}\cdot \sin \vartheta_i} \cdot D(\vartheta_i,E_{sh}) \cdot 2\alpha_i)^j}{j!} \right. \\
& \hspace{1.6cm} \left. \cdot\, \textrm{exp}\{-(N_i \cdot a_i(\gamma_i) \cdot \Lambda_{att} 
\cdot \frac{2\alpha_i}{360^{\circ}\cdot \sin \vartheta_i} \cdot D(\vartheta_i,E_{sh}) \cdot 2\alpha_i)\} \cdot P_g(n_{i,meas},j) \right). 
\end{split}
\end{equation}

\section{The Maximum Likelihood Fit}\label{sec:likelihood}

The shower direction and energy are determined by maximising the likelihood in 
equation~(\ref{eq:prob}). In this process, the {\it log likelihood} $\ln \mathcal{P}$ is considered
instead of $\mathcal{P}$. By multiplying the expression with $-1$, the maximisation problem is
replaced by a minimisation. The final expression to be minimised becomes therefore

\begin{equation}\label{eq:min_like}
\mathcal{L} = - \ln \mathcal{P} = \sum_{i=1}^{900} (- \ln (L_i)).
\end{equation}

To minimise $\mathcal{L}$, the fitting software MINUIT~\cite{minuit} which was developed at CERN
is used. The technical details of the minimisation are explained in appendix~\ref{sec:tech_details}, 
where also a chart of all steps of the reconstruction can be found. 

\subsection{Starting Values for the Fit}\label{sec:scan}

\begin{figure}[h] \centering
\includegraphics[width=7.4cm]{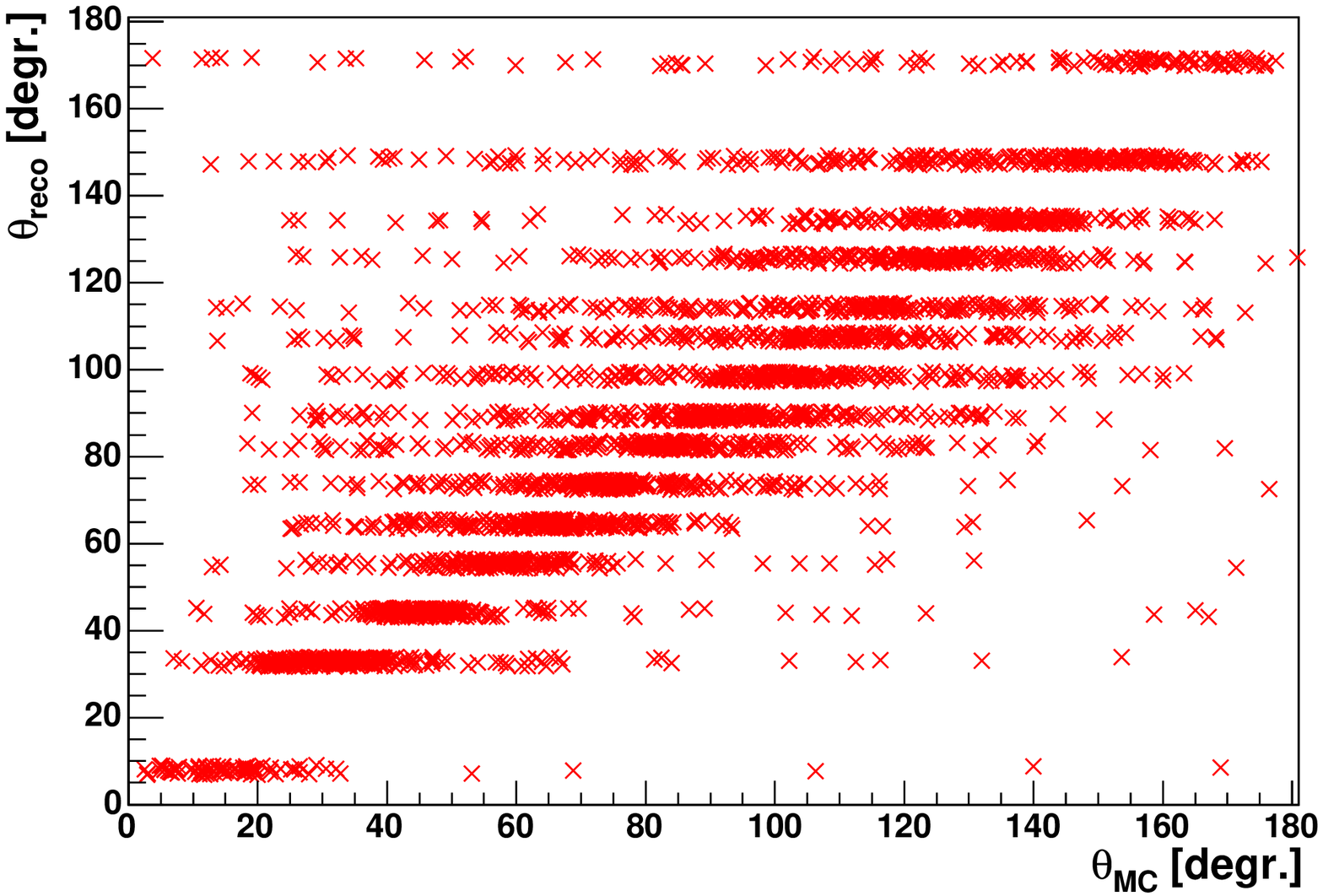}
\includegraphics[width=7.4cm]{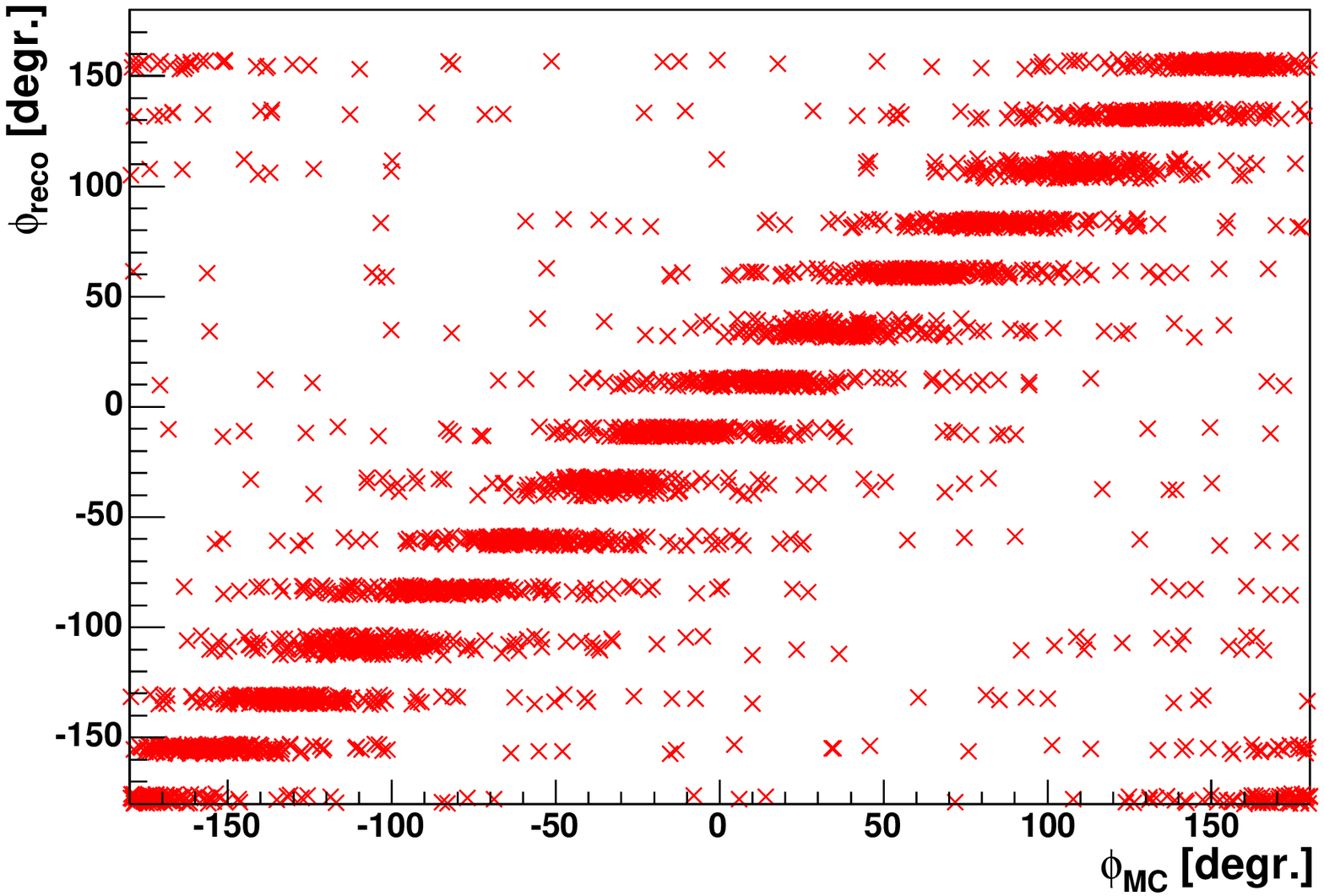}
\caption[$\theta$ and $\phi$ from scan of likelihood parameter space]{Fit variables $\theta$ and
  $\phi$ as determined by a scan of the likelihood parameter space drawn over their respective MC
  values. One can see that the algorithm is sufficient in most cases to find a parameter value
  close to the MC value.} 
\label{fig:theta_phi_scan}
\end{figure}

\begin{figure}[h] \centering
\includegraphics[width=10cm]{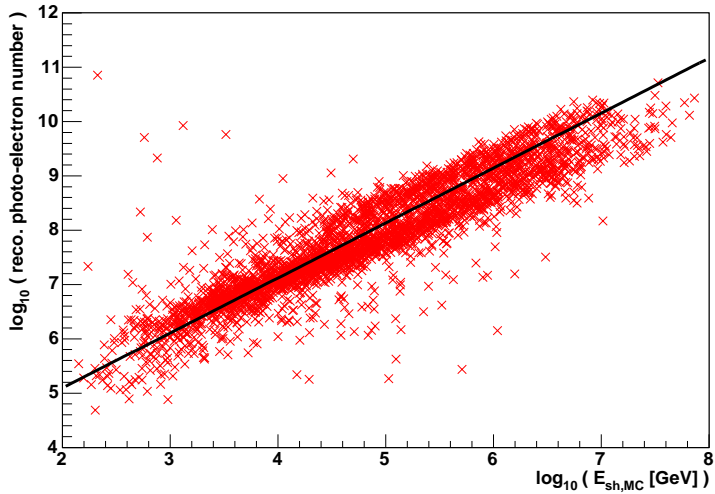}
\caption[Photo-electron number from scan of likelihood parameter space]{The fit variable \lq\lq
  photo-electron number\rq\rq\ as determined by a scan of the likelihood parameter space, vs.~the MC
  shower energy. The linear function (\ref{eq:nphot_E}) which is used to calculate the shower energy
  from the photo-electron number is also shown.} 
\label{fig:nphot_scan}
\end{figure}

To start the fit, suitable values for the three parameters $\theta, \phi$ and $E_{sh}$ are 
needed. They are determined by scanning the whole parameter space. The scan is done by calculating the
value of the likelihood function for a selectable number of data points. For the results shown in the
following, 15 points were used for $\theta$ and $\phi$ each, and 5 for the photo-electron number, so
that the likelihood was calculated at $15 \times 15 \times 5$ points. The 15 points of the directional
variables were equally distributed, between $-\pi$ and $\pi$ in $\phi$ and between $-1$ and $1$ in
$\cos \theta$. The 5 points for the photo-electron number were equally distributed between
$\log_{10}(N_m) - 2$ and $\log_{10}(N_m) + 2$, where $N_m$ is the median of the total photo-electron
numbers calculated for all hits in the event, as described in Section~\ref{sec:n_phot}. The
reconstructed position of the shower, calculated as described in Section~\ref{sec:pos}, and the
shower direction, calculated under the assumption that all photons are emitted from the shower under
the same angle $\vartheta_C$, as described in Section~\ref{sec:prefit_dir1}, were used as input
variables for the calculation of $N_m$. \\  
The parameter values for which the likelihood is smallest are then used as starting values for the
actual fit. These starting values are often already quite
close to the MC values, as can be seen from Figures~\ref{fig:theta_phi_scan}
and~\ref{fig:nphot_scan}, where the values of $\theta, \phi$ and the photo-electron number from the
scan are plotted versus the respective MC values. While the bins for the likelihood calculation are
visible in $\theta$ and $\phi$, for the photo-electron number this is not the case, because $N_m$ is
different in each event. 

\subsection{Calculation of Neutrino Energy from the Photo-Electron Number}\label{sec:calc_nu_energy}

\begin{figure} \centering
\includegraphics[width=10cm]{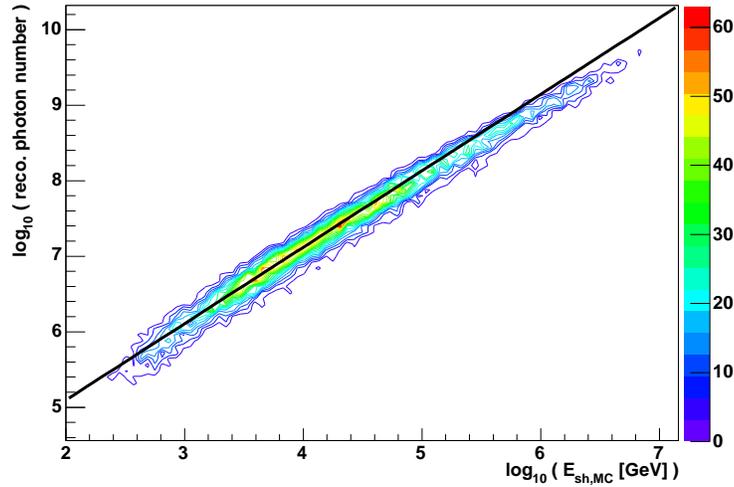}
\caption[MC shower energy over reconstructed photo-electron number]{MC shower energy over reconstructed
  photo-electron number, for events with an angular error $< 10^{\circ}$, together with the fit from
  equation~(\ref{eq:nphot_E}).} 
\label{fig:nphot_fit_E}
\end{figure}

A formula to calculate the energy of the hadronic shower from the reconstructed photo-electron
number was determined from successfully reconstructed events, by comparing the reconstructed
photo-electron number to the MC shower energy. Here, a sample of 140000 NC events between 10\,GeV
and 10\,PeV was used (event sample B, see Appendix~\ref{sec:nc_sample}). The events were
reconstructed by minimising equation~(\ref{eq:min_like}); Figure~\ref{fig:nphot_fit_E} shows the
distribution of those events for which the direction was reconstructed with an error smaller than
$10^{\circ}$, and a parameterisation of the correlation between photo-electron number and MC energy,

\begin{equation}\label{eq:nphot_E}
    \log_{10} (E_{sh}/\textrm{GeV}) = -3.03 + 0.99 \cdot \log_{10} N.
\end{equation}

\begin{figure}[h] \centering
\includegraphics[width=10cm]{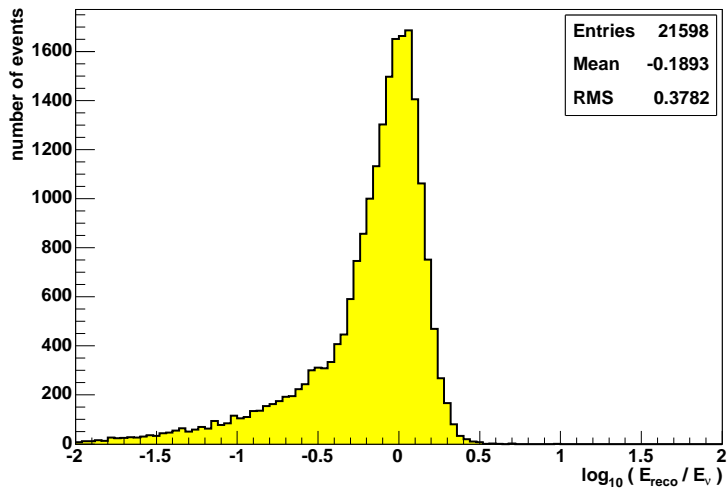}
\caption[Error in neutrino energy for shower-type events]{Error in reconstructed neutrino energy for
  a mixed sample of shower-type events reconstructed with an angular error smaller than $10^{\circ}$.}
\label{fig:reco_energy}
\end{figure}

Function (\ref{eq:nphot_E}) is valid within the energy range of $10^{3}$\,GeV$\le E_{sh} \le
10^{5.2}$\,GeV. For lower and higher shower energies, the distribution is bent towards smaller
photo-electron numbers\footnote{if the aim was to describe the whole distribution as exactly as
  possible, a second degree polynomial would be more suitable. However, it would then be necessary
  to restrict the parameter space to the region where the polynomial is monotonically increasing.
  The usage of constrained parameters is not recommended by the MINUIT authors~\cite{minuit} and was
  therefore avoided.}.
\\
The shower energy can thus be calculated from the photo-electron number. However, to retrieve the
neutrino energy is not straight-forward, as for NC events a fraction of $(1-y)$ of the
primary neutrino energy (where $y$ is the Bjorken variable, see Section~\ref{sec:nu_variables}) goes
into the secondary neutrino. Therefore, the shower energy as calculated from the reconstructed
photo-electron number only yields a lower limit on the primary neutrino 
energy. On the other hand, in the CC reaction of $\nu_e$, all neutrino energy goes into showers, and
the signature of this event type will not be distinguishable from that of NC events, so that in the
end the energy calculated using equation (\ref{eq:nphot_E}) is to be regarded as the neutrino
energy. The comparison with the MC neutrino energy should then, for a combined sample of shower-type
events, lead to a narrow peak at ${E_{sh} = E_{\nu}}$ for the $\nu_e$ CC events, with a long tail
towards the region of a reconstructed energy that is smaller than the MC neutrino energy, as shown
in Figure~\ref{fig:reco_energy}. For this plot, events with an angular error smaller than
$10^{\circ}$ were selected from event sample B (NC events) and from the $\nu_e$ CC event sample (see
Appendix~\ref{sec:nue_sample}). 

\afterpage{\clearpage}

\subsection{Examination of the Likelihood Parameter Space}\label{sec:topologies}

From the description of the likelihood function given in the above sections it is in principle
possible to reconstruct any kind of shower-type event producing hits in the detector. In reality,
however, it is important for the success of the reconstruction where, with respect to the 
instrumented volume, an interaction took place, and whether this position is well reconstructed or
not. In this section, several different event topologies are presented. The list of topologies is by
no means exhaustive, but it gives an overview of the possible problems that might be encountered in
the event reconstruction. 

\subsubsection{The Ideal Event}

\begin{figure}[h] \centering
\includegraphics[width=4.9cm]{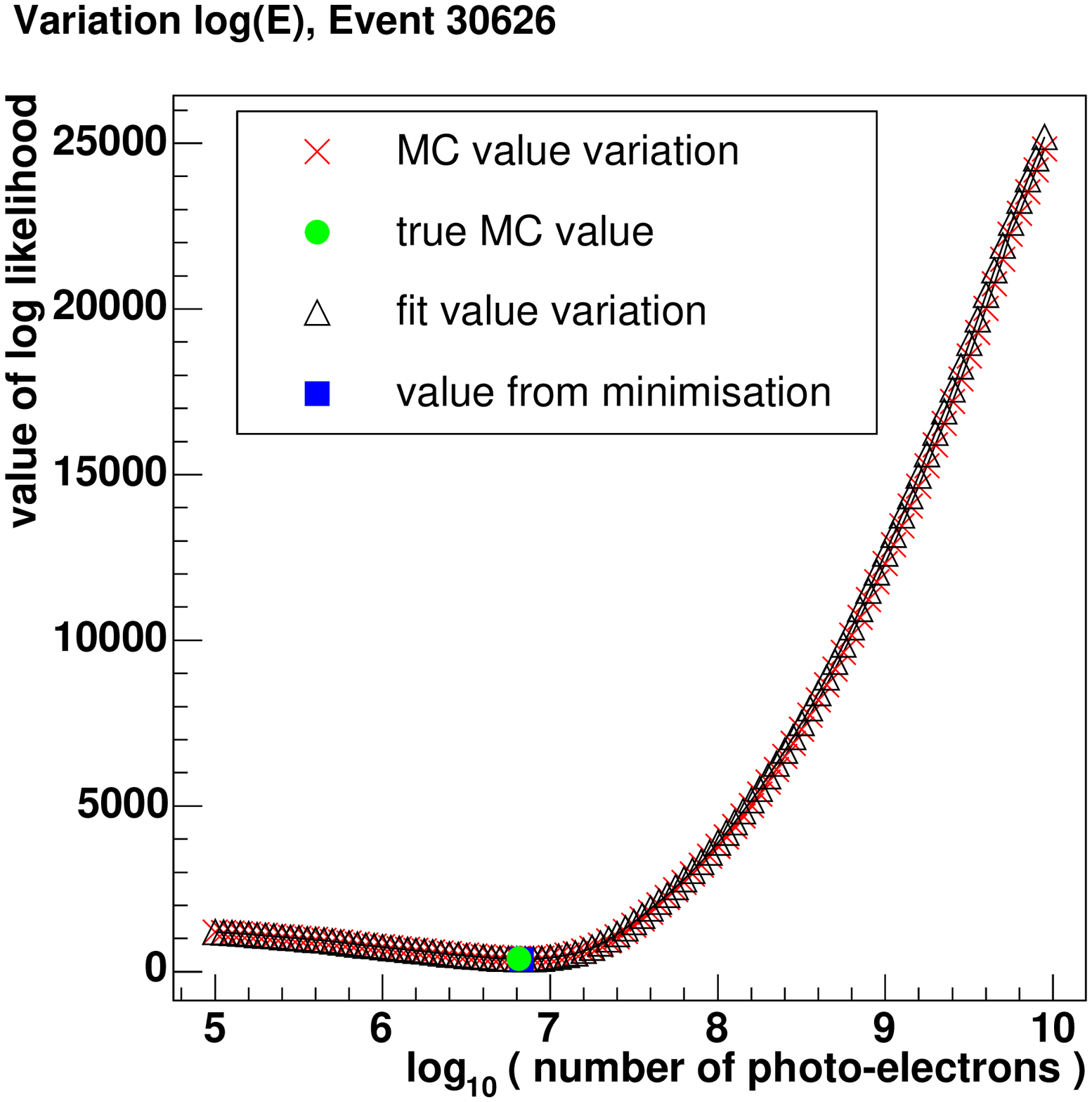}
\includegraphics[width=4.9cm]{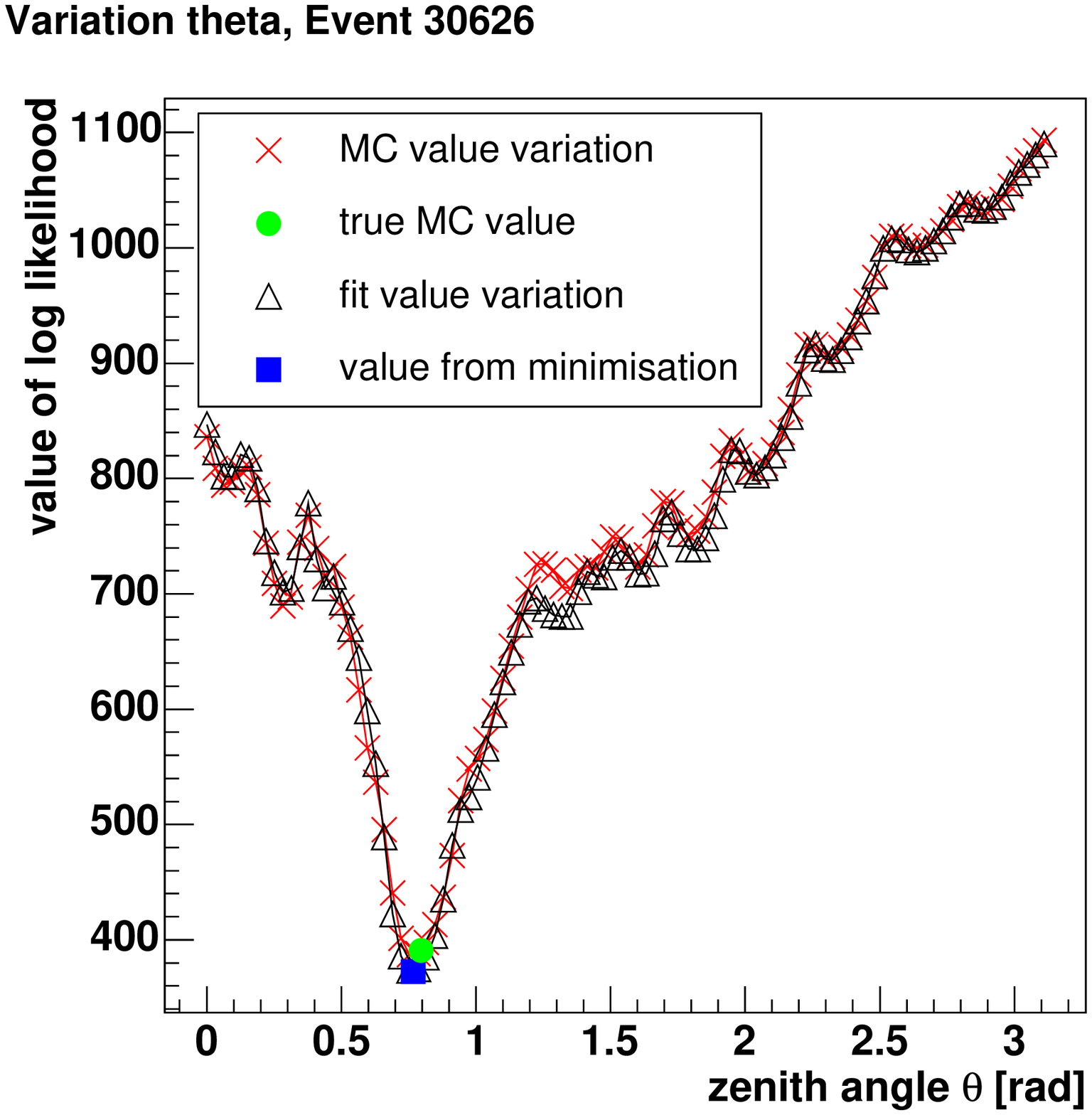}
\includegraphics[width=4.9cm]{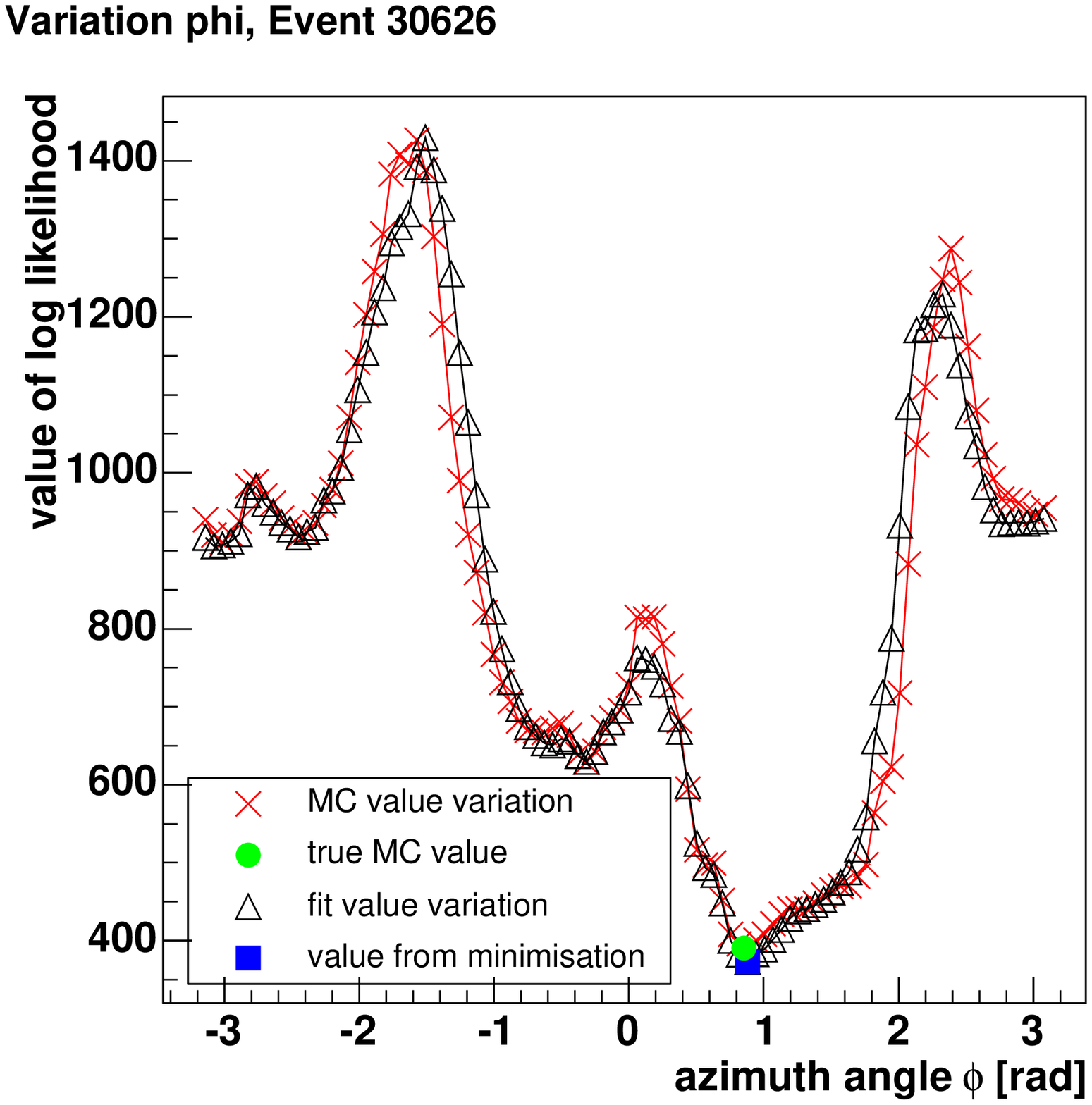}
\caption[Likelihood parameter space for an upgoing event]{Values of the likelihood function for the
  variation of the photo-electron number (left), the zenith angle $\theta$ (middle) and the azimuth
  angle $\phi$ (right), for a well-reconstructible event. The plots show the log likelihood
  calculated for the variation of one of the three fit parameters, keeping the other  
  two parameters constant at their true values (red crosses) or their final values from the fit
  (black triangles). The true MC value (green circle) and the reconstructed value (blue square) are
  also shown.}
\label{fig:good_ev30626}
\end{figure}

\begin{figure}[h] \centering
\includegraphics[width=4.9cm]{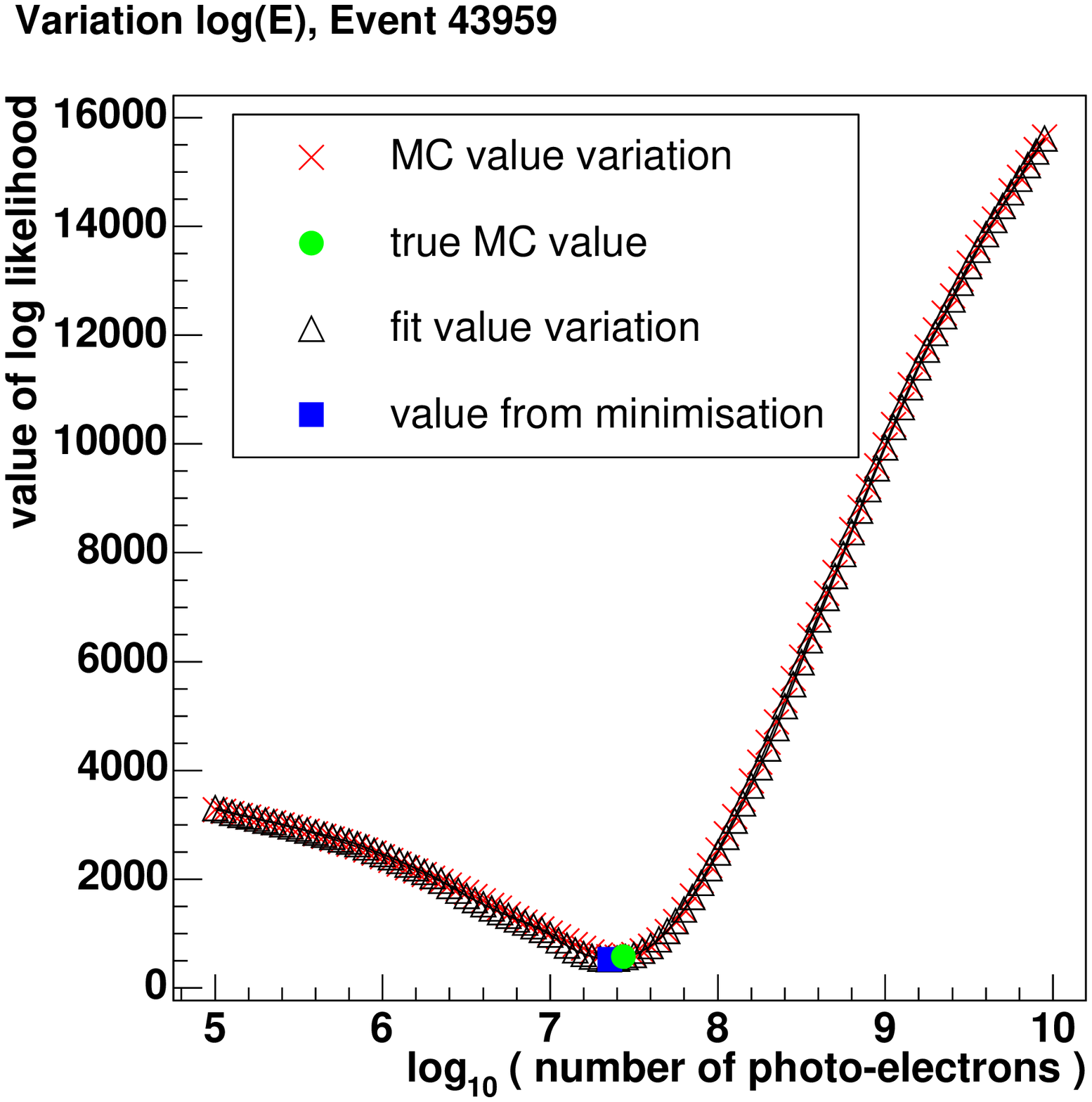}
\includegraphics[width=4.9cm]{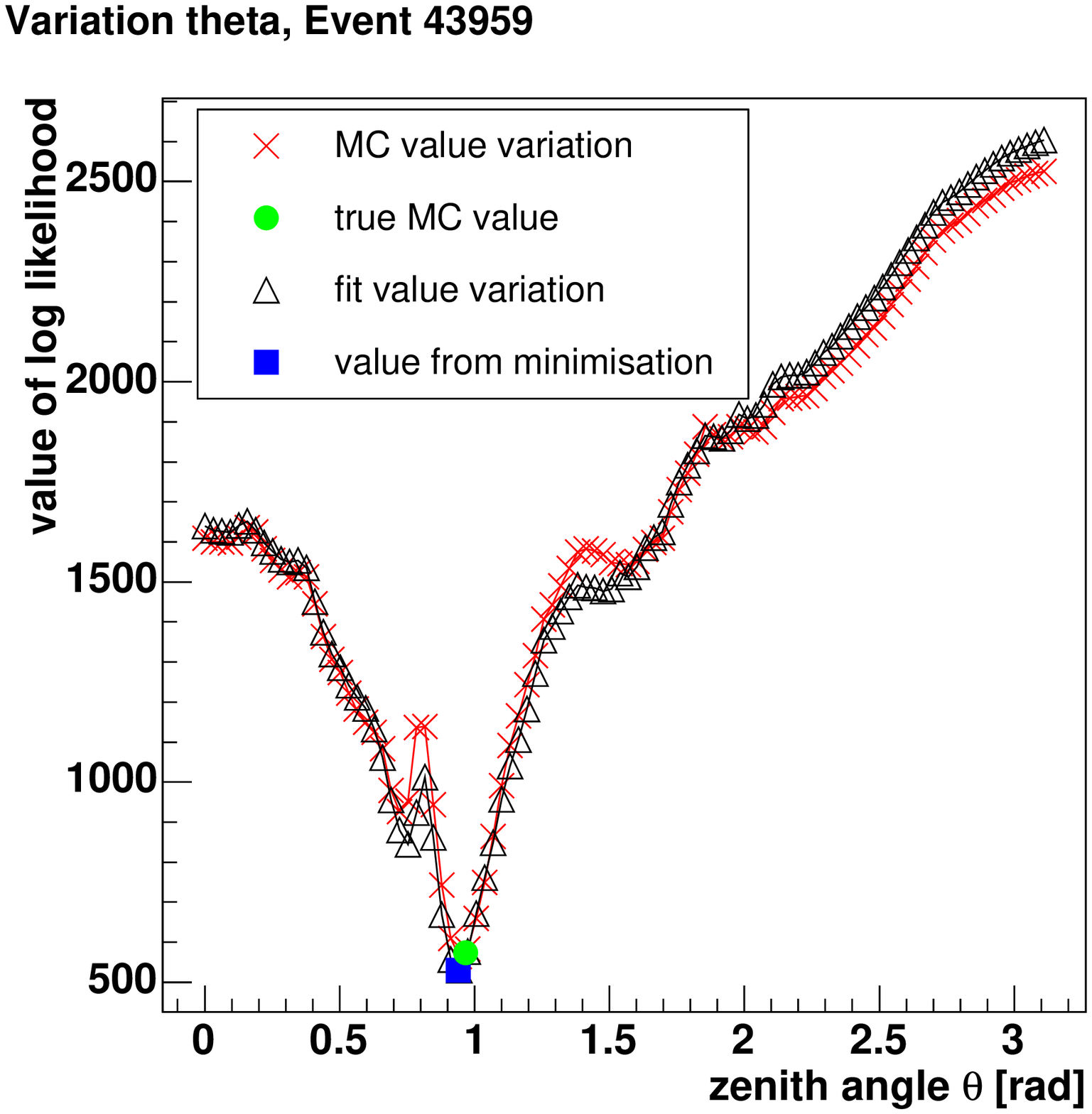}
\includegraphics[width=4.9cm]{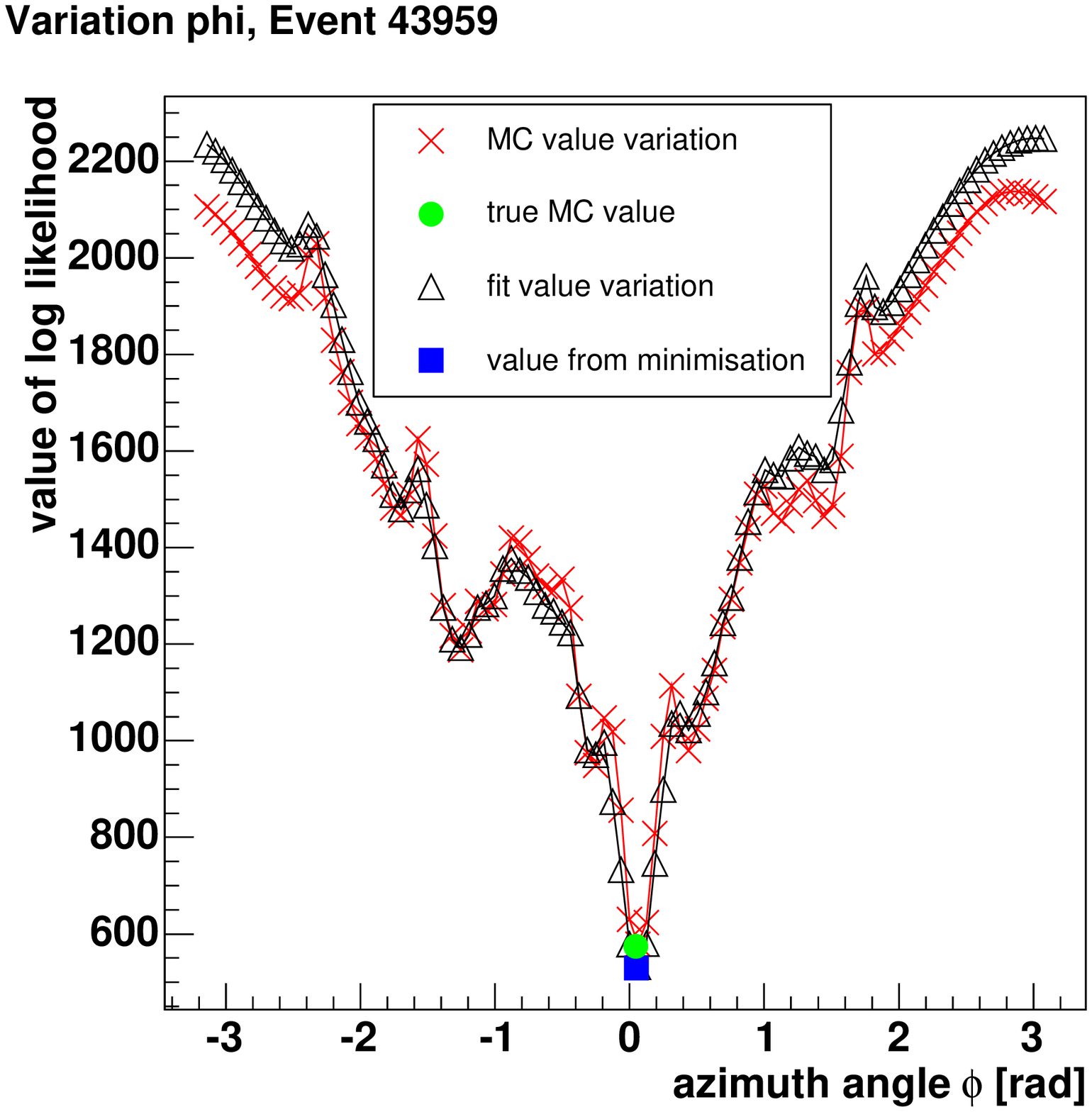}
\caption[Likelihood parameter space for an upgoing event]{Values of the likelihood function for the
  variation of the photo-electron number (left), the zenith angle $\theta$ (middle) and the azimuth
  angle $\phi$ (right), for a well-reconstructible event. See Figure~\ref{fig:good_ev30626} for
  details.} 
\label{fig:good_ev43959}
\end{figure}

\begin{figure}[h] \centering
\includegraphics[width=4.9cm]{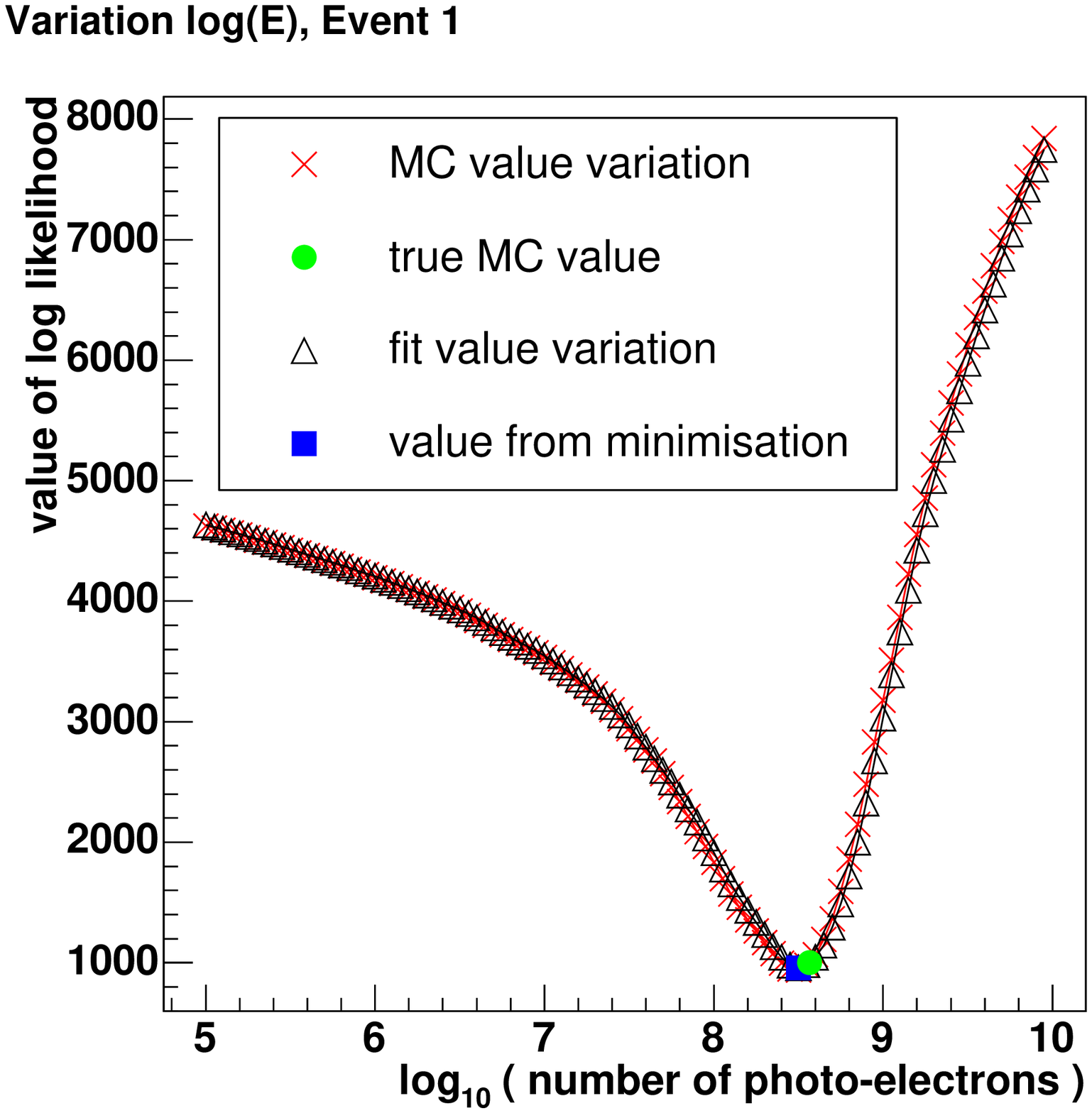}
\includegraphics[width=4.9cm]{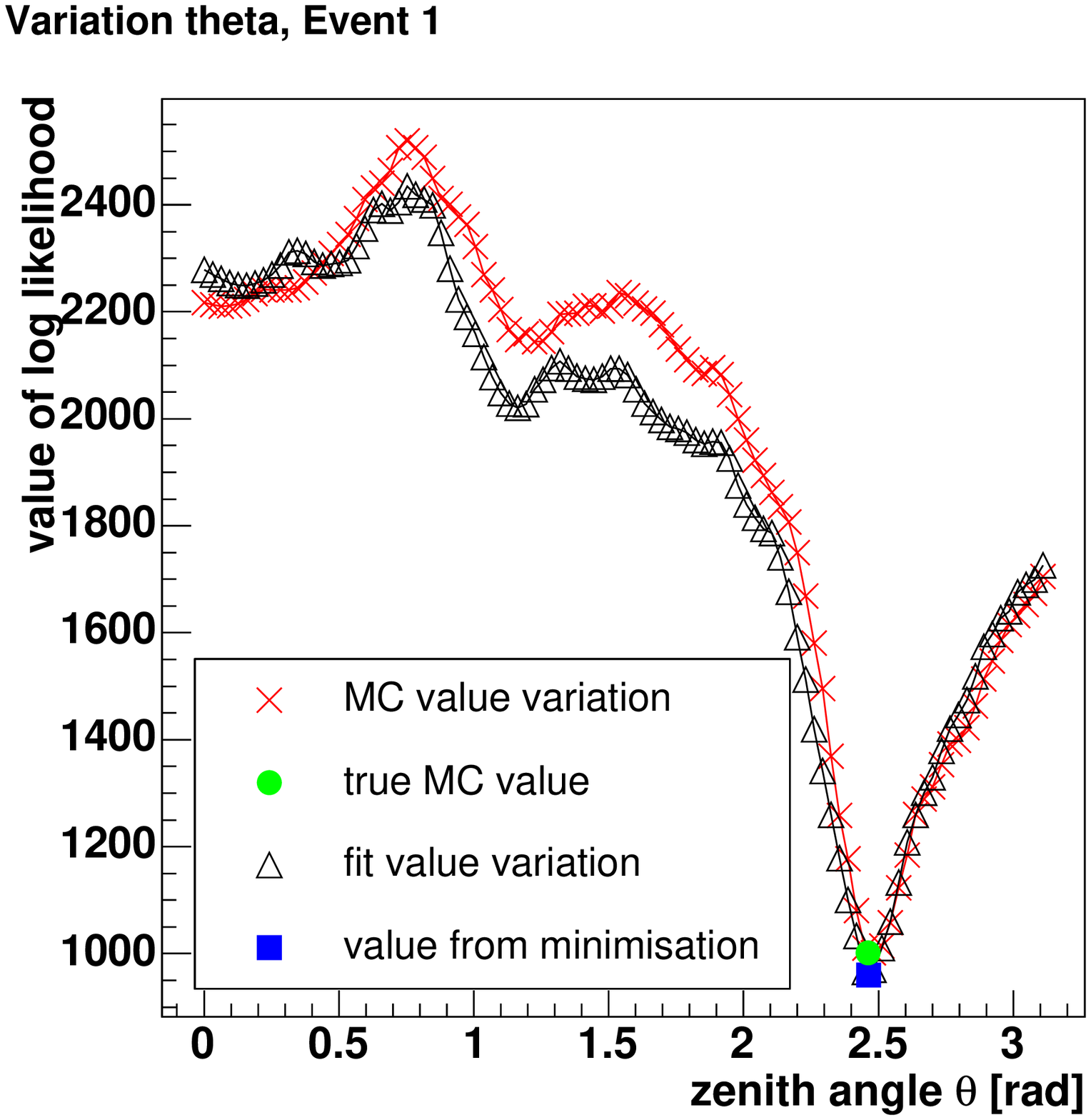}
\includegraphics[width=4.9cm]{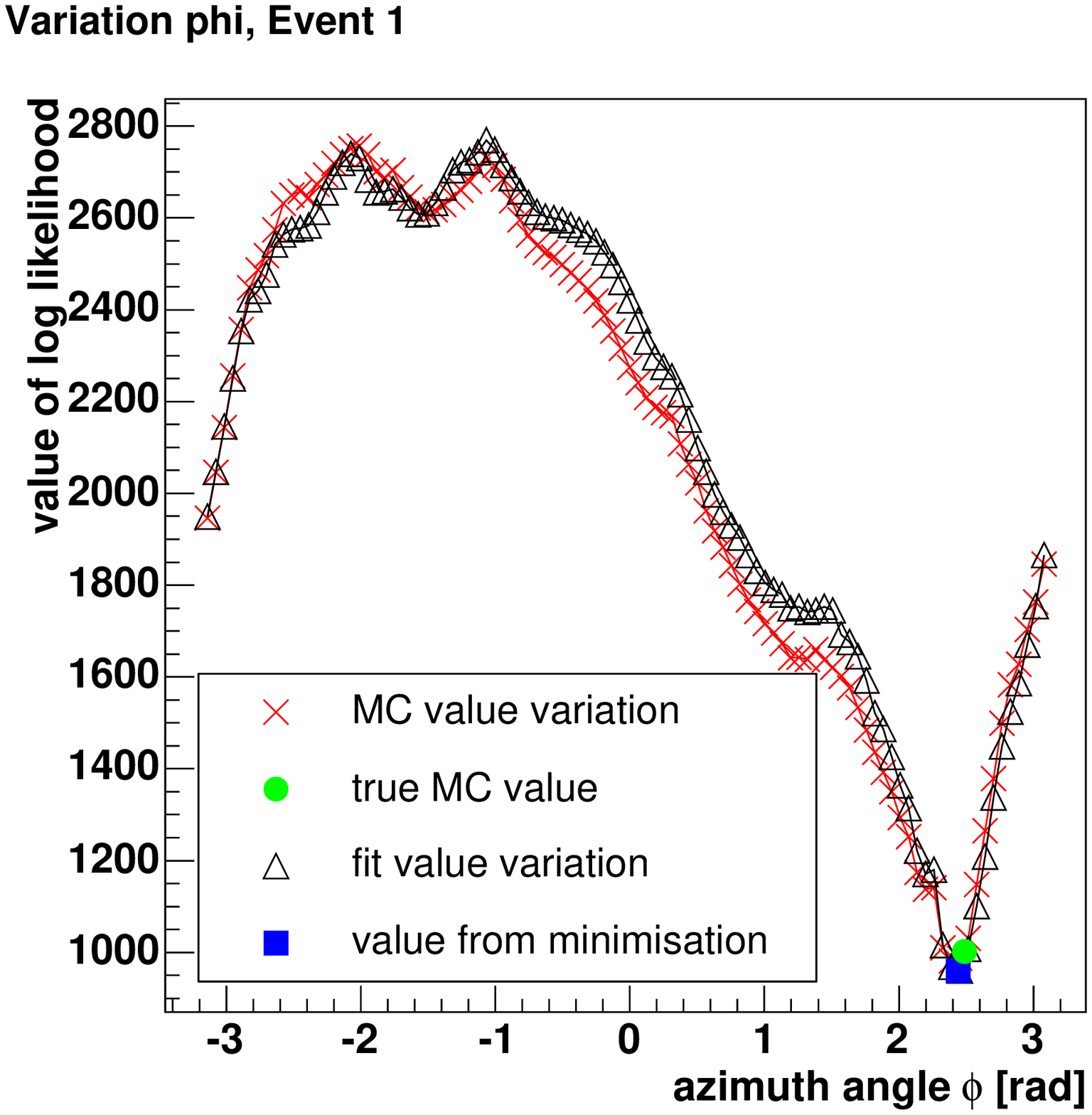}
\caption[Likelihood parameter space for a downgoing event]{Values of the likelihood function for the
  variation of the photo-electron number (left), the zenith angle $\theta$ (middle) and the azimuth
  angle $\phi$ (right), for a well-reconstructible event induced by a downgoing neutrino. See
  Figure~\ref{fig:good_ev30626} for details.}
\label{fig:good_ev51490}
\end{figure}

\begin{figure}[h] \centering
\includegraphics[width=4.9cm]{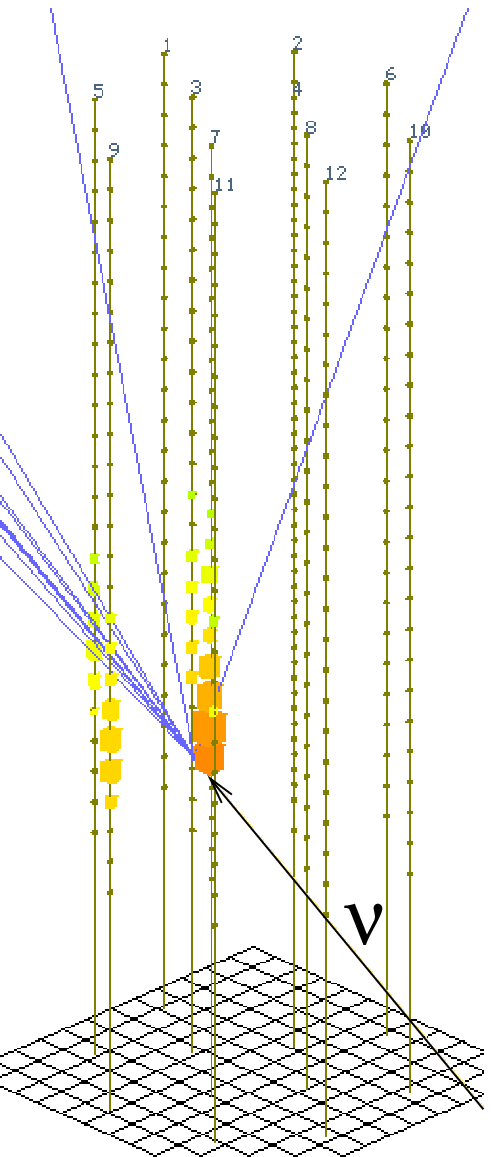}
\includegraphics[width=4.9cm]{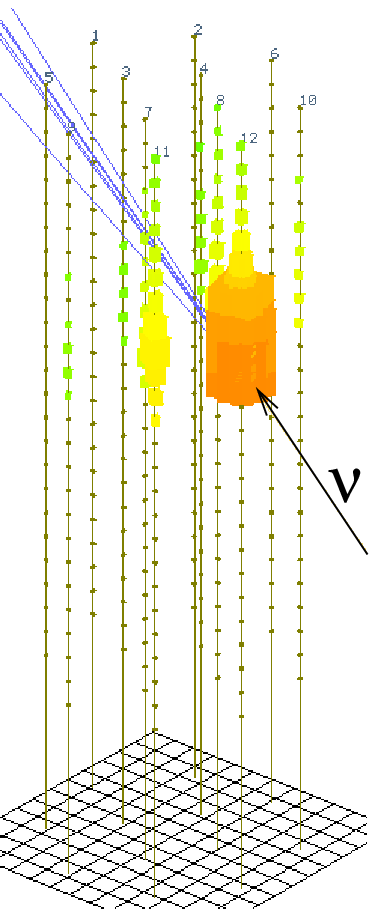}
\includegraphics[width=4.9cm]{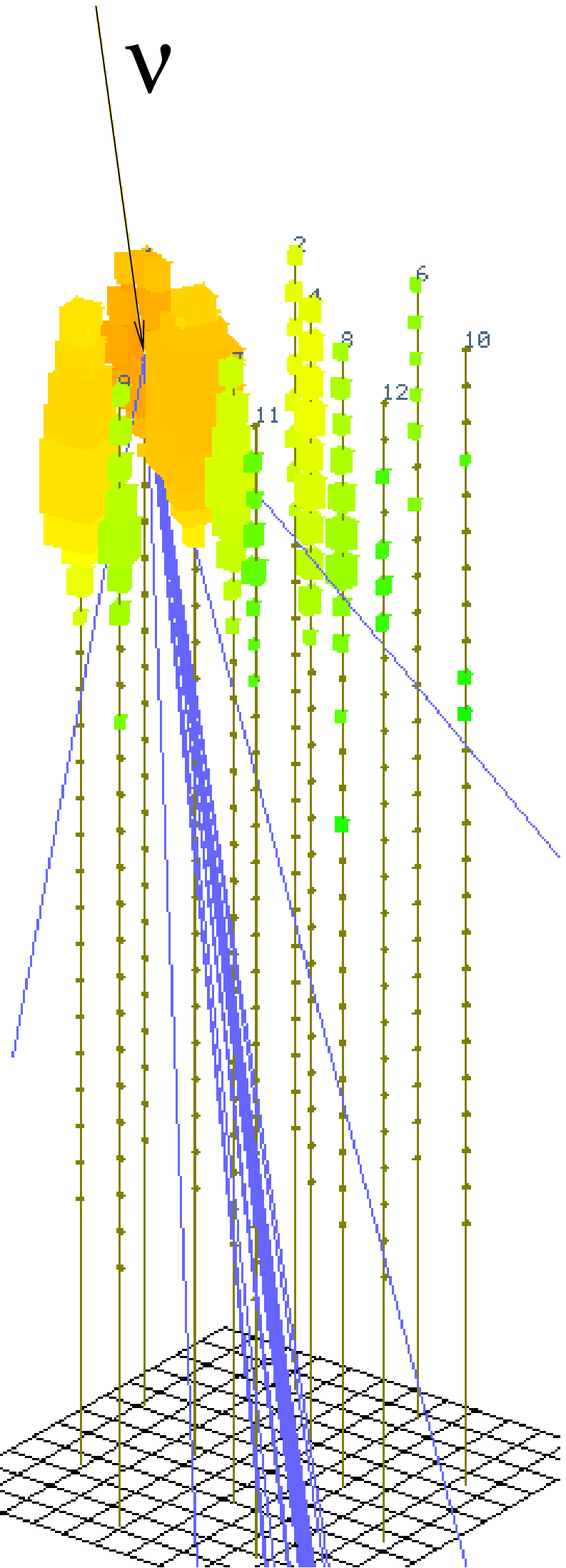}
\caption[Well-reconstructible events in the ANTARES event display]{Events from
  Figures~\ref{fig:good_ev30626},~\ref{fig:good_ev43959} and~\ref{fig:good_ev51490} as
  seen in the ANTARES event display. The detector strings are shown as numbered lines, with dots
  marking the different storeys. The incident neutrino is shown in black, the blue lines mark the
  particle tracks of the shower (not to scale), and the squares show the photo-electron signals,
  integrated over 25\,ns. The size of the squares is proportional to the hit amplitude, and 
  their colour coding is according to the arrival time of the first photon in the photomultiplier,
  from red over yellow to green.} 
\label{fig:good_events}
\end{figure}

The ideal shower-type event is that of a neutrino which interacts inside the detector, so
that the whole shower is right in the middle of the instrumented volume. In
Figures~\ref{fig:good_ev30626},~\ref{fig:good_ev43959} and~\ref{fig:good_ev51490}, the shape of the
likelihood function for three well-reconstructible events is shown. The log likelihood is calculated 
for the variation of one of the three fit parameters, keeping the other two parameters constant at
their true values or at their final values from the fit. The true MC value and the reconstructed value
are also shown. The events shown in Figures~\ref{fig:good_ev30626} and~\ref{fig:good_ev43959} had
similar shower directions, but different shower energies of $E_{sh} = 3.7$\,TeV and $E_{sh} =
19$\,TeV, respectively. The event shown in Figure~\ref{fig:good_ev51490} had a shower energy of
$E_{sh} = 269$\,TeV. For the variation of the photo-electron number, the shape of the 
likelihood is similar in all three events, and very smooth. For the variation of $\theta$ and $\phi$, the 
likelihood shows larger fluctuations; a clear global minimum is visible in all cases, but a number
of local minima is present as well. The neutrino shown in Figure~\ref{fig:good_ev51490} had a large
zenith angle, i.e.~it came from above; it interacted not far away from the upper edge of the
detector. Even though the resolution for the reconstruction of downgoing events is generally worse
than for upgoing events (due to the orientation of the photomultipliers, see also
Section~\ref{sec:upgoing}), this example shows that a good reconstruction is also possible for this
event type (total angular error in this case: $1.65^{\circ}$). \\ 
Snapshots of the three events in the ANTARES event display~\cite{a3d} are shown in
Figure~\ref{fig:good_events}. For better visibility, only hits passing the filter
conditions (see Section~\ref{ch:trigger}) are displayed.


\afterpage{\clearpage}

\subsubsection{Impacts of Shifts in the Shower Position}

\begin{figure}[h] \centering
\includegraphics[width=4.9cm]{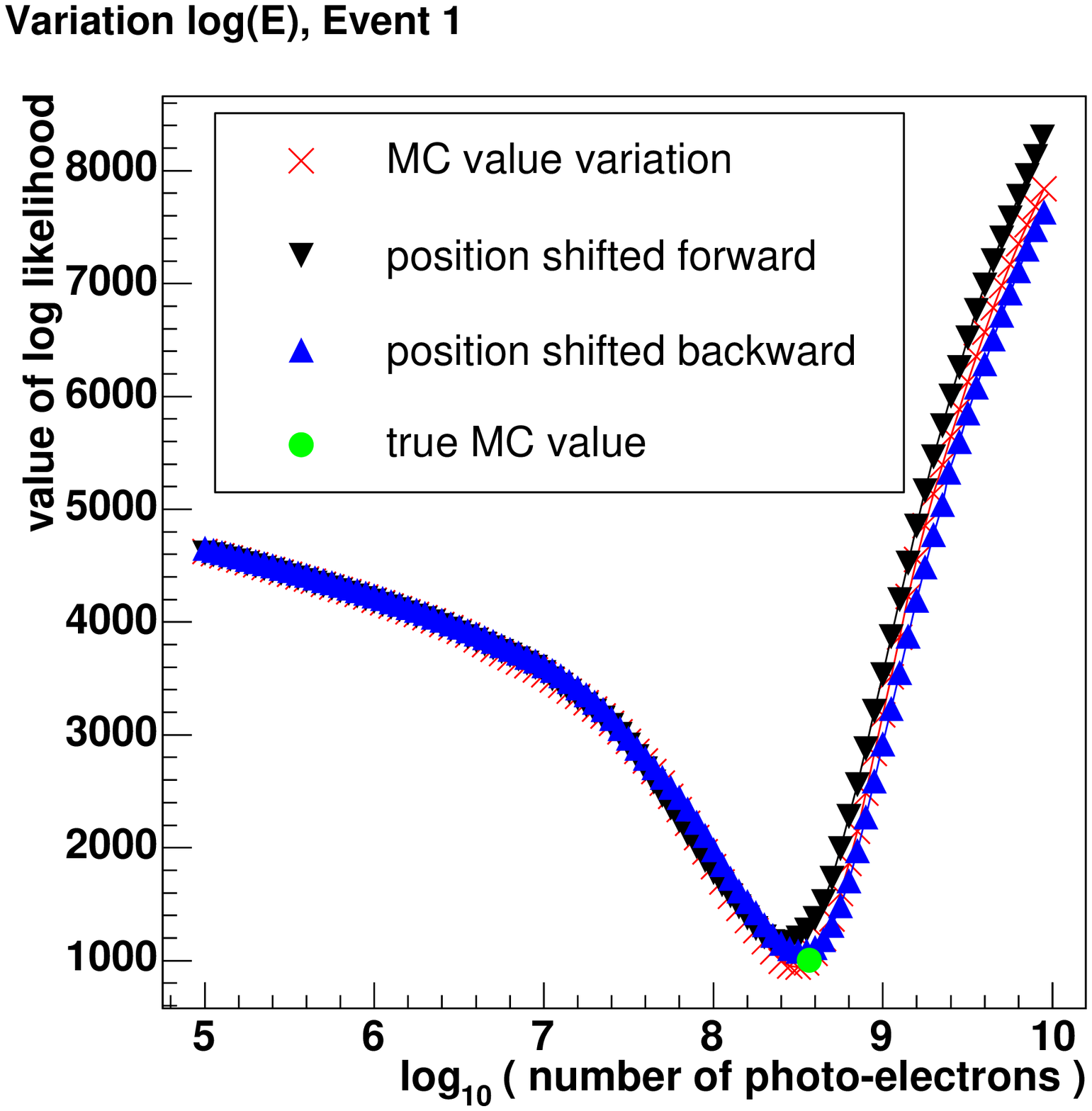}
\includegraphics[width=4.9cm]{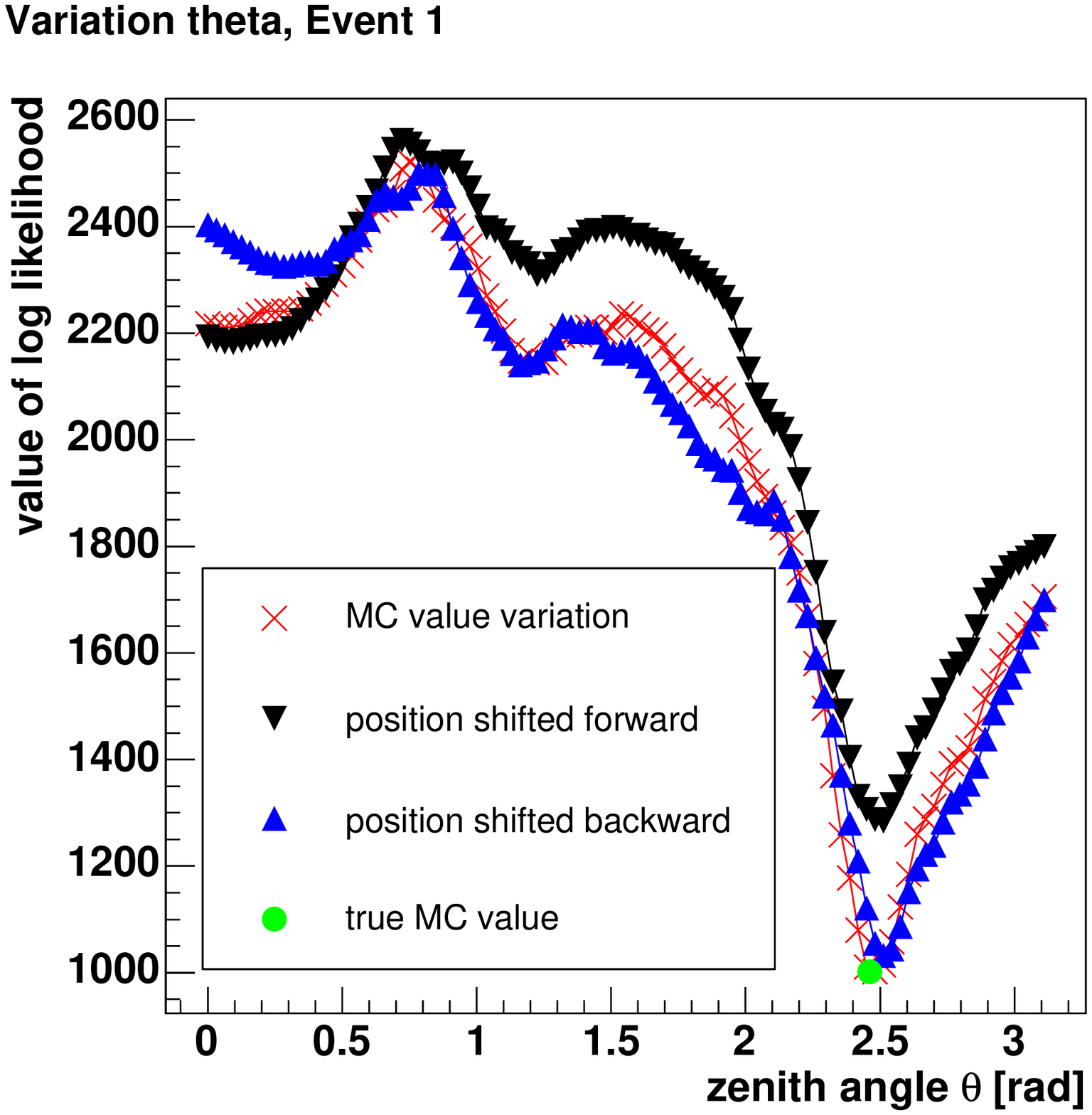}
\includegraphics[width=4.9cm]{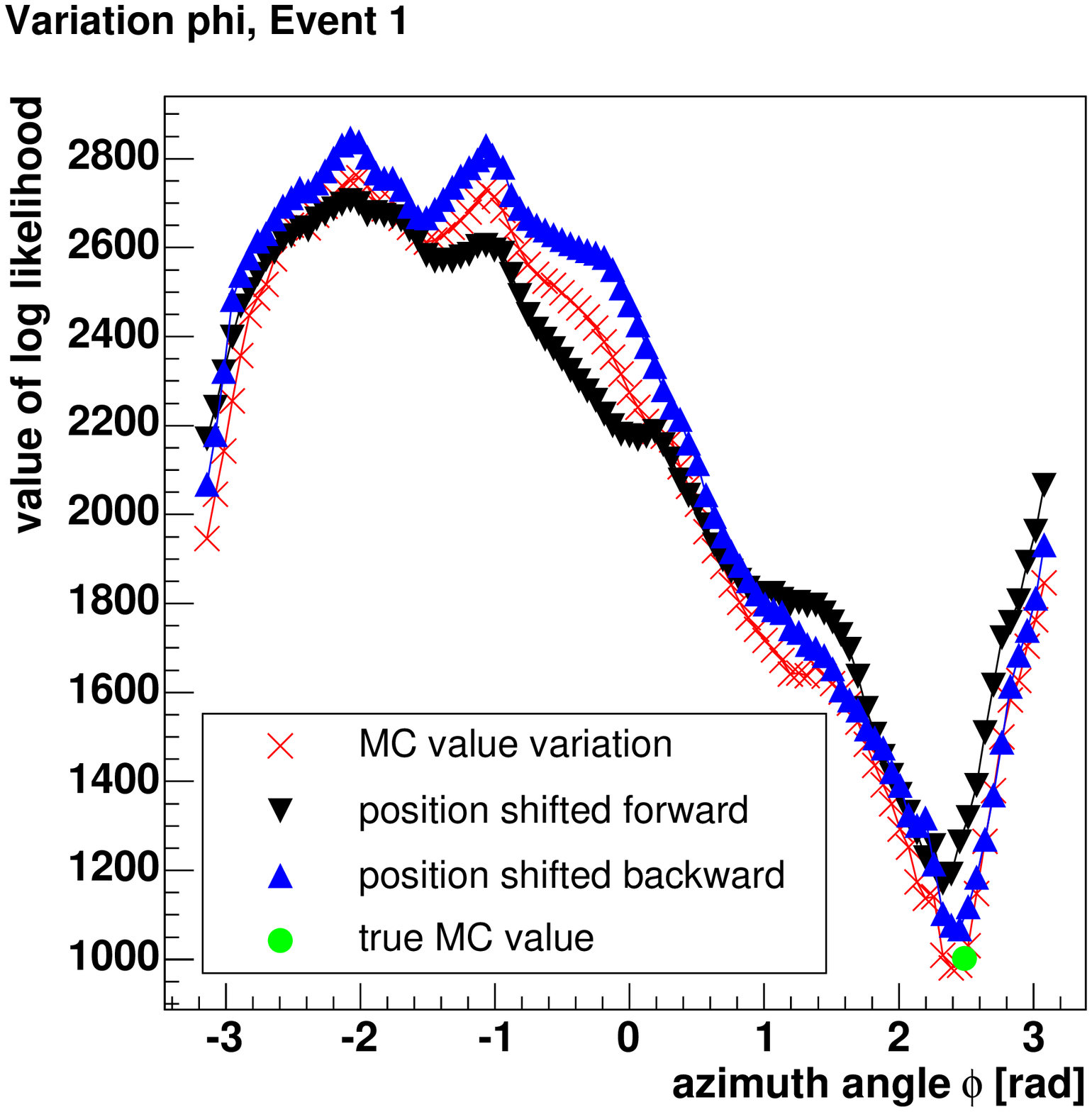}
\caption[Likelihood parameter space for positional variations]{Values of the likelihood function
  using the true position and two positions shifted along the shower axis, varying the
  photo-electron number (left), the zenith angle $\theta$ (middle) and the azimuth angle $\phi$
  (right), for a well-reconstructed event.} 
\label{fig:good_ev39113}
\end{figure}

\begin{figure}[h] \centering
\includegraphics[width=4.9cm]{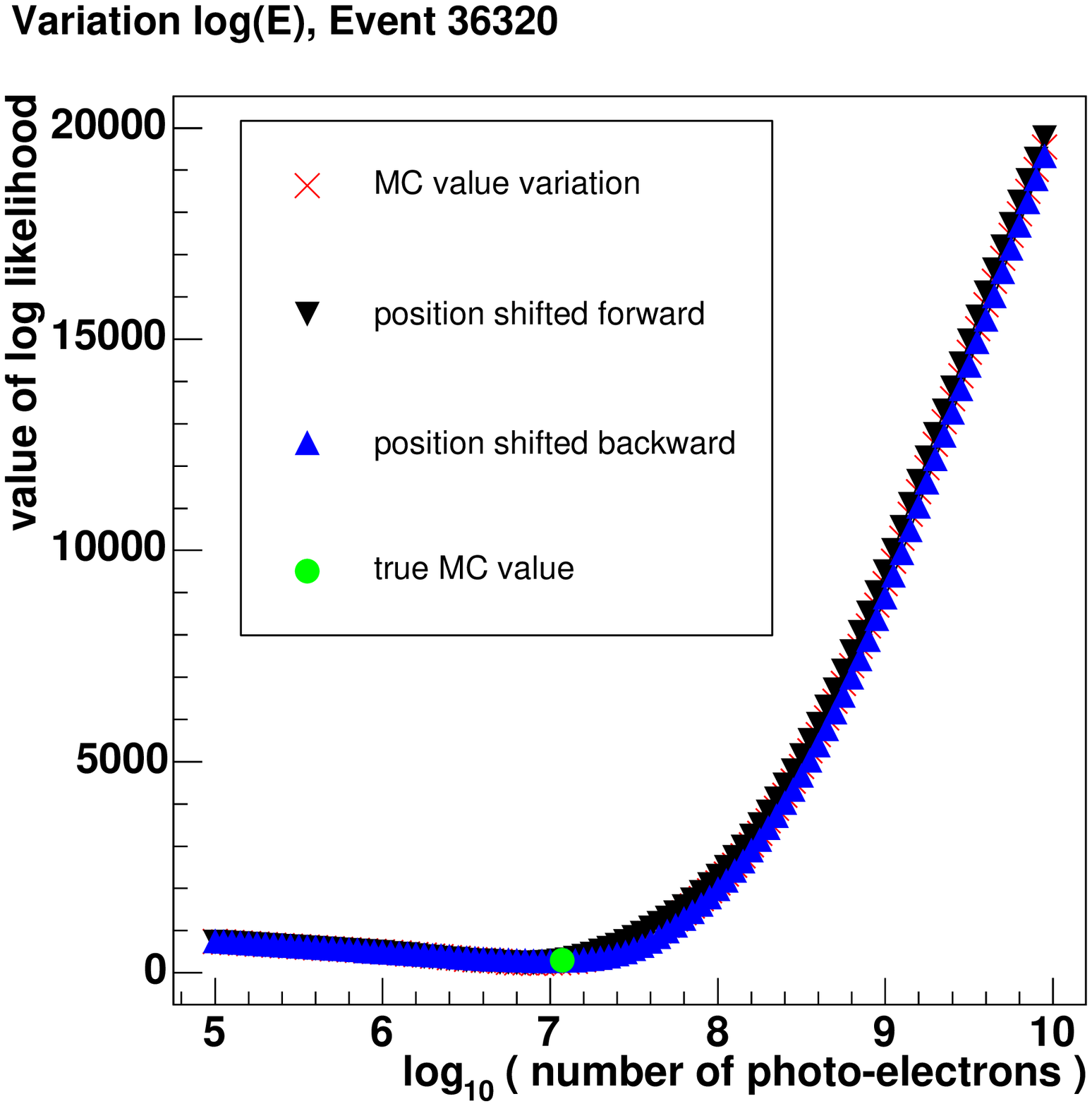}
\includegraphics[width=4.9cm]{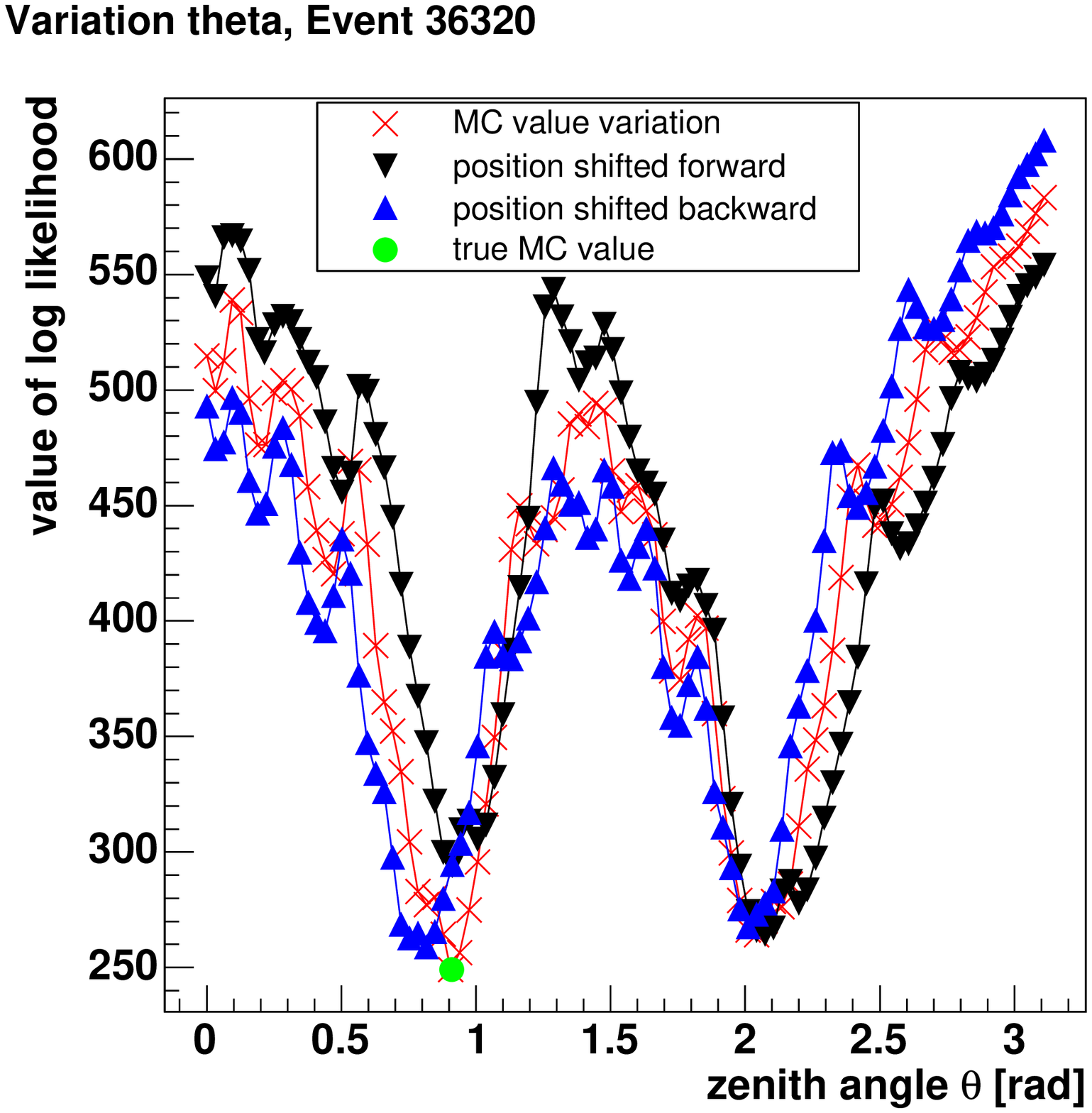}
\includegraphics[width=4.9cm]{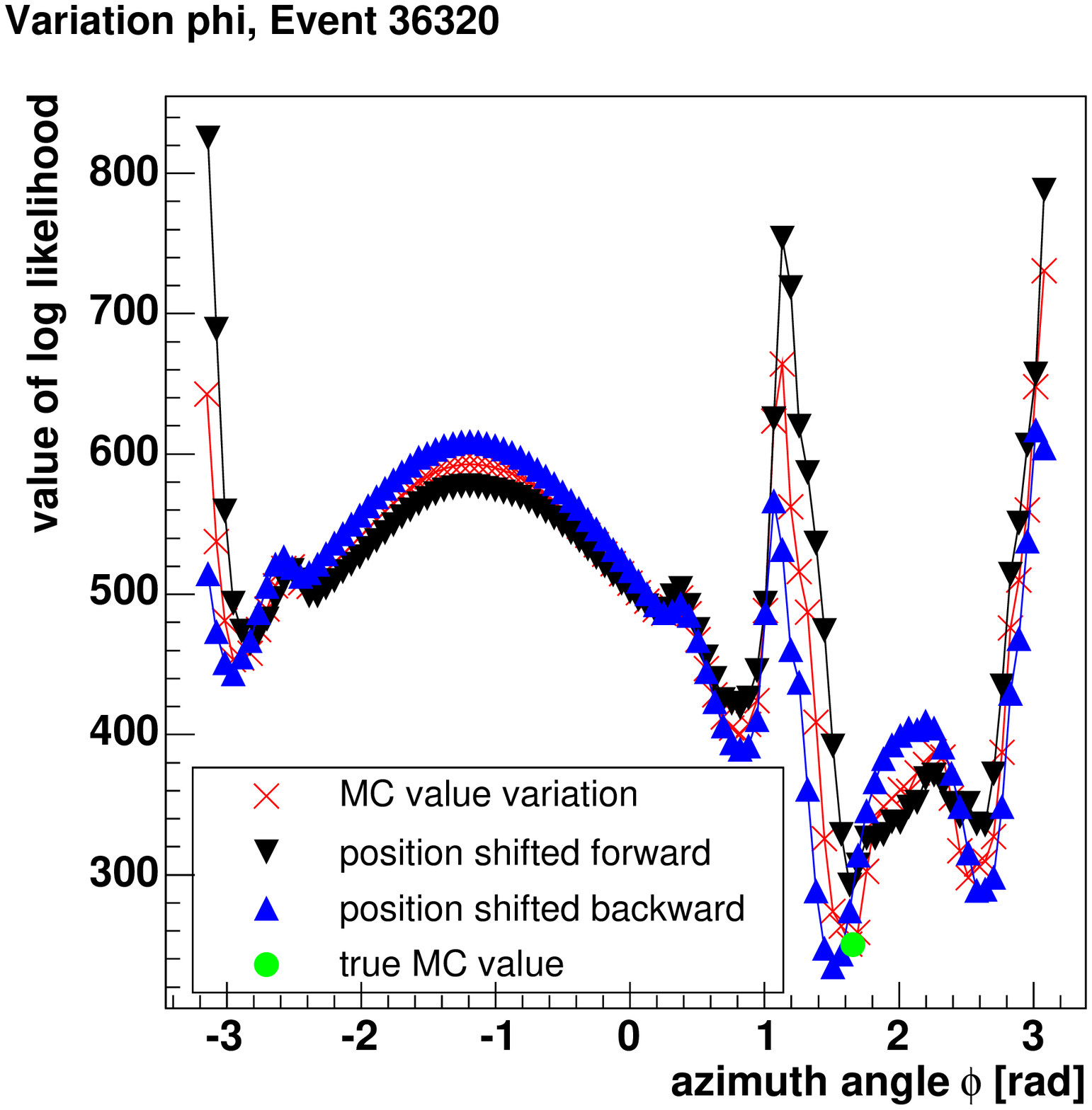}
\caption[Likelihood parameter space for positional variations]{Values of the likelihood function
  using the true position and two positions shifted along the shower axis, varying the
  photo-electron number (left), the zenith angle $\theta$ (middle) and the azimuth angle $\phi$
  (right), for a non-contained event.}  
\label{fig:bad_ev36320}
\end{figure}

The success of the direction and energy depends on the accuracy of the reconstructed position of the
shower. If the position is wrong by several ten metres (e.g.~inside the instrumented volume instead
of outside), the shape of the likelihood function can change significantly, and a correct
reconstruction of direction and energy is not possible. Even if the position is only wrong by a few
metres, the reconstruction can end up with totally different results. As examples, the well
reconstructed event shown above in Figure~\ref{fig:good_ev51490} is displayed again in
Figure~\ref{fig:good_ev39113}, for varied positions, as explained below. The event takes place
inside the instrumented volume (i.e.~it is {\it contained}). Figure~\ref{fig:bad_ev36320}, on the
other hand, displays an event whose interaction vertex is outside the instrumented volume, a so
called {\it non-contained event}. The latter yields a large angular error after reconstruction. For
both events, the likelihood function was calculated varying one of the three parameters, as above,
and keeping the others constant at their MC values. The true value of the respective parameter is
again marked by a green circle. The other two distributions show the values of the likelihood, again
varying one parameter and keeping the others at their MC values, but with a position that is 
{\it shifted 6\,m along the MC shower axis}, once forward in shower direction, and once backward,
against the shower direction. For the contained event, the shape of the likelihood varies slightly
with these shifts, but the position of the global minimum remains relatively stable. For the
non-contained event, however, there is a second minimum in the likelihood for the variation of
$\theta$, which becomes deeper than the one containing the MC value, if the position is shifted in
forward direction. Though the shower reconstruction algorithm describes the event correctly, which
can be seen from the fact that the MC values are located in the global minimum of the parameter
space, the ambiguities are large enough to create a second, deep local minimum which can become
global for slight shifts in the position. It should also be noted that the difference between the
two minima in $\theta$ is approximately twice the Cherenkov angle of $42^{\circ}$.

\subsubsection{Non-contained Events}

\begin{figure}[h] \centering
\includegraphics[width=4.9cm]{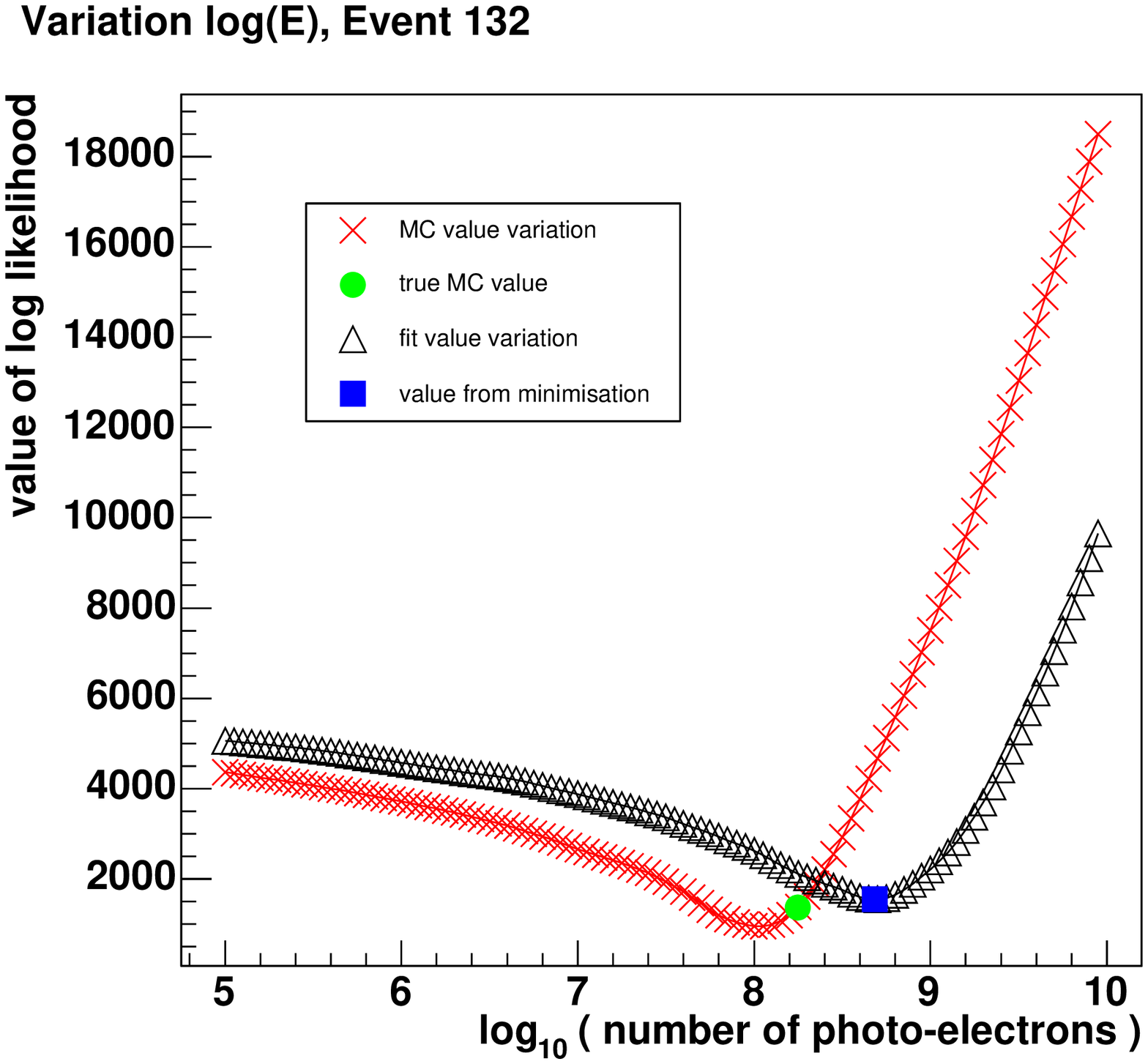}
\includegraphics[width=4.9cm]{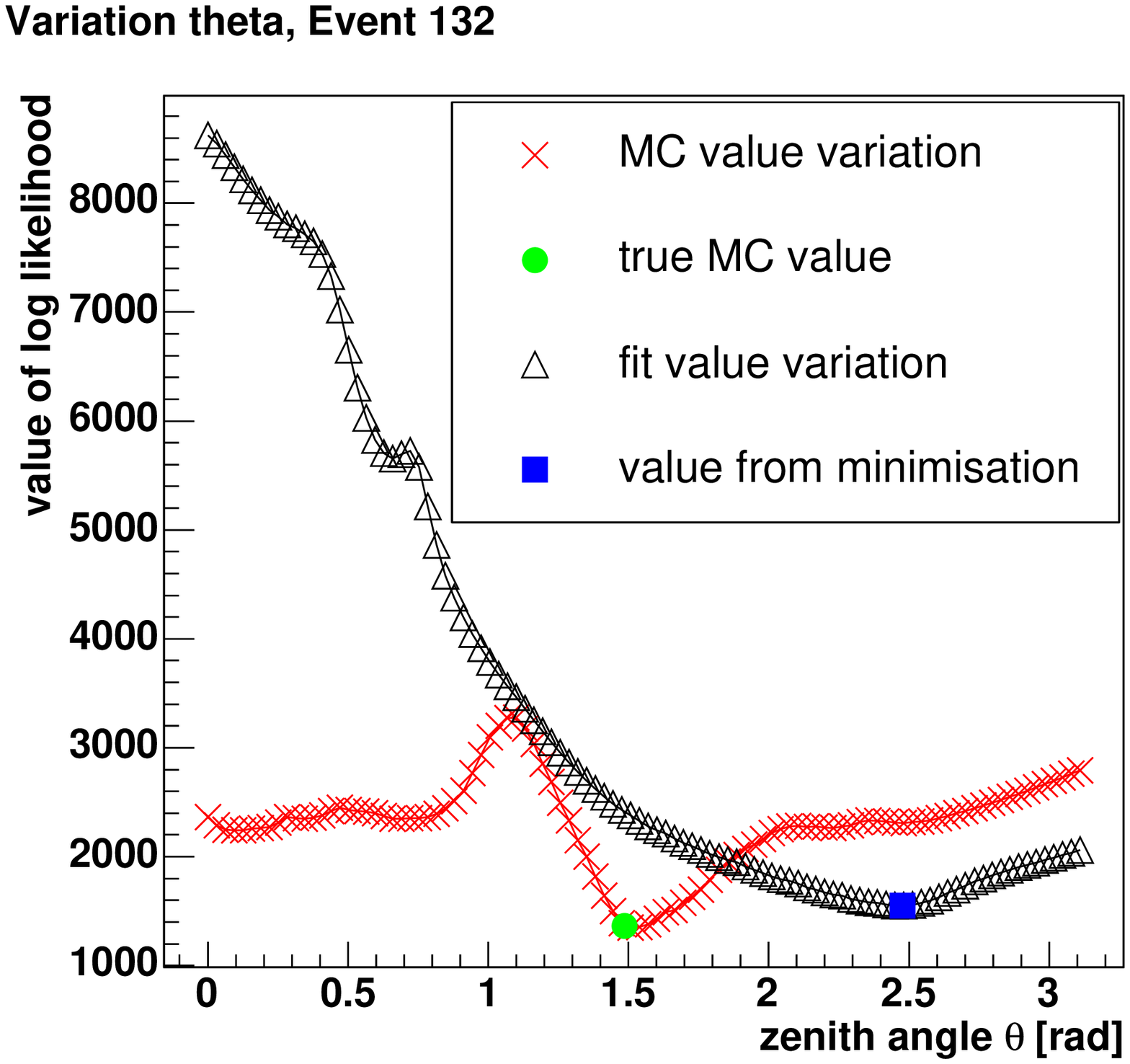}
\includegraphics[width=4.9cm]{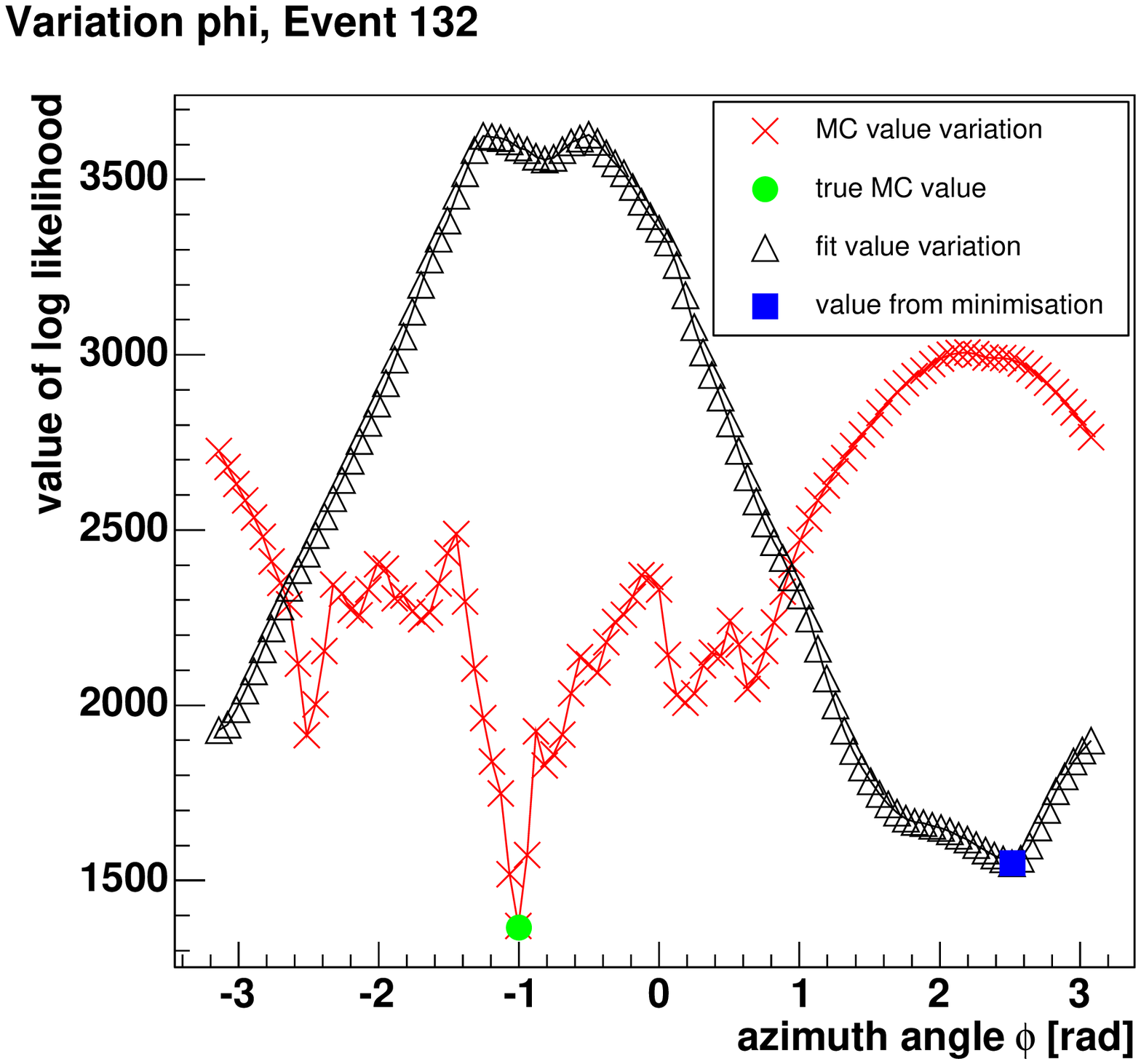}
\caption[Likelihood parameter space for an event with a wrong position]{Values of the likelihood
  function for the variation of the photo-electron number (left), the zenith angle $\theta$ (middle)
  and the azimuth angle $\phi$ (right), for a non-contained, poorly reconstructed event.}
\label{fig:bad_ev47344}
\end{figure}

Because of the directional characteristics of showers, non-contained events, i.e.~events whose
interaction vertex lies outside the instrumented volume, are very difficult to reconstruct, as
the fact that only one side of the detector measures incident photons induces a bias in the
signal. What adds to this is that for these events the position reconstruction 
as described in Section~\ref{sec:pos} is more likely to fail. In such a case it is almost impossible
to get good results from the reconstruction of direction and energy, as the fit then minimises the
likelihood using wrong assumptions. As an example for this, the shape of the likelihood function for
a badly reconstructed event is shown in Figure~\ref{fig:bad_ev47344}. For this event, the
interaction vertex lies outside the instrumented volume at the point (-82, 120, -160) in the ANTARES
coordinate system, where (0, 0, 0) lies right in the centre of the detector, and the detector edge
lies approximately at $(\pm 100,\pm 100,\pm 175)$. The position was reconstructed at (-56, 72, -136), {\it
  inside} the instrumented volume, with a large positional error. For the reconstructed position,
the measured hit amplitudes in the individual OMs do not match the true direction of the shower. The
azimuth angle $\phi$ is flipped to the opposite direction. As the $z$ coordinate is reconstructed
24\,m too large, the shower seems to be downgoing (i.e.~$\theta > \pi/2$); consequently, the zenith
angle $\theta$ is reconstructed too large. Because of the worse acceptance of the OMs for downgoing
photons, also the shower energy is overestimated, as can be seen from the left plot of
Figure~\ref{fig:bad_ev47344}. 

\subsubsection{Events Close to the Detector Edge}

\begin{figure} \centering
\begin{minipage}{6cm}
\epsfig{figure=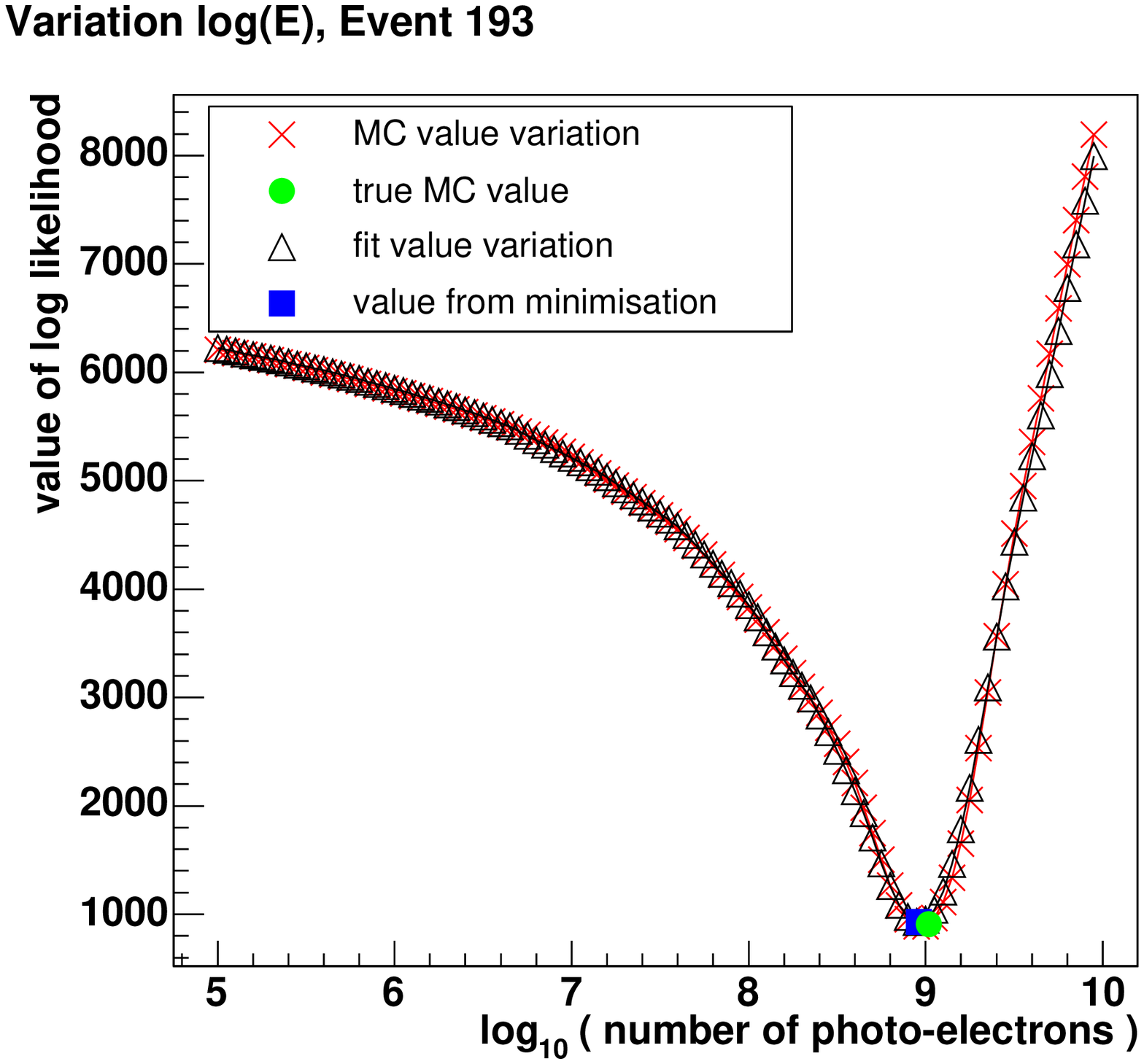, width=4.9cm}
\epsfig{figure=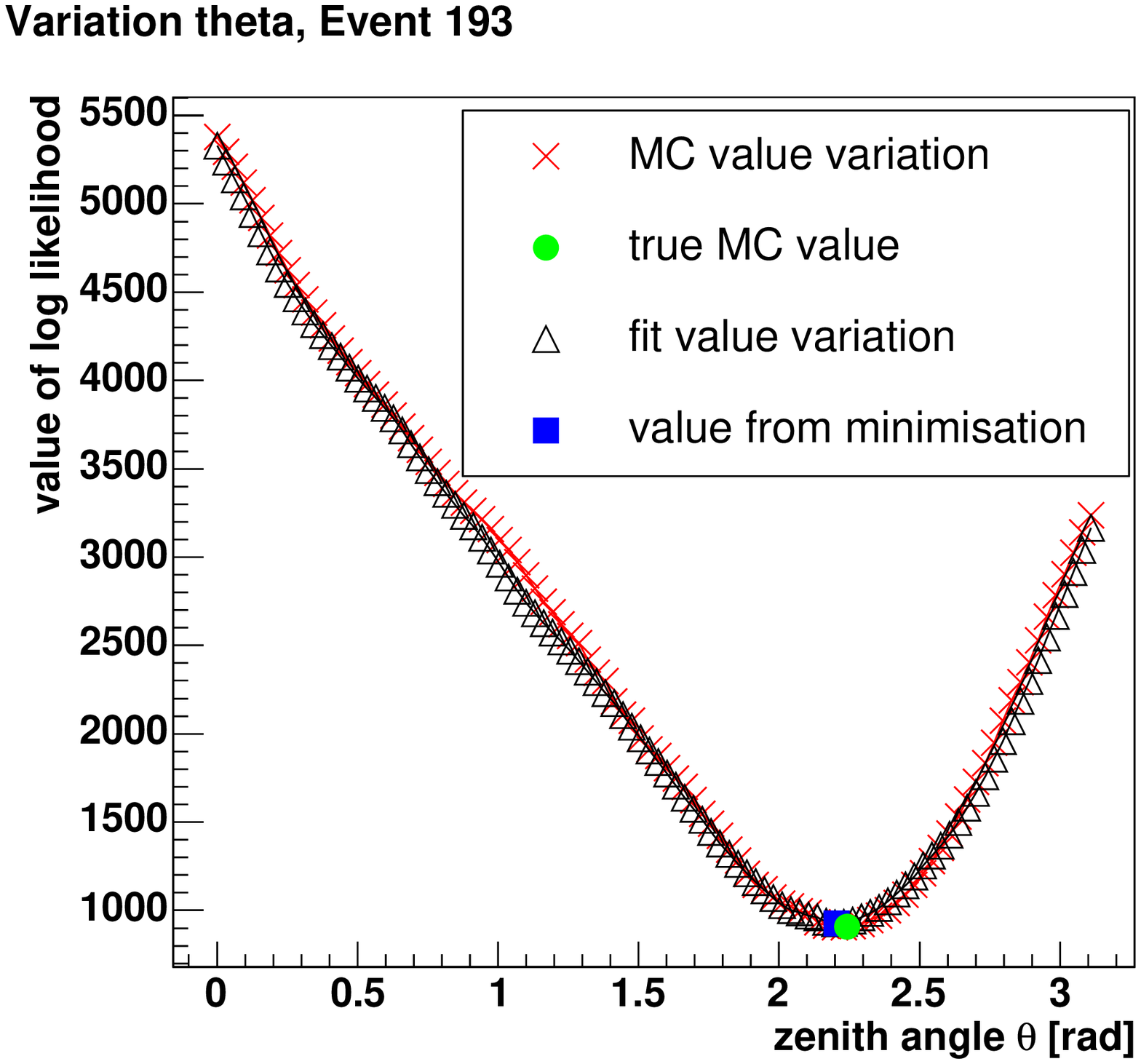, width=4.9cm}
\epsfig{figure=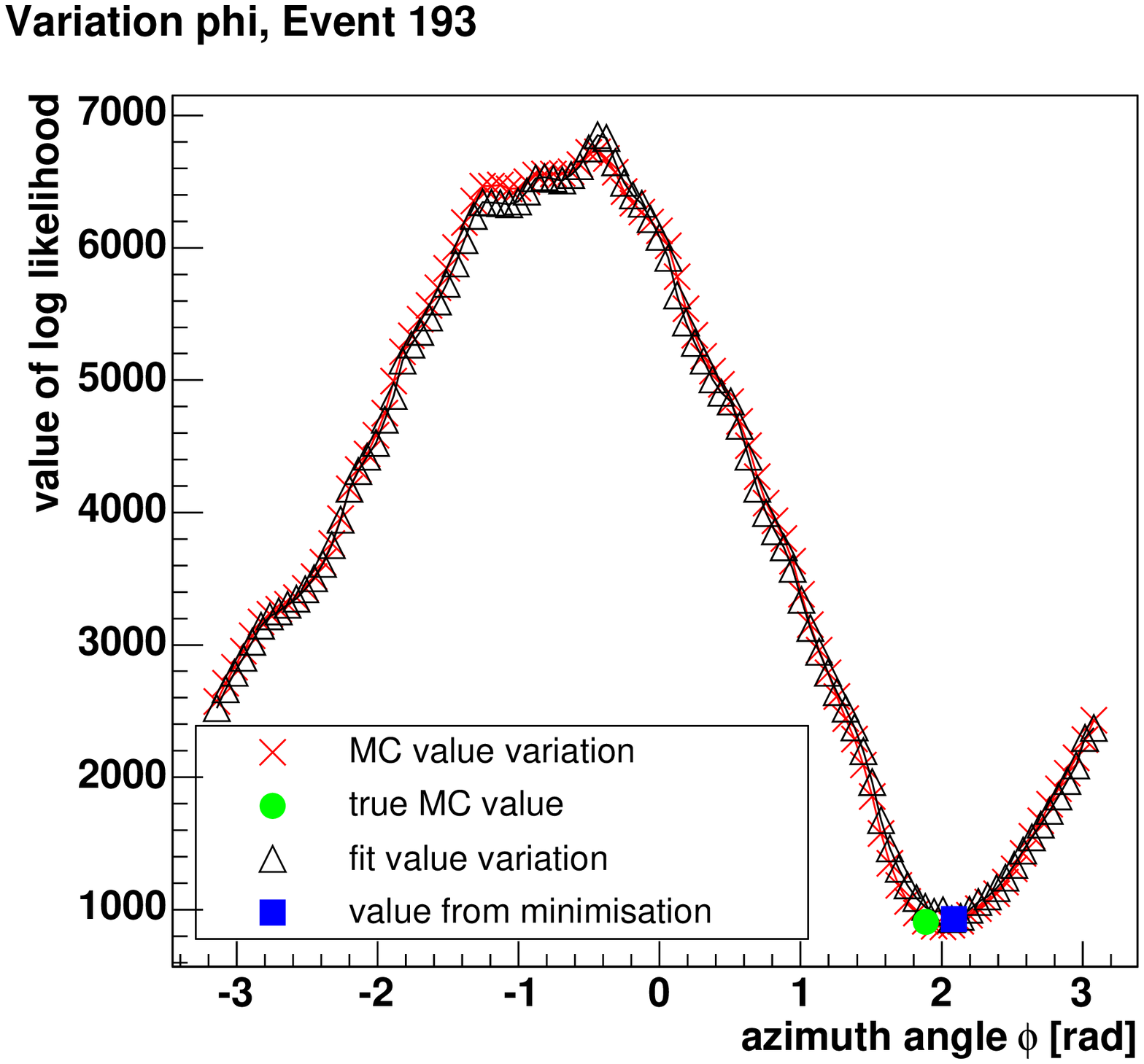, width=4.9cm}
\caption[Likelihood parameter space for an edge event]{Values of the likelihood function for the
  variation of the photo-electron number (top), the zenith angle $\theta$ (middle) and the azimuth
  angle $\phi$ (bottom), for an edge event.}
\label{fig:bad_ev2785}
\end{minipage}
\hspace{5mm}
\begin{minipage}{7.3cm}
\epsfig{width=6cm,figure=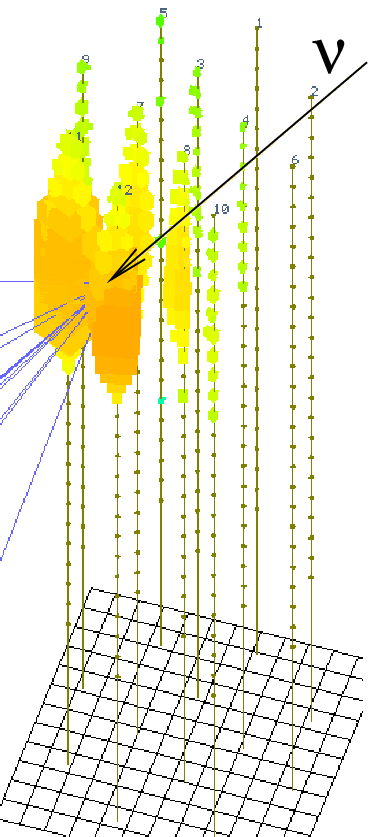}
\caption[Edge event seen in the ANTARES event display]{Edge event from Figure~\ref{fig:bad_ev2785}
  as seen in the ANTARES event display. Only the hits passing the filter conditions are shown.} 
\label{fig:bad_event2785}
\end{minipage}
\end{figure}

Neutrinos which produce showers at the edge of the detector, such that the largest part of the
signal is measured by a few strings at one side of the detector, are not as problematic as
non-contained events. However, also these events cannot always be reconstructed well. The reason for
this is that the likelihood description sometimes lacks the necessary information for a successful
reconstruction, because only one side of the shower or even less is visible for the
detector. However, the fact that the shower is happening close to the 
detector edge does not necessarily mean that this event will be badly reconstructed. In
fact, quite a number of such events are still reconstructed well, so that it is difficult to decide
from the reconstructed position or other criteria if the result of a reconstruction is unreliable
and the event should be discarded. \\
An example for the variation of the likelihood function of a well reconstructed edge event
are shown in Figure~\ref{fig:bad_ev2785}. The event display for this event is shown in
Figure~\ref{fig:bad_event2785}. As before, again the values of the likelihood function is shown in 
Figure~\ref{fig:bad_ev2785}, varying one of the parameters, and keeping the others either at their
MC or at their reconstructed values.  One can see from the upmost plot of
Figure~\ref{fig:bad_ev2785} that the energy of this event is relatively large; this is the reason
for the large detector signal, despite the fact that the neutrino is pointing to the outside the
instrumented volume. The shapes of the likelihood function using the MC or the reconstructed values
are very similar. The distributions are characteristical for an edge event: They display only one
minimum, as there is only one possible orientation of the shower with respect to the measured hit
pattern; and the minimum is relatively broad for $\theta$ and $\phi$, because the information
obtained from the measurement is not as precise as would be the case for an interaction in the
middle of the detector.

\subsection{Numerical Stability Check}
\begin{figure}[h] \centering
\includegraphics[width=4.9cm]{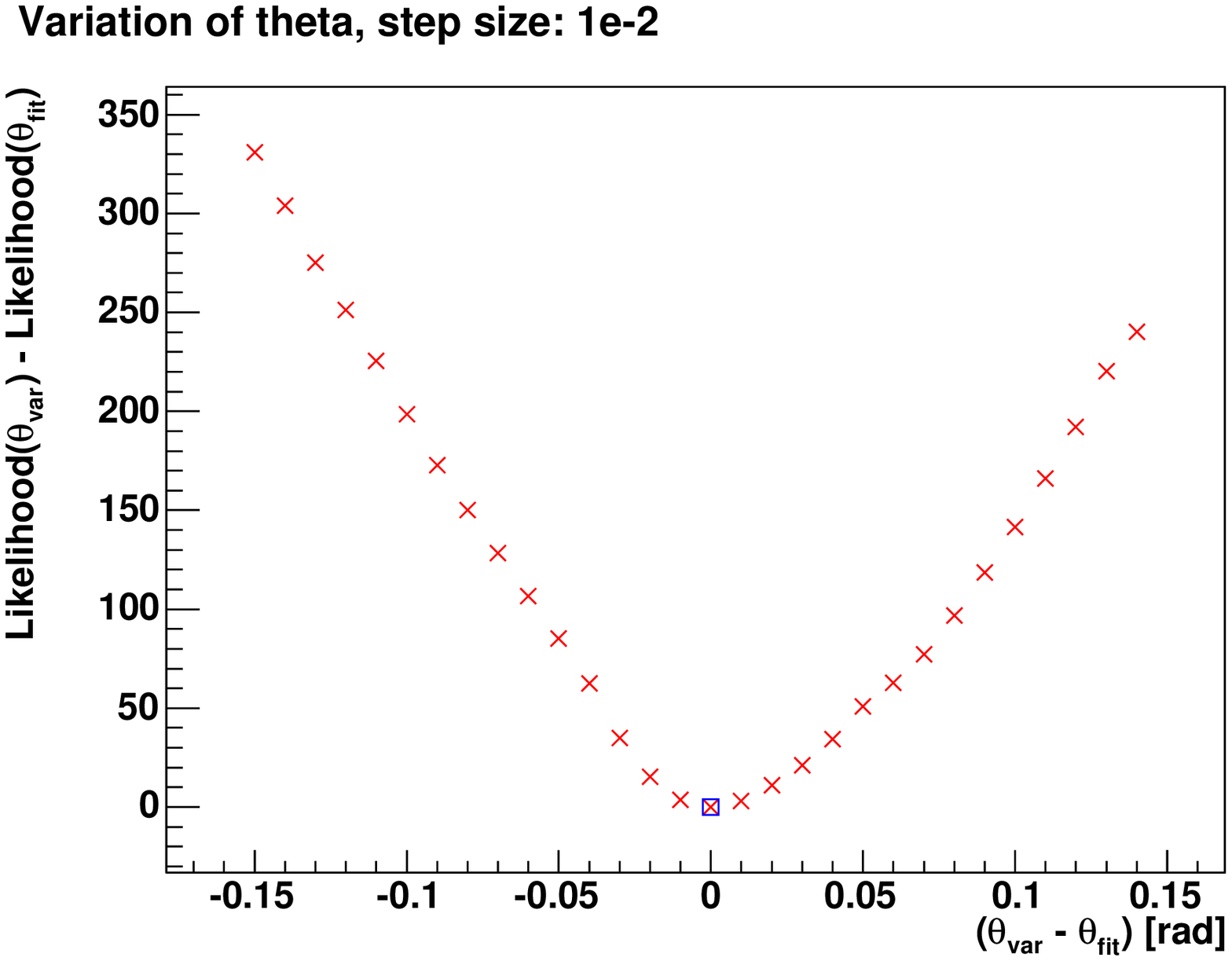}
\includegraphics[width=4.9cm]{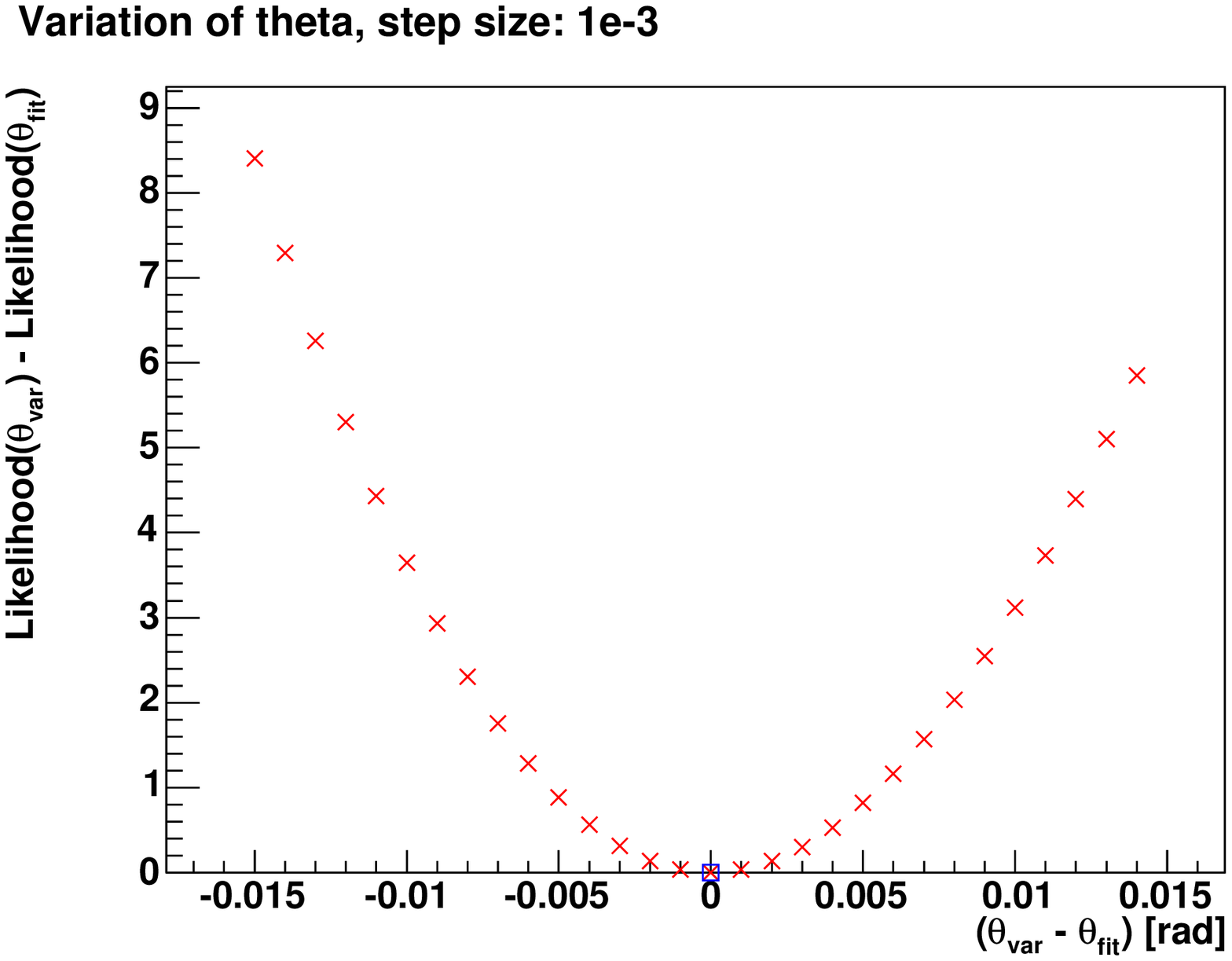}
\includegraphics[width=4.9cm]{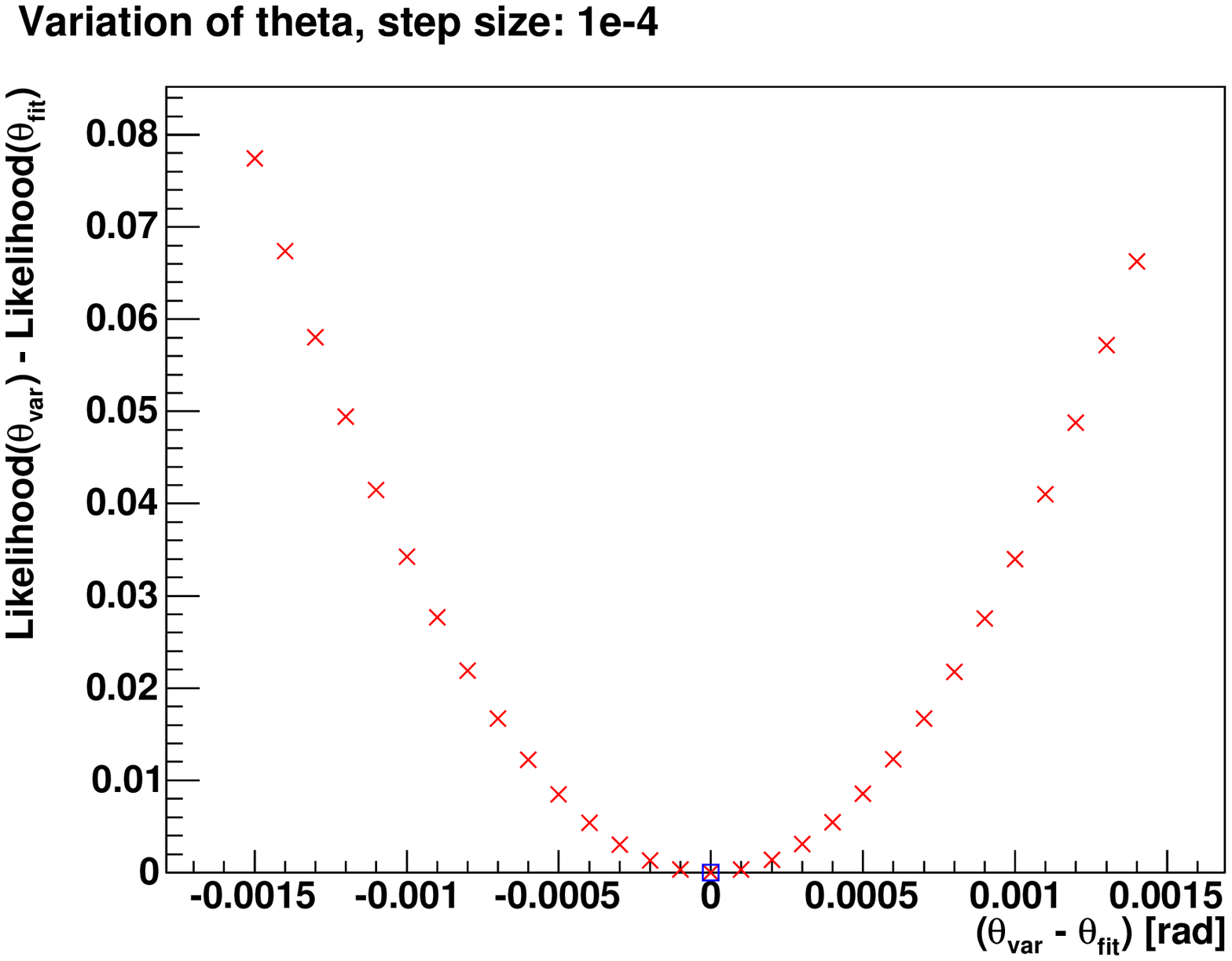}
\includegraphics[width=4.9cm]{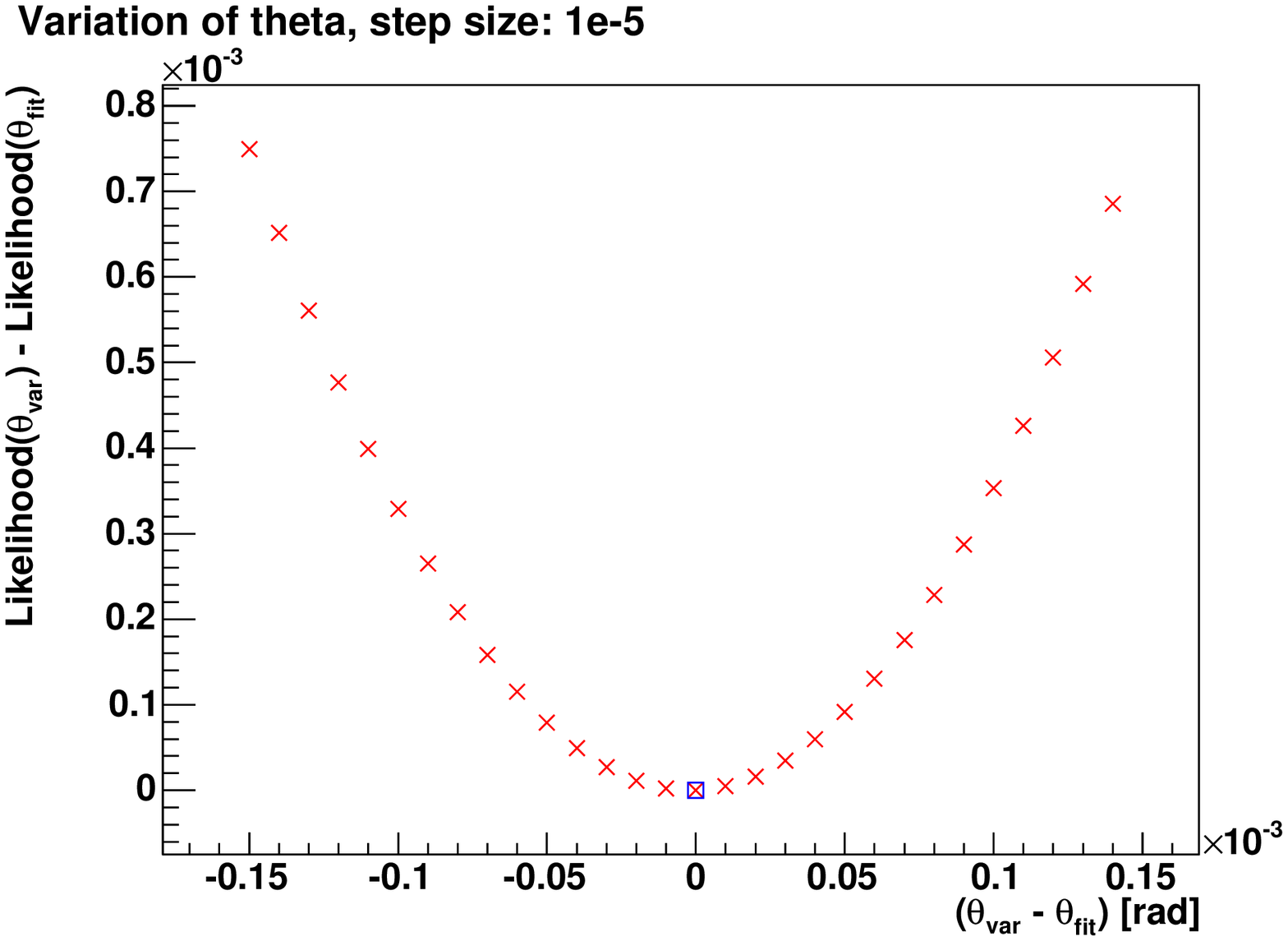}
\includegraphics[width=4.9cm]{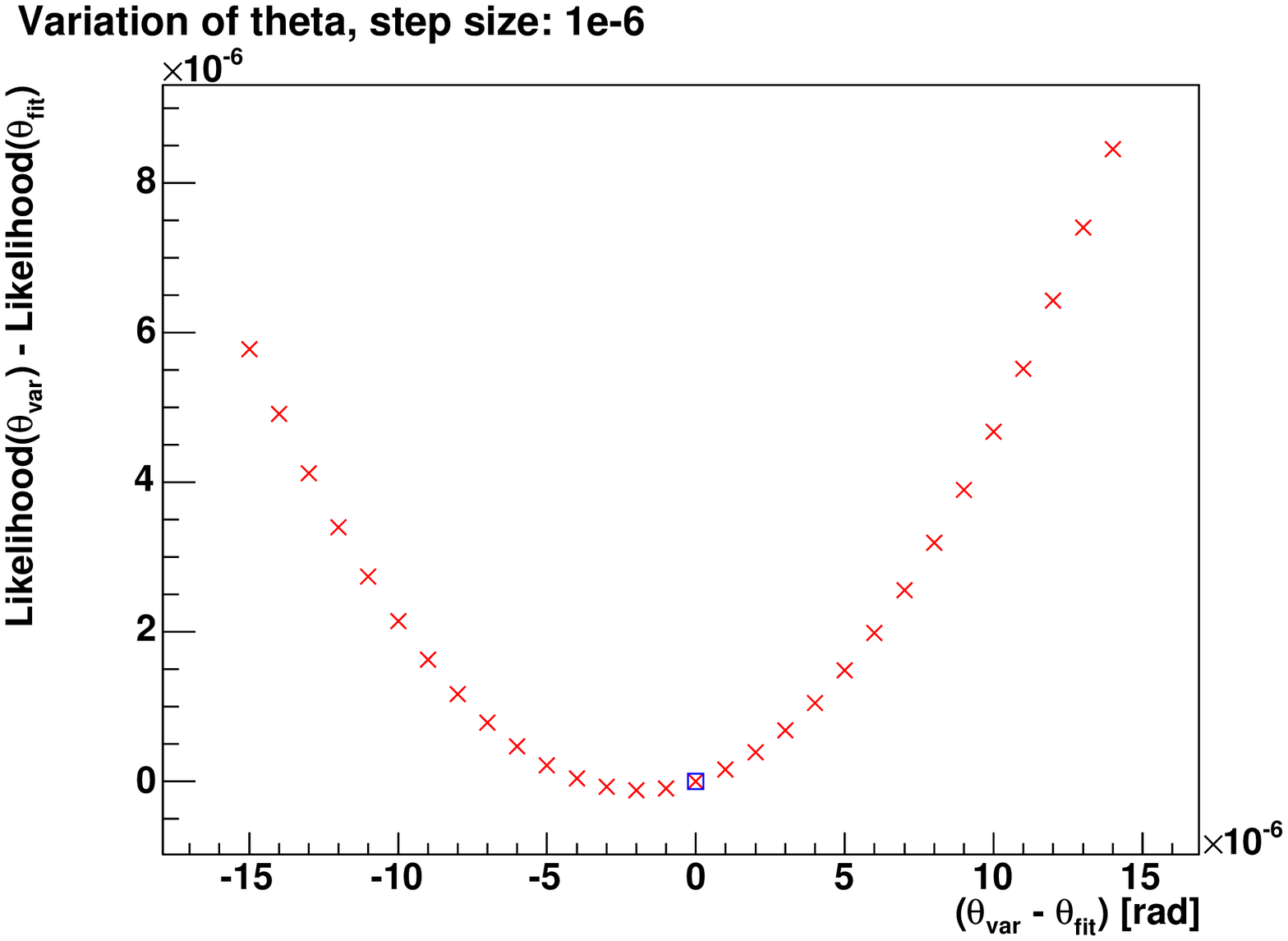}
\includegraphics[width=4.9cm]{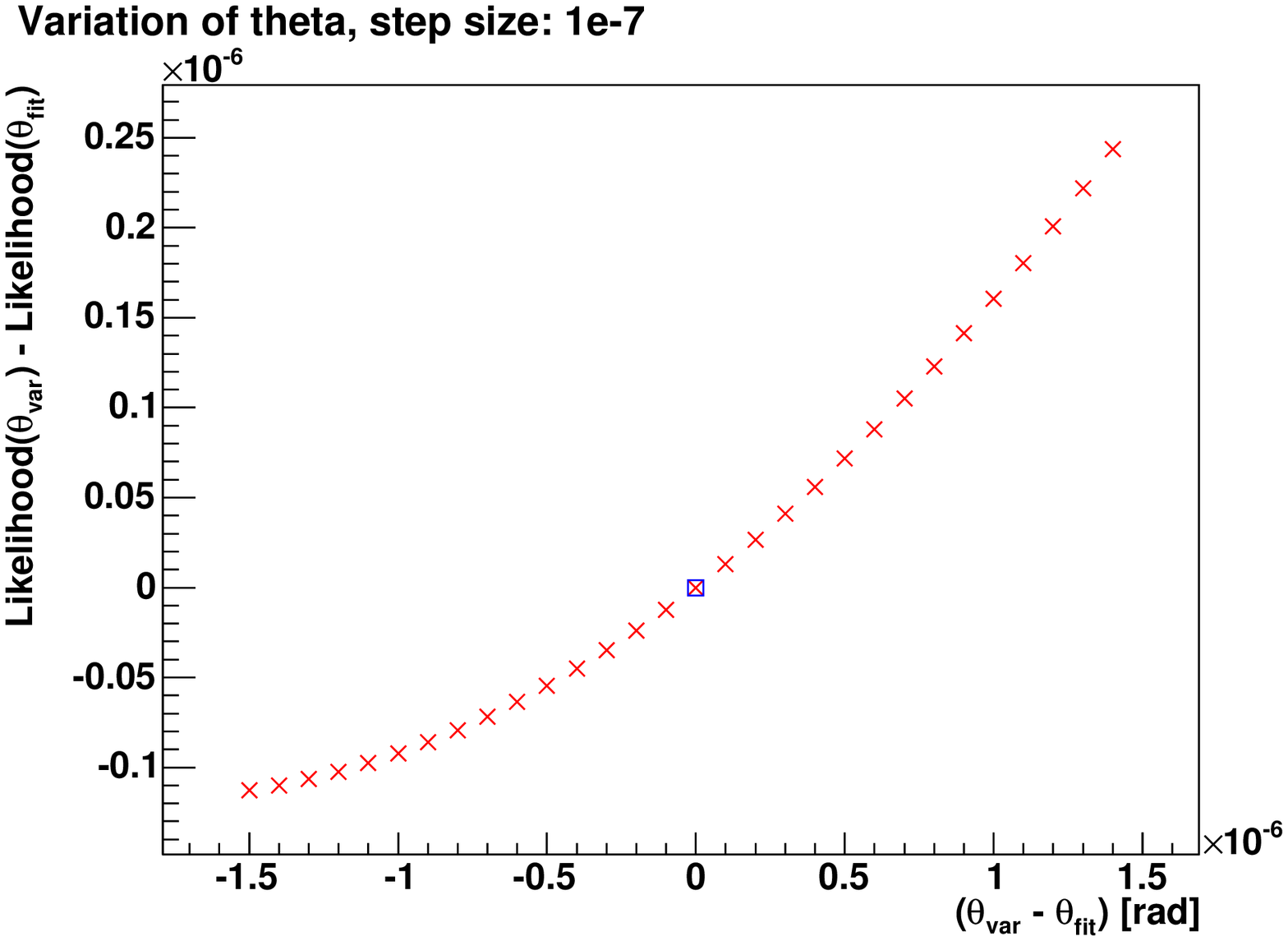}
\includegraphics[width=4.9cm]{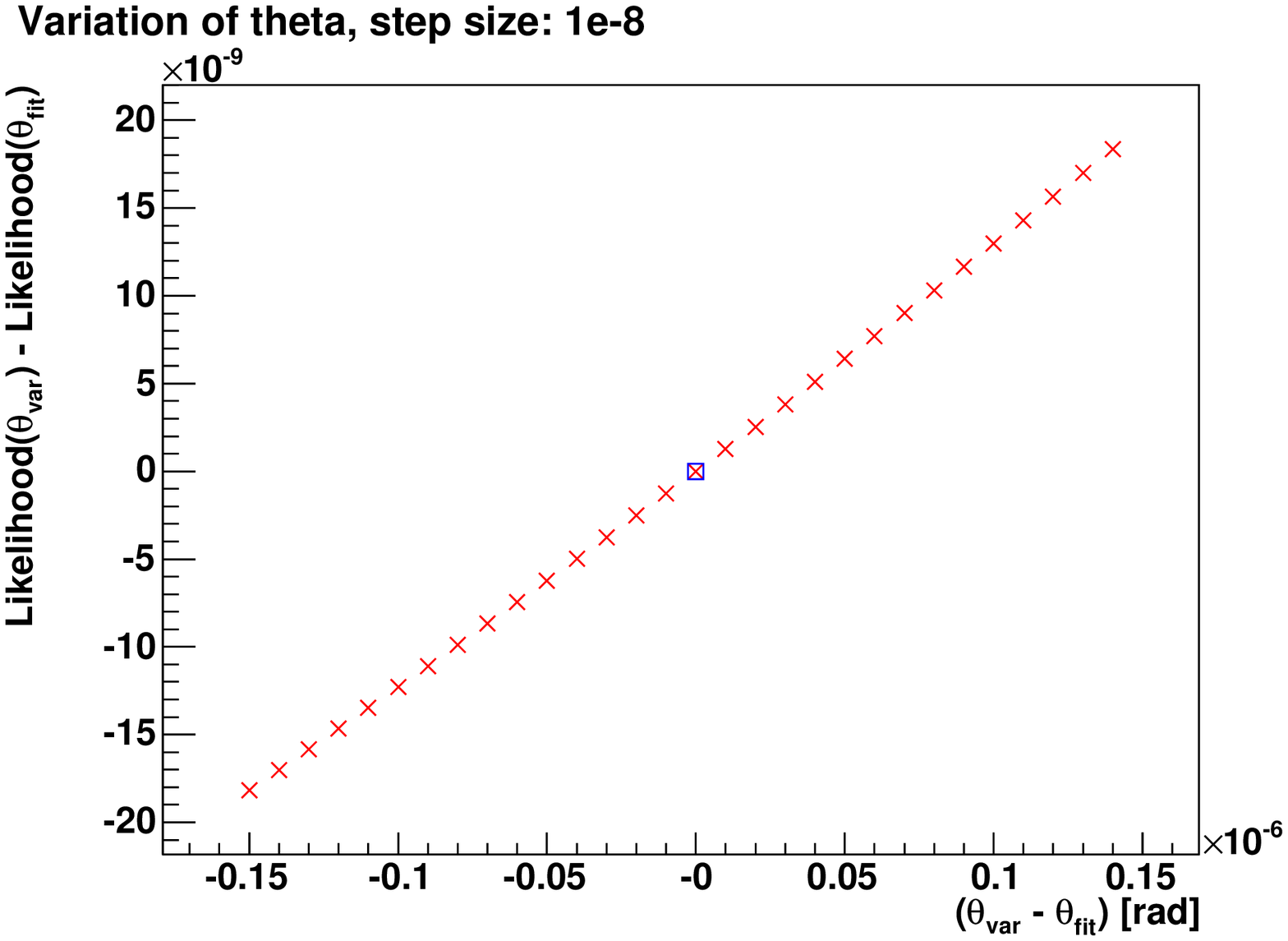}
\includegraphics[width=4.9cm]{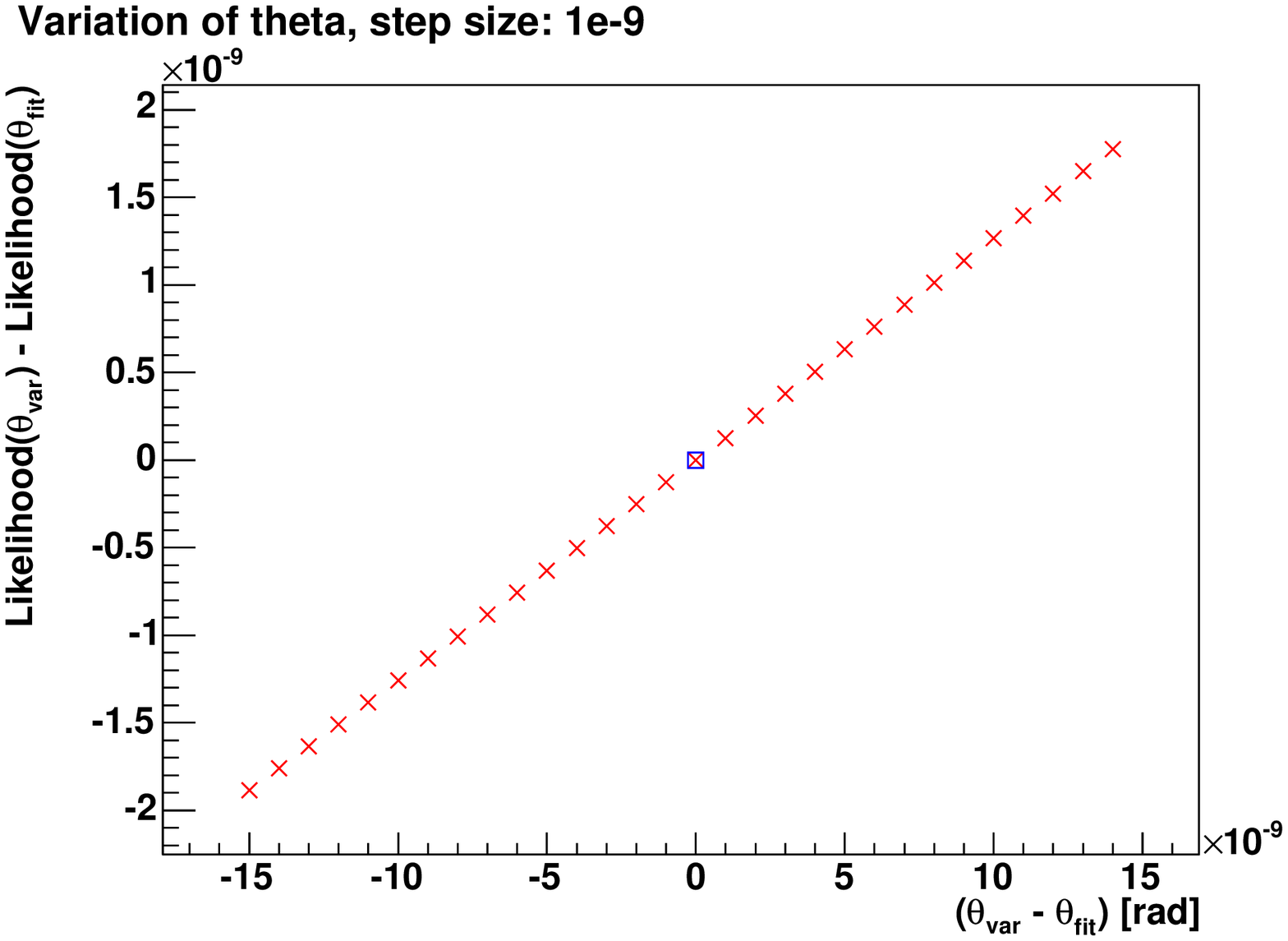}
\includegraphics[width=4.9cm]{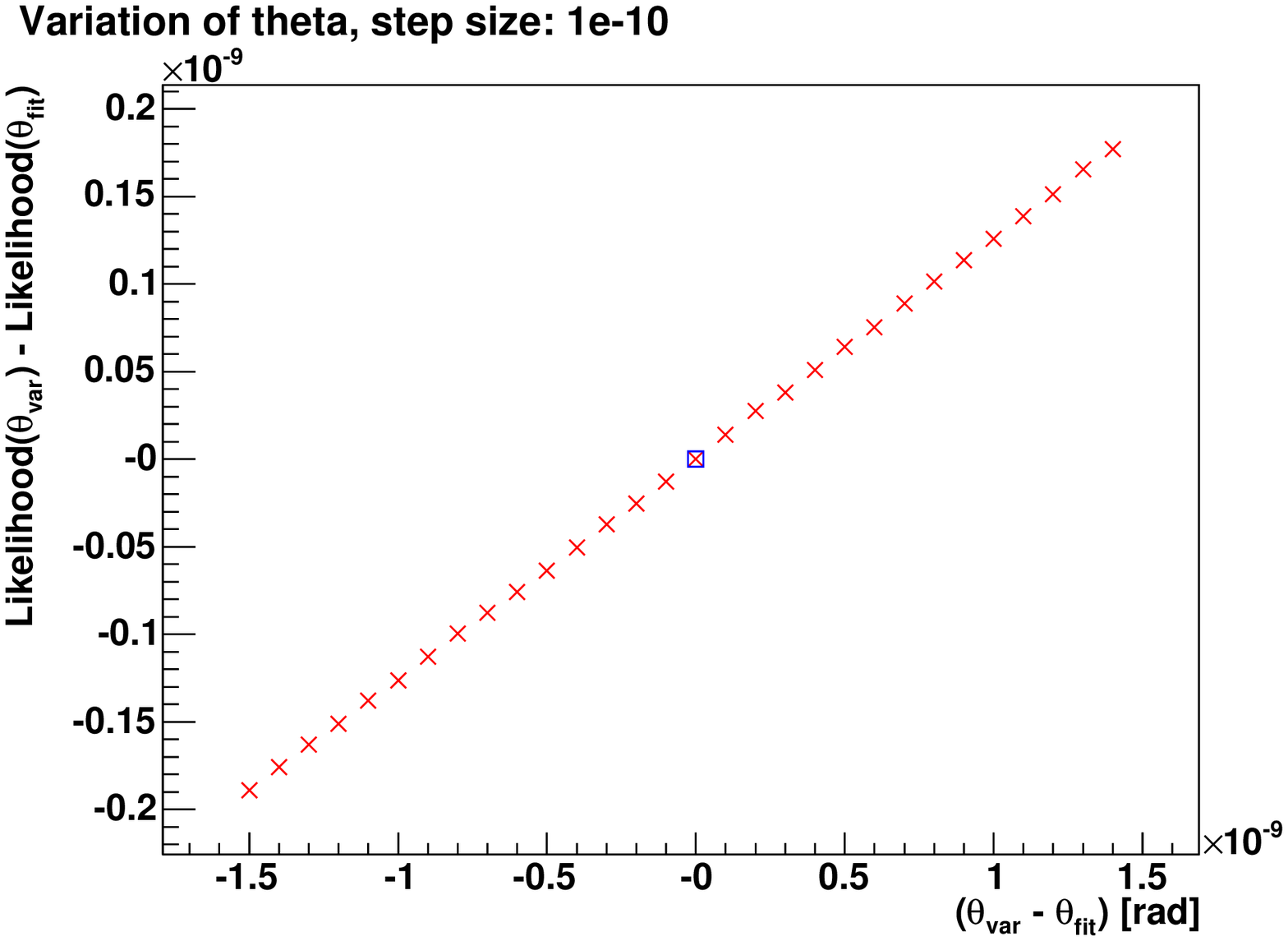}
\includegraphics[width=4.9cm]{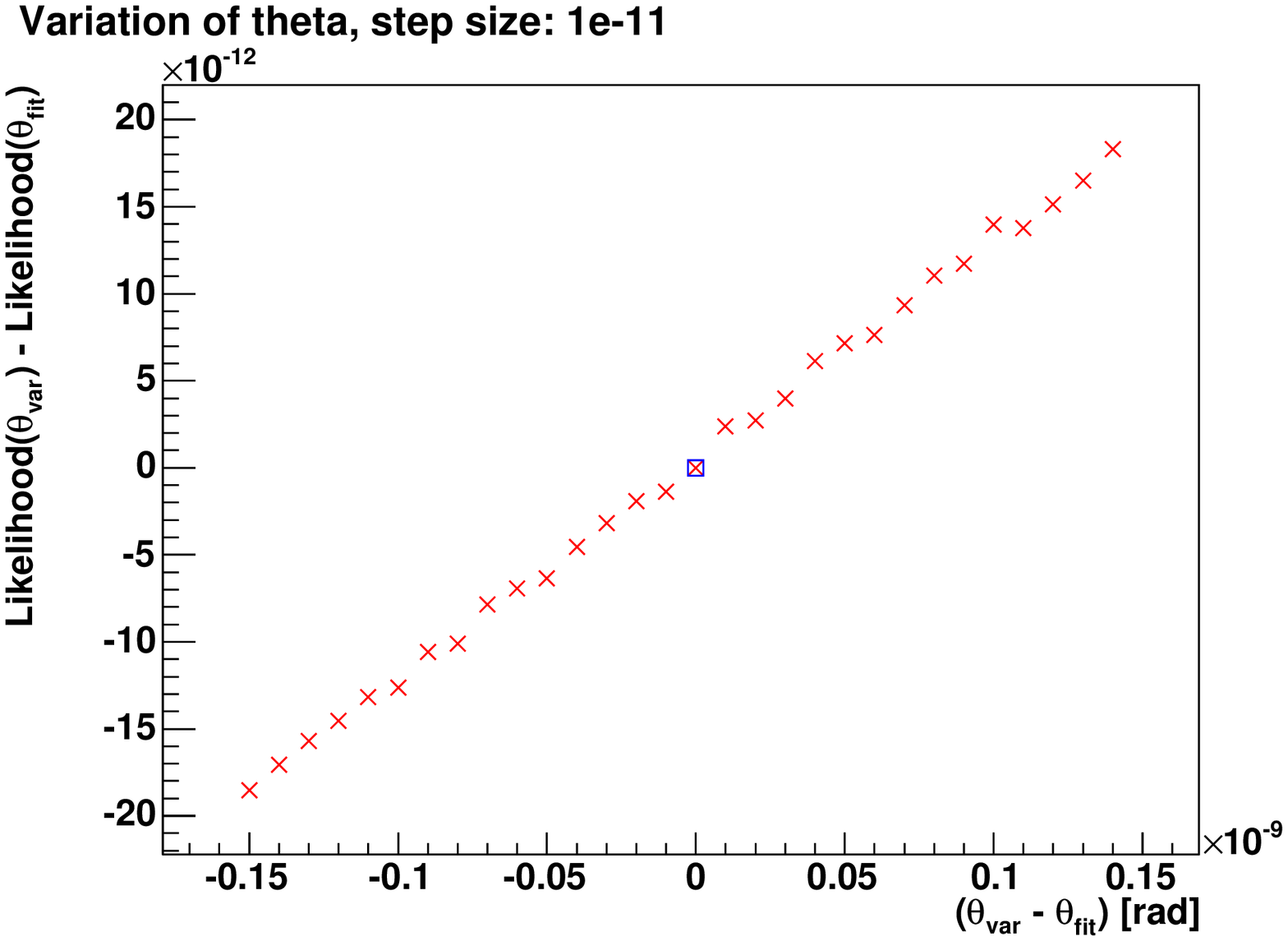}
\includegraphics[width=4.9cm]{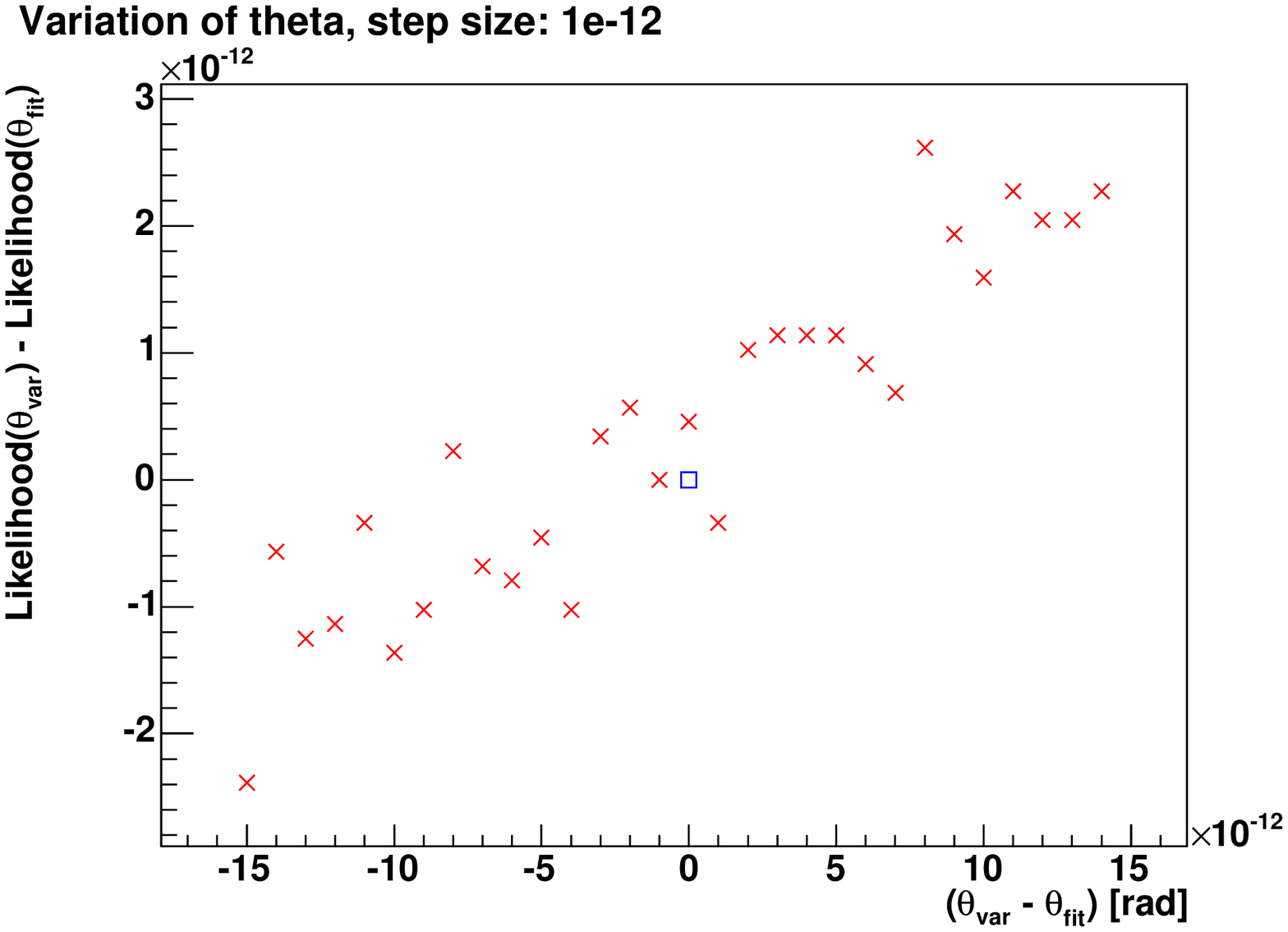}
\includegraphics[width=4.9cm]{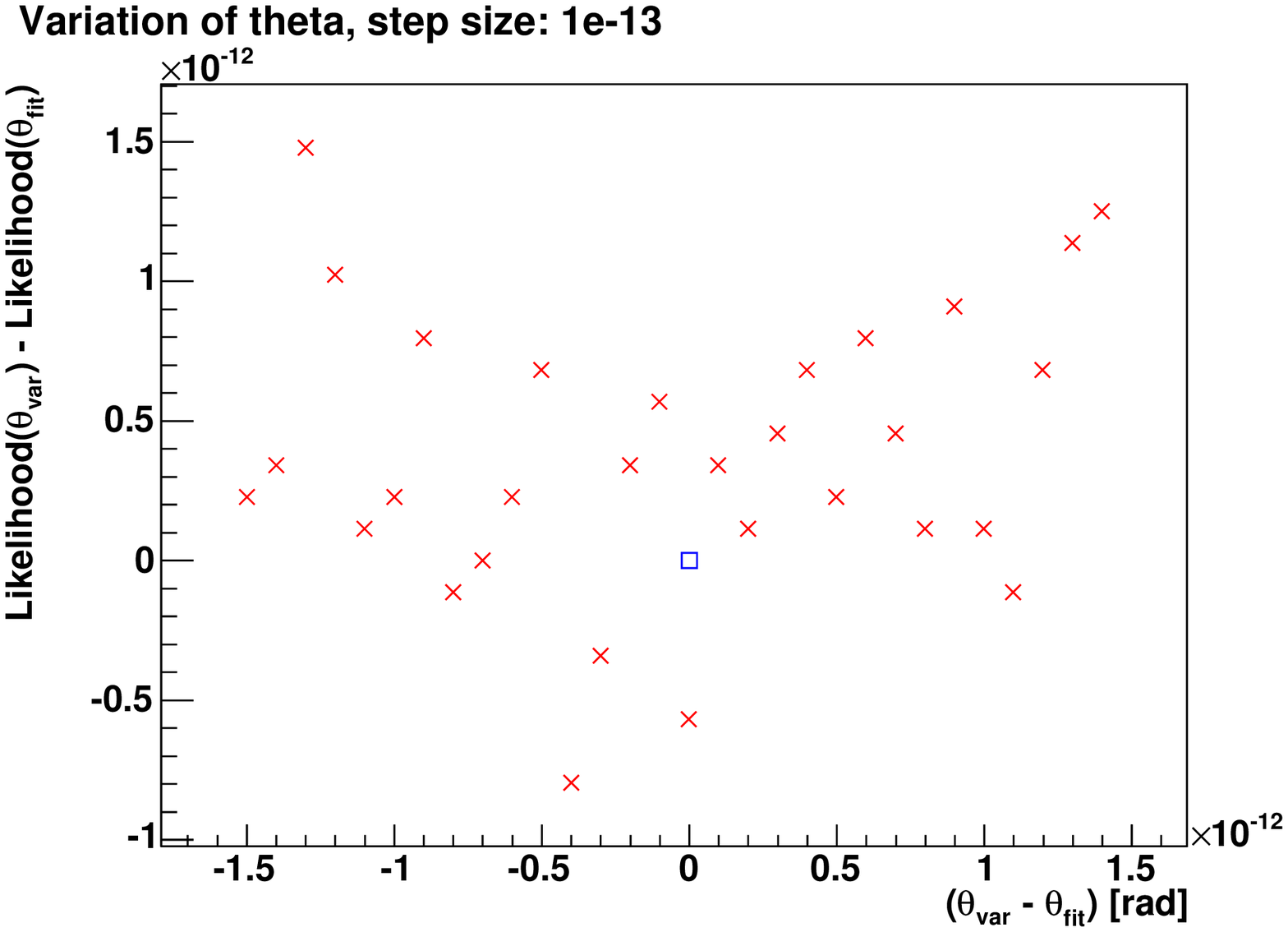}
\caption[Likelihood for variation of $\theta$]{Values of the likelihood function for variations of
  $\theta$ around the fit value.}
\label{fig:var_theta}
\end{figure}

\begin{figure}[h] \centering
\includegraphics[width=4.9cm]{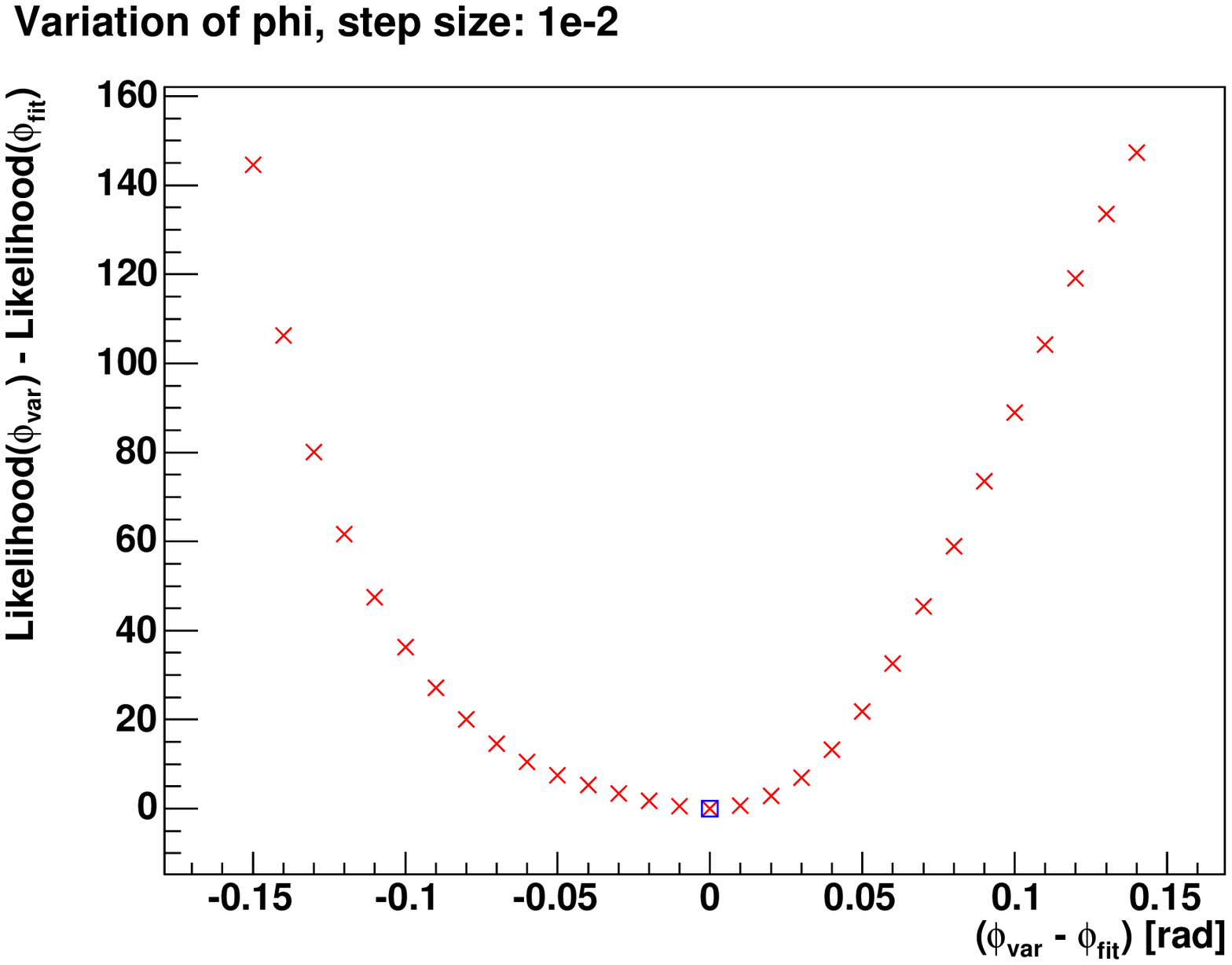}
\includegraphics[width=4.9cm]{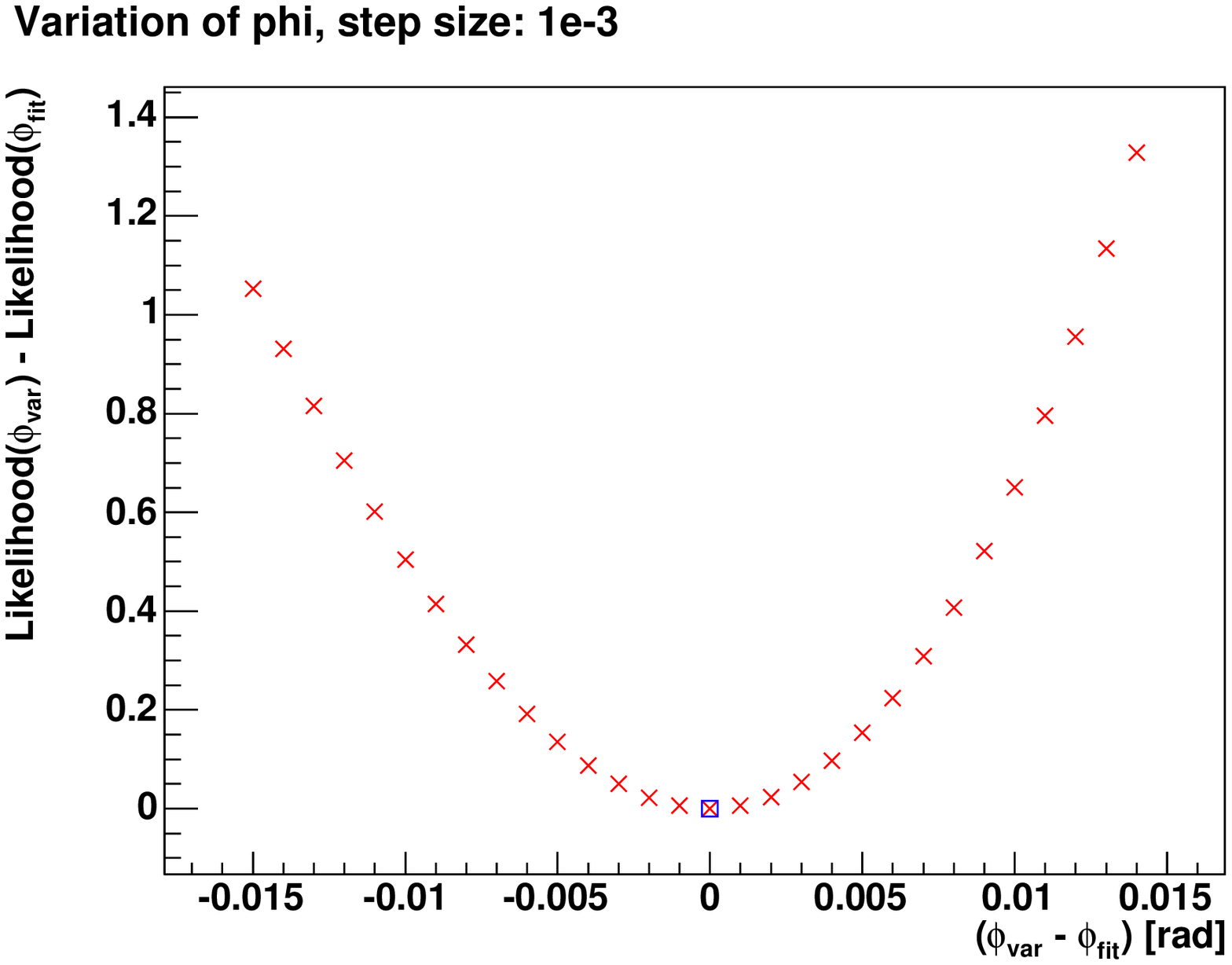}
\includegraphics[width=4.9cm]{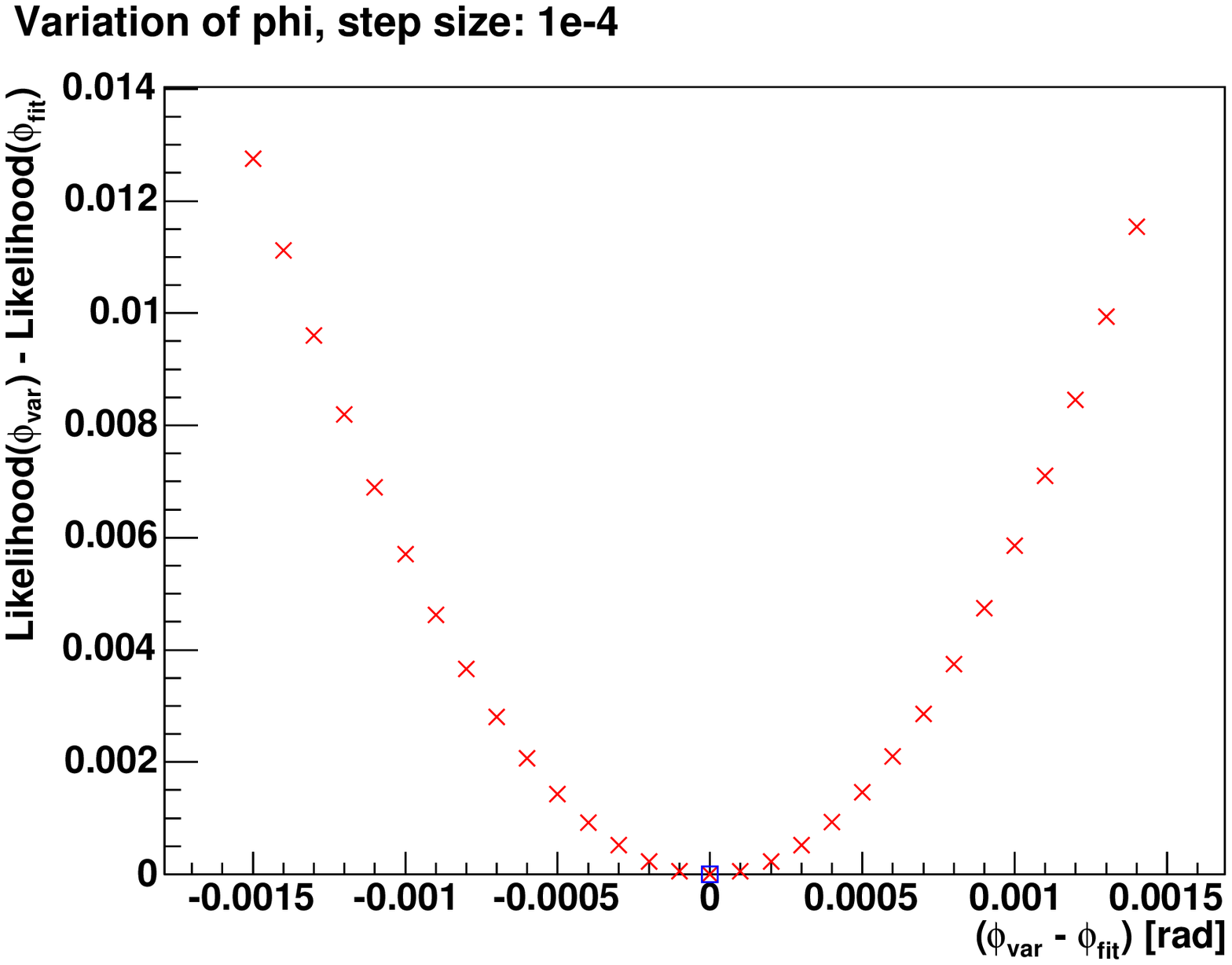}
\includegraphics[width=4.9cm]{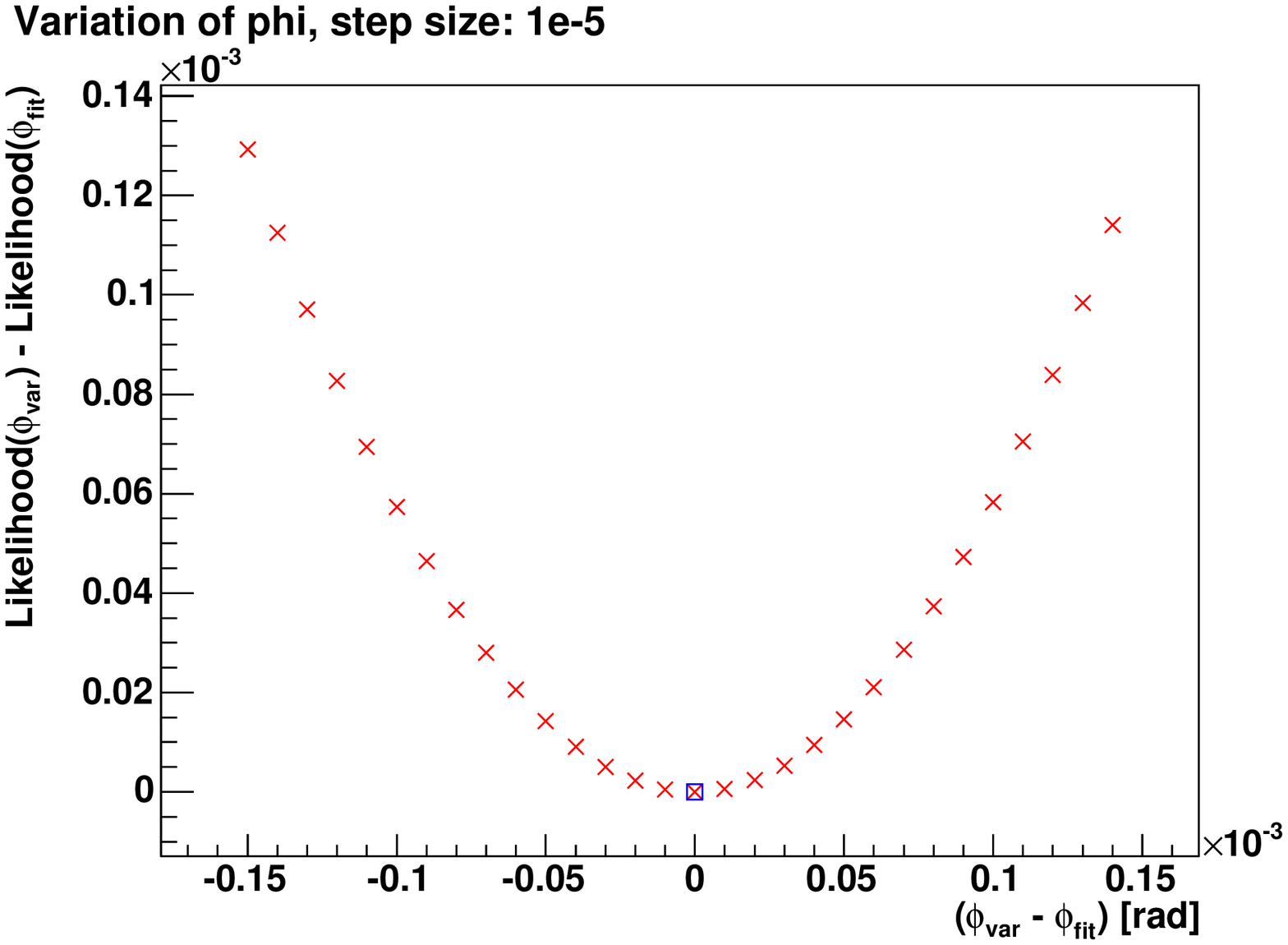}
\includegraphics[width=4.9cm]{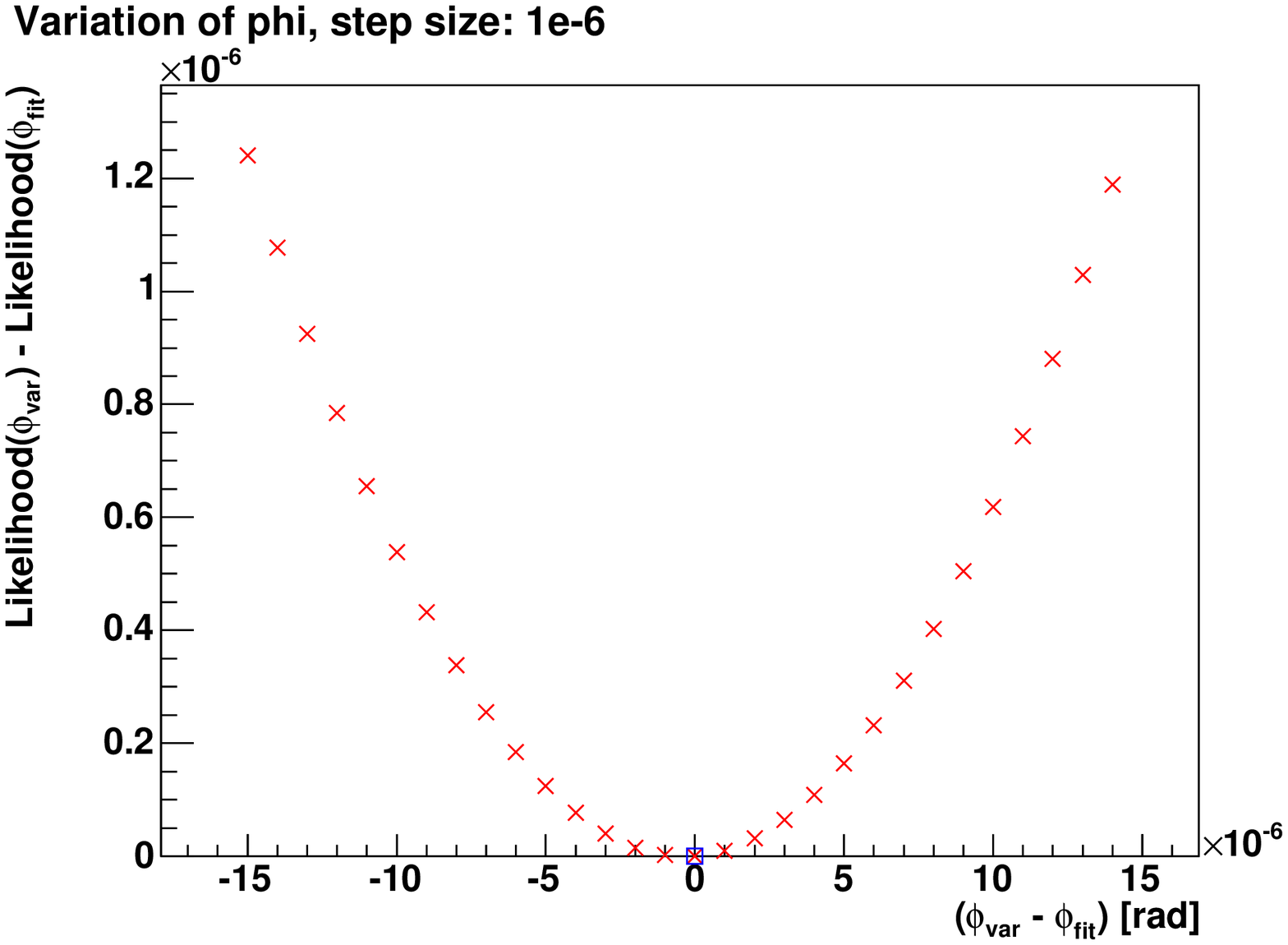}
\includegraphics[width=4.9cm]{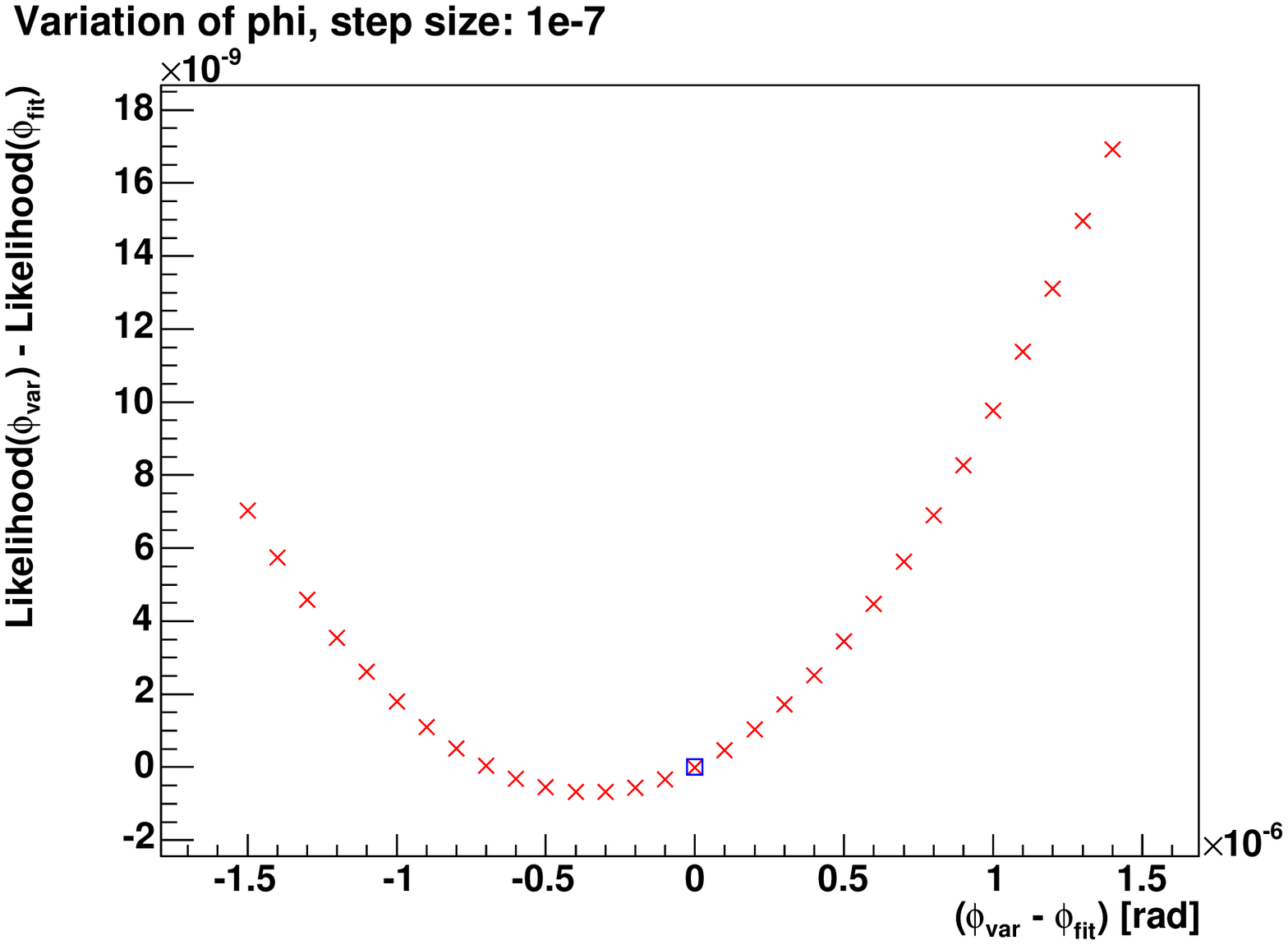}
\includegraphics[width=4.9cm]{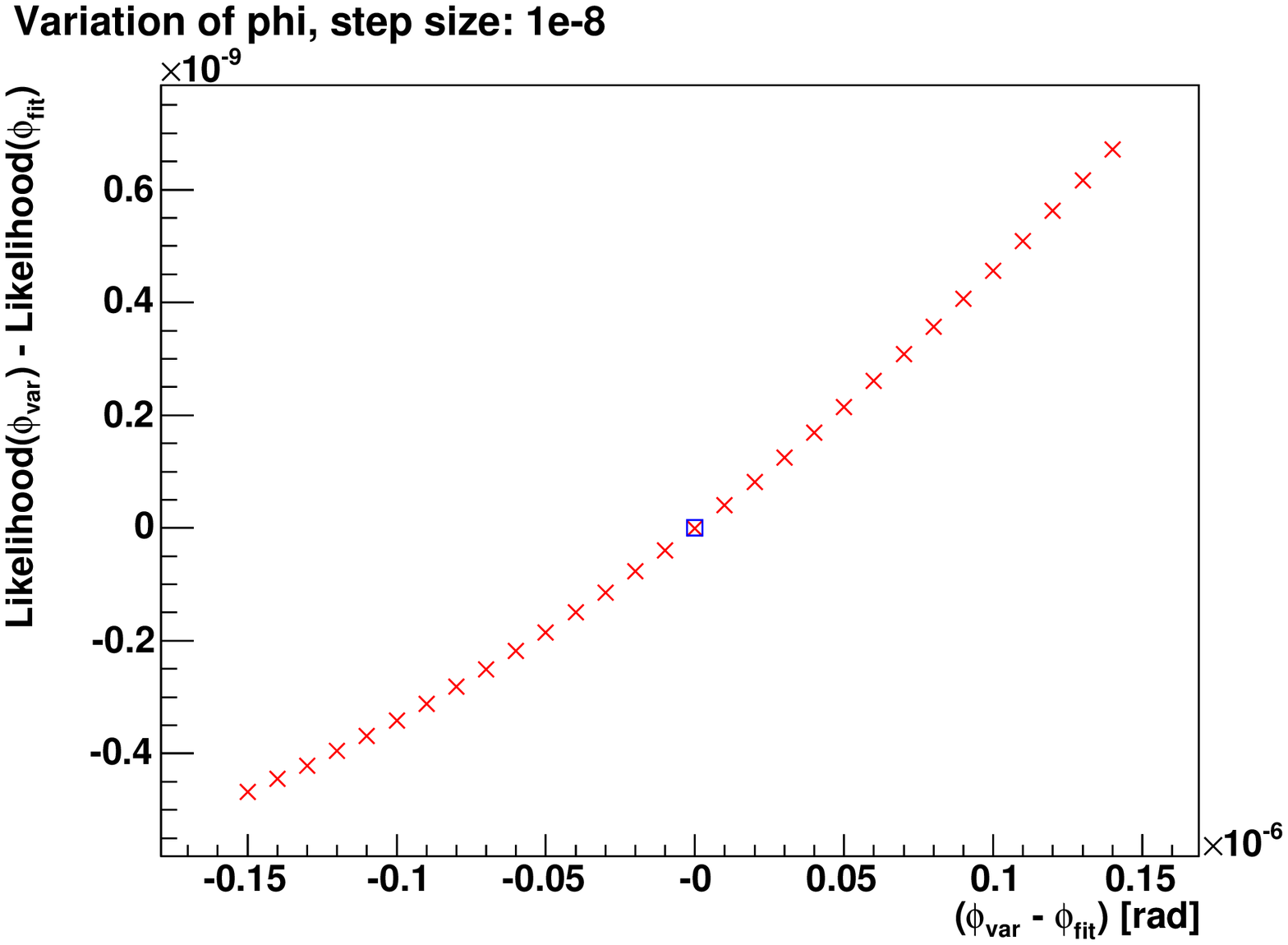}
\includegraphics[width=4.9cm]{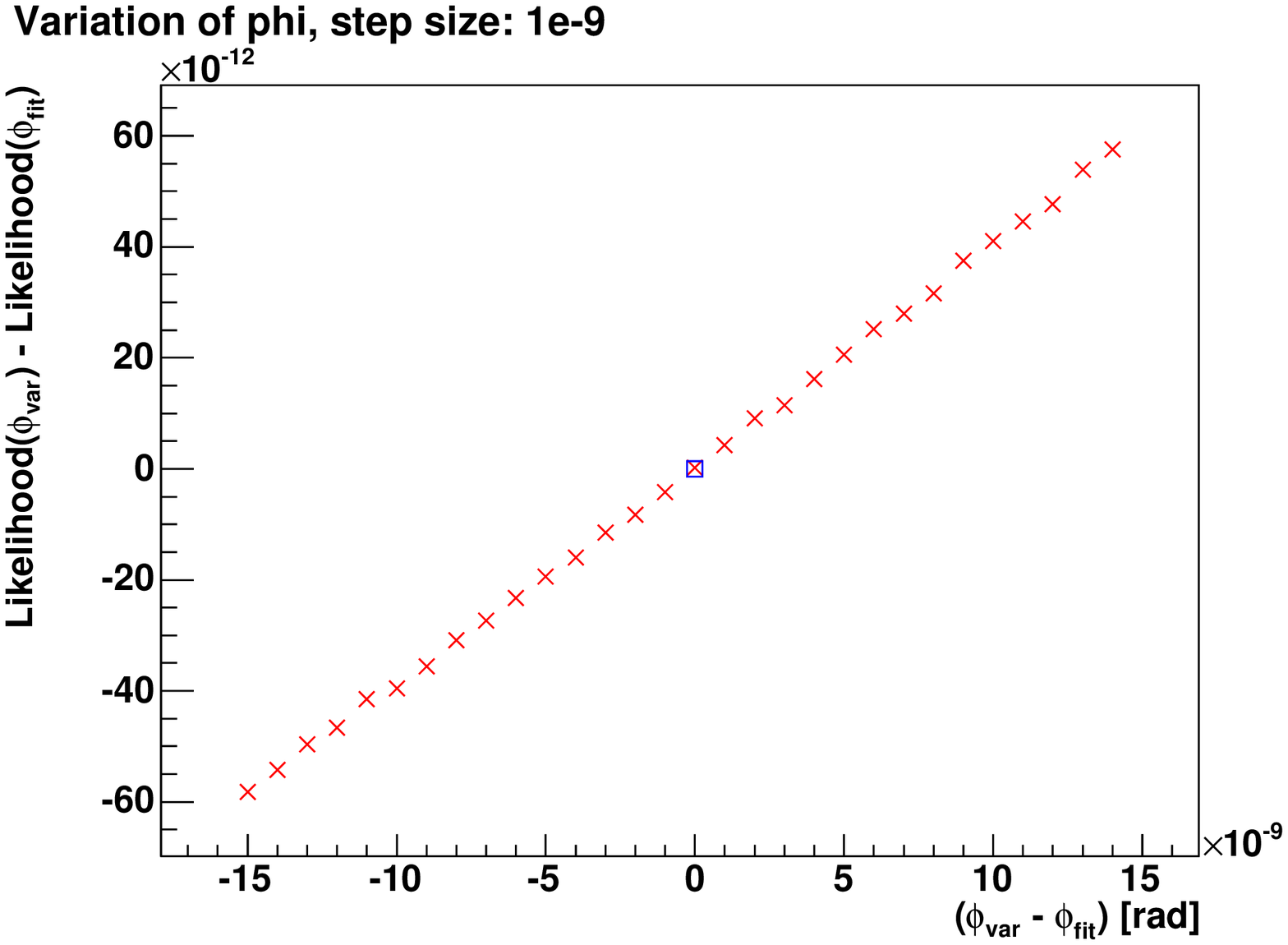}
\includegraphics[width=4.9cm]{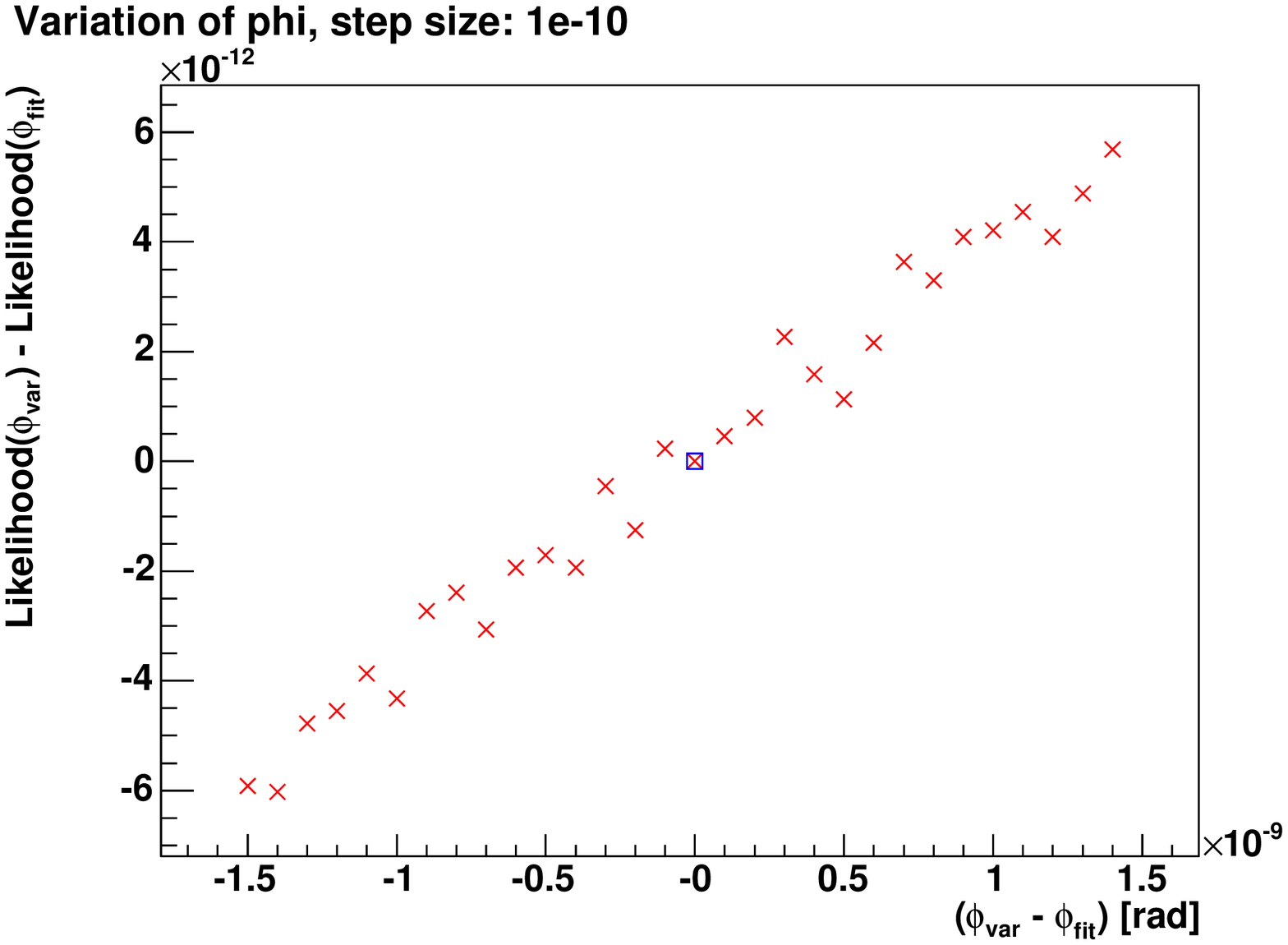}
\includegraphics[width=4.9cm]{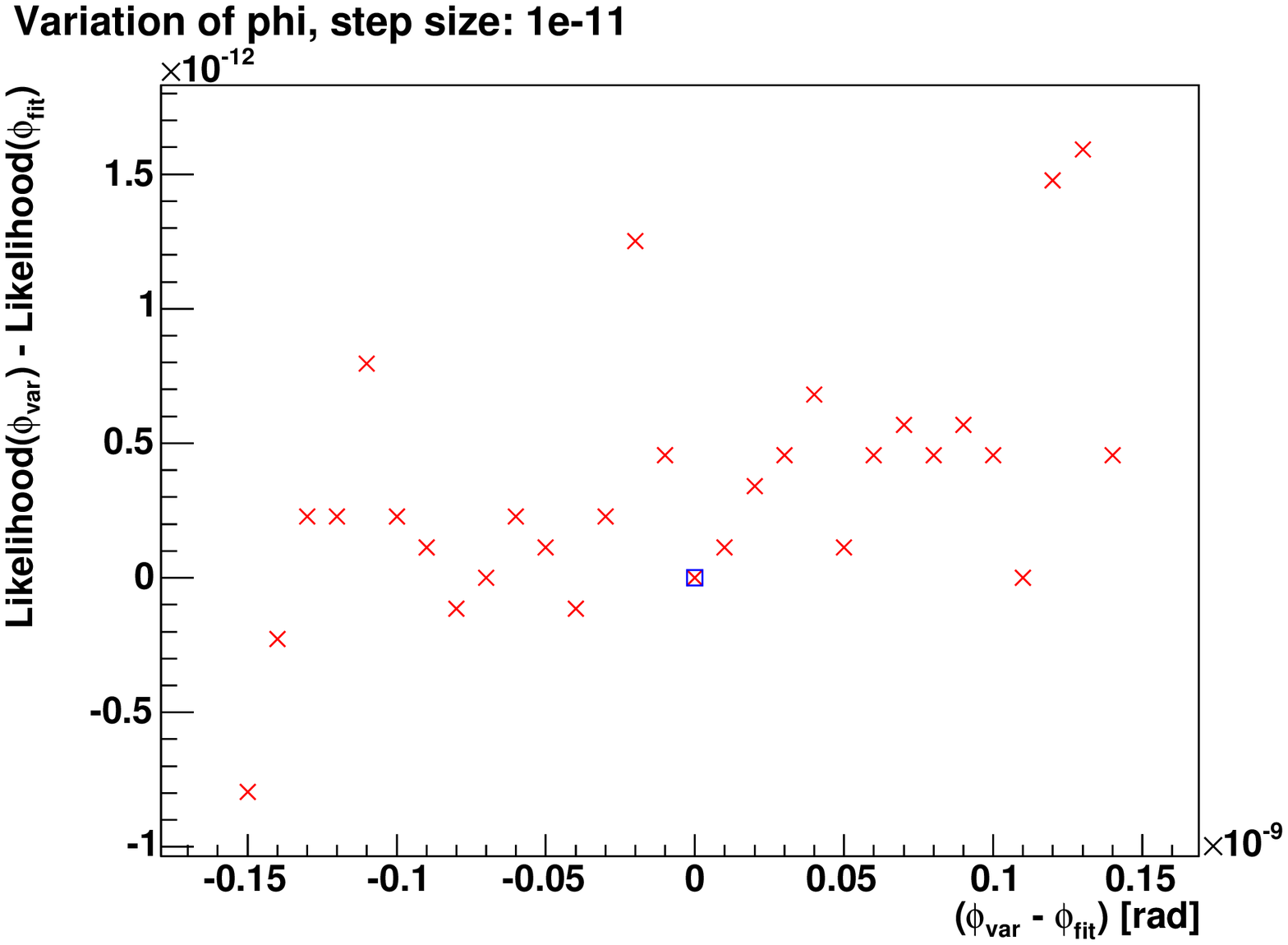}
\includegraphics[width=4.9cm]{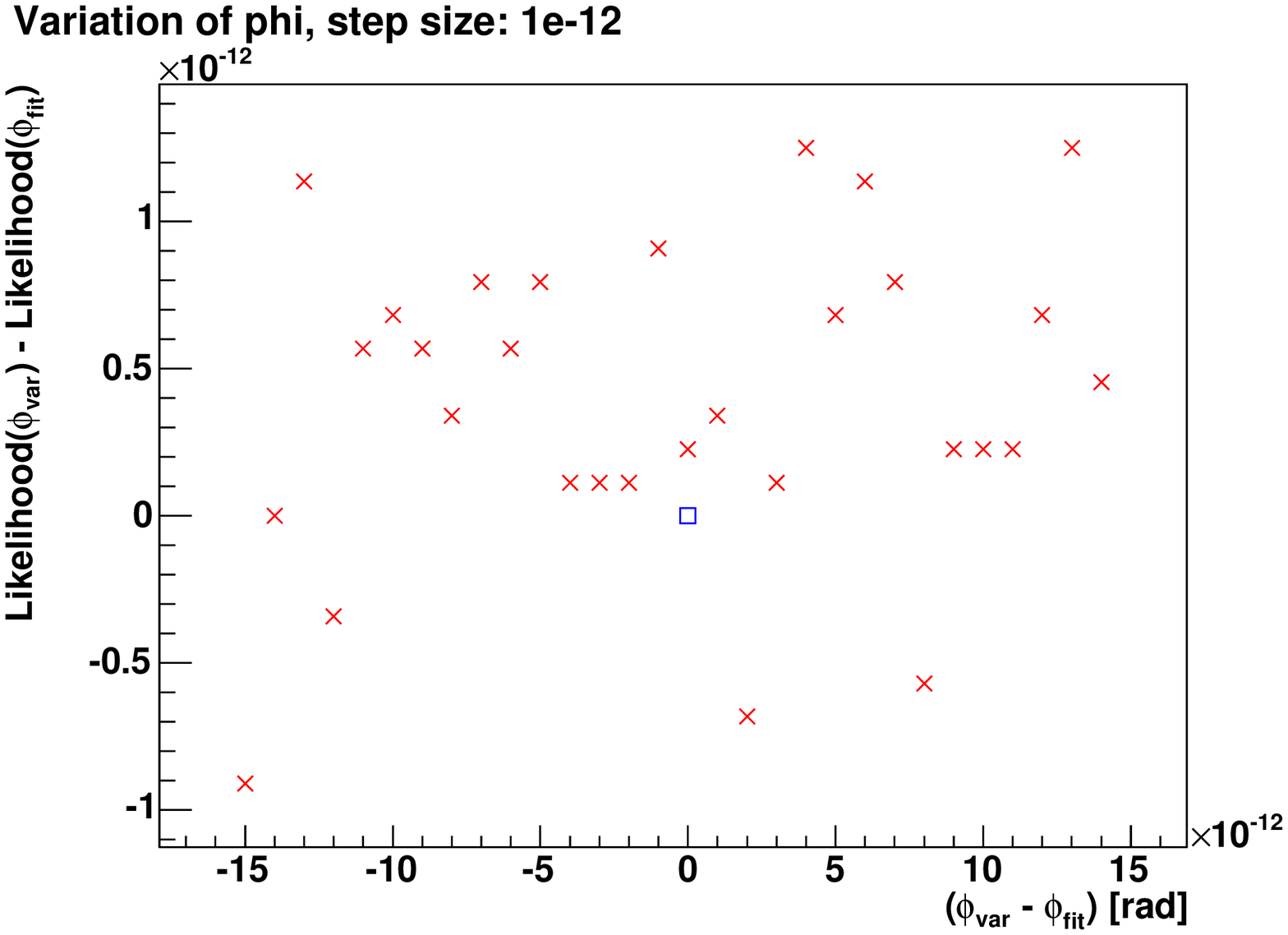}
\includegraphics[width=4.9cm]{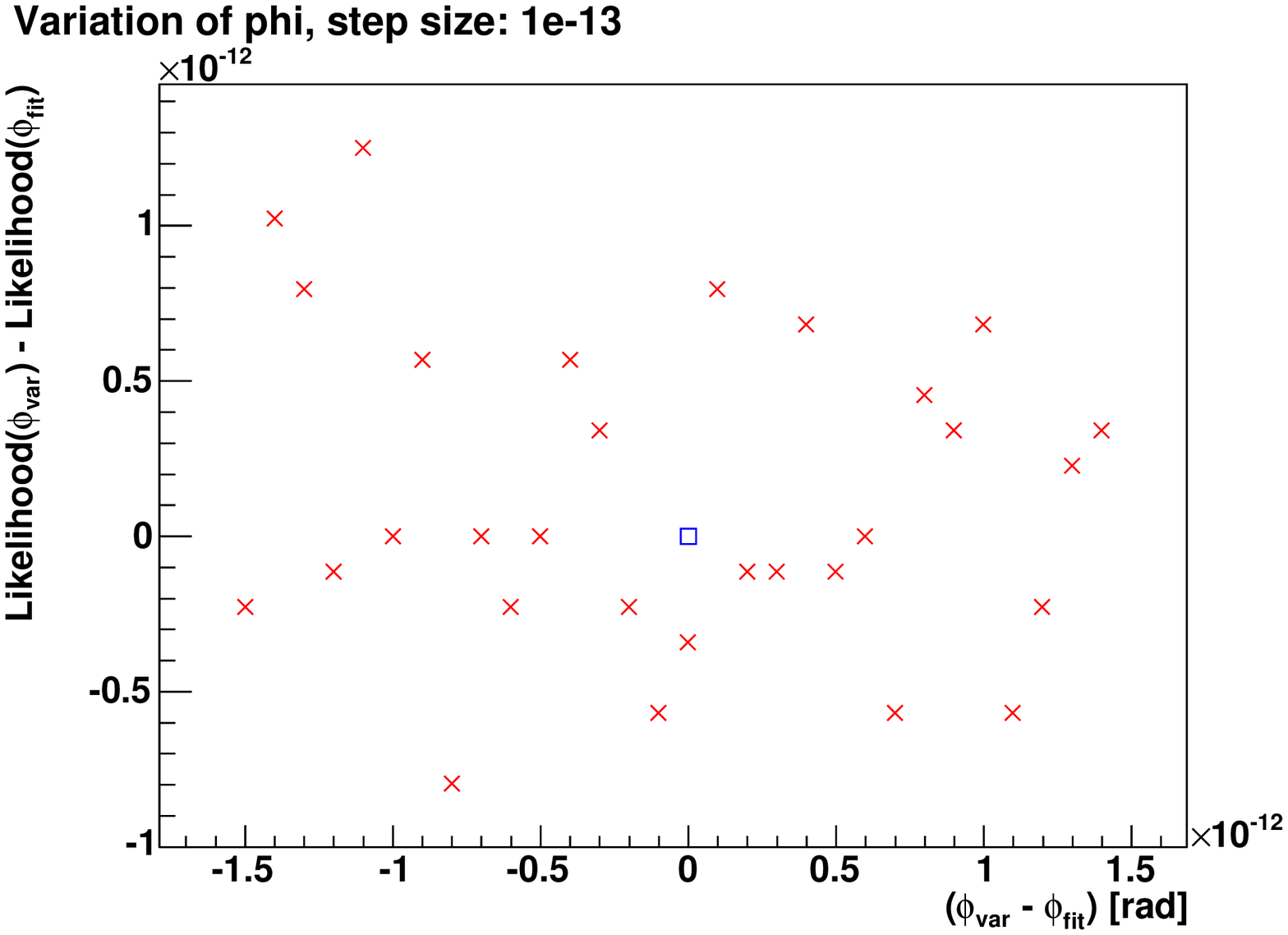}
\caption[Likelihood for variation of $\phi$]{Values of the likelihood function for variations of
  $\phi$ around the fit value.}
\label{fig:var_phi}
\end{figure}

Steps or jumps in the likelihood function can prevent the fit from converging, because the gradient 
cannot be calculated at these likelihood values and therefore the fit does not \lq\lq know\rq\rq \, 
in which direction to continue. The steps can be caused e.g. if fit parameters have a limited
validity range. The likelihood function therefore has to be carefully checked for instabilities.
Additionally, the numerical precision of the calculation is an intrinsic feature which can
cause problems if the precision of the fit parameters becomes equal or higher, which would let 
the fit run into numerical noise, causing uncontrolled, random steps and preventing any further
convergence. It is therefore important to ensure that all values in the function are calculated with
double precision. This is usually not guaranteed just by setting all the user's variables to double
precision, as the single precision could also be used in an external function over which the user
does not have control. \\  
A safe way to ensure that neither steps caused by limits of the likelihood parameter space nor jumps
caused by numerical noise lead to problems in the fit is to examine the likelihood parameter
space by keeping all but one of the fit parameters constant at the minimum of the likelihood
function and varying the remaining parameter in very small intervals around the minimum. \\
Examples for this are shown in figures~\ref{fig:var_theta} and~\ref{fig:var_phi} for the variation
of two variables of the same event. In these figures, the size of the interval inside which the 
likelihood was calculated is decreased by a decade for each subsequent plot. 
In Figure~\ref{fig:var_theta}, the likelihood was calculated at the fixed fit values for $\phi$ and
the photo-electron number, and $\theta$ was varied around the fit value in the steps given in the
header of the respective graph. Plotted in the graphs is the difference between the varied value of
$\theta$ and the value reached as the final result of the fit. Down to a precision of $10^{-6}$,
which was the chosen precision of the minimisation for this example, the fit value, marked by the
blue square, always lies at the lowest point of the graph, as expected. Numerical noise only starts
at a precision of $10^{-11}$, and besides this noise there are no other steps visible in the
graphs. The same is true for Figure~\ref{fig:var_phi}, where $\phi$ was varied instead of
$\theta$. \\
The calculations shown in this subsection have been done for a several dozen events with different 
characteristics, i.e.~different shower directions and energies. It was concluded that 
stability problems do not influence the fit. 

\subsubsection{Conclusion}
In this chapter, a method has been described which allows the combined reconstruction
of the direction and energy of a shower, using the shower position as calculated in 
Section~\ref{sec:pos} together with directional and energy estimates from a scan of the likelihood
parameter space as starting values. Studies of the likelihood parameter space show that the
introduced pattern matching algorithm describes the shower very well (see also
Chapter~\ref{sec:results}, where results both for NC events and for $\nu_e$ CC events are
presented). Not all events can be reconstructed equally successfully; unfortunate event topologies
will sometimes prevent a good result. Quality cuts to distinguish poorly reconstructed events from
well reconstructed ones will be discussed in the following chapter.  

\chapter{Quality Cuts for the Event Selection}\label{sec:cuts}

Clearly, not all shower-type events are reconstructible with the same precision, and some events are
not usable for physics analysis. If there is a problem with the convergence of the minimisation,
the fit program itself will notice this, and the reconstruction will be aborted. However, sometimes
events cannot be reconstructed well because their shape does not agree with the assumed model. This
is the case e.g.~if an event has touched only the edge of the detector, producing enough hits to pass the
filter conditions but not enough information in the hits to allow for a good 
reconstruction, or if the position of the shower has been miscalculated and therefore the likelihood
function does not have a minimum close to the true values of the parameters
(cf.~Section~\ref{sec:topologies}). \\
Quality cuts are used to separate the poorly reconstructed events from the well reconstructed
ones, and to separate off events which are no shower events at all. The latter refers especially to
atmospheric muon events which might be misinterpreted as shower events if they consist of several
muons in a bundle, or if the muon undergoes a catastrophic bremsstrahlung loss and thereby produces
an electromagnetic cascade (see Section~\ref{sec:atm_muons}). \\ 
In Section~\ref{sec:atm_mu}, the suppression of this atmospheric muon background is
discussed. Sections~\ref{sec:phi_diff} to~\ref{sec:pos_cuts} discuss quality cuts that are intended
to allow for the separation of well reconstructed events from poorly reconstructed 
ones. Note that only the cuts introduced in Sections~\ref{sec:likeli_cut},~\ref{sec:E_pre_E_fit} 
and~\ref{sec:large_energies} are used as final quality cuts on the events, as they showed the best 
efficiencies and purities. Efficiencies and purities of all the cuts introduced here are summarised in 
Section~\ref{sec:effpur_cuts} \\ 
Within this chapter, a separation of events according to the angular misreconstruction of the
shower direction with respect to the MC shower axis will be made. Events with a deviation
$\Delta\alpha < 10^{\circ}$ will be considered as {\it good events}, those with ${10^{\circ} \le \Delta\alpha <
30^{\circ}}$ as {\it moderate events} and those with $\Delta\alpha \ge 30^{\circ}$ as {\it bad events}. 
As the quality of the reconstruction also depends on the shower energy, the events will be grouped
in three different energy bins: $E_{sh} < 10$\,TeV, 10\,TeV $\le E_{sh} < 1$\,PeV and $E_{sh} \ge
1$\,PeV, where $E_{sh}$ is the MC shower energy. The shower energy and the neutrino direction are 
strongly correlated; therefore, a badly reconstructed direction will in many cases go along with a badly
reconstructed energy. \\ 
The effects of the cuts will be shown for two different data samples, namely, a sample of
140000 NC events (event sample B, see Appendix~\ref{sec:nc_sample}), and a sample of
$\nu_e$ CC events of about the same size (see Appendix~\ref{sec:nue_sample}). About 18000 events
(13.0\%) of the NC sample, and 31000 events (22.5\%) of the $\nu_e$ CC sample remain in the
sample after the optical background filter and the reconstruction. The number of events passing filter
and reconstruction highly depends on the neutrino energy (see Figure~\ref{fig:event_stages_nue}) and the
generation volume, which exceeded the instrumented volume by one absorption length ($\sim 55$\,m). 
The reason for the higher efficiency for the $\nu_e$ CC events is that for these events the entire
energy of the primary neutrino is transfered into the electromagnetic and hadronic shower, while 
for NC events the average fraction of the primary energy which is transfered to the hadronic shower
is only between $\sim 32\%$ and $\sim 46\%$, depending on the neutrino energy (see Figure~\ref{fig:bjorken}
in Section~\ref{sec:nu_variables}). \\
The numbers of events for the different energy bins and reconstruction qualities, before all cuts,
are shown in Table~\ref{tab:beforecuts}. See Figure~\ref{fig:event_stages_nue} and
Tables~\ref{tab:nc_sample} and~\ref{tab:cc_sample} (Appendix~\ref{sec:data_sample}) for more
details.

\begin{table}[h] \centering
\begin{tabular}{|>{\centering}p{4cm}|>{\raggedleft}p{2cm}|>{\raggedleft}p{2cm}|>{\raggedleft}p{2cm}|>{\raggedleft}p{2cm}|}
\hline
NC events                     & good & moderate & bad  & total \tabularnewline \hline
$E_{sh} < 10$\,TeV            & 3318 &     2527 & 3568 &  9413 \tabularnewline
10\,TeV $\le E_{sh} < 1$\,PeV & 4082 &     2113 & 1943 &  8138 \tabularnewline
$E_{sh} \ge 1$\,PeV           &  382 &      152 &  144 &   678 \tabularnewline \hline
total                         & 7782 &     4792 & 5655 & 18229 \tabularnewline
\hline
\end{tabular}
\begin{tabular}{|>{\centering}p{4cm}|>{\raggedleft}p{2cm}|>{\raggedleft}p{2cm}|>{\raggedleft}p{2cm}|>{\raggedleft}p{2cm}|}
\hline
$\nu_e$ CC events             &  good & moderate & bad  & total \tabularnewline \hline
$E_{sh} < 10$\,TeV            &  6201 &     4148 & 6317 & 16666 \tabularnewline
10\,TeV $\le E_{sh} < 1$\,PeV &  6135 &     3103 & 2759 & 11997 \tabularnewline 
$E_{sh} \ge 1$\,PeV           &  1480 &      523 &  495 &  2498 \tabularnewline \hline 
total                         & 13816 &     7774 & 9571 & 31161 \tabularnewline
\hline
\end{tabular}
\caption[Number of events before all cuts]{Number of well, moderately and badly reconstructed events
  in three energy bins, before all cuts (see text for definitions).}
\label{tab:beforecuts}
\end{table}

\section{Suppression of Atmospheric Muons}\label{sec:atm_mu}

For the investigation of atmospheric muons, a CORSIKA~\cite{corsika} simulation with protons as 
primary cosmic rays was used. Protons are the dominating contributors to the cosmic radiation, 
at least up to the \lq\lq knee\rq\rq\, at about 4\,PeV~\cite{hoerandel05}. Muons produced by the interaction
of the protons in the atmosphere can produce a signal in the ANTARES detector. The event sample 
used for this study was taken from the ANTARES MC data repository (see 
Appendix~\ref{sec:atm_muon_sample} for more details on the data sample). \\
Those events of the muon sample that arrived at the instrumented volume and produced hits in the
detector were processed through the shower reconstruction. For a large number of these atmospheric muon
events, the fit does not converge; those events are thus not reconstructed.

\subsection{Expected and Measured Amplitudes: Quality Cut}\label{sec:likeli_cut}

For the rejection of the atmospheric muons remaining after the reconstruction, it is useful to
consider the correlation between the amplitudes expected in each OM according to the result of the
direction and energy reconstruction, and the measured amplitudes. Even though those atmospheric
muons which survive the reconstruction have an event signature similar to that of shower events, there are
significant differences in the hit pattern, as muons produce hits with small amplitudes along the muon 
trajectory, whereas showers produce relatively localised, large hits. \\
As a measure for the difference between the amplitude $n_{i,c}$ expected for the reconstructed
direction and energy, and the measured amplitude $n_{i,meas}$ in each OM, a variable $\xi$ is defined: 

\begin{equation}\label{eq:xi}
\xi = \frac{1}{N} \sum_{i=1}^{N} \frac{(n_{i,c} - n_{i,meas})^2}{{n_{i,c}}^2}.
\end{equation}

\begin{figure}[h] \centering
\includegraphics[width=7.4cm]{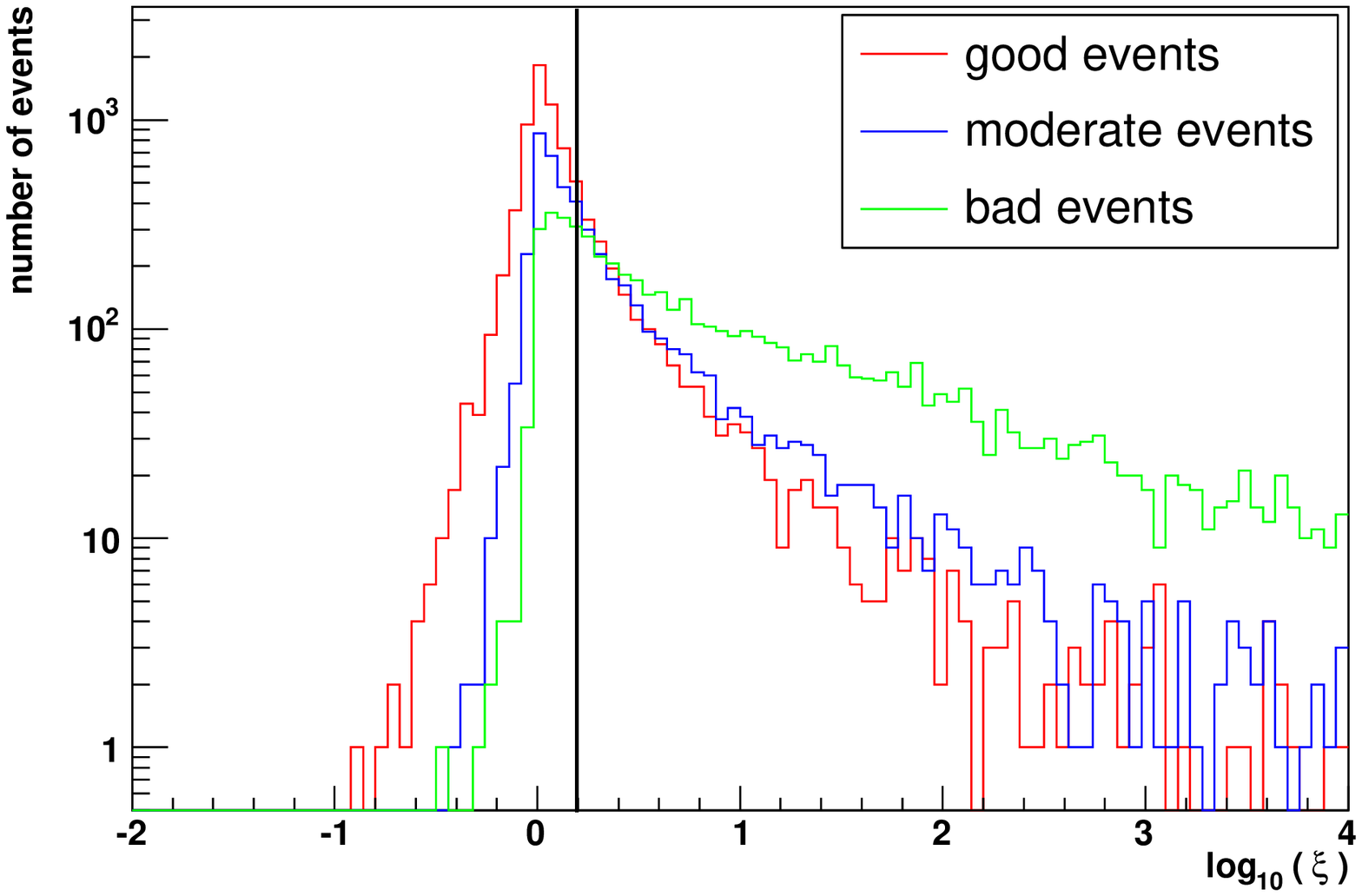}
\includegraphics[width=7.4cm]{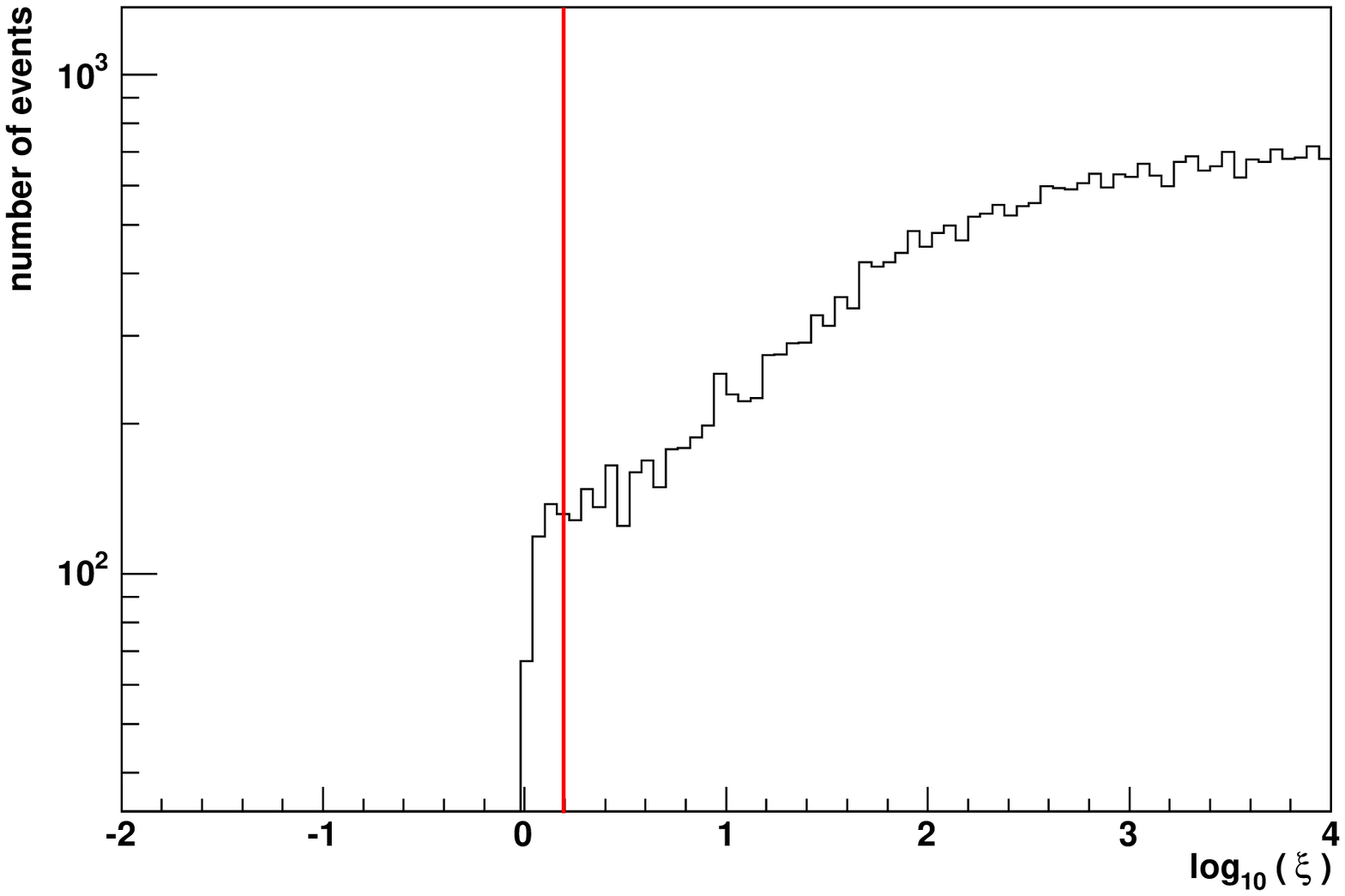}
\caption[Quality variable $\xi$ for NC events and atmospheric muons]{Variable $\xi$ as defined in
  equation~(\ref{eq:xi}), for the NC event sample B (left), and for the atmospheric muons
  (right). The position of the cut is also indicated.} 
\label{fig:chi_square}
\end{figure}

\begin{figure}[h] \centering
\includegraphics[width=7.4cm]{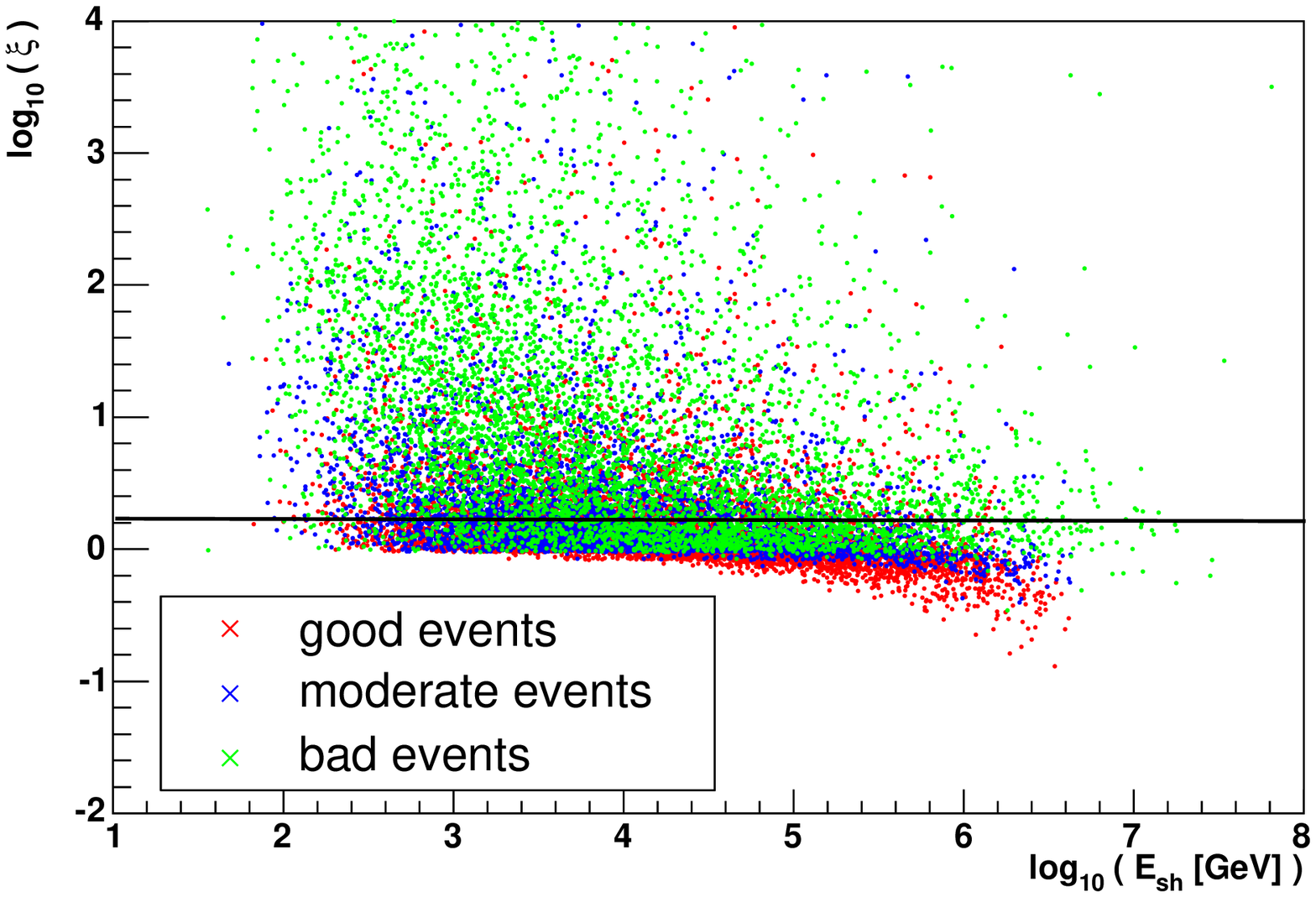}
\includegraphics[width=7.4cm]{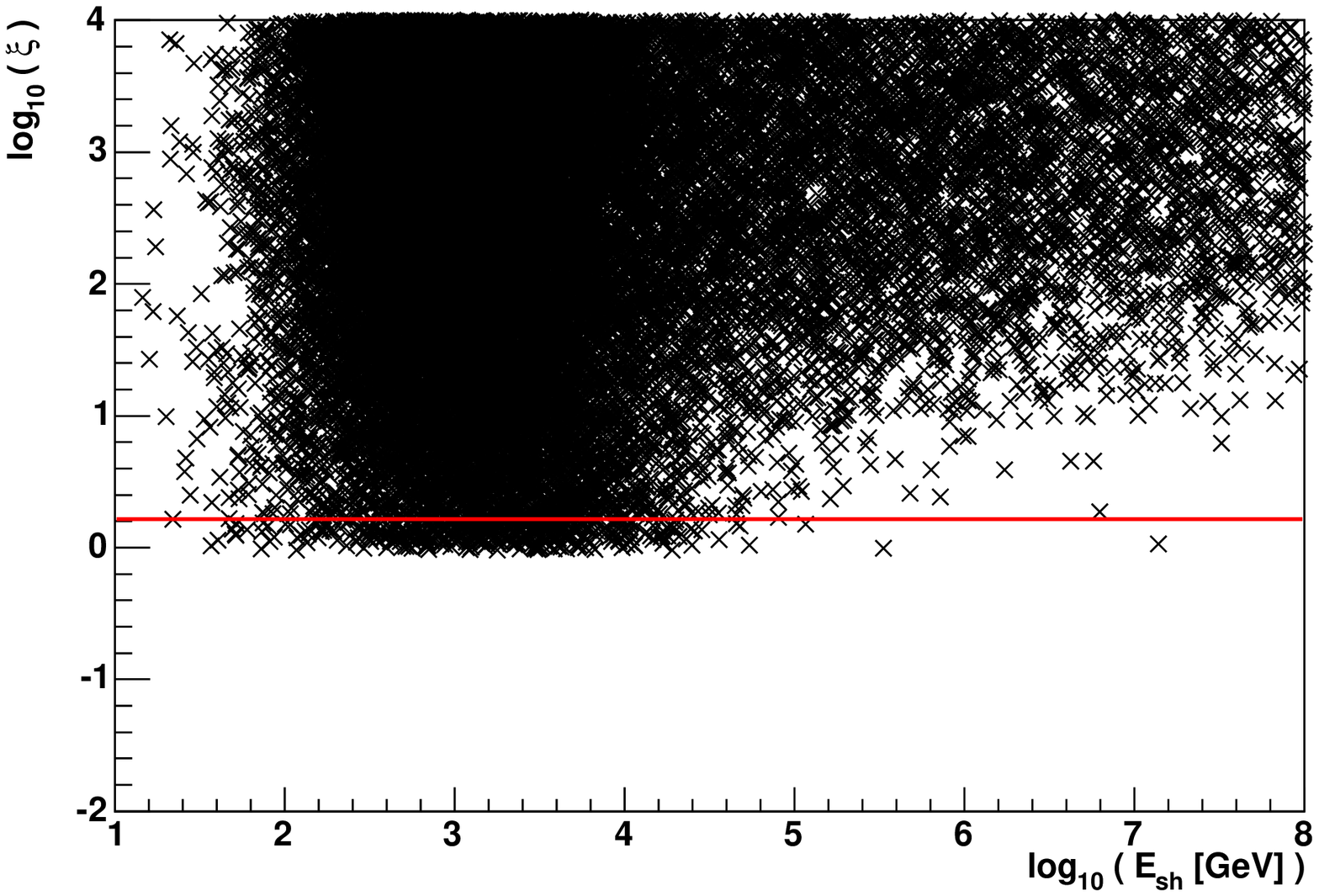}
\caption[Quality variable $\xi$ over energy, for NC events and atmospheric muons]{$\xi$ over the
  reconstructed shower energy, for the NC event sample B (left), and for the atmospheric muons
  (right). The position of the cut is also indicated.} 
\label{fig:chi_square_over_E}
\end{figure}

Here, $N$ is the number of OMs for which $n_{i,c}$ or $n_{i,meas}$ are above the minimum threshold of 
0.3\,pe. The variable $\xi$ is expected to be distributed very differently for muon events than for 
shower events, because of the large number of small amplitude hits that a muon produces  
in OMs where no signal is expected by the shower reconstruction strategy. \\
Figure~\ref{fig:chi_square} shows the distribution of $\xi$ for shower events on the
left hand side, and for muons on the right hand side. The distributions for shower events have their
maximum at ${\log_{10} \xi \approx 0}$, i.e.~$\xi \approx 1$, for all reconstruction qualities. 
However, the distribution of the poorly reconstructed events (green histogram in 
Figure~\ref{fig:chi_square}) extends to larger values of $\xi$, because the agreement between the
calculated and the measured amplitudes is worse for these events, compared to the moderately and
well reconstructed ones. The plot shown here was generated using NC events; for $\nu_e$ CC events,
the distribution is very similar, as can also be seen from the event numbers after a cut on
$\log_{10} \xi < 0.2$ (see Table~\ref{tab:cutchi}). The cut applied at ${\log_{10} \xi < 0.2}$
keeps $66\%$ of the well and moderately reconstructed shower-type events in the data sample,
while $80\%$ of the poorly reconstructed shower events are suppressed.  \\ 
The atmospheric muon events, on the right hand side of Figure~\ref{fig:chi_square}, are distributed
towards much larger values of $\xi$, with a peak around ${\log_{10} \xi \approx 4}$. Most
of the muon events are rejected by the cut at ${\log_{10} \xi < 0.2}$. \\
Figure~\ref{fig:chi_square_over_E} displays the distribution of $\xi$ versus 
the reconstructed shower energy. Due to the increasing number of large hits for higher
energies, the $\xi$ distribution of the shower events decreases with increasing energy. For the
atmospheric muons, the opposite is the case: For the highest energies, almost no events remain in
the sample. \\
Table~\ref{tab:atm_muon_reco} lists the numbers of muon events remaining in different angular
and energy bins after the reconstruction and the cut. A comparison between the remaining atmospheric 
muon event rate and that of the neutrino signal is given in Section~\ref{sec:result_flux}.  

\begin{table}[h] \centering
\begin{tabular}{|>{\centering}p{3.7cm}|>{\raggedleft}p{2.4cm}|>{\raggedleft}p{2.2cm}|>{\raggedleft}p{2.2cm}|>{\raggedleft}p{2.4cm}|}
\hline
NC events                     & good & moderate & bad & total \tabularnewline \hline
$E_{sh} < 10$\,TeV            & 2135 (64.3\%) & 1071 (42.4\%) &  540 (15.1\%) & 3746 (39.8\%) \tabularnewline
10\,TeV $\le E_{sh} < 1$\,PeV & 3356 (82.2\%) & 1424 (67.4\%) &  657 (33.8\%) & 5437 (66.8\%) \tabularnewline
$E_{sh} \ge 1$\,PeV           &  347 (90.8\%) &  120 (78.9\%) &   61 (42.4\%) &  528 (77.9\%) \tabularnewline \hline
total                         & 5838 (75.0\%) & 2615 (54.6\%) & 1258 (22.2\%) & 9711 (53.3\%) \tabularnewline
\hline
\end{tabular}
\begin{tabular}{|>{\centering}p{3.7cm}|>{\raggedleft}p{2.4cm}|>{\raggedleft}p{2.2cm}|>{\raggedleft}p{2.2cm}|>{\raggedleft}p{2.4cm}|}
\hline
$\nu_e$ CC events             & good & moderate & bad & total \tabularnewline \hline
$E_{sh} < 10$\,TeV            &  3758 (60.6\%) & 1576 (38.0\%) &  772 (12.2\%) &  6106 (36.6\%) \tabularnewline
10\,TeV $\le E_{sh} < 1$\,PeV &  4914 (80.1\%) & 2053 (66.2\%) &  863 (31.3\%) &  7830 (65.3\%) \tabularnewline
$E_{sh} \ge 1$\,PeV           &  1385 (93.6\%) &  434 (83.0\%) &  224 (45.3\%) &  2043 (81.8\%) \tabularnewline \hline
total                         & 10057 (72.8\%) & 4063 (52.3\%) & 1859 (19.4\%) & 15979 (51.3\%) \tabularnewline
\hline
\end{tabular}
\caption[Number of shower events after the cut on $\xi$]{Number of good, moderate and bad
  shower events in the three energy bins, after the cut on $\log_{10} \xi < 0.2$; the
  percentages were calculated with respect to the number of events after the reconstruction, as
  displayed in Table~\ref{tab:beforecuts}.}
\label{tab:cutchi}
\end{table}

\begin{table}[h] \centering 
\begin{tabular}{|>{\centering}p{5.2cm}|>{\centering}p{2.cm}|>{\centering}p{2.cm}|>{\centering}p{2.cm}|}
\hline
angle & \multicolumn{3}{c|}{$0^{\circ} - 60^{\circ}$} \tabularnewline
\hline
primary energy [TeV/nucleon]& 1-10   & 10-100 & 100-10$^5$     \tabularnewline \hline
generated events            & $10^8$ & $10^7$ & $10^7$ \tabularnewline
events producing hits       & 917045 & 419280 & 2383537        \tabularnewline
events after reconstruction & 966    & 2002   & 76803          \tabularnewline
events after $\xi$ cut      & 11     &  32    & 472            \tabularnewline 
event rate after cut        & 7.51\,h$^{-1}$ & 7.73\,h$^{-1}$ & 2.187\,h$^{-1}$ \tabularnewline \hline
\end{tabular}
\begin{tabular}{|>{\centering}p{5.2cm}|>{\centering}p{2.cm}|>{\centering}p{2.cm}|>{\centering}p{2.cm}|}
\hline
angle & \multicolumn{3}{c|}{$60^{\circ} - 85^{\circ}$} \tabularnewline
\hline
primary energy [TeV/nucleon]& 1-10   & 10-100 & 100-10$^5$     \tabularnewline \hline
generated events            & $10^8$ & $10^7$ & $10^7$         \tabularnewline
events producing hits       & 8147   & 30831  & 492487         \tabularnewline
events after reconstruction & 66     & 525    & 27676          \tabularnewline
events after $\xi$ cut      & 0      & 3      & 218            \tabularnewline 
event rate after cut        & $\lesssim 20$\,h$^{-1\,*}$ & 0.858\,h$^{-1}$ & 0.737\,h$^{-1}$ \tabularnewline \hline
\end{tabular}
\caption[Atmospheric muon sample]{Numbers of events in the different energy and angular bins of the
  atmospheric muon event sample, after the reconstruction and the quality cut on $\xi$. The event 
  rate after the $\xi$ cut is also shown. The upper limit given in the bin containing 0 events 
  after the cut (marked with an asterisk) corresponds to the 90\% confidence level. }
\label{tab:atm_muon_reco}
\end{table}

\begin{figure}[h] \centering
\begin{minipage}[h]{7.2cm}
\centering \epsfig{figure=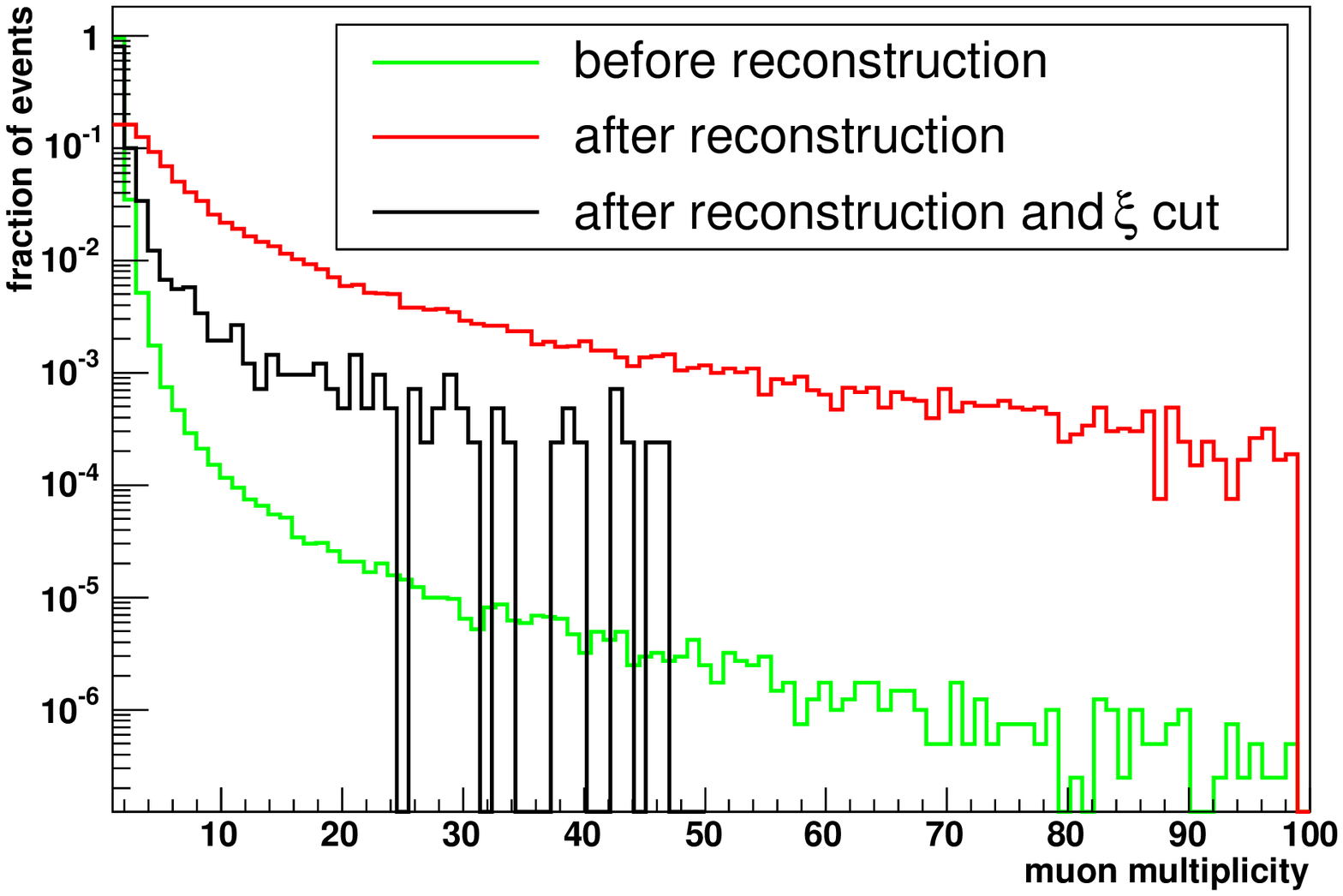, width=7.4cm}
\caption[Muon multiplicity before and after reconstruction]{Muon multiplicity before the
  reconstruction (green), after the reconstruction (red) and after the reconstruction and the cut
  (black). The histograms have been normalised to 1.}  
\label{fig:mult_reco}
\end{minipage}
\hspace{0.2cm}
\begin{minipage}[h]{7.2cm}
\centering \epsfig{figure=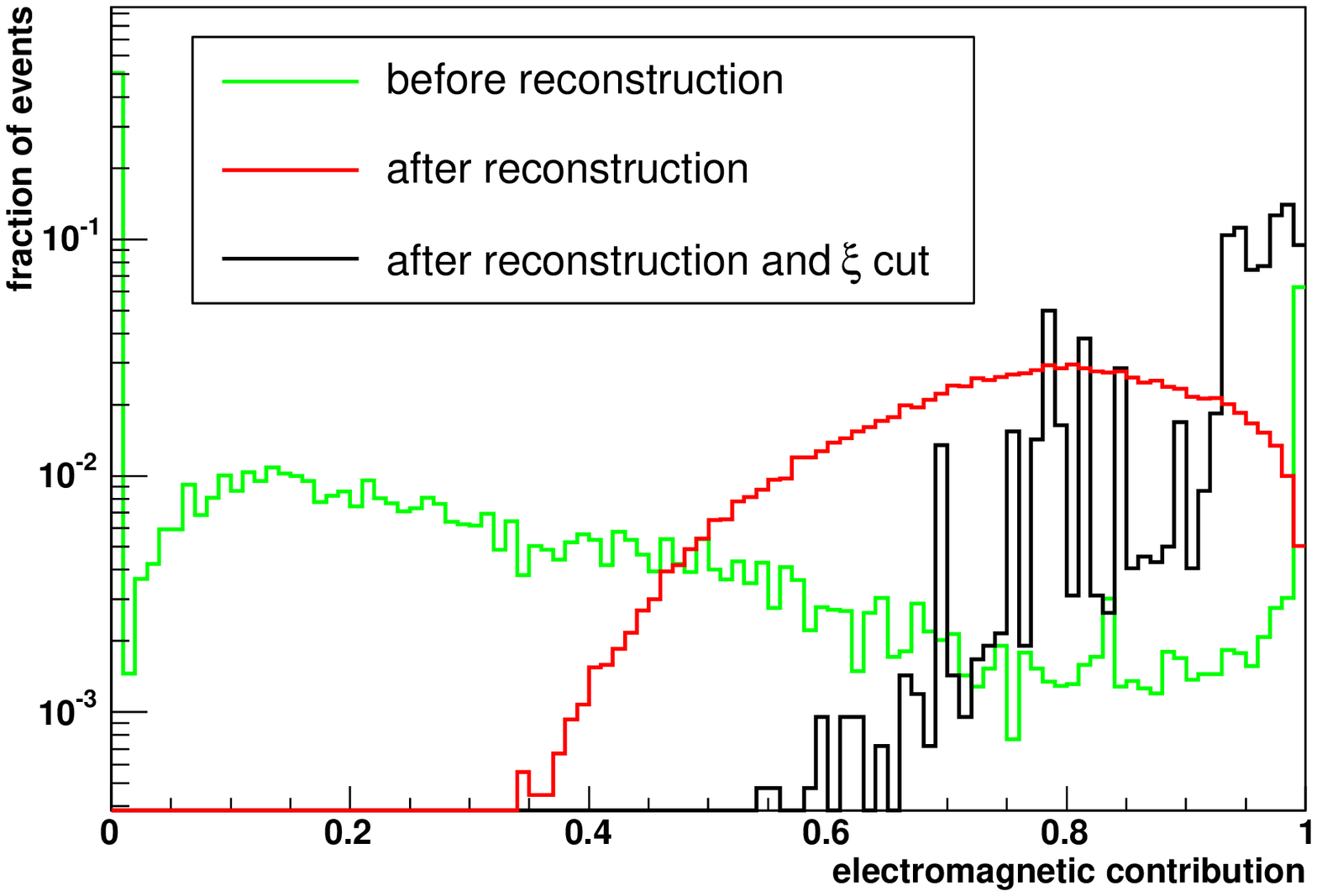, width=7.4cm}
\caption[Electromagnetic contribution before and after reconstruction]{Electromagnetic contribution
  before the reconstruction (green), after the reconstruction (red) and after the reconstruction and
  the cut (black). The histograms have been normalised to 1.} 
\label{fig:em_cont_reco}
\end{minipage}
\end{figure}

The characteristics of the atmospheric muon events remaining in the sample after the shower
reconstruction and the cut on $\xi$ will be discussed in the following. \\
In Figure~\ref{fig:atm_mu_flux} it has been shown that the flux of
single muons at the ANTARES site is about a factor 80 higher than the flux of muon bundles with more
than 4 muons. This can be compared to Figure~\ref{fig:mult_reco}, where the muon multiplicities
(i.e.~number of muons in an event) for the primary proton events that were examined in this study
are shown, in green for those events producing hits in the detector,
in red for those remaining after the reconstruction and in black for those events
remaining after the reconstruction and the $\xi$ cut. The entries of the histograms have been
weighted with to the respective event weights, so as to reproduce the expected energy spectrum
of the primaries. The histograms have both been normalised to 1. As
expected, the relative muon multiplicity has increased after the reconstruction. However, after
the cut on $\xi$, those events with a very large muon multiplicity $\gtrsim 40$ are not present in
the sample anymore; the percentage of single muon events is larger than after the reconstruction,
which points to the fact that those event passing the cut are due to muons undergoing a 
catastrophic energy loss. \\
Figure~\ref{fig:em_cont_reco} shows the fraction of the total hit amplitude which originates from
electromagnetic contributions, again before the reconstruction (green), after the reconstruction
(red) and after the reconstruction and the cut on $\xi$ (black). The entries of the histograms have
been weighted as described above, and the histograms have both been normalised to 1. The 
electromagnetic contribution can be considered as a measure for the energy that has been deposited 
into electromagnetic showers, i.e.~bremsstrahlung losses. Before the reconstruction, the 
electromagnetic contribution shows a very large peak at zero and another peak at 1; in-between, 
the distribution is relatively flat, decreasing towards larger electromagnetic contributions. 
The very large contributions at 0 and 1 are caused by events with very few hits in the detector, 
which were either totally muonic or totally electromagnetic. Such events fail to pass the filter 
conditions which precede the shower reconstruction. \\
After the reconstruction, the electromagnetic contribution is shifted to values of 40\% or higher;
the mean of the distribution is at 0.76. After the cut on $\xi$, the electromagnetic fraction is
distributed to even larger values: All remaining events have an electromagnetic contribution of 55\%
or higher, and the mean is at 0.92. 

\subsection{Catastrophic Losses and Multi-Muons in the Remaining Events}

\begin{figure}[h] \centering
\includegraphics[width=10cm]{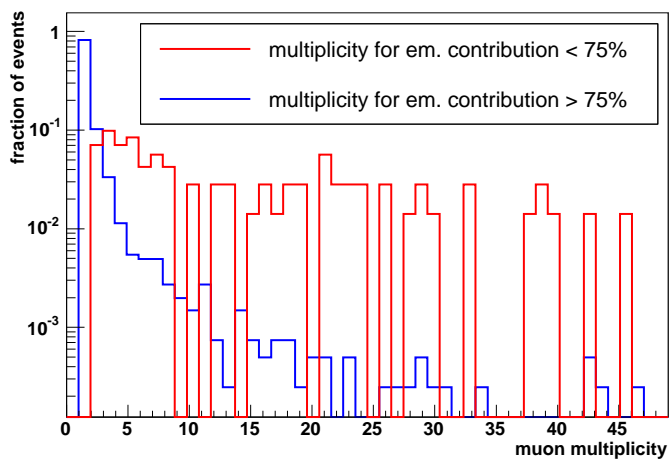}
\caption[Muon multiplicity for bundles and catastrophic events]{Muon multiplicity for those events
  with an electromagnetic contribution larger than 75\% (blue) and for those with a smaller contribution (red). 
 Both histograms have been normalised to 1.}  
\label{fig:mult}
\end{figure}

To determine whether those events with large electromagnetic contributions tend to be caused by muon
bundles with a large multiplicity or not, the sample of events remaining after the cut is divided into two
classes: those with an electromagnetic contribution larger than 75\%, and those with a smaller contribution 
than this value.
653 of the events remaining in the sample had an electromagnetic contribution larger than 75\%; all but
one of the remaining 83 events have a muon multiplicity of at least two and can therefore be considered
multi-muon events. Figure~\ref{fig:mult} shows the muon multiplicity of the two classes of events. 
Again, the entries to both histograms have been weighted according to the primary proton flux, and
both curves have been normalised to 1. One can see that the muon multiplicity for the events with
large electromagnetic contributions is much smaller than for the other events, which leads to the
conclusion that the events of the first category are indeed characterised by catastrophic losses of
single muons.

\vspace{1cm}
Even though the atmospheric muon event rate has been suppressed significantly by the cut
described above, it is still very high compared to the expected neutrino-induced shower
rates. The following sections discusses several cuts for the differentiation between well
reconstructed events and poorly reconstructed ones. It will be shown that with these cuts, also a
suppression of the muon events, at least above a certain energy threshold, is possible. 

\section{Comparison of Azimuth from Calculation and Fit}\label{sec:phi_diff}

\begin{figure}[h] \centering
\includegraphics[width=14cm]{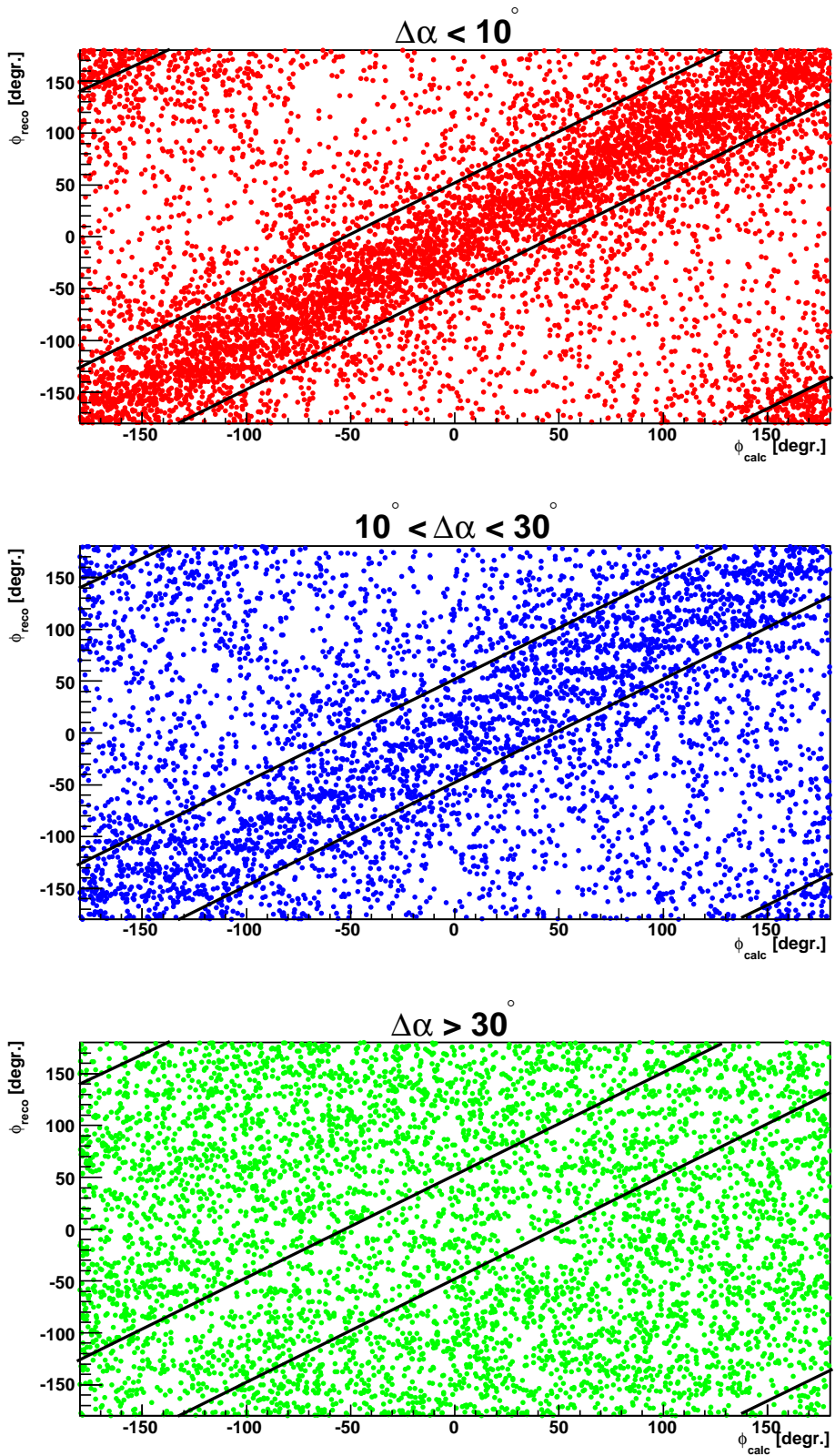}
\caption[Azimuth angle from fit vs.~calculated azimuth angle]{Azimuth angle obtained from the fit,
  $\phi_{reco}$, vs.~the azimuth angle calculated as described in Section~\ref{sec:prefit_dir1},
  $\phi_{calc}$, for well (top), moderately (middle) and poorly (bottom) reconstructed
  events. The region of event selection is marked by the black lines. }
\label{fig:phi_pre_reco}
\end{figure}

An algebraic method to calculate a rough estimate of the neutrino direction was introduced in
Section~\ref{sec:prefit_dir1}. Using the method presented there, the azimuth angle $\phi$ can
be reconstructed with a RMS of $79^{\circ}$. Even though this is a poor resolution, the
results of this calculation and of the fit are correlated, as can be seen from
Figure~\ref{fig:phi_pre_reco}: Here, the azimuth angle resulting from the fit, $\phi_{reco}$, is
plotted versus the azimuth angle from the calculation, $\phi_{calc}$, for well reconstructed events
(upper plot), moderately reconstructed ones (middle plot) and poorly reconstructed ones (lower
plot), for event sample B. The black lines mark the regions of  ${|\phi_{calc} -
  \phi_{reco}| < 50^{\circ}}$ and ${|\phi_{calc} - \phi_{reco}| > 320^{\circ}}$. The 
results for the well reconstructed events are correlated. The correlation between $\phi_{calc}$ 
and $\phi_{reco}$ weakens for the moderate events, and the majority of the poorly reconstructed 
events lie outside the regions marked by the black lines. Events will therefore be discarded if 
they do not fulfil the conditions ${|\phi_{calc} - \phi_{reco}| < 50^{\circ}}$ or 
${|\phi_{calc} - \phi_{reco}| > 320^{\circ}}$. 

In principle, this cut could also be applied for the zenith angle $\theta$. However,
as can be seen from Figure~\ref{fig:theta_phi_pre} in Section~\ref{sec:prefit_dir1}, the algebraic
method tends to shift the zenith angle towards smaller angles, i.e.~to an upward orientation; 
therefore, the correlation between $\theta_{calc}$ and $\theta_{reco}$ is not as clear as for $\phi$. \\
The results for the $\phi$ cut are listed in Table~\ref{tab:phicond}, in absolute numbers and in
percentages with respect to the number of events after the reconstruction. The cut
has a relatively high purity, as can be seen from the small percentages of poorly reconstructed events
remaining in the sample. However, for the highest energy bin also the 
number of suppressed good events is relatively high, which is due to the fact that the
calculation of the azimuth yields its best results for energies below 10\,TeV, as was demonstrated
in Section~\ref{sec:prefit_dir1} in Figure~\ref{fig:delta_over_E}. Due to this poor efficiency, the 
cut is not used for the final event selection. \\
Around 60\% of the atmospheric muon events remaining after the cut on $\xi$ are suppressed by this
cut. The numbers of atmospheric muon events in the individual bins after the different cuts are
listed at the end of the chapter in Table~\ref{tab:atm_muon_cut345}.

\begin{table}[h] \centering 
\begin{tabular}{|>{\centering}p{3.7cm}|>{\raggedleft}p{2.4cm}|>{\raggedleft}p{2.2cm}|>{\raggedleft}p{2.2cm}|>{\raggedleft}p{2.4cm}|}
\hline
NC events         & good & moderate & bad & total \tabularnewline \hline
$E_{sh} < 10$\,TeV            & 2675 (80.6\%) & 1690 (66.9\%) &  950 (26.6\%) & 5315 (56.5\%) \tabularnewline
10\,TeV $\le E_{sh} < 1$\,PeV & 2275 (55.7\%) &  570 (27.0\%) &  326 (16.8\%) & 3171 (39.0\%) \tabularnewline
$E_{sh} \ge 1$\,PeV           &  110 (28.8\%) &   10 (6.58\%) &   32 (22.2\%) &  152 (22.4\%) \tabularnewline\hline
total                         & 5060 (65.0\%) & 2270 (47.4\%) & 1308 (23.1\%) & 8638 (47.4\%) \tabularnewline
\hline
\end{tabular}
\begin{tabular}{|>{\centering}p{3.7cm}|>{\raggedleft}p{2.4cm}|>{\raggedleft}p{2.2cm}|>{\raggedleft}p{2.2cm}|>{\raggedleft}p{2.4cm}|}
\hline
$\nu_e$ CC events & good & moderate & bad & total \tabularnewline \hline
$E_{sh} < 10$\,TeV            & 4921 (79.4\%) & 2690 (64.9\%) & 1795 (28.4\%) &  9406 (56.4\%) \tabularnewline
10\,TeV $\le E_{sh} < 1$\,PeV & 3120 (50.9\%) &  656 (21.1\%) &  471 (17.1\%) &  4247 (35.4\%) \tabularnewline
$E_{sh} \ge 1$\,PeV           &  349 (23.6\%) &   40 (7.65\%) &   68 (13.7\%) &   457 (18.3\%) \tabularnewline\hline
total                         & 8390 (60.7\%) & 3386 (43.6\%) & 2334 (24.4\%) & 14110 (45.3\%) \tabularnewline
\hline
\end{tabular}
\caption[Number of events after $\Delta \phi$ cut]{Number of good, moderate and bad events in the three
  energy bins, after the cut on $|\phi_p - \phi_f|$. The percentages were calculated with respect to
  the number of events after the reconstruction, as displayed in
  Table~\ref{tab:beforecuts}.}
\label{tab:phicond}
\end{table}

\section{Comparison of Energy from Calculation and Fit}\label{sec:E_pre_E_fit}

\begin{figure}[h] \centering
\includegraphics[width=10cm]{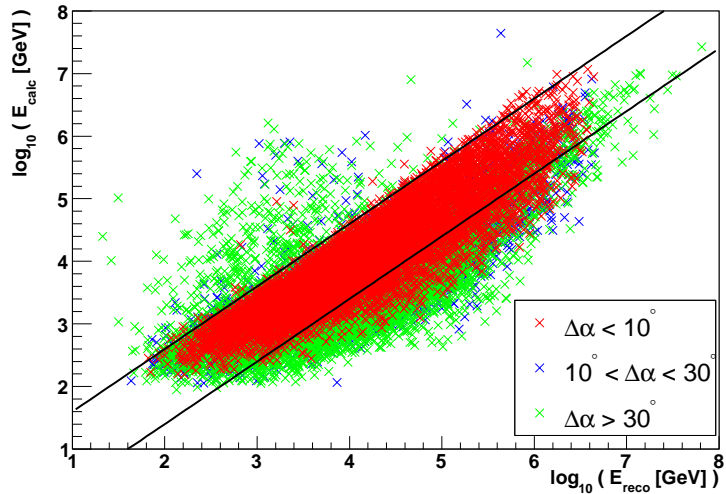}
\caption[$E_{calc}$ vs.~$E_{reco}$ and cut conditions]{$E_{calc}$ vs.~$E_{reco}$ for good, moderate and bad events,
  and the cut conditions (bold lines).}
\label{fig:Epre_E}
\end{figure}

For the energy, an analogous method to that introduced above in Section~\ref{sec:phi_diff}
can be applied. In Section~\ref{sec:e_pre}, an algorithm was described to reconstruct the
shower energy to a factor of about $2.3$ (for WF mode). This method assumes that all photons are
emitted under a fixed angle and uses the position and direction of the shower, calculated 
according to the descriptions in Sections~\ref{sec:pos} and~\ref{sec:prefit_dir1}. \\
The parameterisation of the angular photon distribution $D(\vartheta)$ is not used
for the calculation. Therefore the calculated energy is not directly correlated to the result from
the likelihood minimisation (though the same values for the shower position are used), and thus, a
comparison between the reconstructed energy from the fit, $E_{reco}$, and the energy from the
calculation, $E_{calc}$, can be used to assess the consistency of the results. The correlation between
$E_{calc}$ and $E_{reco}$ is shown in Figure~\ref{fig:Epre_E}, for the NC event sample B. Events pass the cut
if they lie within the region marked by the two bold lines, i.e.~if 
the difference between $\log_{10} E_{calc}$ and $\log_{10} E_{reco}$ is not larger than 0.6. 

\clearpage
The events remaining in the sample after this cut are listed in Table~\ref{tab:Econd}. The
percentages given refer to the number of events that remained in the sample after the
reconstruction, as given in Table~\ref{tab:beforecuts}. The efficiency of this cut is 
high. It was therefore selected as one of the final cuts on the event samples. \\
For the muon suppression, this cut is almost as effective as the cut on the azimuth
angles introduced in the previous section. About 51\% of the atmospheric muons remaining in the
sample after the cut on $\xi$ are suppressed by this cut. Detailed numbers are given 
in Table~\ref{tab:atm_muon_cut345}.

\begin{table}[h] \centering 
\begin{tabular}{|>{\centering}p{3.7cm}|>{\raggedleft}p{2.4cm}|>{\raggedleft}p{2.2cm}|>{\raggedleft}p{2.2cm}|>{\raggedleft}p{2.4cm}|}
\hline
NC events         & good & moderate & bad & total \tabularnewline \hline
$E_{sh} < 10$\,TeV            & 3197 (96.4\%) & 2219 (87.8\%) & 2224 (62.3\%) &  7640 (81.2\%) \tabularnewline
10\,TeV $\le E_{sh} < 1$\,PeV & 3349 (82.0\%) & 1072 (50.7\%) &  817 (42.0\%) &  5238 (64.4\%) \tabularnewline
$E_{sh} \ge 1$\,PeV           &  277 (72.5\%) &   69 (45.4\%) &   91 (63.2\%) &   437 (64.5\%) \tabularnewline\hline
total                         & 6823 (87.8\%) & 3360 (70.1\%) & 3132 (55.4\%) & 13315 (73.0\%) \tabularnewline
\hline
\end{tabular}
\begin{tabular}{|>{\centering}p{3.7cm}|>{\raggedleft}p{2.4cm}|>{\raggedleft}p{2.2cm}|>{\raggedleft}p{2.2cm}|>{\raggedleft}p{2.4cm}|}
\hline
$\nu_e$ CC events & good & moderate & bad & total \tabularnewline \hline
$E_{sh} < 10$\,TeV            & 5982 (96.5\%) & 3599 (86.8\%) & 4078 (64.6\%) & 13659 (82.0\%) \tabularnewline
10\,TeV $\le E_{sh} < 1$\,PeV & 4896 (79.8\%) & 1425 (45.9\%) & 1155 (41.9\%) &  7476 (62.3\%) \tabularnewline
$E_{sh} \ge 1$\,PeV           & 1113 (75.2\%) &  244 (46.7\%) &  288 (58.2\%) &  1645 (65.9\%) \tabularnewline\hline
total                        & 11991 (86.8\%) & 5268 (67.8\%) & 5521 (57.7\%) & 22780 (73.1\%) \tabularnewline
\hline
\end{tabular}
\caption[Number of events after $\Delta E$ cut]{Number of good, moderate and bad events in the three
  energy bins, after the cut on $|\log_{10}(E_{calc}/E_{reco})|$. The percentages were calculated with respect to
  the number of events after the reconstruction, as displayed in
  Table~\ref{tab:beforecuts}.}
\label{tab:Econd}
\end{table}

\section{Large Energies}\label{sec:large_energies}

Due to the larger total number of hits and the larger hit amplitudes, the pattern matching algorithm 
improves for energies in the TeV region or higher. Also, the suppression of the optical
background shows the best efficiency at high energies; and finally, the atmospheric muon background
decreases significantly if only high-energy events are considered. Therefore, a cut is proposed at
a minimum reconstructed shower energy of 5\,TeV. \\
The results of this cut are shown in Table~\ref{tab:large_E}. As expected, nearly all events with a
true shower energy above 10\,TeV pass the cut. Most of the events remaining in the lowest energy bin
have true shower energies above 5\,TeV, though there are a few misreconstructed events with lower
true energies which pass the cut (see Figures~\ref{fig:effpur_NC} and~\ref{fig:effpur_CC} at the end
of the chapter). 

\begin{table}[h] \centering 
\begin{tabular}{|>{\centering}p{3.7cm}|>{\raggedleft}p{2.4cm}|>{\raggedleft}p{2.2cm}|>{\raggedleft}p{2.2cm}|>{\raggedleft}p{2.4cm}|}
\hline
NC events          & good & moderate & bad & total \tabularnewline \hline
$E_{sh} < 10$\,TeV            &  964 (29.1\%) &  603 (23.9\%) & 1155 (32.4\%) &  2722 (28.9\%) \tabularnewline
10\,TeV $\le E_{sh} < 1$\,PeV & 4071 (99.7\%) & 2046 (96.8\%) & 1655 (85.2\%) &  7772 (95.5\%) \tabularnewline
$E_{sh} \ge 1$\,PeV           &  382 (100\%)  &  150 (98.7\%) &  140 (97.2\%) &   672 (99.1\%) \tabularnewline\hline
total                         & 5417 (69.6\%) & 2799 (58.4\%) & 2950 (52.2\%) & 11166 (61.3\%) \tabularnewline
\hline
\end{tabular}
\begin{tabular}{|>{\centering}p{3.7cm}|>{\raggedleft}p{2.4cm}|>{\raggedleft}p{2.2cm}|>{\raggedleft}p{2.2cm}|>{\raggedleft}p{2.4cm}|}
\hline
$\nu_e$ CC events & good & moderate & bad & total \tabularnewline \hline
$E_{sh} < 10$\,TeV            & 2009 (32.4\%) &  991 (23.9\%) & 1827 (28.9\%) &  4827 (29.0\%) \tabularnewline
10\,TeV $\le E_{sh} < 1$\,PeV & 6113 (99.6\%) & 3037 (97.9\%) & 2360 (85.5\%) & 11510 (95.9\%) \tabularnewline
$E_{sh} \ge 1$\,PeV           & 1479 (99.9\%) &  519 (99.2\%) &  490 (99.0\%) &  2488 (99.6\%) \tabularnewline\hline
total                         & 9601 (69.5\%) & 4547 (58.5\%) & 4677 (48.9\%) & 18825 (60.4\%) \tabularnewline
\hline
\end{tabular}
\caption[Number of events after energy cut]{Number of good, moderate and bad events in the three
  energy bins, after the cut on $E_{reco} > 5$\,TeV. The percentages were calculated with respect to
  the number of events after the reconstruction, as displayed in
  Table~\ref{tab:beforecuts}.}
\label{tab:large_E}
\end{table}

\section{Upgoing Events}\label{sec:upgoing}

For energy regions where the  atmospheric muon background cannot be suppressed
sufficiently by the cut on $\xi$ described above, it is necessary to consider only events which have been
reconstructed as upgoing. As the photomultipliers in the ANTARES experiment look downwards, the
resolution for events from above is generally worse than for those from below. However, because of
the photomultiplier orientation, the event vertices tend to be reconstructed too high, i.e.~with a too 
large $z$ coordinate (see Figure~\ref{fig:ortsfehler} in Section~\ref{sec:pos}). In such cases,
a downgoing direction is favoured in the pattern matching. 
Events with a large positional error are therefore often reconstructed as downgoing.
$12\%$ of the upgoing, but only $5\%$ of the downgoing events were reconstructed with a
wrong orientation, i.e.~downgoing instead of upgoing or vice-versa. \\
The results of selecting only events which have been reconstructed as upgoing from the shower event
samples are listed in Table~\ref{tab:upgoing}. As the primary neutrinos were generated isotropically in 
$4 \pi$, one can see that up to 1\,PeV the upgoing events are reconstructed more efficiently than the
downgoing ones: The percentage of remaining good or moderate events is larger than
that of the remaining bad events. \\ 
Considering the atmospheric muon sample, it turns out that 18\% of the events remaining
in the sample after the muon reconstruction are reconstructed as upgoing by the shower reconstruction
strategy (see Table~\ref{tab:atm_muon_cut345}).
Therefore, even if only upgoing events are studied, the background contamination is still 
considerably high, at least below $\sim 5$\,TeV, as will be shown in Chapter~\ref{sec:results}.
Furthermore, the advantage of neutrino-induced showers to be able to study downgoing events, 
would be lost by such a cut.

\begin{table}[h]  \centering
\begin{tabular}{|>{\centering}p{3.7cm}|>{\raggedleft}p{2.4cm}|>{\raggedleft}p{2.2cm}|>{\raggedleft}p{2.2cm}|>{\raggedleft}p{2.4cm}|}
\hline
NC events         & good & moderate & bad & total \tabularnewline \hline
$E_{sh} < 10$\,TeV            & 2182 (65.8\%) & 1463 (57.9\%) & 1696 (47.5\%) & 5341 (56.7\%) \tabularnewline
10\,TeV $\le E_{sh} < 1$\,PeV & 2351 (57.6\%) &  927 (43.9\%) &  768 (39.5\%) & 4046 (49.7\%) \tabularnewline
$E_{sh} \ge 1$\,PeV           &  187 (49.0\%) &   69 (45.4\%) &   68 (47.2\%) &  324 (47.8\%) \tabularnewline\hline
total                         & 4720 (60.7\%) & 2459 (51.3\%) & 2532 (44.8\%) & 9711 (53.3\%) \tabularnewline
\hline
\end{tabular}
\begin{tabular}{|>{\centering}p{3.7cm}|>{\raggedleft}p{2.4cm}|>{\raggedleft}p{2.2cm}|>{\raggedleft}p{2.2cm}|>{\raggedleft}p{2.4cm}|}
\hline
$\nu_e$ CC events & good & moderate & bad & total \tabularnewline \hline
$E_{sh} < 10$\,TeV            & 4143 (66.8\%) & 2410 (58.1\%) & 3054 (48.3\%) &  9607 (57.6\%) \tabularnewline
10\,TeV $\le E_{sh} < 1$\,PeV & 3515 (57.3\%) & 1348 (43.4\%) & 1128 (40.9\%) &  5991 (49.9\%) \tabularnewline
$E_{sh} \ge 1$\,PeV           &  700 (47.3\%) &  255 (48.8\%) &  240 (48.5\%) &  1195 (47.8\%) \tabularnewline\hline
total                         & 8358 (60.5\%) & 4013 (51.6\%) & 4422 (46.2\%) & 16793 (53.9\%) \tabularnewline
\hline
\end{tabular}
\caption[Number of events after cut on neutrino orientation]{Number of good, moderate and bad events
  in the three energy bins, after checking if the neutrino was reconstructed as upgoing. The
  percentages were calculated with respect to the number of events after the reconstruction,
  as displayed in Table~\ref{tab:beforecuts}.}
\label{tab:upgoing}
\end{table}

\section{Cuts Based on Geometrical Considerations}\label{sec:pos_cuts}

In this section, the results on studies of the suppression of poorly reconstructed events by
selecting events by their position and direction with respect to the instrumented volume are presented.
Even though events with an interaction vertex outside the detector, or events pointing to
the outside of the detector, are generally more difficult to reconstruct, there are
actually a lot of events which show very satisfying results. It turns out that, attempting to
suppress events which are geometrically unfavourable for the reconstruction, also a large fraction
of well reconstructed events are rejected. A reason for this is that, due to the usage of the filter
conditions on all events (see Chapter~\ref{ch:trigger}), those events which really produce only a
small signal, because of their disadvantageous position and direction with respect to the detector,
are not considered for the reconstruction in the first place. \\ 
In the following two subsections, selection criteria for geometrically unfavourable events are 
introduced. Due to the unsatisfying efficiency of these cuts, none of them were used in the final
event selection.

\subsection{Events Leaving the Instrumented Volume}\label{sec:leaving}

\begin{figure}[h] \centering
\includegraphics[width=10cm]{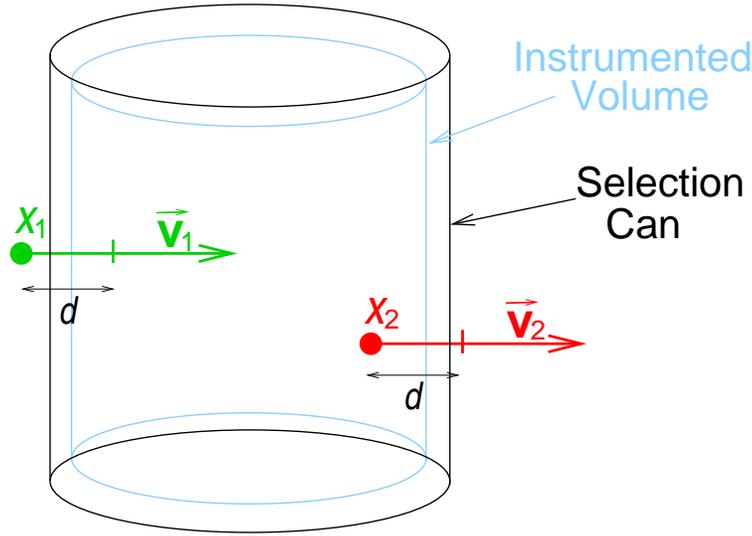}
\caption[Principle of the cut on events leaving the detector]{Geometrical principle of the cut on
  events leaving the detector. The green event labelled 1 is accepted, while the red one labelled 2
  is rejected.} 
\label{fig:leaving}
\end{figure}

As has been discussed in Section~\ref{sec:topologies}, the conditions for a successful 
reconstruction are fulfilled if the shower is contained inside the instrumented volume of the
detector or points right into it. A means to check for these geometrical criteria in an event is
to verify whether a point at a distance $d$ from the reconstructed position $X$, along the reconstructed 
direction $\vec{v}$, lies a can of selectable size around
the instrumented volume. The principle of this selection criterion is demonstrated in
Figure~\ref{fig:leaving}: The can inside which the point must lie is marked in black, the
instrumented volume in light blue: Event 1 in green fulfils the criterion, while event 2 in red
does not.  
 
The results of this cut, for a chosen distance $d = 20$\,m and a can which extends 10\,m beyond the
instrumented volume, are shown in Table~\ref{tab:leaving}. For the chosen sizes of $d$ and the
can, the efficiency of this cut is still above 50\% for the good events, 
but on the other hand, also half of the poorly reconstructed events remains in the sample. This
number could be decreased by decreasing the selected can size, but that would
consequently also lead to a stronger suppression of the well reconstructed events and is therefore
not favourable. 

\begin{table}[h]  \centering
\begin{tabular}{|>{\centering}p{3.7cm}|>{\raggedleft}p{2.4cm}|>{\raggedleft}p{2.2cm}|>{\raggedleft}p{2.2cm}|>{\raggedleft}p{2.4cm}|}
\hline
NC events         & good & moderate & bad & total \tabularnewline \hline
$E_{sh} < 10$\,TeV            & 2785 (83.9\%) & 2098 (83.0\%) & 2276 (63.8\%) &  7159 (76.1\%) \tabularnewline
10\,TeV $\le E_{sh} < 1$\,PeV & 2391 (58.6\%) &  808 (38.2\%) &  751 (38.7\%) &  3950 (48.5\%) \tabularnewline
$E_{sh} \ge 1$\,PeV           &  205 (53.7\%) &   34 (22.4\%) &   64 (44.4\%) &   303 (44.7\%) \tabularnewline\hline
total                         & 5381 (69.1\%) & 2940 (61.4\%) & 3091 (54.7\%) & 11412 (62.6\%) \tabularnewline
\hline
\end{tabular}
\begin{tabular}{|>{\centering}p{3.7cm}|>{\raggedleft}p{2.4cm}|>{\raggedleft}p{2.2cm}|>{\raggedleft}p{2.2cm}|>{\raggedleft}p{2.4cm}|}
\hline
$\nu_e$ CC events & good & moderate & bad & total \tabularnewline \hline
$E_{sh} < 10$\,TeV            & 5162 (83.3\%) & 3437 (82.9\%) & 4164 (65.9\%) & 12763 (76.6\%) \tabularnewline
10\,TeV $\le E_{sh} < 1$\,PeV & 3488 (56.9\%) & 1076 (34.7\%) &  974 (35.3\%) &  5538 (46.2\%) \tabularnewline
$E_{sh} \ge 1$\,PeV           &  714 (48.2\%) &  138 (26.4\%) &  195 (39.4\%) &  1047 (41.9\%) \tabularnewline\hline
total                         & 9364 (67.8\%) & 4651 (59.8\%) & 5333 (55.7\%) & 19348 (62.1\%) \tabularnewline
\hline
\end{tabular}
\caption[Number of events after cut on pointing direction]{Number of good, moderate and bad events in the
  three energy bins, after the cut on the pointing direction of the reconstructed shower. The
  percentages were calculated with respect to the number of events after the reconstruction,
  as displayed in Table~\ref{tab:beforecuts}.}
\label{tab:leaving}
\end{table}

\subsection{Events Inside the Instrumented Volume} \label{sec:instrvol}

A method that is very similar to the one described in the previous subsection is to check if
the reconstructed position of an event lies within the instrumented volume, i.e.~if the event is
{\it contained}. An event passes this cut if its radial distance from the centre of the detector is
smaller than 100\,m, and if its vertical distance to the horizontal plane through the detector
centre is less than 175\,m. \\ 
As this method does not check for the direction of the shower, it is less efficient than the one introduced
above; also the purity is lower (see plots at the end of the chapter). \\
The results of the cut on contained events can be seen from Table~\ref{tab:instrVol}.

\begin{table}[h]  \centering
\begin{tabular}{|>{\centering}p{3.7cm}|>{\raggedleft}p{2.4cm}|>{\raggedleft}p{2.2cm}|>{\raggedleft}p{2.2cm}|>{\raggedleft}p{2.4cm}|}
\hline
NC events         & good & moderate & bad & total \tabularnewline \hline
$E_{sh} < 10$\,TeV            & 2369 (71.4\%) & 1831 (72.5\%) & 2173 (60.9\%) &  6373 (67.7\%) \tabularnewline
10\,TeV $\le E_{sh} < 1$\,PeV & 1978 (48.5\%) &  741 (35.1\%) &  686 (35.3\%) &  3405 (41.8\%) \tabularnewline
$E_{sh} \ge 1$\,PeV           &  174 (45.5\%) &   28 (18.4\%) &   59 (41.0\%) &   261 (38.5\%) \tabularnewline\hline
total                         & 4521 (58.1\%) & 2600 (54.3\%) & 2918 (51.6\%) & 10039 (55.1\%) \tabularnewline 
\hline
\end{tabular}
\begin{tabular}{|>{\centering}p{3.7cm}|>{\raggedleft}p{2.4cm}|>{\raggedleft}p{2.2cm}|>{\raggedleft}p{2.2cm}|>{\raggedleft}p{2.4cm}|}
\hline
$\nu_e$ CC events & good & moderate & bad & total \tabularnewline \hline
$E_{sh} < 10$\,TeV            & 4346 (70.1\%) & 2959 (71.3\%) & 3936 (62.3\%) & 11241 (67.4\%) \tabularnewline
10\,TeV $\le E_{sh} < 1$\,PeV & 2845 (46.4\%) &  999 (32.2\%) &  903 (32.7\%) &  4747 (39.6\%) \tabularnewline
$E_{sh} \ge 1$\,PeV           &  603 (40.7\%) &  114 (21.8\%) &  177 (35.8\%) &   894 (35.8\%) \tabularnewline\hline
total                         & 7794 (56.4\%) & 4072 (52.4\%) & 5016 (52.4\%) & 16882 (54.2\%) \tabularnewline 
\hline
\end{tabular}
\caption[Number of events after cut on containment]{Number of good, moderate and bad events in the
  three energy bins, after the cut on containment. The percentages were calculated with respect to
  the number of events after the reconstruction, as displayed in
  Table~\ref{tab:beforecuts}.}
\label{tab:instrVol}
\end{table}

\section{Efficiencies and Purities of the Cuts}\label{sec:effpur_cuts}

The quality of a cut depends on two factors: How many of the {\it good} events remain in the sample
after the cut, and how many of the {\it bad} events are removed. The efficiency $\mathcal{E}$ of a cut is
defined as the ratio between the number of good events after the cut, $n^{good}_c$, and the number of good
events before the cut, $n^{good}$:
 
\begin{alignat*}{2}
\mathcal{E} & = & \frac{n^{good}_c}{n^{good}}.
\end{alignat*}

The purity $\mathcal{P}$, on the other hand, is defined as the number of good events after the cut,
$n^{good}_c$, divided by the total number of events after the cut, $n_c$:

\begin{alignat*}{2}
\mathcal{P} & = & \frac{n^{good}_c}{n_c}.
\end{alignat*}

\begin{figure} \centering
\includegraphics[width=11cm]{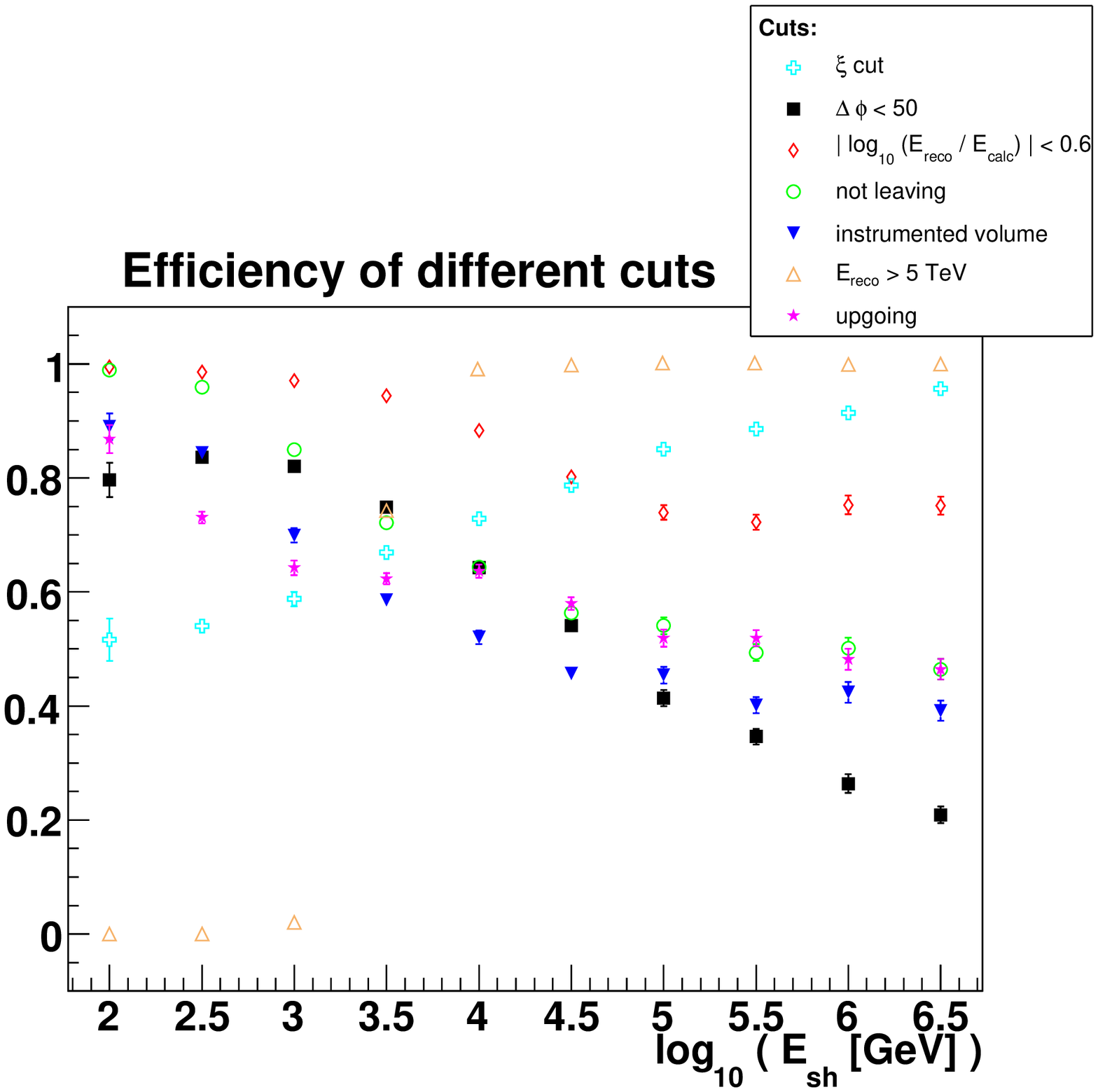}
\includegraphics[width=11cm]{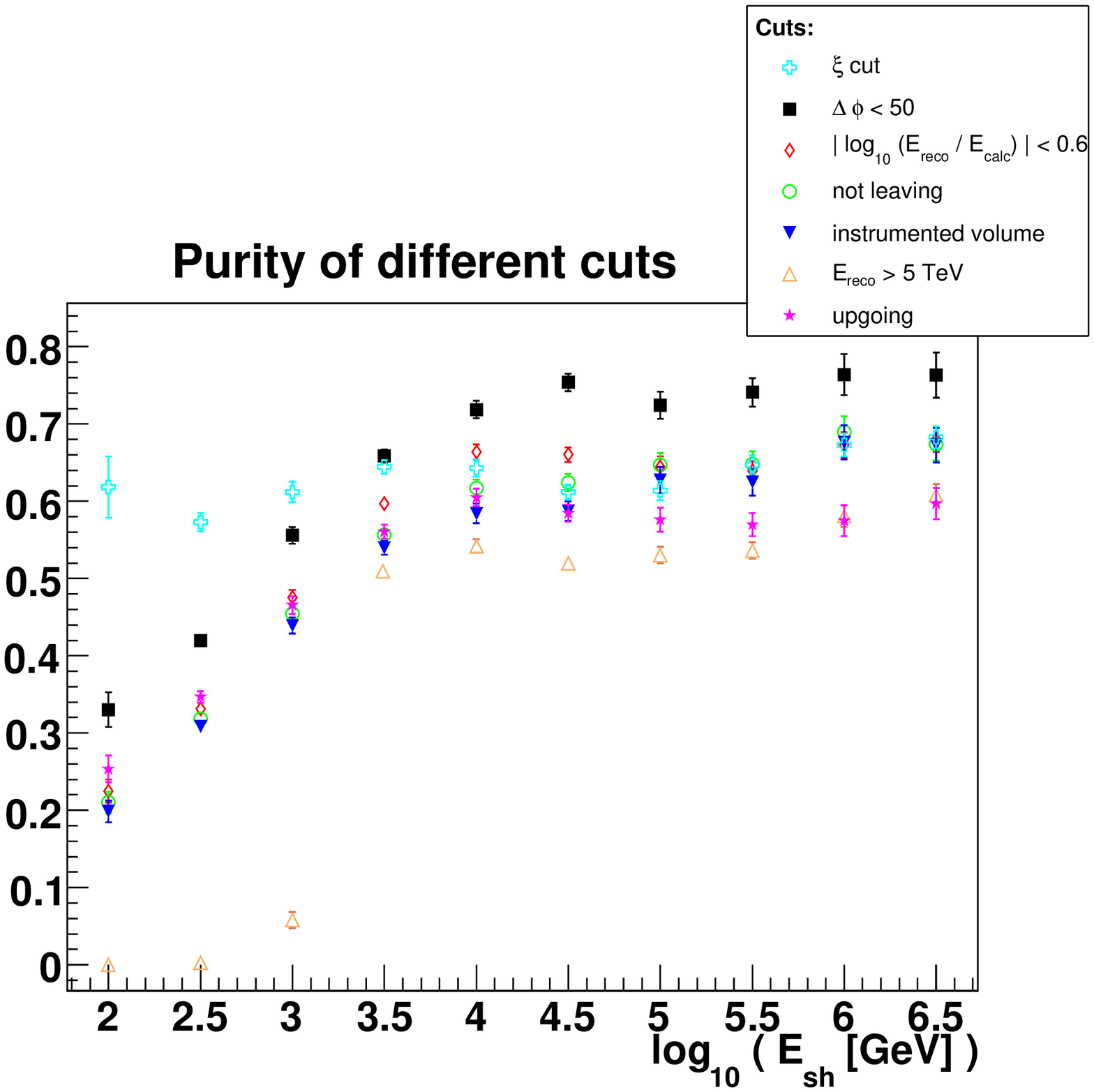}
\caption[Efficiency and purity of the cuts for $\nu_e$ CC events]
{Efficiency (top) and purity (bottom) of the cuts described in the text, for the $\nu_e$ CC sample.}
\label{fig:effpur_CC}
\end{figure}

\begin{figure} \centering
\includegraphics[width=7.4cm]{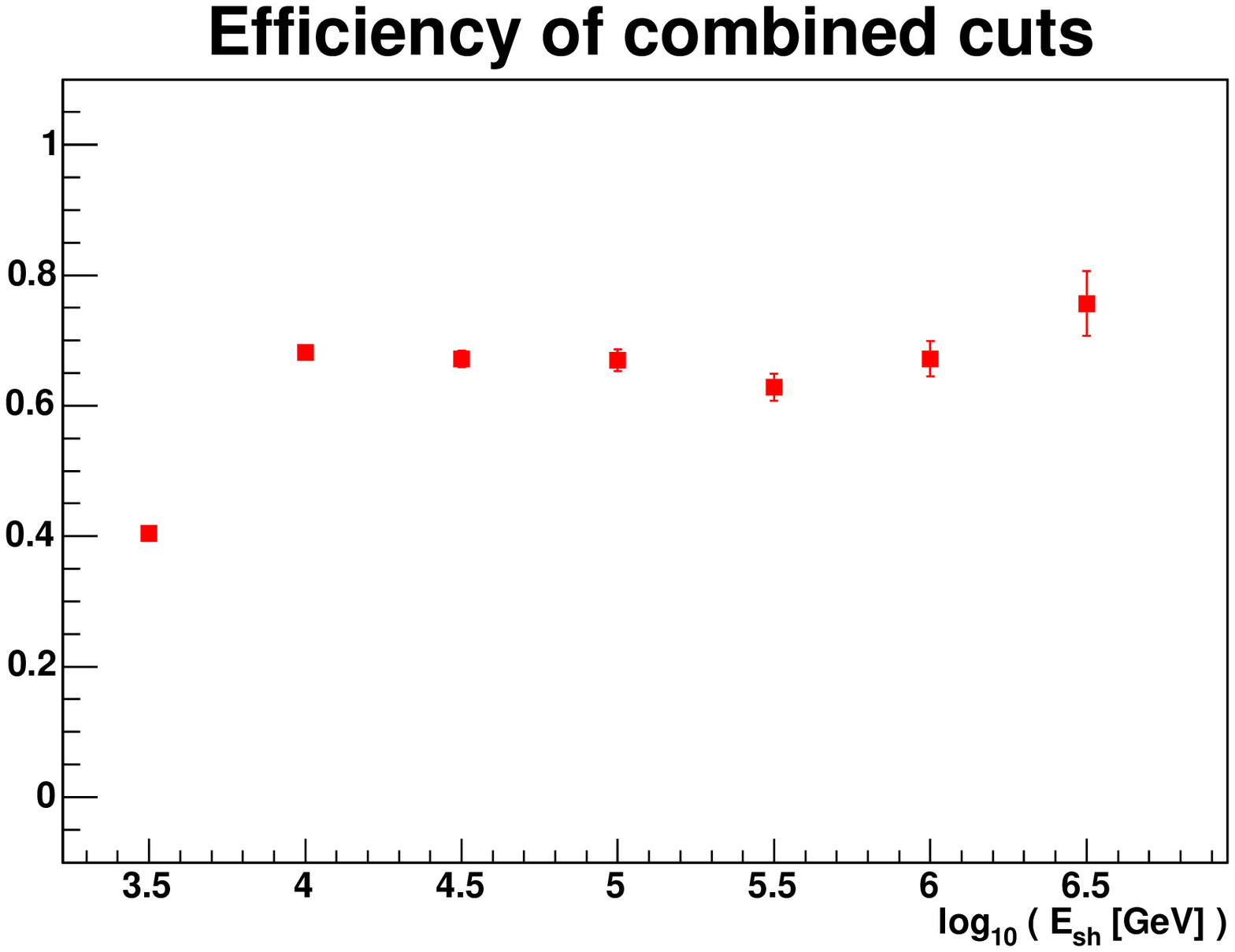}
\includegraphics[width=7.4cm]{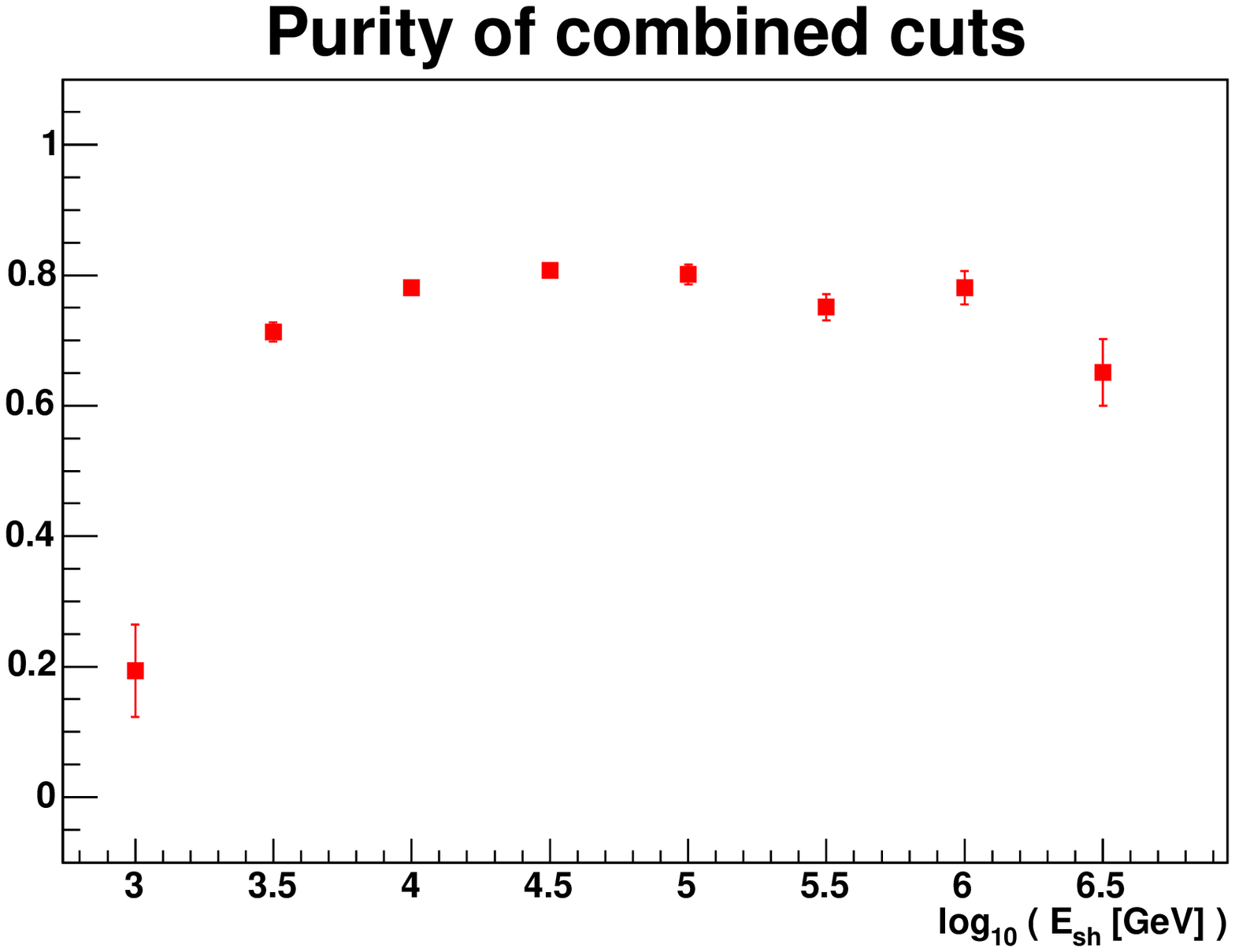}
\caption[Efficiency and purity of the combined cuts, for NC events]
{Combined efficiency (left) and purity (right) of the selected cuts (see text), for the NC sample.}
\label{fig:effpur_comb}
\end{figure}

\begin{figure} \centering
\includegraphics[width=7.4cm]{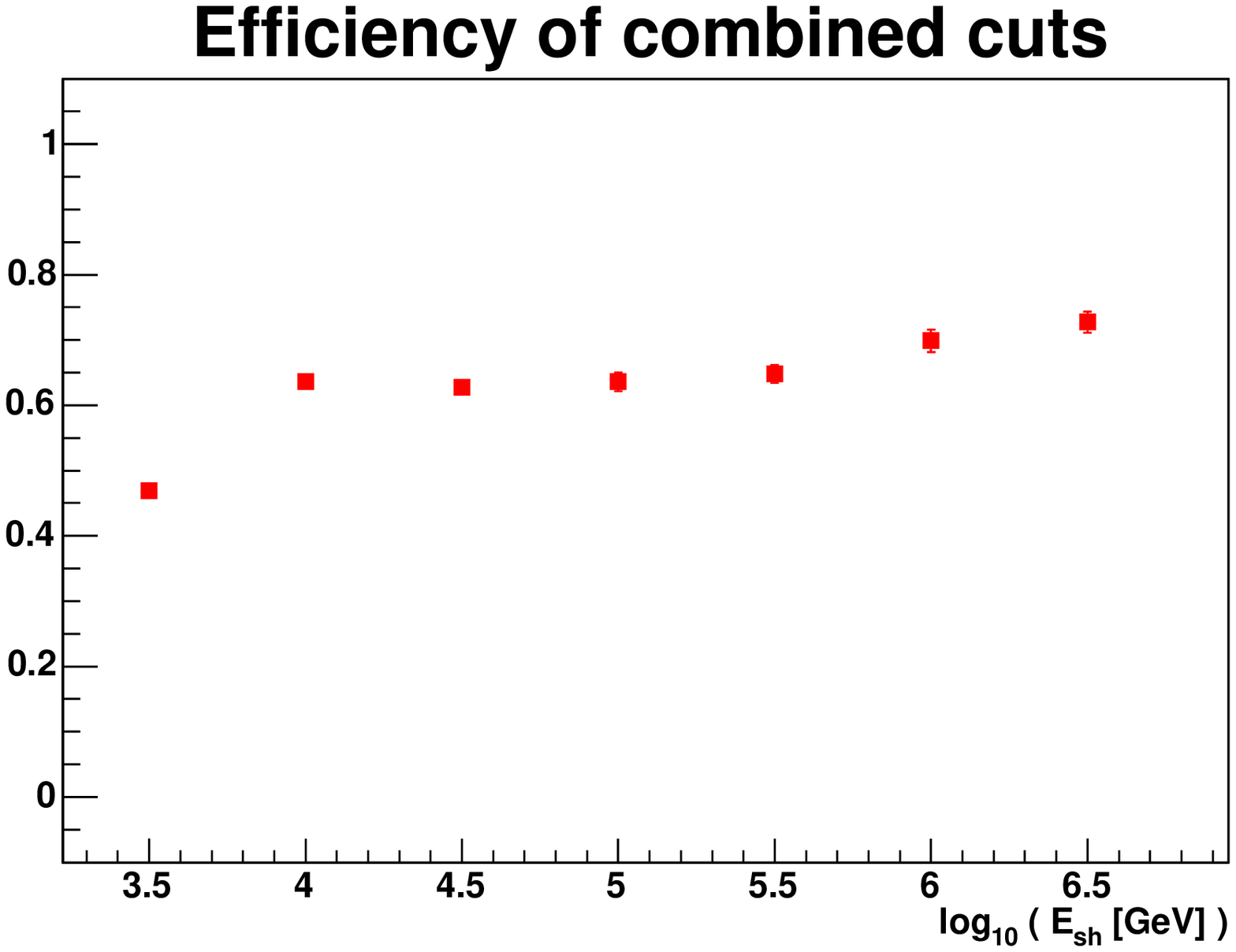}
\includegraphics[width=7.4cm]{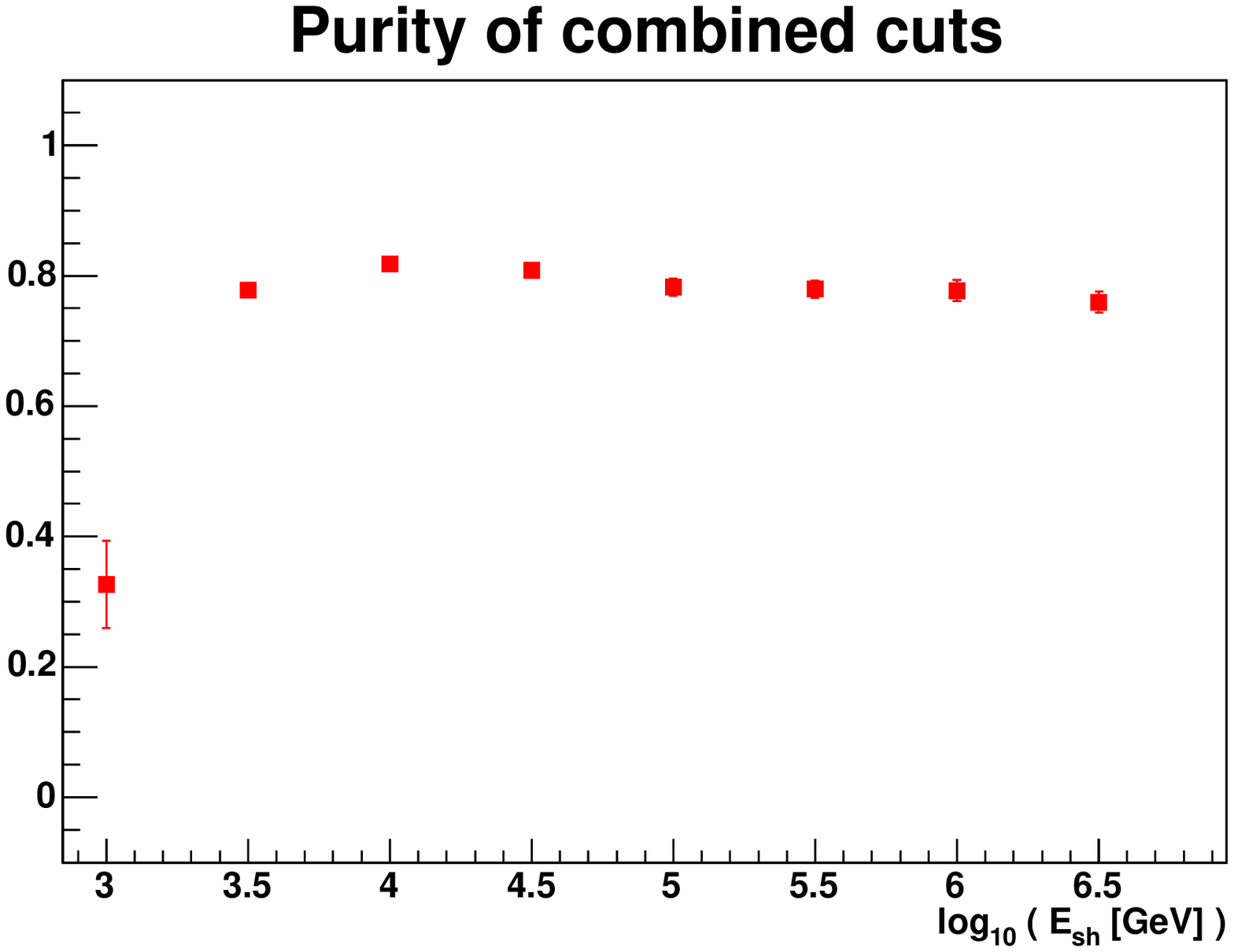}
\caption[Efficiency and purity of the combined cuts, for $\nu_e$ CC events]
{Combined efficiency (left) and purity (right) of the selected cuts (see text), for the $\nu_e$ CC sample.}
\label{fig:effpur_CC_comb}
\end{figure}

For the plots shown in the following, a {\it good} event is defined as described at the beginning
of this chapter: an event reconstructed with a total angular error smaller than $10^{\circ}$. For
this discussion, only {\it good}, and not also the {\it moderate} events, contribute to a high
efficiency and purity, as those events with a small angular error below $10^{\circ}$ are of the 
primary importance for the analysis. \\
For the sample of NC events that was used in this chapter, event sample B, the efficiencies and
purities for the different cuts are given in Figure~\ref{fig:effpur_NC}, together with the
statistical errors (see appendix~\ref{sec:stat_errors} for details on the error calculation). The
efficiencies and purities of the cuts applied to the $\nu_e$ CC sample are shown in
Figure~\ref{fig:effpur_CC}. For the analysis of measured shower events, which will consist of both
types of shower events without a possibility to distinguish between them, the chosen cuts should be
effective on both data samples. All cuts have a very similar effect both to NC and $\nu_e$ CC
events, as can be seen from the figures. \\
Apart from the cut on $\xi$ (see Section~\ref{sec:likeli_cut}), which suppresses the atmospheric 
muon background very efficiently, it was decided to use the cut on the energy comparison 
(see Section~\ref{sec:E_pre_E_fit}), because it shows a very high efficiency throughout the whole 
energy range, combined with a satisfying purity which exceeds 50\% above $\sim 5$~TeV. As the 
energy region above 5\,TeV is the region of main interest, it is also advisable to restrain on 
events with a reconstructed energy above 5\,TeV, as proposed in Section~\ref{sec:large_energies}.  
The combined efficiency and purity of these three cuts are displayed in Figure~\ref{fig:effpur_comb}
for NC events, and in Figure~\ref{fig:effpur_CC_comb} for $\nu_e$ CC events. The low energy region 
has been omitted for these plots. 

\begin{figure} \centering
\includegraphics[width=11cm]{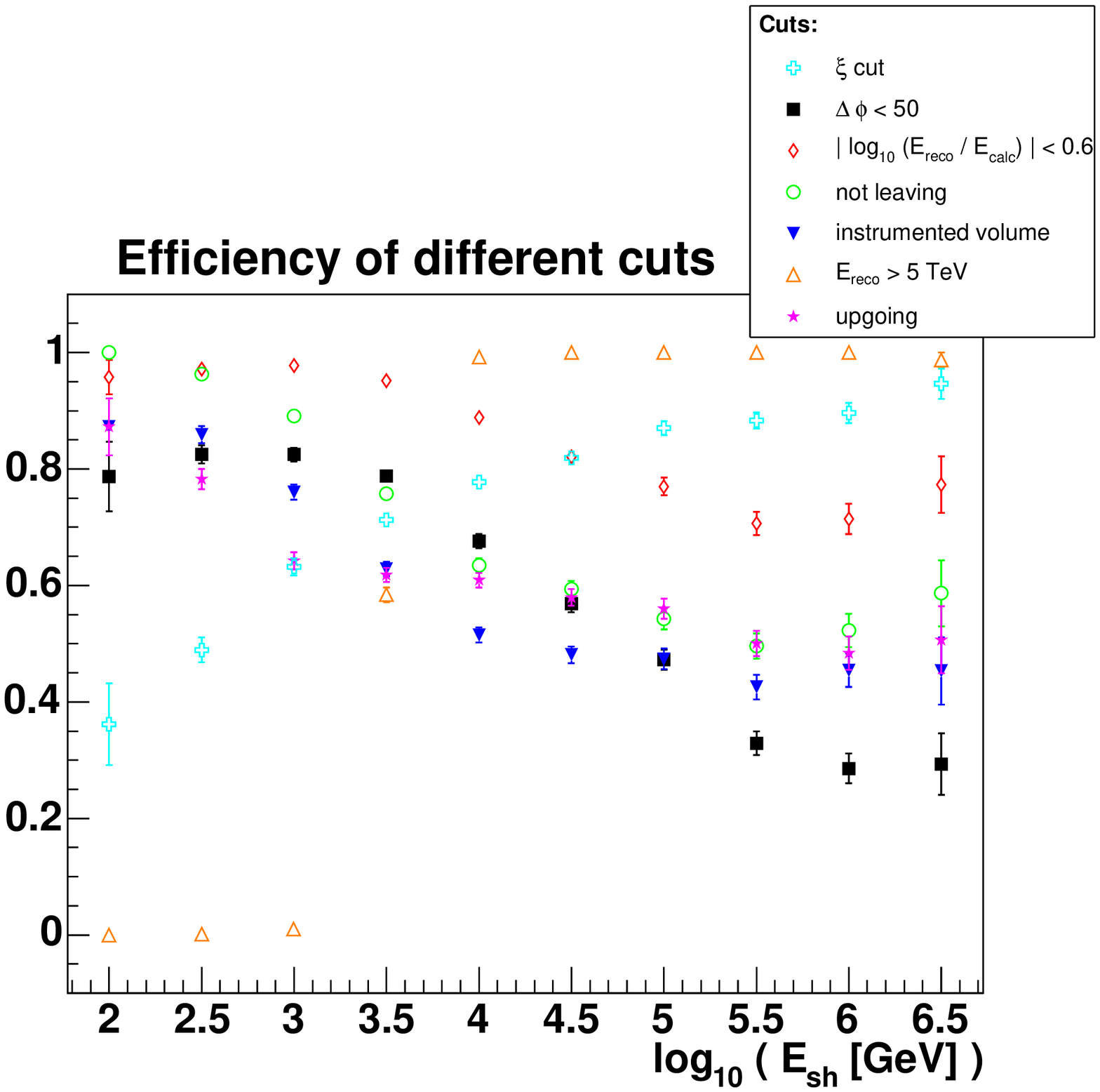}
\includegraphics[width=11cm]{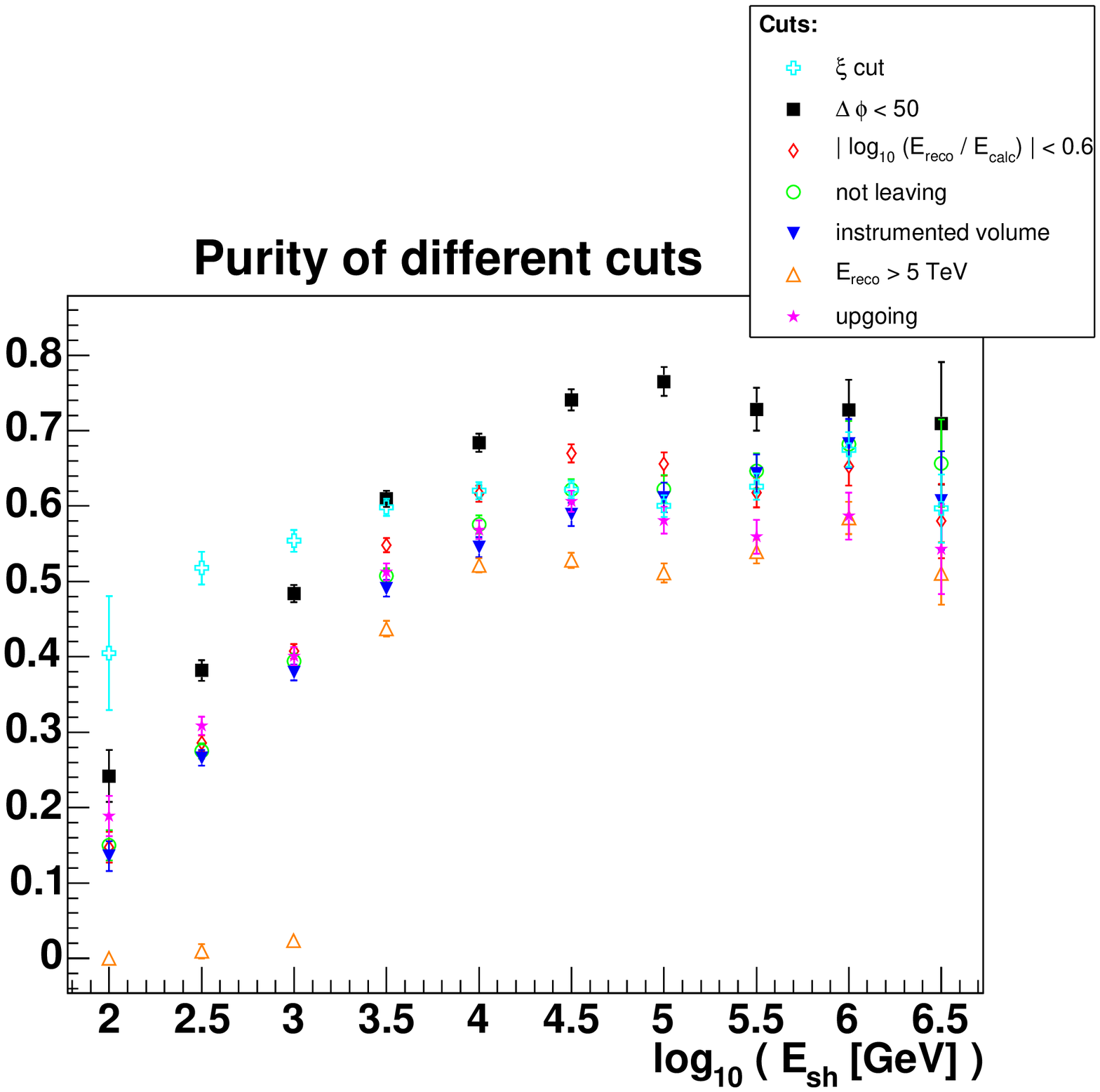}
\caption[Efficiency and purity of the cuts for NC events]
{Efficiency (top) and purity (bottom) of the cuts described in the text, for the NC sample.}
\label{fig:effpur_NC}
\end{figure}

Table~\ref{tab:atm_muon_cut345} lists the numbers of events remaining in the different bins of the
atmospheric muon sample after the cuts described above. The event rates after some of the cuts will
be presented in the next chapter, together with the respective neutrino rates, in
Figure~\ref{fig:event_rates} in Section~\ref{sec:result_flux}. 

\begin{table}[h] \centering 
\begin{tabular}{|>{\centering}p{5.3cm}|>{\centering}p{2.cm}|>{\centering}p{2.cm}|>{\centering}p{2.cm}|}
\hline
angle & \multicolumn{3}{c|}{$0^{\circ} - 60^{\circ}$} \tabularnewline
\hline
primary energy [TeV/nucleon] & 1-10 & 10-100 & 100-10$^5$ \tabularnewline \hline
$|\Delta\phi < 50^{\circ}|$  &  3 (27.3\%) & 11 (34.4\%) &  178 (37.7\%) \tabularnewline \hline  
$|\log_{10}(E_{calc}/E_{reco})| < 0.6$ &  5 (45.5\%) & 14 (43.8\%) & 223 (47.2\%) \tabularnewline \hline  
$E_{reco} > 5$\,TeV          &  0 (0\%) & 6 (18.75\%) & 127 (26.9\%) \tabularnewline \hline
upgoing                      &  2 (18.2\%) &  4 (12.5\%) &  73 (15.5\%) \tabularnewline \hline
not leaving                  & 11 (100\%)  & 28 (87.5\%) & 357 (75.6\%) \tabularnewline \hline
instrumented volume          & 10 (90.9\%) & 27 (84.3\%) & 330 (69.9\%) \tabularnewline \hline
\end{tabular}
\begin{tabular}{|>{\centering}p{5.3cm}|>{\centering}p{2.cm}|>{\centering}p{2.cm}|>{\centering}p{2.cm}|}
\hline
angle & \multicolumn{3}{c|}{$60^{\circ} - 85^{\circ}$} \tabularnewline
\hline
primary energy [TeV/nucleon] & 1-10\,$^{*}$ & 10-100 & 100-10$^5$ \tabularnewline \hline
$|\Delta\phi < 50^{\circ}|$  & 0\,$^{*}$ & 1 (33.3\%) & 95 (43.6\%) \tabularnewline \hline  
$|\log_{10}(E_{calc}/E_{reco})| < 0.6$ & 0\,$^{*}$ & 1 (33.3\%) & 132 (60.6\%) \tabularnewline \hline  
$E_{reco} > 5$\,TeV          & 0\,$^{*}$ & 0 (0\%) & 33 (15.1\%) \tabularnewline \hline 
upgoing                      & 0\,$^{*}$ & 2 (66.7\%) & 42 (19.3\%) \tabularnewline \hline
not leaving                  & 0\,$^{*}$ & 2 (66.7\%) & 178 (81.7\%) \tabularnewline \hline
instrumented volume          & 0\,$^{*}$ & 2 (66.7\%) & 164 (75.2\%) \tabularnewline \hline
\end{tabular}
\caption[Atmospheric muon sample, after cuts]{Numbers of events in the different energy 
  and angular bins of the atmospheric muon event sample, after the cuts described in
  Sections~\ref{sec:phi_diff} to~\ref{sec:instrvol}. The percentages were calculated with respect to
  the number of events after the cut on $\xi$, as displayed in Table~\ref{tab:atm_muon_reco}. The
  lowest energy horizontal bin, marked with an asterisk, contained zero events after the cut on $\xi$.}
\label{tab:atm_muon_cut345}
\end{table}


\chapter{Results of the Reconstruction}\label{sec:results}

In this chapter, the results for the reconstruction of different shower-type event samples are
presented. The results are based on the assumption that the saturation in the photomultiplier
electronics is reached at 200\,pe, i.e.~that data are taken in WF mode, except for the results
presented in Section~\ref{sec:SPE_results}, which were obtained under the assumption of data taking
in SPE mode. In the first section, Section~\ref{sec:nc_results}, results for the reconstruction of
NC events are presented; results for the reconstruction of $\nu_e$ CC events follow in
Section~\ref{sec:nu_e_CC}. The results are shown after the cut on a reconstructed shower energy $>
5$\,TeV (see Section~\ref{sec:large_energies}), and after the additional cuts on the variable $\xi$
comparing expected and measured hit amplitudes (see Section~\ref{sec:likeli_cut}) and on the
comparison between reconstructed and calculated energy (see Section~\ref{sec:E_pre_E_fit}). \\
Sections~\ref{sec:result_effarea} and~\ref{sec:result_flux} show the
effective areas obtained with the reconstruction algorithm presented in this thesis, together with
calculations of the sensitivity of the ANTARES experiment to the diffuse cosmic neutrino fluxes,
using neutrino-induced showers. 

\section{Reconstruction of NC Events}\label{sec:nc_results}

\begin{figure}[h]  
\includegraphics[width=7.4cm]{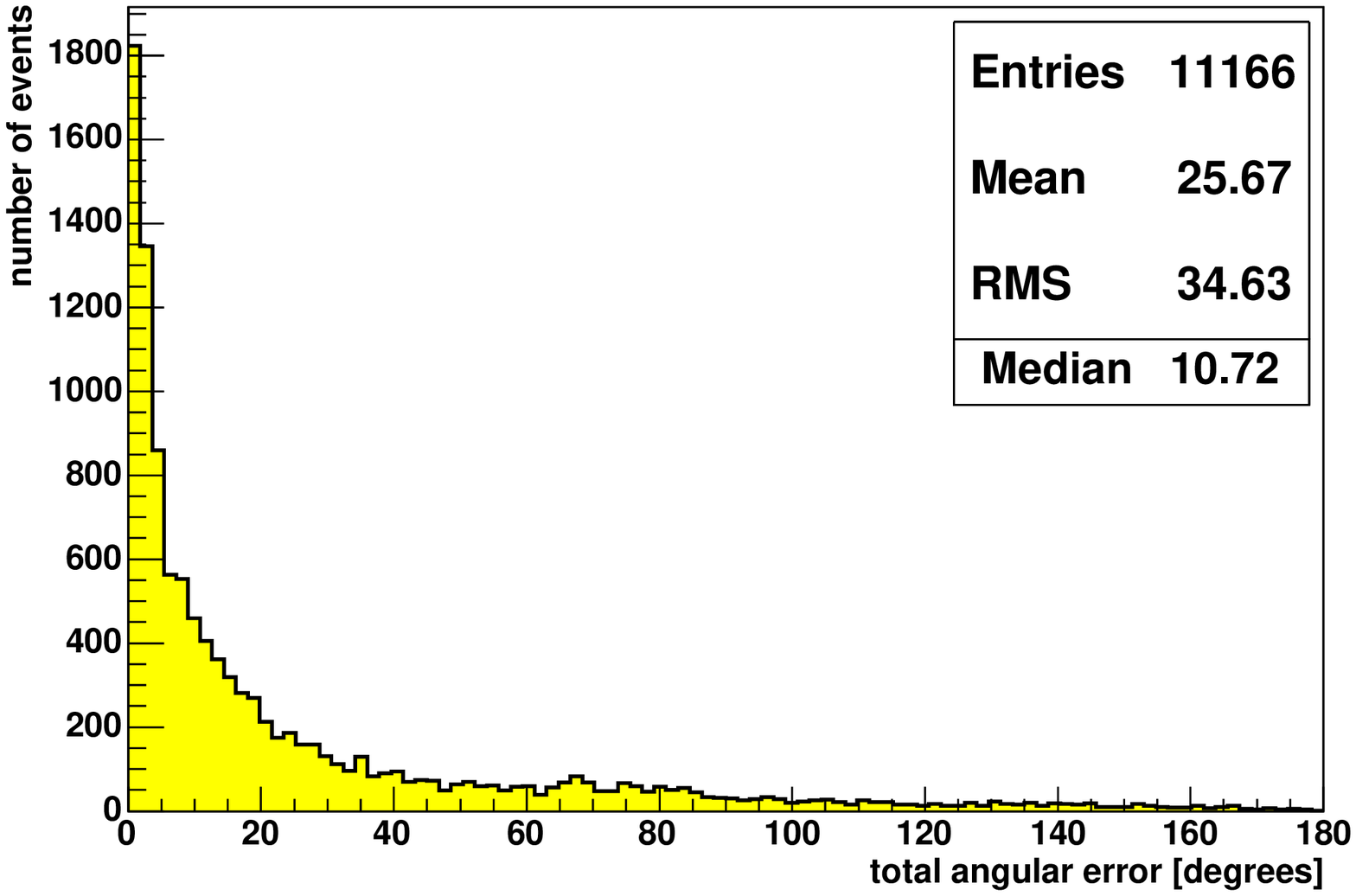}
\includegraphics[width=7.4cm]{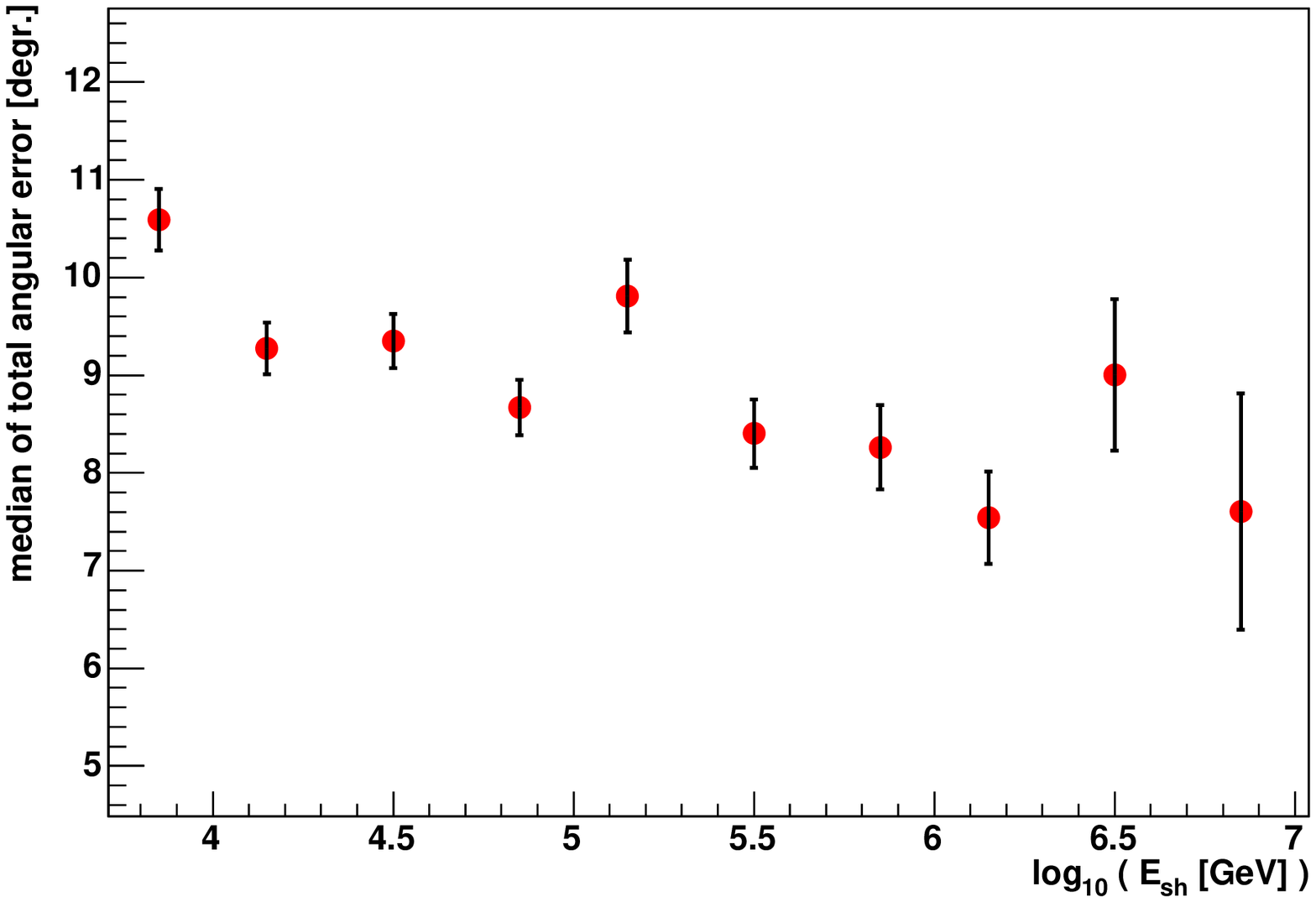}
\caption[Results for direction reconstruction, NC events]{Total angular error $\Delta\alpha$ between
  neutrino and reconstructed direction, after the cut on $E_{reco} > 5$\,TeV: Distribution for all 
  energies (left), and median of total angular error for different MC shower energy bins (right). 
  The error bars mark the statistical standard errors whose calculation is described in
  Appendix~\ref{sec:err_median}.}  
\label{fig:alpha_no_cut}
\end{figure}

\begin{figure}[h] \centering 
\includegraphics[width=7.4cm]{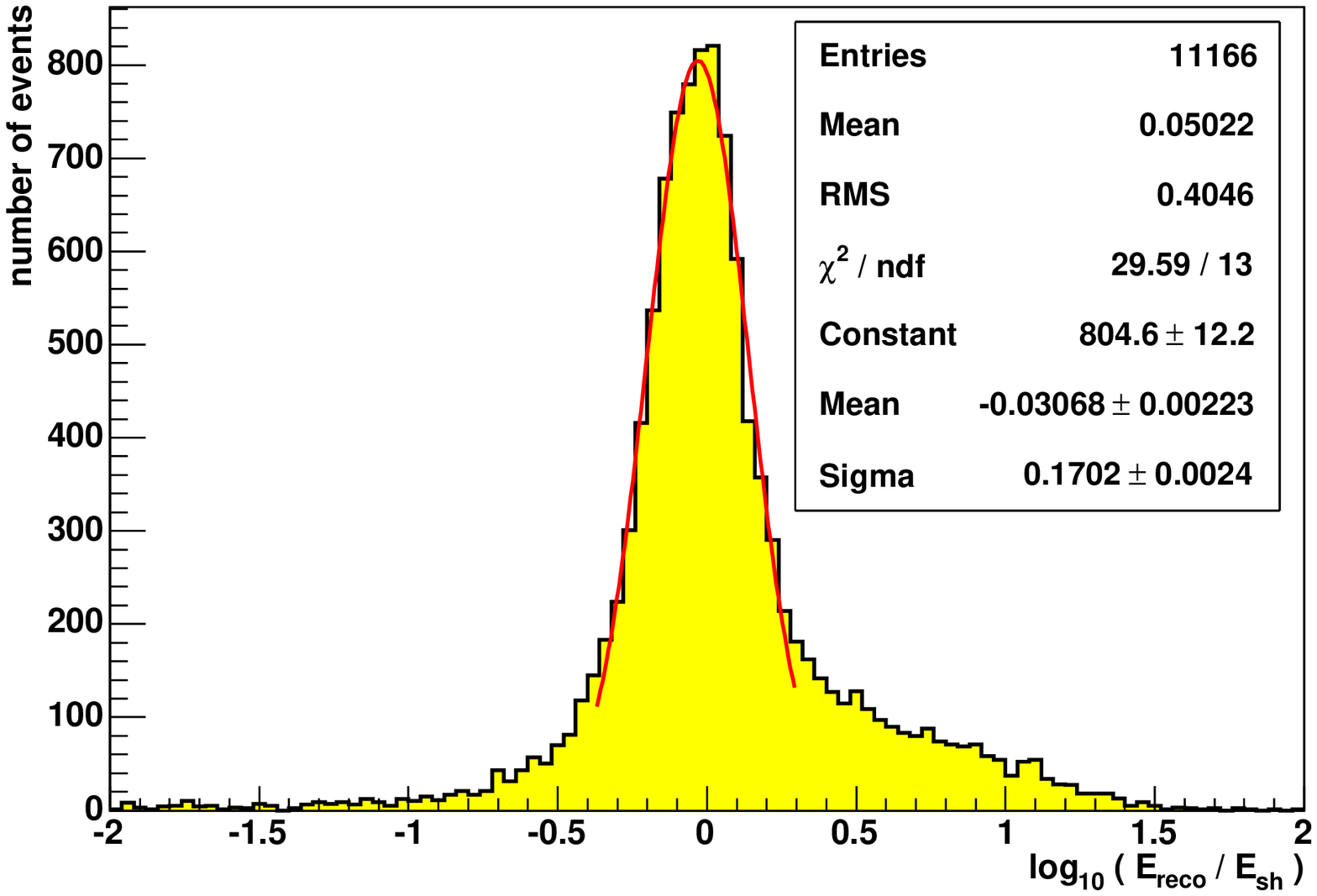}
\includegraphics[width=7.4cm]{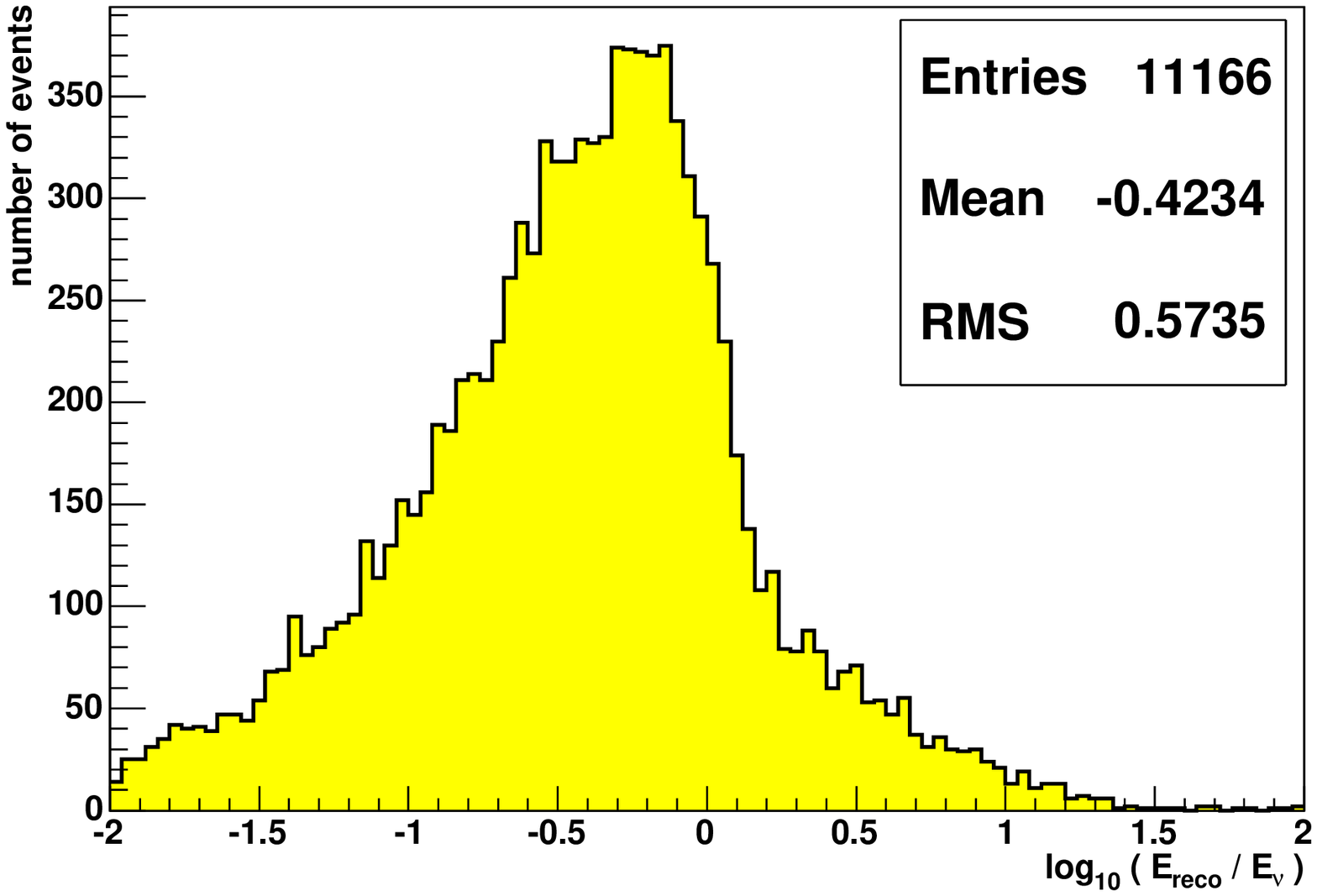}
\caption[Results for energy reconstruction, NC events]{Distribution of the logarithmic difference
    between reconstructed energy and MC shower energy (left), and between reconstructed energy
    and neutrino energy (right), after the cut on $E_{reco} > 5$\,TeV. A Gaussian distribution
    shown in red has been fitted to the peak of the shower energy error.}  
\label{fig:energy_no_cut}
\end{figure}

\begin{figure}[h] \centering  
\includegraphics[width=9cm]{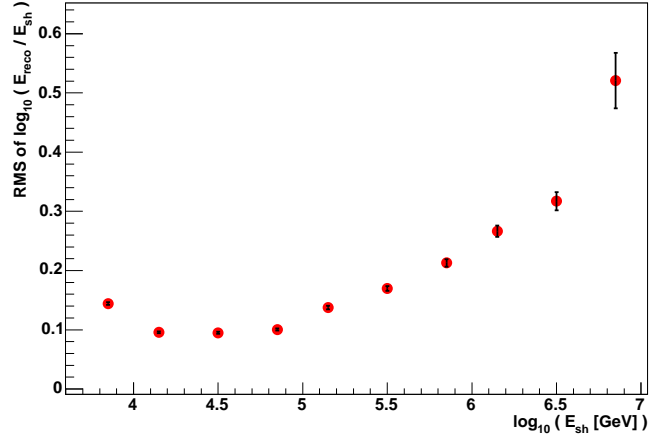}
\caption[Results for energy reconstruction, seperate energy bins]{RMS of the logarithmic error on
  the shower energy, for different bins in the MC shower energy, after the cut on $E_{reco} > 5$\,TeV. }
\label{fig:E_RMS}
\end{figure}

The results shown in this section were obtained reconstructing event sample B which contains 140000
NC events (see Appendix~\ref{sec:nc_sample} for details). Some 18000 events remain in the sample
after the reconstruction. In Figures~\ref{fig:alpha_no_cut} and \ref{fig:energy_no_cut}, results
after the appliance of the cut selecting events with a reconstructed shower energy above 5\,TeV are shown.
Figure~\ref{fig:alpha_no_cut} shows the total angular error $\Delta\alpha$
between the neutrino direction and the reconstructed direction, in degrees, on the left hand side
for all energies, and on the right hand side the median of $\Delta\alpha$ for different bins in the
MC shower energy. \\
In Figure~\ref{fig:energy_no_cut} the distribution of the logarithmic error
on the reconstructed shower energy is shown on the left hand side, and the distribution of the
logarithmic error on the neutrino energy is shown on the right. As the fraction of the primary
neutrino energy which is transferred into the hadronic shower is not known, only a lower limit can
be given for the reconstructed neutrino energy (see comments in
Section~\ref{sec:calc_nu_energy}). The RMS of the logarithmic error on the shower energy 
corresponds to a factor ${10^{0.4} \approx 2.5}$ between the reconstructed and the true shower
energy, though the peak is narrower, with a width of $\sim \ncElog$, obtained from the Gaussian fit
marked by the red line in Figure~\ref{fig:energy_no_cut}. This width corresponds to a factor of
\ncE. The RMS of the logarithmic error on the shower energy, calculated within different MC energy
regions, is displayed in Figure~\ref{fig:E_RMS}. 
 
\begin{figure}[h]  
\includegraphics[width=7.4cm]{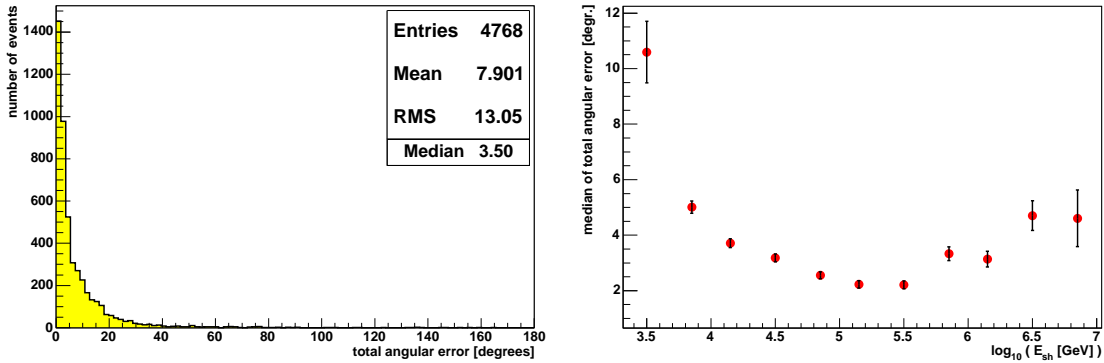}
\includegraphics[width=7.4cm]{alpha_median_cuts.eps}
\caption[Results for direction reconstruction after cuts, NC events]{Total angular error between
  neutrino and reconstructed direction for all energies (left), and median of total angular error
  for different MC shower energy bins (right), with statistical errors, after all three cuts.}
\label{fig:alpha_cut}
\end{figure}

\begin{figure}[h]  
\includegraphics[width=7.4cm]{Ediff_shower_cuts.eps}
\includegraphics[width=7.4cm]{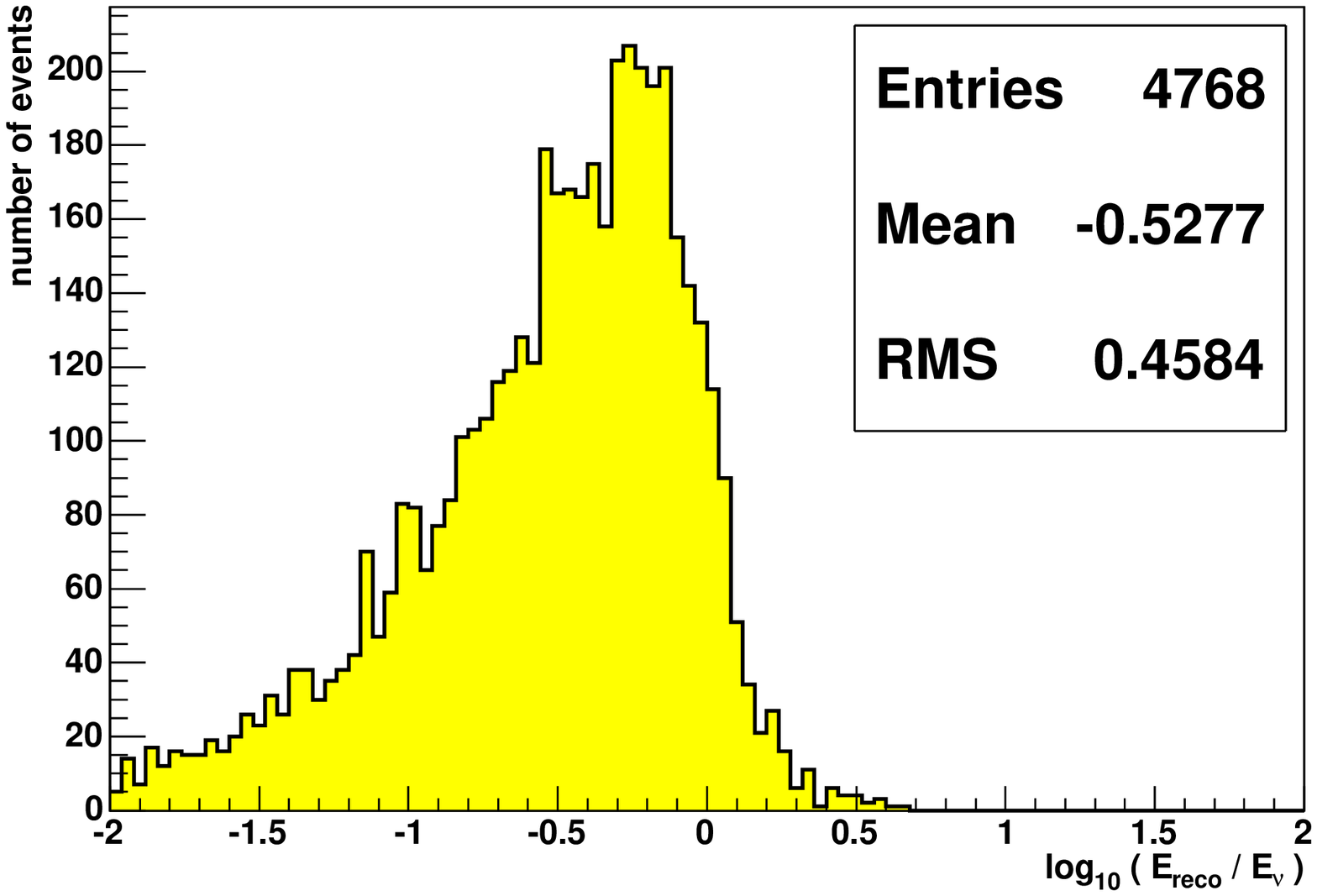}
\caption[Results for energy reconstruction after cuts, NC events]{Distribution of the logarithmic error between
    reconstructed energy and MC shower energy (left), and between reconstructed energy and
    neutrino energy (right), after all three cuts.}  
\label{fig:energy_cut}
\end{figure}

In addition to the cut on the reconstructed shower energy, two more cuts were chosen according to
their efficiencies and the purities, one cutting on a variable $\xi$ (see
Section~\ref{sec:likeli_cut}) and one on a comparison between reconstructed and calculated shower
energy (see Section~\ref{sec:E_pre_E_fit}). 43\% of the events passing the reconstruction and
having a reconstructed shower energy $> 5$\,TeV survive these additional 
cuts. The results for the error on direction and energy after these cuts are 
shown in figures~\ref{fig:alpha_cut} and~\ref{fig:energy_cut}. Both the angular and the energy 
resolution have improved significantly, from a median of \ncalpha~to \ncalphacut~in the
direction, with values of about \alphabest~around 100\,TeV (see right plot of
Figure~\ref{fig:alpha_cut}), and an improvement from 0.40 to 0.20 in the RMS of the logarithmic
shower energy resolution. The peak of this distribution has a width of \ncEcutlog, which means a
resolution correspoding to a factor of $\ncEcut$.  

\subsection{Results for SPE Mode}\label{sec:SPE_results}

\begin{figure}[h]  
\includegraphics[width=7.4cm]{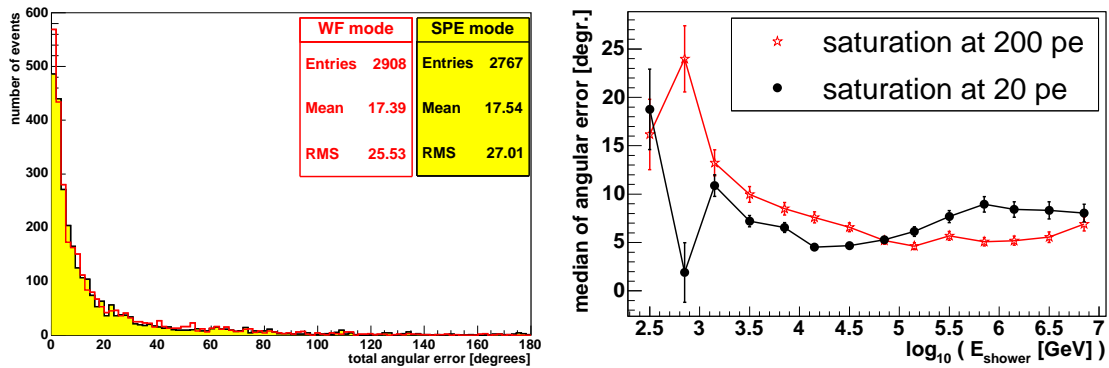}
\includegraphics[width=7.4cm]{alpha_E_SPE_WF.eps}
\caption[Results for direction reconstruction, SPE mode]{Total angular error between neutrino and
  reconstructed direction, for all energies
  (left), and median of total angular error for different MC shower energy bins (right), with statistical
  errors, before all cuts, for SPE and WF mode. On the left plot, the events from WF mode
  are drawn with a red line, as is their statistics, while the events from SPE mode are represented
  by the yellow histogram.} 
\label{fig:alpha_SPE}
\end{figure}

\begin{figure}[h]  
\includegraphics[width=7.4cm]{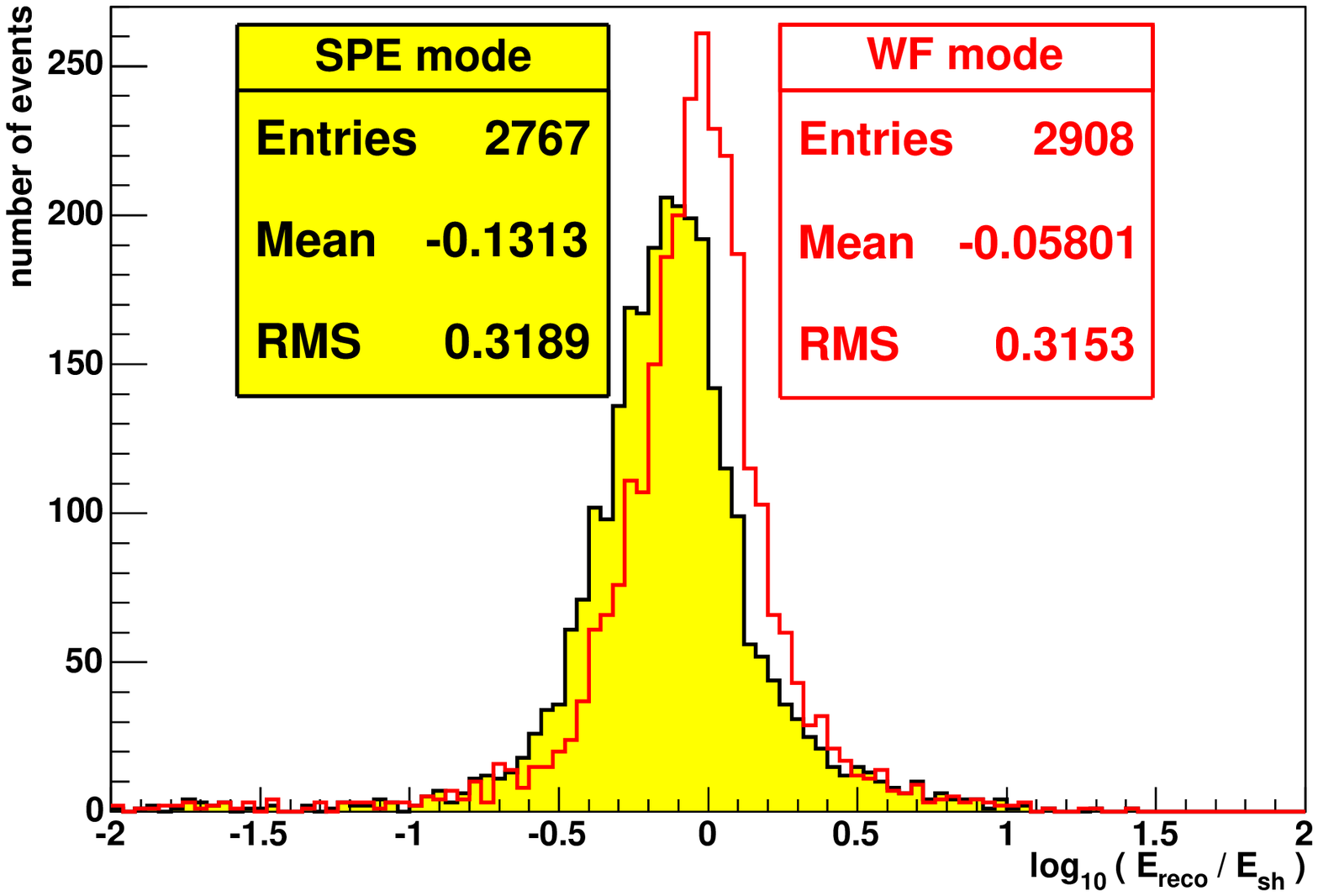}
\includegraphics[width=7.4cm]{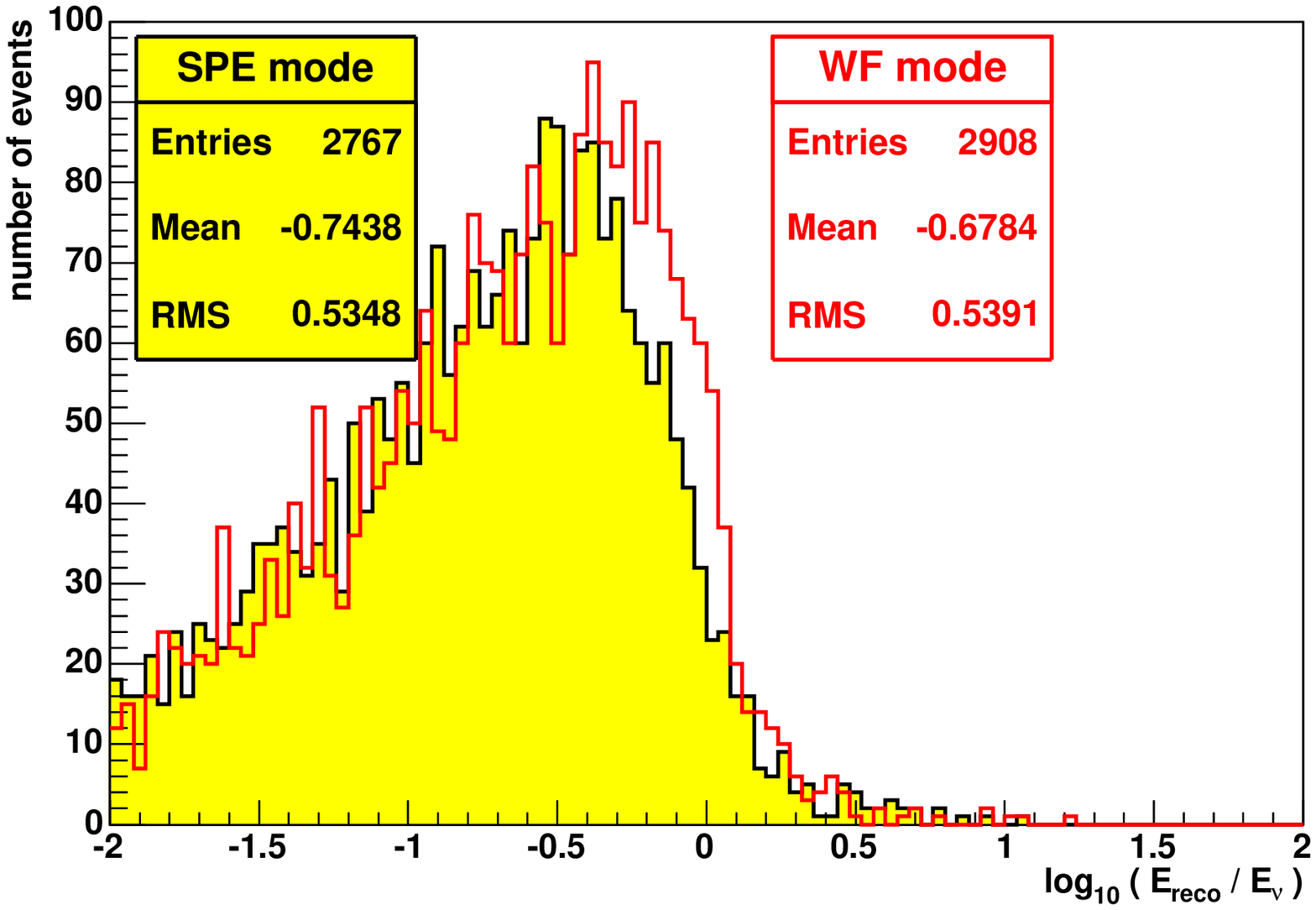}
\caption[Results for energy reconstruction, SPE mode]{Distribution of logarithmic error between
    reconstructed energy and MC shower energy (left), and between reconstructed energy and neutrino
    energy (right), before all cuts, for SPE and WF mode. The events obtained in WF mode are drawn in red,
    those obtained in SPE mode are filled in yellow. The offset for the SPE mode in the
    left plot is caused by the fact that the calculation of the shower energy from the photon number
    is optimised for the WF mode.}  
\label{fig:energy_SPE}
\end{figure}

The default value for the saturation of the photomultiplier electronics in the ShowerFitter
is the value of the WF mode, 200\,pe. However, as discussed in
Section~\ref{sec:digitisation}, data taken in the WF mode need forty times more bandwidth and
storage capacity than in the SPE mode, which has a saturation at 20\,pe. It is therefore not sure if the
ANTARES detector will take a large number of runs in the WF mode. The effect of a lower saturation
level has therefore been examined using the smaller NC event sample A. Note that
in comparison to event sample B, this sample contains only events with interaction vertices
inside the instrumented volume, and is therefore expected to produce better results in the
reconstruction than event sample B. \\
Event sample A has been reconstructed twice, once in WF mode and once in SPE mode. 
As the main principle of the fit is the matching of measured and calculated amplitudes, it is
expected that the results deteriorate when the saturation level is set to a smaller value, since 
this means an information loss. \\
Shown as results for the SPE mode are the total angular error $\Delta\alpha$ in
Figure~\ref{fig:alpha_SPE}, for all energies (left), and the median of $\Delta\alpha$ for the
different energy bins (right). The total number of reconstructed events is smaller than the number
of the reconstructed events in the WF mode, 2767 compared to 2908 events. This is mainly caused by failures
of the fit for high-energy events in the SPE mode, if no satisfying result is found in the minimisation.
The results for the reconstruction of the same sample using the WF mode are shown as
the red lines on the left, and as red stars on the right of Figure~\ref{fig:alpha_SPE}. 
For shower energies below $\sim 70$\,TeV, the resolution for the SPE mode is even slightly better
than for the WF mode, because the statistical fluctuations in the photon distributions which occur
at these smaller energies are damped by lower amplitude saturation. For higher energies, the
resolution for the SPE mode deteriorates, but not much. For the WF mode, a
deterioration of the resolution due to the saturation effects is observed as well, but starting at
higher energies.  \\ 
For the reconstruction of the shower energy, the resolution is equivalent in WF and SPE mode,
see left plot of Figure~\ref{fig:energy_SPE}; the distribution obtained in SPE mode is shifted to
energies reconstructed too low, as expected for a lower saturation (note that the calculation of the
shower energy from the number of photo-electrons in an event was optimised for the WF mode). The
reconstructed neutrino energy shows deteriorations only slightly, see right plot in the same
figure. \\
From this comparison, it can be concluded that, even though the saturation level is significantly
smaller for the SPE mode, the deterioration of the resolution is not large. A good reconstruction 
of shower events is therefore possible also for data taken in the SPE mode. 

\section{Reconstruction of $\nu_e$ CC events}\label{sec:nu_e_CC}

\begin{figure}[h] \centering
\includegraphics[width=7.4cm]{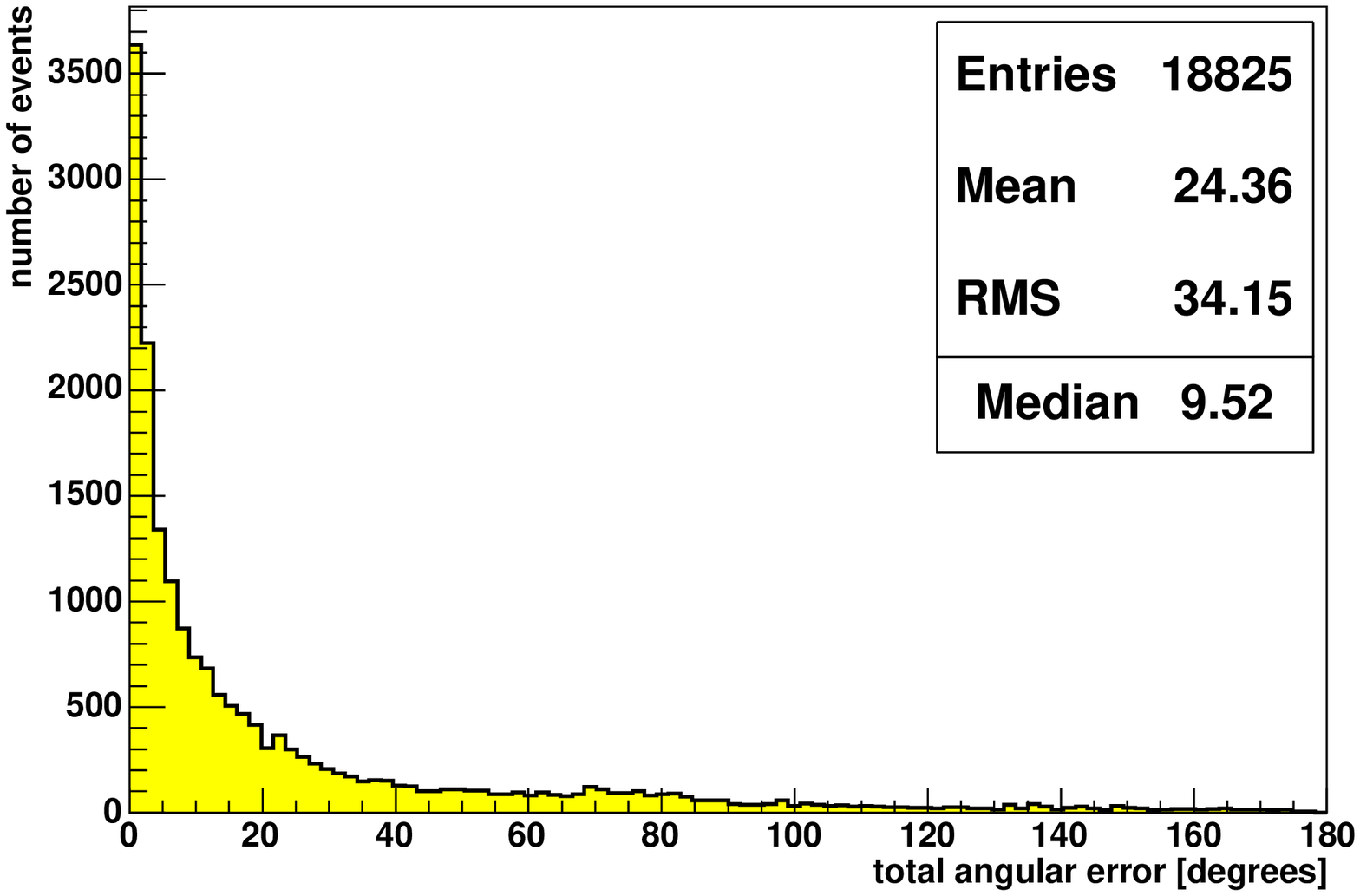}
\includegraphics[width=7.4cm]{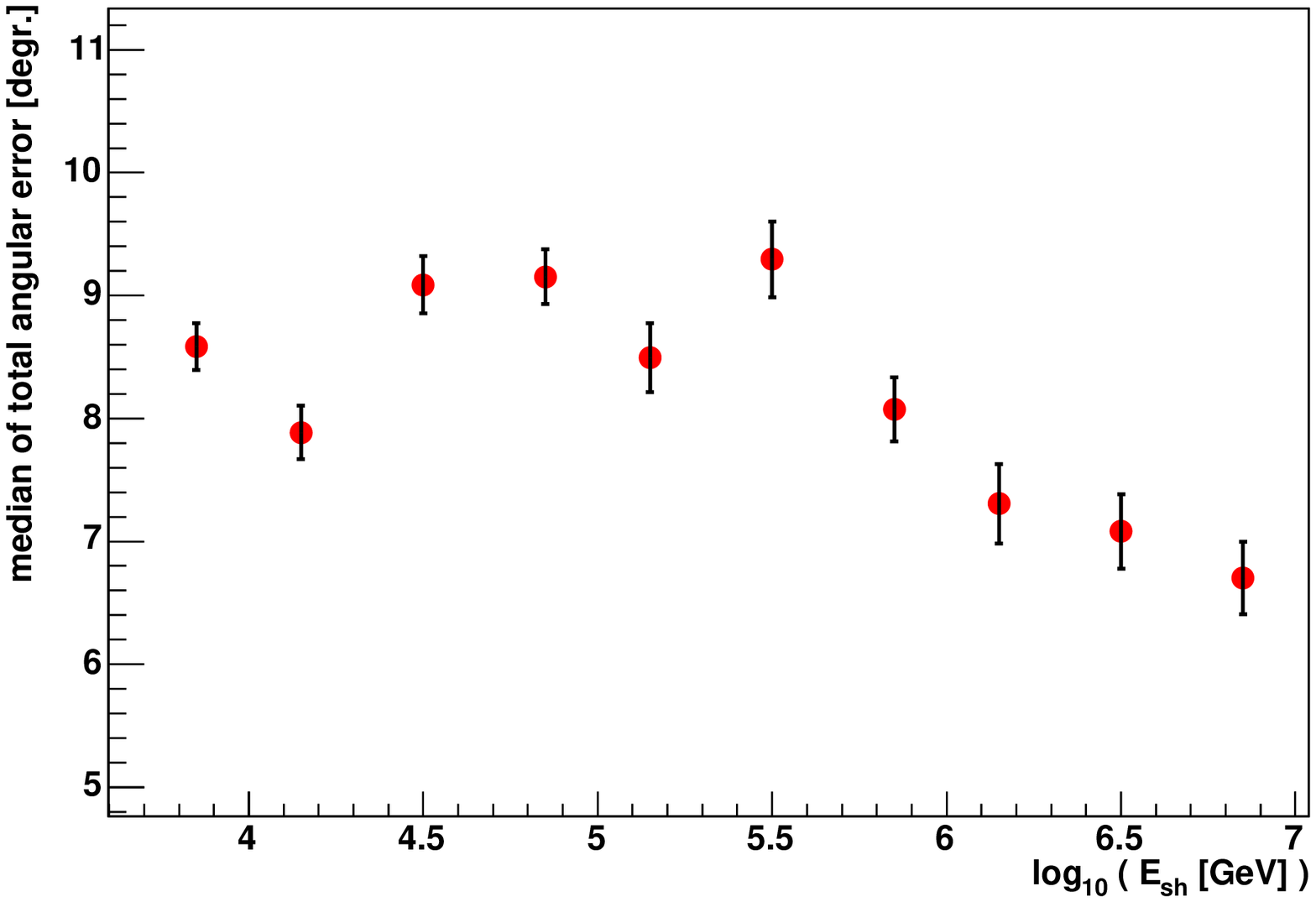}
\caption[Results for direction reconstruction, $\nu_e$ CC events]{Total angular error between
  neutrino and reconstructed direction, for all energies
  (left), and median of total angular error for different MC shower energy bins, with statistical
  errors (right), after the cut on the reconstructed shower energy, for $\nu_e$ CC events.}
\label{fig:nue_alpha_nocuts}
\end{figure}

\begin{figure}[h] \centering
\includegraphics[width=7.4cm]{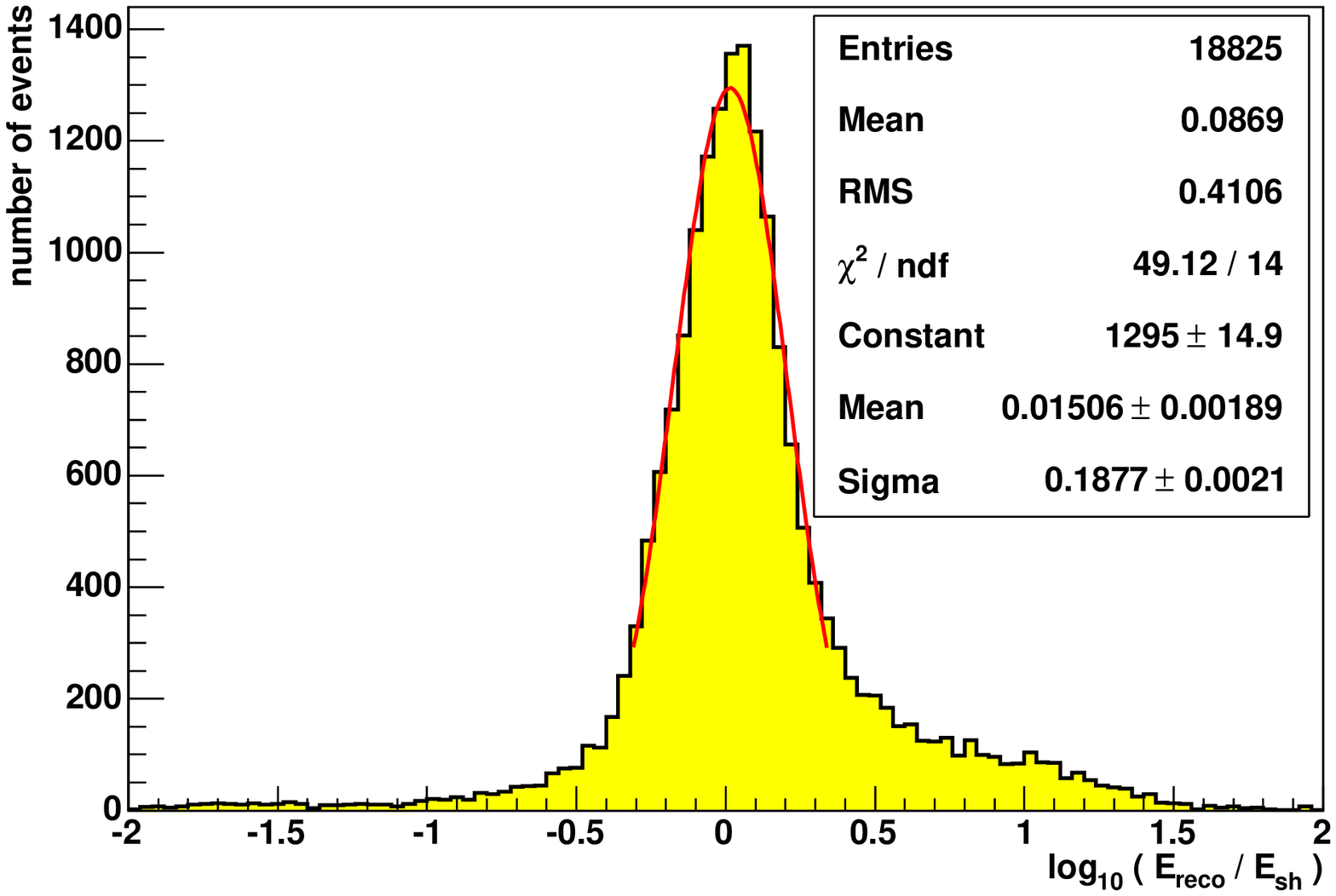}
\includegraphics[width=7.4cm]{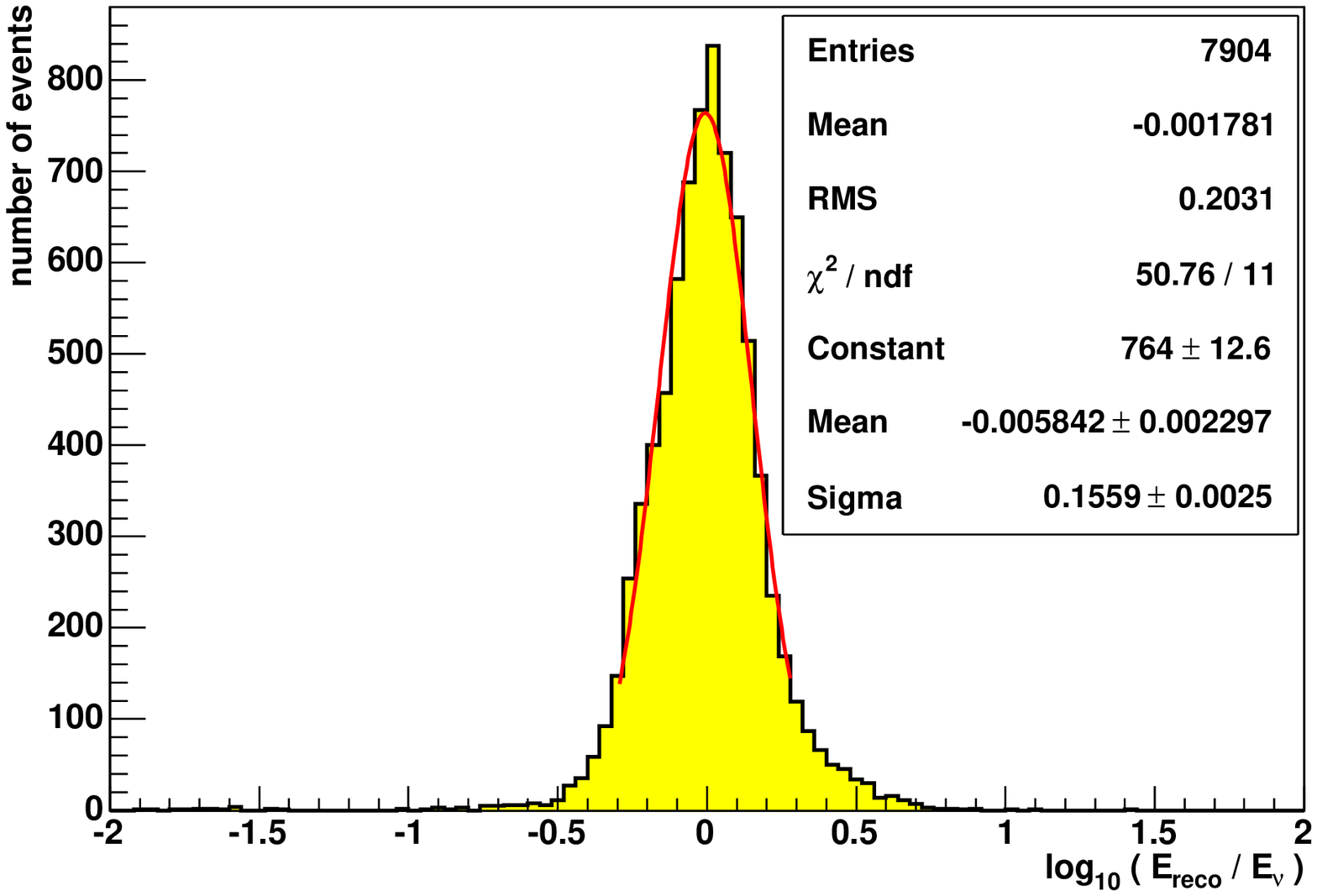}
\caption[Results for energy reconstruction, $\nu_e$ CC events]{Logarithmic error between
    reconstructed and MC shower (=neutrino) energy for $\nu_e$ CC events, after the cut on the
    reconstructed shower energy (left), and after the additional two cuts (right).} 
\label{fig:nue_E}
\end{figure}

\begin{figure}[h] \centering
\includegraphics[width=7.4cm]{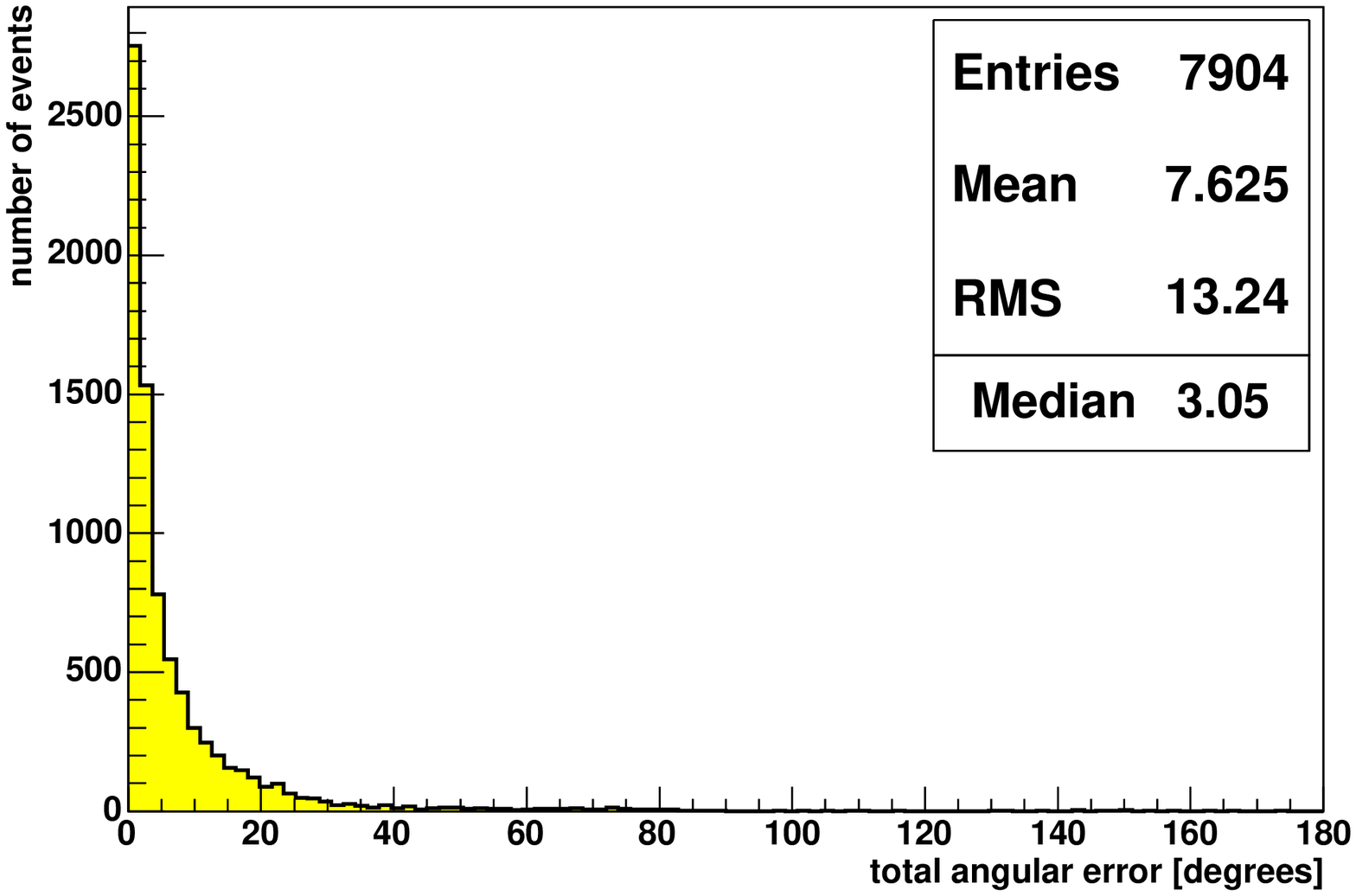}
\includegraphics[width=7.4cm]{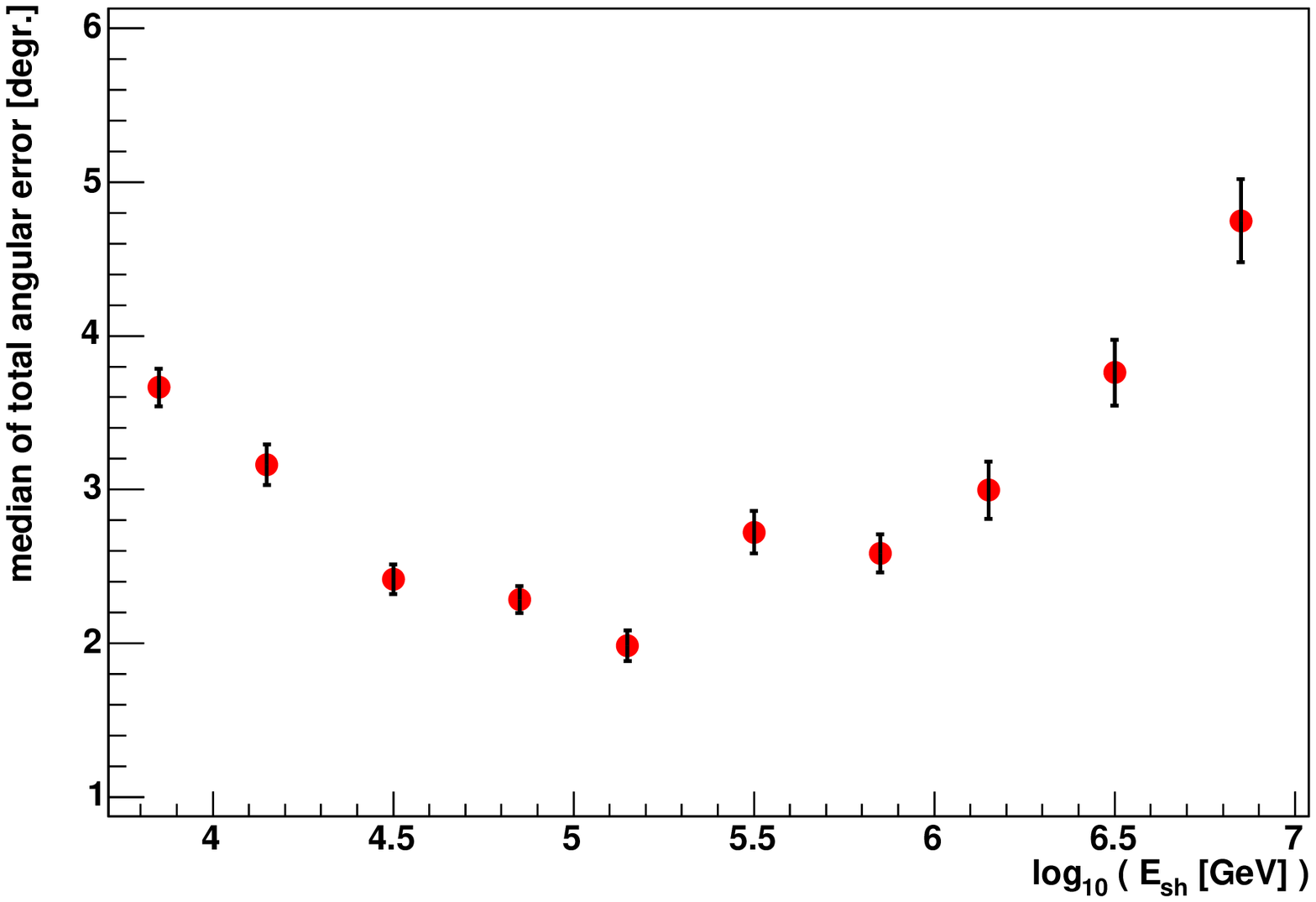}
\caption[Results for direction reconstruction after cuts, $\nu_e$ CC events]{Total angular error
  between neutrino and reconstructed direction,for all
  energies (left), and median of total angular error for different MC shower energy bins (right), with
  statistical errors, after the cuts.}
\label{fig:nue_alpha_cuts}
\end{figure}

In principle, the event topology of $\nu_e$ CC events is different to that of NC events, because of
the higher electromagnetic contribution (see Section~\ref{sec:em_showers}). However, the
electromagnetic component of a hadronic shower grows with increasing shower energy and exceeds 90\%
already at 1\,TeV (cf.~Section~\ref{sec:hadronic_showers}). Therefore, a
large amount of photons and electrons is present in NC events, and additionally, $\nu_e$ CC
events are always accompanied by a hadronic shower. Thus it is well possible
to reconstruct NC and $\nu_e$ CC events with the same reconstruction strategy. In a sparsely
instrumented detector like ANTARES, it will most probably not be possible to distinguish between the
two event types, because of their similar topologies. \\
The crucial difference between the two event types is in the energy reconstruction. While in NC
events a (mostly larger) part of the energy of the primary neutrino is carried away unseen
by the outgoing neutrino, in $\nu_e$ CC events all energy goes into showers and can therefore be
detected. It is therefore expected that for these events the resolution of the neutrino energy
reconstruction is much better than for NC events. \\
In the following, results for a sample of 138500 $\nu_e$ CC events are shown. The energy and angular
distributions of the primary neutrinos are the same as in the NC sample. The $\nu_e$ CC 
data sample is described in more detail in Appendix~\ref{sec:nue_sample}. \\
The results obtained are very similar to those shown for the NC sample. The overall angular
resolution $\Delta\alpha$ after the cut on $E_{reco} > 5$\,TeV, as shown in
Figure~\ref{fig:nue_alpha_nocuts}, has a median of \ccalpha, almost the same as for the NC sample
shown in Figure~\ref{fig:alpha_no_cut}, where the median is \ncalpha. It should be noted that due to
the kinematics, the $\nu_e$ CC events are a factor of 2 -- 3 more energetic than the NC events, with
respect to the shower energy, and therefore the contribution of lower energy events in the sample is
smaller than for the NC events, which leads to a slight overall improvement of the results. \\
Figure~\ref{fig:nue_E} shows the resolution of the shower energy, which in this case is equivalent
to the neutrino energy, after the cut on $E_{reco} > 5$\,TeV, on the left hand side. The RMS of the
distribution is 0.41, with a width of \ccElog\ in the Gaussian fitted peak. This width corresponds to a
factor of \ccE~in the energy resolution, which is again very similar to that of the NC events shown in
Figure~\ref{fig:energy_no_cut} (left). \\ 
The same additional two cuts that were applied to the NC sample above (see
Section~\ref{sec:likeli_cut} and Section~\ref{sec:E_pre_E_fit}) were also applied to this sample. 
37\% of the events which have
passed the reconstruction and the cut on the reconstructed shower energy remain after these
cuts. The right hand side of Figure~\ref{fig:nue_E} shows the energy resolution after the cuts. The
RMS has improved from 0.41 to 0.21 with a width of the peak of around \ccEcutlog, which corresponds
to a factor of ${\ccEcut}$ in the energy reconstruction. The result for the total angular resolution
after the cuts is shown in Figure~\ref{fig:nue_alpha_cuts}. The overall median has improved to
\ccalphacut, and the best resolution, reached at about $100$\,TeV, is below \alphabest. \\[0.1cm] 

It can therefore be concluded that the reconstruction algorithm presented in this thesis is suitable
both for NC and for $\nu_e$ CC events, which is of extreme importance for its applicability to real
data, because a distinction between the two event types may not be possible. From these results the
expected resolution for a shower-type event after the shown cuts is about $3^{\circ}$--$4^{\circ}$, 
with an energy resolution corresponding to a factor of \ccEcut. 

\section{Effective Volume and Effective Neutrino Area}\label{sec:result_effarea}

\begin{figure}[h] \centering
\includegraphics[width=7.4cm]{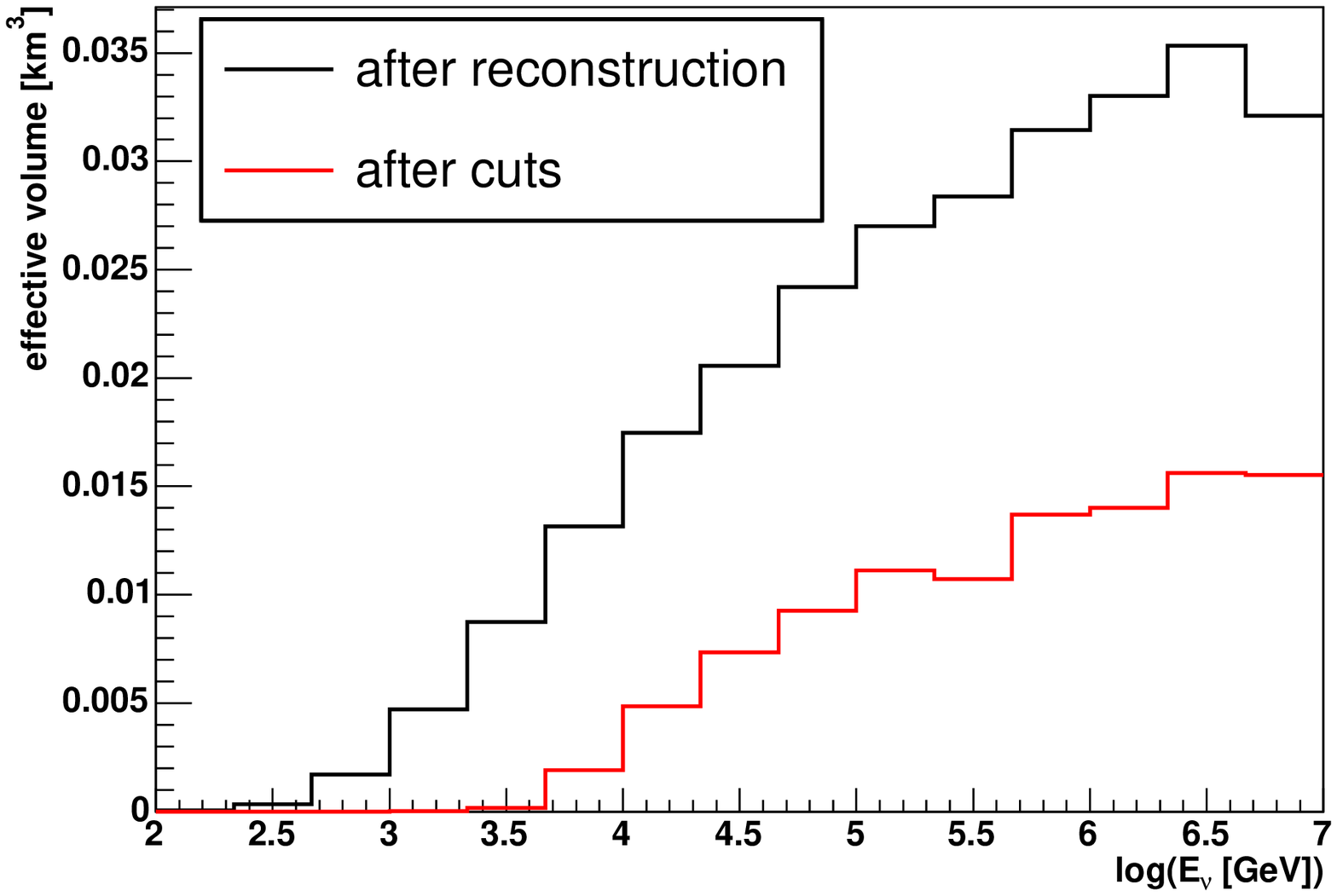}
\includegraphics[width=7.4cm]{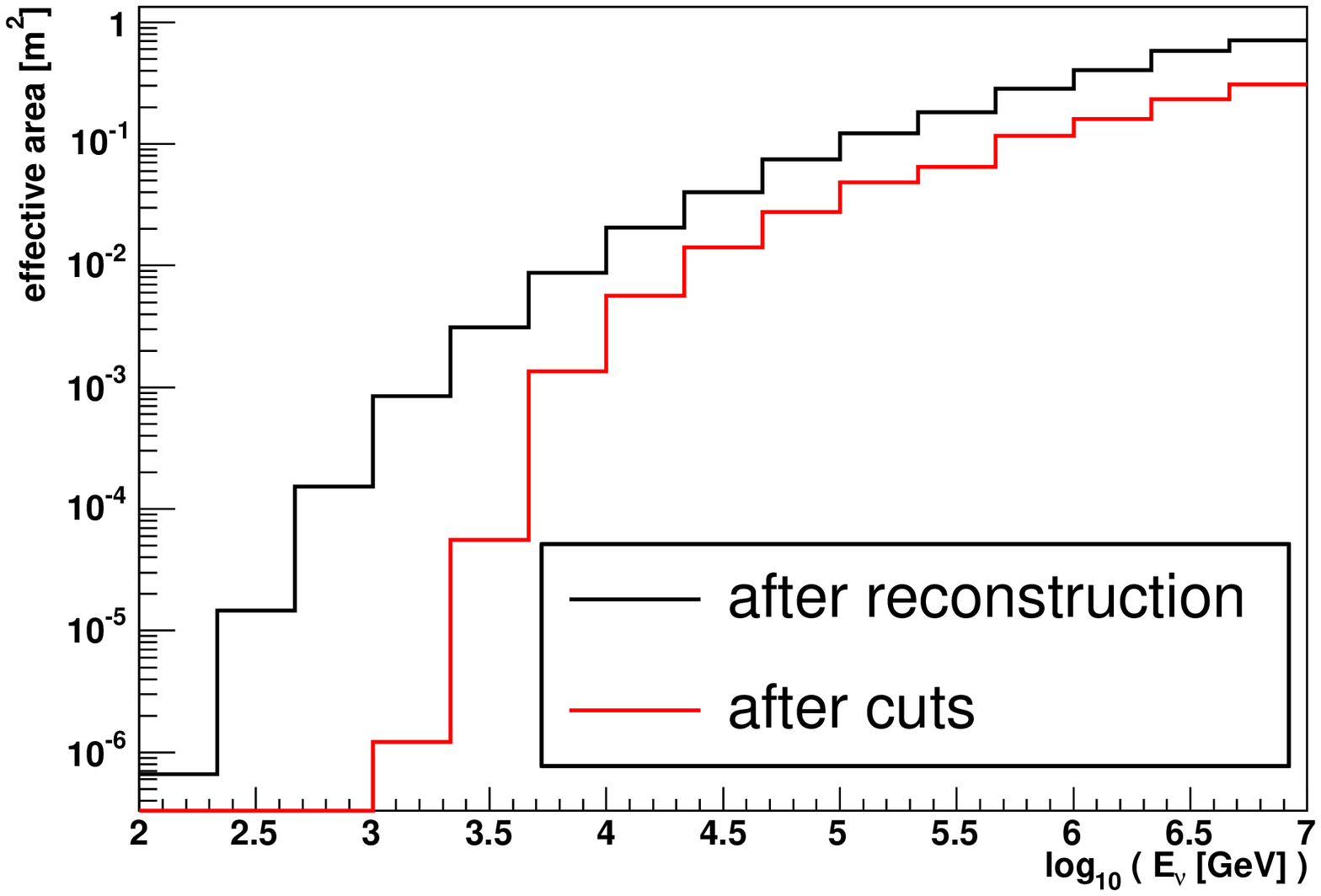}
\caption[Effective volume and effective area for NC events]{Effective volume (left) and effective
  neutrino area (right) for the reconstruction of NC events presented in
  Section~\ref{sec:nc_results}, after the reconstruction (black) and the cuts (red).}
\label{fig:eff_area}
\end{figure}

To give an estimate on the sensitivity of the ANTARES experiment to cosmic neutrino flux, the effective
volume and effective neutrino area of the detector are determined as 
described in Section~\ref{sec:eff_area}. \\
Figure~\ref{fig:eff_area} shows the results for the NC sample. The figure shows effective volume
(left) and effective area (right) over the MC neutrino energy, after the reconstruction (black line) 
and after the three cuts (red line, see Sections~\ref{sec:likeli_cut}, \ref{sec:E_pre_E_fit} 
and~\ref{sec:large_energies}). Before the cuts, the effective area corresponds roughly to 1/10th of the
effective muon neutrino area reached for $\nu_{\mu}$ CC events without cuts~\cite{muon_eff_area}. 

\begin{figure}[h] \centering
\includegraphics[width=7.4cm]{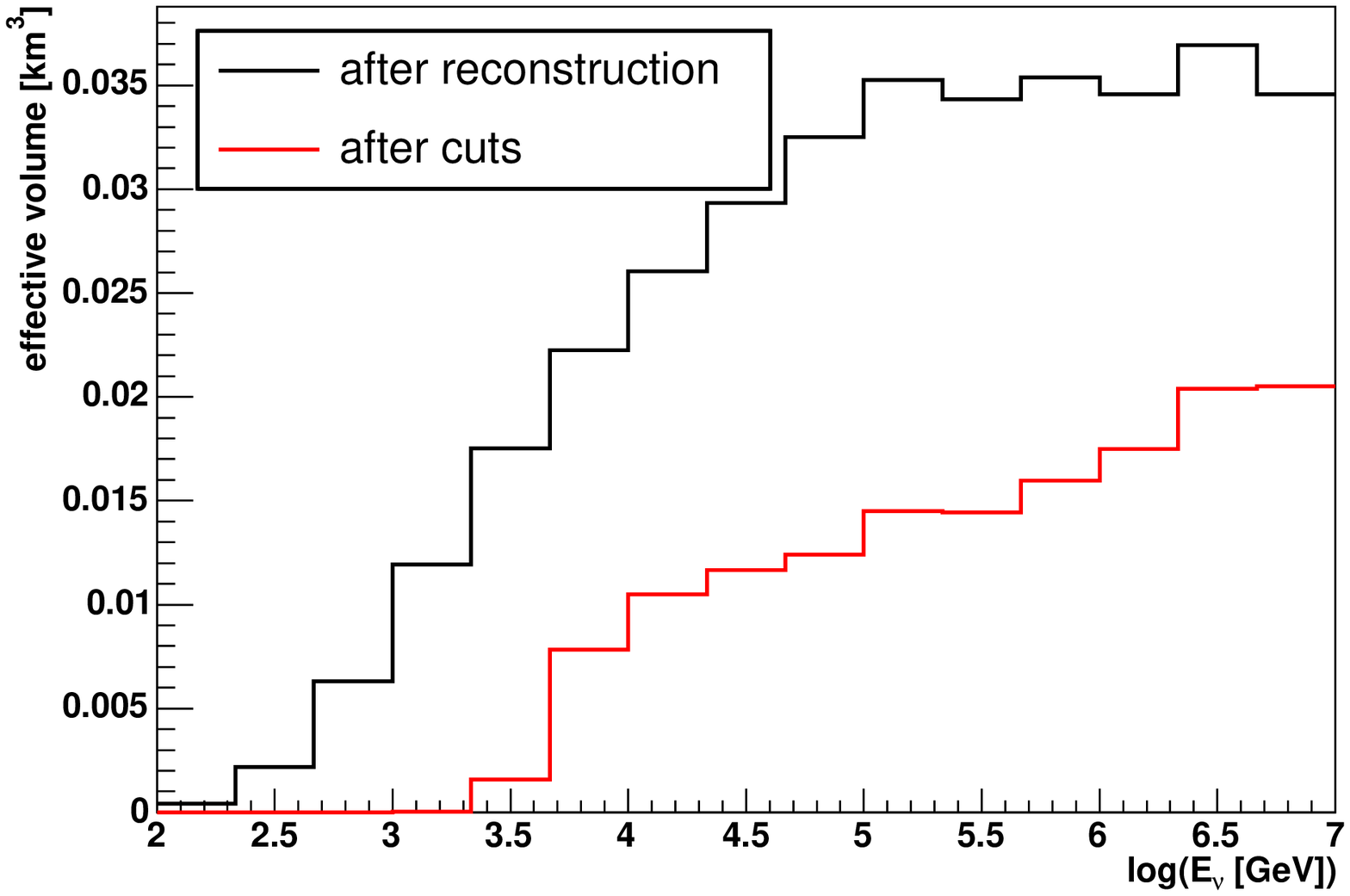}
\includegraphics[width=7.4cm]{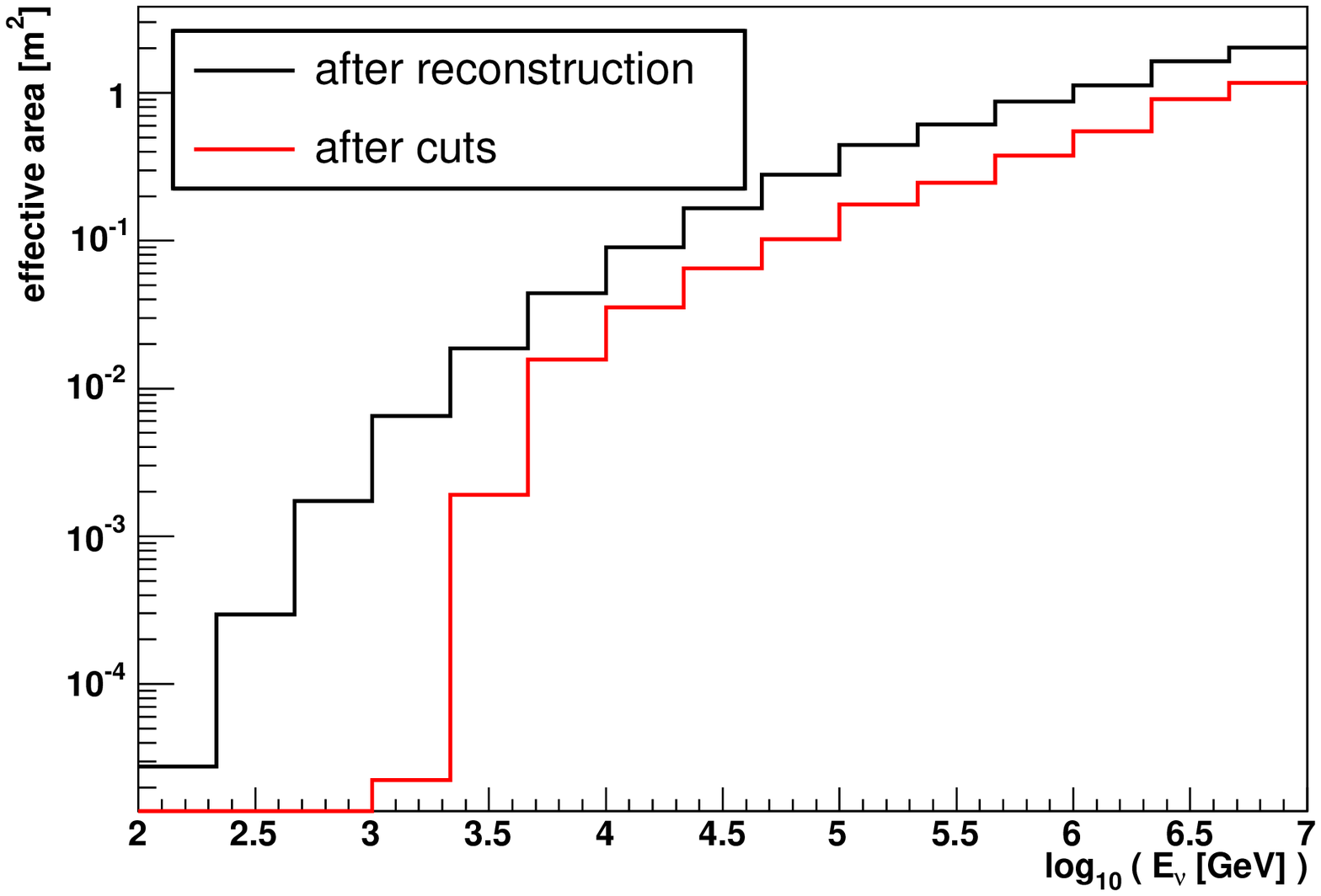}
\caption[Effective volume and effective area for $\nu_e$ CC events]{Effective volume (left) and effective
  neutrino area (right) for the reconstruction of $\nu_e$ CC events presented in
  Section~\ref{sec:nu_e_CC}, after the reconstruction (black) and the cuts (red).}
\label{fig:eff_area_nue}
\end{figure}

In Figure~\ref{fig:eff_area_nue}, the effective volume (left) and effective area (right) for the
$\nu_e$ CC events are shown. Results for these events were presented in
Section~\ref{sec:nu_e_CC}. The colour coding is the same as for Figure~\ref{fig:eff_area}. Above
100\,TeV, the effective volume before the cuts reaches a plateau which corresponds
to the generation volume of the events, pointing at the fact that practically all events above this
energy pass the reconstruction. The values of effective volume and effective area after the cuts
are slightly larger than those reached for the NC events, as the showers induced in $\nu_e$ CC
events are more energetic with respect to the neutrino energy, and therefore more likely to pass the
reconstruction. 

\section{Predictions on Rates and Diffuse Flux Sensitivity}\label{sec:result_flux}

\begin{figure}[h] \centering
\includegraphics[width=14.8cm]{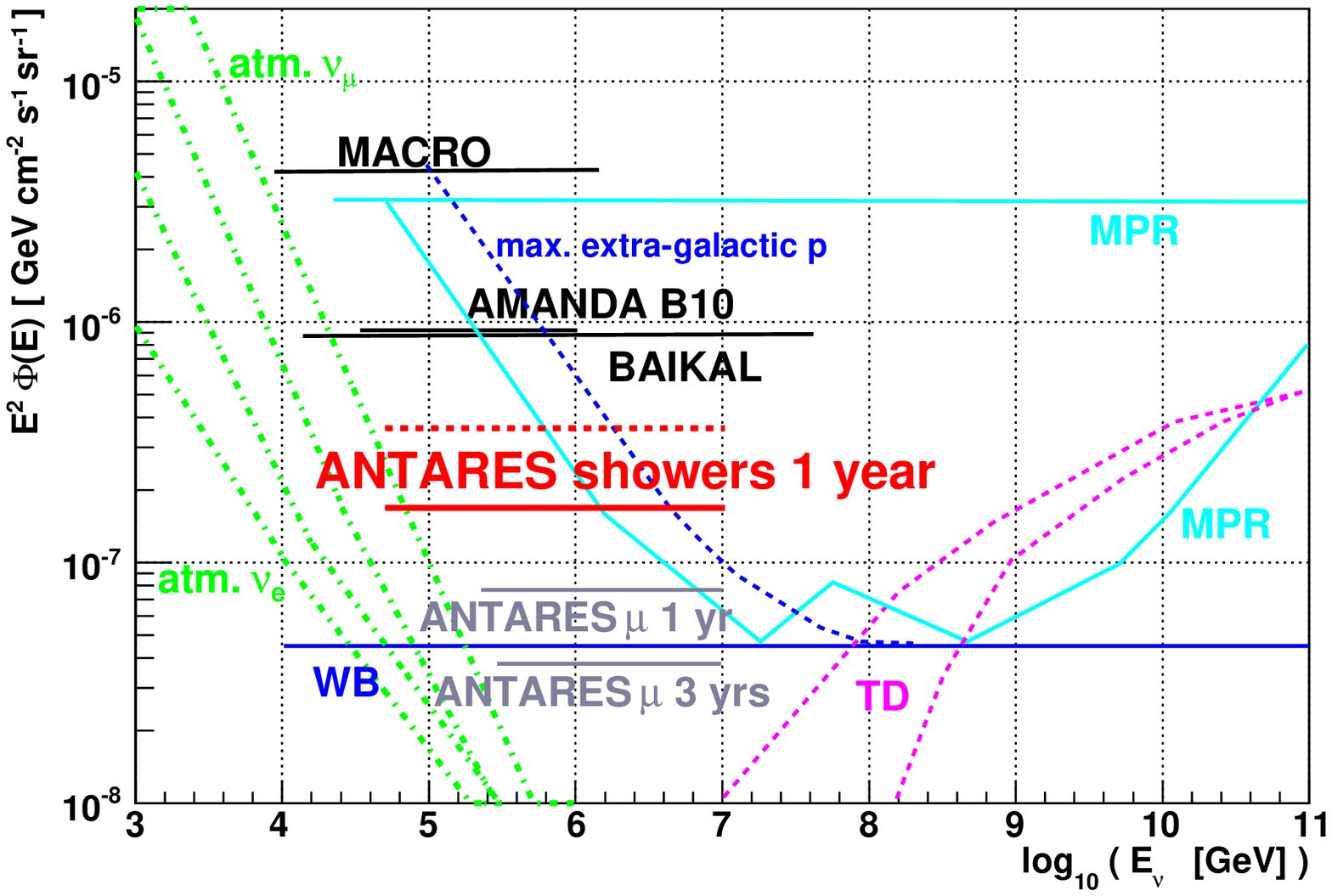}
\caption[Flux limit for isotropic events]{Flux limits for isotropic (drawn red line) and upgoing 
(dashed red line) shower events in ANTARES, as determined in this study, in comparison to results 
  for other event types or
  experiments. See Figure~\ref{fig:flux_limits} in Section~\ref{sec:other_ex} for a description of
  the other limits.}  
\label{fig:sensitivity}
\end{figure}

Expected rates of shower-type events for the ANTARES detector have been estimated from the events
remaining in the NC sample B and the $\nu_e$ CC sample after the reconstruction and the cuts.  
For the cosmic neutrino event rate, a hypothetical flux corresponding to the Waxman-Bahcall 
limit~\cite{waxman-bahcall1} was assumed:

\begin{equation}
E_{\nu}^2 \Phi_{cosmic} = 4.5 \times 10^{-8}\,\textrm{GeV s}^{-1}\,\textrm{sr}^{-1}\,\textrm{cm}^{-2}.
\end{equation}
It was assumed that, due to oscillations, the cosmic neutrino ratio at Earth is the same for all
flavours, $\nu_e : \nu_{\mu} : \nu_{\tau} = 1 : 1 : 1$. As the event rates for the NC sample
correspond to one of six possible channels of the type 
\begin{equation*}
(\nu_l,\bar{\nu}_l) + N \to (\nu_l,\bar{\nu}_l) +  \textrm{hadronic shower},
\end{equation*}
the rates for the sample were multiplied by six to retrieve the event rates for all-flavour NC
interactions, assuming identical cross sections for anti-neutrinos and neutrinos. This simplification
leads to an overestimation of the rates of $\sim 18\%$ at 50\,TeV, decreasing to almost zero 
for 1\,PeV where the cross sections can be regarded as identical (see Section~\ref{sec:cross_sections}).
Likewise, the rates retrieved for the cosmic neutrinos from the $\nu_e$ CC sample were multiplied by
two to take into account the $\bar{\nu}_e$ CC channel as well. Here, the assumption of identical cross sections for
electron neutrinos and anti-neutrinos leads to an overestimation of $\sim 10\%$ at 50\,TeV which 
decreases to almost zero for energies above 300\,TeV. Note that the Glashow resonance for the $\bar{\nu}_e$
CC channel at 6.3\,PeV, which causes an increase of the cross section at the resonance energy, was not
taken into account here. \\
For the atmospheric neutrino background, it was assumed that no tau neutrinos occur. Shower events can
then be induced by muon neutrinos in a NC interaction, or by electron neutrinos in a NC or CC
interaction. To estimate the total rate for NC interactions, the muon neutrino rate was calculated from the event
weights for atmospheric neutrinos in event sample B (see Appendix~\ref{sec:nc_sample});
those event weights were calculated during the production according to the Bartol
flux~\cite{bartol}. The expected rate for atmospheric electron neutrinos was calculated  
estimating that the rate of atmospheric electron neutrinos is 16 times smaller
than that of atmospheric muon neutrinos for energies below 10\,TeV and identical to the atmospheric
muon neutrino rate above 1\,PeV (see e.g.~Figure~\ref{fig:sensitivity}); in between, the difference
$\Delta \Phi_{\nu_{\mu}:\nu_e}$ between the two fluxes was parameterised as 

\begin{equation}
  \log_{10} \Delta \Phi_{\nu_{\mu}:\nu_e} = -0.45 \cdot \log_{10} E_{\nu}/{\textrm{GeV}} 
  + 2.7, \quad  10^4\,{\textrm{GeV}} \leq E_{\nu} \leq 10^6\,{\textrm{GeV}}
\end{equation}

\begin{figure}[h] \centering
\includegraphics[width=11cm]{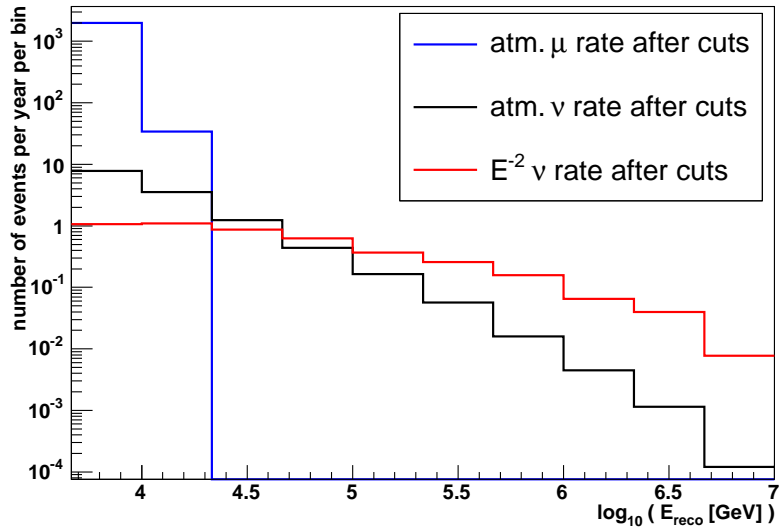}
\caption[Isotropic neutrino event rates]{Rates for isotropic shower events:
  atmospheric muons (blue) and atmospheric (black) and cosmic (red) neutrinos versus the reconstructed energy.}
\label{fig:event_rates}
\end{figure}

\begin{figure}[h] \centering
\includegraphics[width=11cm]{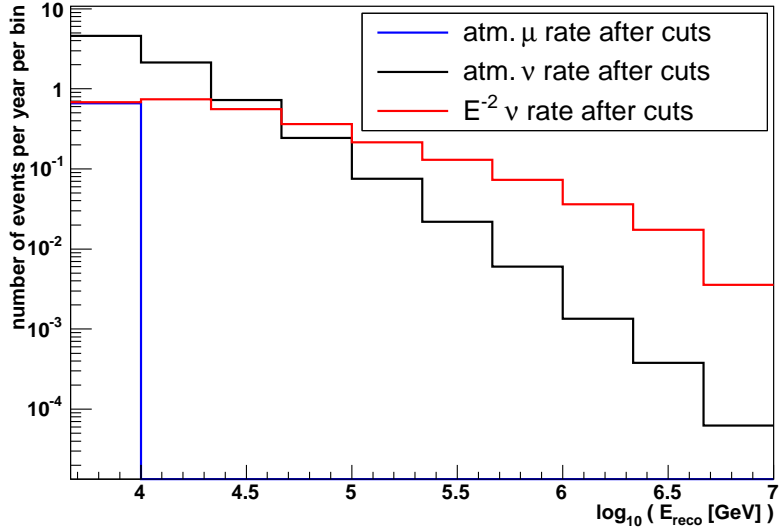}
\caption[Upgoing neutrino event rates]{Rates for upgoing shower events:
  atmospheric muons (blue) and atmospheric (black) and cosmic (red) neutrinos versus the reconstructed energy.}
\label{fig:event_rates_upgoing}
\end{figure}

To this approximated rate for NC events, the rate for $\nu_e$ CC had to be added. The results from
the $\nu_e$ CC event sample (see Appendix~\ref{sec:nue_sample}) were used for this.
As for the cosmic neutrinos, the rates were again considered to be the same for neutrinos and
anti-neutrinos. \\ 
The rates thus determined can be considered as the total expected rates for shower-type events from 
atmospheric and cosmic neutrinos in ANTARES. The resulting rates per year after
reconstruction and cuts are plotted for isotropic fluxes in Figure~\ref{fig:event_rates}, vs.~the 
reconstructed energy. The cosmic neutrino rates exceeds the atmospheric neutrino rate above $\sim 50$\,TeV. \\
The rate of atmospheric muons, again versus the reconstructed energy, is shown in blue, after 
the cuts which were also applied to the neutrino samples. The rate drops to zero above $\sim 20$\,TeV, since 
no events remain in the sample above this energy. This may be due to statistics; note, however, that 
especially the high-energy atmospheric muons are very strongly suppressed by the cut on $\xi$ (see
Figure~\ref{fig:chi_square_over_E}). Nevertheless, the atmospheric muon rate may be underestimated. \\
In Figure~\ref{fig:event_rates_upgoing}, the rates for events which were reconstructed as upgoing 
are shown. For the selected energy region of $E_{reco} > 5$\,TeV, the neutrino flux exceeds the atmospheric 
muon flux in all energy bins shown. It can therefore be concluded that a reliable detection of neutrino-induced 
shower events will be possible in this energy region. \\

\begin{figure}[h] \centering
\includegraphics[width=7.4cm]{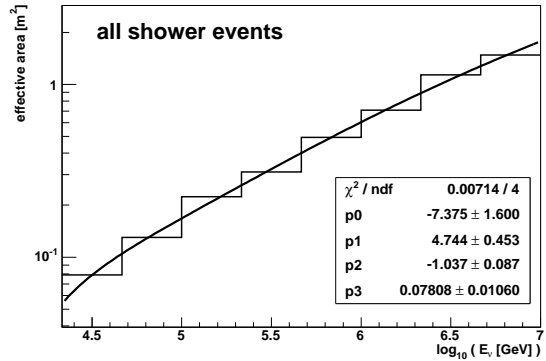}
\caption[Total effective area for isotropic events]{Total effective area for isotropic shower
  events, and fit used for flux limit estimations.} 
\label{fig:eff_area_total}
\end{figure}

From the background rates shown in Figure~\ref{fig:event_rates}, one can deduce an upper 
limit of the experiment for a diffuse flux, as has been described in 
Section~\ref{sec:calc_diffuse_flux}. Above 20\,TeV, no atmospheric muons are present in the sample. 
The cosmic neutrino rate exceeds the atmospheric neutrino rate above 50\,TeV. The atmospheric
neutrino rate above this energy is \nnu/year. 
The number of events to reject the background-only hypothesis at a 90\% confidence level for the
expected background rate is~\cite{feldman} $N = 3.77$. A parameterisation of the total effective
area for shower events as a function of the neutrino energy, needed for the calculation of the flux
limit, is shown in Figure~\ref{fig:eff_area_total}. Assuming a cosmic
neutrino flux proportional to $E^{-2}$, one receives, according to equation~\ref{eq:sensitivity} and
the quoted number of events, a flux limit of  

\begin{equation}\label{eq:sens_result}
E_{\nu}^2 {\Phi_{90\%}} = \sensitivity\,\textrm{GeV cm}^{-2}\,\textrm{s}^{-1}\,\textrm{sr}^{-1}
\end{equation}

after the cuts mentioned above. 

The result from equation~(\ref{eq:sens_result}) can be compared to the flux limit to shower 
events of the AMANDA detector~\cite{amanda_cascades}, where the 
expected background above $\sim 50$\,TeV is $0.90^{+0.69}_{-0.43}$ atmospheric muon events, and
$0.06^{+0.09}_{-0.04}$ atmospheric neutrinos, assuming a different model for the conventional
flux~\cite{lipari} and disregarding the prompt neutrino flux because of its large
uncertainties~\cite{kowalski}. In the analysed data taken in 2000, one event remains after 
the cuts. The authors derive from this an upper limit for the flux of neutrinos of all flavours in
AMANDA of

\begin{equation}
E_{\nu}^2 {\Phi_{90\%}}^{\textrm{AMANDA}} = 8.6 \times 10^{-7}\,\textrm{GeV
  cm}^{-2}\,\textrm{s}^{-1}\,\textrm{sr}^{-1}, 
\end{equation}

at 90\% confidence level, assuming a mean background of 0.96 events. This limit is approximately a 
factor of 5 higher than the result retrieved in this study, which may point to an underestimation of the
atmospheric muon flux; however, also the effective areas presented in this study are larger by about 
a factor of 2 than those shown in~\cite{amanda_cascades}.  \\ 
The flux limit derived here is shown in comparison to other measured and expected rates in 
Figure~\ref{fig:sensitivity}. It is around a factor 2.2 above the flux limit for charged
$\nu_{\mu}$ events expected in ANTARES for the same measurement period~\cite{zornoza} (grey line
marked \lq\lq ANTARES $\mu$ 1\,year\rq\rq). See Figure~\ref{fig:flux_limits} in
Section~\ref{sec:other_ex} for an explanation of the experimental limits, and
Figure~\ref{fig:theo_fluxes} in Section~\ref{sec:diffuse_flux} for a description of the theoretical
predictions.  

\begin{figure}[h] \centering
\includegraphics[width=7.4cm]{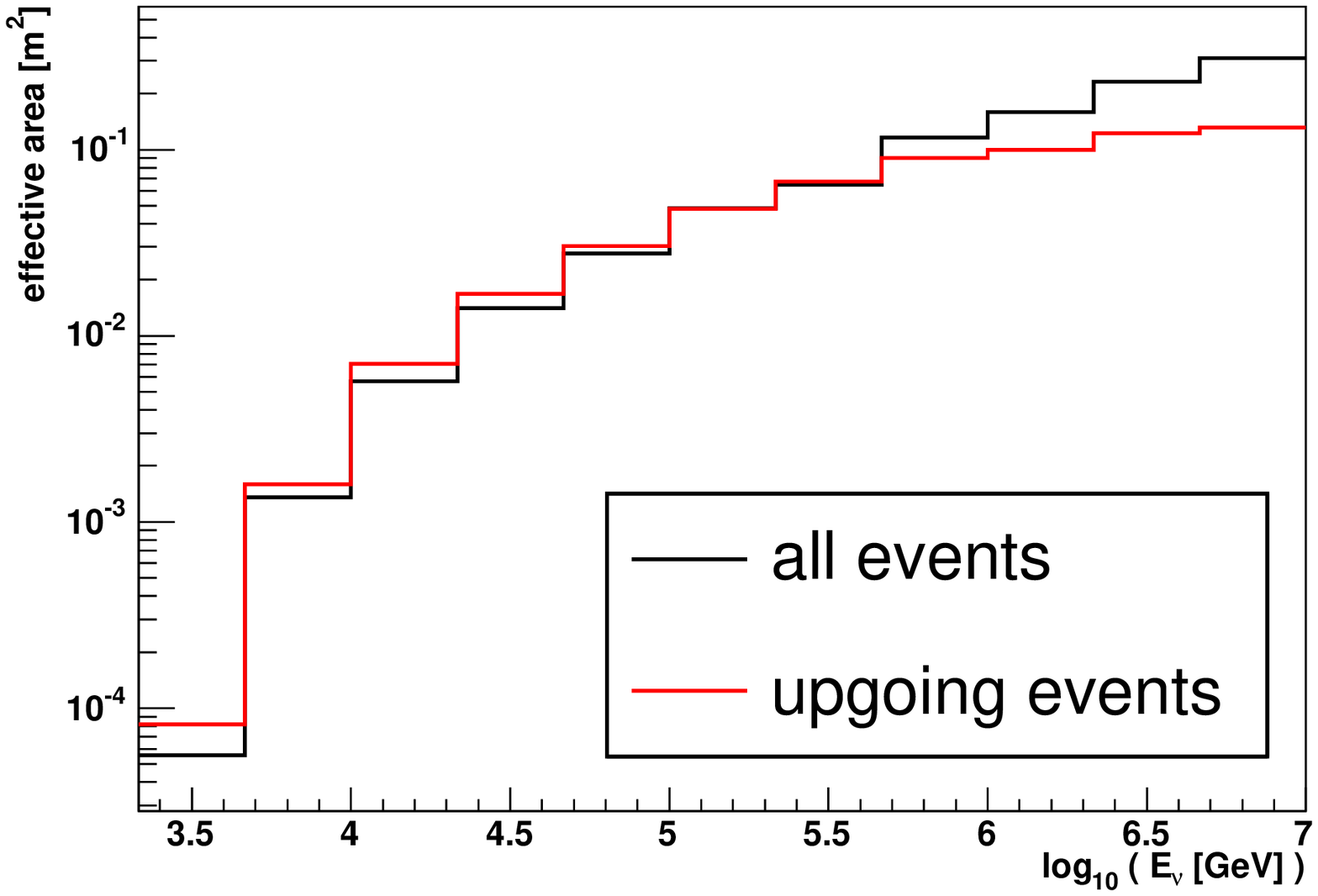}
\includegraphics[width=7.4cm]{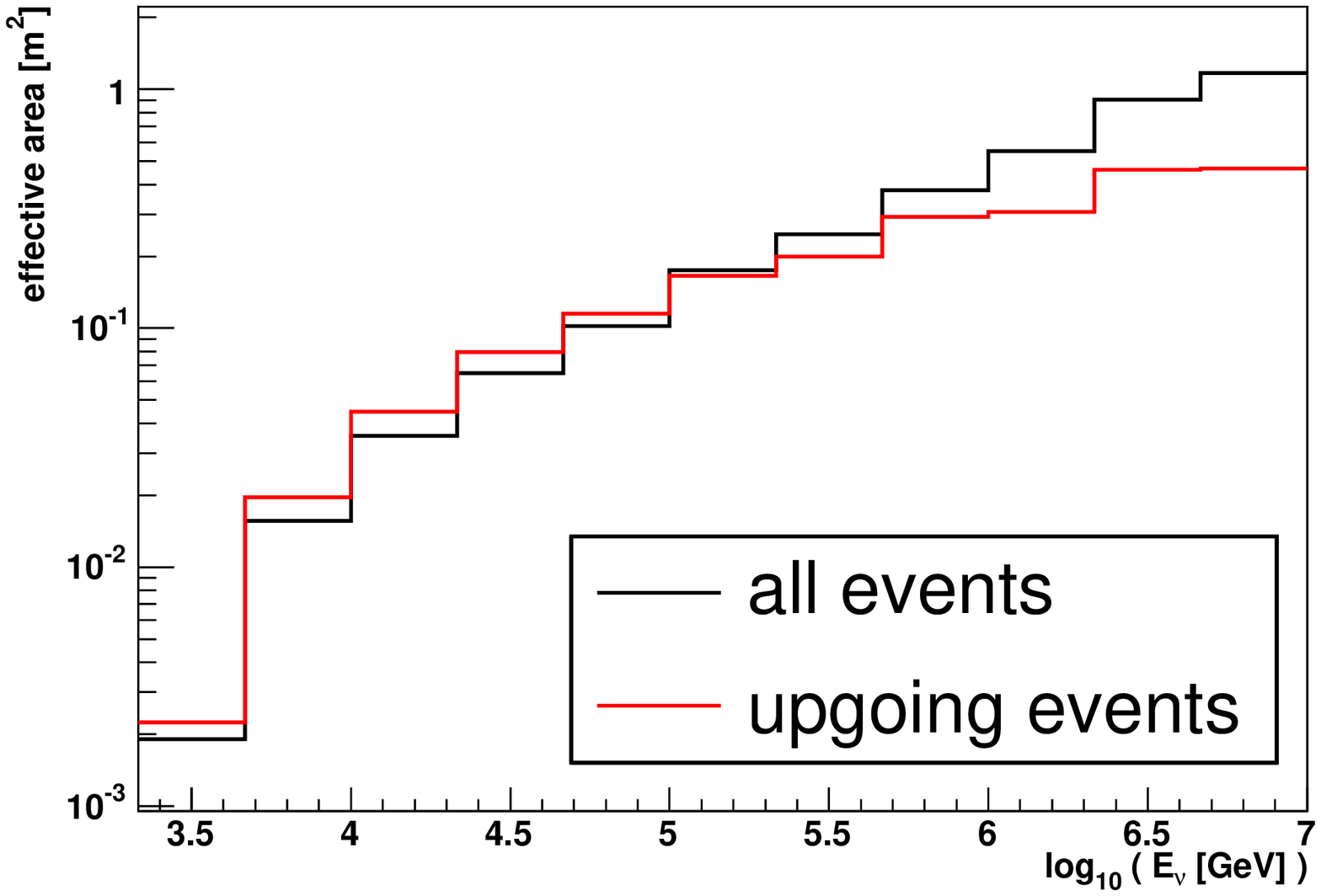}
\caption[Effective areas for isotropic and upgoing events]{Effective areas for NC events (left) and
  $\nu_e$ CC events (right), for isotropic events (black lines) and for upgoing events (red lines).}
\label{fig:eff_area_up}
\end{figure}

As the angular resolution of the shower reconstruction is about \alphabest~above some 10\,TeV after the
cuts, only upgoing events can be considered as well. Both the events from sample B and the $\nu_e$ CC
events were produced with isotropic neutrino distributions, and therefore, the original number of
upgoing neutrinos was considered to be half of the total number of events produced. The
effective areas for upgoing events can then be determined, and from this and from the expected background 
rates, the expected flux limit for upgoing events can be calculated. The effective areas from which the 
upgoing flux limit
was retrieved are shown in Figure~\ref{fig:eff_area_up}, for NC events (left) and $\nu_e$ CC events
(right), in comparison to the effective areas of isotropic events. While the effective area for
neutrino energies below $\sim 100$\,TeV is slightly larger for upgoing events because of their better
reconstructibility, above this value, the effective area flattens because of the 
opaqueness of the Earth (see Section~\ref{sec:nu_passage}). Consequently, also the flux limit for the highest
energy bins, as well as the overall flux limit, will be worse for the upgoing events than for the
events distributed isotropically. 

\begin{figure}[h] \centering
\includegraphics[width=7.4cm]{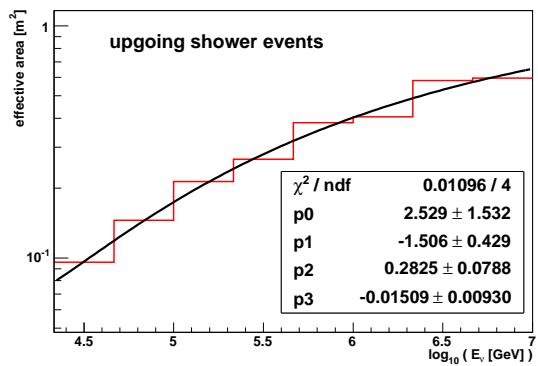}
\caption[Total effective area for upgoing events]{Total effective area for upgoing shower events,
  and fit used for flux limit estimations.} 
\label{fig:eff_area_total_up}
\end{figure}

As for the isotropic events, the energy-independent flux limit of the upgoing events was calculated
by parameterising the effective area in the relevant energy region. The parameterisation function 
of the total effective area for upgoing shower events as a function of the neutrino energy, for
energies above 5\,TeV, is shown in Figure~\ref{fig:eff_area_total_up}.

From Figure~\ref{fig:event_rates_upgoing}, the expected atmospheric neutrino background above
50\,TeV is 0.35 events per year, and the atmospheric muon background above 50\,TeV is negligible,
so that the case of zero measurement can be considered, which yields a maximum number of 2.12
events for the expected background rate at 90\% confidence level~\cite{feldman}. For this case, one
receives an energy-independent flux limit of 

\begin{equation}
E_{\nu}^2 {\Phi_{90\%}}^{up}  = \sensitivup\,\textrm{GeV cm}^{-2}\,\textrm{s}^{-1}\,\textrm{sr}^{-1}, 
\end{equation}

a slightly larger value than for the isotropic neutrino flux, and around a factor 4.7 larger than the
expected ANTARES flux limit for $\nu_{\mu}$ CC events. The flux limit for upgoing events is
shown in Figure~\ref{fig:sensitivity} as dashed lines.

\subsubsection{Conclusion}

The feasibility of reconstructing neutrino-induced showers with a precision of down to \alphabest~in
angular resolution has been demonstrated. A number of cuts had to be applied to 
sufficiently suppress the atmospheric muon background; the flux expected from this background lies
now below the neutrino flux, for energies above 20\,TeV for isotropic, and above 5\,TeV for upgoing
events, so that it is possible to study the diffuse neutrino flux above these
energies. Sensitivities both for isotropic and upgoing shower events have been calculated, and they 
are found to be around a factor 2--5 above the flux limit for charged $\nu_{\mu}$ events expected in
ANTARES for the same measurement period~\cite{zornoza} (grey line marked \lq\lq $\mu$ one year\rq\rq\ in
Figure~\ref{fig:sensitivity}). This ratio agrees well with the rough flux limit estimation given in
Section~\ref{sec:point_sources}, where a deterioration of a factor 5 was predicted. It can therefore
be concluded that the shower reconstruction strategy adds an important, independent contribution to the
measurable neutrino flux in ANTARES.

\chapter{Summary and Outlook}\label{ch:conclusion}

This thesis is devoted to the reconstruction of shower-type events with the ANTARES detector, a
neutrino telescope that is currently being installed in a depth of 2400\,m in the Mediterranean
Sea. The main objective of ANTARES is the detection of high-energy cosmic neutrinos that are
presumably produced in astrophysical objects like Supernova Remnants, Active Galactic Nuclei or
Gamma Ray Bursts. The detection of these neutrinos will permit a new insight into the physics of
these objects. Neutrinos are detected by measuring the Cherenkov light signal that is caused by the
charged secondary particles of a neutrino interaction with matter, in this case, the water of the
deep sea. The amplitudes and timings of the light signals are measured in photomultiplier tubes
inside Optical Modules, which are installed covering a volume of $\sim 0.03$\,km$^3$. From the
signals, the direction and energy of the neutrinos are to be reconstructed. \\ 
Depending on the type of neutrino interactions, different secondaries can be generated, and
different strategies must be used to reconstruct them. This thesis deals with the reconstruction of
neutrino-induced showers, both hadronic and electromagnetic. Showers are present in all neutrino
interactions, and there are two types of reactions for which they are the only detectable
objects: Neutral current (NC) interactions and charged current (CC) electron neutrino
interactions. It has been shown in this thesis that both event types can be reconstructed using the
same strategy. \\ 
Neutrino telescopes are usually optimised on the reconstruction of muons from CC muon neutrino
interactions. In these events, a muon and a hadronic shower are produced in the interaction of the
neutrino with the ambient matter. For the considered energies above a few 100\,GeV, the muon has a path
length of 1\,km or more and its direction can be reconstructed to a precision of a few tenth of a
degree. Showers, on the other hand, occur on a relatively small scale with respect 
to the spacing of the Optical Modules in the detector -- the typical length of a shower is 10\,m,
while the distance between the Optical Modules of the detector is 14.5\,m in vertical, and 60 to 75\,m
in horizontal direction. Therefore, the directional characteristics of a shower cannot be resolved
well in the ANTARES detector, so that a sub-degree angular resolution, as reached for muons, is not
achievable for showers. On the other hand, since the showers have considerably small dimensions,
almost the entire amount of light produced in the shower can be detected, which
makes the reconstruction of the shower energy much less problematic. \\
A strategy to reconstruct the shower direction together with the shower energy has been developed in
the context of this thesis. The first reconstruction step is the determination of the position of
the shower maximum, which is calculated assuming an isotropic emission of light. Direction and
energy are then determined using a pattern matching algorithm: The signal amplitudes that have been
measured in the Optical Modules of the detector are compared to those that are calculated
assuming specific values for shower direction and energy, taking into account the
angular efficiency of the photomultipliers, the absorption of the photons in water and the energy
dependent angular distribution of the photons with respect to the shower axis. Direction and energy
are then determined using log-likelihood optimisation. \\
The reconstruction of the shower events yields an overall median of the directional error of the
shower about $10^{\circ}$, both for NC and $\nu_e$ CC events, for reconstructed energies above 5\,TeV. 
The width of the peak in the logarithmic shower energy resolution is $\sim \ccElog$ which means 
that the shower energy can be determined to a factor of ${10^{\ccElog} \approx  \ccE}$. Some additional 
quality cuts which proved to have a high combined efficiency and purity have been applied to the data 
samples. The combined efficiency of all cuts it at about 70\%. After the cuts, the reconstruction of 
the shower energy improves to a factor of ${10^{\ccEcutlog} \approx \ccEcut}$. The directional 
resolution improves to $3.5^{\circ}$, with values at \alphabest\ for 
energies of about 100\,TeV. Since the background dominates the neutrino signal below $\sim 50$\,TeV (see
below), the reconstruction algorithm is optimised to the energy region above this value. \\
When deducing the neutrino energy from the shower energy, it must be taken into account that
for NC events, a part of the primary neutrino energy is carried away by the outgoing neutrino. This
fraction could be determined statistically, but as NC events will not be distinguishable in the
experimental data from $\nu_e$ CC (for which the shower energy is equivalent to the neutrino
energy), it was preferred for this study to regard the reconstructed shower energy {\it a priori} as
a lower limit for the primary neutrino energy.
\\ 
The results presented here were obtained assuming the saturation of the photomultiplier electronics at
200\,photo-electrons. However, the data taking mode which would enable this high saturation level is
very bandwidth-consuming. It has been found that under the assumption of a saturation at
20\,photo-electrons, which is the saturation level corresponding to the default data taking mode in
ANTARES, the resulting resolution below 70\,TeV is slightly better than for a higher saturation,  
because the large fluctuation effects that occur at these smaller energies are damped by the lower
amplitude saturation. For higher energies, the lower saturation leads to a slight deterioration in the
angular resolution, which becomes about $3^{\circ}$ larger than the resolution achieved for
the saturation at 200\,pe. The energy resolution shows very similar results for both saturation levels. \\
As the ANTARES detector is not built in a sterile surrounding, but in natural salt water,
the Optical Modules are exposed to constant optical noise from $^{40}$K decays and from
bacteria bioluminescence. The background can be suppressed by applying filter conditions on the signals
in the individual Optical Modules, concerning the causal connection of the signal with the signals
in the other Optical Modules, and the size of the signal amplitude. The filter works
most efficiently for events with shower energies above $\sim 10$\,TeV. \\
A second type of background originates from high-energy particle interactions caused by atmospheric
muons and atmospheric neutrinos. \\
Atmospheric muons are a dangerous background to neutrino-induced showers if they cannot be
clearly identified as muons, either because they undergo strong bremsstrahlung losses from which an
electromagnetic cascade is produced, or because they occur as multi-muon bundles. Over 99\% of the
atmospheric muon background are suppressed by some quality cuts; the background becomes negligible
above 20\,TeV. \\
Atmospheric neutrinos are an insuppressible background to cosmic neutrinos; the latter can only be
identified by detecting an excess of events over the atmospheric neutrino spectrum. \\ 
Due to the background of atmospheric neutrinos, which dominates the signal of shower
events from cosmic neutrinos below $\sim 50$\,TeV, the detection of an isotropic, diffuse cosmic
flux is only possible above this energy. The expected isotropic rate for atmospheric neutrinos above
this energy was determined as \nnu/year. At a 90\% confidence level and under the assumption of the
detection of one background event, this yields an energy-independent upper limit on the cosmic
neutrino flux of  

\begin{equation*}
E_{\nu}^2 {\Phi_{90\%}} = \sensitivity\,\textrm{GeV cm}^{-2}\,\textrm{s}^{-1}\,\textrm{sr}^{-1}
\end{equation*}

for one year of data taking in ANTARES, assuming that the cosmic neutrino flux is proportional to
$E^{-2}$. If only upgoing events are considered, the atmospheric muon background (consisting of
downgoing events which have been misreconstructed as upgoing) above 5\,TeV is negligible, and
the atmospheric neutrino rate is \nnuup/year. The corresponding upper limit on the cosmic neutrino
flux for one year of data taking in ANTARES, assuming no event detection at all, becomes then

\begin{equation*}
E_{\nu}^2 {\Phi_{90\%}}^{up} = \sensitivup\,\textrm{GeV cm}^{-2}\,\textrm{s}^{-1}\,\textrm{sr}^{-1}.
\end{equation*}

These flux limits are about a factor 2--5 larger than those predicted for the CC muon
neutrinos in ANTARES, due to the smaller effective area for shower events. \\
While all other reconstruction strategies for the ANTARES detector are intended for
CC muon neutrino events, the strategy introduced here allows now to also reconstruct shower
events. It therefore closes an important gap in the ANTARES event reconstruction, as now the
reconstruction of all event topologies is possible, and the access to events which were
unreconstructible until now is permitted. The detectable neutrino flux for the ANTARES detector has
therefore been significantly increased. \\
The shower reconstruction could also lead to an important improvement in the reconstruction of the
neutrino energy for $\nu_{\mu}$ CC events, if the event has taken place close enough to the
instrumented volume for the hadronic shower to be detected as well. In this case, the muon and the
hadronic shower could be reconstructed separately and the results could be combined afterwards. \\ 
In principle, the strategy introduced here is adaptable to other experiments as well; the
parameterisations used have been tuned to the ANTARES detector, but could be tuned to other
detectors as well. For experiments at sites with a significantly shorter
scattering length, like IceCube, the strategy is probably not applicable, because the directional
characteristics of the Cherenkov photons of the shower are deteriorated too strongly to permit a 
reconstruction.

\begin{appendix}
\chapter{ANTARES Event Simulation Packages}\label{sec:software}

In this appendix, the software chain for the production of ANTARES Monte Carlo (MC) events is
described, as far as it was used for the presented study. The MC production is usually divided into 
four parts: The interaction of the neutrino,
which is simulated with {\it genhen}, see Section~\ref{sec:genhen}; the propagation of secondaries and
the production of Cherenkov light, for which {\it geasim} is used in the case of non-muonic events,
see Section~\ref{sec:geasim}; the selection of physics hit from noise hits, for which a software
filter is used, see Section~\ref{sec:trigger_soft}; and finally the event reconstruction, which is
done with {\it reco}, see Section~\ref{sec:reco}. \\
The chapter closes with a description of the different data samples used for this study in
Section~\ref{sec:data_sample}. \\ 
A more detailed summary of the ANTARES simulation tools can be found in~\cite{brunner_software}. 
As the main part of this work, the {\it ShowerFitter}, a reconstruction
package for showers, was developed within the reco frame. The physical and mathematical background
of this shower reconstruction is described separately in Chapter~\ref{sec:shower_fitter}. 

\section{Monte Carlo Event Generation: genhen}\label{sec:genhen}

Genhen~\cite{genhen,genhenv6r3} is an event generator for ANTARES events written in Fortran. From
version v6 upwards, the generation of events from all neutrino flavours is supported. To generate
NC or $\nu_e$ CC events using genhen, the user defines a cylindrical object called {\it can} within 
which the events are generated\footnote{The can is also defined by the user for $\nu_{\mu}$ CC events;
  however, for this event type the neutrino interactions take place outside the can and the muons
  are propagated until they are either stopped or have reached the can.}. Depending on the event
type, the can extends up to a few hundred metres beyond the instrumented volume, but for the
generation of shower events, which have a relatively short range, an extension of a few absorption 
lengths is sufficient. If a $\nu_{\mu}$ CC event is generated, the shower that accompanies the $\mu$
is only written into the genhen output if it occurs close enough to the detector to be measured; else, 
only the single muon is displayed. \\
The neutrino interactions are simulated using the LEPTO~\cite{lepto} package for the deep
inelastic scattering and the RSQ~\cite{rsq} package for resonant and quasi-elastic events. The CTEQ
parton distribution functions~\cite{cteq} are used as default; the most recent version which can be
used is CTEQ6D. The hadronisation is done using PYTHIA/JETSET~\cite{pythia}.
The neutrino energy range in genhen is valid from the lower energy
threshold of the detector, a few dozen GeV, up to 1\,EeV. However, due to uncertainties
in the structure functions, the total uncertainty in the neutrino cross section exceeds 10\% at 100
PeV; and for neutrino energies above 1 PeV, also the results from PYTHIA and JETSET, which are
originally intended for particle accelerator physics, and therefore tuned for lower energies, might
be wrong.   
\\ 
The output of genhen consists of the primary neutrino, its kinetic properties and the position of
the interaction vertex, plus the position, direction and energy of the long-lived secondaries that 
were produced by the interaction. For secondaries with a lifetime $\lesssim 10^{-11}$\,s, the
respective decay products are displayed in the output instead. \\
The program also calculates and displays event weights which take
into account the absorption probability of neutrinos on their passage through the Earth, and gives
event rates for an assumed neutrino flux. 

\section{Particle Tracking and Light Generation: geasim}\label{sec:geasim}

\begin{figure}[h]\centering
\includegraphics[width=7cm]{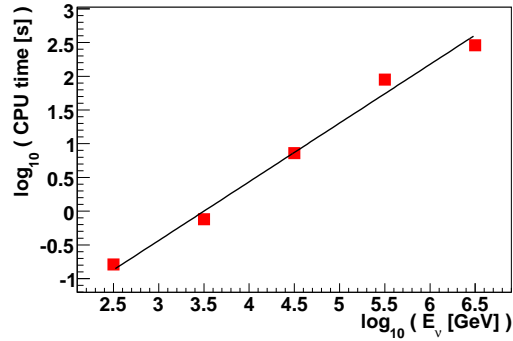}
\caption[Geasim CPU usage]{CPU usage of one event in geasim as a function of the neutrino
  energy.}
\label{fig:geasim_CPU}
\end{figure}

In the ANTARES software chain, the tracking of particles other than muons is done with the geasim
package~\cite{geasim} which is based on GEANT 3.21~\cite{geant}. All relevant physics processes that
can occur during the passage of particles through a medium are taken into account. At each tracking
step of each particle, a Cherenkov cone is produced; if an OM lies within the rays of light from the
cone, a hit is produced, whose amplitude is calculated according to the photon density on the cone,
using Poissonian statistics. The the characteristics of the PMTs, like quantum efficiency, angular
efficiency or the transmission coefficient of the glass sphere are taken into account as well, as is
the wavelength-dependent attenuation of light in water (see Section~\ref{sec:properties}). However,
scattering effects are not simulated; the number of photons is only damped according to the
attenuation length.  \\
The radial size of the shower is neglected in geasim as it is only a few cm. 
Monte Carlo studies have shown that the angular distribution of the Cherenkov light with respect to
the shower axis is not energy-dependent for electromagnetic showers (see
Section~\ref{sec:em_showers}). Therefore, this distribution has been parameterised to save CPU
time. For hadronic showers, however, the distribution is energy-dependent; it is subject to large
fluctuations and therefore no parameterisation is done. It is thus necessary to track every single
particle in the shower down to the Cherenkov level. This is very time-consuming, and the CPU time
rises exponentially with  energy, as can be seen in Figure~\ref{fig:geasim_CPU}, where the average
CPU time for the processing of one event is shown. The average CPU time was determined by processing
10 mono-energetic events for each decade of energy, so fluctuations are possible, but the trend is
clearly visible.  

This high consumption of time makes the production of events above $\sim 10$\,PeV a tedious
business. There are also some known errors in the cross sections from GEANT 3.21. The authors
of geasim therefore do not recommend the usage above 100\,TeV, although events have been produced
within this work up to energies of 100\,PeV without displaying any obvious flaws other than technical
ones\footnote{One technical problem was, for example, that the maximum number of tracking steps was
  reached for some of the showers produced at the highest energies, which in consequence made the
  program abandon the tracking, so that not all tracks were followed down to the Cherenkov level and
  the event had to be discarded afterwards.}. \\
After the event processing, the output of geasim is added to the event file generated by genhen. The 
geasim output consists of a
list of all OMs which have been hit, the amplitude of the hits, their timing, particle of origin and
track of origin. The hit digitisation in the PMTs is simulated by producing the so
called {\it raw hits} from the hits. For this, data taking characteristics (see
Section~\ref{sec:digitisation}) like the ARS integration time of 25\,ns, the number of ARSs and the
dead times are taken into account. The timing information can be smeared by a Gaussian (as was done
for the data used in this study) or a Laplace
distribution. The raw hits do not contain any MC information any more, as they are intended to be in
the same format as measured signal hits. \\
In geasim, it is also possible to add optical background. To do this, the user defines the required
background frequency; geasim will then add random single photon hits to the events according to that
frequency. 

\section{Hit Selection: Software Filter}\label{sec:trigger_soft}

For real data, a software trigger will be used to decide whether the detector has
measured a physics event or only noise. This trigger collects hits that are causally connected and/or
have a sufficiently large amplitude; when the trigger conditions are fulfilled, a physics event is
built. The selection criteria can also be used as a filter to eliminate noise hits in a MC event
sample, as has been described in the chapter on different sources of background, in
Section~\ref{ch:trigger}. In this case, only the hits passing the filter conditions will be used for
further analysis. If an event, after the filter has been applied to it, does not contain a
sufficient number of hits for reconstruction, it will be discarded. In the final version of the
event simulation, the filter conditions will be applied to 
the events in a stage between the hit generation and the reconstruction. For this study, the filter
conditions were part of the reconstruction code that is described in the next section.

\section{Event Reconstruction: reco}\label{sec:reco}

Reco is the ANTARES reconstruction package within which different reconstruction algorithms, the
so called {\it strategies}, can be placed and run. The software in its current form was re-written
from the previously existing code in 1999 by~\cite{recov2r0}. Information on the code structure and
a collection of documentation can be found on the reco home-page~\cite{reco}. There exists also a
handbook on different reco strategies~\cite{reco_dev} which describes some of the algorithms and
their implementation in the code. \\
Almost all reconstruction algorithms which have been developed for the ANTARES detector until now
are intended for the reconstruction of $\nu_{\mu}$ CC events. There is one strategy called {\it
  BrightPointFitter} which reconstructs the position of a point source of light by calculating the
centre-of-gravity of all hits in an event. The same algorithm is also used for the reconstruction of
the vertex in a shower event, as described in Section~\ref{sec:pos}. A reconstruction
strategy for showers has been developed in 2000 by~\cite{bernard} but it was never
integrated into the official ANTARES reconstruction code.

\subsubsection{General Remark on the Data Format}
At the writing of this thesis, the standard format of the event files was in ASCII code. However,
work is going on to change this into the ROOT~\cite{root} format, and this process will probably be
completed 
within the first half of 2006. The new data format forsees a stricter separation between physics
characteristics of an event and characteristics that are caused by the detector properties. For
example, the conversion from {\it hits} to {\it raw hits}, and the addition of optical 
background, will not be done any more within the simulation of the particle tracking and light
production, but in a separate program devoted for digitisation and triggering. 

\section{The Monte Carlo Data Samples}\label{sec:data_sample}

This section describes the different MC data samples that were used for this thesis. Both NC events 
and $\nu_e$ CC events have been studied. Atmospheric muon background was studied as well, using 
events which were produced within the ANTARES collaboration. This section summarises the 
characteristics of the used samples: the neutrino energy range and spectrum, the angular distribution 
of the neutrino, the optical background that was added and the remaining event numbers after the 
different steps of the reconstruction.

\subsection{The NC Events}\label{sec:nc_sample}

\subsubsection{Event Sample A}

The Monte Carlo distributions and some of the results shown in this note are based on a $\nu_e$ NC data
sample of 4776 events between $10^2$ and $10^8$ GeV, called {\it event sample A} within this
thesis. The neutrino energy spectrum of this production is $E^{-1}$ and the angular distribution is
isotropic over $4\pi$. The sample is available without optical background and with different rates
of optical background, between 60\,kHz and 140\,kHz, as well. Note that 60\,kHz is the default
background rate used with this sample, as long as no other background rate is implicitly mentioned
in the text. The can inside which the events were produced
was of about the size of the instrumented volume of the ANTARES detector, thus the events can be 
considered to be {\it contained events}. 

\subsubsection{Event Sample B}

Because the statistics of the above-mentioned event sample A are rather limited, $\nu_{\mu}$ NC
events that were part of a $5 \times 10^5$ events mass production of shower-type events were used
where larger statistics were required. These events are denoted as {\it event sample B}. The events
were produced with a neutrino energy between $10^2$ and $10^7$\,GeV, an energy spectrum of $E^{-1}$ 
and inside a generation volume that exceeds the instrumented volume by one absorption length ($\sim
55$\,m). The primary neutrinos were distributed isotropically in the whole $4\pi$ solid angle. An 
optical background of 60\,kHz was added
to these events as well. The numbers of events in event sample B, after the different stages of
event production as described in the first four sections of this chapter, are given in
Table~\ref{tab:nc_sample}. Figure~\ref{fig:event_stages_nue} (top) shows these numbers in relation
to the number of originally generated events, again for the different energy bins. As expected, the
percentage of events that survive all production stages increases with increasing energy, because
the number of produced photons is proportional to the shower energy, and the filter and
reconstruction stages become more efficient for larger numbers of hits. It should also be noted that
the filter efficiency is worse than for the events shown in Figure~\ref{fig:eff_trig} in
Section~\ref{sec:event_purity}, because the events shown there are {\it contained events}.

\begin{table}[h] \centering 
\begin{tabular}{|>{\centering}p{3.2cm}|>{\raggedleft}p{2.2cm}|>{\raggedleft}p{2.6cm}|>{\raggedleft}p{2.4cm}|>{\raggedleft}p{2.4cm}|}\hline
neutrino energy [GeV] & generated (Section~\ref{sec:genhen}) & producing hits (Section~\ref{sec:geasim}) & filtered (Section~\ref{sec:trigger_soft}) & reconstructed (Section~\ref{sec:reco}) \tabularnewline \hline
$10^2 - 10^3$ & 99633 &  67266 (67.5\%) &  1935 (1.94\%) &  1747 (1.75\%) \tabularnewline
$10^3 - 10^4$ & 22623 &  20000 (88.4\%) &  5541 (24.5\%) &  5142 (22.7\%) \tabularnewline
$10^4 - 10^5$ & 10318 &  10000 (96.9\%) &  5939 (57.6\%) &  5486 (53.2\%) \tabularnewline
$10^5 - 10^6$ &  5040 &   5000 (99.2\%) &  4085 (81.1\%) &  3739 (74.2\%) \tabularnewline
$10^6 - 10^7$ &  2465 &   2457 (99.7\%) &  2288 (92.8\%) &  2115 (85.8\%) \tabularnewline \hline
total events & 140079 & 104723 (74.8\%) & 19788 (14.1\%) & 18229 (13.0\%) \tabularnewline
\hline 
\end{tabular}
\caption[Numbers of events in event sample B]{Numbers of events in event sample B, after the
  different production stages. The percentages given refer to the original number of events (first column).}
\label{tab:nc_sample}
\end{table}

The sample also contains event weights for the atmospheric neutrino flux, according to the Bartol 
model~\cite{bartol}, which were calculated during the simulation (see
Section~\ref{sec:genhen}). Beside the conventional neutrino 
flux, prompt neutrinos were taken into account as well, according to the studies of~\cite{naumov}
and using the recombination quark-parton model (RQPM)~\cite{RQPM}.

\subsection{The $\nu_e$ CC Events}\label{sec:nue_sample}

\begin{figure}[h]\centering
\includegraphics[width=10cm]{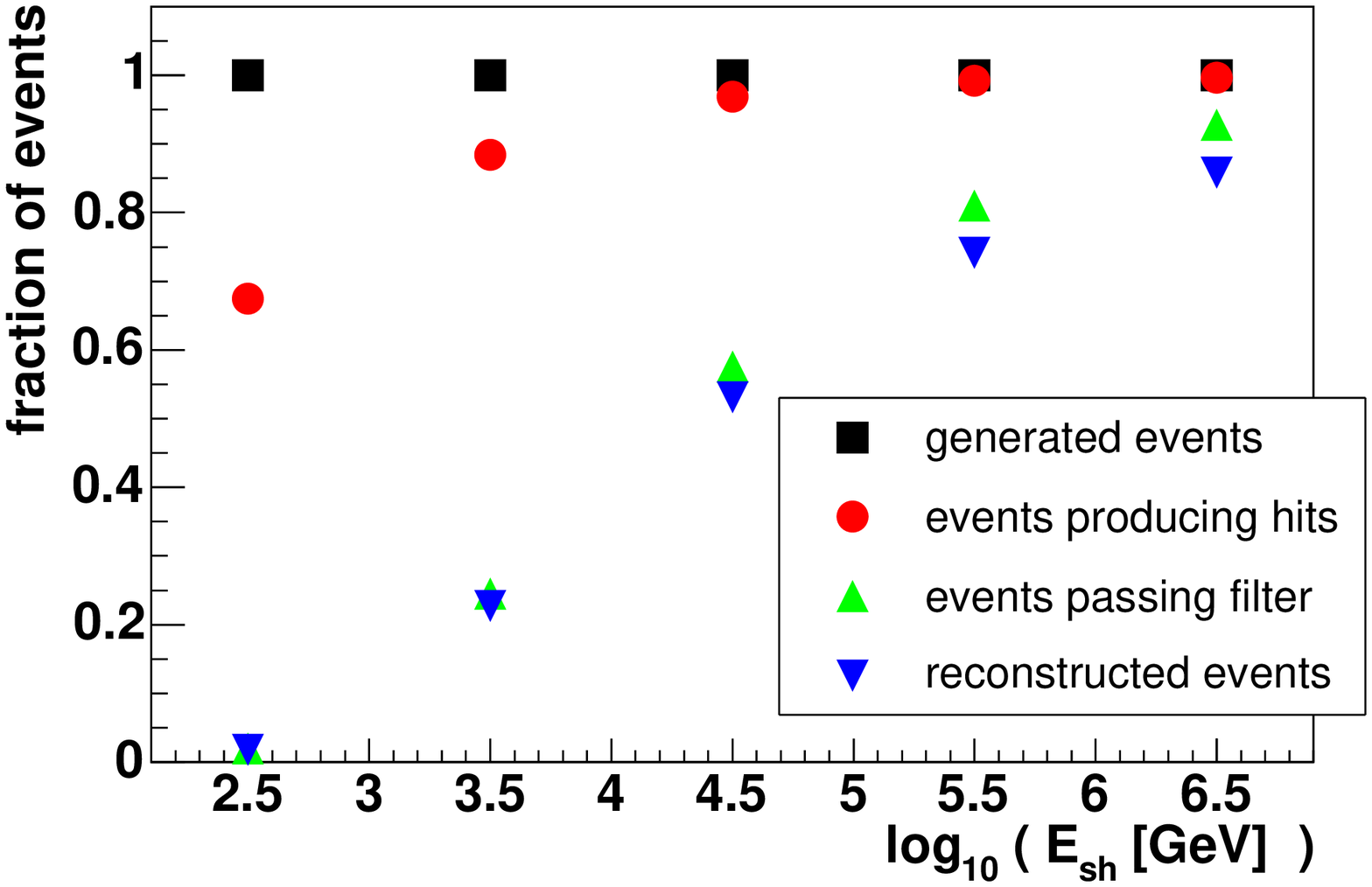}
\includegraphics[width=10cm]{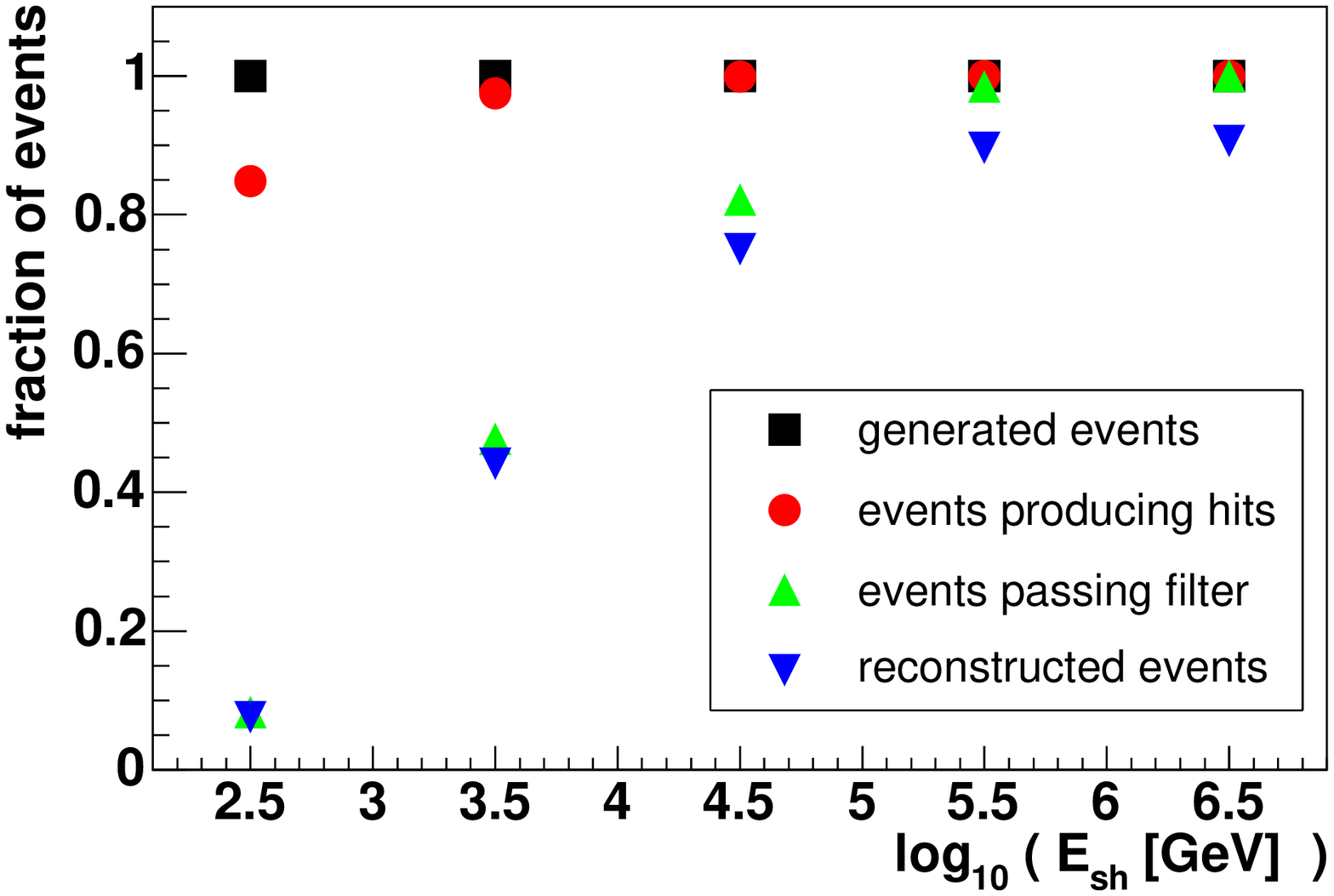}
\caption[Number of events after different production stages]{Number of NC events from sample B
  (top) and number of $\nu_e$ CC events (bottom) after the different stages of production as listed
  in tables~\ref{tab:nc_sample} and \ref{tab:cc_sample}, over the shower energy. The event numbers
  have been normalised to the respective numbers of generated events.} 
\label{fig:event_stages_nue}
\end{figure}

The $\nu_e$ CC events which were used for this study were produced within the same mass production
and with the same simulation parameters 
as event sample B of the NC events (see above). The energy of the primary neutrino lies between
$10^2$ and $10^7$\,GeV. The energy spectrum used for the production is $E^{-1}$, and the events were
produced inside a generation volume that exceeds the instrumented volume by one absorption length
($\sim 55$\,m). The primary neutrino was distributed isotropically in the whole $4\pi$ solid angle. A
optical background of 60\,kHz was added. As for event sample B, event weights for the atmospheric
neutrino flux according to the Bartol model~\cite{bartol} and including prompt
neutrinos~\cite{naumov} were added as information to each event. \\
The event sample was chosen from the mass production to be originally of about the same size as
event sample B of the NC events; as all energy of the primary neutrino goes into showers for this
event type, the showers in these events are actually more energetic than the those in the NC events 
from the same mass production, and thus, a larger number of events than for the NC events remains in 
the sample after the different production stages. The numbers of events for the different energy 
bins and the different 
stages of production are given in Table~\ref{tab:cc_sample}. Figure~\ref{fig:event_stages_nue}
(bottom) shows these numbers in relation to the number of originally generated events over the shower
energy of the respective bin. The percentage of events that survive all production stages rises with
increasing energy; however, the fraction of actually reconstructed events is lower in the highest
energy bin than than in the second-highest one, an evidence for the fact that for some of the events,
saturation effects in the OMs keep the reconstruction from converging.

\begin{table}[h] \centering 
\begin{tabular}{|>{\centering}p{3.2cm}|>{\raggedleft}p{2.2cm}|>{\raggedleft}p{2.6cm}|>{\raggedleft}p{2.4cm}|>{\raggedleft}p{2.4cm}|}\hline
neutrino energy [GeV] & generated (Section~\ref{sec:genhen}) & producing hits (Section~\ref{sec:geasim}) & filtered (Section~\ref{sec:trigger_soft}) & reconstructed (Section~\ref{sec:reco}) \tabularnewline \hline
$10^2 - 10^3$ & 100270 &  85068 (84.8\%) &  8326 (8.30\%) &  7610 (7.59\%)\tabularnewline
$10^3 - 10^4$ &  20506 &  20000 (97.5\%) &  9774 (47.7\%) &  9056 (44.2\%)\tabularnewline
$10^4 - 10^5$ &  10007 &  10000 (99.9\%) &  8224 (82.2\%) &  7514 (75.1\%)\tabularnewline
$10^5 - 10^6$ &   5000 &   5000 (100\%)  &  4926 (98.5\%) &  4483 (89.7\%)\tabularnewline
$10^6 - 10^7$ &   2757 &   2757 (100\%)  &  2757 (100\%)  &  2499 (90.6\%)\tabularnewline \hline
total events  & 138540 & 122825 (88.7\%) & 34007 (24.5\%) & 31162 (22.5\%)\tabularnewline
\hline 
\end{tabular}
\caption[Numbers of events in the $\nu_e$ CC sample]{Numbers of events in the $\nu_e$ CC event
  sample, after generation, hit production with geasim, the filter conditions, and the
  reconstruction. The percentages given refer to the original number of events (first column).}  
\label{tab:cc_sample}
\end{table}

\subsection{The Atmospheric Muon Background}\label{sec:atm_muon_sample}

For the studies on the suppression of the atmospheric muon background that are discussed in
Section~\ref{sec:atm_mu}, a sample of atmospheric muons produced in 2005 at Bologna for the
ANTARES collaboration was used. The whole data sample is described in more detail
in~\cite{corsika_muons}. It contains events with protons, helium, nitrogen, magnesium or
iron as primaries interacting in the atmosphere, produced within different energy and zenith angle 
ranges. For the results shown here, only the proton events were used. \\
For the higher energies, due to the high amount of CPU usage, only about 10\% of the whole production were used. \\
Table~\ref{tab:atm_muon_sample} shows the numbers of events in the different energy and angular bins
for the used data sample.

\begin{table}[h] \centering 
\begin{tabular}{|c|c|c|c|c|c|c|}
\hline
zenith angle & \multicolumn{3}{c|}{$0^{\circ} - 60^{\circ}$} & \multicolumn{3}{c|}{$60^{\circ} - 85^{\circ}$} \\
\hline
energy [TeV/nucleon] & 1-10 & 10-100 & 100-10$^5$ & 1-10 & 10-100 & 100-10$^5$ \\ \hline
primary protons    & $10^8$ & $10^7$ & $10^7$ & $10^8$ & $10^7$ & $10^7$ \\ \hline
approx.~equiv.~lifetime    & $\sim 1$\,h  & $\sim 2.6$\,h  & $\sim 6$\,days & $\sim 1$\,h & $\sim
2.6$\,h & $\sim 6$\,days \\ \hline
\end{tabular}
\caption[Atmospheric muon sample]{Numbers of events in the different energy and angular bins of the
  atmospheric muon event sample. }
\label{tab:atm_muon_sample}
\end{table}

The primary spectrum used was $E^{-2}$; the air showers from the primaries were produced using the
CORSIKA package~\cite{corsika}, version 6.203. After the production of secondaries in CORSIKA, the
secondary muons were propagated to the detector with MUSIC~\cite{music}, and the detector response
was simulated using km3~\cite{km3}. A optical background of 60\,kHz was also added to the data
sample. 

\chapter{Technical Details of the ShowerFitter Package}\label{sec:tech_details}

\section{Parameters Used for the Reconstruction}
 
Table~\ref{tab:reco_params} lists the fixed parameters that were used in the reconstruction, and their
values.

\begin{table}[h] 
\begin{tabular}{|l|l|l|}
\hline
Parameter          & Value     & Source/comment \\ \hline\hline
refraction index $n$ & 1.3499    & \cite{antcc} \\
photon wavelength $\lambda$ & 475\,nm    & minimal absorption \\
absorption length $\tau$ & 54.9451\,m & for $\lambda = 475$\,nm,~\cite{antcc}  \\
PM saturation      & 200\,pe / 20\,pe & WF mode / SPE mode  \\ \hline
minimum pe threshold & 0.3\,pe & \\
{\it filter parameters (see Section~\ref{sec:trigger}):} &  &  \\
causality time window $\delta t$ & 100\,ns & condition 1 \\
minimum amplitude $A_{min}$      & 3\,pe   & condition 2 \\
coincidence time window & 20\,ns  & condition 2 \\
number of coincidences on one string & 2 & condition 3 \\
minimum amplitude $A_{min,2}$ & 1.5\,pe & condition 3 \\
minimum number of hit strings &  3 & filter condition \\
\hline
\end{tabular}
\caption[Parameters in the reconstruction]{List of the values of environmental and data acquisition 
  parameters used in the ShowerFitter.}
\label{tab:reco_params}
\end{table}

\section{Schematic Description of the Fit}

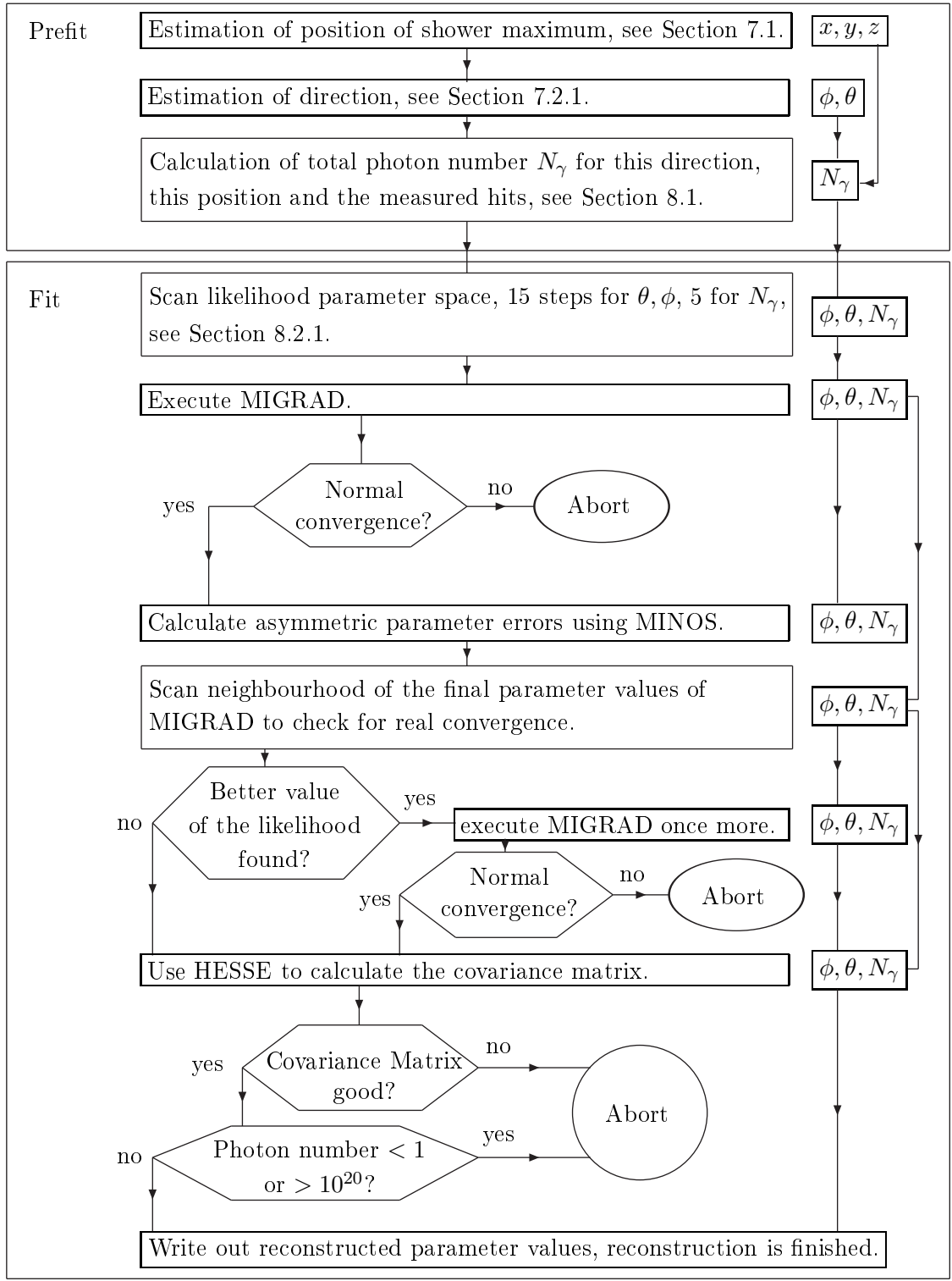
\begin{figure}[h] \centering 
\begin{picture}(0,570)(250,0)

\Line(40,560)(460,560)
\Line(40,560)(40,450)
\Line(460,560)(460,450)
\Line(40,450)(460,450)

\Text(50,550)[l]{Prefit}

\Text(100,550)[l]{\fbox{Estimation of position of shower maximum, see Section~\ref{sec:pos}.}}
\ArrowLine(245,540)(245,525)
\Text(400,550)[l]{\fbox{$x,y,z$}}
\Line(428,542)(428,480)
\ArrowLine(428,480)(420,480)
\Text(100,520)[l]{\framebox[290pt][l]{Estimation of direction, see Section~\ref{sec:prefit_dir1}.}}
\ArrowLine(245,510)(245,500)
\Text(400,520)[l]{\fbox{$\phi, \theta$}}
\ArrowLine(410,510)(410,491)

\Line(100,500)(390,500)
\Line(100,463)(390,463)
\Line(100,500)(100,463)
\Line(390,463)(390,500)
\Text(104,490)[l]{Calculation of total photon number $N_{\gamma}$ for this direction, }
\Text(104,475)[l]{this position and the measured hits, see Section~\ref{sec:pattern_matching}.}
\ArrowLine(245,463)(245,440)
\Text(400,483)[l]{\fbox{$N_{\gamma}$}}
\ArrowLine(410,472)(410,429)

\Line(40,445)(460,445)
\Line(40,-8)(460,-8)
\Line(460,445)(460,-8)
\Line(40,445)(40,-8)

\Text(50,430)[l]{Fit}

\Line(100,440)(390,440)
\Line(100,403)(390,403)
\Line(100,440)(100,403)
\Line(390,403)(390,440)
\Text(104,430)[l]{Scan likelihood parameter space, 15 steps for $\theta,\phi$, 5 for $N_{\gamma}$,}
\Text(104,415)[l]{see Section~\ref{sec:scan}.}
\ArrowLine(245,403)(245,390)
\Text(400,422)[l]{\fbox{$\phi, \theta, N_{\gamma}$}}
\ArrowLine(410,412)(410,392)

\Text(100,385)[l]{\framebox[290pt][l]{Execute MIGRAD.}}
\ArrowLine(198,377)(198,355)
\Text(400,385)[l]{\fbox{$\phi, \theta, N_{\gamma}$}}
\ArrowLine(410,375)(410,293)
\Line(441,385)(446,385)
\ArrowLine(446,385)(446,250)
\Line(441,250)(446,250)

\Line(175,355)(220,355)
\Line(150,336)(175,355)
\Line(150,336)(175,318)
\Line(245,336)(220,355)
\Line(245,336)(220,318)
\Line(175,318)(220,318)
\Text(200,345)[c]{Normal}  
\Text(200,330)[c]{convergence?}
\Line(150,336)(130,336)
\ArrowLine(130,336)(130,292)
\Text(110,336)[l]{yes}
\ArrowLine(245,336)(275,336)
\Text(255,345)[l]{no}
\Oval(305,336)(16,30)(0)
\Text(305,338)[c]{Abort}

\Text(100,285)[l]{\framebox[290pt][l]{Calculate asymmetric parameter errors using MINOS.}}
\Text(400,285)[l]{\fbox{$\phi, \theta, N_{\gamma}$}}
\ArrowLine(245,276)(245,265)

\Line(100,265)(390,265)
\Line(100,228)(390,228)
\Line(100,265)(100,228)
\Line(390,228)(390,265)
\Text(104,255)[l]{Scan neighbourhood of the final parameter values of}
\Text(104,240)[l]{MIGRAD to check for real convergence.}
\ArrowLine(155,228)(155,220)
\Text(400,248)[l]{\fbox{$\phi, \theta, N_{\gamma}$}}
\Line(441,245)(446,245)
\ArrowLine(446,245)(446,130)
\Line(441,130)(446,130)
\ArrowLine(410,238)(410,203)

\Line(130,220)(190,220)
\Line(105,195)(130,220)
\Line(105,195)(130,170)
\Line(215,195)(190,220)
\Line(215,195)(190,170)
\Line(130,170)(190,170)
\Text(160,210)[c]{Better value}
\Text(160,195)[c]{of the likelihood}
\Text(160,180)[c]{found?}
\ArrowLine(105,195)(105,136)
\Text(95,195)[c]{no}
\ArrowLine(215,195)(240,195)
\Text(225,205)[c]{yes}

\Text(240,195)[l]{\framebox[150pt][l]{execute MIGRAD once more.}}
\ArrowLine(262,188)(262,182)
\Text(400,195)[l]{\fbox{$\phi, \theta, N_{\gamma}$}}
\ArrowLine(410,186)(410,138)

\Line(240,182)(285,182)
\Line(215,163)(240,182)
\Line(215,163)(240,144)
\Line(310,163)(285,182)
\Line(310,163)(285,144)
\Line(240,144)(285,144)
\Text(265,172)[c]{Normal} 
\Text(265,157)[c]{convergence?} 
\ArrowLine(215,163)(215,136)
\Text(205,160)[c]{yes}
\ArrowLine(310,163)(335,163)
\Text(320,172)[c]{no}
\Oval(365,163)(16,30)(0)
\Text(365,163)[c]{Abort}

\Text(100,130)[l]{\framebox[290pt][l]{Use HESSE to calculate the covariance matrix.}}
\ArrowLine(197,123)(197,105)
\Text(400,130)[l]{\fbox{$\phi, \theta, N_{\gamma}$}}
\ArrowLine(410,121)(410,13)

\Line(170,105)(225,105)
\Line(145,86)(170,105)
\Line(145,86)(170,67)
\Line(250,86)(225,105)
\Line(250,86)(225,67)
\Line(170,67)(225,67)
\Text(200,90)[c]{Covariance Matrix}
\Text(200,75)[c]{good?}
\ArrowLine(145,86)(145,60)
\Text(130,86)[c]{yes}
\ArrowLine(250,86)(300,86)
\Text(260,95)[c]{no}

\Line(140,60)(215,60)
\Line(105,46)(140,60)
\Line(105,46)(140,27)
\Line(250,46)(215,60)
\Line(250,46)(215,27)
\Line(140,27)(215,27)
\Text(180,50)[c]{Photon number $< 1$}
\Text(180,35)[c]{or $> 10^{20}$?}
\ArrowLine(105,46)(105,13)
\Text(95,46)[c]{no}
\ArrowLine(250,46)(300,46)
\Text(260,55)[c]{yes}
\Oval(322,66)(30,30)(0)
\Text(322,66)[c]{Abort}

\Text(100,5)[l]{\fbox{Write out reconstructed parameter values, reconstruction is finished.}}

\end{picture}
\caption[Reconstruction Scheme]{Reconstruction Scheme of the ShowerFitter. The parameters that are
  determined in the respective reconstruction step are listed in the boxes on the right of the figure.}
\label{fig:reco_scheme}
\end{figure}

A schematic process chart of the whole ShowerFitter strategy, its parameters and the references to
the descriptions of the respective steps in the chapters of this work, is provided in 
Figure~\ref{fig:reco_scheme}.

\newpage

\chapter{Error Calculations}

\section{Calculation of the Errors on Efficiency and Purity}\label{sec:stat_errors}

Efficiency and purity are both functions of the form $f(n,n_1) = \frac{n_1}{n}$. As $n_1$ is a
subsample of $n$, and therefore not independent from it, the error propagation rules cannot be 
used directly. Instead, one defines $n \equiv n_1 + n_2$. 
The function $f$ becomes then $f(n_1,n_2) = \frac{n_1}{n_1 + n_2}$ with the two independent variables
$n_1$ and $n_2$. The error on $f, \, \sigma_{\f}$, can then be calculated following the Gaussian
error propagation rules: 

\begin{alignat}{2}
\sigma_{\f} &= \sqrt{\left(\frac{df}{dn_1}\right)^2 \sigma_{n_1}^2 
                + \left(\frac{df}{dn_2}\right)^2 \sigma_{n_2}^2}\label{eq:error} \\
         &= \sqrt{\left(\frac{n_2}{(n_1+n_2)^2}\right)^2 \cdot n_1 
                + \left(\frac{-n_1}{(n_1+n_2)^2}\right)^2 \cdot n_2}
         \notag \\
         &= \sqrt{\frac{(n_2)^2 n_1}{(n_1+n_2)^4} + \frac{(n_1)^2 n_2}{(n_1+n_2)^4}} \notag \\
         &= \sqrt{\frac{n_1 n_2 (n_2 + n_1)}{n^4}}  \notag \\
         &= \frac{\sqrt{n_1 (n-n_1) n}}{n^2}.
\end{alignat}

If the subsample $n_1$ is equal to $n$ (and therefore, $n_2 = 0$), one can no longer assume that the
statistical error of the 
variable $n_2$ is $\sqrt{n_2}$ (this would always result in $\sigma_{\f} = 0$, no matter how large or
small $n$ is). Instead, one has to take into account Poissonian statistics:
For a $1 \sigma$ error, (100 - 68 = 32)\% of the events lie outside the $1
\sigma$ area. For the case of $n_1 = n$, only the error in one direction has to be considered; the
other one is really zero. Therefore, one has to define the Poisson distribution such that 32\% / 2 =
16\% of the events lie outside of the curve. 84\% should thus lie inside. The normalised Poisson
distribution thus should solve to

\begin{gather}
\int_0^t P(x,0) = \int_0^t \frac{x^0 e^{-x}}{0!} = \int_0^t e^{-x} = -e^{-t} + e^0 = 1-e^{-t} = 0.84
\\
\leadsto t = - \ln 0.16 \approx 1.83 = \sigma_{n_2}.
\end{gather}

Equation (\ref{eq:error}) therefore becomes

\begin{gather}
\sigma_{\f} = \sqrt{\left(\frac{-n}{n^2}\right)^2 \sigma_{n_2}^2} = \frac{\sigma_{n_2}}{n} = 1.83 / n. 
\end{gather}

\section{Standard Error on the Median}\label{sec:err_median}

For the calculation of the standard error on the median, the derivation from the book of
Kendall and Stuart~\cite[pp.236-237]{kendall} was used: \\
A frequency distribution can be described by {\it quantiles}, i.e.~the $(m-1)$ values which divide a
frequency distribution into $m$ equal parts (i.e.~with an equal number of entries). The most common
quantile is the {\it median}, which divides a frequency distribution into 2 equal parts. \\
If a sample of $n$ values distributed according to a frequency distribution $f(x)$ is considered,
with a fraction of $pn$ 
values below and a fraction of $qn$ values above the quantile $x_1$, it can be shown that the {\it
  variance} $V(x_1)$ is given by 

\begin{equation}
V(x_1) = \frac{pq}{nf(x_1)^2}.
\end{equation}

In the case of the median, $p = q = \frac{1}{2}$, and therefore $V(x_m) =  \frac{1}{4nf(x_m)^2}$,
where $x_m$ is the value of the median. Assuming that the distribution is normal with a variance
$\sigma^2$, the standard error on the median $\sigma_m$ becomes

\begin{equation}
\sigma_m = \sigma \cdot \sqrt{V(x_m)} = \sigma \cdot \frac{1}{2\sqrt{n}f(x_m)}.
\end{equation}

As $x_m = 0$ for the normal distribution, ${f(x_m) = \frac{1}{\sqrt{2 \pi}} \exp\{-\frac{1}{2}x_m^2\} =
\frac{1}{\sqrt{2 \pi}} \approx 0.399}$. The standard error on the median becomes therefore

\begin{equation}
\sigma_m = 1.253 \cdot \frac{\sigma}{\sqrt{n}}.
\end{equation}
 
\end{appendix}

\chapter*{Danksagung}
\selectlanguage{ngerman}

Eine Vielzahl von Menschen hat zum Gelingen dieser Arbeit beigetragen, und ich bin allen
zu gro\ss em Dank verpflichtet. \\
Zuallererst danke ich meinem Betreuer, Prof.~Dr.~U.~Katz, daf\"ur, dass er es mir erm\"oglichte, meine
Arbeit zu einem so interessanten Thema zu schreiben, und mir w\"ahrend meiner Zeit als Doktorandin
unerm\"udlich mit Rat und Tat zur Seite stand. \\
Prof.~Dr.~E.~Steffens hat sich freundlicherweise bereit erkl\"art, das Zweigutachten zu meiner
Arbeit zu schreiben, wof\"ur ihm herzlicher Dank geb\"uhrt. \\
I would like to thank the ANTARES collaboration for the friendly and fruitful atmosphere that I 
have experienced while being a member of the collaboration. I especially thank J\"urgen Brunner for
his prompt answers to my emails and his help in all technical and software questions. \\
Ich danke allen Kollegen der Erlanger ANTARES-Gruppe f\"ur die angehme Arbeitsatmosph\"are und die
gute Zusammenarbeit, und auch den Nicht-ANTARESianern von Abteilung 1 und 
4 f\"ur die gute Gesamtstimmung im Institut, ob beim gemeinsamen Gang zur Mensa (die dort
gereichte Kost w\"are ohne Gesellschaft jedenfalls nicht auf Dauer auszuhalten!), bei der
Kaffeerunde oder auf diversen Festivit\"aten. \\
Besonderer Dank geb\"uhrt Alexander Kappes, der nicht nur die Arbeit gr\"undlich Korrektur
gelesen hat, sondern sich auch immer die Zeit genommen hat, physikalische Fragen zu diskutieren
und technische Probleme zu l\"osen, sowie 
Felix Fehr und Kay Graf f\"ur das sorgf\"altige Korrekturlesen meiner Arbeit. Bei
Kay Graf bedanke ich mich au\ss erdem f\"ur den seelischen und moralischen Beistand insbesondere
w\"ahrend der arbeitsreichen Zeit des Zusammenschreibens \Heart.

\newpage
\thispagestyle{empty}
\hspace{3cm}
\newpage

\thispagestyle{empty}
\begin{center} \begin{large}LEBENSLAUF 
\end{large} \end{center}

\noindent\underline{\bf{Pers\"onliches}}\\[0.3cm]
\begin{tabular}{p{5cm}p{9cm}}
  Name: & Bettina Diane Hartmann \\
  Adresse:        & Gerhart-Hauptmann-Stra\ss e 11 \\
                  & 91058 Erlangen  \\ 
  Telefon:        & 09131 -- 16 06 79  \\
  E-Mail:         & hartmann@physik.uni-erlangen.de \\
  Geburtsdatum:   & 14.~Februar 1977 \\
  Geburtsort:     & Stuttgart \\
\end{tabular}

\vspace{0.5cm}

\noindent\underline{\bf{beruflicher Werdegang}}  \\[0.3cm]
\begin{tabular}{p{5cm}p{9cm}}
  \emph{Bildung} & \\[0.2cm]
  1987 -- 1996 & H\"olderlingymnasium Stuttgart \\
  August -- Oktober 1993 & Bexley Highschool, Columbus, Ohio, USA \\
  Oktober 1996 -- Juni 2002 & Universit\"at Karlsruhe (TH): Physik (Diplom) \\
  September 2002 -- M\"arz 2006 & Universit\"at Erlangen--N\"urnberg: Promotion 
  \\[0.5cm]
  \emph{Pr\"ufungen} & \\[0.2cm]
  Juni 1996 & Abitur, Note: 1,3 \\
  Oktober 1998 & Vordiplom, Note: 2,5 \\
  Juni 2002 & Diplom, Note: 1,4 \\
  Juni 2006 & Promotionspr\"ufung 
  \\[0.5cm]
  Diplomarbeit: & \\
  Mai 2001 -- Juni 2002 & am Institut f\"ur Experimentelle Kernphysik Karlsruhe \\ 
                       & bei Prof.~Dr.~W.~de Boer \\  
                       & Thema: \emph{Suche nach einem leichten Higgsboson im} \\ 
                       & \emph{$WH$-Kanal bei CDF mit neuronalen Netzen} \\
  Doktorarbeit: & \\
  seit September 2002 & am Physikalischen Institut I der Universit\"at \\
                      & Erlangen-N\"urnberg bei Prof.~Dr.~U.~Katz
  \\[0.5cm]
  \emph{Praktika} & \\[0.2cm]
  September \& Oktober 1999 & Praktikum am Institut f\"ur angewandte Physik an der \\ 
                            & Technischen Universit\"at Kaunas, Litauen \\ 
  Juli \& August 2000 & Sommerstudentin am Forschungszentrum CERN \\
                      & (Genf, Schweiz)\\
  Oktober 2000 & Wissenschaftliche Ausstellungsbetreuerin am \\
               & Deutschen Elektronen-Synchrotron DESY, Hamburg \\

\end{tabular}

\newpage
\thispagestyle{empty}
\hspace{3cm}
\newpage

\end {document}